\documentclass[10pt,a4paper]{report}
\usepackage[utf8]{inputenc}
\usepackage[english]{babel}
\usepackage{amsmath}
\usepackage{amsfonts}
\usepackage{amssymb}
\usepackage{graphicx}
\usepackage{color}
\usepackage{rotating}
\usepackage{float}

\usepackage{appendix}
\usepackage{geometry}
\everymath{\displaystyle}
\usepackage{cite}

\usepackage{slashed}

\usepackage{axodraw}

\usepackage{breqn}
\usepackage{booktabs}

\usepackage{makeidx}
\usepackage{lmodern}

\newcommand\ea{{\em et al.}}

\usepackage{pgf}
\usepackage{multirow}

\def\CKM {\texttt{CKMfitter}}
\def\NP {{Neyman-Pearson}}

\def\noi {\noindent}

\def\tHF {{'t Hooft-Feynman }}

\usepackage[absolute]{textpos} 
\usepackage{multicol} 
\usepackage{calc} 
\usepackage{tikz} 

\label{The thesis}
\newcommand{\PhDTitleFR}{Ph\'{e}nom\'{e}nologie de Mod\`{e}les \`{a} Sym\'{e}trie Droite-Gauche \\ dans le secteur des quarks} 

 \label{Personnal}
\newcommand{\PhDname}{M. Luiz Henrique \textsc{Vale Silva}} 
\newcommand{\NNT}{2016SACLS249} 
\newcommand{\ecodocnum}{576} 
\newcommand{\ecodoctitle}{Particules, hadrons, \'{e}nergie, noyau, \\ instrumentation, image, cosmos et simulation (PHENIICS)} 
\newcommand{\PhDspeciality}{Physique des particules} 
\newcommand{\PhDworkingplace}{\`{a} l'Universit\'{e} Paris-Sud} 
\newcommand{\defenseplace}{Orsay} 
\newcommand{\defensedate}{20 Septembre 2016} 
\newcommand{\logoEt}{\includegraphics[scale=1.2]{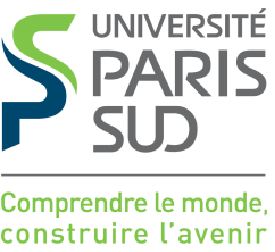}} 
\newcommand{\vpos}{1} 

\label{Jury}
\newcommand{\jurynameA}{Hagop Sazdjian}
\newcommand{\jurygenderA}{M.} 
\newcommand{\juryadressA}{\it IPN Orsay}
\newcommand{\jurygradeA}{Professeur \'{e}m\'{e}rite}
\newcommand{\juryroleA}{Pr\'{e}sident du jury} %
\newcommand{\jurynameB}{Ulrich Nierste}
\newcommand{\jurygenderB}{M.} 
\newcommand{\juryadressB}{\it TTP Karlsruhe}
\newcommand{\jurygradeB}{Prof. Dr.}
\newcommand{\juryroleB}{Rapporteur}
\newcommand{\jurynameC}{Renata Zukanovich Funchal}
\newcommand{\jurygenderC}{Mme} 
\newcommand{\juryadressC}{\it IF S\~ao Paulo}
\newcommand{\jurygradeC}{Prof.}
\newcommand{\juryroleC}{Rapporteur}
\newcommand{\jurynameD}{J\'{e}r\^{o}me Charles}
\newcommand{\jurygenderD}{M.} 
\newcommand{\juryadressD}{\it CPT Marseille}
\newcommand{\jurygradeD}{Charg\'{e} de Recherche}
\newcommand{\juryroleD}{Examinateur}
\newcommand{\jurynameE}{V\'{e}ronique Bernard}
\newcommand{\jurygenderE}{Mme} 
\newcommand{\juryadressE}{\it IPN Orsay}
\newcommand{\jurygradeE}{Directeur de Recherche}
\newcommand{\juryroleE}{Directrice de th\`{e}se}
\newcommand{\jurynameF}{S\'{e}bastien Descotes-Genon}
\newcommand{\jurygenderF}{M.} 
\newcommand{\juryadressF}{\it LPT Orsay}
\newcommand{\jurygradeF}{Directeur de Recherche}
\newcommand{\juryroleF}{Directeur de th\`{e}se}
\label{Document}

\begin{document}
\renewcommand{\arraystretch}{1.3}


\begingroup
\pagenumbering{gobble}
\fontsize{12pt}{14pt}\selectfont
\newgeometry{textheight=150ex,textwidth=40em,top=30pt,headheight=30pt,headsep=30pt,inner=80pt}
\label{cotutelle}

\begin{tikzpicture}[remember picture,overlay,color=blue!20!red!45!black!75!]
		([yshift=-160pt,xshift=55pt]current page.north west)--     
		([yshift=-160pt,xshift=-15pt]current page.north east)--    
		([yshift=35pt,xshift=-15pt]current page.south east)--      
		([yshift=35pt,xshift=55pt]current page.south west)--cycle; 
\end{tikzpicture}

\begin{textblock}{13}(1.35,3.3)
  NNT : \NNT
\end{textblock}

\begin{textblock}{1}(1.15,1)
\includegraphics[height=2.4cm]{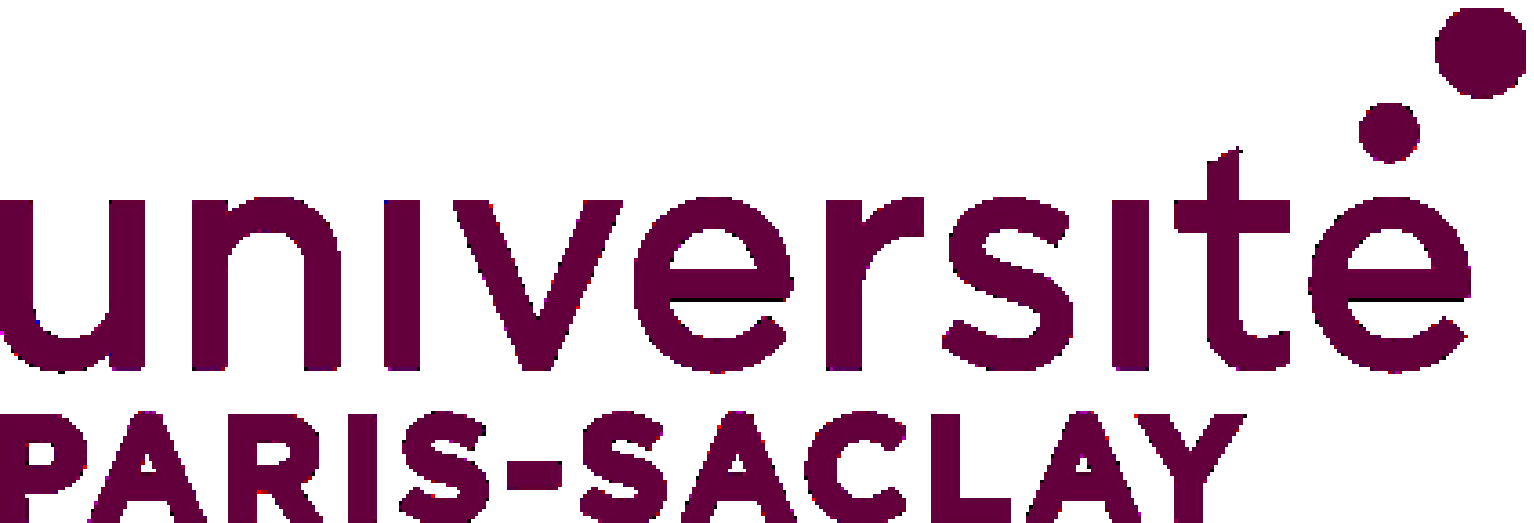} 
\label{Logo Paris Saclay}
\end{textblock}

\begin{textblock}{1}(12,\vpos)
\logoEt 
\label{Logo Etablissement}
\end{textblock}

\vspace{6cm}
\color{blue!20!red!45!black} 
  \begin{center}    
    \LARGE\textsc{Th\`{e}se de doctorat\\ de l'Universit\'{e} Paris-Saclay} \\
    \LARGE{\textsc{pr\'{e}par\'{e}e \PhDworkingplace}} \\ \bigskip
  \color{black} 
	\vfill
    \Large{Ecole doctorale n$^{\circ}\ecodocnum$}\\ 
     \Large{\ecodoctitle}  \\

     \Large{Sp\'{e}cialit\'{e} de doctorat: \PhDspeciality} 
    \vfill  
   \Large{par}
   \vfill
   \LARGE{\textbf{\PhDname}} 
    \vfill
    \Large{\PhDTitleFR} 
    \vfill
    \bigskip
\end{center}
\color{black}
\begin{flushleft}
Th\`{e}se pr\'{e}sent\'{e}e et soutenue \`{a} \defenseplace, le \defensedate. \\
\bigskip
Composition du Jury :
\end{flushleft}

\begin{center}
\begin{tabular}{llll}

    \jurygenderA & \textsc{\jurynameA}  & \jurygradeA & (\juryroleA) \\
    \null & \null & \juryadressA &\\   
   
    \jurygenderB & \textsc{\jurynameB}  & \jurygradeB & (\juryroleB) \\
    \null & \null & \juryadressB &\\ 
    
    \jurygenderC & \textsc{\jurynameC}  & \jurygradeC & (\juryroleC) \\
    \null & \null & \juryadressC &\\ 
    
    \jurygenderD & \textsc{\jurynameD}  & \jurygradeD & (\juryroleD) \\
    \null & \null & \juryadressD &\\ 
    
    \jurygenderE & \textsc{\jurynameE}  & \jurygradeE & (\juryroleE) \\
    \null & \null & \juryadressE &\\ 
    
    \jurygenderF & \textsc{\jurynameF}  & \jurygradeF & (\juryroleF) \\
    \null & \null & \juryadressF &\\ 
   
  \end{tabular}    
\end{center}
\restoregeometry
\endgroup



\newpage\null\thispagestyle{empty}\newpage

\begin{center}
{\bf {\Large Phenomenology of Left-Right Models in the quark sector } }

\vspace{0.5cm}

\vspace{3\baselineskip}

\vspace*{0.5cm}
\textbf{Abstract}\\
\vspace{1\baselineskip}
\parbox{0.9\textwidth}{A natural avenue to extend the Standard Model (SM) is to embed it into a more symmetric framework. Here, I focus in Left-Right (LR) Models, which treat left- and right-handed chiralities on equal footing. Important information about the structure of LR Models comes from meson-mixing observables. Due to the impact of the new contributions to these processes, I consider the calculation of the short-distance QCD effects correcting the LR Model contributions to meson-mixing observables at the Next-to-Leading Order. I then revisit the phenomenology of a simple realization of LR Models, containing doublet scalars, and combine in a global fit electroweak precision observables, direct searches of the new gauge bosons and meson oscillation observables, a task performed within the \texttt{CKMfitter} statistical framework. Finally, I also cover a different issue, namely the modeling of theoretical uncertainties, a class of uncertainties specially important for flavour physics. Different frequentist schemes are compared, and their differences are illustrated with flavour-oriented examples.}
\end{center}

\thispagestyle{empty}

\clearpage
\setcounter{page}{1}


\newpage\null\thispagestyle{empty}\newpage

\newpage

\pagenumbering{arabic}

\tableofcontents

\newpage

\chapter*{Introduction}
\addcontentsline{toc}{chapter}{Introduction}



The Standard Model (SM) of particle physics is built on top of basic requirements such as Lorentz covariance and renormalizability, and offers a common framework for the description of all known microscopic interactions in terms of local, gauged symmetries. This theory was the fruit of the work of many generations of talented physicists, and is certainly one of the most impressive achievements of sciences. For example, one has tracked a long road to achieve the formulation of the theory of electroweak interactions, unifying Electromagnetism and Weak processes. Indeed, first the weak interactions were introduced as a new fundamental interaction in the 30's by Fermi \cite{Fermi:1934hr}, formulated at that time as a contact interaction. Later, exactly 60 years ago, parity violation in weak decays was suggested \cite{Lee:1956qn}, triggering doubts about charge-conjugation and time reversion symmetries \cite{Lee:1957qq}. The observation of parity violation \cite{Wu:1957my,Garwin:1957hc,Goldhaber:1958nb} in the following year confirmed such a hypothesis and was of utmost importance for the understanding of the weak interactions (see \cite{HistoryParity,Nishijima} for historical details). Following the discovery of parity violation, they were conceived as a $ V - A = \gamma^\mu - \gamma^\mu \gamma_5 $ (i.e. a coupling to left-handed fields) interaction \cite{Feynman:1958ty,Sudarshan:1958vf}, suggesting that the exchange of vector bosons was the underlying reason for the weak force.


The short-distance character of the weak interactions, related to the exchange of heavy gauge bosons, was elegantly interpreted as the low-energy limit of a more fundamental and symmetric theory, the Electroweak interaction of Glashow-Salam-Weinberg. The Electroweak symmetry is spontaneously broken by the vacuum expectation value of a scalar field $ \phi $, which also introduces the Higgs boson of the SM. At the same time that this mechanism, named Brout-Englert-Higgs (BEH) \cite{BEH}, explains the short-distance nature of weak interactions by giving masses to the $ W^\pm $ and $ Z^0 $ bosons, the particles responsible for the weak forces, it also offers an origin for the masses of the quarks and charged leptons, through their interaction with the same scalar field $ \phi $.

Together with the strong interactions, this overall picture has been verified in an accurate way by measurements coming from different sectors, two important examples being EW Precision Observables (EWPO) \cite{ALEPH:2005ab,Baak:2014ora,Arbuzov:2005ma} and Flavour Observables \cite{CKMfitterStandard,UTfit,Scan}, which test very different aspects of the theory, including $ Z $ boson couplings and the $ \mathcal{C P} $ violation encoded in the CKM matrix. More recently, the historical discovery of the Higgs boson \cite{Aad:2012tfa,Chatrchyan:2012xdj} has been made, the only remaining block of the SM not directly observed until then.

It is interesting to note that the SM gives hints towards the possibility of having something more fundamental beyond itself. Indeed, the hierarchical structure of the CKM matrix, together with the strong hierarchy of the masses of the quarks and leptons, claim for a deeper understanding and questioning. Moreover, the values of the gauge couplings $ g_s, g_L, g_Y $ are roughly similar: following a successful tradition of unifying interactions (gravity effects on the ground and celestial movement, electric and magnetic forces, electromagnetic and weak interactions, etc.), one may be tempted to do the same for the known quantum fundamental interactions.

On top of that, though very successful in explaining a wide variety of particle physics phenomena, the SM leaves unexplained some properties of nature. Here we will focus on the different behaviours of left- and right-handed chiralities of the known fermions, or in other words the violation of parity symmetry. A possible and somewhat natural avenue to explain this feature is to embed the SM into a more symmetric model, which treats the chiralities on equal footing. Looking for new symmetries (e.g. supersymmetry, gauge unification, etc.) or to reasons why we do not see them (e.g. explaining lepton flavour violation in neutrino oscillations, or why the Yukawa matrices in the SM break the symmetries between different generations in the way they do, etc.) can improve crucially our understanding of fundamental phenomena, and indeed it has proven to be the case over the past, the SM itself being an example.

The class of models restoring parity, the Left-Right (LR) Models, has been first conceived in the seventies \cite{Pati:1974yy,Mohapatra:1974hk,Senjanovic:1975rk,Senjanovic:1978ev}, and since then it is at the origin of fruitful investigations. This is certainly due to the flexibility it has concerning its specific realization, a property exploited for addressing a wide variety of phenomenological problems, including the smallness of neutrino masses \cite{Mohapatra:1980yp} and strong $ \mathcal{C P} $ violation \cite{Babu:1989rb,Barr:1991qx}. At the same time, the LR Model may result from Grand Unified gauge groups \cite{GUT}, as part of their spontaneous breaking pattern. From these perspectives then, investigating the violation of parity symmetry may be a window for dealing with other questions in particle physics.




The first point concerning their formulation is that the LR Models introduce a new weak interaction which couples preferentially to right-handed fields, analogously to the situation found in the SM for left-handed currents. This is encoded in the gauge group

\begin{equation}
SU(3)_c \times SU(2)_L \times SU(2)_R \times U(1)_{B-L} \, , \nonumber
\end{equation}
where $ B-L $ states for baryonic minus leptonic number. At an energy scale beyond the EW symmetry breaking, LR symmetry is spontaneously broken giving origin to the SM framework and to parity violating phenomena. Following the spontaneous breaking of the LR gauge group, the spectrum of gauge bosons includes heavy $ {W'}^\pm $ and $ {Z'}^0 $ bosons, which are associated to a rich phenomenology: for instance, the $ {W'}^\pm $ couples to right-handed fields with  a strength in the quark sector given by a CKM-like mixing matrix, thus introducing the mixing of different generations and new sources of $ \mathcal{C P} $ violation beyond the one of the CKM matrix. Moreover, the $ {Z'}^0 $ and the $ {W'}^\pm $ mix respectively with the $ {Z}^0 $ and the $ {W}^\pm $ bosons thus changing the way in which the known weak gauge bosons couple to fermions, a situation that can be tested by EWPO. Note also that, more recently, the potential for observing the LR particles $ {W'}^\pm $ and $ {Z'}^0 $ in high-energy colliders has triggered new activities in the domain, e.g. \cite{Maiezza:2010ic}.

The specific way in which the spontaneous breaking of the LR gauge group happens depends on the scalar content of the model. It is usual to consider triplet representations since they give rise to a see-saw mechanism for the light neutrinos. We would like here to revisit a simpler realization of the LR Models containing doublets instead of triplets, which is less constrained from the point of view of the spontaneous breaking pattern of the LR symmetry. Indeed, the way the masses of the known gauge bosons $ W, Z $ are related, satisfying $ M_W \approx M_Z \cdot g_L / \sqrt{g_Y^2 + g_L^2}  $, constrains the vacuum expectation value of one of the triplet representations, left unconstrained in the case of doublet representations. This explains our choice for the title (``Phenomenology of Left-Right Models \textit{ in the quark sector}''): we focus here on the more fundamental aspects of LR Models, namely the pattern of its Spontaneous Symmetry Breaking and the properties of its minimal scalar sector, while the question of the smallness of the neutrino masses may require further additional elements to be integrated on top of the doublet scenario under investigation here.

We consider the study of this doublet scenario based on the phenomenology of the new gauge bosons and the new scalar sector. The latter includes new neutral scalars that have Flavour Changing Neutral Couplings (FCNC) at tree level, which are strongly suppressed in the SM, where they first arrive at the one-loop order. FCNCs typically provide extremely powerful constraints on New Physics models, and therefore deserve close attention. They provide new contributions to meson-mixing observables, which have been intensively studied in the context of LR Models \cite{Beall:1981ze,Frere:1991db,Barenboim:1996nd,Barenboim:1996wz,Ball:1999mb,Kiers:2002cz,Zhang:2007da,Maiezza:2010ic,Blanke:2011ry,Bertolini:2014sua}, and point towards lower bounds for the $ W' $ mass of a few TeV and the much constraining lower bound of $ \mathcal{O} (10) $~TeV or even higher for the masses of the extended scalar sector.

When computing the LR contributions to meson-mixing processes, trustful predictions require the knowledge of QCD effects. If in one hand long-distance QCD effects have been addressed by several groups \cite{Carrasco:2015pra,Hudspith:2015wev,Jang:2015sla,Bazavov:2016nty,Carrasco:2013zta} and one expects their accuracy to increase in the near future, on the other one would like to improve the accuracy in the calculation of the short-distance QCD effects. In order to achieve such a task, we have therefore considered their calculation at the NLO, and compared the methods used previously in the literature \cite{Vainshtein:1976eu,Vysotsky:1979tu,Gilman:1982ap,Herrlich:1993yv,Herrlich:1996vf,Ecker:1985vv,Bigi:1983bpa,Blanke:2011ry}.

By studying the constraints LR Models are submitted to, we aim at having a clearer picture of their structure, namely energy scales and couplings. To this effect, we perform a combined analysis of EWPO, direct searches for the new gauge bosons and meson-mixing observables. Their combination is provided by \texttt{CKMfitter}, a powerful statistical analysis framework which has proven for instance very useful in the extraction of the Wolfenstein parameters in the context of the SM \cite{CKMfitterStandard,Charles:2015gya}. 







Shifting to a different issue, the QCD effects mentioned above as well as other theoretical parameters are subjected to systematic uncertainties, in many cases the main source of uncertainty concerning their true values. The combination of different classes of observables should in principle take into account the particularity of theoretical uncertainties, which are different in nature compared to random statistical errors. In fact, their very interpretation is subject to some ambiguity, since they do not fit straightforwardly in the usual statistical framework. We will then discuss possible models of theoretical errors, an issue particularly important for flavour physics in which this class of uncertainties are usually large.









\vspace{1cm}

In Chapter~\ref{ch:SM} we are going to introduce the basic elements of the SM, and test its basic features based on two classes of observables, EWPO and flavour observables in the quark sector. While the first test aspects of the SM such as the EW Symmetry Breaking, the second includes phenomena of $ \mathcal{C P} $ violation, which in the SM come from a unique complex phase from the CKM matrix. Then in Chapter~\ref{ch:LRM} we are going to introduce the LR Model, discussing various aspects of its gauge, scalar, quark a leptonic sectors. In Chapter~\ref{ch:EWPO} we consider EWPO in the context of the LR Model as a first test of its viability, and for further constraining the LR Model structure we also consider bounds on the mass of the $ W' $ coming from its direct search. Then we move to a second class of observables, consisting of meson-mixing observables. In order to constrain and test the LR Model with accuracy we consider addressing one fundamental element necessary for phenomenological analyses which are the short-distance QCD corrections. The basic elements necessary for the computation are given in Chapter~\ref{ch:generalEFT}, while the computation in the LR Model is left for Chapter~\ref{ch:technicalEFT}. We combine the set of the previous observables, namely EWPO, bounds from direct searches for the LR spectrum and meson-mixing observables in an exploratory study in Chapter~\ref{ch:PHENO}. Then in Chapter~\ref{ch:theo} we compare different models of theoretical uncertainties for dealing with flavour observables. This last chapter corresponds to a prospective study of the \texttt{CKMfitter} Collaboration, for the further improvement of the extraction of the parameters characterizing the CKM matrix from global fits. Finally, some points are complemented in a series of Appendices.

\chapter{The Standard Model}\label{ch:SM}



At the present state of accuracy, the SM (Standard Model) theory of particle physics succeeds in the description of a wide variety of observations, such as the weak interactions, which lead to generation mixing in the quark sector. These interactions are formulated as chiral phenomena where for instance the $ W $ boson, whose exchange is responsible for the generation mixing, couples exclusively to left-handed fermions. The coupling strengths of the $ W $ boson to quarks are the elements of a unitary matrix called Cabibbo-Kobayashi-Maskawa (CKM) matrix \cite{Cabibbo:1963yz,Kobayashi:1973fv}, introduced in the diagonalization of the quark mass matrices issued from Yukawa-type interactions. These same couplings introduce the only (sizeable) source of $ \mathcal{C} \mathcal{P} $ violation in the SM. 

The goal of the present chapter is to render explicit the success of the SM in the description of flavour processes in the quark sector. We are going to compare the SM predictions with the most solid and accurate flavour observables, which will lead us to the extraction of the strengths of the $ W $ couplings.



Since quarks are confined into hadronic states and weak interactions are formulated in terms of quark states, one cannot ``decouple'' weak interactions from QCD effects (particularly those non-perturbative in $ \alpha_s $). Therefore, in order to test the weak sector of the SM, a good knowledge of the strong dynamics of the theory, ubiquitous in quark processes, is mandatory. Moreover, to match the experimental precision, accurate predictions must be made, requiring one to compute perturbative QCD effects in $ \alpha_s $ (apart from other corrections such as QED radiative effects). We will thus pay some attention to parameters originating from QCD, both from its short- (perturbative) and long-range (non-perturbative) dynamics.

It should be kept in mind that the extraction of SM parameters is meaningful only if the formulation of the SM as a whole, described briefly in the next section, is adapted to correctly model nature. After reviewing the SM and considering EWPO, we are going to briefly discuss the inputs and the statistical framework we are going to employ in order to draw a picture of the SM prediction of $ \mathcal{C P}- $violating processes. 





\section{Introduction to the SM}

In what follows, we are going to gradually introduce the necessary elements to build the SM. We do not intend to be comprehensive or self-contained, and therefore we are going to jump many steps in its formulation (they can be found in detail in e.g. \cite{AlvarezGaume:2012zz,Branco:1999fs,Ramond:1981pw}).



\subsection{Gauge symmetries}



We start with the gauge symmetries of the SM

\begin{equation}\label{eq:SM}
SU(3)_c \times SU(2)_L \times U(1)_Y \, ,
\end{equation}
where the conserved charge of the first symmetry is color, the second is weak isospin and the third is hypercharge, and each group has a characteristic coupling strength: $ g_s, g_L $ and $ g_Y $ denote respectively the gauge couplings of the three groups $ SU(3)_c $, $ SU(2)_L $ and $ U(1)_Y $.

The generators of the symmetry groups $ SU(3)_c $ and $ SU(2)_L $ ($ Y $ is an operator proportional to the identity) are Hermitian operators obeying to the commutation rules

\begin{equation}
[T^s_a, T^s_b] = i f^{a b c}_s T^s_c \quad {\rm and} \quad [T^L_a, T^L_b] = i f^{a b c}_L T^L_c \, ,
\end{equation}
respectively. Above, $ f^{a b c}_s $ and $ f^{a b c}_L $ are the \textit{structure constants} of the groups $ SU(3)_c $ and $ SU(2)_L $ (in the latter case we have $ f^{a b c}_L = \epsilon^{a b c} $, $ \epsilon $ being the antisymmetric symbol -- more generally, $ f^{a b c} $ is antisymmetric and proportional to $ {\rm tr} \{ [T_a , T_b] \, T_c \} $).

To each of these symmetries it is associated a distinct vector field $ X_\mu = A_{\mu}, W_{\mu} $ or $ B_\mu $ (all massless at this stage) satisfying the gauge transformations

\begin{equation}
X_\mu \rightarrow X'_\mu = U X_\mu U^{-1} - \frac{1}{i g} U \partial_\mu U^{-1} \, , \quad U = \exp \{ i \chi (x) \} \, ,
\end{equation}
where $ g = g_s, g_L $ or $ g_Y $, $ \chi (x) $ is called \textit{gauge function}, and $ X^\mu \equiv X^{a \mu} T_a $, $ \chi \equiv \chi^a T_a $. Under the gauge transformation, $ G^{\mu \nu}_a, F^{\mu \nu}_a $ and $ F^{\mu \nu}_Y $, defined as

\begin{eqnarray}
G^{\mu \nu}_a &=& \partial^\mu A^{a \nu} - \partial^\nu A^{a \mu} + g_s f^{a b c}_s A^{b \mu} A^{c \nu} \, , \\
F^{\mu \nu}_a &=& \partial^\mu W^{a \nu} - \partial^\nu W^{a \mu} + g_L f^{a b c}_L W^{b \mu} W^{c \nu} \, , \\
F^{\mu \nu}_Y &=& \partial^\mu B^\nu - \partial^\nu B^\mu \, ,
\end{eqnarray}
transform as $ F^{\mu \nu} \rightarrow U F^{\mu \nu} U^{-1} $. Therefore, the following Yang-Mills Lagrangian

\begin{equation}\label{eq:YangMills}
\mathcal{L}_{gauge} = - \frac{1}{4} G^{\mu \nu}_a G_{a \mu \nu} - \frac{1}{4} F^{\mu \nu}_a F_{a \mu \nu} - \frac{1}{4} F^{\mu \nu}_Y F_{Y \mu \nu} \, , 
\end{equation}
where $ a $ runs over the total number of generators, is invariant under the gauge transformations.

Apart from the terms in $ \mathcal{L}_{gauge} $, in full generality when building the SM we should consider a term like

\begin{equation}
\frac{1}{2} \epsilon_{\mu \nu \alpha \beta} G^{\mu \nu}_a G^{\alpha \beta}_a \, ,
\end{equation}
where $ \epsilon_{\mu \nu \alpha \beta} $ is the anti-symmetric tensor. This is a $ \mathcal{P}- $ and $ \mathcal{T}- $violating term, where $ \mathcal{P} $ states for parity transformation, $ (t, \overrightarrow{r}) \rightarrow (t, -\overrightarrow{r}) $, and $ \mathcal{T} $ for time reversion, $ (t, \overrightarrow{r}) \rightarrow (-t, \overrightarrow{r}) $. Despite being an important issue in the SM (and its extensions), we are not going to discuss it in detail (see Ref.~\cite{Branco:1999fs} for an introduction into this problem). 


\subsection{Matter fields}

The kinetic term for a spinor field $ f $ is given by\footnote{The irreducible representations of the Lorentz group are \textit{Weyl} spinors, which are complex two-component objects of definite chiralities (right- or left-handed). A \textit{Dirac} spinor $ f $ is built out of two Weyl spinors $ u_{R,L} $, $ f = \begin{pmatrix}
u_R \\ u_L
\end{pmatrix} $, where $ u_{R,L} $ have definite parity transformations

\begin{equation}
\mathcal{P} : u_{R,L} \rightarrow u_{L,R} \, .
\end{equation}

A Weyl spinor is then a Dirac spinor that is an eigenstate of $ \gamma_5 $: $ \gamma_5 f = \pm f \Leftrightarrow \gamma_5 f^c = \mp f^c $, where $ f^c $ is defined so that $ f^c = C A^T {f^\dagger}^T $ -- $ A $ and $ C $ are matrices satisfying $ A \gamma_\mu = \gamma^\dagger_\mu A $ and $ \gamma_\mu C = - C \gamma_\mu^T $ (for the usual Dirac representations, $ A = \gamma^0 $). By definition, a \textit{Majorana} spinor is a Dirac spinor such that $ f^c = f $, up to an arbitrary phase.}

\begin{equation}
\mathcal{L}^{(free)}_{matter} = \bar{f} i \gamma^\mu \partial_\mu f \, , \quad \bar{f} = f^\dagger \gamma^0 \, ,
\end{equation}
where we assume for the time being a massless field $ f $, and $ \gamma^\mu $ are $ 4 \times 4 $ matrices satisfying the Dirac algebra

\begin{equation}
\{ \gamma^\mu, \gamma^\nu \} = 2 g^{\mu \nu} \mathbf{1}_{4 \times 4} \, .
\end{equation}

The matter content of the Standard Model is made of three copies or generations of the following fields (all three equivalents at this stage):

\begin{center}
\begin{tabular}{ll}
left-handed quarks: & $ Q_L = \begin{pmatrix}
u_L \\ d_L
\end{pmatrix} = (\mathbf{3},\mathbf{2},1/3) $ \, , \\
right-handed quarks: & $ u_R = (\mathbf{3},\mathbf{1},4/3) \, , \quad d_R = (\mathbf{3},\mathbf{1},-2/3) \, , $ \\
left-handed leptons: & $ L_L = \begin{pmatrix}
\nu_L \\ \ell_L
\end{pmatrix} = (\mathbf{1},\mathbf{2},-1) $ \, , \\
right-handed leptons: & $ \ell_R = (\mathbf{1},\mathbf{1},-2) \, , $ \\
\end{tabular}
\end{center}
\noindent
where the quantum numbers refer to the gauge group in Eq.~\eqref{eq:SM}, and $ Y = 2 (Q - T^L_{3}) $ is chosen for left and right-handed chiralities such that up-type quarks $ u $ have electric charge $ Q = +2/3 $ and down-type quarks $ d $ have electric charge $ Q = -1/3 $, while we have neutral leptons $ \nu $ with no electric charge and charged leptons $ \ell $ with electric charge $ Q = -1 $. Now it should be clear the subscript ``$ L $'' in the $ SU(2)_L $ gauge group: the right- and left-handed fields are charged differently under this symmetry, the former being singlets and the latter being doublets.



When considering interactions with the gauge sector we substitute $ \partial \rightarrow D $, where the \textit{covariant derivative} $ D $ includes the gauge transformations of the fermionic fields:

\begin{equation}\label{eq:covariantDerivativeSM}
D^\mu = \partial^\mu - i ( g_s A^{a \mu} T^s_{a} + g_L W^{a \mu} T^L_{a} + g_Y B^\mu Y / 2 ) \, ,
\end{equation}
where $ T^s_{a} $ and $ T^L_{a} $ are the generators of the symmetry groups $ SU(3)_c $ and $ SU(2)_L $. The free Lagrangian $ \mathcal{L}^{(free)}_{matter} $ is then replaced by

\begin{equation}\label{eq:LmatterNotFree}
\mathcal{L}_{matter} = \bar{f} i \gamma^\mu D_\mu f \, .
\end{equation}
The full Lagrangian is now $ \mathcal{L}_{gauge} + \mathcal{L}_{matter} $ and it leads to parity and charge-conjugation violation (i.e. the discrete symmetry relating a particle and its anti-particle), a consequence of the gauging of $ V-A $.

Note that the mass term $ - m \bar{f} f = - m (\overline{f_R} f_L + h.c.) $ is not allowed by symmetry reasons, since it cannot be derived from the matter content (right-handed singlets and left-handed doublets) of the SM. At low energies, where the local symmetries $ SU(2)_L \times U(1)_Y $ are (spontaneously) broken, a mass term is compatible with the remaining symmetry, which is the electromagnetism. We will come back to the question of the masses of the fermions later on.

\subsection{EW symmetry breaking}

A further piece of the Lagrangian, necessary to implement symmetry breaking in the SM, includes a complex scalar field $ \phi $ in the following way

\begin{equation}\label{eq:LScalarFree}
\mathcal{L}^{(free)}_{scalar} = (\partial_\mu \phi) (\partial^\mu \phi)^\dagger - V(\phi) \, ,
\end{equation}
where the first term is the kinetic energy density of $ \phi $, and the second its self-interaction

\begin{equation}\label{eq:potentialSM}
V(\phi) = -\mu^2 \phi^\dagger \phi + \frac{\lambda}{2} (\phi^\dagger \phi)^2 \, .
\end{equation}
The scalar field $ \phi $ has the following quantum numbers

\begin{equation}
\phi = (\mathbf{1},\mathbf{2},1) \, , 
\end{equation}
i.e. a colourless weak isodoublet

\begin{equation}
\phi = \begin{pmatrix}
\varphi^+ \\ \varphi^0
\end{pmatrix} \, ,
\end{equation}
where the superscripts $ +, 0 $ refer to the values of the operator $ T^L_3 + Y/2 $, which remains an unbroken symmetry at low energies, identified with the electromagnetism.

If $ \mu^2 $ from Eq.~\eqref{eq:potentialSM} is positive, the field $ \phi $ acquires a vacuum expectation value (VEV) given by

\begin{equation}\label{eq:VEVSM}
\langle \phi \rangle = \begin{pmatrix}
0 \\ (\mu^2 / \lambda)^{1/2}
\end{pmatrix} \, ,
\end{equation}
which does not introduce a source of $ \mathcal{C P} $ violation in the SM: though one could choose to have a complex VEV, its phase can be eliminated by a non-physical field redefinition $ \phi \rightarrow {\rm e}^{i \alpha} \phi $, i.e. by a global phase shifting. From the expansion of $ \phi $ around its VEV

\begin{equation}
\phi = \langle \phi \rangle + \begin{pmatrix}
\varphi^+ \\ \frac{(H^0 + i \chi^0)}{\sqrt{2}}
\end{pmatrix} \, ,
\end{equation}
one has

\begin{equation}\label{eq:brokenPotential}
V = - \frac{\mu^4}{2 \lambda} + \mu^2 (H^0)^2 + \ldots \, ,
\end{equation}
where the ellipsis contains cubic and quartic interactions. The first term is related to the vacuum energy density, while the second states that there is a massive scalar field of mass $ \sqrt{2} \mu $, which is the SM Higgs particle. The full spectrum contains still massless particles, as attested by the remaining degrees of freedom $ \chi^0 $ and $ \varphi^\pm $, which are the (would-be) Goldstone bosons \cite{Goldstone}.

Interactions between the scalar $ \phi $ and the gauge sector are implemented through the substitution $ \partial \rightarrow D $, which follows by charging $ \phi $ under Eq.~\eqref{eq:SM}, and requiring $ \mathcal{L}^{(free)}_{scalar} \rightarrow \mathcal{L}_{scalar} $ to be invariant under the gauge transformations of $ \phi $: 

\begin{equation}\label{eq:LscalarNotFree}
\mathcal{L}_{scalar} = (D_\mu \phi) (D^\mu \phi)^\dagger - V(\phi) \, .
\end{equation}


Eq.~\eqref{eq:LscalarNotFree} implies one of the major achievements of particle physics: through the interplay of the two sectors, the gauge bosons acquire masses at low energies, thus explaining why the weak interactions have a short range (in other words, the corresponding forces are described by a Yukawa potential). Let us now see how it happens, a mechanism called Brout-Englert-Higgs (BEH) \cite{BEH}. By expanding $ (D^\mu \phi) (D_\mu \phi)^\dagger $ around the vacuum expectation value of $ \phi $, we have the mass terms

\begin{equation}
M^2_W W^- W^+ + \frac{1}{2} M^2_Z Z^2 \, , \quad M_W = \frac{g_L v}{2} \, , \quad M_Z = \frac{M_W}{c_\theta} \, ,
\end{equation}
where $ v = \sqrt{2} (\mu^2 / \lambda)^{1/2} $ and the fields $ W^\pm, Z $ are

\begin{equation}
W^\pm = \frac{W^1 \mp W^2}{\sqrt{2}} \, , \quad \begin{pmatrix}
A \\ Z
\end{pmatrix} = \begin{pmatrix}
c_\theta & s_\theta \\ - s_\theta & c_\theta
\end{pmatrix} \begin{pmatrix}
B \\ W^3
\end{pmatrix} \, ,
\end{equation}
and 

\begin{equation}
e = (g_L^{-2} + g_Y^{-2})^{-1/2} \, , \quad s_\theta = \sin \theta = e / g_L \, , \quad c_\theta = \cos \theta = e / g_Y \, .
\end{equation}
The mass terms above are not compatible with the electroweak gauge symmetries stated in the previous section: while there still remains a massless field $ A $, the electromagnetic vector field, the remaining symmetries are said to be \textit{spontaneously broken}, due to the VEV of $ \phi $. Note that out of $ A $ we have two physical degrees of freedom, corresponding to the possible transverse polarizations, while the massive fields $ W^\pm $ and $ Z $ have extra degrees of freedom corresponding to the Goldstone bosons discussed after Eq.~\eqref{eq:brokenPotential}, which are $ \varphi^\pm $ and $ \chi^0 $, respectively.





It is an amazing fact that the same phenomenon we are discussing, leading to the spontaneous breaking of local (gauge) symmetries, also implies a mechanism for mass generation in the fermionic sector through (primes are used previous to going to the eigenmass basis)

\begin{eqnarray}\label{eq:LYukawa}
&& \mathcal{L}_{Yukawa} = - (\overline{Q'_L} Y \phi d'_R + \overline{Q'_L} \tilde{Y} \tilde{\phi} u'_R + \overline{L'_L} Y^{lept} \phi \ell'_R) + h.c. \, , \quad \overline{f_L} = \bar{f} P_R \, , \nonumber\\
&& \qquad \quad \tilde{\phi} = i \tau_2 {\phi^\dagger}^T \, , \quad \tau_2 = \begin{pmatrix}
0 & - i \\ i & 0
\end{pmatrix} \, ,
\end{eqnarray}
where $ Y, \tilde{Y}, Y^{lept} $ are $ 3 \times 3 $ matrices over generation indices, called Yukawa matrices. Plugging the vacuum expectation value of the Higgs field, we obtain the following mass term

\begin{equation}
\mathcal{L}_{Yukawa} \ni - (\overline{d'_L} M_d d'_R + \overline{u'_L} M_u u'_R + \overline{\ell'_L} M_{\ell} \ell'_R) + h.c. \, ,
\end{equation}
where $ M_d = \frac{v}{\sqrt{2}} Y $, $ M_u = \frac{v}{\sqrt{2}} \tilde{Y} $ and $ M_{\ell} = \frac{v}{\sqrt{2}} Y^{lept} $. The Yukawa matrices (not necessarily Hermitian) are diagonalized by the bi-unitary transformations $ U^{u,d}_{L,R} $

\begin{eqnarray}
&& u'_R = U^u_R u_R \, , \quad u'_L = U^u_L u_L \, , \\
&& d'_R = U^d_R d_R \, , \quad d'_L = U^d_L d_L \, ,
\end{eqnarray}
under which we have, by definition,

\begin{equation}
U^{u \dagger}_L M_u U^u_R = {\rm diag} (m_u, m_c, m_t) \, , \quad U^{d \dagger}_L M_d U^d_R = {\rm diag} (m_d, m_s, m_b) \, ,
\end{equation}
where $ m_u, m_d, m_s, m_c, m_b, m_t $ are real and positive. Following this discussion, $ \mathcal{L}_{Yukawa} $ leads to symmetry breaking among the generations: previous to considering the Yukawa interactions, the Lagrangian was invariant under the interchange of generations, but now generations are differentiated by their masses, $ m_t \simeq 170 $~GeV, $ m_b \simeq 4 $~GeV, $ m_c \simeq 1 $~GeV, etc.


\subsection{Full model}

The full SM theory is described by

\begin{equation}
\mathcal{L}_{SM} = \mathcal{L}_{gauge} + \mathcal{L}_{scalar} + \mathcal{L}_{matter} + \mathcal{L}_{Yukawa} \, ,
\end{equation}
where the individual terms are defined in Eqs.~\eqref{eq:YangMills}, \eqref{eq:LscalarNotFree}, \eqref{eq:LmatterNotFree}, \eqref{eq:LYukawa}, \eqref{eq:covariantDerivativeSM}, and \eqref{eq:potentialSM}. The full Lagrangian leads to the phenomenon of \textit{generation mixing}, as we now see. For a fermionic multiplet $ f $, the gauge interactions are derived from $ \overline{f} \gamma^\mu i D_\mu f $, from which one has for the weak charged current

\begin{equation}
\frac{g_L}{\sqrt{2}} (W^+_\mu \overline{u'_L} \gamma^\mu d'_L + W^-_\mu \overline{d'_L} \gamma^\mu u'_L) \, ,
\end{equation}
which can be put into the mass basis for quarks

\begin{equation}\label{eq:genMixVector}
\frac{g_L}{\sqrt{2}} (W^+_\mu \overline{u_L} \gamma^\mu V d_L + W^-_\mu \overline{d_L} \gamma^\mu V^\dagger u_L) \, .
\end{equation}
Above, the matrix

\begin{equation}
V \equiv U^{u \dagger}_L U^d_L
\end{equation}
is called the Cabibbo-Kobayashi-Maskawa (CKM) matrix. Its non-diagonal structure, as we measure it, leads to generation mixing, as seen from Eq.~\eqref{eq:genMixVector}. Note that the analogous unitary matrix $ U^{u \dagger}_R U^d_R $ is not observable in the SM.\footnote{A comment about notation: when shifting to the Left-Right Model, $ V $ is going to be replaced by $ V^L $.}

The phenomenon of generation mixing is exclusive of the weak charged coupling, i.e. in the SM the couplings of the $ Z $, the photon and the gluon are diagonal over family indices. Another important property concerning the matrix $ V $ is that it has a complex phase which leads to the only sizable source of $ \mathcal{C P} $ violation in the SM.


\section{EWPO for testing the SM}\label{sec:EWtestsSM}

A crucial way of testing the picture described in the previous section, i.e. the couplings of the gauge bosons to fermions and the way in which the local gauged symmetries of the SM are broken at low energies, has been provided by precise measurements made at LEP ($ e^- e^+ $ collider, including ALEPH, DELPHI, OPAL, and L3) and the Tevatron ($ p \bar{p} $ collider) in the 80's, among others. These experiments were able to collect a large amount of data and build many different observables involving the production and decay of $ Z $ bosons, and -- to a lesser extent -- $ W^\pm $ bosons. In the case of the $ Z $ boson, these observables were specially designed to test the Lorentz structure of its couplings to fermions, i.e. $ g_V^f $ and $ g_A^f $ defined as follows

\begin{eqnarray}
&& \frac{g_L}{2 \cos \theta} Z^\mu \overline{f} \gamma_\mu (g_V^f - g_A^f \gamma_5) f = \frac{g_L}{2 \cos \theta} Z^\mu \frac{1}{2} \overline{f_L} \gamma_\mu (g^f_V - g^f_A \gamma^5) f_L + (L \leftrightarrow R) \, , \\
&& g^f_V = T^L_3 (f) - 2 Q (f) \sin^2 \theta , \qquad g^f_A = T^L_3 (f) \, ,
\end{eqnarray}
for a flavour $ f $ (a lepton or a quark). The different values of these couplings are given in Table~\ref{tab:SMquantumNumbers}.

Taking into account the set of the most precise measurements and predictions, one can perform a global fit and test the overall agreement of the SM picture with data. At the same time, if this agreement is good enough, it is possible to extract the values of the underlying quantities of the theory, as it was the case for the mass of the Higgs boson \cite{Baak:2012kk}. Note that this is an indirect extraction, much different in the case of the Higgs mass from modern direct measurements. In what follows, we perform a global fit of the EW Precision Observables (EWPO).

\subsection{Inputs}\label{sec:inputsforEWPO}

The couplings $ g^f_V $ and $ g^f_A $ are the basic ingredients to define many EWPO. Among these observables we have left-right asymmetries of the couplings


\begin{equation}
\mathcal{A}_f = 2 \frac{g^f_V g^f_A}{(g^f_V)^2 + (g^f_A)^2} \, , \quad f = e, \mu, \tau, b, c \, ,
\end{equation}
forward-backward asymmetries in the $ Z $ boson production

\begin{equation}
A_{FB} (f) = \frac{3}{4} \mathcal{A}_e \mathcal{A}_f \, , \quad f = e, \mu, \tau, b, c \, ,
\end{equation}
total cross section of the $ Z $ boson into hadrons

\begin{equation}
\sigma_{had} = \sum_{f \in \{ u,d,s,c,b \}} \frac{12 \pi}{M^2_Z} \frac{\Gamma_{e \bar{e}} \Gamma_{f \bar{f}}}{\Gamma^2_Z} \, ,
\end{equation}
ratios of partial widths for quarks

\begin{equation}
R_q = \frac{\Gamma_{q \bar{q}}}{\Gamma_{had}} \, , \quad q = b, c \, ,
\end{equation}
etc., defined at the pole of the $ Z $ boson from $ e^+ e^- $ collision. The full set of EWPO, including Atomic Parity Violation measurements obtained at low-energies, is defined explicitly in Appendix~\ref{sec:EWPOTreeLevel}. 

The set of observables we use in our fit is given in Table \ref{tab:tab1}. Correlations were taken into account and can be found in the quoted references. The inputs are dominated by statistical uncertainties, an important difference with respect to flavour observables that will be introduced later.

\begin{table}[t]
	\centering
	\begin{tabular}{|c|c|c|c|c|c|}
		\hline
		left-handed & $ T_3^L (f) $ & $ Q (f) $ & $ g^f_V $ & $ g^f_A $ & $ g^f_V / g^f_A $ \\
		\hline
		$ \nu_{e L}, \nu_{\mu L}, \nu_{\tau L} $ & $ + 1/2 $ & 0 & $ + 1/2 $ & $ + 1/2 $ & $ 1 $ \\
		$ e_L, \mu_L, \tau_L $ & $ - 1/2 $ & $ - 1 $ & $ - 1/2 + 2 \sin^2 \theta \sim - 0.04 $ & $ - 1/2 $ & $ \sim 0.08 $ \\
		$ u_L, c_L, t_L $ & $ + 1/2 $ & $ + 2/3 $ & $ + 1/2 - 4/3 \sin^2 \theta \sim 0.19 $ & $ + 1/2 $ & $ \sim 0.38 $ \\
		$ d_L, s_L, b_L $ & $ - 1/2 $ & $ - 1/3 $ & $ - 1/2 + 2/3 \sin^2 \theta \sim - 0.35 $ & $ - 1/2 $ & $ \sim 0.70 $ \\
		\hline
		\hline
		right-handed & $ T_3^L (f) $ & $ Q (f) $ & $ g^f_V $ & $ g^f_A $ & $ g^f_V / g^f_A $ \\
		\hline
		$ e_R, \mu_R, \tau_R $ & $ 0 $ & $ - 1 $ & $ + 2 \sin^2 \theta \sim 0.46 $ & $ 0 $ & - \\
		$ u_R, c_R, t_R $ & $ 0 $ & $ + 2/3 $ & $ - 4/3 \sin^2 \theta \sim - 0.31 $ & $ 0 $ & - \\
		$ d_R, s_R, b_R $ & $ 0 $ & $ - 1/3 $ & $ + 2/3 \sin^2 \theta \sim 0.15 $ & $ 0 $ & - \\
		\hline
	\end{tabular}
	\caption{\it EW quantum numbers of the different SM fermions.}\label{tab:SMquantumNumbers}
\end{table}



Thanks to the huge effort from the theoretical side to match the experimental accuracy of these observables, see \cite{ALEPH:2005ab} and references therein, which required going beyond the tree level order, one was able to probe the structure of the SM in detail. The higher order effects are sensitive to the underlying parameters of the model, such as the masses of the top-quark, $ Z^0 $ boson and Higgs boson, and apart from that these corrections are also sensitive to the coupling constants $ \alpha_s $ and $ \alpha $, at the energy scale $ M_Z $ for $ Z^0 $ observables. 


Beyond the loop-corrections alluded in the last paragraph, Initial State Radiation (ISR) and Final State Radiation (FSR) need also to be taken into account. They correspond to high-order corrections in QED where one (or more) photon(s) are emitted by the initial electron or/and positron states, in the case of ISR, or by the final states, in the case of FSR. Taking into account this class of corrections is of capital importance: they reduce the value of $ \sigma_{had} $ by about $ 36~\% $, and the value of $ A^{\mu, \tau}_{FB} $ by as much as $ 80~\% $. In what follows, they have already been subtracted from the experimental values \cite{ALEPH:2005ab}, quoted in Table~\ref{tab:tab1}. To distinguish the initial and the ISR/FSR-subtracted values, the latter are referred to as ``pseudo-observables,'' and indicated with a superscript 0 (but we do not systematically use this notation). Note an important point: their extraction is made in a model-independent way, since these are pure QED effects, and therefore do not depend on the EW sector which is under test here.


We give a few more comments about one of the inputs. The value for the strong coupling, $ \alpha_{s} (M_{Z}) = 0.1185 \pm 0.0005_{syst} $ \cite{Beringer:1900zz}, takes into account the four following classes of inputs: $ \tau- $decay, Lattice, DIS, and $ e^+ e^- $. More recent information on $ \alpha_s $ from hadronic collider studies \cite{CMSalphas} are not included, because this extraction has a more important uncertainty when compared with the other four classes. Note that the $ \tau- $decay provides a value for the strong coupling at $ m_\tau $, and needs to be evolved from low-energy scales up to $ M_Z $. This can be done up to the $ {\rm N^3LO} $ \cite{Chetyrkin:2000yt}, and requires the inclusion of thresholds for the charm- and bottom-quarks. The strong constant extracted from $ \tau- $decays is the only input used by \texttt{Gfitter} in the SM fit, presumably because the full set of them has a negligible impact on the prediction for the Higgs mass, see Ref.~\cite{Flacher:2008zq}.


\subsection{Parameterization}\label{sec:secSparameters}


The corrections beyond tree level of any observable $ \mathcal{X} $ can be written in terms of the parameters 

\begin{equation}\label{eq:parametersEWPOSM}
\mathcal{S} \equiv \{ m^{pole}_{top} , \alpha_{s} (M_{Z}) , M_{Z} , M_H , \Delta \alpha (M_{Z}) \} \, .
\end{equation}
\noindent
One can thus write

\begin{eqnarray}\label{eq:parametersForX}
& \mathcal{X} = \mathcal{X}_{0} + c_{1} \cdot L_{H} + c_{2} \cdot \Delta_{t} + c_{3} \cdot \Delta_{\alpha_{s}} \\
& + c_{4} \cdot \Delta^{2}_{\alpha_{s}} + c_{5} \cdot \Delta_{\alpha_{s}} \Delta_{t} + c_{6} \cdot \Delta_{\alpha} + c_{7} \cdot \Delta_{Z} \, , \nonumber\\
& L_{H} = \log \frac{M_{H}}{125.7 \, \operatorname{GeV}} \, , \\
& \Delta_{t} = \left( \frac{m^{pole}_{top}}{173.2 \, \operatorname{GeV}} \right)^{2} - 1 \, , \\
& \Delta_{\alpha_{s}} = \frac{\alpha_{s} (M_{Z})}{0.1184} - 1 \, , \\
& \Delta_{\alpha} = \frac{\Delta \alpha (M_{Z})}{0.059} - 1 \, , \\
& \Delta_{Z} = \frac{M_{Z}}{91.1876 \, \operatorname{GeV}} - 1 \, .
\end{eqnarray}
The parameterization in Eq.~\eqref{eq:parametersForX} describes very accurately the ensemble of the observables we consider, and higher-order terms in $ L_{H}, \Delta_{t}, \Delta_{\alpha_{s}}, \Delta_{\alpha}, \Delta_{Z} $ compared to those already shown can be safely neglected. We must now determine the coefficients of the parameterization, and we have used \texttt{Zfitter} 6.42 \cite{Arbuzov:2005ma,Riemann} for determining them. \texttt{Zfitter} consists in a set of codes integrating higher-order corrections to a variety of observables, mainly those defined at the pole of the $ Z $ boson. For a given set of values of $ \{ m^{pole}_{top} , \alpha_{s} (M_{Z}) , M_{Z} , M_H , \Delta \alpha (M_{Z}) \} $, \texttt{Zfitter} provides the numerical value of $ \mathcal{X} $, and we are then able to determine the values of $ \mathcal{X}_0, c_{1,2,3,4,5,6,7} $ seen in Appendix \ref{sec:AppParameters}.

Reference \cite{Freitas:2014hra} includes further corrections for the observables $ \Gamma_{Z}, \sigma_{had}, R_{b, c} $ (two-loop EW diagrams, not included in the version 6.42 of \texttt{Zfitter}) and gives their parameterization in the same way seen in Eq.~\eqref{eq:parametersForX}. We have then considered its results for the coefficients of the parameterization of $ \Gamma_{Z}, \sigma_{had}, R_{b, c} $.



A last point concerning the parameters of Eq.~\eqref{eq:parametersForX}: it will be more useful to parameterize $ \alpha $ in a different way. First we write

\begin{equation}
\alpha (s) = \frac{\alpha (0)}{1 - \Delta \alpha (s)} \, , \quad \alpha (0) = 1/137.035999074 \, ,
\end{equation}
for $ \alpha $ calculated at the energy $ s $ (when not stated otherwise, $ \alpha $ is calculated at $ M_Z $), and then we use the following decomposition into quark (except the top), charged leptonic and top contributions

\begin{equation}
\Delta \alpha (s) = \Delta \alpha^{(5)}_{had} (s) + \Delta \alpha_{e \mu \tau} (s) + \Delta \alpha_{top} (s) \, ,
\end{equation}
where \cite{Steinhauser:1998rq, Kuhn:1998ze}

\begin{equation}
\Delta \alpha_{e \mu \tau} (M_{Z}) = 0.031497686 \, , \quad \Delta \alpha_{top} (M_{Z}) = - 0.000072 \, ,
\end{equation}
with negligible uncertainties. In the following, we use the parameter $ \Delta \alpha^{(5)}_{had} (M_Z) $ instead of $ \Delta \alpha (M_{Z}) $. Note that we do not use an input for $ \Delta \alpha_{had}^{(5)} $ due to the wild spread of central values and uncertainties found in the literature (see the ``EW model and constraints on NP'' review in \cite{Beringer:1900zz}).

\subsection{Global fit}

\begin{table}
\begin{center}
\begin{tabular}{|cc|c|c|c|}
\hline
Observable &Ref.& input & SM fit (1 $ \sigma $) & pull \\
\hline
$ \Delta \alpha_{had}^{(5)} $ & & - & $ 0.02736^{+0.00041}_{-0.00042} $ & - \\
$ M_H $ [GeV] & \cite{Aad:2012tfa} \cite{Chatrchyan:2012xdj} & $ 125.7 \pm 0.4 $ & $ 125.7 \pm 0.4 $ & 0.97 \\
$ m_{top}^{pole} $ [GeV] & \cite{ATLAS:2014wva} & $ 173.34 \pm 0.36 \pm 0.67 $ & $ 174.04^{+0.36}_{-1.48} $ & 0.95 \\
$ M_Z $ [GeV] & \cite{ALEPH:2005ab} & $ 91.1876 \pm 0.0021 $ & $ 91.1876 \pm 0.0021  $ & 0.44 \\
$ \alpha_s $ & \cite{Beringer:1900zz} & $ 0.1185 \pm 0 \pm 0.0005 $ & $ 0.11900^{+0.00012}_{-0.00109} $ & 0.23 \\
\hline
$ \Gamma_Z $ [GeV] & \cite{ALEPH:2005ab} & $ 2.4952 \pm 0.0023 $ & $ 2.49493^{+0.00042}_{-0.00084} $ & 0.56 \\
$ \sigma_{had} $ [nb] & \cite{ALEPH:2005ab} & $ 41.541 \pm 0.037 $ & $ 41.4857^{+0.0067}_{-0.0023} $ & 1.42 \\
$ R_b $ & \cite{ALEPH:2005ab} & $ 0.21629 \pm 0.00066 $ & $ 0.215762^{+0.000055}_{-0.000022} $ & 0.56 \\
$ R_c $ & \cite{ALEPH:2005ab} & $ 0.1721 \pm 0.0030 $ & $ 0.172256^{+0.000019}_{-0.000037} $ & 0.14 \\
$ R_e $ & \cite{ALEPH:2005ab} & $ 20.804 \pm 0.050 $ & $ 20.7445^{+0.0029}_{-0.0088} $ & 0.77 \\
$ R_\mu $ & \cite{ALEPH:2005ab} & $ 20.785 \pm 0.033 $ & $ 20.7445^{+0.0029}_{-0.0088} $ & 1.08 \\
$ R_\tau $ & \cite{ALEPH:2005ab} & $ 20.764 \pm 0.045 $ & $ 20.7915^{+0.0030}_{-0.0088} $ & 0.87 \\
$ A_{FB} (b) $ & \cite{ALEPH:2005ab} & $ 0.0992 \pm 0.0016 $ & $ 0.10363 \pm 0.00079 $ & 2.89 \\
$ A_{FB} (c) $ & \cite{ALEPH:2005ab} & $ 0.0707 \pm 0.0035 $ & $ 0.07409 \pm 0.00061 $ & 0.62 \\
$ A_{FB} (e) $ & \cite{ALEPH:2005ab} & $ 0.0145 \pm 0.0025 $ & $ 0.01639^{+0.00024}_{-0.00025} $ & 0.39 \\
$ A_{FB} (\mu) $ & \cite{ALEPH:2005ab} & $ 0.0169 \pm 0.0013 $ & $ 0.01639 \pm 0.00024 $ & 0.29 \\
$ A_{FB} (\tau) $ & \cite{ALEPH:2005ab} & $ 0.0188 \pm 0.0017 $ & $ 0.01639 \pm 0.00024 $ & 1.41 \\
$ \mathcal{A}_b $ & \cite{ALEPH:2005ab} & $ 0.923 \pm 0.020 $ & $ 0.93464 \pm 0.00011 $ & 0.41 \\
$ \mathcal{A}_c $ & \cite{ALEPH:2005ab} & $ 0.670 \pm 0.027 $ & $ 0.66823 \pm 0.00050 $ & 0.15 \\
$ \mathcal{A}_e^{SLD} $ & \cite{ALEPH:2005ab} & $ 0.1516 \pm 0.0021 $ & $ 0.1478 \pm 0.0011 $ & 2.15 \\
$ \mathcal{A}_e (P_{\tau}) $ & \cite{ALEPH:2005ab} & $ 0.1498 \pm 0.0049 $ & $ 0.1478 \pm 0.0011 $ & 0.42 \\
$ \mathcal{A}_\mu^{SLD} $ & \cite{ALEPH:2005ab} & $ 0.142 \pm 0.015 $ & $ 0.1478 \pm 0.0011 $ & 0.40 \\
$ \mathcal{A}_\tau^{SLD} $ & \cite{ALEPH:2005ab} & $ 0.136 \pm 0.015 $ & $ 0.1478 \pm 0.0011 $ & 0.82 \\
$ \mathcal{A}_\tau (P_\tau) $ & \cite{ALEPH:2005ab} & $ 0.1439 \pm 0.0043 $ & $ 0.1478 \pm 0.0011 $ &  0.95 \\
\hline
$ M_W $ [GeV] & \cite{Aaltonen:2013iut} \cite{Awramik:2003rn} & $ 80.385 \pm 0.015 \pm 0.004 $ & $ 80.3694^{+0.0075}_{-0.0081} $ & 0.89 \\
$ \Gamma_W $ [GeV] & \cite{ALEPH:2010aa} & $ 2.085 \pm 0.042 $ & $ 2.09151^{+0.00062}_{-0.00093} $ &  0.15 \\
\hline
$ Q_{W} (Cs) $ & \cite{Wood:1997zq} \cite{Guena:2004sq} & $ - 73.20 \pm 0.35 $ & $ - 72.959 \pm 0.036 $ & 0.69 \\
$ Q_{W} (Tl) $ & \cite{Edwards:1995zz} \cite{Vetter:1995vf} & $ - 116.4 \pm 3.6 $ & $ - 116.457^{+0.059}_{-0.057} $ & 0.01 \\
\hline
\end{tabular}
\caption{\it Inputs and results for the SM global fit.}\label{tab:tab1}
\end{center}
\end{table}

We now comment on the results of the global fit. The observables were combined using \texttt{CKMfitter}, a statistical tool which will be described in the next section. The value for the $ \chi^{2}_{\min} $ of the global fit is 22.24, and with 22 degrees of freedom we have the p-value $ \sim 45~\% $, good enough for a meaningful extraction of the fundamental parameters. The extracted best fit points and intervals are seen in Table~\ref{tab:tab1}.

\begin{figure}
	\centering
	\includegraphics[scale=0.45]{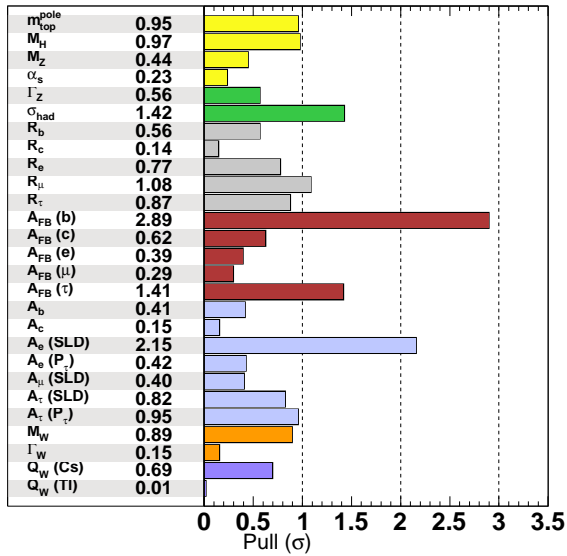}
	\caption{\it The pulls defined in Eq.~\eqref{eq:flavourPull} indicate the tensions between the measured and the (indirectly) predicted values given by the SM.}\label{fig:PullsEW}
\end{figure}

The overall conclusion we can draw is that EWPO are well described in the context of the SM. There are some tensions, however, indicated by the pulls defined for an observable $ o $ as

\begin{equation}\label{eq:flavourPull}
{\rm pull} = \sqrt{\chi^2_{min} - \chi^2_{min,o!}} \, ,
\end{equation}
where ``$ o! $" means that the second $ \chi^2 $ is built without the experimental information on $ o $. This is a different definition from the one used used in EWPO: in the context of EWPO, a different definition is usually found in the literature \cite{ALEPH:2005ab}. As can be seen from Figure~\ref{fig:PullsEW}, the main tensions are found for $ \sigma_{had}, A_{FB} (b, \tau), \mathcal{A}_e^{SLD} $, which are left unexplained at this stage in the SM context. We now move to a much different class of observables, consisting of flavour physics observables.

\section{CKM matrix phenomenology and fit}

The strengths of the $ W $ couplings to quarks are left free in the SM and must be extracted from the observation of nature. One may determine these couplings from a well-measured set of observables, performing a global combination of them. The exercise of combining a large set of observables in the extraction of the CKM matrix has been executed by many different collaborations \cite{CKMfitterStandard,UTfit,Scan}, and they have all pointed towards a consistent description of flavour processes made by the SM. This is usually indicated by Unitarity Triangle (UT) fits, showing that at the current level of accuracy the observables agree about the true values of the fundamental quantities parameterizing the CKM matrix, in particular the one related to $ \mathcal{C P}- $violating phenomena. Previous to performing a global fit, we now introduce the most relevant facts about the CKM matrix. 


\subsection{The CKM matrix}\label{sec:CKMmatrixBasics}

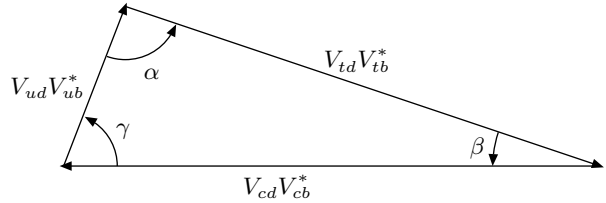
\begin{figure}
\centering
\begin{picture}(210,80)(-10,-20)
	\LongArrow(200,0)(0,0)
	\Text(80,-12)[b]{\small $ V^{}_{cd} V^{*}_{cb} $}
	\LongArrow(0,0)(23,60)
	\Text(-20,30)[l]{\small $ V^{}_{ud} V^{*}_{ub} $}
	\LongArrow(23,60)(200,0)
	\Text(122,40)[r]{\small $ V^{}_{td} V^{*}_{tb} $}
	\LongArrowArc(23,60)(20,-110,-19.6)
	\Text(33,33)[b]{\small $ \alpha $}
	\LongArrowArc(200,0)(40,161.5,180)
	\Text(152,7)[l]{\small $ \beta $}
	\LongArrowArc(0,0)(20,0,67)
	\Text(25,13)[r]{\small $ \gamma $}
\end{picture}
	\caption{\it Example of Unitarity Triangle. In this specific case, the sides of the (un-normalized) triangle have lengths of order $ \lambda^3 $.}\label{fig:UTBdBdbar}
\end{figure}

The elements of the CKM matrix

\begin{equation}
V = \begin{pmatrix}
V_{ud} & V_{us} & V_{ub} \\
V_{cd} & V_{cs} & V_{cb} \\
V_{td} & V_{ts} & V_{tb} \\
\end{pmatrix}
\end{equation}
describe the coupling strengths of the $ W $ boson to a pair of up- and down-type quarks. Being a unitary matrix, the product of two different columns of $ V $ satisfies

\begin{equation}\label{eq:unitarityCKM}
V_{u \alpha} V^{*}_{u \beta} + V_{c \alpha} V^{*}_{c \beta} + V_{t \alpha} V^{*}_{t \beta} = 0 \, ,
\end{equation}
where $ \{ \alpha = d, \beta = s \} $ corresponds to products of elements of $ V $ found in the $ K \overline{K} $ meson-mixing system, while $ \{ \alpha = d, \beta = b \} $ ($ \{ \alpha = s, \beta = b \} $) is found in the $ B_d \overline{B}_d $ ($ B_s \overline{B}_s $ respectively) system. Graphically, Eq.~\eqref{eq:unitarityCKM} can be represented by a triangle whose sides correspond to the following three vectors in the complex plane: $ V_{u \alpha} V^{*}_{u \beta} $, $ V_{c \alpha} V^{*}_{c \beta} $, and $ V_{t \alpha} V^{*}_{t \beta} $, see Figure~\ref{fig:UTBdBdbar}.

These same elements can be parameterized in terms of mixing angles $ \theta_{12} $, $ \theta_{13}$, $ \theta_{23} $, also called Cabibbo angles, between different generations (Chau-Keung parameterization \cite{Chau:1984fp})

\begin{eqnarray}\label{eq:VChauKeung}
V &=& \begin{pmatrix}
1 & 0 & 0 \\
0 & c_{23} & s_{23} \\
0 & -s_{23} & c_{23} \\
\end{pmatrix}
\begin{pmatrix}
c_{13} & 0 & s_{13} {\rm e}^{-i \delta} \\
0 & 1 & 0 \\
-s_{13} {\rm e}^{i \delta} & 0 & c_{13} \\
\end{pmatrix}
\begin{pmatrix}
c_{12} & s_{12} & 0 \\
-s_{12} & c_{12} & 0 \\
0 & 0 & 1 \\
\end{pmatrix} \nonumber\\
&=& \begin{pmatrix}
c_{12} c_{13} & s_{12} c_{13} & s_{13} {\rm e}^{-i \delta} \\
- s_{12} c_{23} - c_{12} s_{23} s_{13} {\rm e}^{i \delta} & c_{12} c_{23} - s_{12} s_{23} s_{13} {\rm e}^{i \delta} & s_{23} c_{13} \\
s_{12} s_{23} - c_{12} c_{23} s_{13} {\rm e}^{i \delta} & -c_{12} s_{23} - s_{12} c_{23} s_{13} {\rm e}^{i \delta} & c_{23} c_{13} \\
\end{pmatrix} \, , \nonumber\\
\end{eqnarray}
where $ s_{ij} = \sin \theta_{ij} $ and $ c_{ij} = \cos \theta_{ij} $ for $ i,j = 1,2,3 $, with $ i < j $. The complex phase $ \delta $ is, as stated previously, the only sizeable source of $ \mathcal{C} \mathcal{P} $ violation in the SM. The Chau-Keung parameterization (and the Wolfenstein parameterization that will be introduced afterwards) is based on a phase convention for the CKM matrix, i.e. the value of $ \delta $ depends on an arbitrary choice. However, it is shown that \cite{Jarlskog} 

\begin{equation}
J \equiv {\rm Im} (V^{}_{us} V^{}_{cb} V^{*}_{ub} V^{*}_{cs}) = c_{12} s_{12} c^2_{13} s_{13} c_{23} s_{23} \sin \delta \, ,
\end{equation}
called the Jarlskog invariant, is independent on the phase convention. Since mixing angles are small, as we will see, $ J $ and therefore $ \mathcal{C P}- $violating processes are naturally suppressed: $ J \simeq 10^{-5} $, much smaller than its maximum allowed value, \textit{viz.} $ 1/(6 \sqrt{3}) \simeq 0.1 $.


It is experimentally observed that the structure of the CKM matrix is hierarchical, i.e. mixings between different generations are suppressed. It is then useful to exploit the hierarchy of the CKM elements and introduce the Wolfenstein parameterization \cite{Wolfenstein:1983yz}, where to all orders in the small parameter $ \lambda \simeq 0.2 $ we have:


\begin{eqnarray}
&& s_{12} = \frac{\vert V_{us} \vert}{(\vert V_{ud} \vert^2 + \vert V_{us} \vert^2)^{1/2}} \equiv \lambda  \, , \\
&& s_{23} = \frac{\vert V_{cb} \vert}{(\vert V_{ud} \vert^2 + \vert V_{us} \vert^2)^{1/2}} \equiv A \lambda^2 \, , \\
&& s_{13} {\rm e}^{-i \delta} = V_{ub} \equiv A \lambda^3 (\rho - i \eta) \, .
\end{eqnarray}
Therefore, a non-vanishing value for $ \eta $ is equivalent to having a complex phase in the CKM matrix.


Note that, though the elements $ V_{cd}, V_{ts} $ (and $ V_{cs} $) are complex, their phases are suppressed by extra powers of $ \lambda $ if compared to $ V_{ub}, V_{td} $ (which is in fact a convention dependent statement). This can be more immediately seen from a systematic expansion in powers of $ \lambda $ (up to order $ \mathcal{O} (\lambda^6) $)


\begin{eqnarray}
V_{ud} &=& 1 - \frac{\lambda^2}{2} - \frac{\lambda^4}{8} \, , \label{eq:VudWolfenstein}\\
V_{us} &=& \lambda \, , \label{eq:VusWolfenstein}\\
V_{ub} &=& A \lambda ^3 (\rho -i \eta ) \, , \label{eq:VubWolfenstein}\\
V_{cd} &=& -\lambda +\frac{\lambda ^5}{2} A^2 (1-2 (\rho +i \eta )) \, , \label{eq:VcdWolfenstein}\\
V_{cs} &=& 1 - \frac{\lambda ^2}{2} -\frac{\lambda ^4}{8} \left(1+4 A^2\right) \, , \label{eq:VcsWolfenstein}\\
V_{cb} &=& A \lambda ^2 \, , \label{eq:VcbWolfenstein}\\
V_{td} &=& A \lambda ^3 (1 -\rho-i \eta)+\frac{\lambda^5}{2} A (\rho +i \eta ) \, , \label{eq:VtdWolfenstein}\\
V_{ts} &=& -A \lambda ^2 +\frac{\lambda ^4}{2} A (1-2 (\rho +i \eta )) \, , \label{eq:VtsWolfenstein}\\
V_{tb} &=& 1-\frac{\lambda ^4}{2}A^2 \, . \label{eq:VtbWolfenstein}
\end{eqnarray}

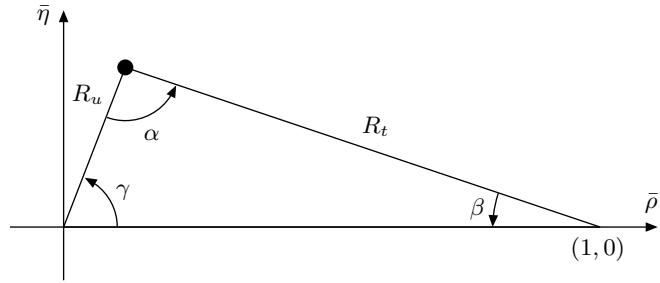
\begin{figure}
\centering
\begin{picture}(210,80)(-10,-20)
	\Vertex(23,60){3}
	\Line(200,0)(0,0)
	\Text(200,-13)[b]{\small $ (1,0) $}
	\Line(0,0)(23,60)
	\Text(3,50)[l]{\small $ R_u $}
	\Line(23,60)(200,0)
	\Text(122,38)[r]{\small $ R_t $}
	\LongArrowArc(23,60)(20,-110,-19.6)
	\Text(33,33)[b]{\small $ \alpha $}
	\LongArrowArc(200,0)(40,161.5,180)
	\Text(152,7)[l]{\small $ \beta $}
	\LongArrowArc(0,0)(20,0,67)
	\Text(25,13)[r]{\small $ \gamma $}
	\LongArrow(-20,0)(220,0)
	\Text(220,13)[t]{\small $ \bar{\rho} $}
	\LongArrow(0,-20)(0,80)
	\Text(-10,80)[l]{\small $ \bar{\eta} $}
\end{picture}
	\caption{\it Example of Unitarity Triangle. In this specific case, the sides of the triangle have lengths of order 1. The coordinates $ (\bar{\rho}, \bar{\eta}) $ determine the position of the apex of the triangle in the $ \bar{\rho} \, {\rm vs.} \, \bar{\eta} $ plane, indicated by the black blob, and therefore its shape, i.e. the angles $ \alpha, \beta, \gamma $.}\label{fig:UTBdBdbarNorm}
\end{figure}

It turns out that the three contributions relevant for the $ B_d \overline{B}_d $ system in Eq.~\eqref{eq:unitarityCKM} have roughly the same size, order $ \lambda^3 $, cf. Figure~\ref{fig:UTBdBdbar}. Its internal angles, which are invariant under phase redefinitions of the quark fields, can be easily checked to be given by

\begin{equation}
\alpha = {\rm arg} \left( - \frac{V_{td} V^*_{tb}}{V_{ud} V^*_{ub}} \right) \, , \quad \beta = {\rm arg} \left( - \frac{V_{cd} V^*_{cb}}{V_{td} V^*_{tb}} \right) \, , \quad \gamma = {\rm arg} \left( - \frac{V_{ud} V^*_{ub}}{V_{cd} V^*_{cb}} \right) \, .
\end{equation}
Once all the sizes have roughly the same length, we see that

\begin{equation}\label{eq:usefulTriangle}
1 + \frac{V_{u d} V^{*}_{u b}}{V_{c d} V^{*}_{c b}} + \frac{V_{t d} V^{*}_{t b}}{V_{c d} V^{*}_{c b}} = 0
\end{equation}
is a useful relation in graphical representations, as a triangle whose sides are:

\begin{equation}
1 \, , \quad  R_u = \vert V_{u d} V^{}_{u b} / V_{c d} V^{}_{c b} \vert \sim 1 \quad {\rm and} \quad R_t = \vert V_{t d} V^{}_{t b} / V_{c d} V^{}_{c b} \vert \sim 1 \ ,
\end{equation}
and the internal angles are $ \alpha, \beta $ and $ \gamma $, see Figure~\ref{fig:UTBdBdbarNorm}. For this same triangle, $ (\bar\rho , \bar\eta) $ defined as follows 

\begin{equation}\label{eq:rhobaretabarDef}
\bar\rho+i \bar\eta = - \frac{V^{}_{ud} V^{*}_{ub}}{V^{}_{cd} V^{*}_{cb}}
\end{equation}
gives the coordinate of the apex of the triangle equivalent to Eq.~\eqref{eq:usefulTriangle} in the $ \bar{\rho} \, {\rm vs.} \, \bar{\eta} $ plane, of which the side opposite to the apex is the basis and has length $ 1 $. Note that $ \bar{\rho}, \bar{\eta} $ differ from $ \rho, \eta $ by $ \mathcal{O} (\lambda^2) $ corrections, which can be seen from the relation

\begin{equation}
\rho+i \eta = \frac{\sqrt{1-A^2 \lambda ^4} (\bar\rho+i \bar\eta)}{\sqrt{1-\lambda ^2} \left(1-A^2 \lambda ^4 (\bar\rho+i\bar\eta)\right)} \, ,
\end{equation}
valid to all orders in $ \lambda $.

Since our goal is to show the success of the SM to describe the structure of the weak interactions in the quark sector through a UT fit, we are going to perform a global fit to determine precisely the shape of the triangle defined from Eq.~\eqref{eq:usefulTriangle}. Let us now see which classes of observables are going to be considered to perform this task. 

\subsection{Observables}


We collect in Table~\ref{tab:expinputs} the full set of inputs we use in the global fit: from the experimental side, we have gained much precision in the determination of the Unitarity Triangle in the last years, thanks to which it was possible to confirm the mechanism of Kobayashi and Maskawa \cite{Kobayashi:1973fv} for the origin of $ \mathcal{C} \mathcal{P} $ violation in $ K $ decays, as we will see. This was in particular due to the $ B- $factories Belle and BaBar (based on the decays of $ \Upsilon(4S) $ into $ b \bar{b} $ states) \cite{Bfactories}, where $ B $ related observables are collected. 

The individual inputs in Table~\ref{tab:expinputs} have very different impacts when constraining the Unitarity Triangle, so that the different categories of inputs deserve a dedicated discussion. The following scheme contains the basic information one needs in the global fit: 

\begin{eqnarray}\label{eq:schemeV}
V &=& \begin{pmatrix}
           & \mathbf{d} & \mathbf{s} & \mathbf{b} \\
\mathbf{u} & n \rightarrow p + e \nu & K \rightarrow \pi + \ell \nu & B \rightarrow \pi + \ell \nu \\
\mathbf{c} & D \rightarrow \pi + \ell \nu & D \rightarrow K + \ell \nu & B \rightarrow D + \ell \nu \\
\mathbf{t} & B^0 \bar{B}^0 & B^0_s \bar{B}^0_s & t \rightarrow b + W \\
\end{pmatrix} \\
&=& \, \begin{pmatrix}
           & \mathbf{d} & \mathbf{s} && \mathbf{b} \\
\mathbf{u} & 1 - \frac{\lambda^2}{2} & \lambda && A \lambda ^3 (\rho -i \eta) \\
\mathbf{c} & -\lambda & 1 - \frac{\lambda ^2}{2} && A \lambda ^2 \\
\mathbf{t} & A \lambda ^3 (1 -\rho-i \eta) & -A \lambda ^2 && 1 \\
\end{pmatrix} + \mathcal{O} (\lambda^4) \, , \nonumber
\end{eqnarray}
\noindent
where, as indicated in the first line, $ \vert V_{ud} \vert $ is mainly determined from nuclear transitions, while $ \vert V_{us} \vert, \vert V_{cd} \vert, \vert V_{cs} \vert, \vert V_{ub} \vert, \vert V_{cb} \vert $ come from leptonic and semileptonic decays, and $ \vert V_{td} \vert, \vert V_{ts} \vert $ come from $ B $ meson-mixing observables (mass differences). The knowledge of the (semi-)leptonic decay rates and mixing observables are limited by the theoretical uncertainties coming from hadronic parameters discussed in the next section. We further add that the element $ \vert V_{tb} \vert $ can be in principle probed by high-energy processes where a top is produced in association with a $ W $, but the resulting accuracy is not yet competitive with the global fit prediction coming from low-energy physics, see e.g. Ref.~\cite{Lacker:2012ek}. 

In the Wolfenstein parameterization, the parameters $ \lambda $ and $ A $ are basically fixed by the $ s \rightarrow u $ and $ b \rightarrow c $ semileptonic decays, as seen from the part of Eq.~\eqref{eq:schemeV}. The constraints on $ \bar{\rho}, \bar{\eta} $ come from different sources, among which we have semileptonic decays $ b \rightarrow u $ and $ b \rightarrow c $:

\begin{eqnarray}
&& b \rightarrow u, c \Rightarrow \quad\quad \vert V_{ub} \vert \simeq A \lambda^3 R_u \quad\quad {\rm and \; the \; ratio} \quad\quad \vert V_{ub} / V_{cb} \vert \simeq \lambda R_u \, , \nonumber\\
&& \qquad\qquad {\rm with} \quad\quad R_u = \sqrt{\bar{\rho}^2 + \bar{\eta}^2} \, ,
\end{eqnarray}
and $ B $ meson-mixing observables

\begin{eqnarray}
&& B^0_{d,s} \overline{B}^0_{d,s} \Rightarrow \quad\quad \vert V_{td} \vert \simeq A \lambda^3 R_t \quad\quad {\rm and \; the \; ratio} \quad\quad \vert V_{td} / V_{ts} \vert \simeq \lambda R_t \, , \nonumber\\
&& \qquad\qquad {\rm with} \quad\quad R_t = \sqrt{(1 - \bar{\rho})^2 + \bar{\eta}^2} \, ,
\end{eqnarray}
where ratios are considered in order to have a better control over uncertainties. Note that we have given the expressions of the last two classes of observables as functions of $ R_{u,t} $: since both are individually compatible with $ \bar{\eta} = 0 $ (cf. Figure~\ref{fig:globalFit}), they are referred to as $ \mathcal{C P}- $conserving observables.

The $ K $ meson-mixing also provides information on $ \bar{\rho}, \bar{\eta} $, of the form

\begin{equation}
| \varepsilon_K | \;\; \Rightarrow \;\; \eta (a_1 - \rho) = a_2 \, ,
\end{equation}
where $ a_{1,2} > 0 $ do not need to be given at the moment. We see then that the $ K \overline{K} $ system provides a constraint of a different sort compared to the $ B $ systems, since $ \eta = 0 $ is not allowed, i.e. its observation alone is a clear sign of $ \mathcal{C P} $ violation.


\begin{sidewaystable}
\renewcommand\arraystretch{1.2}
{\footnotesize \begin{tabular}{c|c|cccc|ccc}
CKM  & Process  & \multicolumn{4}{c|}{Observables}  & \multicolumn{3}{c}{Theoretical inputs}\\
\hline
$|V_{ud}|$ & $0^+\to 0^+$ $\beta$
                  & $|V_{ud}|_{\rm nucl}$&=& $0.97425\pm 0\pm 0.00022$
                  & \cite{TownerHardy} & \multicolumn{3}{c}{Nuclear matrix elements} \\
                    \hline
$|V_{us}|$ & $K\to\pi\ell\nu$ 
                  & $|V_{us}|_{\rm SL}f_+^{K\to\pi}(0)$&=& $ 0.2163\pm0.0005 $ & \cite{Agashe:2014kda}
                  & $f_+^{K\to\pi}(0)$&=& $0.9645\pm 0.0015\pm 0.0045$\\
                 &  $K\to e\nu_e$ 
                 & ${\cal B}(K\to e\nu_e)$&=&$(1.581\pm0.008)\cdot 10^{-5}$ & \cite{Agashe:2014kda}
                 &  $f_K$&=& $155.2\pm0.2\pm0.6 $ MeV \\
                &  $K\to \mu\nu_\mu$ 
                &  ${\cal B}(K\to \mu\nu_\mu)$&=& $0.6355 \pm 0.0011$
                & \cite{Agashe:2014kda}\\
                 &  $\tau \to K \nu_\tau$ 
                 & ${\cal B}(\tau \to K\nu_\tau)$&=&$(0.6955 \pm 0.0096)\cdot 10^{-2}$
                 & \cite{Agashe:2014kda}\\
                 \hline
$\frac{|V_{us}|}{|V_{ud}|}$                 &  $K\to \mu\nu/\pi\to\mu\nu$ & 
                 $\displaystyle \frac{{\cal B}(K\to \mu\nu_\mu)}{{\cal B}(\pi \to \mu\nu_\mu)}$
                         &=&$1.3365 \pm 0.0032$
                 & \cite{Agashe:2014kda} &
                 $f_K/f_\pi$&=&$1.1952 \pm 0.0007\pm0.0029$ 
                  \\
                 &  $\tau\to K\nu/\tau \to \pi\nu$ &   
                 $\displaystyle \frac{{\cal B}(\tau \to K\nu_\tau)}{{\cal B}(\tau \to \pi\nu_\tau)}$
                        &=& $(6.431 \pm 0.094)\cdot 10^{-2}$
                 & \cite{Agashe:2014kda} \\
                 \hline
$|V_{cd}|$   & $\nu N$ & $|V_{cd}|_{\rm not \; lattice}$ &=& $0.230\pm 0.011$ & \cite{Agashe:2014kda}\\
                   & $D\to \mu\nu $ & ${\cal B}(D\to \mu\nu)$ &=& $(3.74\pm0.17)\cdot 10^{-4}$ 
                   & \cite{Amhis:2014hma}
                   &$f_{D_s}/f_D$&=&$1.175 \pm 0.001\pm0.004$\\
                   & $D\to \pi\ell\nu $ & $|V_{cd}|f_+^{D\to \pi}(0)$ &=& $0.148 \pm 0.004$ 
                   & \cite{DtopiandK}
                   &$f_+^{D\to \pi}(0)$&=&$0.666\pm 0.020\pm 0.048 $\\
                   \hline
$|V_{cs}|$   &  $W\to c\bar{s}$ &   $|V_{cs}|_{\rm not \; lattice}$ &=& $0.94^{+0.32}_{-0.26}\pm 0.13$ & \cite{Agashe:2014kda}\\ 
                   & $D_s\to \tau\nu$ 
                   & ${\cal B}(D_s\to \tau\nu)$&=& $(5.55\pm0.24) \cdot 10^{-2}$ 
                   &  \cite{Amhis:2014hma} 
                   & $f_{D_s}$ &=& $248.2\pm 0.3 \pm 1.9$ MeV\\
                   & $D_s\to \mu\nu$ 
                   & ${\cal B}(D_s\to \mu\nu_\mu)$&=& $(5.57\pm0.24)\cdot 10^{-3}$ 
                   &  \cite{Amhis:2014hma}\\
                   & $D\to K\ell\nu $ & $|V_{cs}|f_+^{D\to K}(0)$ &=& $0.712 \pm 0.007$ 
                   & \cite{DtopiandK,Babar-DtoK}
                   &$f_+^{D\to K}(0)$&=&$ 0.747\pm0.011\pm0.034$\\
                   \hline
$|V_{ub}|$ & semileptonic $B$
                  & $|V_{ub}|_{\rm SL}$ &=& $(4.01 \pm 0.08 \pm 0.22)\cdot 10^{-3}$ 
                  &  \cite{Amhis:2014hma}&
                     \multicolumn{3}{c}{form factors, shape functions}\\
                  & $B\to \tau\nu$ 
                  & ${\cal B}(B\to\tau\nu)$ &=& $(1.08\pm0.21) \cdot 10^{-4}$ & \cite{btaunu,Amhis:2014hma}
                  &   $f_{B_s}/f_B$&=& $1.205\pm 0.003 \pm 0.006 $\\
                  \hline
$|V_{cb}|$ & semileptonic $B$
                 & $|V_{cb}|_{\rm SL}$ &=& $(41.00\pm 0.33 \pm 0.74)\cdot 10^{-3}$ &  \cite{Amhis:2014hma}
                 &  \multicolumn{3}{c}{form factors, OPE matrix elements}\\               
\hline
$|V_{ub}/V_{cb}|$ & semileptonic $\Lambda_b$
                 & $\frac{{\cal B}(\Lambda_p\to p\mu^-\bar\nu_\mu)_{q^2>15}}{{\cal B}(\Lambda_p\to \Lambda_c\mu^-\bar\nu_\mu)_{q^2>7}}$ &=& $(1.00\pm 0.09 ) \cdot 10^{-2}$ &  \cite{Aaij:2015bfa}
                 &  \multicolumn{3}{c}{$\frac{\zeta(\Lambda_p\to p\mu^-\bar\nu_\mu)_{q^2>15}}{\zeta(\Lambda_p\to \Lambda_c\mu^-\bar\nu_\mu)_{q^2>7}}=1.471\pm 0.096\pm 0.290$}
                 \\
\hline
$\alpha$ & $B\to\pi\pi$, $\rho\pi$, $\rho\rho$ 
                & \multicolumn{3}{c}{branching ratios, $CP$ asymmetries} & \cite{Amhis:2014hma} 
                & \multicolumn{3}{c}{isospin symmetry}\\
                \hline
$\beta$   & $B\to (c\bar{c}) K$ 
               & $\sin(2\beta)_{[c\bar{c}]}$ &=& $0.691 \pm 0.017$ 
               & \cite{Amhis:2014hma} & \multicolumn{3}{c}{subleading penguins neglected}\\
\hline
$\gamma$ & $B\to D^{(*)} K^{(*)}$ 
                 & \multicolumn{3}{c}{inputs for the 3 methods}
                 &  \cite{Amhis:2014hma}& \multicolumn{3}{c}{GGSZ, GLW, ADS methods} \\
                   \hline
$\phi_s$ & $B_s\to J/\psi (KK, \pi\pi)$ & $(\phi_s)_{b \rightarrow c \bar{c} s}$ &=& $-0.015\pm 0.035$
             & \cite{Amhis:2014hma}&
 \\
  \hline  
$V_{tq}^*V_{tq'}$       & $\Delta m_d$ 
                & $\Delta m_d$ &=& $0.510 \pm 0.003$ ps${}^{-1}$
                & \cite{Amhis:2014hma}
                &  $\hat{B}_{B_s}/\hat{B}_{B_d}$ &=& $1.023 \pm 0.013\pm0.014$\\ 
                 & $\Delta m_s$ & $\Delta m_s$ &=& $17.757\pm0.021$ ps${}^{-1}$ 
                 & \cite{Amhis:2014hma} 
                 & $\hat{B}_{B_s}$&=& $1.320\pm0.016\pm0.030$\\
                                 & $B_s\to \mu\mu$ & ${\cal B}(B_s\to\mu\mu)$ &=& $(2.8^{+0.7}_{-0.6})\cdot 10^{-9}$
                 & \cite{Bsmumu} & $f_{B_s}$ &=& $224.0\pm1.0\pm2.0 $ MeV \\
                 \hline
$V_{td}^*V_{ts}$ and  
      & $\varepsilon_K$ & $|\varepsilon_K|$ &=& $(2.228\pm0.011)\cdot 10^{-3}$
       & \cite{Agashe:2014kda} 
       &$\hat{B}_K$&=& $0.7615\pm0.0027\pm0.0137$\\
$V_{cd}^*V_{cs}$      &   &&&&& $\kappa_\varepsilon$&=& $0.940\pm0.013\pm0.023 $\\
\end{tabular}}
\caption{\it Constraints used for the global fit, and the main inputs involved (more information can be found in ref.~\cite{CKMfitterStandard}). When two errors are quoted, the first one is statistical, and the second one systematic. In the cases of $ \alpha $ and $ \gamma $ angles, many different channels or methods are used to extract their values, and the full resulting p-value profiles are used as the inputs for the global fit \cite{CKMfitterStandard}.\label{tab:expinputs}}
\end{sidewaystable}

Such as $ \varepsilon_K $, $ \mathcal{C P} $ asymmetries also indicate clear signs of $ \mathcal{C P} $ violation, by measuring the difference in rates of a process and its $ \mathcal{C P}- $conjugated one. The ratio \cite{Bigi:1981qs}

\begin{equation}\label{eq:CPaymmetriesBigi}
\frac{\Gamma (X \rightarrow f) - \Gamma (\bar{X} \rightarrow \bar{f})}{\Gamma (X \rightarrow f) + \Gamma (\bar{X} \rightarrow \bar{f})} \equiv \frac{\Gamma - \overline{\Gamma}}{\Gamma + \overline{\Gamma}} \, ,
\end{equation}
where $ \overline{\Gamma} $ denotes the rate of the $ \mathcal{C P}- $conjugated process of rate $ \Gamma $, defines schematically this class of observables. Taking the ratio has a clear advantage, because it permits to have a good control over hadronic uncertainties: since QCD effects are $ \mathcal{CP}- $conserving, they drop in Eq.~\eqref{eq:CPaymmetriesBigi}. As seen in Table~\ref{tab:expinputs}, such asymmetries directly measure the angles $ \alpha, \beta, \gamma $ through decays of $ B_d $ mesons: for instance, $ B_d \rightarrow J/\psi K $ gives a clean measure of $ \sin (2 \beta) $ -- see e.g. Refs.~\cite{Branco:1999fs,Grinstein:2015nya} for the related calculations.

Apart from the role played by $ \mathcal{C P}- $conserving and violating quantities, there is also a clear distinction between tree level (e.g. $ B \rightarrow X + \ell \nu $ decays, $ X = \pi, D $) and loop-level (e.g. $ B^0_q \bar{B}^0_q $ mixing, $ q= d, s $) dominated observables. While tree level dominated processes give a cleaner and safer extraction of SM parameters, the question is less straightforward for loop-level processes. Indeed, if New Physics effects are large enough to be observed, it is generally expected that relatively to tree level amplitudes they are much suppressed, but compared to loop-induced SM amplitudes they can have a similar size. 

Apart from the observables discussed in this section, other observables are not included because they do not have a sufficient accuracy yet, such as $ \mathcal{B} (B_d \rightarrow \mu \mu) $, or because the theoretical uncertainties deserve still some improvement or at least more discussion, e.g. $ \Delta \Gamma_s $ and $ \varepsilon'_K / \varepsilon_K $. Therefore, it would be premature to use them in a precision fit.


\subsection{Theoretical inputs}\label{sec:theoInputs}



From the elements of the SM we calculate the amplitudes for low-energy processes. The most efficient way to build a Lagrangian specially designed for low-energy observables is to build an effective description which keeps only the most relevant set of physical operators, neglecting others which are too much suppressed by the high-energy scales \cite{Wilson:1969zs}. Then, for low enough energies, i.e. for long enough wavelengths, QCD effects become non-perturbative. Since the characteristic wavelength of weak interactions ($ \sim 1/M_W $) is much smaller, everything happens in two steps: first we build an effective Lagrangian of weak interactions including short-range QCD corrections, and then this picture is completed at long distances by taking into account the hadronic environment, where the low-energy effective quantities can be extracted from experiment.

Nowadays, non-perturbative effects are most of the time computed numerically from first principles by a technique called Lattice QCD. It consists in the computation of correlation functions defined on a lattice of discrete points (in the Euclidean space). Physical quantities are then determined from the extrapolation of the results in the cases where the lattice inter-space goes to zero and its total size goes to infinite, apart from other possible limits related to the physical masses of the particles simulated. This effective description has been enormously developed in the last decades, thanks to the increasing computing capacities, and has acquired a great degree of sophistication \cite{LATTICE}.




Let us see a few examples where hadronic quantities must be known in order to predict an observable. In the leptonic decay $ \pi \rightarrow \ell \nu $, we need the value of the $ \pi $ \textit{decay constant}, called $ f_\pi $ and defined as

\begin{equation}
- p_\mu f_\pi = \langle 0 \vert (\bar{d} \gamma_\mu \gamma_5 u) \vert \pi (p) \rangle \, ,
\end{equation}
up to an arbitrary phase, correcting the coupling of the $ W $ boson to a pair of $ d $ and/or $ u $ in the hadronic environment. Similarly, when calculating the decay rate of the semi-leptonic decay $ K \rightarrow \pi + \ell \nu $, one faces the hadronic amplitude of $ K \rightarrow \pi $, whose non-perturbative effects are contained in the \textit{form factor} $ f^{K \rightarrow \pi} $, defined from


\begin{equation}
f^{K \rightarrow \pi}_+ (q^2) (p + p')_\mu + f^{K \rightarrow \pi}_- (q^2) (p - p')_\mu = \langle \pi (p') \vert (\bar{s} \gamma_\mu P_L u) \vert K (p) \rangle \, ,
\end{equation}
where $ q_\mu \equiv (p - p')_\mu $ gives a term proportional to the mass $ m_\ell $ ($ \sim 0 $ for $ \ell = e, \mu $) when the contraction with the leptonic current is taken. Both $ f_\pi $ and $ f^{K \rightarrow \pi}_+ $ must be determined from elsewhere in order to predict the SM values of the related processes.

Another class of non-perturbative parameters concerns $ \vert \Delta F \vert = 2 $ mixing processes, which require long-distance parameters called \textit{bag parameters} $ B_P $, $ P = K, B_d, B_s $. These are defined as the ratio of the $ \vert \Delta F \vert = 2 $ hadronic amplitude over its Vacuum Insertion Approximation (VIA) estimate (e.g. see the Appendix C of \cite{Branco:1999fs}). For definiteness, consider the $ K^0 \bar{K}^0 $ system


\begin{equation}
B_K \propto \frac{\langle \overline{K} \vert (\bar{s} \gamma^\mu P_L d) \, (\bar{s} \gamma_\mu P_L d) \vert K \rangle}{\langle \overline{K} \vert (\bar{s} \gamma^\mu P_L d) \vert 0 \rangle \, \langle 0 \vert (\bar{s} \gamma_\mu P_L d) \vert K \rangle} \, ,
\end{equation}
\noindent
where the vacuum is indicated by the state $ \vert 0 \rangle $. Note that $ \langle 0 \vert (\bar{s} \gamma_\mu P_L d) \vert K \rangle $ is proportional to the decay constant $ f_K $: we thus have

\begin{equation}
\frac{2}{3} m^2_K f^2_K B_K = \langle \overline{K} \vert (\bar{s} \gamma^\mu P_L d) \, (\bar{s} \gamma_\mu P_L d) \vert K \rangle \, ,
\end{equation}
\noindent
where $ m_K $ is the average of the eigenmasses of the $ K \overline{K} $ system. Note that $ B_K $ depends on the scale $ \mu_{had} $ where $ \langle \overline{K} \vert (\bar{s} \gamma^\mu P_L d) \, (\bar{s} \gamma_\mu P_L d) \vert K \rangle $ is determined: by factorizing out this dependence, one defines a scale independent quantity usually indicated by a hat, $ \hat{B}_K $.

The full set of theoretical inputs needed in our analysis is contained in the third column of Table~\ref{tab:expinputs}. A different class of inputs is necessary when computing meson-mixing observables, which corresponds to short-distance QCD corrections that can be computed by a perturbative expansion in $ \alpha_s $ (more on this subject will come in Chapters~\ref{ch:generalEFT} and \ref{ch:technicalEFT}). We have employed the following values, which collect these perturbative effects \cite{Brod:2011ty,Buras:1990fn,Buchalla:1995vs,Brod:2010mj}

\begin{eqnarray}\label{eq:differentEtasSM}
&& \eta_{cc} = 1.87 \pm 0 \pm 0.76 \, , \; \eta_{tt} = 0.5765 \pm 0 \pm0.0065 \, , \; \eta_{ct} = 0.497 \pm 0 \pm 0.047 \, , \nonumber\\
&& \eta_B = 0.5510 \pm 0 \pm 0.0022 \, ,
\end{eqnarray}
where the first uncertainty is statistical and the second theoretical. The values correspond respectively to short-distance QCD corrections to $ K \overline{K} $ mixing (first line) and $ B \overline{B} $ mixing (second line). For $ \eta_{cc} $, subjected to the largest uncertainty, an important fraction of the uncertainty comes from the poor convergence behaviour of the series: indeed, the shift NLO $ \rightarrow $ NNLO enhances the central value by $ \sim 30~\% $.

For completeness, the remaining parameters used in the fit include

\begin{equation}
\alpha_s (M_Z) = 0.1185 \pm 0 \pm 0.0006 \, , \\
\end{equation}
for the strong coupling \cite{Beringer:1900zz} (including information from the $ Z $ pole, coming from an electroweak precision fit \cite{Baak:2014ora}), and \cite{ATLAS:2014wva}

\begin{equation}
\bar{m}_t (\bar{m}_t) = 165.95 \pm 0.35 \pm 0.64 \; {\rm GeV} \, , \quad \bar{m}_c (\bar{m}_c) = 1.286 \pm 0.013 \pm 0.040  \; {\rm GeV} \, ,
\end{equation}
for the masses of the top- and charm-quarks running inside the loops of the meson-mixing amplitudes. Above, the input for the top-quark mass is determined from $ m^{pole}_{top} $ by using the following relation, valid around the value $ m^{pole}_{top} = 173.34 $~GeV:

\begin{equation}
\bar{m}_t (\bar{m}_t) \simeq 0.9626 \times m^{pole}_{top} - 0.90 \, {\rm GeV} \, ,
\end{equation}
which comes from a one-loop calculation described in Ref.~\cite{Chetyrkin}.

\subsection{\texttt{CKMfitter}}\label{sec:CKMfitterPhilo}

We combine the observables in a pure \textit{frequentist} approach based in a $ \chi^2 $ analysis. In the context of flavour physics, theoretical uncertainties, mainly introduced by the theoretical inputs from Section~\ref{sec:theoInputs}, deserve a special attention. Contrarily to EWPO, they are very much significant in the comparison between measurement and theoretical prediction. When considering this class of uncertainties, one quotes a range $ [-\Delta, \Delta] $, and we will admit that it contains the true value $ \delta $ of the theoretical uncertainty: if $ \delta $ was known, instead of quoting $ X \pm \sigma_{stat} \pm \Delta $, we would quote $ (X + \delta) \pm \sigma_{stat} $ (further discussion follows in Chapter~\ref{ch:theo}). 

\begin{figure}
	\begin{center}
	\includegraphics[scale=0.45]{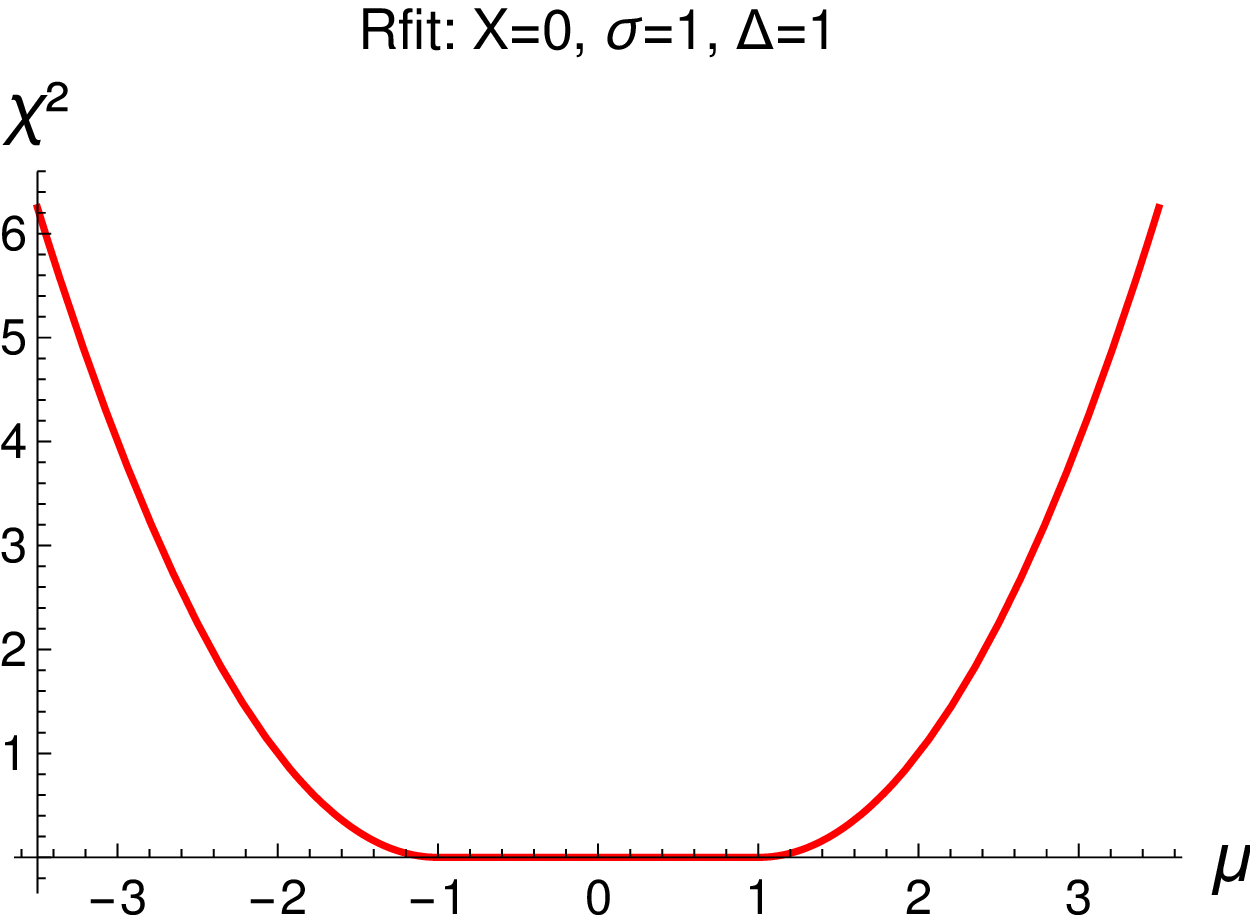}
	\hspace{1.5cm}
	\includegraphics[scale=0.45]{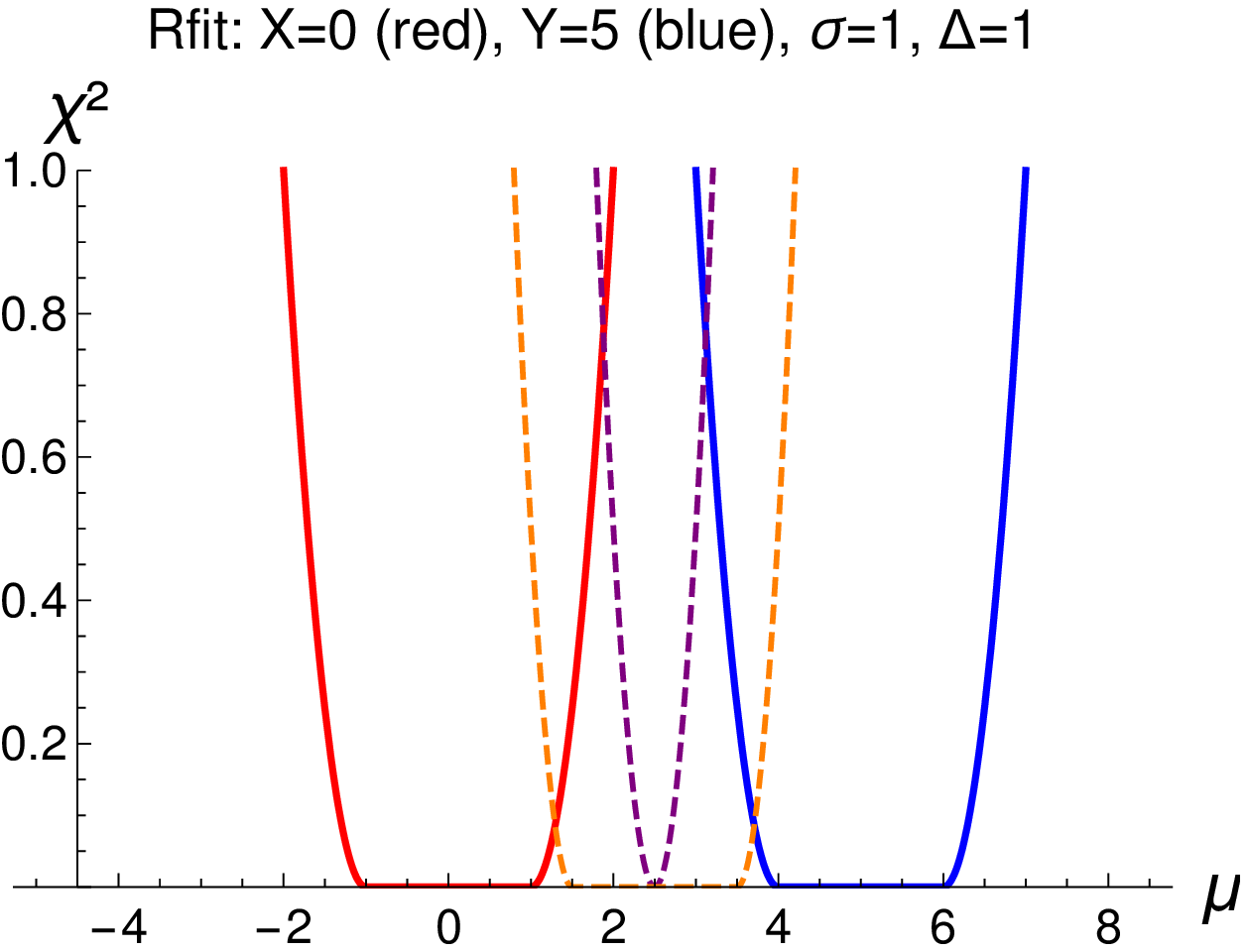}
	\end{center}
	\caption{\it (Left) $ \chi^2 $ for a random variable of variance $ \sigma^2 = 1 $ whose measure is $ X_{mes} = 0 \equiv X $. The theoretical uncertainty is modelled in the \textit{Rfit} scheme with range $ [-\Delta, \Delta] = [-1, 1] $: this is seen as a \textit{plateau} at $ \chi^2 = 0 $ for $ [-1, 1] $, while outside the plateau we have $ \chi^2 = (X_{mes} - \mu)^2 / \sigma^2 $, where $ \mu \equiv X_{theo} $ is the parameter we want to extract. (Right) Combination of two $ \chi^2 $ profiles, similar to the one described in the left figure, one for which $ X_{mes} = 0 \equiv X $ (red) and one for which $ X_{mes} = 5 \equiv Y $ (blue); in the \textit{naive Rfit} procedure, the combined profile results in the dashed, purple curbe, with no theoretical uncertainty, while in this case the \textit{educated Rfit} method would consider a resulting theoretical uncertainty equals to the smallest of the theoretical uncertainties (dashed orange).}\label{fig:chiSquaredRfit}
\end{figure}

We adopt a \textit{Rfit} scheme for dealing with theoretical uncertainties. In practice, this means that one can vary freely the true value of the theoretical uncertainty inside the quoted uncertainty $ \delta \in [- \Delta, \Delta] $ without any penalty from the $ \chi^2 $. This implies a \textit{plateau} for the preferred value of the parameter we would like to extract, see Figure~\ref{fig:chiSquaredRfit} (Left), a shape that will be further discussed in Chapter~\ref{ch:theo}. Subsequently, Confidence Level intervals are determined from varying the $ \chi^2 $ around the best fit point, and the \textit{goodness of the fit} is calculated by supposing that $ \chi^2_{min} $ is distributed as a $ \chi^2- $distribution.

Different extractions of the same quantity are combined previous to the fit (more on that in Chapter~\ref{ch:theo}), and when different inputs point towards a tension, such as in $ \vert V_{xb} \vert $, $ x = u, c $, for inclusive and exclusive extractions, we adopt a special procedure. To illustrate this point, Table~\ref{tab:educatedRfit} contains the inclusive and exclusive inputs for $ \vert V_{ub} \vert $ and $ \vert V_{cb} \vert $ previous to their average, which are substantially in tension given the size of their uncertainties. A procedure we call \textit{naive Rfit} average leads to no theoretical uncertainty if the data are barely compatible, i.e. the resulting average is not affected by the unknown true value of the theoretical uncertainties $ \delta $. This is illustrated in Figure~\ref{fig:chiSquaredRfit} (Right), where it is shown that one has a vanishing theoretical uncertainty: the resulting dotted-purple curve has not a shape presenting a flat bottom, a feature symptomatic of theoretical uncertainties in the Rfit scheme. Since it creates a very aggressive situation where two inputs in disagreement end up implying a very precise average, without any theoretical uncertainty related, we adopt a conservative procedure in which the final theoretical uncertainty is equal to the smallest of the individual theoretical uncertainties. This procedure is called \textit{educated Rfit}, and it is represented by the dotted-orange curve in Figure~\ref{fig:chiSquaredRfit} (Right). Both methods are compared in Table~\ref{tab:educatedRfit}.


\begin{table}[ht]
\begin{center}
	\begin{tabular}{cccc}
	$ \vert V_{ub} \vert \times 10^3 $ & central value & stat. uncertainty & theo. uncertainty \\
	\hline
	exclusive & 3.28 & $ \pm $ 0.15 & $ \pm $ 0.26 \\
	inclusive & 4.36 & $ \pm $ 0.18 & $ \pm $ 0.44 \\
	\textit{naive Rfit} & 3.70 & $ \pm $ 0.12 & $ \pm $ 0.00 \\
	\textit{educated Rfit} & 3.70 & $ \pm $ 0.11 & $ \pm $ 0.26 \\
	\\
	$ \vert V_{cb} \vert \times 10^3 $ & central value & stat. uncertainty & theo. uncertainty \\
	\hline
	exclusive & 38.99 & $ \pm $ 0.49 & $ \pm $ 1.17 \\
	inclusive & 42.42 & $ \pm $ 0.44 & $ \pm $ 0.74 \\
	\textit{naive Rfit} & 41.00 & $ \pm $ 0.33 & $ \pm $ 0.00 \\
	\textit{educated Rfit} & 41.00 & $ \pm $ 0.33 & $ \pm $ 0.74 \\
	\hline	
	\end{tabular}
\end{center}
\caption{\it Inclusive and exclusive inputs for $ \vert V_{ub} \vert $ and $ \vert V_{cb} \vert $, and their averages under two different procedures: \textit{naive Rfit} and \textit{educated Rfit}.}\label{tab:educatedRfit}
\end{table}


For form factors, bag parameters and decay constants, previous to the global fit we average over the extractions made by different groups (the individual references are found in \cite{CKMfitterStandard}). Moreover, if for a quantity we have many sources of systematic uncertainty $ \Delta_1, \ldots, \Delta_n $, they are treated at the same footing and summed linearly, $ \Delta_1 + \ldots + \Delta_n $. The resulting averages are seen in the last column of Table~\ref{tab:expinputs}.\footnote{Note that, although different numbers of dynamical flavours on the Lattice are simulated, $ N_f = 2, 2+1, 2+1+1 $, we suppose that they extract the same underlying quantity, and therefore they can all be used in the average, without preference for a particular $ N_f $.}

The procedure depicted in the last paragraphs is adopted by the \texttt{CKMfitter} Collaboration 	\cite{CKMfitterStandard}. Apart from its main statistical lines we have briefly described, found in more details in \cite{CKMfitterStandard}, at a more computational level \texttt{CKMfitter} is a modularized set of files of code where each class of observables is defined in terms of the underlying relevant parameters: these are the Wolfenstein parameters, for factors, bag parameters, etc. To make the computational work more efficient, the derivatives of the  observables are calculated symbolically in order to optimize the extremization procedure, necessary in the determination of the best fit point.

Having discussed the observables, the theoretical inputs and the statistical treatment, we now shift to the results of our analysis, obtained through the \texttt{CKMfitter} framework.


\subsection{Results}

\begin{figure}
	\centering
	\includegraphics[scale=0.5]{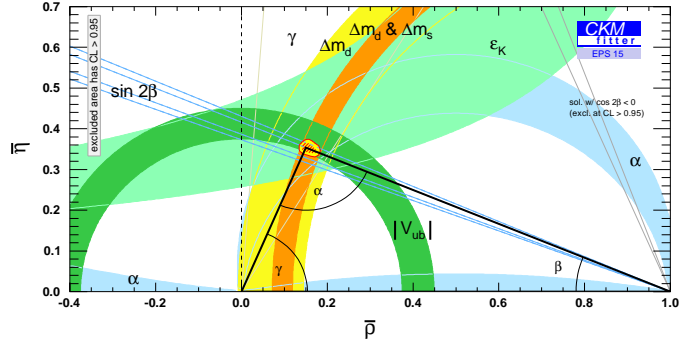}
	\caption{\it Plot in the $ \bar{\rho} \, {\rm vs.} \, \bar{\eta} $ plane showing the constraints of the individual inputs (not shown are, for example, constraints from $ V_{us} $ and $ V_{cs} $, which are relevant only for $ A, \lambda $, cf. Eqs.~\eqref{eq:VusWolfenstein} and \eqref{eq:VcsWolfenstein}). The set of observables point towards an apex at $ \sim (0.15, 0.35) $, with $ 68~\% $ CL ($ 95~\% $ CL) represented in hashed red (yellow with red contour). (Note that there is a discrete ambiguity for $ \beta $ coming from the $ \sin 2 \beta $ constraint, indicated by the gray region for which $ \cos 2 \beta < 0 $. However, this ambiguity is excluded at $ 95~\% $ CL.)}\label{fig:globalFit}
\end{figure}

The results of the global fit for $ \bar{\rho}, \bar{\eta} $ are summarized in Figure~\ref{fig:globalFit}, where we represent the Unitarity Triangle for the $ B_d \overline{B}_d $ system -- corresponding to the orthogonal relation between the first and the third columns of the CKM matrix (divided by $ V_{c d} V^{*}_{c b} $), as discussed in Section~\ref{sec:CKMmatrixBasics}. Of course, other Unitarity Triangles could also be represented, cf. Ref.~\cite{CKMfitterStandard}. In the same figure, we indicate the individual $ 68~\% $ CL (Confidence Levels): in order to determine these confidence intervals in the $ \bar{\rho} \, {\rm vs}. \, \bar{\eta} $ plane, theoretical inputs are obviously employed, apart from the necessary experimental information.

The outcome of the fit points towards a unique region in the $ (\bar{\rho}, \bar{\eta}) $ plane. For the goodness of the fit we find a p-value equals to $ 66~\% $, or $ 0.4 \; \sigma $ in units of sigma: the good agreement implies that the extraction of the fundamental parameters of the SM is meaningful and we have

\begin{eqnarray}\label{eq:resultsCKMparams}
A= 0.8227^{+0.0066}_{-0.0136}\,, &\qquad&
\lambda=0.22543^{+0.00042}_{-0.00031}\,, \label{eq:resultsVCKM}\\
\bar\rho = 0.151^{+0.012}_{-0.006}\,, &\qquad&
\bar\eta = 0.354^{+0.007}_{-0.008}\,. \nonumber
\end{eqnarray}

\begin{figure}
	\centering
	\includegraphics[scale=0.45]{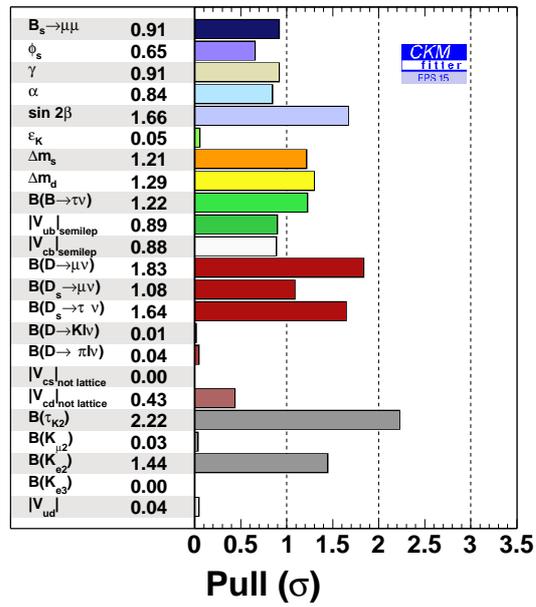}
	\caption{\it Pulls for the individual observables in Table~\ref{tab:expinputs} in units of $ \sigma $. Since correlations are present, the number of observables which have a pull larger than $ n \times \sigma $ is not a meaningful information.}\label{fig:pullsGlobalFit}
\end{figure}


Individually, each observable provides a test of the validity of the SM, as indicated by the pulls defined in Eq.~\eqref{eq:flavourPull}, which are normally distributed with mean zero and a dispersion of one.\footnote{The inputs $ \mathcal{B} (K^- \rightarrow \mu^- \bar{\nu}_\mu) / \mathcal{B} (\pi^- \rightarrow \mu^- \bar{\nu}_\mu) $ and $ \mathcal{B} (\tau^- \rightarrow K^- \bar{\nu}_\tau) / \mathcal{B} (\tau^- \rightarrow \pi^- \bar{\nu}_\tau) $ are not shown because they are correlated with $ \mathcal{B} (K^- \rightarrow \mu^- \bar{\nu}_\mu) $ and $ \mathcal{B} (\tau^- \rightarrow K^- \bar{\nu}_\tau) $.} The values of the pulls seen in Figure~\ref{fig:pullsGlobalFit} tell us that each single observable has a suitable SM prediction compared to the experimental value. Since the observables have correlated fits, which is for instance the case for $ \sin 2 \beta $ and $ \mathcal{B} (B \rightarrow \tau \nu) $  \cite{CKMfitterStandard}, the distribution of the pulls is not normal. Note that the presence of a \textit{plateau} in the \textit{Rfit} model for theoretical uncertainties may lead to a vanishing pull for some quantities even in cases where the predicted and the observed values are not identical.



\begin{figure}
	\hspace{-1.2cm}\includegraphics[scale=0.42]{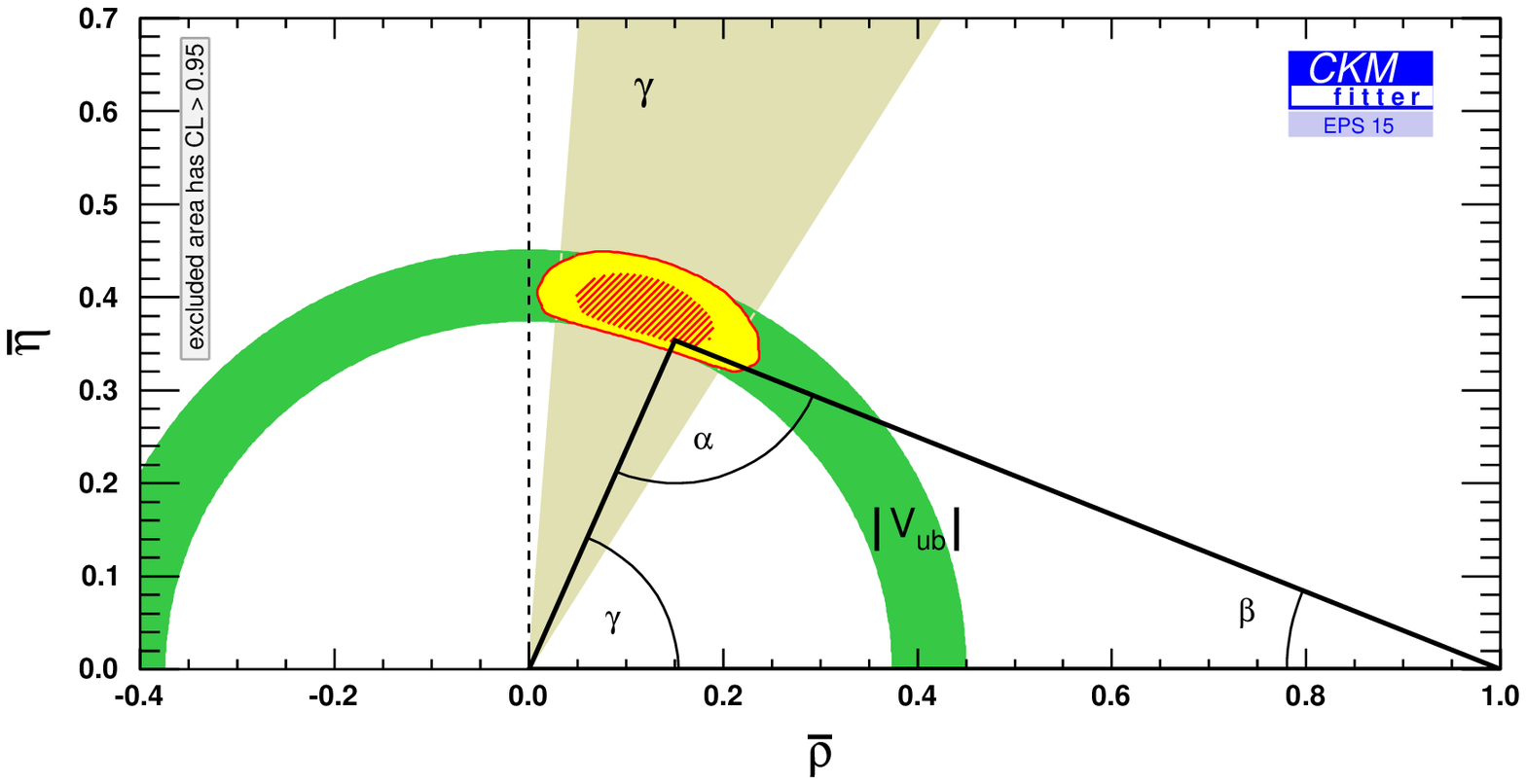}
	\includegraphics[scale=0.42]{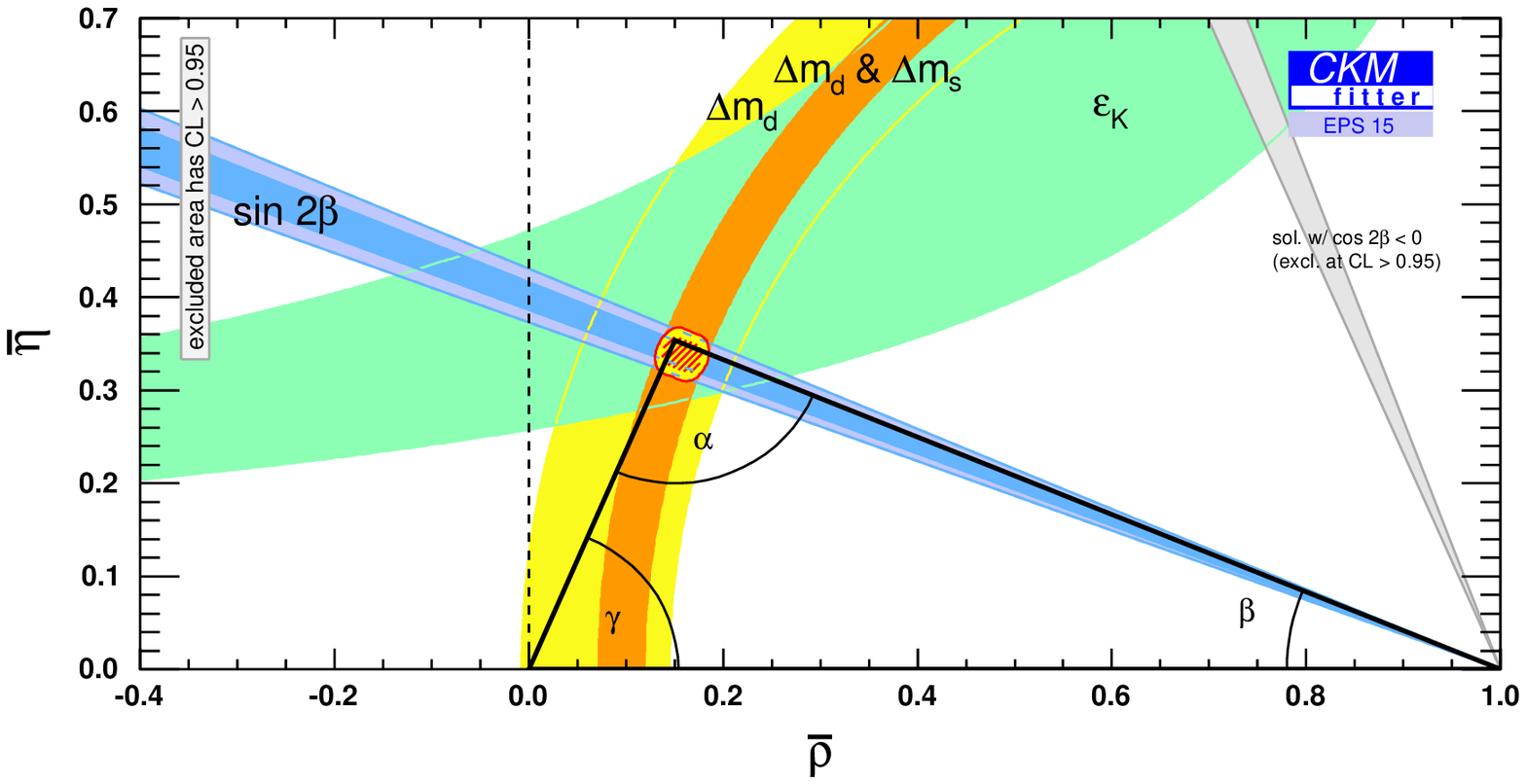}
	\caption{\it (Left) combination of observables dominated by tree level amplitudes (only $ \gamma (DK) $ is used \cite{CKMfitterStandard}), (Right) combination of processes dominated by loop-induced amplitudes. $ 68~\% $ confidence level intervals ($ 95~\% $ CL) are represented in dashed red (yellow with red contour). The apex of the unitarity triangle is determined from the global fit.}\label{fig:individualTreeLoop}
\end{figure}

\begin{figure}
	\hspace{-1.2cm}\includegraphics[scale=0.42]{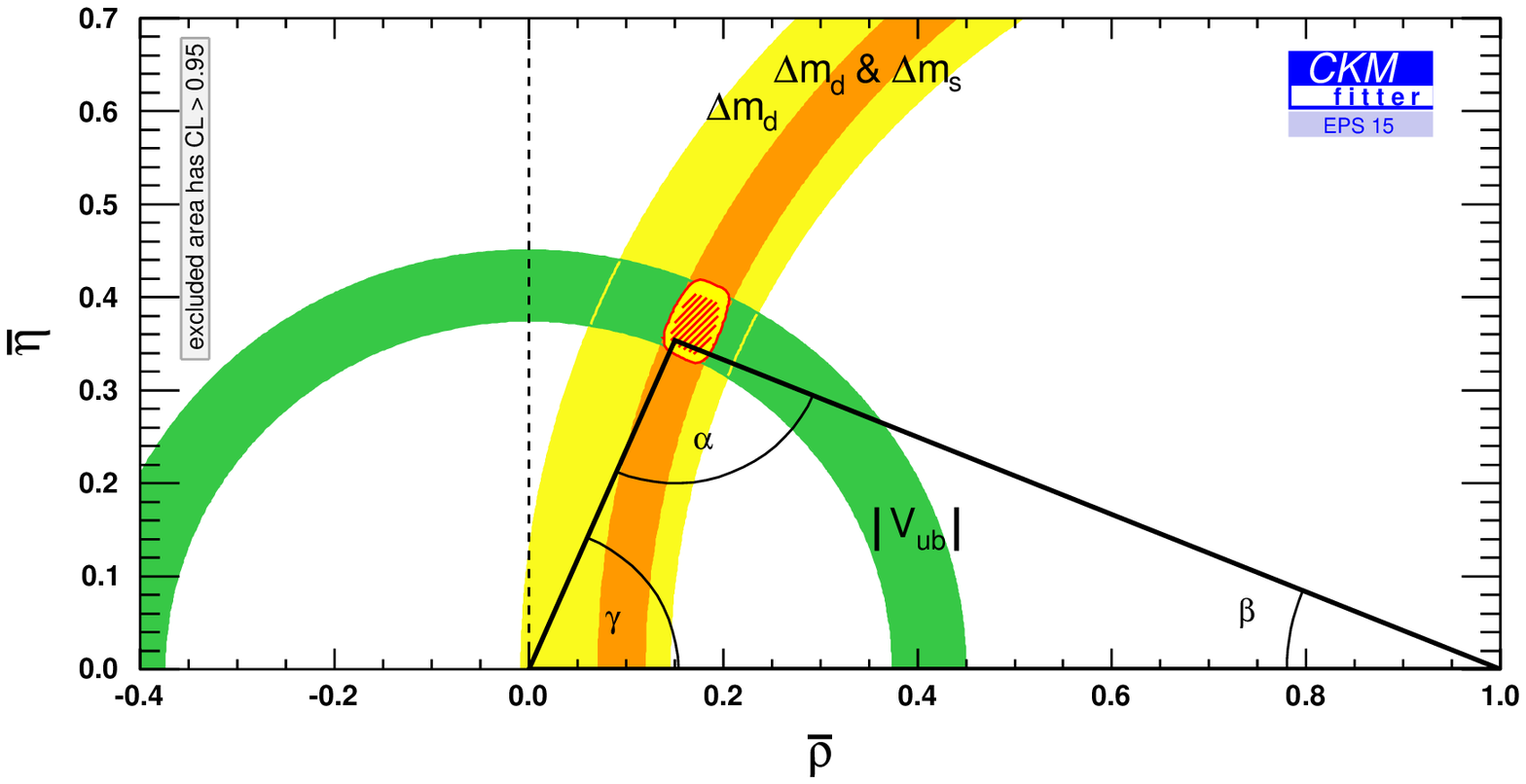}
	\includegraphics[scale=0.42]{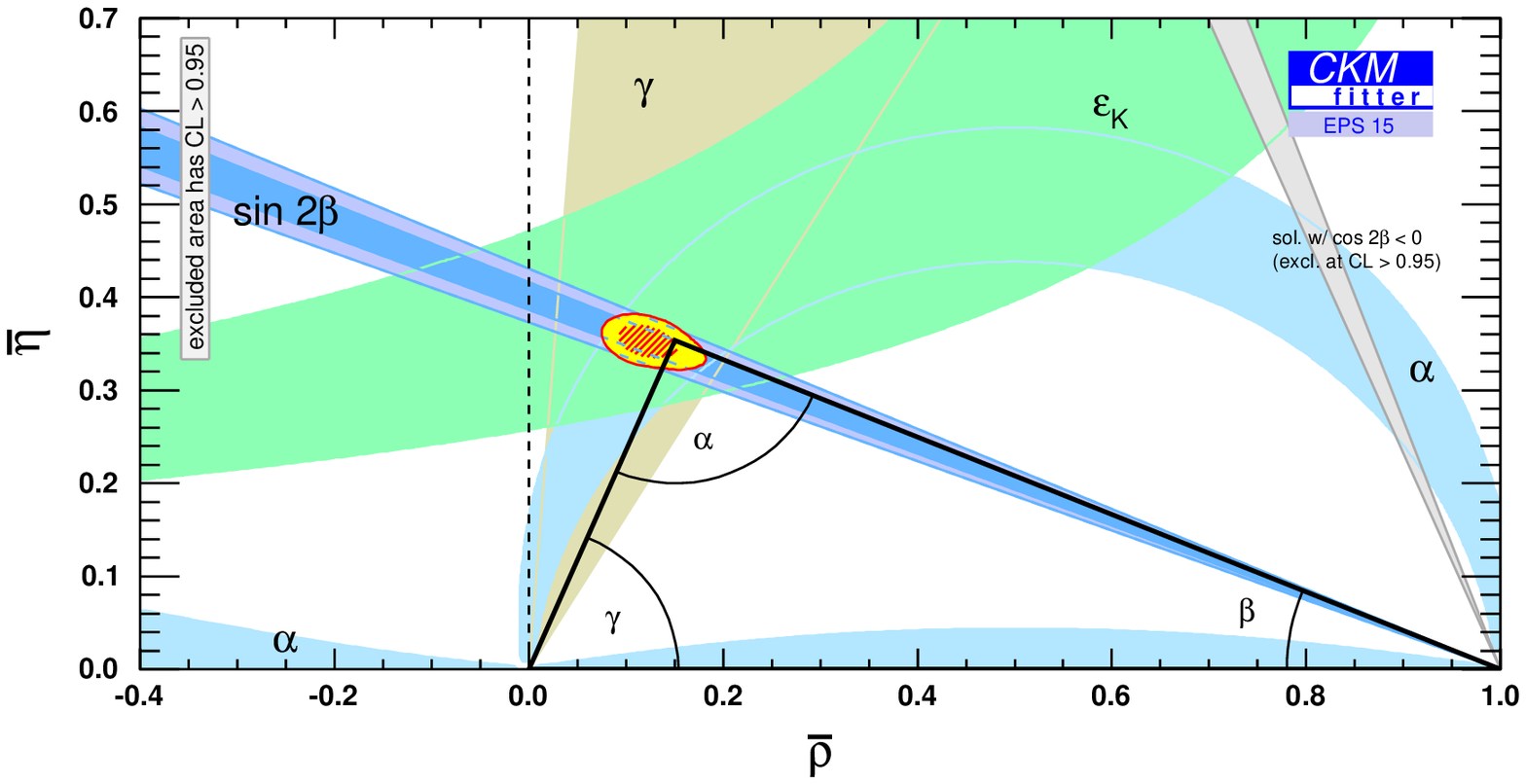}
	\caption{\it (Left) fit of observables that can be individually explained by $ \bar{\eta} = 0 $ (called $ \mathcal{C P}- $conserving), (Right) combination of processes that individually exclude $ \bar{\eta} = 0 $ (except for the ambiguity in the $ \alpha $ constraint). $ 68~\% $ confidence level intervals ($ 95~\% $ CL) are represented in dashed red (yellow with red contour). The apex of the unitarity triangle is determined from the global fit.}\label{fig:individualCPconsviol}
\end{figure}

In Figure~\ref{fig:individualTreeLoop} we show the outcome of the fit when only tree level or loop-induced processes are considered. Both fits lead to values of $ \bar{\rho}, \bar{\eta} $ in agreement with the global extraction of $ \bar{\rho}, \bar{\eta} $ shown in Figure~\ref{fig:globalFit}. We also consider a plot, Figure~\ref{fig:individualCPconsviol}, containing only $ \mathcal{C P}- $conserving quantities, i.e. observables which individually do not exclude a vanishing $ \mathcal{C P}- $violating phase, or equivalently $ \bar{\eta} = 0 $. Its outcome, and the one from a global fit considering only $ \mathcal{C P}- $violating observables, also agrees with the global fit combining all classes of observables at once.


\section{Conclusions and what comes next}


The SM succeeds in explaining a wide variety of classes of observables: in the context of the underlying gauge structure of the model, the EWPO are of particular interest, and have shown a great predictivity for the Higgs mass and top-quark mass before they were discovered or determined with accuracy \cite{Baak:2012kk}. As we have seen, there are some unexplained tensions that we have found in Section~\eqref{sec:EWtestsSM}, such as $ A_{FB} (b, \tau) $, $ \mathcal{A}^{SLD}_e $ and $ \sigma_{had} $, which have the largest pulls.

We have also seen over this chapter the success of the SM in describing the set of flavour observables shown in Table~\ref{tab:expinputs}. In particular, this success implies that the mechanism of $ \mathcal{C P} $ violation in the SM gives an accurate picture of nature, and we were able to extract very accurate values for the Wolfenstein parameters describing the CKM matrix.

However, this is not the full story: extensions of the SM are typically evoked at high energy scales to explain poorly understood features of the SM, or some tensions when comparing predictions and measurements. The great success of the SM to describe many observables helps us to test the very existence of these extensions and to probe their structure.


Starting from the next chapter, we are going to analyse a particular extension of the Standard Model which introduces weak charged right-handed currents, to be contrasted with the pure weak charged left-handed currents of the SM. We now discuss this extension of the SM, called \textit{Left-Right Model}. 

\chapter{Left-Right Models}\label{ch:LRM}












Parity violation is an experimental fact used in the formulation of the SM, where left- and right-handed fields have different gauge structures. Therefore, the SM incorporates this feature rather than explains its origin. One possibility for a better understanding of why one observes parity violation is to assume it at high enough energies, and then assume it is spontaneously broken when one goes down in energy, leading to parity violating phenomena.

To accomplish such a mechanism we introduce a second $ SU(2) $ local symmetry under which left-handed fields transform as singlets and right-handed fields transform as doublets \cite{Pati:1974yy,Mohapatra:1974hk,Senjanovic:1975rk,Senjanovic:1978ev}:

\begin{equation}\label{eq:theLRM}
SU(3)_c \times SU(2)_L \times SU(2)_R \times U(1)_{\tilde{Y}} \, ,
\end{equation}
where $ \tilde{Y} $ is a hypercharge in principle different from the one of the SM. This therefore adds up a new term in the Yang-Mills Lagrangian compared to the SM case

\begin{equation}
\mathcal{L}_{gauge} = - \frac{1}{4} G^{\mu \nu}_a G_{a \mu \nu} - \frac{1}{4} L^{\mu \nu}_a L_{a \mu \nu} - \frac{1}{4} R^{\mu \nu}_a R_{a \mu \nu} - \frac{1}{4} F^{\mu \nu}_{\tilde{Y}} F_{\tilde{Y} \mu \nu} \, , 
\end{equation}
replacing the notation of the field tensor $ F $ of the SM by $ L $ and adding up the Lagrangian density that corresponds to the new local symmetry $ SU(2)_R $. This is the basic starting point of the class of models called \textit{Left-Right (LR) Model}, and another feature we can point out is the existence of new gauge bosons:

\begin{equation}
W' \, \quad {\rm and} \, \quad Z' \, ,
\end{equation}
which acquire masses as a result of the symmetry breaking pattern, when the extended gauge group of the LR Models breaks down into the local symmetries of the SM:

\begin{equation}
SU(3)_c \times SU(2)_L \times SU(2)_R \times U(1)_{\tilde{Y}} \rightarrow SU(3)_c \times SU(2)_L \times U(1)_{Y} \, ,
\end{equation}
thus breaking the parity symmetric structure spontaneously. We are going to discuss at length in Section~\ref{sec:EWSBSMLRM} this pattern and the scalar field content necessary to realize it.\footnote{Of course, at each step of the symmetry breaking, the Lagrangian built out of renormalization and symmetry arguments is supplemented by higher-dimensional operators which are induced by the previous (itself effective or UV complete) Lagrangian, after that some of its degrees of freedom are integrated out.} 

Apart from the issue of parity, the SM raises some other intriguing questions that we could attempt to solve by restoring parity

\begin{itemize}
	\item[(a)] in the way the SM is introduced, the hypercharges are in principle completely arbitrary; they are then fixed in order to accommodate the electric charges, 
	\item[(b)] the strong $ \mathcal{C P}- $violating phase, if it exists at all, is ``unexpectedly'' small,
	\item[(c)] in the SM, neutrinos are massless, and there is no mechanism for the description of the observed neutrino oscillations.
\end{itemize}
\noindent
Possible answers or ways to deal with these problems in the LR Models are: (a) the hypercharges in the LR Model are naturally interpreted; first the electric charge $ Q $ is given by

\begin{equation}
Q = T^L_3 + T^R_3 + \frac{\tilde{Y}}{2} \, ,
\end{equation}
and then $ \tilde{Y} $ results being equal to $ B-L $, where $ B $ is the baryonic number and $ L $ is the leptonic number. (b) The strong $ \mathcal{C P} $ problem can be naturally investigated in the LR Model framework, since the $ \mathcal{C P}- $violating term $ - \frac{1}{4} \tilde{G}_{\mu \nu} G^{\mu \nu} $ is also parity-violating \cite{Babu:1989rb,Barr:1991qx,Lavoura:1996iu}. (c) In LR Models, right-handed neutrinos must be introduced, thus leading to mechanisms for mass generation.

At a practical level, many different ways of implementing a Left-Right Model have been investigated in the literature: a possible avenue is to consider extra fermions \cite{alternatveLRM}, which may serve as candidates for Dark Matter \cite{LRDM}; otherwise, in the effort to solve the naturalness problem (the hierarchy between the EW scale and whatever comes beyond), LR supersymmetric models have been considered \cite{SUSYLR,Mohapatra:1986uf}; different scalar sectors, with two bi-doublets, with triplets and/or doublets, etc. have been discussed in order to realize a series of features, see e.g. \cite{Mohapatra:1977mj,Chang:1983fu,Branco:1985ng}. All of these realizations of LR Models have their own motivations. Here, we would like to address a particular realization of Left-Right Models, called \textit{doublet scenario}, that will be described in the next sections. 

\section{Matter content}

Before explaining the spontaneous symmetry breaking in LR Models, let us introduce its matter content, which is a generic feature found in many of its realizations. As we have already stated, right-handed fermions are put into doublets, and right-handed neutrinos are introduced:

\begin{center}
{\renewcommand{\arraystretch}{1.5}
\begin{tabular}{ll}
left-handed quarks: & $ Q_L = \begin{pmatrix}
u_L \\ d_L
\end{pmatrix} = (\mathbf{3},\mathbf{2},\mathbf{1},1/3) $ , \\
right-handed quarks: & $ Q_R = \begin{pmatrix}
u_R \\ d_R
\end{pmatrix} = (\mathbf{3},\mathbf{1},\mathbf{2},1/3) $ , \\
left-handed leptons: & $ L_L = \begin{pmatrix}
\nu_L \\ \ell_L
\end{pmatrix} = (\mathbf{1},\mathbf{2},\mathbf{1},-1) $ , \\
right-handed leptons: & $ L_R = \begin{pmatrix}
\nu_R \\ \ell_R
\end{pmatrix} = (\mathbf{1},\mathbf{1},\mathbf{2},-1) $ , \\
\end{tabular}}
\end{center}
\noindent
showing an elegant and self-speaking symmetry between left- and right-handed fields.

The part of the full Lagrangian describing the interactions with the fields $ f $ shown above is

\begin{equation}\label{eq:LmatterNotFreeLRM}
\mathcal{L}_{matter} = \bar{f} i \gamma^\mu D_\mu f \, ,
\end{equation}
where the covariant derivative $ D $ includes the new gauge symmetries introduced by the LR Models

\begin{equation}\label{eq:fullDerivativeCovariantLRM}
D^\mu = \partial^\mu - i \left( g_s A^{a \mu} T^s_{a} + g_L W^{a \mu}_L T^L_{a} + g_R W^{a \mu}_R T^R_{a} + g_{B-L} B^\mu \frac{B-L}{2} \right) \, ,
\end{equation}
where $ g_s $, $ g_L $, $ g_R $ and $ g_{B-L} $ are the gauge couplings.

\section{Spontaneous symmetry breaking in the LR Model}\label{sec:EWSBSMLRM}

The LR Models must reproduce the observations we have made so far, where weak right-handed currents are suppressed compared to weak left-handed ones, and where parity $ \mathcal{P} $ and charge-conjugation $ \mathcal{C} $ are both violated over the energy scales we have had access so far. This set of characteristics are achieved in the same way as in the SM, where $ W, Z- $mediated interactions are weak due to the large masses of these gauge bosons (and a relatively small gauge coupling strength), generated through the BEH mechanism. 

In the LR Model case, a way to implement the spontaneous symmetry breaking (SSB) pattern of Eq.~\eqref{eq:theLRM} is to have a doublet under $ SU(2)_R $ whose vacuum expectation value leads to the SSB

\begin{equation}
SU(2)_R \times U(1)_{B-L} \rightarrow U(1)_Y \, .
\end{equation}
Therefore, we introduce the $ SU(2)_R $ doublet

\begin{eqnarray}
\chi_{R} = \begin{pmatrix} \chi^{+}_{R} \\ \chi^{0}_{R} \end{pmatrix} \, ,
\end{eqnarray}
which breaks $ SU(2)_R \times U(1)_{B-L} $ when $ \chi^{0}_{R} $ develops the VEV $ \kappa_R / \sqrt{2} $ as follows

\begin{equation}\label{eq:VEVsDoubletR}
\chi^{0}_{R} = (\chi^{0r}_{R} + i \chi^{0i}_{R} + \kappa_R) / \sqrt{2} \, ,
\end{equation}
where $ \chi^{0r}_{R} $ and $ \chi^{0i}_{R} $ are two distinct real fields of null VEV. The VEV $ \kappa_R $ is expectedly larger than the one responsible for the SSB of the EW group

\begin{equation}
SU(2)_L \times U(1)_Y \rightarrow U(1)_{EM} \, ,
\end{equation}
as argued in the first paragraph.

To further discuss the Brout-Englert-Higgs mechanism in LR Models, we first shift to the mass generation mechanism for fermions. Of course, the primary interest is to implement the BEH mechanism, but out of this discussion we will have picked up the scalar field that has the good quantum numbers for both phenomena, BEH and the mass generation of fermions. In the SM, masses come from the coupling of a scalar doublet to fermionic singlets and fermionic doublets, which introduces squared $ n_g- $dimensional matrices called Yukawa matrices, where $ n_g $ is the number of generations (3 in the SM and in the LR Model). In LR Models, a Yukawa term invariant under $ SU(3)_c \times SU(2)_L \times SU(2)_R \times U(1)_{B-L} $ requires a scalar bi-doublet, $ \Phi = (\mathbf{1},\mathbf{2},\mathbf{2},0) $, which transforms like\footnote{Note that, due to the extra $ SU(2)_R $ symmetry, the bi-doublet cannot be mapped onto two doublets, and therefore we cannot trivially think about the LR Model as a two-Higgs doublet model (THDM) extension of the SM. For further discussion on the parallel between LR Model and THDM, see Ref.~\cite{Olness:1985bg} in the context of flavour physics.}

\begin{equation}
\Phi \rightarrow U_L \Phi U_R^\dagger \, .
\end{equation}
The scalar bi-doublet has the expression

\begin{eqnarray}
\Phi = \begin{pmatrix} \varphi^{0}_{1} & \varphi^{+}_{2} \\ \varphi^{-}_{1} & \varphi^{0}_{2} \end{pmatrix} \, ,
\end{eqnarray}
and from the same degrees of freedom of $ \Phi $ one defines

\begin{equation}
\tilde{\Phi} = \tau_{2} {\Phi^{\dag}}^{T} \tau_{2} = \begin{pmatrix} \varphi^{0 \dag}_{2} & -\varphi^{+}_{1} \\ -\varphi^{-}_{2} & \varphi^{0 \dag}_{1} \end{pmatrix} \, ,
\end{equation}
which transforms in the same way as $ \Phi $, namely $ \tilde{\Phi} \rightarrow U_L \tilde{\Phi} U_R^\dagger $. The fields $ \varphi^0_{1,2} $ acquire the vacuum expectation values

\begin{eqnarray}
\varphi^{0}_{1} &=& (\varphi^{0r}_{1} + i \varphi^{0i}_{1} + \kappa_{1}) / \sqrt{2} , \label{eq:VEVsBiDoublet}\\
\varphi^{0}_{2} &=& (\varphi^{0r}_{2} + i \varphi^{0i}_{2} + \kappa_{2}) \operatorname{e}^{i \alpha} / \sqrt{2} \, , \nonumber
\end{eqnarray}
and $ \kappa_{1,2} $ trigger EWSB and generate the masses of the fermions. Since the bi-doublet is also charged under $ SU(2)_R $, it corrects slightly the picture of the first symmetry breaking, given at leading order by $ \kappa_R $, and corrected by $ \kappa_{1,2} $.

So far, we have introduced a doublet under $ SU(2)_R $ and a bi-doublet under $ SU(2)_L \times SU(2)_R $, and we have seen why they are required: in order to break the LR Model gauge group down to the SM one, to implement the SSB in the SM and to generate masses for fermions. However, our final goal is to build a model which is invariant under parity, i.e. to introduce the required degrees of freedom to define a $ \mathcal{P} $ symmetric model. Therefore, we introduce a doublet under $ SU(2)_L $, $ \chi_{L} = (\mathbf{1},\mathbf{2},\mathbf{1},1) $, i.e.

\begin{eqnarray}
\chi_{L} = \begin{pmatrix} \chi^{+}_{L} \\ \chi^{0}_{L} \end{pmatrix} \, .
\end{eqnarray}
In full generality, it acquires a non-vanishing VEV\footnote{Here, the phases of $ \varphi^{0}_{1} $ and $ \chi^{0}_{R} $ have been rotated away, and $ \langle \varphi^{0}_2, \chi^{0}_L \rangle $ are the only complex VEVs.}

\begin{equation}\label{eq:VEVsDoubletL}
\chi^{0}_{L} = (\chi^{0r}_{L} + i \chi^{0i}_{L} + \kappa_L) \operatorname{e}^{i \theta_{L}} / \sqrt{2} \, ,
\end{equation}
which corrects both the first ($ SU(2)_R \times U(1)_{B-L} \rightarrow U(1)_Y $, since $ \chi_L $ is charged under $ U(1)_{B-L} $) and the second symmetry breakings. Concerning the latter, it is a combination of $ \kappa_{1,2,L} $, more precisely the combination

\begin{equation}
\sqrt{\kappa^2_1 + \kappa^2_2 + \kappa^2_L} \simeq 246 \operatorname{GeV}
\end{equation}
that characterizes the energy scale of the EWSB. It will be useful in the following to characterize the SSB in terms of the parameters that follow

\begin{equation}\label{eq:parametersDefinedVEVs}
\epsilon \equiv \frac{\kappa_1}{\kappa_R} \sqrt{1 + r^2 + w^2} \, , \quad r \equiv \frac{\kappa_2}{\kappa_1} \, , \quad w \equiv \frac{\kappa_L}{\kappa_1} \, .
\end{equation}
Together with the angles $ \theta $ and $ \phi $ introduced below, they reflect the way in which the gauge symmetries are spontaneously broken: in particular, due to the expected hierarchy of SSB scales, $ \epsilon $ is a small parameter, and we are generally going to keep only first order corrections in $ \epsilon $.

Summarizing the previous discussion, we have the following symmetry breaking pattern

\begin{center}
	\begin{tabular}{lcc}
	&&$ SU(3)_c \times SU(2)_L \times SU(2)_R \times U(1)_{Y} $ \\
	\textbf{Stage 1:} && $ \downarrow $ \\ 
	&&$ SU(3)_c \times SU(2)_L \times U(1)_Y $ \\
	\textbf{Stage 2:} && $ \downarrow $ \\ 
	&&$ SU(3)_c \times U(1)_{EM} $ \\
	\end{tabular} .
\end{center}

\noindent
At both stages, we have $ SU(2) \times U(1) \rightarrow U(1) $: for the first stage, there is a ``weak'' angle $ \phi $, analogous to $ \theta $ from Chapter~\ref{ch:SM}, which describes the direction in which the first breaking occurs, see Table~\ref{tab:gaugeCouplingsLRM}.


\begin{table}[t]
\centering
{\def\arraystretch{2.0}
\begin{tabular}{cccc}
\hline
first breaking: & $ \tan \phi = t_\phi \equiv \frac{g_{B-L}}{g_R} $ & $ \frac{1}{g^2_Y} \equiv \frac{1}{g^2_R} + \frac{1}{g^2_{B-L}} $ \\
\hline
second breaking: & $ \tan \theta = t_{\theta} \equiv \frac{g_Y}{g_L} $ & $ \frac{1}{e^2} \equiv \frac{1}{g^2_L} + \frac{1}{g^2_R} + \frac{1}{g^2_{B-L}} $ \\
\hline
\hline
$ g_L = \frac{e}{s_{\theta}} $ & $ g_R = \frac{e}{c_{\theta} s_\phi} $ & $ g_{B-L} = \frac{e}{c_{\theta} c_\phi} $ \\
\hline
\end{tabular}
}
\caption{\it Couplings and relevant angles of the symmetry breaking pattern.}\label{tab:gaugeCouplingsLRM}
\end{table}

In the BEH mechanism, when generating the masses for the gauge bosons $ W, Z, W', Z' $, an equivalent number of scalars become their longitudinal degrees of freedom. It is clear from Eqs.~\eqref{eq:VEVsDoubletR}, \eqref{eq:VEVsBiDoublet}, \eqref{eq:VEVsDoubletL} that there are eight remaining neutral degrees of freedom, the SM-like Higgs plus three $ \mathcal{C P}- $even and two $ \mathcal{C P}- $odd scalars, while in the charged sector we have two remaining charged degrees of freedom. The scenario we describe here corresponds, however, to the \textit{minimal} possible scalar content: we have already stated the necessity of $ \chi_L $ for a structure which is symmetric under parity, while a bi-doublet, necessary for the generation of masses in the fermionic sector, cannot fully break the $ SU(2)_L \times SU(2)_R \times U(1)_{B-L} $ gauge group down to $ U(1)_{EM} $, but only produce a partial breaking $ SU(2)_L \times SU(2)_R \rightarrow SU(2)_{L+R} $, i.e. simultaneous and identical transformations under both $ SU(2) $ symmetries.

\section{Gauge boson spectrum}\label{sec:spectrumGaugeBosonsLRM}

Each step of the symmetry breaking is represented in the neutral sector in the following way

\begin{equation}\label{eq:breakingSymmetries}
\underbrace{ \begin{pmatrix}
s_\theta & c_\theta & 0 \\
c_\theta & - s_\theta & 0 \\
0 & 0 & 1 \\ 
\end{pmatrix} }_{\mathbf{\rm Stage \; 2}} \;\;
\underbrace{ \begin{pmatrix}
1 & 0 & 0 \\
0 & s_\phi & c_\phi \\
0 & c_\phi & - s_\phi \\
\end{pmatrix}  }_{\mathbf{\rm Stage \; 1}}
\begin{pmatrix} W_{L}^3 \\ W_{R}^3 \\ B \end{pmatrix} = \begin{pmatrix} A \\ X_1 \\ X_2 \end{pmatrix} \, ,
\end{equation}
where $ c_\theta \equiv \cos \theta $, etc. On the other hand, we have for the charged gauge bosons

\begin{equation}
W^\pm_L = \frac{W^1_L \mp W^2_L}{\sqrt{2}} \, , \qquad W^\pm_R = \frac{W^1_R \mp W^2_R}{\sqrt{2}} \, .
\end{equation}
The fields $ X_1, X_2 $ and $ W_L, W_R $ can mix depending on the specific Spontaneous Symmetry Breaking occurring in LR Models, and therefore the physical states $ Z, Z' $ and $ W, W' $ are linear combinations of $ X_1, X_2 $ and $ W_L, W_R $, respectively.



The physical states can be determined by diagonalizing the mass matrix of the gauge bosons, which results from the couplings to the scalars. First then, we have the covariant derivative $ D $ as in Eq.~\eqref{eq:fullDerivativeCovariantLRM}, from which one has the following gauge invariant part of the full Lagrangian

\begin{equation}
\mathcal{L}_{scalar} = (D_\mu \chi_R) (D^\mu \chi_R)^\dagger + (D_\mu \chi_L) (D^\mu \chi_L)^\dagger + (D_\mu \Phi) (D^\mu \Phi)^\dagger - V \, ,
\end{equation}
where the potential $ V \equiv V(\chi_R, \chi_L, \Phi) $ is going to be discussed later. (Note that, since $ \Phi $ transforms as $ \Phi \rightarrow U_L \Phi U_R^\dagger $, the part of the covariant derivative on $ W^{a}_R $ gets an opposite sign compared to that of $ W^{a}_L $.)

Now, expliciting the VEVs of the scalars fields, we have the following mass term


\begin{eqnarray}
& \mathcal{L}_{mass} = \begin{pmatrix}
W^{+}_{L} & W^{+}_{R} \\
\end{pmatrix}
\begin{pmatrix}
\tilde{M}_{W}^2 & \delta \tilde{M}_{W}^2 \\
\delta \tilde{M}_{W}^2 & \tilde{M}_{W'}^2 + \Delta \tilde{M}_{W'}^2 \\
\end{pmatrix}
\begin{pmatrix}
W^{-}_{L} \\
W^{-}_{R} \\
\end{pmatrix} \nonumber\\
& + \frac{1}{2} \begin{pmatrix}
A & X_1 & X_2 \\
\end{pmatrix}
\begin{pmatrix}
0 & 0 & 0 \\
0 & \tilde{M}_{Z}^2 & \delta \tilde{M}_{Z}^2 \\
0 & \delta \tilde{M}_{Z}^2 & \tilde{M}_{Z'}^2 + \Delta \tilde{M}_{Z'}^2 \\
\end{pmatrix}
\begin{pmatrix}
A \\
X_1 \\
X_2 \\
\end{pmatrix} \, ,
\end{eqnarray}
where the mass terms at the tree level are given in Table~\ref{tab:massesSMandLRM}.

\begin{table}[t]
\centering
\begin{tabular}{c@{\hskip 0.15in}c@{\hskip 0.15in}c@{\hskip 0.15in}c@{\hskip 0.15in}c@{\hskip 0.15in}c@{\hskip 0.15in}c@{\hskip 0.15in}c}
	\midrule
	$ \tilde{M}_{Z}^2 $ & $ \tilde{M}_{Z'}^2 $ & $ \Delta \tilde{M}_{Z'}^2 $ & $ \delta \tilde{M}_{Z}^2 $ \\
	\midrule
	$ \frac{1}{4} (g_{L}^2 + g_{Y}^2) \kappa^2 $ & $ \frac{1}{4} (g_{R}^2 + g_{B-L}^2) \kappa_{R}^2 $ & $ \frac{c^2_\phi}{4} g_{R}^2 \left[ \kappa^2-\kappa_{L}^2 \frac{1 - 2 c^2_\phi}{c^4_\phi} \right] $ & $ - \frac{c^2_\phi}{4 e} g_{L} g_{R} g_{B-L} \left[ \kappa^2-\frac{\kappa_{L}^2}{c^2_\phi} \right] $ \\
	\midrule
	\midrule
	$ \tilde{M}_{W}^2 $ & $ \tilde{M}_{W'}^2 $ & $ \Delta \tilde{M}_{W'}^2 $ & $ \delta \tilde{M}_{W}^2 $ \\
	\midrule
	$ \frac{1}{4} g_{L}^2 \kappa^2 $ & $ \frac{1}{4} g_{R}^2 \kappa_{R}^2 $ & $ \frac{1}{4} g_{R}^2 (\kappa_{1}^2 + \kappa_{2}^2) $ & $ - \frac{s_{2 \beta}}{4} g_{L} g_{R} (\kappa_{1}^2 + \kappa_{2}^2) $ \\
	\midrule
\end{tabular}
\caption{\it Mass terms at the tree level. Above, $ \kappa^2 \equiv \kappa^2_1 + \kappa^2_2 + \kappa^2_L $, and $ s_{2 \beta} = 2 r / (1 + r^2) $}\label{tab:massesSMandLRM}
\end{table}


The off-diagonal terms lead to mixing between $ X_1 $ and $ X_2 $, or between $ W_L^\pm $ and $ W_R^\pm $: for instance, the $ \sim 80 $~GeV charged gauge boson $ W^\pm $ results from the mixing of the fields associated to the $ T_{\pm}^L $ and $ T_{\pm}^R $ weak isospin generators, which is suppressed by the ratio of vacuum expectation values, i.e. $ \epsilon^2 $. If we now diagonalize the gauge boson matrix, the eigenmasses are given by:

\begin{eqnarray}
M_{Z}^2 &=& \tilde{M}_{Z}^2 - \frac{\delta \tilde{M}_{Z}^4}{\tilde{M}_{Z'}^2} , \qquad M_{Z'}^2 = \tilde{M}_{Z'}^2 + \Delta \tilde{M}_{Z'}^2 = \frac{M_{W'}^2}{c_\phi^2} (1 + \mathcal{O} (\epsilon^2)) \, , \label{eq:massZequation}\\
M_{W}^2 &=& \tilde{M}_{W}^2 - \frac{\delta \tilde{M}_{W}^4}{\tilde{M}_{W'}^2} , \qquad M_{W'}^2 = \tilde{M}_{W'}^2 + \Delta \tilde{M}_{W'}^2 = \frac{g_R^2 \kappa_R^2}{4} (1 + \mathcal{O} (\epsilon^2)) \, , \nonumber\\ 
\end{eqnarray}
keeping only the first corrections on $ \tilde{M}_{Z,W}^2 $, $ \delta \tilde{M}_{Z,W}^2 $, $ \Delta \tilde{M}_{Z',W'}^2 $ over $ \tilde{M}_{Z',W'}^2 $.





To conclude this section, we would like to discuss the chosen scalar representations in connection with the gauge boson spectrum. If a triplet representation $ \Delta_{L} = (\mathbf{1},\mathbf{3},\mathbf{1},2) $ was considered instead of $ \chi_L $, the VEV $ \kappa^{triplet}_L  $ would be very much suppressed. This is due to the different ways in which this VEV contributes to the masses of the $ W^\pm, Z $ gauge bosons: since the so-called $ \rho- $parameter

\begin{equation}\label{eq:rhodef}
\rho = \frac{M^{2}_{W}}{M^{2}_{Z} \cos^{2} \theta}
\end{equation}
is very close to 1, $ \langle \Delta^0_L \rangle $ would be small, a dozen GeV at most (for $ \rho = 0.99 $ for instance, we find $ \kappa^{triplet}_L \simeq 17 $~GeV, see Appendix~\ref{sec:rhoParameter}). On the other hand, since in the doublet LR Model case the spontaneous symmetry breaking mechanism is triggered by doublets and a bi-doublet, $ \rho $ does not provide a very important constraint at tree level, see Appendix \ref{sec:rhoParameter}. 




\section{Masses in the fermionic sector}\label{sec:quarkSectorMasses}

When the scalar doublet of the SM develops a VEV, it gives masses to the fermions (apart from the neutrinos), which are equal to the VEV times the diagonal elements of the diagonalized Yukawa matrices. The unitary matrices that diagonalize the mass matrices lead to the CKM matrix, discussed in the last chapter, responsible for mixing among different generations in the context of the SM.

In LR Models, when the bi-doublet develops VEVs, it generates masses to the fermions (including neutrinos) and the diagonalization of the mass matrix leads to non-diagonal mixing matrices responsible for the flavour phenomenology in LR Models. In this context, we have two possible structures

\begin{eqnarray}\label{eq:YukawaQuarksLRM}
\mathcal{L}^{(quarks)}_{Yukawa} = - \overline{Q'_{L}} (\Phi Y + \tilde{\Phi} \tilde{Y}) Q'_{R} + h.c. \, ,
\end{eqnarray}
where $ Y, \tilde{Y} $ are the two Yukawa matrices. Given the vacuum expectation values seen in Eqs.~\eqref{eq:VEVsBiDoublet}, we have the mass matrices

\begin{eqnarray}\label{eq:massesLRM}
M_{u} = \frac{\sqrt{\kappa^2_{1} + \kappa^2_{2}}}{\sqrt{2}} (c Y + s \tilde{Y}) \, , \qquad M_{d} = \frac{\sqrt{\kappa^2_{1} + \kappa^2_{2}}}{\sqrt{2}} (s Y + c \tilde{Y}) \, ,
\end{eqnarray}
where $ c \equiv 1 / (\sqrt{1 + r^2}) $ and $ s \equiv r / (\sqrt{1 + r^2}) $. We now diagonalize the mass matrices $ M_{u,d} $ as in the previous chapter, resulting in

\begin{equation}
U^{u \dagger}_L M_u U^u_R = \widehat{M}_u = {\rm diag} (m_u, m_c, m_t) \, , \quad U^{d \dagger}_L M_d U^d_R = \widehat{M}_d = {\rm diag} (m_d, m_s, m_b) \, ,
\end{equation}
and the same unitary transformations $ U^{u,d}_R, U^{u,d}_L $ introduce the mixing matrices $ V^{L,R} $, which are defined as the unitary matrices

\begin{equation}
V^{L,R} = U^{u \dag}_{L,R} U^{d}_{L,R} \, .
\end{equation}

We now discuss an important point concerning $ \widehat{M}_{u, d} $. Note that when $ s = \mathcal{O} (1)\, c \Leftrightarrow r = \mathcal{O} (1) $, a linear combination of $ Y, \tilde{Y} $ gives a large mass in one case (the top-quark mass in the LHS of Eq.~\eqref{eq:massesLRM}), and not in the other (down-type quark masses in the RHS of Eq.~\eqref{eq:massesLRM}). It is usual then to ask for $ r \ll 1 $ (see e.g. Ref.~\cite{Maiezza:2010ic}, where large spontaneous $ \mathcal{C P}- $violating phases are considered) and $ Y $ alone would account for the very large mass of the top. However, in order to remain as general as possible, we will investigate arbitrary values for $ r $, but we will keep in mind that, even if not impossible, having $ m_t $ in one case and $ m_b $ in the other would state a somewhat unexpected relation among the VEVs $ \kappa_{1,2} $, related to EWSB, and the components of the Yukawas $ Y, \tilde{Y} $.


Shifting to the leptonic sector, we first note that in the SM neutrinos are massless. A ``trivial" extension of the SM, however, could be to consider three copies of right-handed neutrinos \linebreak $ \nu_R = (\mathbf{1},\mathbf{1},0) $, thus leading to a Dirac mass term for $ \nu_{L,R} $ just like for the other fermions. Due to their quantum numbers, $ \nu_R $ are sterile, i.e. they do not interact with the other particles, and their existence would at our present stage of knowledge remain based on the observation of neutrino masses and generation mixing (the PMNS matrix).


In the LR Models, right-handed neutrinos must be introduced to establish a LR symmetry in the leptonic sector and their existence could be tested by means of right-handed currents. In the doublet scenario under investigation here we have

\begin{eqnarray}
\mathcal{L}^{(leptons)}_{Yukawa} = - \overline{L'_{L}} (\Phi Y^{lept} + \tilde{\Phi} \tilde{Y}^{lept}) L'_{R} + h.c. \, ,
\end{eqnarray}
similarly to Eq.~\eqref{eq:YukawaQuarksLRM}. Once plugging the VEV of the bi-doublet, we have the following non-diagonalized leptonic mass matrices

\begin{equation}
M_{\nu} = \frac{\sqrt{\kappa^2_{1} + \kappa^2_{2}}}{\sqrt{2}} (c Y^{lept} + s \tilde{Y}^{lept}) , \qquad M_{e} = \frac{\sqrt{\kappa^2_{1} + \kappa^2_{2}}}{\sqrt{2}} (s Y^{lept} + c \tilde{Y}^{lept}) \, ,
\end{equation}
the diagonalization of them leading to the mixing-matrices $ V^{L,R}_{lept} $, analogously to the quark sector. Note that when sending the masses of the neutrinos to zero one does not always gain in symmetry as in the SM for charged fermions \cite{'tHooft:1979bh}, because $ \nu_L $ and $ \nu_R $ in the LR model can be related by the exchange of physical $ W $ or $ W' $ gauge bosons when $ r \neq 0 $ \cite{Branco:1978bz}. This results in finite radiative contributions to the masses of the neutrinos proportional to the masses of the charged leptons. Whether or not this situation results in fine-tuned masses for the neutrinos at low energies in the doublet case when $ r $ is arbitrary must still be checked.



\section{Discrete symmetries and the scalar potential}


In LR Models, we may define the following parity transformation (where global phases are absorbed into the fields)

\begin{eqnarray}
\mathcal{P} : \left\{ \begin{matrix}
Q_{L} \leftrightarrow Q_{R} \\
\Phi \rightarrow \Phi^{\dag} \\
\chi_{L} \leftrightarrow \chi_{R}
\end{matrix} \right.
\end{eqnarray}
(we omit $ \overrightarrow{x} \rightarrow - \overrightarrow{x} $ or $ \overrightarrow{p} \rightarrow - \overrightarrow{p} $). We also define the charge conjugation transformation\footnote{See Refs.~\cite{Ecker:1980at,Branco:1985ng} for other transformations in the context of LR Models, and see Ref.~\cite{Lee:1966ik} for a more general discussion.}


\begin{eqnarray} \label{eq:eq1}
\mathcal{C} : \left\{ \begin{matrix}
Q_{L} \leftrightarrow (Q_{R})^{c} \\
\Phi \rightarrow \Phi^{T} \\
\chi_{L,R} \leftrightarrow \chi^{\dagger T}_{R,L}
\end{matrix} \right. .
\end{eqnarray}

We have not yet mentioned the case $ \mathcal{C} $ so far: after all, parity and charge-conjugation violations are closely related in the SM, and LR Models alternatively offer the possibility of restoring $ \mathcal{C} $ at high energies. Therefore, one could as well rewrite the statements we have made so far about the restoration of $ \mathcal{P} $ by considering $ \mathcal{C} $ instead of $ \mathcal{P} $. Moreover, restoring $ \mathcal{C} $ is motivated by possible extensions of the LR Model \cite{Maiezza:2010ic}.









We now move to a discussion of the scalar potential. Its most general form, symmetric under the parity transformation defined above is


\begin{dmath}\label{eq:potentialSymPdoublet}
V = - \mu^{2}_{1} \operatorname{tr} (\Phi^{\dag} \Phi) - \mu^{2}_{2} \operatorname{tr} (\tilde{\Phi}^{\dag} \Phi +  \tilde{\Phi} \Phi^{\dag}) - \mu^{2}_{3} (\chi^{\dag}_{L} \chi_{L} + \chi^{\dag}_{R} \chi_{R}) + {\mu'}_{1} (\chi^{\dag}_{L} \Phi \chi_{R} + \chi^{\dag}_{R} \Phi^{\dag} \chi_{L}) + {\mu'}_{2} (\chi^{\dag}_{L} \tilde{\Phi} \chi_{R} + \chi^{\dag}_{R} \tilde{\Phi}^{\dag} \chi_{L}) + \lambda_{1} [\operatorname{tr} (\Phi^{\dag} \Phi)]^{2} + \lambda_{2} \left( [\operatorname{tr} (\tilde{\Phi}^{\dag} \Phi)]^{2} + [\operatorname{tr} (\tilde{\Phi} \Phi^{\dag})]^{2} \right) + \lambda_{3} \operatorname{tr} (\tilde{\Phi}^{\dag} \Phi) \operatorname{tr} (\tilde{\Phi} \Phi^{\dag}) + \lambda_{4} \operatorname{tr} (\Phi^{\dag} \Phi) \operatorname{tr} (\tilde{\Phi}^{\dag} \Phi +  \tilde{\Phi} \Phi^{\dag}) + \rho_{1} [ (\chi^{\dag}_{L} \chi_{L})^{2} + (\chi^{\dag}_{R} \chi_{R})^{2} ] + \rho_{3} (\chi^{\dag}_{L} \chi_{L})(\chi^{\dag}_{R} \chi_{R}) + \alpha_{1} (\chi^{\dag}_{L} \chi_{L} + \chi^{\dag}_{R} \chi_{R}) \operatorname{tr} (\Phi^{\dag} \Phi) + \frac{\alpha_{2}}{2} \{ \operatorname{e}^{i \delta_{2}} [\chi^{\dag}_{L} \chi_{L} \operatorname{tr} (\tilde{\Phi} \Phi^{\dag}) + \chi^{\dag}_{R} \chi_{R} \operatorname{tr} (\tilde{\Phi}^{\dag} \Phi)] + \operatorname{e}^{-i \delta_{2}} [\chi^{\dag}_{L} \chi_{L} \operatorname{tr} (\tilde{\Phi}^{\dag} \Phi) + \chi^{\dag}_{R} \chi_{R} \operatorname{tr} (\tilde{\Phi} \Phi^{\dag})] \} + \alpha_{3} (\chi^{\dag}_{L} \Phi \Phi^{\dag} \chi_{L} + \chi^{\dag}_{R} \Phi^{\dag} \Phi \chi_{R}) + \alpha_{4} (\chi^{\dag}_{L} \tilde{\Phi} \tilde{\Phi}^{\dag} \chi_{L} + \chi^{\dag}_{R} \tilde{\Phi}^{\dag} \tilde{\Phi} \chi_{R}),
\end{dmath}
\noindent
where $ \delta_{2} $ is a $ \mathcal{C P}- $violating phase, and $ \mu^{2}_{1, 2, 3} $, $ {\mu'}_{1,2} $, $ \lambda_{1,2,3,4} $, $ \rho_{1,3} $ and $ \alpha_{1,2,3,4} $ are all real. Even if there are $ \mathcal{O} (20) $ new parameters in the potential, not all of them are relevant, and only a few show up in the expressions of couplings and masses of the heavy sector, as seen in their expressions given in Appendix~\ref{sec:spectrumScalars}.

For simplicity reasons, we focus specifically in the case where no complex phase is present in the potential, namely $ \sin \delta_2 = 0 $. As argued in Appendix~\ref{sec:stableConds}, the phases of the VEVs are related to $ \delta_2 $, and we are going as well to set all of them to zero, i.e. $ \sin \alpha = \sin \theta_L = 0 $.

The basic features of the scalar potential symmetric under $ \mathcal{C} $ can be found in Appendix~\ref{sec:tripletCaseStability} in a slightly different context, where it is argued that the $ \mathcal{C} $ case introduces extra complex phases. However, since we are interested in the simplified case where the complex phases in the potential are set to zero, both cases $ \mathcal{P} $ and $ \mathcal{C} $ lead to the same discussion. Moreover, neither one or the other implies the vanishing of a possible coupling between the fields $ \Phi, \chi_{L,R} $: all the possible combinations are already present in Eq.~\eqref{eq:potentialSymPdoublet}, and what changes is the relation between the coefficients of the different structures and the number of possible complex phases. Then, we expect for all our purposes to have essentially the same discussion when $ \mathcal{P} $ or $ \mathcal{C} $ invariant scalar potentials are considered, and perhaps even when neither of them is assumed.

By minimizing the potential, one determines the VEVs in terms of the parameters of $ V $. At the minimum, the stability conditions which must be satisfied are

\begin{equation}
\frac{\partial V}{\partial x} \Bigg|_{ \{ \kappa_{1,2,L,R}, \alpha, \theta_L \} } = 0 \, ,
\end{equation}
where $ x \in \{ \varphi^{0r}_{1}, \varphi^{0r}_{2}, \chi^{0r}_{R}, \chi^{0r}_{L}, \varphi^{0i}_{1}, \varphi^{0i}_{2}, \chi^{0i}_{R}, \chi^{0i}_{L} \} $ (out of them, two equations are redundant, i.e. there are six independent equations relating the set of the six parameters $ \{ \kappa_{1,2,L,R}, \alpha, \theta_L \} $ characterizing the minimum). The mass matrix in the scalar sector is determined at the minimum by

\begin{equation}\label{eq:formalMassSquared}
\mathcal{M}^2 = \frac{\partial^2 V}{\partial x \, \partial y} \Bigg|_{\{ \kappa_{1,2,L,R}, \alpha, \theta_L \}} \, ,
\end{equation}
which is an $ 8 \times 8 $ matrix in the neutral sector ($ x $ as above, and with $ y $ assuming values over the same set), or a $ 4 \times 4 $ matrix in the charged sector ($ x, y \in \{ \varphi^\pm_{1} , \varphi^\pm_{2} , \chi^\pm_{R} , \chi^\pm_{L} \} $).

In agreement with the discussion at the end of Section~\ref{sec:EWSBSMLRM}, the resulting scalar spectrum calculated from Eq.~\eqref{eq:formalMassSquared} consists of:

\begin{itemize}
	\item one light SM-like scalar $ h^0 $, of mass $ \sim \kappa $, and
	\item three heavy $ \mathcal{C P}- $even scalars $ H^0_{1,2,3} $,
	\item two heavy $ \mathcal{C P}- $odd scalars $ A^0_{1,2} $,
	\item two heavy charged scalars $ H^\pm_{1,2} $,
\end{itemize}

\noindent
where the heavy particles have masses $ \sim \kappa_R $. As seen in Appendix~\ref{sec:spectrumScalars}, some of the physical scalars are degenerate at leading order in $ \epsilon $: thus $ \{ H^0_i, A^0_i ; H^\pm_i \} $, $ i = 1, 2 $, have the same masses. Of course, two other $ \mathcal{C P}- $odd scalars and two other charged scalars correspond to the would-be Goldstone bosons and are absorbed as the longitudinal degrees of freedom of the massive gauge bosons.  



\section{Couplings to fermions}

Here, we focus on the basic characteristics of the gauge boson and physical scalar couplings to the fermionic sector. More information and technical details can be found in Appendix~\ref{sec:spectrumScalars}.


\subsection{Couplings of the gauge bosons to the fermions}\label{sec:couplingsTreeLevel}

As in the SM, the couplings of the fermions to the gauge bosons can be determined from their charges. Then, these couplings are corrected when going to the mass basis of the gauge bosons due to the mixing of the known and new gauge bosons. This mixing is suppressed by the hierarchy of the SSB energy scales, and comes at order $ \mathcal{O} (\epsilon^2) $. For the sake of readability, we provide in this chapter the couplings at leading order in $ \epsilon $, while the full expressions are given in Appendix~\ref{sec:spectrumScalars}. It is straightforward that at this order the couplings to the charged vector bosons are given by

\begin{eqnarray}\label{eq:couplingsWWpr}
&& \frac{g_L}{\sqrt{2}} W^-_{L \mu} \overline{u_L} \gamma^\mu V^L d_L + \frac{g_R}{\sqrt{2}} {W}^-_{R \mu} \overline{u_R} \gamma^\mu V^R d_R \\
&& + \frac{g_L}{\sqrt{2}} W^-_{L \mu} \overline{\nu_L} \gamma^\mu V^L_{lept} \ell_L + \frac{g_R}{\sqrt{2}} {W}^-_{R \mu} \overline{\nu_R} \gamma^\mu V^R_{lept} \ell_R + h.c. \, , \nonumber
\end{eqnarray}
which has a symmetric structure for right- and left-handed fields.

In the SM, the CKM matrix is the mixing matrix corresponding to $ V^{L} = U^{u \dag}_{L} U^{d}_{L} $, while $ V^R $ is not relevant. Concerning the structure of $ V^L $, since it is unitary there are $ \frac{n_{g} (n_{g} + 1)}{2} $ possible complex phases, where $ n_{g} $ is the number of generations. However, some of them are non-physical: one can redefine the relative phases of the quark fields, thus eliminating $ 2 n_{g} - 1 $ phases. With $ n_{g} = 3 $, we have one physical complex phase, which alone introduces $ \mathcal{C P} $ violation in the quark sector of the SM. When there is a second mixing matrix, the one corresponding to right-handed quark currents, one has a total of $ n_{g} (n_{g} + 1) $ possible complex phases distributed over the two mixing matrices, but as before one can eliminate $ 2 n_{g} - 1 $ phases by redefining the quark fields (in what follows, one eliminates as much as possible the phases of $ V^{L} $). Hence, there is a total of $ n^{2}_{g} - n_{g} + 1 $ complex phases in $ V^{L,R} $. When $ n_{g} = 3 $, the total number of complex phases introduced by $ V^{L,R} $ is seven, i.e. six extra phases compared to the SM, constituting new sources of $ \mathcal{C P} $ violation in the quark sector.

\begin{figure}
	\hspace{1cm} \SetScale{0.5} \SetWidth{2} \begin{picture}(40,40)(-30,0)
		\Photon(-60,0)(0,0){4}{4}
		\ArrowLine(20,-40)(0,0)
		\ArrowLine(0,0)(20,40)
		\Text(-12,12)[t]{\small $ W $}
		\Text(-12,-5)[t]{\small $ V^L $}
		\Text(16,-20)[t]{\small $ u_L $}
		\Text(17,20)[t]{\small $ d_L $}
	\end{picture}
	\hspace{1.0cm} \SetScale{0.5} \SetWidth{2} \begin{picture}(40,40)(-30,0)
		\Photon(-60,0)(0,0){4}{4}
		\ArrowLine(20,-40)(0,0)
		\ArrowLine(0,0)(20,40)
		\Text(-12,12)[t]{\small $ W' $}
		\Text(-12,-5)[t]{\small $ V^R $}
		\Text(16,-20)[t]{\small $ u_R $}
		\Text(17,20)[t]{\small $ d_R $}
	\end{picture}
	\hspace{1.5cm} \SetScale{0.5} \SetWidth{2} \begin{picture}(40,40)(-30,0)
		\Photon(-80,0)(0,0){4}{5}
		\ArrowLine(20,-40)(0,0)
		\ArrowLine(0,0)(20,40)
		\Text(-16,12)[t]{\small $ Z $}
		\Text(-23,-5)[t]{\small $ g^f_V \mathbf{1} - g^f_A \gamma_5 $}
		\Text(16,-17)[t]{\small $ f $}
		\Text(16,20)[t]{\small $ f $}
	\end{picture}
	\hspace{1.5cm} \SetScale{0.5} \SetWidth{2} \begin{picture}(40,40)(-30,0)
		\Photon(-80,0)(0,0){4}{5}
		\ArrowLine(20,-40)(0,0)
		\ArrowLine(0,0)(20,40)
		\Text(-16,12)[t]{\small $ Z' $}
		\Text(-23,-5)[t]{\small $ d^f_V \mathbf{1} + d^f_A \gamma_5 $}
		\Text(16,-17)[t]{\small $ f $}
		\Text(16,20)[t]{\small $ f $}
	\end{picture}
	\vspace{1cm}
	\caption{\it Main couplings of the $ W, W', Z $ and $ Z' $ particles (which are further corrected by their mixing).}
\end{figure}
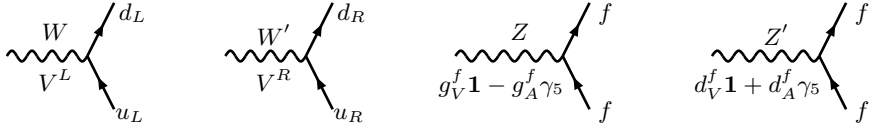

The same exercise for deriving Eq.~\eqref{eq:couplingsWWpr} can also be made for the couplings to the neutral gauge bosons $ Z \sim X_1 $ and $ Z' \sim X_2 $. From Eq.~\eqref{eq:breakingSymmetries} we have


\begin{eqnarray}\label{eq:WeakInteractionsLRM}
&& \bar{f} \gamma^\mu \Bigg[ \left( g_L T^L_3 s_\theta + g_R T^R_3 s_\phi c_\theta + g_{B-L} \frac{B - L}{2} c_\phi c_\theta \right) A_\mu \nonumber\\
&& \quad + \left( g_L T^L_3 c_\theta - g_R T^R_3 s_\phi s_\theta - g_{B-L} \frac{B - L}{2} c_\phi s_\theta \right) X_{1 \mu} \\
&& \quad \quad \left( g_R T^R_3 c_\phi - g_{B-L} \frac{B - L}{2} s_\phi \right) X_{2 \mu} \Bigg] f \, , \nonumber
\end{eqnarray}
where $ f = Q_{L,R} $ or $ L_{L,R} $. The reader should note that the $ Z' $ boson introduced here, which is an admixture of $ X_2 $ (the main component) and $ X_1 $, does not introduce flavour changing neutral currents, as it may happen in other models which have $ Z' $ bosons. The reason for this is the same found in the SM for explaining the absence of FCNC of the $ Z $: up-like quarks or down-like quarks of the same chirality and different generations have exactly the same quantum numbers (except for flavour), assuring flavour diagonal couplings (see \cite{Glashow:1970gm,Paschos:1976ay,Chanowitz:1977ye,Gripaios:2015gxa}).

The weak interaction part of the expression~\eqref{eq:WeakInteractionsLRM} can be compactly written as

\begin{equation}
Z^\mu_1 J^0_\mu + (Z')^\mu_2 K^0_\mu \, ,
\end{equation}
which is corrected at order $ \mathcal{O} (\epsilon^2) $ by the mixing between $ X_1 $ and $ X_2 $, and where

\begin{eqnarray}
J^0_\mu &=& \frac{g_L}{c_\theta} ( T^L_{3} (f) \overline{f_L} \gamma_\mu f_L - s^2_{\theta} Q (f) \overline{f} \gamma_\mu f ) \\
&=& \frac{g_L}{c_\theta} \frac{1}{2} \bar{f} \gamma_\mu (g^f_V - g^f_A \gamma^5) f \, , \nonumber\\
K^0_\mu &=& g_L \frac{t_\theta}{t_\phi} \left( T^R_{3} (f) \overline{f_R} \gamma_\mu f_R - t^2_\phi \, \frac{(B-L) (f)}{2} \overline{f} \gamma_\mu f \right) \\
&=& g_L \frac{t_\theta}{t_\phi} \frac{1}{2} \left( d^f_V \overline{f} \gamma_\mu f + d^f_A \overline{f} \gamma_\mu \gamma_5 f \right) \, , \nonumber
\end{eqnarray}
and

\begin{eqnarray}
&& g^f_V = T^L_3 (f) - 2 Q (f) s^2_\theta , \qquad g^f_A = T^L_3 (f) \, , \\
&& d^f_V = T^R_{3} (f) / c^2_\phi + t^2_\phi \, ( T^L_{3} (f) - 2 Q (f) ) \, , \\
&& d^f_A = T^R_{3} (f) / c^2_\phi - t^2_\phi \, T^L_{3} (f) \, ,
\end{eqnarray}
given in Table~\ref{tab:LRMquantumNumbers}. The first line of Eq.~\eqref{eq:WeakInteractionsLRM} can be simplified as

\begin{equation}
A_\mu J^\mu_{EM} \, , \qquad J^\mu_{EM} = e Q (f) \overline{f} \gamma^\mu f \, ,
\end{equation}
where we have employed

\begin{equation}
Q (f) = T^R_3 (f) + T^L_3 (f) + \frac{(B-L) (f)}{2} \, . \nonumber
\end{equation}
Just like the last expression is only a consistency check (the electric charge must be equal to the unbroken $ U(1) $ charge), the couplings to gluons have the obvious expression

\begin{equation}
g_s \bar{f} \gamma^\mu G^a_{\mu} T_a f \, .
\end{equation}

\begin{table}[t]
	\centering
	\begin{tabular}{|c|c|c|c|c|c|}
		\hline
		left-handed & $ T_3^L (f) $ & $ T_R^L (f) $ & $ Q (f) $ & $ d^f_V $ & $ d^f_A $ \\
		\hline
		$ \nu_{e L}, \nu_{\mu L}, \nu_{\tau L} $ & $ + 1/2 $ & 0 & 0 & $ t^2_\phi ( + 1/2 ) $ & $ - t^2_\phi ( + 1/2 ) $ \\
		$ e_L, \mu_L, \tau_L $ & $ - 1/2 $ & 0 & $ - 1 $ & $ t^2_\phi 3/2 $ & $ - t^2_\phi ( - 1/2 ) $ \\
		$ u_L, c_L, t_L $ & $ + 1/2 $ & 0 & $ + 2/3 $ & $ - t^2_\phi 5/6 $ & $ - t^2_\phi ( + 1/2 ) $ \\
		$ d_L, s_L, b_L $ & $ - 1/2 $ & 0 & $ - 1/3 $ & $ t^2_\phi/6 $ & $ - t^2_\phi ( - 1/2 ) $ \\
		\hline
		\hline
		right-handed & $ T_3^L (f) $ & $ T_R^L (f) $ & $ Q (f) $ & $ d^f_V $ & $ d^f_A $ \\
		\hline
		$ \nu_{e R}, \nu_{\mu R}, \nu_{\tau R} $ & 0 & $ + 1/2 $ & 0 & $ + 1 / (2 c^2_\phi) $ & $ + 1 / (2 c^2_\phi) $ \\
		$ e_R, \mu_R, \tau_R $ & $ 0 $ & $ - 1/2 $ & $ - 1 $ & $ - 1 / (2 c^2_\phi) + 2 t^2_\phi $ & $ - 1 / (2 c^2_\phi) $ \\
		$ u_R, c_R, t_R $ & $ 0 $ & $ + 1/2 $ & $ + 2/3 $ & $ + 1 / (2 c^2_\phi) - 4 t^2_\phi / 3 $ & $ + 1 / (2 c^2_\phi) $ \\
		$ d_R, s_R, b_R $ & $ 0 $ & $ - 1/2 $ & $ - 1/3 $ & $ - 1 / (2 c^2_\phi) + 2 t^2_\phi / 3 $ & $ - 1 / (2 c^2_\phi) $ \\
		\hline
	\end{tabular}
	\caption{\it Quantum numbers of the different fermions in the LR Model.}\label{tab:LRMquantumNumbers}
\end{table}

\subsection{Couplings of the scalars to the fermions}\label{sec:couplingsHfermionsLRM}

Similarly to the vectorial couplings, the couplings of the fermions to the would-be Goldstones can also be evaluated. For the sake of readability, their expressions are given in Appendix~\ref{sec:spectrumScalars}.




Here, we focus on the physical scalars. Developing the Yukawa terms around the VEV of the scalar fields, beyond the quark mass terms one gets the couplings


\begin{eqnarray}
\overline{u_L} \frac{1}{\kappa_1 (1 - r^2)} \left[ \widehat{M}_u (\varphi^{0r}_{1} - r \varphi^{0r}_{2}) + V^L \widehat{M}_d V^{R \dagger} (\varphi^{0r}_{2} - r \varphi^{0r}_{1}) \right] u_R + h.c. \, ,
\label{eq:couplingsSMHiggs}
\end{eqnarray}
\noindent
where a similar expression holds for the down-type sector. Above, the first term inside brackets is a diagonal coupling to the up-type generations, while the second introduces Flavour Changing Neutral Currents (FCNC). To further interpret this result, one needs to express the set of unphysical fields $ \{ \varphi^{0r}_{1}, \varphi^{0r}_{2}, \chi^{0r}_L, \chi^{0r}_R \} $ in terms of the physical $ \mathcal{C P}- $even scalar fields, $ h^0, H^0_{1,2,3} $. In Appendix~\ref{sec:spectrumScalars}, we show the expressions for $ h^0, H^0_{1,2,3} $ in terms of $ \{ \varphi^{0r}_{1}, \varphi^{0r}_{2}, \chi^{0r}_L, \chi^{0r}_R \} $ up to leading order in $ \epsilon $. Here we pay special attention to the light SM-like Higgs field:

\begin{equation}\label{eq:combinationLightHiggs}
h^0 = \frac{1}{\sqrt{1 + r^2 + w^2}} (\varphi^{0r}_{1} + r \varphi^{0r}_{2} + w \chi^{0r}_L) + \mathcal{O} (\epsilon) \chi^{0r}_R \, .
\end{equation}
It then follows from Eq.~\eqref{eq:couplingsSMHiggs} that $ h^0 $ does not couple non-diagonally, since $ h^0 $ is orthogonal to the combination $ \varphi^{0r}_{2} - r \varphi^{0r}_{1} $ and the $ h^0- $coupling vanishes. Clearly, this is a feature at leading order in $ \epsilon $, but it is also valid when considering $ \mathcal{O} (\epsilon) $ corrections, thanks to the structure of the next-to-leading order term, proportional to $ \chi^{0r}_R $ as seen from Eq.~\eqref{eq:combinationLightHiggs}.

On the other hand, the diagonal couplings of the Higgs are given by

\begin{equation}\label{eq:SMLikeHiggsCoupling}
- \left( \frac{\widehat{M}_d}{\sqrt{\kappa^2_1 + \kappa^2_2 + \kappa^2_L}} \overline{d_L} d_R + \frac{\widehat{M}_u}{\sqrt{\kappa^2_1 + \kappa^2_2 + \kappa^2_L}} \overline{u_L} u_R \right) h^0 + \mathcal{O} (\epsilon^2) + h.c.
\end{equation}
Note that in the SM we have

\begin{equation}\label{eq:SMHiggsCoupling}
- \left( \frac{\widehat{M}_d}{v} \overline{d_L} d_R + \frac{\widehat{M}_u}{v} \overline{u_L} u_R \right) H^0 + h.c.
\end{equation}
Both expression are identical up to $ \mathcal{O} (\epsilon^2) $ corrections since $ \sqrt{\kappa^2_1 + \kappa^2_2 + \kappa^2_L} $ is equivalent to $ v $ found in the SM (up to $ \mathcal{O} (\epsilon^2) $ corrections), and $ H^0 $ from the SM is identified with $ h^0 $, by construction. The comparison between Eq.~\eqref{eq:SMLikeHiggsCoupling} and \eqref{eq:SMHiggsCoupling} implies that the measurement of the intensity of the couplings of the Higgs to the fermions does not set alone bounds on $ \kappa_L $, as it could be naively thought once the bi-doublet alone couples to the fermions.


Of course, other combinations of the fields $ \{ \varphi^{0r}_{1}, \varphi^{0r}_{2}, \chi^{0r}_L, \chi^{0r}_R \} $ will have couplings which introduce FCNC in the model. This is the main reason why one generally has very strong bounds on the scalar sector of the model, pushing the neutral scalar masses beyond many TeV (see Chapter~\ref{ch:PHENO}).




\section{Structure of the mixing matrix $ V^R $}\label{sec:VRStructure}



The new mixing matrix $ V^R $ describes the couplings of the new charged gauge boson $ {W'}^\pm $ to quarks and introduces new free parameters in generic versions of the model.\footnote{The mass spectrum $ \widehat{M}_{u,d} $ and the mixing matrices $ V^{L,R} $ are all related to the underlying $ Y, \tilde{Y} $. Knowing $ \{ m_{u,d,s,c,b,t} \} $ and the Cabibbo angles of $ V^L $, namely $ \{ \sin \theta_{12}, \sin \theta_{13}, \sin \theta_{23} \} $, one has the possibility to constrain the structure of $ V^R $. This is studied in \cite{Kiers:2002cz}, where constraints sensitive to $ \mathcal{C P} $ violation are also employed.} However, by considering the explicit realization of a discrete symmetry, parity $ \mathcal{P} $ or charge-conjugation $ \mathcal{C} $, one may constrain $ V^R $. 
Let us see the structure the Yukawa matrices must have under these discrete symmetries. Under $ \mathcal{P} $, we have \cite{Branco:1999fs}

\begin{dmath}
\overline{Q_{L}} \Phi Y Q_{R} \stackrel{\mathcal{P}}{\rightarrow} \overline{Q} \gamma^0 P_R \Phi^\dagger Y P_R \gamma^0 Q = \overline{Q_R} \Phi^\dagger Y Q_L \, ,
\end{dmath}
and to fit the Hermitian conjugate

\begin{equation}
(\overline{Q_{L}} \Phi Y Q_{R})^\dagger = \overline{Q_R} \Phi^\dagger Y^\dagger Q_L \, ,
\end{equation}
we require $ Y = Y^\dagger $ (and $ \tilde{Y} = \tilde{Y}^\dagger $). On the other hand, under charge-conjugation

\begin{dmath}
\overline{Q_{L}} \Phi Y Q_{R} \stackrel{\mathcal{C}}{\rightarrow} \overline{(Q_{R})^{c}} \Phi^{T} Y (Q_{L})^{c} = \overline{Q_{L}} \Phi Y^{T} Q_{R} \, .
\end{dmath}


\noindent
Hence, $ Y = Y^{T} $ (and $ \tilde{Y} = \tilde{Y}^{T} $) for charge-conjugation invariance.






The particular structures for $ V^R $ in each of these two cases are: (a) under a parity-symmetric model

\begin{eqnarray}
V^R &=& S_u V^L S_d \, , \label{eq:manifestVLVR}\\
S_q &=& {\rm diag} (s_{q_i}) \, , \quad s_{q_i} = \pm \, , \quad i = 1,2,3 \, , \nonumber
\end{eqnarray}
which is valid in the case where there is no complex phase from the bi-doublet vacuum expectation values, as we assume here for simplicity reasons (for a more general discussion, see \cite{Maiezza:2010ic} and \cite{GeneralVRmatrixParity}). (b) under charge-conjugation


\begin{eqnarray}
V^R &=& K_u (V^L)^* K_d \, , \label{eq:pseudomanifestVLVR}\\
K_q &=& {\rm diag} (k_{q_i}) \, , \quad k_{q_i} = {\rm e}^{i \theta_{q_i}} \, , \quad i = 1,2,3 \, , \nonumber
\end{eqnarray}
where one of the phases $ \theta_{q_i} $ is fixed by the others. In this case, there are extra $ 2 n_g - 1 = 5 $ complex phases in total.

We call the expression stated by Eq.~\eqref{eq:manifestVLVR} a \textbf{manifest} relation between the two mixing matrices, while the relation stated in Eq.~\eqref{eq:pseudomanifestVLVR} can be called \textbf{pseudo-manifest}, which are two common patterns investigated in LR Model studies. 

Eqs.~\eqref{eq:manifestVLVR} and \eqref{eq:pseudomanifestVLVR} may be seen as further constraints to the model in order to have more predictive versions of LR Models, and follow from $ \mathcal{P} $ or $ \mathcal{C} $ that we have till now advocated for. However, we would like to test the more general case where $ V^L, V^R $ are not related by a manifest or a pseudo-manifest relation. The reader may be worried by the fact that, therefore, we would have no exact discrete symmetry at all. Nonetheless, the exact equality of \textit{couplings}, Eqs.~\eqref{eq:manifestVLVR} and \eqref{eq:pseudomanifestVLVR} (and moreover the gauge \textit{couplings} $ g_L = g_R $), should be the concern of more elaborate Grand Unified Theories, while LR Models introduce the required degrees of freedom for a $ \mathcal{P} $ or $ \mathcal{C} $ symmetric theory. In this picture, differences between left and right couplings would be explained by the running from the scale of unification of the couplings down to the scale $ \sim \kappa_R $, inducing $ g_L \neq g_R $ and $ V^R \nsim V^L $ (up to signs or complex phases) in the LR Model.\footnote{For this reason, we have avoided so far the use of the term ``Left-Right \textit{Symmetric} Model.''} 



\section{Triplet model}\label{sec:TripletModelDiscussion}

While the model with doublets we have described has been the first one to be considered in the end of the `70s when LR Models were conceived, the model with two triplets and one bi-doublet has been extensively considered in the literature, starting from Ref.~\cite{Mohapatra:1980yp}. It has the interesting feature of implementing a mechanism where the known neutrinos have very light masses due to very heavy counterparts (i.e. a see-saw mechanism). 

For further discussion, let us investigate over the remaining of this section the triplet scenario, in particular the interactions with leptons, where we find the most important differences. To this effect, we have the two triplets $ \Delta_{R} = (\mathbf{1},\mathbf{1},\mathbf{3},2) $ and $ \Delta_{L} = (\mathbf{1},\mathbf{3},\mathbf{1},2) $, which introduce the following degrees of freedom 

\begin{eqnarray}
\Delta_{R} = \begin{pmatrix} \delta^{+}_{R}/\sqrt{2} & \delta^{++}_{R} \\ \delta^{0}_{R} & -\delta^{+}_{R}/\sqrt{2} \end{pmatrix} \, , \quad \Delta_{L} = \begin{pmatrix} \delta^{+}_{L}/\sqrt{2} & \delta^{++}_{L} \\ \delta^{0}_{L} & -\delta^{+}_{L}/\sqrt{2} \end{pmatrix}.
\end{eqnarray}
(A discussion concerning the scalar potential in this case, together with the Higgs spectrum is found in Appendix~\ref{sec:tripletCaseStability}.)

In the triplet case, beyond the Yukawa terms

\begin{eqnarray}
\mathcal{L}^{(leptons)}_{Yukawa} \ni - \overline{L'_{L}} (\Phi Y^{lept} + \tilde{\Phi} \tilde{Y}^{lept}) L'_{R} + h.c. \, ,
\end{eqnarray}
we also have

\begin{eqnarray}	
\mathcal{L}^{(leptons)}_{Yukawa} \ni - \frac{1}{2} (\overline{(L'_L)^c} i \tau_{2} \Delta_{L} Y^{lept}_{L} L'_{L} + \overline{(L'_R)^c} i \tau_{2} \Delta_{R} Y^{lept}_{R} L'_{R}) + h.c. \, ,
\end{eqnarray}
thus leading to Majorana masses. The full mass matrix is then

\begin{eqnarray}
\mathcal{L}^{(leptons)}_{mass} = - \frac{1}{2} \left( \overline{(\nu_{L})^{c}} \quad \overline{\nu_{R}} \right) \left( \begin{array}{cc} h^L_{M} \kappa_L & h_{D} \kappa_{+} \\ h^{T}_{D} \kappa_{+} & h^R_{M} \kappa_R \end{array} \right) \left( \begin{array}{c} \nu_{L} \\ (\nu_{R})^{c} \end{array} \right) + h.c. \, , \label{eq:tripletLeptonMatrix}
\end{eqnarray}
where $ h_{D} = (c Y^{lept} + s \tilde{Y}^{lept}) / \sqrt{2} $, $ h^L_{M} = Y^{lept}_{L} / \sqrt{2} $, $ h^R_{M} = Y^{lept}_{R} / \sqrt{2} $ and $ \kappa_+ \equiv \sqrt{\kappa^2_1 + \kappa^2_2} $.

Note that the mass matrix of Eq.~(\ref{eq:tripletLeptonMatrix}) has a $ 6 \times 6 $ structure: when writing the physical interactions with light neutrinos, one would find a non-unitary $ 3 \times 3 $ matrix, implying a more involved analysis of neutrino processes compared to the doublet case.

For simplicity, let us consider the case of only one generation \cite{Branco:1999fs}. When going to the mass basis, one finds (we choose conveniently the global phases of the fields so that $ h^R_M > 0 $ and assume $ h^R_{M} \kappa_R \gg h_{D} \kappa_+ \gg h^L_{M} \kappa_L $)

\begin{eqnarray}\label{eq:MassesNeutrinosLRMT}
m_{\nu} \simeq \frac{h^{2}_{D}}{h^R_{M}} \frac{\kappa^{2}_{+}}{\kappa_R} - h^L_{M} \kappa_L, \qquad m_{N} \simeq h^R_{M} \kappa_R \, ,
\end{eqnarray}
\noindent
and
\begin{eqnarray}
\left( \begin{array}{c} \nu \\ N \end{array} \right) =
\begin{pmatrix}
+ i \cos \Theta & - i \sin \Theta \\
\sin \Theta & \cos \Theta \\
\end{pmatrix}
\left( \begin{array}{c} \nu_{L} \\ (\nu_{R})^{c} \end{array} \right),
\end{eqnarray}
\noindent
where $ \cos \Theta \simeq 1 - \frac{1}{2} \left( \frac{h_{D}}{h^R_{M}} \frac{\kappa_{+}}{\kappa_R} \right)^{2} $. Then, the left-handed field $ \nu_L $ is basically light, while the right-handed counterpart $ \nu_R $ is much heavier. To compare the two contributions to $ m_\nu $ in Eq.~\eqref{eq:MassesNeutrinosLRMT}, one has the following relation between the VEVs $ \kappa_{L,R} $, valid in the triplet case

\begin{eqnarray}
&& (2 \rho_{1} - \rho_{3}) \frac{\kappa_L}{\kappa_{+}} \simeq \left[ \beta_{1} \frac{\kappa_{1} \kappa_{2}}{\kappa^{2}_{+}} \cos (\theta_{L} - \alpha) + \beta_{2} \frac{\kappa^{2}_{1}}{\kappa^{2}_{+}} \cos \theta_{L} \right. \label{eq:eqSeeSaw}\\
&& \left. + {\beta_{3} \frac{\kappa^{2}_{2}}{\kappa^{2}_{+}} \cos (\theta_{L} - 2 \alpha)} \right] \frac{\kappa_{+}}{\kappa_R} \Rightarrow \frac{\kappa_L}{\kappa_{+}} \equiv \gamma \frac{\kappa_{+}}{\kappa_R} \, , \nonumber
\end{eqnarray}
at leading order in $ \epsilon = \kappa_{1}/\kappa_R $, which is one of the stability conditions from the scalar potential (see Appendix~\ref{sec:tripletCaseStability}). Therefore, Eq.~(\ref{eq:eqSeeSaw}) leads to

\begin{eqnarray}\label{eq:finalSeeSaw}
m_{\nu} = \left( h^{2}_{D} - \gamma h^L_M h^R_{M} \right) \kappa^{2}_{+} / (h^R_{M} \kappa_R) \, , 
\end{eqnarray}
where $ \gamma $, defined implicitly from Eq.~(\ref{eq:eqSeeSaw}), is of any sign. This relation is called a see-saw mechanism: $ \kappa_{+} $ being fixed by the EWSB, the bigger $ \kappa_R $ (setting the scale for $ m_{N} $) is, the smaller $ m_{\nu} $ becomes. Note that the see-saw mechanism, and therefore the smallness of neutrinos masses, is closely related to the spontaneous breaking of parity in a very elegant way, through $ \kappa_L \ll \kappa_{R} $.

Let us have a closer look to the spectrum. Given

\begin{equation}
\kappa_R \lesssim 10 \operatorname{TeV} \quad {\rm and} \quad \kappa_{+} \simeq 246 \operatorname{GeV} \, ,
\end{equation}
if all the involved parameters of the potential and Yukawas were of order $ \mathcal{O}(1) $, one would have $ m_\nu \gtrsim 6 \operatorname{GeV} $, implying far too big neutrino masses. Therefore,

\begin{equation}\label{eq:orders}
h_D^2, \quad \gamma h^L_M \qquad {\rm of \; order} \quad 10^{-9}
\end{equation}
are usually required, while $ h^R_M \sim \mathcal{O} (1) $ controls the mass of the heavy neutrino. To accomplish Eq.~\eqref{eq:orders}, one may imagine having

\begin{equation}
h^L_M \quad {\rm of \; order} \quad \mathcal{O} (1) \, ,
\end{equation}
just like $ h^R_M $, and $ \gamma $ much suppressed. At the same time,

\begin{equation}
h_D \sim 10^{-4} \, ,
\end{equation}
which is not surprising given the sizes of the Dirac Yukawa couplings in the other sectors, e.g. $ m_{e} / \kappa_+ $. As noted in \cite{Mohapatra:1980yp,Mohapatra:1986uf}, asking for small $ \beta_{1,2,3} $ in the definition of $ \gamma $ is not a problem, since quantum corrections to these same coefficients would not spoil their smallness.\footnote{Note that the suppression of $ \gamma $ implies the suppression of $ \kappa_L $ through Eq.~\eqref{eq:eqSeeSaw}.} It has also been argued over the literature that it is possible to evoke approximate symmetries to suppress the coefficient of $ m_{\nu} \propto \kappa^{2}_{+}/\kappa_R $ (e.g. Refs.~\cite{Deshpande:1990ip,Khasanov:2001tu,Kiers:2005gh}).

Note that Eq.~\eqref{eq:finalSeeSaw} is the usual motivation for considering the triplet scenario. However, in view of the requirement of further suppressions for explaining the smallness of $ \nu $ masses, Eq.~\eqref{eq:orders}, we see that the original interest for introducing triplets in the model (i.e. \textit{predicting} small neutrino masses), is less compelling than usually advocated. On the other hand, though in the doublet scenario no prediction is made concerning the smallness of the neutrino masses, it is a simpler scenario in the sense that no Yukawas $ Y^{lept}_{L,R} $ are present, thus implying that the mixing among leptonic generations is described by $ 3 \times 3 $ unitary matrices, and the number of physical scalar fields is lower (there are no doubly charged fields $ \delta^{\pm \pm}_{L,R} $).


\section{Conclusions}

Over this chapter, we have introduced LR Models, and we have seen their main features:

\begin{itemize}
	\item They consist in the gauge group $ SU(3)_c \times SU(2)_L \times SU(2)_R \times U(1)_{B-L} $, offering a framework where one is in principle able to implement parity $ \mathcal{P} $ or charge-conjugation $ \mathcal{C} $ symmetry, 
	\item In order to spontaneously break the new local symmetries, $ SU(2)_R \times U(1)_{B-L} \rightarrow U(1)_Y $, to implement the EW spontaneous breaking, $ SU(2)_L \times U(1)_Y \rightarrow U(1)_{EM} $, and to generate masses for the fermions, we include two scalar doublets $ \chi_R = (\mathbf{1},\mathbf{1},\mathbf{2},1) $ and $ \chi_L = (\mathbf{1},\mathbf{2},\mathbf{1},1) $, and a bi-doublet $ \Phi = (\mathbf{1},\mathbf{2},\mathbf{2},0) $, this being the minimal possible set of fields,
	\item The vacuum expectation value $ \langle \chi^0_R \rangle = \kappa_R / \sqrt{2} $ breaking the LR Model gauge group down to the SM one is expected to be much larger than $ \kappa \simeq 246 $~GeV, a way to explain why LR Model effects have not been observed so far,
	\item After the first symmetry breaking, new gauge bosons are present: a new charged gauge boson $ {W'}^\pm $ and a new neutral gauge boson $ {Z'}^0 $, whose masses are proportional to $ \kappa_R $,
	\item $ {W'}^\pm $ couples to quarks through a mixing matrix $ V^R $ analogous to the CKM matrix: its structure can be constrained by considering $ \mathcal{P} $ or $ \mathcal{C} $ to be exactly restored at the energy scale $ \sim \kappa_R $, but the full restoration of $ \mathcal{P} $ or $ \mathcal{C} $ may be achieved \textit{a priori} only at higher energy scales,
	\item The matrix $ V^R $, which has in principle an arbitrary structure, introduces new sources of $ \mathcal{C P} $ violation,
	\item Beyond the SM-like Higgs field $ h^0 $, other physical scalars are present. These are: three $ \mathcal{C P}- $even $ H^0_{1,2,3} $, two $ \mathcal{C P}- $odd $ A^0_{1,2} $, and two charged $ H^\pm_{1,2} $, which have masses proportional to $ \kappa_R $. The scalars $ H^0_{1,2} $ and $ A^0_{1,2} $ have Flavour Changing Neutral Couplings, a feature not present in the SM. This then implies contributions to meson-mixing observables that are relevant for phenomenology.
\end{itemize}

Starting from the next chapter, we aim at constraining the doublet scenario of the LR Model. First we will consider EWPO, in order to constrain the specific way in which local symmetries are spontaneously broken in the LR Model, and then we will shift to meson-mixing observables.

\chapter{Testing the SSB pattern of LR Models through EWPO and direct searches}\label{ch:EWPO}

The first class of observables of interest to constrain LR Models includes EWPO \cite{Bernard:2007cf,Silva:2015tha}. In the Standard Model, they constitute constraints of utmost importance for testing the consistency of one of its most salient features, namely the spontaneous breaking of the Electroweak symmetry $ SU(2)_L \times U(1)_Y $, as seen in Chapter~\ref{ch:SM}. Being very sensitive to the precise way in which the Brout-Englert-Higgs mechanism operates in the SM, one expects to gain valuable information concerning the way in which the LR Model gauge group is spontaneously broken.

A meaningful computation of these observables in the SM framework requires radiative corrections, as discussed in Section~\ref{sec:inputsforEWPO}. On top of that, we add tree level contributions introduced by the LR Model with respect to the pure SM (i.e. the SM with no extra fields or parameters), and from this point the LR Model can be constrained - we postpone the discussion of radiative corrections in the LR Model framework to \cite{Bernard}. We reconsider the global fit of the full set of observables already studied in the context of the SM in Section~\ref{sec:EWtestsSM}, see also \cite{Erler:1999ub} \cite{Gonzalez:2011he} \cite{Eberhardt:2012gv} \cite{Wells:2014pga}. Moreover, we also consider the impact of direct searches for $ W' $ bosons.

It should be stressed that throughout this chapter we are not interested in flavour-sensitive observables (this being the concern of Chapter \ref{ch:PHENO}): the main goal here is to probe one simple and in some sense new realization of Left-Right Models (the doublet LR Model) from the point of view its Spontaneous Symmetry Breaking features.

\section{Corrections from the Left-Right Model to the EWPO}

We will correct the SM predictions by including the contributions from the LR Model at tree level. Then, from these corrections, we will be able to constrain this particular extension of the SM. We consider the calculation of a low-energy effective theory at tree level by integrating out the (heavy) gauge bosons $ Z', W' $. 

In our case, contributions from the extended Higgs sector at tree level are not important, since they are expectedly heavy (due to Flavour Changing Neutral Current related processes that we will discuss later) and we work with energies around $ M_Z $ or lower. Moreover the physical scalars couple with strengths proportional to the masses of the quarks and leptons, which are the light degrees of freedom for the tree level processes under discussion.

\subsection{Effective Lagrangians}

Here we will closely follow \cite{Hsieh:2010zr}. Having $ \kappa_L \neq 0 $, however, is the essential difference between our analysis and the one found in reference \cite{Hsieh:2010zr}, where $ \kappa_L $ is set to zero as a simplification.

The Lagrangian including gauge boson interactions with fermions and gauge boson mass terms is

\begin{eqnarray}
\mathcal{L} &=& \frac{1}{2} M^2_Z Z^\mu Z_\mu + M^2_W W^{+ \mu} W^-_\mu + \frac{1}{2} M^2_{Z'} {Z'}^\mu {Z'}_\mu + M^2_{W'} {W'}^{+ \mu} {W'}^-_\mu \nonumber\\
&+& Z_\mu \left( J^{0 \mu} - \frac{\delta \tilde{M}^2_Z}{\tilde{M}^2_{Z'}} K^{0 \mu} \right) + {Z'}_\mu \left( K^{0 \mu} + \frac{\delta \tilde{M}^2_Z}{\tilde{M}^2_{Z'}} J^{0 \mu} \right) + A_\mu J^\mu_{EM} \\
&+& \left[ W^+_\mu \left( J^{+ \mu} - \frac{\delta \tilde{M}^2_W}{\tilde{M}^2_{W'}} K^{+ \mu} \right) + {W'}^+_\mu \left( K^{+ \mu} + \frac{\delta \tilde{M}^2_W}{\tilde{M}^2_{W'}} J^{+ \mu} \right) + (+ \leftrightarrow -) \right] \, , \nonumber
\end{eqnarray}
where the currents $ J^{0}, K^{0}, J^{\pm}, K^{\pm}, J_{EM} $ are read from Section~\ref{sec:couplingsTreeLevel}. Note that, in the expressions above weak charged currents are in the flavour basis: going to the mass basis introduces the mixing matrices $ V^{L,R} $ and $ V^{L,R}_{lept} $, which are of no relevance here, since the masses of the kinematically allowed fermions are too small compared to $ M_W $ to be significant for the determination of $ \Gamma_{W} $, at the precision we need to calculate LR Model corrections.

At energy scales $ \sqrt{s} $ much below the masses $ M_{Z',W'} $, one can integrate out the $ W' $ and the $ Z' $ leading to

\begin{eqnarray}\label{eq:oneEffectiveLag}
\mathcal{L}^{s \ll M^2_{Z',W'}}_{\rm eff} &=& \frac{1}{2} M^2_Z Z^\mu Z_\mu + M^2_W W^{+ \mu} W^-_\mu \\
&+& Z_\mu \left( J^{0 \mu} - \frac{\delta \tilde{M}^2_Z}{\tilde{M}^2_{Z'}} K^{0 \mu} \right) + \left[ W^+_\mu \left( J^{+ \mu} - \frac{\delta \tilde{M}^2_W}{\tilde{M}^2_{W'}} K^{+ \mu} \right) + (+ \leftrightarrow -) \right] \nonumber\\
&+& A_\mu J^\mu_{EM} - \frac{1}{2 \tilde{M}^2_{Z'}} K^{0 \mu} K^0_\mu - \frac{1}{\tilde{M}^2_{W'}} K^{+ \mu} K^-_\mu \, , \nonumber
\end{eqnarray}
where one recognizes the SM Lagrangian corrected by terms suppressed by $ (\tilde{M}_{Z'})^{-2} $ or $ (\tilde{M}_{W'})^{-2} $.

At energies $ \sqrt{s} $ much smaller than $ M_{Z,W} $ one can integrate out the $ W $ and the $ Z $, and we are left with

\begin{eqnarray}\label{eq:lastLagrangian}
\mathcal{L}^{s \ll M^2_{Z,W}}_{\rm eff} = &-& \frac{1}{2 \tilde{M}^2_Z} \left[ J^{0 \mu} J^0_\mu + \frac{\tilde{M}^2_Z}{\tilde{M}^2_{Z'}} \left( \frac{\delta \tilde{M}^4_Z}{\tilde{M}^4_Z} J^{0 \mu} J^0_\mu - 2 \frac{\delta \tilde{M}^2_Z}{\tilde{M}^2_Z} J^{0 \mu} K^0_\mu + K^{0 \mu} K^0_\mu \right) \right] \nonumber\\
&-& \frac{1}{\tilde{M}^2_W} \Bigg[ J^{+ \mu} J^-_\mu + \frac{\tilde{M}^2_W}{\tilde{M}^2_{W'}} \Bigg( \frac{\delta \tilde{M}^4_W}{\tilde{M}^4_W} J^{+ \mu} J^-_\mu - \frac{\delta \tilde{M}^2_W}{\tilde{M}^2_W} ( J^{+ \mu} K^-_\mu + J^{- \mu} K^+_\mu ) \nonumber\\
&+& K^{+ \mu} K^-_\mu \Bigg) \Bigg] + A_\mu J^\mu_{EM} \, .
\end{eqnarray}
It is from the two last expressions that we have calculated the LR Model corrections used in the fit.

As a parenthesis, compared to the triplet case, where $ SU(2)_R \times U(1)_{B-L} $ is triggered by a triplet $ \Delta_R $ under $ SU(2)_R $ instead of a doublet $ \chi_R $ as in our case, the only difference we have is (taking $ \langle \Delta_R^0 \rangle = \langle \chi_R^0 \rangle $, which may not be the result of phenomenological studies)

\begin{equation}
( \tilde{M}^2_{Z'} )_{doublet} \rightarrow 4 ( \tilde{M}^2_{Z'} )_{triplet} \, , \qquad ( \tilde{M}^2_{W'} )_{doublet} \rightarrow 2 ( \tilde{M}^2_{W'} )_{triplet} \, ,
\end{equation}
while the other mass parameters ($ \tilde{M}^2_Z $, etc.) remain the same when $ \kappa_L = 0 $. 


\subsection{Parameters used in the fit}

Neglecting electroweak corrections, functions of $ \alpha, m_{top} $ and $ M_H $, as well as terms from $ J^+_\mu K^{- \mu} $ and $ K^+_\mu K^{- \mu} $, that do not interfere with the SM $ J^+_\mu J^{- \mu} $ in the limit where the mass of the muon is neglected, one gets the following expression for the Fermi constant

\begin{equation}\label{eq:GFexpressionEWPO}
\frac{G_{F}}{\sqrt{2}} = \frac{g_{L}^2}{8 \tilde{M}_{W}^2} \left( 1 + \frac{\delta \tilde{M}_{W}^4}{\tilde{M}_{W}^2 \tilde{M}_{W'}^2} \right) = \frac{1}{2 \kappa^2} \left( 1 + \epsilon^2 R^4 s^2_{2 \beta} \right) \, ,
\end{equation}
\noindent
where one employs

\begin{equation}\label{eq:R2definition}
R^2 = \left(1 + \frac{w^2}{1 + r^2} \right)^{-1} \leq 1 \, , \quad\quad R^2 \geq 0 \, ,
\end{equation}
and
\begin{eqnarray}\label{eq:sin2betadefinition}
s^{}_{2 \beta} = \frac{2 r}{1 + r^2} \, .
\end{eqnarray}
The Fermi constant is measured from the $ \mu $ lifetime. Note that its value is extremely well measured, and it is given by
\begin{equation}
G_F = ( 1.1663787 \pm 0.0000006 ) \cdot 10^{-5} \operatorname{GeV^{-2}} \qquad ({\rm PDG}) \, .
\end{equation}
Due to the accuracy compared to other experimental inputs, we are going to neglect the uncertainty in the last expression.

To further continue, we recall the expression for the mass of the $ Z $ boson

\begin{equation}\label{eq:MZexpressionEWPO}
M^2_{Z} = \tilde{M}^2_{Z} - \frac{\delta \tilde{M}^4_{Z}}{\tilde{M}^2_{Z'}} = \frac{\alpha (M_Z) \pi}{s^2_{\theta} c^2_{\theta}} \kappa^2 \left[ 1 - \epsilon^2 c^4_\phi \left( 1 - \frac{1 - R^2}{c^2_\phi} \right)^2 \right] \, ,
\end{equation}
where

\begin{equation}\label{eq:cRexpression}
c^2_\phi = 1 - \frac{s^{2}_{\theta}}{1 - s^{2}_{\theta}} \left( \frac{g_{L}}{g_{R}} \right)^{2} \simeq 1 - \frac{1}{3} \left( \frac{g_{L}}{g_{R}} \right)^{2}
\end{equation}
has already been defined and is a parameter indicating the value of the ratio $ g_L / g_R $.

In EW precision tests, it is standard to use $ M_Z $ and $ G_F $ as parameters of the fit, since they are very precisely measured. We will do the same here, trading $ s_{\theta}, \kappa $ by $ G_F, M_Z $, thus leading to the following relations between the SM and the LR Model


\begin{eqnarray}
\kappa^2 &=& (\kappa^2)_{SM} \left( 1 + \epsilon^2 R^4 s^2_{2 \beta} \right) \, , \\
s^2_{\theta} c^2_{\theta} &=& (s^2_{\theta} c^2_{\theta})_{SM} \left[ 1 + \epsilon^2 R^4 s^2_{2 \beta} - \epsilon^2 c^4_\phi \left( 1 - \frac{1 - R^2}{c^2_\phi} \right)^2 \right] \, , \nonumber
\end{eqnarray}
where

\begin{equation}
(\kappa^2)_{SM} \equiv \frac{1}{\sqrt{2} G_{F}} \, , \quad (s^2_{\theta} c^2_{\theta})_{SM} \equiv \frac{\alpha (M_Z) \pi}{\sqrt{2} G_{F} M^2_{Z}} \, .
\end{equation}



Note that there are no tree level corrections to the fine structure constant, as can be seen from the previous Lagrangians. Therefore

\begin{equation}
\alpha = ( \alpha )_{SM} = \frac{e^2}{4 \pi} \, .
\end{equation}

\subsection{Expressions for LR Model corrections}

We now use the definitions given in Appendix~\ref{sec:EWPOTreeLevel} to calculate the tree level expression of an observable in the SM, $ \mathcal{X}^{tree}_{SM} $, and the tree level expression in the LR Model, $ \mathcal{X}^{tree}_{LRM} $, from the effective Lagrangian in Eq.~\eqref{eq:oneEffectiveLag} or \eqref{eq:lastLagrangian}.

Following the discussion in Chapter~\ref{sec:EWtestsSM}, we will thus use the following set of parameters

\begin{equation}\label{eq:parametersSprime1}
\mathcal{S}' = \{ m^{pole}_{top} , \alpha_{s} (M_{Z}) , M_{Z} , M_H , \Delta \alpha^{(5)}_{had} (M_{Z}), \epsilon, c_\phi, r, w \} \, ,
\end{equation}
cf. Eq.~\eqref{eq:parametersEWPOSM}.

We have then the following expressions for $ \frac{1}{\epsilon^2} \cdot \frac{\mathcal{X}^{tree}_{LRM}}{\mathcal{X}^{tree}_{SM}} $ (numerical values are truncated for compactness), where one employs the numerical approximation $ ( s^2_{\theta} )_{SM} \simeq 0.234 $, sufficient for the precision required here:  

\begin{itemize}
	\item Asymmetries
\begin{eqnarray}
&& \frac{1}{\epsilon^2} \cdot \delta \mathcal{A}_{b} / \mathcal{A}_{b} = -0.113 - 0.649 \, c^2_\phi + 0.762 \, c^4_\phi \\ && \quad + (-0.119 + 0.994 \, c^2_\phi) \, R^2 + R^4 \, 0.232 \, (1 - s^{2}_{2 \beta}) \, , \nonumber\\
&& \frac{1}{\epsilon^2} \cdot \delta \mathcal{A}_{c} / \mathcal{A}_{c} = 3.12 - 8.97 \, c^2_\phi + 
 5.86 \, c^4_\phi \\ && \quad + (-4.90 + 7.65 \, c^2_\phi) \, R^2 + 
 R^4 \, 1.79 \, (1 - s^{2}_{2 \beta}) \, , \nonumber\\
&& \frac{1}{\epsilon^2} \cdot \delta \mathcal{A}_{e, \mu, \tau} / \mathcal{A}_{e, \mu, \tau} = 50.7 - 118. \, c^2_\phi + 
 66.9 \, c^4_\phi \\ && \quad + (-71.1 + 87.3 \, c^2_\phi) \, R^2 + 
 R^4 \, 20.4 \, (1 - s^{2}_{2 \beta}) \, , \nonumber\\
&& \frac{1}{\epsilon^2} \cdot \delta A_{FB} (b) / A_{FB} (b) = 50.6 + 67.6 \, c^4_\phi - 71.2 \, R^2 \\ && \quad + 
 c^2_\phi \, (-118. + 88.2 \, R^2) + 
 R^4 \, 20.6 \, (1 - s^{2}_{2 \beta}) \, , \nonumber\\
&& \frac{1}{\epsilon^2} \cdot \delta A_{FB} (c) / A_{FB} (c) = 53.8 + 72.7 \, c^4_\phi - 76.0 \, R^2 \\ && \quad + 
 c^2_\phi \, (-127. + 94.9 \, R^2) + 
 R^4 \, 22.2 \, (1 - s^{2}_{2 \beta}) \, , \nonumber\\
&& \frac{1}{\epsilon^2} \cdot \delta A_{FB} (e, \mu, \tau) / A_{FB} (e, \mu, \tau) = 101. + 134. \, c^4_\phi - 142. \, R^2 \\ && \quad + 
 c^2_\phi \, (-235. + 174. \, R^2) + 
 R^4 \, 40.8 \, (1 - s^{2}_{2 \beta}) \, , \nonumber
\end{eqnarray}
	\item $ Z $ total width
\begin{eqnarray}
&& \frac{1}{\epsilon^2} \cdot \delta \Gamma_{Z} / \Gamma_{Z} = 0.783 + 0.141 \, c^4_\phi - 2.13 \, R^2 \\ && \quad + 
 c^2_\phi \, (-0.924 + 1.49 \, R^2) + 
 R^4 \, 1.35 \, (1 - s^{2}_{2 \beta}) \, , \nonumber
\end{eqnarray}
	\item Rations of $ Z $ widths
\begin{eqnarray}
&& \frac{1}{\epsilon^2} \cdot \delta R_{b} / R_{b} = -0.422 - 0.194 \, c^4_\phi + 
 c^2_\phi \, (0.616 - 0.253 \, R^2) \\ && \quad + 0.481 \, R^2 - 
 R^4 \, 0.0591 \, (1 - s^{2}_{2 \beta}) \, , \nonumber\\
&& \frac{1}{\epsilon^2} \cdot \delta R_{c} / R_{c} = 0.814 + 0.374 \, c^4_\phi + 
 c^2_\phi \, (-1.19 + 0.488 \, R^2) \\ && \quad - 0.928 \, R^2 + 
 R^4 \, 0.114 \, (1 - s^{2}_{2 \beta}) \, , \nonumber\\
&& \frac{1}{\epsilon^2} \cdot \delta R_{e, \mu, \tau} / R_{e, \mu, \tau} = -0.218 + 0.993 \, c^4_\phi - 0.0850 \, R^2 \\ && \quad + 
 c^2_\phi \, (-0.775 + 1.30 \, R^2) + 
 R^4 \, 0.303 \, (1 - s^{2}_{2 \beta}) \, , \nonumber
\end{eqnarray}
	\item Cross-section of the $ Z $ into hadrons
\begin{eqnarray}
&& \frac{1}{\epsilon^2} \cdot \delta \sigma_{had} / \sigma_{had} = 1.09 - 0.140 \, c^4_\phi + 
 c^2_\phi \, (-0.948 - 0.183 \, R^2) \nonumber\\ && \quad - 1.05 \, R^2 - 
 R^4 \, 0.0428 \, (1 - s^{2}_{2 \beta}) \, ,
\end{eqnarray}
where a kinematic correction to the $ Z \rightarrow \overline{b} b $ process \cite{Rosner:2001zy} was included in the calculation of $ \sigma_{had} $ in the denominator of $ \delta \sigma_{had} / \sigma_{had} $:

\begin{equation}\label{eq:equation4}
\left( 1 - \frac{4 m^2_{b}}{M_{Z}^2} \right)^{1/2} \left[ g^b_{V} \left( 1 + \frac{2 m^2_{b}}{M_{Z}^2} \right) + g^b_{A} \left( 1 - \frac{4 m^2_{b}}{M_{Z}^2} \right) \right] \simeq 0.99 .
\end{equation}
	\item $ W $ mass and total width
\begin{eqnarray}
&& \frac{1}{\epsilon^2} \cdot \delta M_{W} / M_{W} = 0.719 - 1.44 \, c^2_\phi + 0.719 \, c^4_\phi \\ && \quad - 1.44 \, R^2 + 
 1.44 \, c^2_\phi \, R^2 + 0.719 \, R^4 (1 - s^{2}_{2 \beta}) \, , \nonumber \label{eq:paramforMW}\\
&& \frac{1}{\epsilon^2} \cdot \delta \Gamma_{W} / \Gamma_{W} = 2.16 - 4.32 \, c^2_\phi + 2.16 \, c^4_\phi \\ && \quad - 4.32 \, R^2 + 
 4.32 \, c^2_\phi \, R^2 + 2.16 \, R^4 (1 - s^{2}_{2 \beta}) \, , \nonumber
\end{eqnarray}
	\item Atomic Parity Violation
\begin{eqnarray}
&& \frac{1}{\epsilon^2} \cdot \delta Q_{W} (p) / Q_{W} (p) = 52.1 + 20.6 \, c^4_\phi - 73.7 \, R^2 \label{eq:paramforQWp}\\ && \quad + 
 c^2_\phi \, 88.0 \, (1 - R^2) + 
 R^4 \, 21.6 \, (1 - s^{2}_{2 \beta}) \, , \nonumber\\
&& \frac{1}{\epsilon^2} \cdot \delta Q_{W} (n) / Q_{W} (n) = -1 + R^4 \, (1 - s^{2}_{2 \beta}) \, . \label{eq:paramforQWn}
\end{eqnarray}
\end{itemize}

Together with the expressions in Appendix~\ref{sec:EWPOTreeLevel}, this set of expressions therefore gives the SM and LR Model tree level contributions $ \mathcal{X}^{tree}_{SM} $ and $ \mathcal{X}^{tree}_{LRM} = \mathcal{O} (\epsilon^2) $. The Standard Model $ \mathcal{X}_{SM} $ can be considered up to the highest order known

\begin{equation}
\mathcal{X}_{SM} = \mathcal{X}^{tree}_{SM} + \mathcal{X}^{radiative}_{SM} \, .
\end{equation}

\noindent
Finally, the expressions corrected by the LR Model are given by

\begin{equation}
\mathcal{X} = \mathcal{X}_{SM} \Bigg( 1 + \underbrace{ \frac{\mathcal{X}^{tree}_{LRM}}{\mathcal{X}^{tree}_{SM}} }_{\mathcal{O} (\epsilon^2)} \Bigg) + \mathcal{O} (\epsilon^4) \, ,
\end{equation}
where corrections coming at order $ \mathcal{O} (\epsilon^4) $ are not taken into account, and $ \mathcal{X}^{radiative}_{SM} \mathcal{X}^{tree}_{LRM} / \mathcal{X}^{tree}_{SM} $ has negligible impact.

\section{Direct searches for LR Model particles}\label{sec:directSearches}




There has been an intensive program to search for a $ W' $ coupling exclusively to right-handed currents. One of the analysed decay channels is

\begin{equation}
{W'}^{+} \rightarrow t \bar{b}
\end{equation}
(or $ {W'}^{-} \rightarrow b \bar{t} $), where the final hadronic pair is on-shell. The pair $ t \bar{b} $ then decays into $ \ell^+ \nu b \bar{b} $ \cite{Aad:2014xea,CMS:2015aal,Chatrchyan:2014koa}, where $ \ell = e, \mu $, or $ q \bar{q} b \bar{b} $ \cite{Aad:2014xra,Khachatryan:2015edz}. Exclusion limits at $ 95~\% $ are set on the mass of the $ W' $ for the special case $ g_R = g_L $, generally excluding masses below $ \sim 2 $~TeV (or excluding large values of $ g_R $, when $ g_R \neq g_L $, for a given value of $ M_{W'} $). In many of these analyses, it is assumed that the right-handed neutrino is heavier than the $ W' $, and therefore the $ W' $ decays exclusively hadronically. The analyses may also consider a $ V^R $ matrix proportional to the identity $ \mathbf{1}_{3 \times 3} $ as a simplification.

Another channel, interesting when right-handed neutrinos are allowed to be heavy, is

\begin{equation}
W' \rightarrow \ell_1 N \rightarrow \ell_1 \ell_2 {W'}^{*} \rightarrow \ell_1 \ell_2 q \bar{q} \, ,
\end{equation}
where $ N $ is a right-handed neutrino. This channel, known as Keung-Senjanovi\'c \cite{Keung:1983uu}, leads to a same-sign pair of leptons in the final state, or in other words lepton number violation (however, the actual analyses may consider both, same- and different-sign, final states). Following a series of simplifications to guarantee the predictivity of the analysis, one is able to set lower bounds on the mass of the $ W' $ of $ \sim 3 $~TeV \cite{Khachatryan:2014dka} (more generally, an exclusion region in the plan $ M_{W'}, M_N $ is quoted). Since the mass of the right-handed neutrino could be light as well, the decay $ W'_L \rightarrow \ell \nu $, where $ \nu $ is a light neutrino, has also been considered, leading to similar bounds on the mass of the $ W'_L $, of $ \sim 2 $~TeV \cite{CMS:2011yqa}, and the more recent stronger lower bound of $ \sim 4.74 $~TeV \cite{ATLAS:2016ecs}. This however corresponds to searches of a $ W'_L $ which is interpreted as a heavier version of the $ W $, coupling exclusively to left-handed currents (and excluding couplings to the SM bosons $ W, Z, h $).


Lower bounds are also found from diboson resonance searches, such as \linebreak $ WW / W Z \rightarrow \ell \nu j j $ \cite{Aad:2013pdy}. In this respect, an excess observed in the invariant mass range $ 1.8 - 2.0 $~TeV \cite{Aad:2015owa} has triggered a relatively large number of theoretical studies, specially when other excesses are considered simultaneously (such as an excess of $ eejj $ in the invariant mass range $\approx 2.1 $~TeV), see e.g. \cite{Brehmer:2015cia} (for a $ W' $ interpretation), \cite{Dobrescu:2015jvn} (for a $ W', Z' $ interpretation) and also \cite{Cheung:2015nha,Deppisch:2015cua}. However, more recent experimental data from the Run-II of LHC have not confirmed such excess, see e.g. \cite{CMS:2015nmz}.

We could still comment on other channels, such as $ W' \rightarrow tb \rightarrow \ell \nu j j $ \cite{Aad:2012ej}, or $ W' \rightarrow t t j $ \cite{Chatrchyan:2012su}. Our point here, however, is that as a matter of fact LR Models have many parameters ($ g_R $, $ V^R $, etc.) or realizations (with heavy or light Dirac and/or Majorana right-handed neutrinos, etc.) and analyses usually consider particular cases. That being said, we consider that it is sufficient for the analysis presented here to assume a strict lower bound of $ 2 $~TeV for the mass of the $ M_{W'} $.

Apart from searching for the $ W' $, $ Z' $ resonances are also intensively looked for in the channels $ \ell \ell $, $ \tau \tau $, $ b b $, $ t t $, etc. As for the $ W' $, many of these analyses consider the Sequential Standard Model (SSM) picture \cite{Altarelli:1989ff}, where new gauge bosons $ Z' $ and $ W'_L $ are assumed to have couplings equal to the ones in the SM. In the SSM, lower bounds of order $ 2 - 3 $~TeV are set on the mass of the $ Z' $,\footnote{Experimental bounds are communicated in terms of production ratio times branching ratio, $ \sigma \times \mathcal{B} $. It does not look obvious, at least at this point, that a simple rescaling of $ \mathcal{B} $ (equal to the ratio of the couplings in the SSM and the LR Model frameworks) would be enough to reinterpret bounds derived in the SSM, in part because the widths of the $ Z' $ in both two cases are different.} while some less constraining limits also exist for LR Models, $ M_{Z'} \gtrsim 1050 $~GeV \cite{Chatrchyan:2011wq} or even higher $ M_{Z'} \gtrsim 2 - 2.5 $~TeV \cite{Patra:2015bga}. In view of that, and due to the weaker impact on the analysis of a lower limit for $ M_{Z'} $ (we remind the reader that in our version of the LR Model, the relation $ M_{Z'} = M_{W'} / c_\phi $ holds), we do not include any \textit{a priori} lower bound on the mass of the $ Z' $.

There has also been interest for the search of LR Model scalar particles, such as the doubly charged scalar when triplets under $ SU(2)_R $ instead of doublets are considered. In view of the very strong bounds coming from indirect processes (that we will consider later on), direct searches are not competitive yet (see Ref.~\cite{Dev:2016dja} for future collider searches).



\section{Results of the global fits}



\begin{figure}
\centering
\includegraphics[scale=0.3]{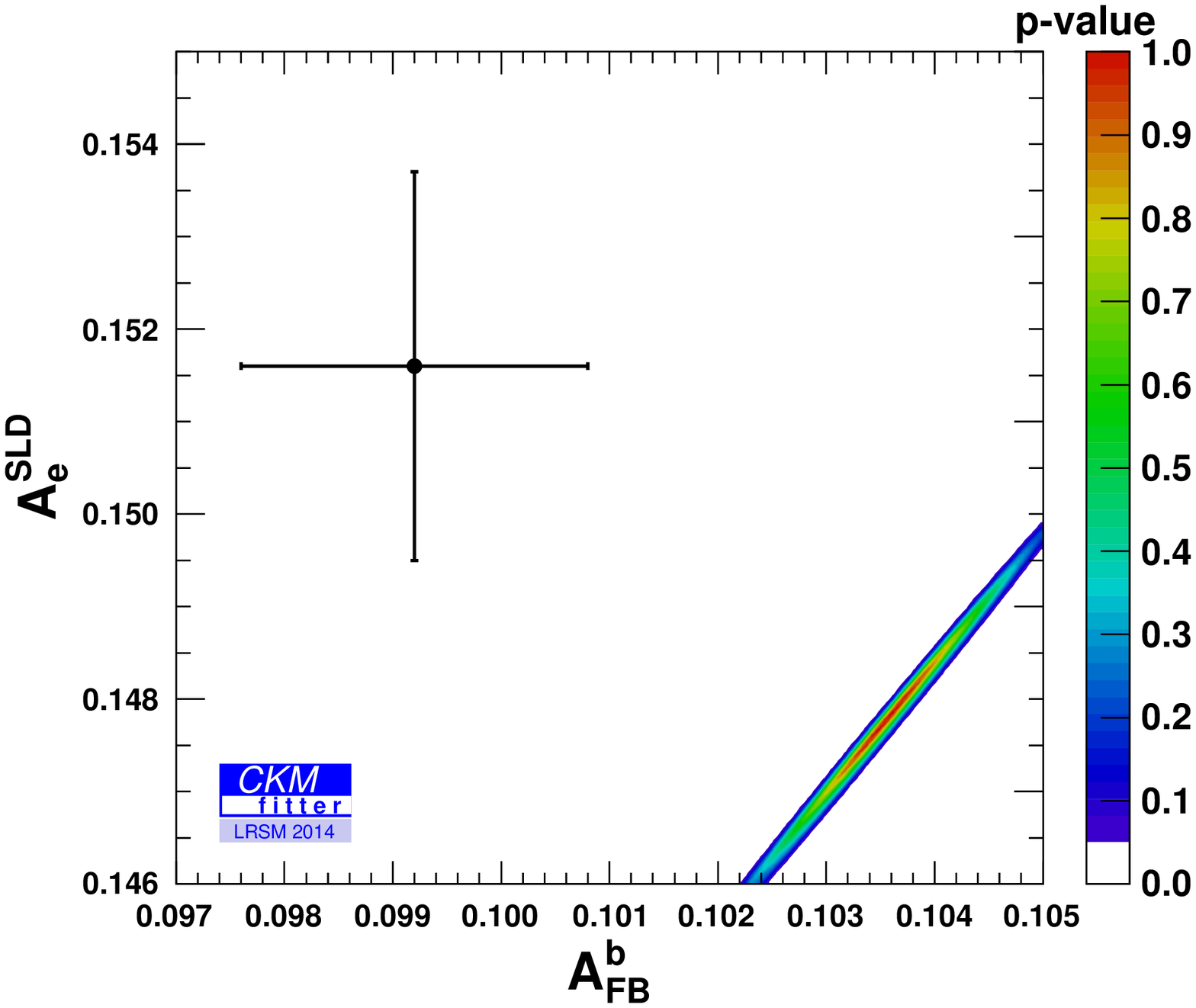}
\hspace{1cm}
\includegraphics[scale=0.3]{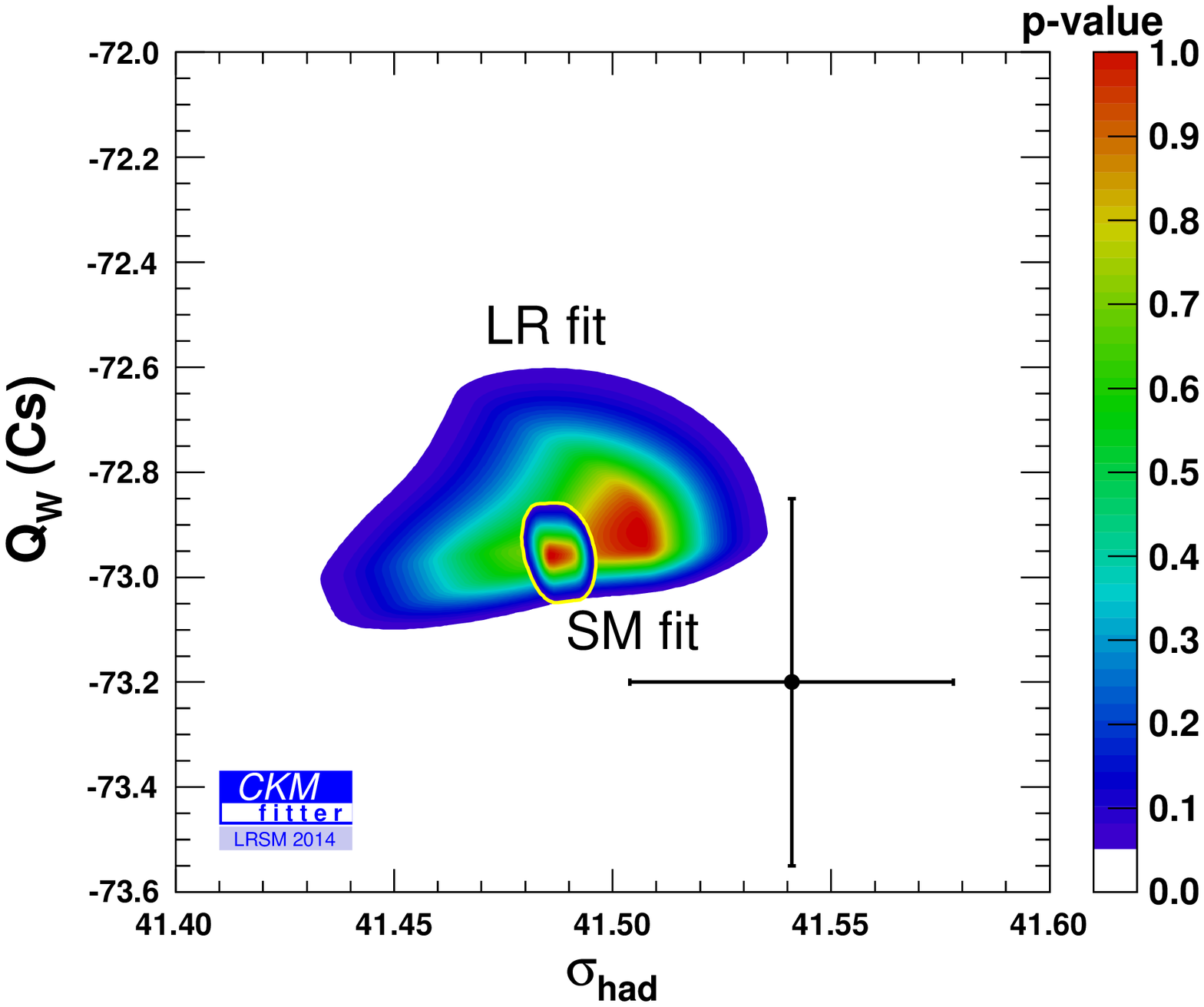}
\caption{\it (Left) Correlation among $ A^{b}_{FB} $ and $ A^{SLD}_e $ in the LR Model showing a strong tension between the two measurements (the SM fit has essentially the same p-value curve and is not shown). (Right) Correlation among $ \sigma_{had} $ and $ Q_W (Cs) $ in both SM and LR Model.}\label{fig:QWCs}
\end{figure}

The set of observables we use in our fit is given in Table~\ref{tab:tab2}, and they were combined using \texttt{CKMfitter}, as in Chapter~\ref{ch:SM}. In our analysis, the parameter $ c_\phi $ is allowed to vary over the range $ [0.1, 0.99] $, resulting from the perturbativity requirements $ g_R^2 / 4 \pi < 1 $ and $ g_{B-L}^2 / 4 \pi < 1 $ (symmetrically, $ c_\phi $ could be considered over the range $ [- 0.99, - 0.1] $, but the observables are not sensitive to the sign of $ c_\phi $). The ratio $ r = \kappa_2 / \kappa_1 $ (where both VEVs are positive) can be taken over the range $ [0, 1] $, otherwise we could redefine $ r $ as $ 1/r $ altogether with $ w/r \rightarrow w $, see Eqs.~\eqref{eq:R2definition}-\eqref{eq:sin2betadefinition}. On the other hand, the range for $ w $ can be larger, and we consider $ w \in [0, 3] $. Having too large values for $ w $ would imply small $ \kappa_{1,2} $, which set the scale of the masses of the fermions: $ w \leqslant 3 $ is therefore required in order to avoid too large Yukawa couplings of the top-quark.

The value for $ \chi^{2}_{\min} $ is 21.47, and for 19 d.o.f. we have the resulting p-value of $ \sim 31~\% $, allowing for a meaningful extraction of the physical parameters. One can consider the Standard Model as a limit case of the LR Model for which $ \epsilon \rightarrow 0^+ $, thus loosing all the dependences on the other LR Model parameters, i.e. $ c_\phi, r, w $. It then follows that the quantity $ \chi^2_{min} |_{SM} - \chi^2_{min} |_{LRM} $ is distributed as a $ \chi^2 $ with 1 degree of freedom. Therefore, the SM hypothesis in the context of the LR Model has a pull of

\begin{equation}
{\rm pull} = \sqrt{\chi^2_{min} |_{SM} - \chi^2_{min} |_{LRM}} = 0.88 \, \sigma \, ,
\end{equation}
interpreted as a $ 0.88 \; \sigma $ deviation, which at this stage is not large enough for substituting the SM hypothesis for the LR Model one.

The results for the best fit point and $ 68~\% $ CL intervals are given in Table~\ref{tab:tab2}. As seen from the predicted values of the different observables, the global fit of the LR Model is rather similar to the SM one discussed in Section~\ref{sec:EWtestsSM}. One sees that the agreement with the experimental values is improved for some observables (e.g. $ \sigma_{had}, M_W $) at the expense of others (e.g. $ R_{e, \mu}, Q_W (Cs) $). Note as well from the pulls shown in Table~\ref{tab:tab2} that, similarly to the SM case, under the LR Model hypothesis the experimental inputs for $ A_{FB} (b, \tau) $, $ \mathcal{A}^{SLD}_e $ and $ \sigma_{had} $ show sizable tensions with the underlying model, which are still left unexplained. Of course, different observables (and their pulls) are correlated: for example, Figure \ref{fig:QWCs} (Left) shows the correlation of $ A_{FB} (b) $ and $ \mathcal{A}^{SLD}_e $, whereas Figure \ref{fig:QWCs} (Right) shows the correlation of $ Q_W (Cs) $ and $ \sigma_{had} $ in both SM and LR Model fits, indicating the possibility of decreasing the tension in $ \sigma_{had} $ at the cost of $ Q_W (Cs) $.






Moreover, one sees in Table~\ref{tab:tab2} that the observables are not powerful enough to constrain $ c_\phi $, $ r $ and $ w $ independently at $ 1 \sigma $ (a situation indicated by ``flat" in that table). One also notes that the true value of $ w $ is poorly constrained, $ w $ preferring the highest value possible we allow it to have. This preference is also illustrated in Table~\ref{tab:tab3} (note that in this table we have fixed $ M_{W'} $ to 1.5~TeV, whereas in the rest of this chapter we have considered the lower bound of 2~TeV), indicated by the smaller values of $ \chi^2_{min} $ one has for larger $ w $. Moreover, when $ w = 0 $, $ g_{B-L} $ reaches its perturbativity limit, $ g^{2}_{B-L} = 4 \pi $.



\begin{table}
\begin{center}
\begin{tabular}{|cc|c|c|c|}
\hline
Observable & & input & LRM fit (1 $ \sigma $) & pull \\
\hline
  $ \Delta \alpha_{had}^{(5)} $ && - & $ 0.02812^{+0.00085}_{-0.00107} $ & - \\
  $ M_H $ [GeV] & \cite{Aad:2012tfa} \cite{Chatrchyan:2012xdj} & $125.7 \pm 0.4$ & $ 125.70 \pm 0.40 $ & 0.81 \\
  $ m_{top}^{pole} $ [GeV] & \cite{ATLAS:2014wva} & $ 173.34 \pm 0.36 \pm 0.67 $ & $ 174.03^{+0.36}_{-1.56} $ & 0.64 \\
  $ M_Z $ [GeV] & \cite{ALEPH:2005ab} & $91.1876 \pm 0.0021$ & $ 91.1875 \pm 0.0021 $ & 0.49 \\
  $ \alpha_s $ & \cite{Beringer:1900zz} & $0.1185 \pm 0 \pm 0.0005$ & $ 0.11900^{+0.00010}_{-0.00101} $ & 0.81 \\
\hline
  $ c_\phi $ && $ [0.1, 0.99] $ & 0.49/flat & - \\
  $ \epsilon $ && $ \geq 0 $ & $ 0.025^{+0.033}_{-0.025} $ & - \\
  $ r $ && $ [0, 1] $ & 0.0075/flat & - \\
  $ w $ && $ [0, 3] $ & large, read text & - \\
\hline
  $ \Gamma_Z $ [GeV] & \cite{ALEPH:2005ab} & $2.4952 \pm 0.0023$ & $ 2.49485^{+0.00082}_{-0.00096} $ & 0.47 \\
  $ \sigma_{had} $ [nb] & \cite{ALEPH:2005ab} & $41.541 \pm 0.037$ & $ 41.5067^{+0.0083}_{-0.0333} $ & 1.78 \\
  $ R_b $ & \cite{ALEPH:2005ab} & $0.21629 \pm 0.00066$ & $ 0.215737^{+0.000067}_{-0.000031} $ & 0.60 \\
  $ R_c $ & \cite{ALEPH:2005ab} & $0.1721 \pm 0.0030$ & $ 0.172292^{+0.000030}_{-0.000077} $ & 0.13 \\
  $ R_e $ & \cite{ALEPH:2005ab} & $20.804 \pm 0.050$ & $ 20.7356^{+0.0136}_{-0.0085} $ & 1.01 \\
  $ R_\mu $ & \cite{ALEPH:2005ab} & $20.785 \pm 0.033$ & $ 20.7356^{+0.0136}_{-0.0085} $ & 1.35 \\
  $ R_\tau $ & \cite{ALEPH:2005ab} & $20.764 \pm 0.045$ & $ 20.7826^{+0.0136}_{-0.0085} $ & 0.65 \\
  $ A_{FB} (b) $ & \cite{ALEPH:2005ab} & $0.0992 \pm 0.0016$ & $ 0.10356^{+0.00081}_{-0.00080} $ & 2.87 \\
  $ A_{FB} (c) $ & \cite{ALEPH:2005ab} & $0.0707 \pm 0.0035$ & $ 0.07401 \pm 0.00062 $ & 0.60 \\
  $ A_{FB} (e) $ & \cite{ALEPH:2005ab} & $0.0145 \pm 0.0025$ & $ 0.01637 \pm 0.00025 $ & 0.29 \\
  $ A_{FB} (\mu) $ & \cite{ALEPH:2005ab} & $0.0169 \pm 0.0013$ & $ 0.01637 \pm 0.00025 $ & 0.30 \\
  $ A_{FB} (\tau) $ & \cite{ALEPH:2005ab} & $0.0188 \pm 0.0017$ & $ 0.01637 \pm 0.00025 $ & 1.42 \\
  $ \mathcal{A}_b $ & \cite{ALEPH:2005ab} & $0.923 \pm 0.020$ & $ 0.93435^{+0.00046}_{-0.00017} $ & 0.40 \\
  $ \mathcal{A}_c $ & \cite{ALEPH:2005ab} & $0.670 \pm 0.027$ & $ 0.66775^{+0.00077}_{-0.00053} $ & 0.16 \\
  $ \mathcal{A}_e^{SLD} $ & \cite{ALEPH:2005ab} & $0.1516 \pm 0.0021$ & $ 0.1478 \pm 0.0011 $ & 2.20 \\
  $ \mathcal{A}_e (P_\tau) $ & \cite{ALEPH:2005ab} & $0.1498 \pm 0.0049$ & $ 0.1478 \pm 0.0011 $ & 0.43 \\
  $ \mathcal{A}_\mu^{SLD} $ & \cite{ALEPH:2005ab} & $0.142 \pm 0.015$ & $ 0.1478 \pm 0.0011 $ & 0.40 \\
  $ \mathcal{A}_\tau^{SLD} $ & \cite{ALEPH:2005ab} & $0.136 \pm 0.015$ & $ 0.1478 \pm 0.0011 $ & 0.82 \\
  $ \mathcal{A}_\tau (P_\tau) $ & \cite{ALEPH:2005ab} & $0.1439 \pm 0.0043$ & $ 0.1478 \pm 0.0011 $ & 0.94 \\
\hline
  $ M_W $ [GeV] & \cite{Aaltonen:2013iut} \cite{Awramik:2003rn} & $ 80.385 \pm 0.015 \pm 0.004 $ & $ 80.3718^{+0.0075}_{-0.0093} $ & 0.72 \\
  $ \Gamma_W $ [GeV] & \cite{ALEPH:2010aa} & $2.085 \pm 0.042$ & $ 2.09170^{+0.00066}_{-0.00084} $ & 0.16 \\
\hline
  $ Q_W (Cs) $ & \cite{Wood:1997zq} \cite{Guena:2004sq} & $-73.20 \pm 0.35$ & $ -72.915^{+0.133}_{-0.070} $ & 0.89 \\
  $ Q_W (Tl) $ & \cite{Edwards:1995zz} \cite{Vetter:1995vf} & $-116.4 \pm 3.6$ & $ -116.39^{+0.22}_{-0.12} $ & 0.00 \\
\hline
  $ M_{W'}^2 $ [$ \operatorname{TeV}^2 $] && $ \geq 4 \; {\rm TeV}^2 $ & $ \geq 4 \, {\rm TeV}^2 $ & - \\
\hline
\end{tabular}
\end{center}
\caption{\it Results for the LR Model global fit. We use the same inputs as for the SM fit, except that we include bounds on the mass of $ W' $ coming from direct searches. The term ``flat" referring to the confidence intervals of $ c_\phi $ and $ r $ means that no bounds at $ 1 \sigma $ are set ($ c_\phi = 0.49 $ and $ r = 0.0075 $ refer to the best fit point). The definition of a pull is given in Eq.~\eqref{eq:flavourPull}.}\label{tab:tab2}
\end{table}

\begin{figure}
\centering
\includegraphics[scale=0.3]{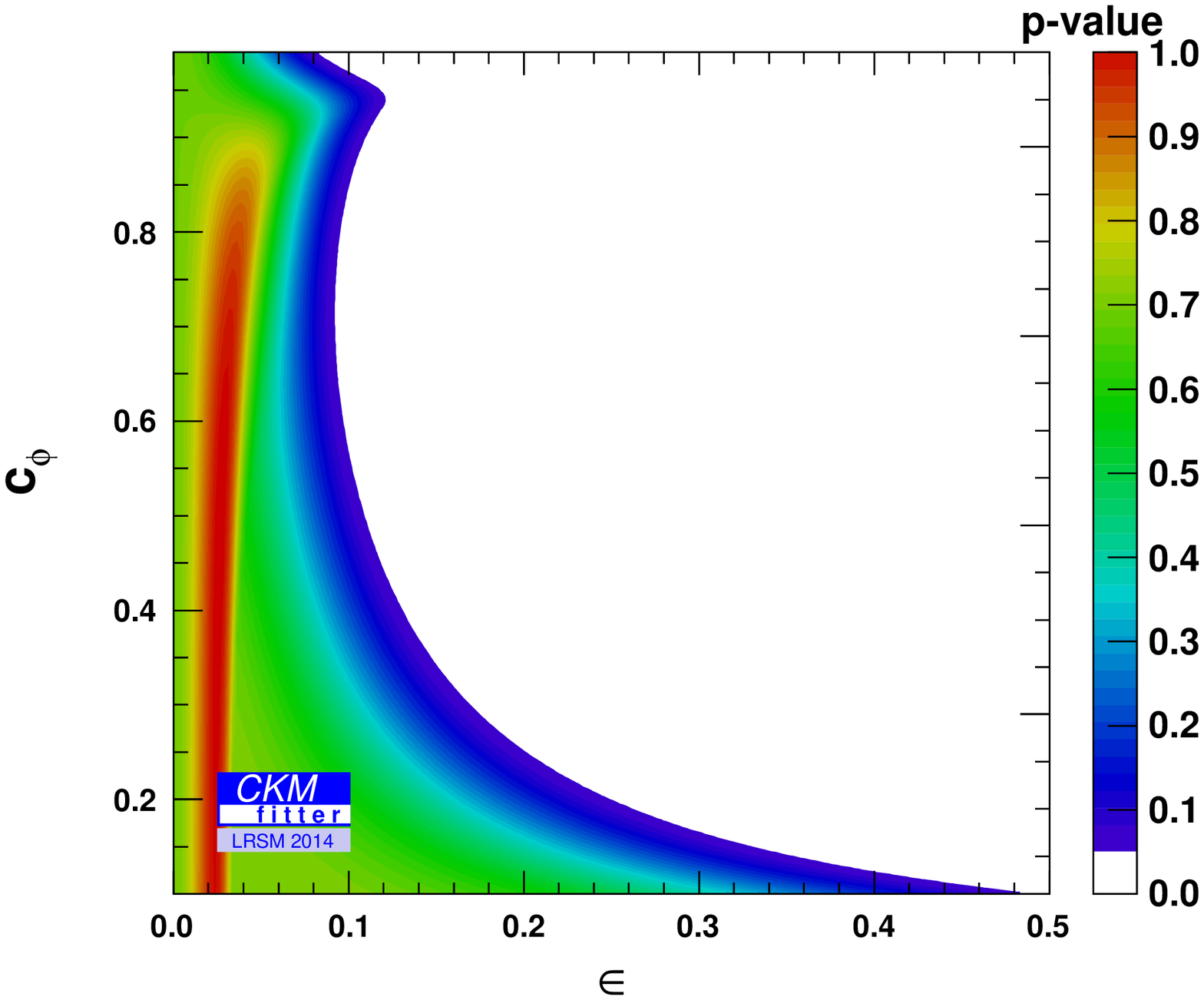}
\hspace{1cm}
\includegraphics[scale=0.3]{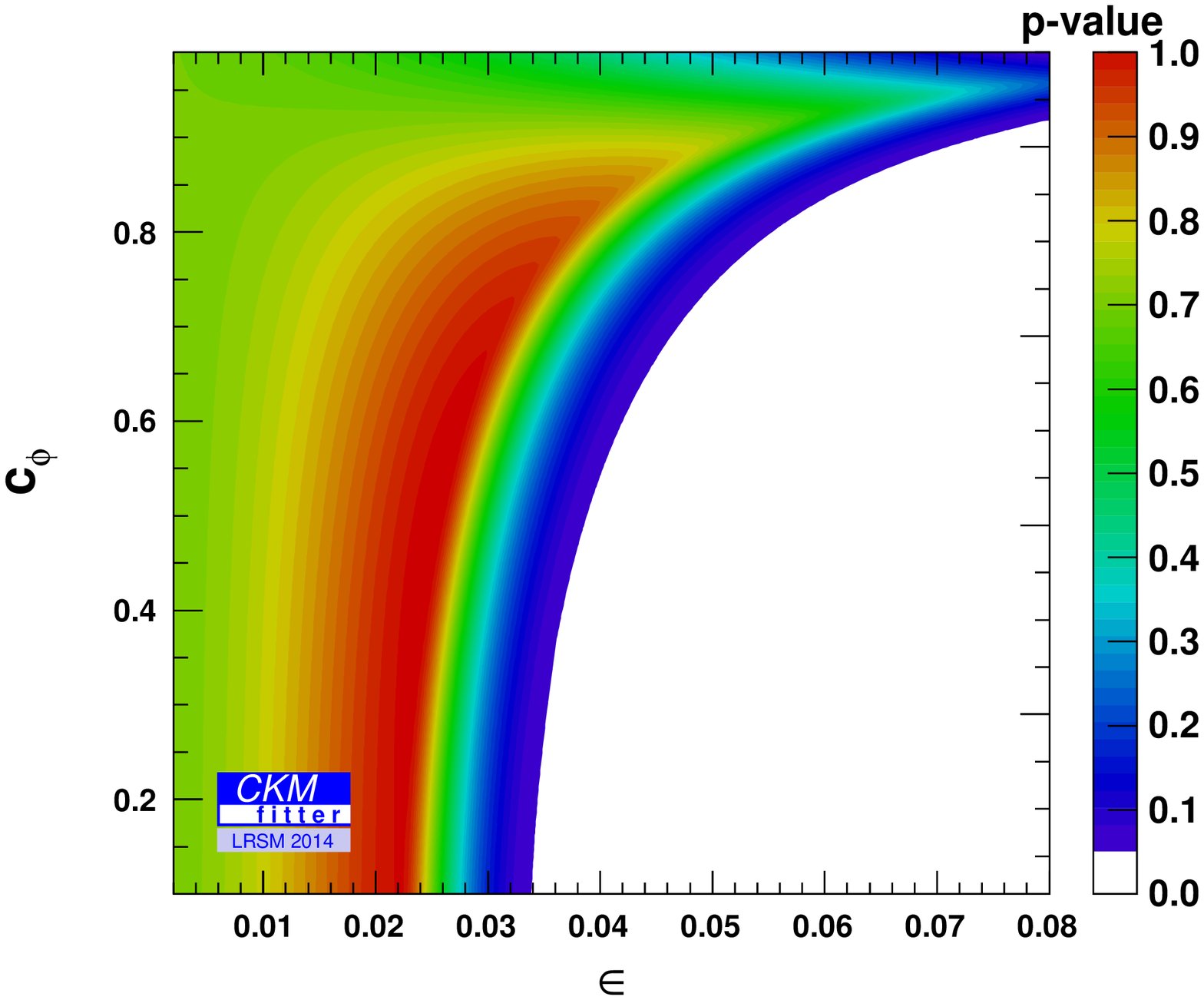}
\caption{\it Correlation among $ \epsilon $ and $ c_\phi $ in two situations: (Left) no constraint from $ M_{W'} $ and (Right) including this direct bound. Note the different scales in $ \epsilon $ used in the two figures.}\label{fig:fig1}
\end{figure}



\begin{figure}
\centering
\includegraphics[scale=0.6]{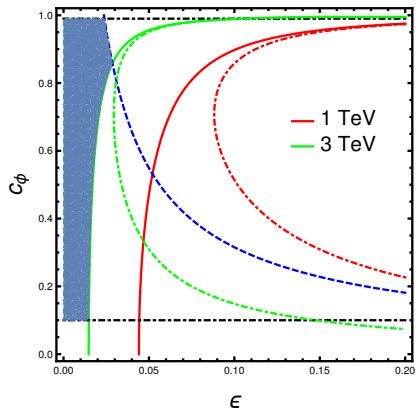}
\caption{\it Illustrative plot showing the impact of different constraints on the $ \epsilon, c_\phi $ plane: the dashed blue line is representative of the EWPO constraints; the solid lines correspond to $ M_{W'} = 1 $~TeV (red) and $ M_{W'} = 3 $~TeV (green), while the red and green dot-dashed lines correspond to $ M_{Z'} = 1 $~TeV and $ M_{Z'} = 3 $~TeV, respectively; finally, the dot-dashed black lines come from the (theoretical) requirements $ g_R^2 / 4 \pi < 1 \Rightarrow \vert c_\phi \vert \lesssim 0.99 $ and $ g_{B-L}^2 / 4 \pi < 1 \Rightarrow \vert c_\phi \vert \gtrsim 0.1 $. The blue region corresponds to the area satisfying simultaneously the representative EWPO constraints, $ M_{W'} > 3 $~TeV and $ 0.1 < \vert c_\phi \vert < 0.99 $.}\label{fig:individualBounds}
\end{figure}

Though $ c_\phi, r, w $ are not all constrained at $ 1 \; \sigma $, we can still have access to their correlations. Figure \ref{fig:fig1} shows the correlation between $ c_\phi $ and $ \epsilon $ in two different cases: (Left) without the information about direct searches for $ W' $, and (Right) when bounds on the mass of $ W' $ from direct searches are included. In the last case, as expected from

\begin{equation}
\tilde{M}_{W'}^2 \simeq \frac{\alpha (M_Z) \pi}{\sqrt{2} G_F (c^{2}_\theta)_{SM}} \frac{1}{s^{2}_\phi \epsilon^{2}} \simeq \left( \frac{44.1 \; {\rm TeV}}{10^3} \right)^2 \frac{1}{s^{2}_\phi \epsilon^{2}}
\end{equation}
($ M_{W'}^2 $ and $ \tilde{M}_{W'}^2 $ differ by $ \mathcal{O} (\epsilon^{0}) $ corrections, and we ignore the latter for the expression of $ M_{W'}^2 $), there is no allowed point in the phase space for high values of $ \epsilon $ and fixed $ c_\phi $, thus ``killing the tail" of the graph in the left. This is better illustrated in Figure~\ref{fig:individualBounds}, where the bounds from direct searches for the $ W' $ boson (setting bounds on $ M_{W'} \propto 1/(s_\phi \epsilon) $) and EWPO in the $ \epsilon, c_\phi $ plane, together with representative limits on the mass of the $ Z' $ (which has the form $ M_{Z'} = M_{W'} / c_\phi \propto 1/(c_\phi s_\phi \epsilon) $) boson and the requirements $ g_R^2 / 4 \pi, g_{B-L}^2 / 4 \pi < 1 $ are shown. 

\begin{table}
\begin{center}
\begin{tabular}{|c|ccccccc|}
\hline
$ w $ & $ \epsilon^{2} $ & $ \vert c_\phi \vert $ & $ \chi^{2}_{\min} \vert_{SM} - \chi^{2}_{\min} \vert_{LR} $ & $ M_{Z'} $ [TeV] & $ g_{R} $ & $ g_{L} $ & $ g_{B-L} $ \\
\hline
0 & 0.88 & 0.11 & 0.01 & 13.1 & 0.36 & 0.65 & 3.57 \\
1 & 1.04 & 0.40 & 0.99 & 3.77 & 0.39 & 0.65 & 0.90 \\
2 & 1.43 & 0.63 & 2.07 & 2.4 & 0.46 & 0.65 & 0.56 \\
\hline
\end{tabular}
\caption{\it Best fit point results for two parameters of the EWPO fits, the $ Z' $ mass, the couplings as well as the $ \chi^{2}_{\min} $ for $ M_{W'} = 1.5 $ TeV and $ w $ fixed as given by the first column. $ \epsilon^{2} $ is in units $ 10^{-3} $. The fit prefers $ w > 0 $, though $ \chi^{2}_{\min} $ does not change by large amounts (a less important decrease of the $ \chi^2_{min} $ is seen for larger values of $ M_{W'} $).}\label{tab:tab3}
\end{center}
\end{table}




\section{Conclusions}

We have considered over this chapter EWPO, which were parameterized in the SM in terms of the mass of the top-quark, the SM-like Higgs mass, the $ Z $ boson mass and the couplings $ \alpha_s $ and $ \alpha $. Their expressions are then corrected by the LR Model, giving contributions suppressed by $ \epsilon $, which is the ratio of EW and LR symmetry breaking scales ($ \sqrt{\kappa^2_L + \kappa^2_1 + \kappa^2_2} $ and $ \kappa_R $, respectively). We have as well considered bounds on the mass of the $ W' $ boson coming from direct searches, which are however not optimized for the most general case under consideration here.


We would like to summarize some results of the outcome of our analysis:

\begin{itemize}
	\item The SM and the LR Model both lead to similar qualities of the global fit and similar predictions for the EWPO,
	\item EWPO plus direct searches mainly set the bound $ \epsilon \lesssim 0.04 $, while we do not have strong constraints on the parameters $ w = \kappa_L / \kappa_1, r = \kappa_2 / \kappa_1 $ and $ g_R $ introduced in the LR Model framework,
	\item There is, however, an intriguing suggestion for $ \kappa_{R} \gg \kappa_{1,2} \sim \kappa_{L} $, i.e. large $ w $. The fact that $ w $, constrained to be essentially zero in the triplet LR Model, is pushed towards non-vanishing values is an interesting feature of the doublet scenario, but it remains to be seen if the other sectors of the theory agree with this tendency. 
\end{itemize}

In what follows, we will discuss a further set of inputs, consisting in meson-mixing observables, in order to further test the LR Models.

\chapter{Overview of meson-mixing observables}\label{ch:generalEFT}


We now shift to a different class of observables, made of meson-mixing observables. As in the case of EWPO, our goal here is to probe the possible structure of LR Models. The difference compared to EWPO is that new parameters show up in the prediction of meson-mixing observables, thus offering the opportunity to constrain them. The new parameters are the elements of the mixing-matrix $ V^R $, which describe the couplings of the $ W' $ boson to quarks, and the masses of the extended scalar sector.

In Section~\ref{sec:DiagramsSMLRM}, we discuss the contributions to meson-mixing in LR Models, which are diagrams including $ W', H^{\pm, 0} $ exchanges, beyond the $ W W $ box already found in the SM. Then, reliable predictions of LR Model rates require the computation of short-distance QCD corrections. Indeed, as we have seen in Chapter~\ref{ch:SM} in the SM these are important corrections, shifting the individual contributions found in the SM framework by factors of 2, cf. Eqs.~\eqref{eq:differentEtasSM}. Over the Sections~\ref{sec:generalEFT}, \ref{sec:running}, \ref{sec:MRforPedestrians} we are going to introduce the basic elements necessary in order to discuss these short-distance QCD corrections in two different approaches, Effective Field Theory (EFT) and Method of Regions (MR), trying to be as general as possible in our description. Then, in Section~\ref{sec:comparisonMREFT} we briefly compare both approaches in order to validate the MR, which is meant to be an approximation to the more complete calculation done in the EFT approach. This calculation is going to be considered in detail in the next chapter, dedicated to more technical elements.

\section{Contributions to meson-mixing}\label{sec:DiagramsSMLRM}

Formally, from the Lagrangian $ \mathcal{L} $ of the theory one builds all the possible contributions to the meson-mixing amplitude from the generating Green's function, $ \langle {\rm \mathbf{T}} \exp \left[ i \int d^4 x \, \mathcal{L} (x) \right] \rangle_{\vert \Delta F \vert = 2} $. Of these, we aim at keeping the set of diagrams compatible with gauge invariance at first order in

\begin{equation}
\beta = M^2_W / M^2_{W'} = \mathcal{O} (\epsilon^2) \, .
\end{equation}
Apart from the $ W W $ box already found in the SM, one important class of contributions includes the exchange of a single $ W' $ in a box together with a $ W $, which turns out not to be gauge invariant by itself. The set of diagrams necessary for the gauge invariance of the $ W W' $ box includes loop corrections of the Higgs self-energy and Higgs $ \vert \Delta F \vert = 1 $ coupling \cite{Basecq:1985cr,Hou:1985ur}. The other classes of diagrams we consider consist of charged Higgs and tree level neutral Higgs exchanges, cf. Figure~\ref{fig:tabNPdiagrams}.

\subsection{SM}

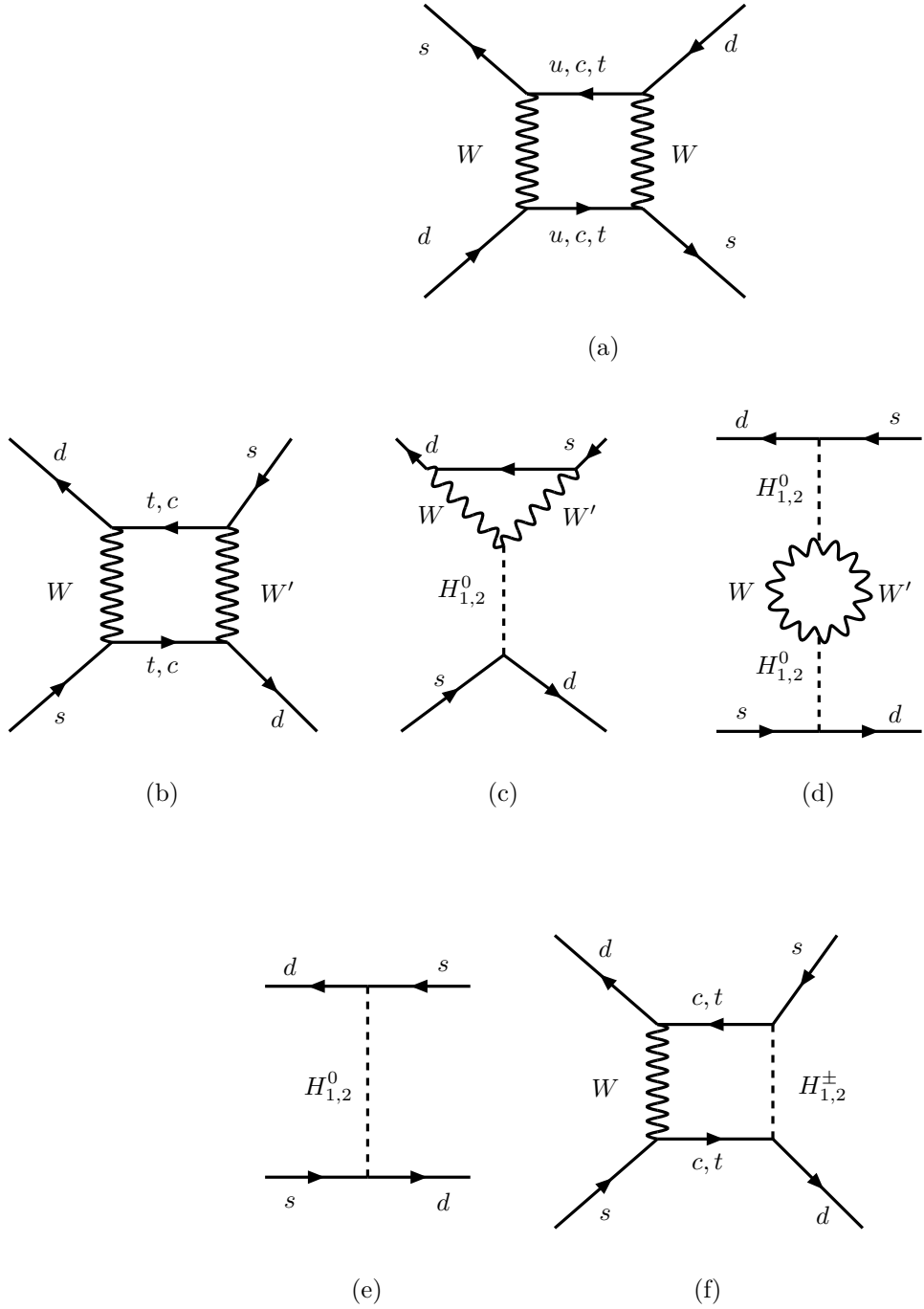
\begin{figure}
	\centering
		\SetWidth{1.2}
		\vspace{-2cm} \begin{picture}(125,115)(-10,0)
		\ArrowLine(40,80)(0,115)
		\Text(0,100)[t]{$ s $}
		\ArrowLine(0,0)(40,35)
		\Text(0,28)[t]{$ d $}
		\ArrowLine(85,80)(40,80)
		\Photon(40,80)(40,35){4}{8}
		\ArrowLine(125,115)(85,80)
		\Text(120,104)[t]{$ d $}
		\ArrowLine(85,35)(125,0)
		\Text(120,24)[t]{$ s $}
		\ArrowLine(40,35)(85,35)
		\Photon(85,80)(85,35){4}{8}
		\Text(60,95)[t]{$ u, c, t $}
		\Text(60,28)[t]{$ u, c, t $}
		\Text(18,60)[t]{$W$}
		\Text(102,60)[t]{$W$}
		\Text(70,-20)[]{(a)}
		\end{picture}
		\vspace{2cm} 
\begin{center}
\begin{picture}(180,50)(90,0)
\ArrowLine(80,25)(40,60)
\Text(60,55)[]{$d$}
\ArrowLine(40,-55)(80,-20) 
\Text(60,-50)[]{$s$}
\ArrowLine(125,25)(80,25) 
\Photon(80,25)(80,-20){4}{8}
\ArrowLine(150,60)(125,25) 
\Text(135,55)[]{$s$}
\ArrowLine(125,-20)(160,-55) 
\ArrowLine(80,-20)(125,-20) 
\Photon(125,25)(125,-20){4}{8}
\Text(145,-50)[]{$d$}
\Text(60,0)[]{$W$}
\Text(100,-30)[]{$t,c$}
\Text(100,35)[]{$t,c$}
\Text(100,-80)[]{(b)}
\Text(145,0)[]{$W'$}
\end{picture}
\begin{picture}(150,50)(120,0)
\ArrowLine(50,48)(38,60)
\Text(52,58)[]{$d$}
\ArrowLine(40,-55)(80,-25)
\ArrowLine(120,60)(108,48)
\Text(64,0)[]{$H^0_{1,2}$}
\ArrowLine(108,48)(54,48)
\Photon(80,18)(50,48){3}{5}
\Photon(108,48)(80,18){3}{5}
\Text(52,30)[]{$ W $}
\Text(110,30)[]{$ W' $}
\DashLine(80,18)(80,-25){3}
\ArrowLine(80,-25)(120,-55)
\Text(106,58)[]{$s$}
\Text(55,-35)[]{$s$}
\Text(106,-35)[]{$d$}
\Text(80,-80)[]{(c)}
\end{picture}
\begin{picture}(140,50)(-90,-51)
\ArrowLine(80,60)(40,60)
\ArrowLine(120,60)(80,60)
\DashLine(80,60)(80,20){3}
\PhotonArc(80,0.225)(17.5,90,270){3}{7}
\PhotonArc(80,0.225)(17.5,270,450){3}{7}
\DashLine(80,-18)(80,-55){3}
\ArrowLine(40,-55)(80,-55)
\ArrowLine(80,-55)(120,-55)
\Text(80,-80)[]{(d)}
\Text(65,40)[]{$H^0_{1,2}$}
\Text(65,-30)[]{$H^0_{1,2}$}
\Text(50,0)[]{$W$}
\Text(110,0)[]{$W'$}
\Text(50,-48)[]{$s$}
\Text(110,-48)[]{$d$}
\Text(50,68)[]{$d$}
\Text(110,68)[]{$s$}
\end{picture}

\vspace{4cm}

\begin{picture}(90,50)(20,-20)
\ArrowLine(80,40)(40,40)
\ArrowLine(120,40)(80,40)
\DashLine(80,40)(80,-35){3}
\ArrowLine(40,-35)(80,-35)
\ArrowLine(80,-35)(120,-35)
\Text(80,-80)[]{(e)}
\Text(65,0)[]{$H^0_{1,2}$}
\Text(50,-45)[]{$s$}
\Text(110,-45)[]{$d$}
\Text(50,48)[]{$d$}
\Text(110,48)[]{$s$}
\end{picture}
\begin{picture}(180,50)(0,-20)
\ArrowLine(80,25)(40,60)
\Text(60,55)[]{$d$}
\ArrowLine(40,-55)(80,-20)
\Text(60,-50)[]{$s$}
\ArrowLine(125,25)(80,25)
\Photon(80,25)(80,-20){4}{8}
\ArrowLine(150,60)(125,25)
\Text(135,55)[]{$s$}
\ArrowLine(125,-20)(160,-55)
\ArrowLine(80,-20)(125,-20)
\DashLine(125,25)(125,-20){3}
\Text(145,-50)[]{$d$}
\Text(60,0)[]{$W$}
\Text(100,-30)[]{$c,t$}
\Text(100,35)[]{$c,t$}
\Text(100,-80)[]{(f)}
\Text(145,0)[]{$H^{\pm}_{1,2}$}
\end{picture}
\end{center}
\vspace{2.5cm}
\caption{\it Figure (a): contribution already found in the SM. Figures (b)-(f): set of diagrams giving the main new contributions to kaon mixing in Left-Right Models. The set (b), (c), (d) and a contribution from (e), tree level neutral Higgs exchanges, forms a gauge invariant set of diagrams. Another important class of new contributions includes charged Higgs exchanges in a box, diagram (f). Instead of the neutral $ \mathcal{CP}- $even Higgs $ H^0_i $, $ i=1,2 $, we could consider as well the $ \mathcal{CP}- $odd ones, $ A^0_i $, $ i=1,2 $. Diagrams where the $ W $ and $ W' $ are replaced by their respective Goldstone bosons have also to be taken into account.}\label{fig:tabNPdiagrams}
\end{figure}

In the SM, the only contribution we have is shown in Figure~\ref{fig:tabNPdiagrams} (a), where two $ W $ bosons are exchanged and the internal flavours can be $ u, c, t $. Of course, the full set of diagrams also includes the exchange of the corresponding Goldstone boson $ G $ instead of $ W $, i.e. $ W G $ and $ G G $ boxes. We note that the set of boxes is gauge invariant by itself, and that it is finite (no renormalization is needed at this stage). The final expression is found for instance in \cite{Buchalla:1995vs}, and is given by

\begin{eqnarray}\label{eq:SMbox}
H^{\rm SM} &=&\frac{G_F^2 M^2_W}{4\pi^2} Q^{VLL}_1 \sum_{U,V=u,c,t}\lambda_U^{LL}\lambda_V^{LL} \\
& \times & [(1+x_U x_V / 4)I_2(x_U,x_V)-2 x_U x_V I_1(x_U,x_V)]\, + h.c. ,
\nonumber
\end{eqnarray}
\noindent
where the operator $ Q^{VLL}_1 $ is

\begin{equation}
Q^{VLL}_1 = \bar{d} \gamma^\mu P_L s \cdot \bar{d} \gamma_\mu P_L s \, .
\end{equation}
The combinations of CKM matrix elements are given by

\begin{equation}
\lambda^{AB}_U = V^{A *}_{U q_1} V^{B}_{U q_2}
\end{equation}
($ q_{1,2} $ are the external flavours) and $ I_{1,2} (x_U, x_V) $ are the Inami-Lim functions \cite{Inami:1980fz}

\begin{eqnarray}\label{eq:InamiLimSMcase}
I_1 (x_U,x_V) & = & \frac{x_U\log x_U}{(1-x_U)^2 (x_U-x_V)} + (U\leftrightarrow V) + \frac{1}{(1-x_U)(1-x_V)} \, , \nonumber\\
I_2 (x_U,x_V) & = & \frac{x_U^2\log x_U}{(1-x_U)^2 (x_U-x_V)} + (U\leftrightarrow V) + \frac{1}{(1-x_U)(1-x_V)} \, .
\end{eqnarray}

We further apply the unitarity of the CKM matrix, i.e. $ \sum_{U=u,c,t} \lambda^{LL}_U = 0 $, to rewrite Eq.~(\ref{eq:SMbox}) under a different form

\begin{eqnarray}\label{eq:SMboxUnitarity}
&& H^{\rm SM} = \frac{G_F^2 M_W^2}{4 \pi^2}
\Bigg[ \lambda_c^{LL} \lambda_c^{LL} S^{LL}(x_c) \\
&& + \lambda_t^{LL} \lambda_t^{LL} S^{LL}(x_t) +2 \lambda_t^{LL} \lambda_c^{LL} S^{LL}(x_c,x_t) \Bigg] Q^{VLL}_1 + h.c. , \nonumber
\end{eqnarray}
where we have defined the following \textit{loop functions}

\begin{eqnarray}\label{eq:SMloopfunctions}
S^{LL}(x_c)&=& x_c+ {\cal O}(x_c^2)\,,\\
S^{LL}(x_t)&=&x_t\biggl(\frac{1}{4}+\frac{9}{4}\frac{1}{1-x_t}
-\frac{3}{2}\frac{1}{(1-x_t)^2}\biggr) -\frac{3}{2}\biggl(\frac{x_t}{1-x_t}
 \biggr)^3 \log x_t \,,
\nonumber \\
S^{LL}(x_c,x_t)&=& -x_c \log x_c+ x_c F(x_t)+ {\cal O}(x_c^2 \log x_c)  \, , \nonumber
\end{eqnarray}
with

\begin{equation}\label{eq:nonLogContribution}
F(x_t)=\frac{x_t^2 - 8 x_t +4}{4(1-x_t)^2} \log x_t +\frac{3}{4} \frac{x_t}{(x_t-1)} \, .
\end{equation}

We refer to the above three contributions as charm-charm, top-top and charm-top, respectively, which are proportional to $ \lambda^{LL}_U \lambda^{LL}_V $, $ U, V = c, t $. Note that, thanks to the unitarity of the CKM matrix, $ \mathcal{O} (\log x_c) $ in the charm-charm contribution vanishes, while due to the different masses in the charm-top case this factor is present.

\subsection{LR Models}

We start by discussing the contributions from box diagrams. Due to the $ W, W' $ mixing, which modifies the structure of the $ W $ coupling at order $ \epsilon^2 $ by the introduction of right-handed couplings, we need in principle to reconsider the SM-like $ W W $ box, which is also present in the LR Model. Moreover, since right-handed couplings are also present, we distinguish the following cases: zero, one, two, three and four right-handed couplings. Each time there is a right-handed coupling, the whole contribution gets suppressed by $ \epsilon^{2} $, and the combination of a left-handed coupling with a right-handed one implies a chiral flip of the intermediate up-type quark. Now, diagrams with an odd number of chirality flips give no contribution when the momenta of the external quarks are set to zero, since the integral on the loop momentum is odd. Therefore, we consider only the left-handed coupling of the $ W $, since the next contribution involving two right-handed couplings is suppressed by $ \epsilon^4 $, thus implying that $ W W $ boxes in LR Models have the same expressions found in the SM, Eqs.~\eqref{eq:SMboxUnitarity}-\eqref{eq:SMloopfunctions}.

Apart from the $ W W $ box, we also have $ W W' $ boxes. The heavy character of the $ W' $ implies that we need to consider only its right-handed coupling. The $ WW' $ box diagrams are usually calculated in the \tHF gauge ($ \xi_{W,W'} = 1 $), see e.g. \cite{Ecker:1985vv}, and we have

\begin{eqnarray}\label{eq:boxWWprimePrevious}
A^{({\rm box})} &=&\frac{G_F^2 M^2_W}{4\pi^2} 2 \beta h^2 \langle Q_2^{LR} \rangle \sum_{U,V=u,c,t}\lambda_U^{LR}\lambda_V^{RL} \sqrt{x_Ux_V} \\
& \times & [(4+x_Ux_V \beta)I_1(x_U,x_V,\beta)-(1+\beta)I_2(x_U,x_V,\beta)] \,,
\nonumber
\end{eqnarray}
\noindent
where the operator $ Q^{LR}_{2} $ is

\begin{equation}
Q^{LR}_{2} = \bar{d} P_R s \cdot \bar{d} P_L s \, .
\end{equation}
Note that this operator has a very different structure when compared to the SM operator, $ Q^{VLL}_1 $.

In Eq.~\eqref{eq:boxWWprimePrevious}, gauge couplings are contained in $ h = g_R / g_L $, and

\begin{eqnarray}\label{eq:InamiLimObetaLRM}
I_1 (x_U,x_V,\beta) & = & \frac{x_U\log x_U}{(1-x_U)(x_U-x_V)} + (U\leftrightarrow V) + \mathcal{O}(\beta) , \\
I_2 (x_U,x_V,\beta) & = & \frac{x_U^2\log x_U}{(1-x_U)(x_U-x_V)} + (U\leftrightarrow V) - \log\beta + \mathcal{O}(\beta) \nonumber
\end{eqnarray}
(corrections in $ \beta $ are found for example in \cite{Ecker:1985vv}). In the \tHF gauge, the contributions to $ A^{({\rm box})} $ seen in the second line of Eq.~\eqref{eq:boxWWprimePrevious} come from the four following diagrams:

\begin{itemize}
	\item $ WW' \Rightarrow I_1(x_U,x_V,\beta) $ term, \\
	\item $ GW' \Rightarrow I_2(x_U,x_V,\beta) $ term, \\
	\item $ WG' \Rightarrow \beta I_2(x_U,x_V,\beta) $ term, of higher order in $ \beta $, and \\
	\item $ GG' \Rightarrow x_U x_V \beta I_1(x_U,x_V,\beta) $ term, also of higher order in $ \beta $, \\
\end{itemize}
where $ G $ ($ G' $) is the Goldstone boson associated to the light (heavy) gauge boson $ W $ ($ W' $).

The different handednesses of the main couplings of the $ W, W' $ imply chiral flips leading to the overall mass term, seen in the first line of Eq.~\eqref{eq:boxWWprimePrevious}. Expanding $ A^{({\rm box})} $ in $ \beta = M^{2}_{W}/M^{2}_{W'} $ and $ x_c = m^{2}_{c}/M^{2}_{W} $ one has

\begin{eqnarray}\label{eq:boxWWprime}
A^{({\rm box})} &=& \frac{G_F^2 M_W^2}{4 \pi^2} 2 \beta h^2 \langle Q_2^{LR} \rangle \biggl[ \lambda_c^{LR}\lambda_c^{RL}S^{({\rm box})}(x_c, x_c, \beta) \\
& + & \lambda_t^{LR}\lambda_t^{RL}S^{({\rm box})}(x_t, x_t, \beta)+  (\lambda_c^{LR}\lambda_t^{RL}+\lambda_t^{LR}\lambda_c^{RL})S^{({\rm box})}(x_c, x_t, \beta)\biggr]\,,\nonumber
\end{eqnarray}
where the loop functions are

\begin{eqnarray}
S^{({\rm box})} (x_c, x_t, \beta) \!\!&=& \!\! \sqrt{x_c x_t}\left[\frac{x_t - 4}{x_t - 1} \log (x_t) + \log (\beta) \right] + \mathcal{O} (\beta, x^{3/2}_c)\,, \\
S^{({\rm box})} (x_t, x_t, \beta) \!\!&=&\!\! x_t\left( \frac{x_t^2 - 2 x_t + 4}{(x_t - 1)^2} \log (x_t) + \frac{x_t - 4}{x_t - 1} + \log (\beta) \right) + \mathcal{O} (\beta)\,, \nonumber\\
&& \\
S^{({\rm box})} (x_c, x_c, \beta) \!\!&=&\!\! x_c  \left( 4 \log (x_c) + 4 + \log (\beta) \right) + \mathcal{O} (\beta, x^2_c)\,. \label{eq:boxWWprimecc}
\end{eqnarray}
Note that due to the overall mass factors, $ x_U $, diagrams involving an up-quark are very much suppressed and can be ignored.

It has been shown in \cite{Kenmoku:1987fm} that the $ W W' $ box diagram in LR Models forms a gauge invariant set only if neutral Higgs exchanges induce no FCNC at tree level. However, we have seen that this is not the case in the class of LR Models we are considering, cf. Section~\ref{sec:couplingsHfermionsLRM}. The question of gauge invariance has been addressed in several papers \cite{Basecq:1985cr,Hou:1985ur,Kenmoku:1987fm,GagyiPalffy:1997hh,Bernard}, and the required diagrams for gauge invariance are shown in Figure~\ref{fig:tabNPdiagrams} (c, d, e). Note that, in order to be consistent with the order of these four point Green's functions, we should consider the renormalization of the Higgs mass and couplings: for this reason, we consider the on-shell subtractions as described in \cite{Basecq:1985cr}.

Below, we give the expressions of the vertex and self-energy diagrams in the \tHF gauge. They depend on the Higgs mass through $ \omega_i = M^2_{W'} / M^2_{H_i} = \mathcal{O} (1) $:

\begin{eqnarray}\label{eq:selfWWprime}
A^{\rm (self)} =
-2\beta \sum^2_{i=1} \mathcal{F}^2_i \omega_i h^2\frac{G_F^2 M_W^2}{4\pi^2}  \langle Q_2^{LR}\rangle S_S(\omega_i) \sum_{U,V=c,t} \lambda_U^{LR}\lambda_V^{RL}  \sqrt{x_U x_V} \,,
\end{eqnarray}

\begin{eqnarray}
S_S(\omega) = -2+\frac{(1-\omega^2)}{\omega}\log\left|\frac{1-\omega}{\omega}\right| +\mathcal{O}(\beta)\,,
\end{eqnarray}
and

\begin{equation}\label{eq:vertexWWprime}
A^{\rm (vertex)} =
-32\beta \sum^2_{i=1} \mathcal{F}_i \omega_i h^2\frac{G_F^2 M_W^2}{4\pi^2} \langle Q_2^{LR}\rangle S_V(\omega_i)  \sum_{U,V=c,t} \lambda_U^{LR}\lambda_V^{RL}  \sqrt{x_U x_V} \,,
\end{equation}

\begin{equation}
S_V(\omega) = -1+(1-\omega)\log\left|\frac{1-\omega}{\omega}\right| + \mathcal{O}(\beta^{1/2}) \, .
\end{equation}
The above expressions include a $ \mathcal{CP}- $even and a $ \mathcal{CP}- $odd Higgs. Note that the mixed propagator of a $ \mathcal{CP}- $even and a $ \mathcal{CP}- $odd Higgs gives no contribution, due to their real and pure imaginary couplings, respectively.

The functions $ \mathcal{F}_i = k F_i G_i $ correct the limiting case $ w \rightarrow 0^+ $, explicitly calculated in e.g. \cite{Bertolini:2014sua}, and are calculated from the corrections to the couplings of the scalar sector to the gauge bosons (proportional to $ F_i $) and quarks (proportional to $ G_i $) when $ w \neq 0 $. Their expressions are given by

\begin{eqnarray}\label{eq:functionF1}
\mathcal{F}_1  &=&\frac{1}{ 2  (1 - r^2)(1 + \beta(x) w^2) (1 - \delta^2)} 
 \biggl( \bigl(-k^2  + (k^2 -2(1 +  \nu(x) )) X \bigr) (1+ \delta^2) \biggr.
\nonumber \\
&&\biggl.+2 \bigl(1 + \nu(x) + \bigl(r^2 -\beta(x) w^2 (1 - r^2)  +  \nu(x)\bigr) \delta^2 \bigr) \biggr)
\end{eqnarray}
while $ \mathcal{F}_2 = 1 - \mathcal{F}_1 $ is calculated from $ \mathcal{F}_1 $ by changing $\delta \to 1/\delta$, where $\delta=M_{H_2}/M_{H_1}$ is the ratio of the mass of the scalars $ H_2, A_2 $ over the one of the scalars $ H_1, A_1 $. The other functions seen in the expression of $ \mathcal{F}_i $ are

\begin{eqnarray}\label{eq:Xnubetak2defs}
&& X=\sqrt{1- \frac{4 \delta^2}{(1+ \delta^2)^2} \frac{\left(1+r^2\right)\left(1+ \beta(x) w^2\right)}{k^2}} \, , \nonumber\\
&& \beta(x)=(1+x^2)/(1+r x)^2  \, , \quad \quad \nu(x) = w^2/(1+ r x ) \, , \\
&& k^2 = 1 + r^2 + w^2 \, . \nonumber
\end{eqnarray}
In the above expressions, we have indicated the dependence on the parameter $ x $ defined as $x=\mu'_1 /\mu'_2$, which is the ratio of the two trilinear coupling constants seen in the scalar potential of Eq.~\eqref{eq:potentialSymPdoublet}. Its origin here amounts to the diagonalization of the mass matrix in the scalar sector, introducing eigenvectors whose coefficients in the original basis depend on $ x $, thus introducing $ x $ in the couplings of the physical scalar particles. 

The remaining contribution necessary for gauge invariance comes from a tree level diagram, cf. Figure~\ref{fig:tabNPdiagrams} (e). It originates from the mixing between $ H^{0}_{1} $ and $ H^{0}_{2} $, or between $ A^{0}_{1} $ and $ A^{0}_{2} $, through $ W W' $ loops leading to a gauge dependence of the scalar couplings, just like the diagram in Figure~\ref{fig:tabNPdiagrams} (d) is not gauge invariant by itself. We give in the following expression both the gauge independent (first line) and the gauge dependent (second line) contributions from the tree level diagram in the \tHF gauge:

\begin{eqnarray}\label{eq:allTheTreeLevelContributions}
A^{(H^0)} &=& \biggl( - \frac{4 G_F \beta u}{\sqrt{2}} \frac{k^2}{1+r^2} \sum^2_{i=1} \omega_i \tilde G_i^2 \nonumber\\
&& + h^2 \frac{G_F^2 M_W^2}{2 \pi^2} \beta {\cal F}^r \sqrt{\omega_1 \omega_2}   S_S\bigl(\sqrt{\omega_1 \omega_2}\bigr) \biggr) \\
&& \sum_{U, V=c, t} \lambda_U^{LR}\lambda_V^{RL} \sqrt{x_U x_V} \langle Q_2^{LR}\rangle \, , \nonumber
\end{eqnarray}
where $u=(1+r^2)^2/(1-r^2)^2$. In this expression, other operator structures such as $ \bar{s} P_L d \cdot \bar{s} P_L d $ or $ \bar{s} P_R d \cdot \bar{s} P_R d $ come at a higher order in $ \beta $ and are neglected. The function $ \mathcal{F}^r $ is defined as $ \mathcal{F}^r = \sum^2_{i=1} \mathcal{F}^2_i - 1 $, where $ \mathcal{F}_i $ is found in Eq.~\eqref{eq:functionF1}. The second line of Eq.~\eqref{eq:allTheTreeLevelContributions} combines in the sum

\begin{equation}
A^{({\rm box})} + A^{({\rm vertex})} + A^{({\rm self})} + A^{({\rm tree})} \vert_{\mathcal{F}^r} \, ,
\end{equation}
to form a gauge invariant expression. A discussion concerning gauge invariance in the general case where $ w \neq 0 $ is found in reference \cite{Bernard}.

The function $ \tilde G_i^2 $ seen in the gauge independent part is calculated from $\tilde G_i^2= (1+r^2) G_i^2/u$ where $ G_i $ corrects the couplings of the scalars to the quarks when $ w \neq 0 $.  Its expression is given by

\begin{equation}
\tilde G_2^2=\frac{k^2(1- \delta^2  ( 1 -\left(1+1/\delta^2\right)X)) -2w^2(1-\beta(x)(1+r^2))}{2 (1-\delta^2)
(1+r^2)  (1+ \beta(x) w^2) } \, ,
\end{equation}
while $ \tilde G_1^2 $ is determined from the relation $\tilde G_1^2=1- \tilde G_2^2$.


Apart from the contributions given in Eqs.~\eqref{eq:SMboxUnitarity}, \eqref{eq:boxWWprime}, \eqref{eq:selfWWprime}, \eqref{eq:vertexWWprime}, \eqref{eq:allTheTreeLevelContributions}, the last set of diagrams, shown in Figure~\ref{fig:tabNPdiagrams} (f) consists of boxes $ W H^{\pm}_i $ and $ G H^{\pm}_i $, $ i = 1,2 $, where $ H^{\pm}_i $ is a heavy, electrically charged Higgs which couples as $ G' $, cf. Appendix~\ref{sec:tableOfCouplings}. As discussed in \cite{GagyiPalffy:1997hh}, it alone does not form a gauge invariant set, but the other diagrams necessary for gauge invariance (vertex and self-energy diagrams) contribute at a higher order in $ \beta $ in the \tHF gauge. In this case, we have

\begin{eqnarray}
H^{(H^{\pm} \, {\rm box})} &=&\frac{G_F^2 M^2_W}{4\pi^2} \frac{k^2}{1+r^2} Q_2^{LR} \sum^2_{i=1} \tilde{G}^2_i 2 \omega_i \beta u (1 - \beta) \sum_{U,V=u,c,t}\lambda_U^{LR} \lambda_V^{RL} \sqrt{x_U x_V} \nonumber\\
& \times & [-I_2(x_U,x_V,\omega_i \beta)+x_U x_V I_1(x_U,x_V,\omega_i \beta)]\, + h.c. \, ,
\end{eqnarray}
\noindent
or, expanding in $ \beta $ and $ x_c $,

\begin{eqnarray}\label{eq:chargedHiggsContribution}
H^{(H^{\pm} \, {\rm box})} &=&
 \frac{G_F^2 M_W^2}{4\pi^2} \frac{k^2}{1+r^2} Q_2^{LR} \sum^2_{i=1} \tilde{G}^2_i 2 \omega_i \beta u \sum_{U,V=c,t} \lambda_U^{LR}\lambda_V^{RL} \nonumber\\
&& \quad \sqrt{x_U x_V} S^H_{LR}(x_U,x_V, \beta \omega_i)\, + h.c. \, ,
\end{eqnarray}
with

\begin{eqnarray}\label{eq:loopFunctionsBoxH}
S^H_{LR} (x_c, x_t, \omega \beta) &=& \left( \frac{x_t}{x_t - 1} \log (x_t) + \log (\omega \beta) \right) + \beta \cdot \mathcal{O} (\beta, x_c) , \\
S^H_{LR} (x_t, x_t, \omega \beta) &=& \left( \frac{2 x_t}{x_t - 1} \log (x_t) - x_t + \log (\omega \beta) \right) + \mathcal{O} (\beta^2), \\
S^H_{LR} (x_c, x_c, \omega \beta) &=& \log (\omega \beta) + \beta \cdot \mathcal{O} (\beta, x_c) .
\end{eqnarray}
When calculating the expressions above, we have considered only the $ m_u^i $ term seen in the third and sixth lines (involving the coupling $ \bar{u}^i_L d^j_R H^+_{1,2} $) of Table~\ref{tab:chargedHiggsContributions} in Appendix~\ref{sec:spectrumScalars}. The $ m_d^j $ terms seen in the different lines of this same table do not contribute in the system of kaons. On the other hand, $ m_u^i $ terms from the second and fourth lines (involving the coupling $ \bar{u}^i_R d^j_L H^+_{1,2} $) provide contributions similar to the $ W G $ and $ G G $ box diagrams found in the SM, i.e. contributions proportional to $ \lambda^{LL}_U \lambda^{LL}_V Q_1^{VLL} $ for $ (U,V) \in \{ (t,t), (c,t), (t,c) \} $. Since these contributions do not carry the same enhancements of the LR operators (i.e. the values of the short-distance QCD corrections for $ tt $ and $ ct $, and the chiral enhancement $ m_K^2 / (m_s + m_d)^2 $ seen in Eq.~\eqref{eq:BagParametersEnhancement} below), we will not further consider them in our analysis.

A last comment is in order. In all cases including physical scalars, when taking the limit where $ w $ goes to zero the contributions from the particles $ H^0_2, A^0_2, H^\pm_2 $ go to zero and this scalar sector decouples from the meson-mixing  phenomenology. In such a case, the expressions given above reduce to those found for example in \cite{Bertolini:2014sua}. 

\subsection{Including short-distance QCD corrections in the LR Model}\label{sec:ExpressionsLRMwetaparameters}

The expressions we have given above correspond to the main contributions up to higher order corrections in $ \beta $ of the full LR Model. On top of that, we must consider QCD corrections, which may shift considerably the individual contributions. These contributions are factorized at a low-energy scale $ \mu_{h} $ into short-distance and long-distance corrections, and the former are calculated by perturbative methods while non-perturbative methods are able to take into account hadronic effects in $ \langle K \vert Q^{LR}_2 \vert \bar{K} \rangle (\mu_{h}) $. Short-distance QCD corrections from the high energy scales $ \mu_W $ down to the low energy scale $ \mu_{h} $ are collected into the $ \bar{\eta} $ parameters seen in the following compact expressions: first, we have

\begin{eqnarray}\label{eq:Fullsetwneq0}
&& \langle H^{(W W')} \rangle = \frac{G_F^2 M_W^2}{4 \pi^2} 8 \beta h^2 \langle Q^{LR}_{2} \rangle (\mu_h) \sum_{U,V=c,t} \bar{\eta}^{(LR)}_{U V} (\mu_h) \lambda_U^{LR} \lambda_V^{RL} \\
&& \sqrt{x_U x_V} S^{LR} (x_U, x_V, \beta, \omega) + h.c. , \nonumber
\end{eqnarray}
with

\begin{equation}\label{eq:finalLoopFunct}
S^{LR} (x_U, x_V, \beta, \omega) = S^{({\rm box})} (x_U, x_V, \beta) / (4 \sqrt{x_U x_V}) + F(\omega_1, \omega_2) / 4 ,
\end{equation}
and

\begin{eqnarray}\label{eq:TheFunctioFomega1omega2}
F(\omega_1, \omega_2) = - \sum^2_{i=1} \mathcal{F}_i \omega_i \left( \mathcal{F}_i S_S (\omega_i) + 16 \; S_V (\omega_i) \right) + \mathcal{F}^r \sqrt{\omega_1 \omega_2} S_S (\sqrt{\omega_1 \omega_2}) \, ,
\end{eqnarray}
where in Eq.~\eqref{eq:Fullsetwneq0} we include the gauge dependent part of Eq.~\eqref{eq:allTheTreeLevelContributions}, necessary for canceling the gauge dependence coming from Eqs.~\eqref{eq:boxWWprime}, \eqref{eq:selfWWprime}, \eqref{eq:vertexWWprime}. Then, for the other contributions

\begin{eqnarray}\label{eq:HiggsBoxFactEta}
\langle H^{(H^{\pm} \, {\rm box})} \rangle &=& 2 \omega \beta u \frac{G_F^2 M_W^2}{4\pi^2} \langle Q_2^{LR} \rangle (\mu_h) \sum_{U,V=c,t} \bar{\eta}^{(H^{\pm} \, {\rm box})}_{U V} (\mu_h) \lambda_U^{LR}\lambda_V^{RL} \nonumber\\
&& \sqrt{x_U x_V} S^H_{LR}(x_U,x_V, \beta\omega) + h.c. \,,
\end{eqnarray}

\begin{equation}\label{eq:HiggsTreeGaugeFree}
\langle H^{(H)} \rangle =
 -\frac{4 G_F}{\sqrt{2}} u\beta\omega \langle Q_2^{LR} \rangle (\mu_h) \sum_{U,V=c,t} \bar{\eta}^{({H})}_{U V} (\mu_h) \lambda_U^{LR}\lambda_V^{RL} \sqrt{x_U x_V} + h.c.
\end{equation}
Above, the masses $ m_U $ are understood to be calculated at $ m_U $, i.e. $ m_U (m_U) $, $ U = u, c, t $.

In the rest of this chapter, we are going to discuss strategies for computing the $ \bar \eta $ in the LR Model. Two approaches are going to be considered, the Effective Field Theory \cite{Gilman:1982ap,Herrlich:1993yv,Herrlich:1996vf} and the Method of Regions \cite{Vainshtein:1976eu,Vysotsky:1979tu,Ecker:1985vv,Bigi:1983bpa}. After comparing them in the SM, where the differences will be clearer, these methods are going to be employed in Chapter~\ref{ch:technicalEFT} for the LR Models.

\section{EFT for meson-mixing in LR Models}\label{sec:generalEFT}

Here we discuss the steps for building the effective theory valid at low energies describing meson-mixing in LR Models, which will be important for the computation of short-distance corrections. We are going to point out some particular features of LR Models, while a comprehensive discussion of the SM case may be found in \cite{Buchalla:1995vs}. For definiteness, we orient our discussion to the kaon system, but a similar discussion also applies in the system of $ B $ mesons.

\subsection{Operator Product Expansion}

The way to formalize an EFT includes an Operator Product Expansion (OPE) \cite{Wilson:1969zs,Witten:1976kx}. Performing an OPE amounts to factorizing short range physics in coupling constants, called Wilson coefficients, and long distance physics corresponding to the dynamical degrees of freedom including any dependence on the external states (supposedly light). This is particularly important when discussing QCD corrections: one collects short-distance, perturbative effects in the Wilson coefficients, while the QCD behaviour at long distances such as hadronization is factorized out and treated at a different step, through appropriate non-perturbative methods.

In order to build an EFT, the first step is precisely to perform an OPE, keeping only those operators which have the lowest power on the high energy scale $ M $, i.e. we keep only the leading power of $ 1/M^2 $ (further corrections in the case of meson-mixing are discussed for example in \cite{Niyogi:1979wm}). The usefulness of this procedure is to simplify the description of the problem by using a limited set of operators. Indeed, in this way suppressed operators are not present from the very beginning.

In few words, building an EFT amounts to defining a new (possibly non-renormalizable) field theory below a certain energy scale $ \mu $ called the \textbf{matching scale}: such a field theory collects the effects of the heavy particles through coupling constants, the Wilson coefficients mentioned above. Each time one builds an effective theory the most general set of operators up to a certain order is taken, and their coefficients or coupling constants are defined by comparison with the full theory.


\subsection{Integrating out heavy particles}

In LR Models, one disposes of the following spectrum of particles

\begin{equation}
H^{0, \pm}_{\alpha}, \, A^{0}_{\alpha}, \, W', \, W, \, t, \, b, \, c, \, s, \, u, \, d ,
\end{equation}
\noindent
where in our case $ \alpha \in \{ 1, 2 \} $, but one can imagine a larger scalar content in other realizations of LR Models. Note that there is a large spread of masses among these particles: $ W' $ for instance is not expected to have a mass below a few TeV, cf. Chapter~\ref{ch:EWPO}, while some of the fermions have masses around a few GeV, or even below. Since some of them are so heavy compared to the other particles, one can consider ``integrating out" $ W', H $ or $ W $ \cite{Buras:1998raa,Donoghue:1992dd}.

In the cases of $ W, W' $, the procedure of integrating out a massive gauge boson implies that the longitudinal degree of freedom, i.e. the non-physical (would-be) Goldstone boson, is also absent in the effective theory. This is consistent with going to a unitary gauge and then integrating out a particle of mass $ M $ whose propagator is

\begin{equation}\label{eq:eqPropW}
\frac{-i}{k^2 - M^2} \left( g_{\mu \nu} - \frac{k_\mu k_\nu}{M^2} \right) .
\end{equation}
Whenever the energy scale $ k^2 $ is much inferior than the mass $ M $ of the propagating vector particle, one can perform the expansion

\begin{equation}\label{eq:expansionWprop}
\frac{i}{M^2} g_{\mu \nu} + \mathcal{O} \left( \frac{k^2}{M^2} \right) \, ,
\end{equation}
meaning that the propagating particle has no more dynamics. In this expansion, the $ \mathcal{O} \left( \frac{k^2}{M^2} \right) $ terms correspond to infinitely many higher dimensional operators in the Fourier transformed space.

\subsection{Set of operators}\label{sec:setOfOpsEFT}

Comparatively to $ W $ and $ t $, the new scalar fields and the $ W' $ boson are much heavier. However, we choose to integrate out all of them at the same energy scale, referred to as $ \mu_W $. We will discuss more on this point at the end of Section~\ref{sec:running}. We note here that we keep only the lowest order corrections in $ \beta $, where $ \beta = M^2_W / M^2_{W'} $, and we do not suppose that $ \omega_i = M^2_{W'} / M^2_{H_i} $ is negligible. Below $ \mu_W $, the most general effective Lagrangian describing meson-mixing in LR Models is given by

\begin{equation}\label{eq:lagrangian5flavours}
\mathcal{L}^{(5)}_{\rm eff} = - \frac{4 G_F}{\sqrt{2}} \mathbf{V}^2 \sum_k C_k Q_k - 2 G^2_F \mathbf{V}^4 \sum_l \tilde{C}_l \tilde{Q}_l ,
\end{equation}
\noindent
where $ \mathbf{V}^n $ indicates the number of powers on the mixing matrices, $ V^{L,R} $. $ Q_k $, $ \tilde{Q}_l $ represent local $ \vert \Delta F \vert = 1 $ and $ \vert \Delta F \vert = 2 $ operators, respectively, while $ C_k $ and $ \tilde{C}_l $ are their corresponding Wilson coefficients. The superscript in parenthesis, ``$ (5) $'' in Eq.~\eqref{eq:lagrangian5flavours}, indicates the number of dynamical flavours. The sum over $ k $ above includes in addition to $ Q^{LR}_{2} $ another LR operator which will be relevant in our discussion afterwards, which is

\begin{equation}
Q^{LR}_{1} = \bar{d} \gamma^\mu P_R s \cdot \bar{d} \gamma_\mu P_L s \, .
\end{equation}

We now discuss the $ \tilde{Q}_l $ operators. Due to chiral flips and Higgs couplings, the expressions in the full theory are proportional to $ m_U \times m_V $, $ U, V = u, c, t $, and we do not need to worry about the up-quark, whose mass is very small and thus set to zero. Therefore, in Eq.~\eqref{eq:lagrangian5flavours} above there is no $ \vert \Delta F \vert = 1 $ operator with an up-quark, and we are left with charm internal flavours only in the EFT. This is different from the SM effective Lagrangian, $ \mathcal{L}^{{\rm SM} \, (5)}_{\rm eff} $, for which the up-quark is present as $ uc, cu $ boxes.

Before further discussion, note that the $ W $ and $ W' $ couple to fermions proportionally to the gauge couplings $ g_L $ and $ g_R $, respectively, while this is not the case for the scalar sector. We therefore deal with (1) the SM contributions, (2) the $ W  \, W' $ box, vertex and self-energy, (3) the $ W  \, H^\pm $ box, and (4) the tree level neutral Higgs exchange separately, which corresponds to distinguishing different powers on $ g_L, g_R $ and respecting gauge invariance. Following the same reasoning, we also distinguish the parts of the effective Lagrangian containing different powers on $ \lambda^{LR}_c $, i.e. we distinguish the three sectors $ cc, ct, tt $. Let us inspect the operators we need in each sector, after integrating out the top-quark, the gauge bosons $ W $ and $ W' $, and the scalars. Below the scale $ \mu_{W} $, we dispose of the light mass $ m_c $ (the masses $ m_t, M_{W,W',H} $ are absorbed into the Wilson coefficients):

\begin{itemize}
\item[$ \mathbf{tt} \, $] We can have only local dimension 6, $ \vert \Delta F \vert = 2 $ operators $ Q^{LR}_{1,2} $.

\item[$ \mathbf{cc} \, $] Here, $ \vert \Delta F \vert = 1 $ local operators are present, and the dimension 8 operators $ m_c^2 \, Q^{LR}_{1,2} $ are needed in order to renormalize the contraction of two $ \vert \Delta F \vert = 1 $ operators, where $ m_c^2 $ appears explicitly since it is related to the light spectrum. Indeed, the $ \log (x_c) $ contribution found in Eq.~\eqref{eq:boxWWprimecc} already indicates the need to include local operators in the EFT.

\item[$ \mathbf{ct} \, $] We have local dimension 7, $ \vert \Delta F \vert = 2 $ operators of the form $ m_c \, Q^{LR}_{1,2} $ ($ m_c $ coming from the matching of the LR Model onto the effective description).
\end{itemize}

Note that there is also the possibility of having $ \vert \Delta F \vert = 1 $ penguin operators coming from the contraction of top propagators, and giving a contribution proportional to $ \lambda^{LL}_t $, or $ \lambda^{RR}_t $. They contribute to the mixing of mesons when contracted with a current-current operator, $ \bar{d} \gamma^\mu P_L q \bar{q} \gamma_\mu P_L s $ or $ \bar{d} \gamma^\mu P_R q \bar{q} \gamma_\mu P_R s $, $ q = u, c $, leading to terms proportional to $ \lambda^{LL}_t \lambda^{RR}_c $, or $ \lambda^{LL}_c \lambda^{RR}_t $. This does not correspond to any of the above mentioned contributions, $ \{ tt, cc, ct \} $, and has not been taken into account in the phenomenological analysis that will follow in Chapter~\ref{ch:PHENO}: by analogy with the SM where penguin operators imply a small effect of the order of $ 1~\% $ \cite{Herrlich:1996vf}, we do not expect a large contribution in LR Models from penguin operators.

Also note that, we have widely ignored operators proportional to the light masses $ m_{d,s} $ so far in the discussion of the effective theory. Accordingly, in the full theory we have neglected corrections which go like $ m^2_{d,s} / M^2_W $, or in other words we have set the external momenta to zero, which consistently corresponds to neglecting higher dimensional operators. (In fact, masses $ m_{d,s} $ may appear, but for a different reason, as off-shell IR regulators.)

To continue, when going down in energy, one goes from $ \mathcal{L}^{(5)}_{\rm eff} $ to $ \mathcal{L}^{(4)}_{\rm eff} $ by integrating out the bottom, thus changing the way the strong coupling $ \alpha_s $ evolves (and possibly having a different set of penguin operators \cite{Herrlich:1996vf}). A further step in the EFT program is to consider the threshold $ \mu_c $ where the charm-quark is integrated out from the theory, through the definition of $ \mathcal{L}^{(3)}_{\rm eff} $, fully described by a set of $ \vert \Delta F \vert = 2 $ local operators $ Q^{LR}_{1,2} $.

A comment concerning $ B $ systems is in order here: in the SM, due to the structure of the CKM matrix and the masses of the up-type quarks, the $ tt $ contribution (proportional to $ m^2_t $) is largely dominant. In LR Models, however, $ ct $ contributions (proportional to $ m_c m_t $) can also be important, given the arbitrary structure of the mixing matrix $ V^R $.\footnote{Following this comment, note that one should as well compute contributions from $ H^\pm_{1,2} $ proportional to $ m_b m_t $, cf. the $ m^j_d $ couplings in Table~\ref{tab:chargedHiggsContributions} in Appendix~\ref{sec:spectrumScalars}, when general structures of $ V^R $ are considered. However, in the phenomenological analysis of Chapter~\ref{ch:PHENO} we consider the special case where $ V^R = V^L $, for which both $ m_c m_t $ and $ m_b m_t $ contributions can be neglected.} For completeness, we also discuss the $ cc $ case, though this contribution is very suppressed even in non-manifest scenarios for $ V^R $. In the $ c t $ and $ c c $ cases, one could formally follow a Heavy Quark Expansion, in which the bottom-quark degree of freedom is decomposed into two pieces, one light and another one heavy, based on a $ 1 / m_b $ expansion. Next, we would consider integrating out the heavy degree of freedom, and neglecting the suppressed corrections which go as $ 1 / m_b $, corresponding to new operators that are usually neglected. In this way, the discussion concerning the operator basis in the $ B $ case follows in exactly the same way the discussion made above for the system of kaons.


\section{Renormalization Group Equations}\label{sec:running}

In the previous section, we have built effective theories by defining effective operators and couplings at the energy scales $ \mu_W $ or $ \mu_c $. These effective couplings, or Wilson coefficients, are not directly observable. Indeed, they generally depend on the energy scale of the renormalization, i.e. the scale at which we choose to subtract possible divergences, which is an arbitrary scale. However, the Lagrangian of the theory, as physical observables, is independent of this choice (there is in fact a residual dependence of the Lagrangian on this scale, which will be discussed at the appropriate moment). This simple remark implies a very beautiful formalism through the introduction of the Renormalization Group Equations (RGE), which tell us how to evolve the couplings of the theory, or other blocks such as matrix elements, from one energy scale to another.

At this moment, we discuss what the running means in practice. To pick an example, the strong coupling $ \alpha_s \equiv 4 \pi \, a $ is not an observable and depends on the energy scale $ \mu $ at which we probe its effects. Its \textit{running} is given by the following RGE

\begin{equation}\label{eq:RunDec}
\frac{d a}{d \log \mu} \stackrel{\rm NLO}{=} - 2 \beta_0 a^2 - 2 \beta_1 a^3 .
\end{equation}
Given a reference scale $ \mu_1 $, the solution to Eq.~\eqref{eq:RunDec} at a scale $ \mu_2 $ is

\begin{equation}
\alpha_s (\mu_2) = \frac{\alpha_s (\mu_1)}{v (\mu_2 ; \mu_1)} \left[ 1 - \frac{\beta_1}{\beta_0} \frac{\alpha_s (\mu_1)}{4 \pi} \frac{\log v (\mu_2 ; \mu_1)}{v (\mu_2 ; \mu_1)} \right] ,
\end{equation}
\noindent
where

\begin{equation}
v (\mu_2 ; \mu_1) = 1 - \beta_0 \frac{\alpha_s (\mu_1)}{4 \pi} \log \left( \frac{\mu_1^2}{\mu_2^2} \right) ,
\end{equation}
and, for the sake of clarity, we have ignored the thresholds at which a heavy quark flavour is integrated out.

Consider now any coupling constant $ C $ (in the absence of mixing between operators). Its running provides multiplicative contributions of the generic form

\begin{equation}\label{eq:eqLOWC}
C (\mu_1) \stackrel{\rm LO}{=} C (\mu_2) \left( \frac{\alpha_s (\mu_1)}{\alpha_s (\mu_2)} \right)^{d} ,
\end{equation}
\noindent
for a certain power $ d $, which can be expanded as follows

\begin{eqnarray}\label{Eq:TaylorforMass}
\left( \frac{\alpha_s (\mu_1)}{\alpha_s (\mu_2)} \right)^{d} & = & v^{d} (\mu_2 ; \mu_1) \cdot \left[ 1 - \frac{\beta_1}{\beta_0} \frac{\alpha_s (\mu_1)}{4 \pi} \frac{\log v (\mu_2 ; \mu_1)}{v (\mu_2 ; \mu_1)} \right]^{-d} \\
& = & 1 - \gamma^{(0)} a (\mu_1) \log \left( \frac{\mu_1}{\mu_2} \right) - \gamma^{(0)} \, \frac{\beta_1}{\beta_0} a^2 (\mu_1) \log \left( \frac{\mu_1}{\mu_2} \right) + \ldots \nonumber ,
\end{eqnarray}
where in the last line we have traded $ d $ by $ \gamma^{(0)} / (2 \beta_0) $. Therefore, by evolving a mass from $ \mu_2 $ to $ \mu_1 $, one collects factors of the form $ a (\mu_1) \log \left( \mu_1 / \mu_2 \right) $, called Leading Order (LO), and $ a^2 (\mu_1) \log \left( \mu_1 / \mu_2 \right) $, called Next-to-Leading Order (NLO), etc.\footnote{To be consistent, at the NLO we should employ

\begin{equation}
C (\mu_1) = C (\mu_2) \left( 1 + \frac{\alpha_s (\mu_2)}{4 \pi} J \right) \left( \frac{\alpha_s (\mu_1)}{\alpha_s (\mu_2)} \right)^{d} \left( 1 - \frac{\alpha_s (\mu_1)}{4 \pi} J \right) ,
\end{equation}
where $ J $ corrects the running at the NLO. Note that the NLO running of $ \alpha_s $ was used even when the running of $ C $ was considered up to the LO: we may say in this case that the calculation is done at the LO with $ \alpha_s $ improved to the NLO.
}

The last Taylor expansion has a meaning only if the terms contained in the ellipsis are much smaller than one. To further discuss this point, let us consider the example of the running of masses and give some numerical values. First

\begin{equation}
\gamma^{(0)}_m = 6 C_F , \quad \beta_0 = \frac{11 N - 2 f}{3} , \quad \beta_1 = \frac{34}{3} N^2 - \frac{10}{3} N f - 2 C_F f , \quad C_F = \frac{N^2 - 1}{2 N} ,
\end{equation}
where $ f $ is the number of dynamical flavours, and $ N $ the number of colors. Then

\begin{equation}
\beta_1 / \beta_0 \sim 6, 5, 3, 1 \qquad {\rm for} \qquad f = 3, 4, 5, 6 \, .
\end{equation}
For $ \mu_1 \rightarrow m_c $ and $ \mu_2 \rightarrow M_W $ one has

\begin{equation}
\alpha_s (m_c) \simeq 0.3 \qquad {\rm and} \qquad \log (M_W / m_c) \simeq 4 \, .
\end{equation}
It is then clear that the Taylor expansion (for $ f = 3, 4, 5 $) made above cannot be justified, because $ \gamma^{(0)}_m a (\mu_1) \vert \log ( \mu_1 / \mu_2 ) \vert \sim 0.8 $ is too large and does not allow for an expansion. This illustrates that the solution to the RGE \textit{resums large logarithms to all orders in perturbation theory}. Therefore, RGE prove to be a very efficient, and even mandatory, way to improve perturbative calculations in $ \alpha_s $.

In this context, we discuss the issue of a unique energy scale for integrating out $  W, W', H^{0, \pm} $ and $ t  $. Alternatively, $ W', H $ could be integrated out at $ \mathcal{O} ( M_{W'}, M_H ) $, and we would be left with a theory containing dynamical $ W $ bosons and top-quarks. At the end of the day, we would have a resummation of $ \alpha_s (M_W) \log \beta $ to all orders in perturbation theory when running the EFT defined at $ \mathcal{O} (M_{W'}, M_H) $ down to $ \mathcal{O} (M_{W}, m_t) $. This then justifies the procedure of considering a single scale $ \mu_W $ for integrating out $ \{ W, W', H^{0, \pm}, t \} $: these resummed factors are small in the interesting phenomenological range of LR Models, namely $ M_{W'} \lesssim 10 $~TeV, for which $ \alpha_s (M_W) \log \beta / \pi \lesssim 0.3 $, and therefore do not require such a precise calculation. Conversely, the need for precision in the resummation of $ \log \beta $ for large $ \log \beta $ with an EFT between EW and LR scales gets damped by an overall $ \beta $ suppression factor. (Similar comments would also apply to the logarithm $ \log (\beta \omega) $, provided the difference of masses of the $ W' $ and the extended Higgs sector is not too large.) Note that, on the other hand, we cannot do the same for contributions which go like $ \log x_c $, cf. Eq.~\eqref{eq:SMloopfunctions}. This can be seen by the comparison between both resummations:

\begin{table}[h!]
\centering
\begin{tabular}{c|c}
$ \alpha_s (m_c) \vert \log (x_c) \vert / \pi $ & $ \alpha_s (M_W) \vert \log (\beta) \vert / \pi $ \\
\hline
$ \sim 0.8 $ & $ \sim [0.2, 0.3] $ \\
\end{tabular}
\end{table}

\noindent
showing that the resummation of $ \alpha_s (m_c) \log (x_c) $ is mandatory, while higher orders in $ \alpha_s (M_W) \log (\beta) $ can be neglected in a first approximation (but will contribute to the error budget).

We consider now the counting of QCD corrections in the situation of having $ \log (x_c) $ in the Wilson coefficient, a discussion that will be relevant in what will follow. As seen from Eq.~\eqref{eq:boxWWprimecc}, the loop-function corresponding to the $ cc $ contribution contains a large logarithm $ \log (x_c) $. In this case, when running from $ \mu_W $ down to $ \mu_c $ at LO we will have the overall factor $ \log (\mu_c / \mu_W) $ \cite{Herrlich:1996vf}. Therefore, when a large logarithm is present in the loop-functions we have the following counting: factors of the form

\begin{equation}
\log \left( \mu_1 / \mu_2 \right) \times \left[ a (\mu_1) \log \left( \mu_1 / \mu_2 \right) \right]^n \, , \quad n \geq 0 \, ,
\end{equation}
are called LO, while

\begin{equation}
\left[ a (\mu_1) \log \left( \mu_1 / \mu_2 \right) \right]^n \, , \quad n \geq 0 \, ,
\end{equation}
are called NLO, and so on. Another interesting aspect of Eq.~\eqref{eq:boxWWprimecc}, combined with the remaining pieces necessary for gauge invariance, is the need to go beyond the LO for computing short-distance corrections: as can be seen from Table~\ref{tab:goingBeyondLO}, the non-logarithmic term $ 1 + \frac{\log (\beta) + F (\omega_1, \omega_2)}{4} $ may be large, thus indicating a sizeable correction coming from the NLO.

\begin{table}[h!]
\centering
\begin{tabular}{|c|cccc|}
\hline
LO & $ \log (x_c) $ & $ \sim $ & $ -8.2 $ & \\
\hline
    & $ 1 $ & & & \\
NLO & $ \frac{\log (\beta)}{4} $ & $ \sim $ & $ [-1.3, -2.4] $, & for $ M_{W'} \in [1,10] $~TeV \\
    & $ \frac{F (\omega_1, \omega_2)}{4} $ & $ \sim $ & $ [-0.9, 5] $, & for $ \omega_1 \in [0.1,1], w = 0 $ \\
\hline
\end{tabular}
\caption{\it Contributions from the charm-charm case: compared to the other factors from Eq.~\eqref{eq:finalLoopFunct}, the factor $ \log (x_c) $ comes at LO, while the remaining terms come at the NLO. When $ w=0 $ as in the simplified case shown here, the function $ F (\omega_1, \omega_2) $ reduces to a function on $ \omega_1 $ only.}\label{tab:goingBeyondLO}
\end{table}


\section{Method of Regions}\label{sec:MRforPedestrians}

Chronologically, before the EFT approach was used to calculate short-distance QCD corrections for $ K \overline{K} $ meson-mixing in \cite{Gilman:1982ap}, a way to estimate them was discussed in Refs.~\cite{Vainshtein:1976eu} and \cite{Vysotsky:1979tu}, by employing a method we call Method of Regions (MR). The idea behind this method is to resum potentially large logarithms of the form $ \alpha_s \cdot \log $ by using RGE, thus providing an estimate for the short-distance QCD corrections. With the development of the EFT formalism, see \cite{Buchalla:1995vs}, the MR may be seen nowadays as an approximation to the complete calculation, made by a systematic use of EFT, which employs RGE in effective descriptions of the full LR Model.

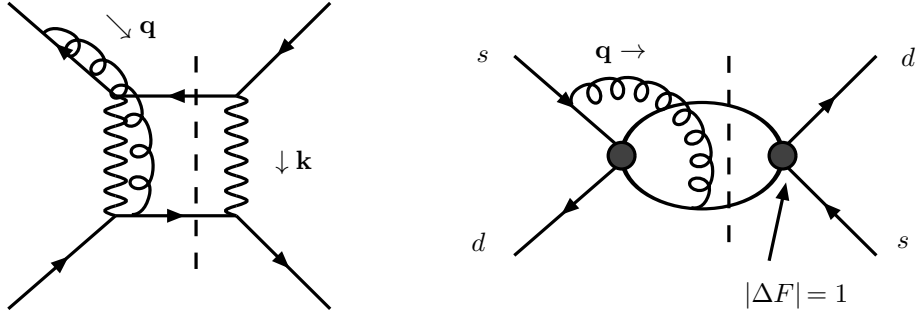
\begin{figure}
	\hspace{1.0cm} \SetWidth{1.2}
	\begin{picture}(200,50)(35,0)
		\ArrowLine(80,25)(40,60)
		\ArrowLine(40,-55)(80,-20)
		\ArrowLine(125,25)(80,25)
		\Photon(80,25)(80,-20){4}{6}
		\GlueArc(40,0)(50,-23,75){4}{6}
		\ArrowLine(160,60)(125,25)
		\ArrowLine(125,-20)(160,-55)
		\ArrowLine(80,-20)(125,-20)
		\Photon(125,25)(125,-20){4}{6}
		\Text(154,0)[r]{$ \mathbf{\downarrow k} $}
		\Text(86,56)[t]{$ \mathbf{\searrow q} $}
		\DashLine(110,-40)(110,40){6}
	\end{picture}
	\SetWidth{1.2}
	\begin{picture}(200,50)(50,20)
		\ArrowLine(40,60)(80,25)
		\Text(26,60)[l]{$ s $}
		\ArrowLine(80,20)(40,-15)
		\Text(24,-10)[l]{$ d $}
		\GOval(110,22.5)(20,30)(360){1}
		\GCirc(80,22.5){5}{0.25}
		\ArrowLine(140,25)(175,60)
		\Text(190,60)[r]{$ d $}
		\ArrowLine(175,-15)(140,20)
		\Text(188,-10)[r]{$ s $}
		\GCirc(140,22.5){5}{0.25}
		\Text(165,-30)[r]{$ \vert \Delta F \vert \, = 1 $}
		\GlueArc(80,18)(30,-30,128){4}{8}
		\Text(80,64)[t]{$ \mathbf{q \rightarrow} $}
		\DashLine(120,-10)(120,60){6}
		\LongArrow(135,-17)(141,11)
	\end{picture}
	\vspace{2cm}
	\caption{\it Example of short-distance corrections to meson-mixing processes in the Method of Regions. (Left) The momentum $ k $ running in the original loop is fixed, and one distinguishes the relevant range of momenta of the gluon  $ q $ giving $ \alpha_s \cdot \log (k^2/M_W^2) $ factors. (Right) Over this range, we integrate out the $ W $ boson and identify the relevant anomalous dimension necessary to resum the potentially large $ \alpha_s \cdot \log (k^2/M_W^2) $ contribution.}\label{fig:MRexample}
\end{figure}

We now explain how the MR operates. Consider dressing the set of Electroweak one-loop diagrams in Figure~\ref{fig:tabNPdiagrams} with the exchange of a gluon in all possible ways. One of the resulting two-loop diagrams is given in Figure~\ref{fig:MRexample}. Based on this particular example, the prescription given by the MR to estimate short-distance QCD corrections is the following:

\begin{itemize}
	\item Fix the momentum $ k $ running in the EW loop. Then, consider the ``two sides'' of the diagram separately, i.e. consider the contraction of two $ \vert \Delta F \vert \, = 1 $ diagrams, one of them containing the exchange of a gluon.
	\item Determine the range of the momentum $ q $ of the gluon implying terms of the generic form $ \alpha_s \cdot \log $. In the present example, we have that the integration of $ q^2 $ over the range $ [k^2, M^2_W] $ results in a term proportional to $ \alpha_s \cdot \log (k^2 / M^2_W) $.
	\item Over the last range, integrate out the $ W $ being exchanged, and identify the anomalous dimensions $ \gamma $ of the four-quark operators. Schematically, the running given by RGE of an operator of anomalous dimension $ \gamma $ from $ M^2_W $ to $ k^2 $ results in a factor $ (\alpha_s (M_W^2) / \alpha_s (k^2))^{\gamma / 2 \beta_0} $. It is precisely this factor that we aim at extracting from all the possible gluon exchanges.
	\item It follows from the last step that the method can be applied if the relevant anomalous dimensions are already known: in the example we have shown, only $ \vert \Delta F \vert \, = 1 $ are required, but for other diagrams anomalous dimensions describing how $ \vert \Delta F \vert \, = 2 $ operators evolve are also necessary.
	\item Finally, perform the integration over $ k^2 $. The dominant range of $ k^2 $ is determined from the loop functions. One distinguishes two possible cases:
	\begin{itemize}
		\item[(a)] the loop function is dominated by a single mass scale, $ m $, for which $ k^2 $ is replaced by $ m^2 $;
		\item[(b)] the loop function is dominated by the range $ [ m_1 , m_2 ] $ coming from the logarithm $ \log (m_1 / m_2) $, in which case one performs the average $ \int^{m_1^2}_{m_2^2} \frac{d k^2}{k^2} \alpha_s^{\gamma / 2 \beta_0} (k^2) $.
	\end{itemize}
\end{itemize}

The last step requires more precision. Generalizing Ref.~\cite{Vainshtein:1976eu} in case (b) at NLO, we define the following averaging function

\begin{eqnarray}
R^{NLO}_{\log}(\gamma,U,J;m_1,m_2) &=& \log^{-1}\frac{m_2^2}{m_1^2}  \left(\frac{\alpha_s(m_1)}{\alpha_s(\mu)}\right)^{-\gamma} \\
&& \qquad \times \int_{m_1^2}^{m_2^2} \frac{dk^2}{k^2}  \left(\frac{\alpha_s(k)}{\alpha_s(\mu)}\right)^{\gamma} \left[U+\frac{\alpha_s(k)}{4\pi}J\right] \, , \nonumber
\label{eq:rnlo}
\end{eqnarray}
where $U, J$ are related to the matching and renormalization of the $ \vert \Delta F \vert \, = 1, 2 $ operators and do not depend on $k$, yielding for $ \gamma \neq 0, 1 $
\begin{eqnarray}
&& R^{NLO}_{\log}(\gamma,U,J;m_1,m_2)=\frac{1}{\log ( m_2^2/m_1^2)}\frac{4\pi}{\beta_0 \alpha_s(m_1)}\\ \nonumber
&&\qquad\qquad\times\biggl[ \frac{1}{1-\gamma}
\biggl\{ \biggl( 
\frac{\alpha_s(m_2)}{\alpha_s(m_1)} 
\biggr)^{\gamma-1} -1
 \biggr\}U \nonumber\\
&&\qquad\qquad\qquad+\frac{\alpha_s(m_1)}{4\pi}\frac{1}{\gamma}\left[\frac{\beta_1}{\beta_0}U-J\right]\biggl\{\biggl (\frac{\alpha_s(m_2)}{\alpha_s(m_1)}\biggr)^{\gamma} -1\biggr\} 
\biggr]\,. \nonumber
\end{eqnarray}
The function defined above includes contributions from the NLO proportional to the factor $ J $, and from the running of the strong coupling constant at the NLO (term proportional to $ \beta_1 $). The case (a), where a single mass scale $m_1$ dominates the integral, is a limit of case (b) and is given by

\begin{equation}\label{eq:rnlo1}
R^{NLO}_1(\gamma,U,J;m_1,m_2)=U+\frac{\alpha_s(m_1)}{4\pi}J \, .
\end{equation}


When moving to the NLO, a feature not present in the original works, we employ the known anomalous dimensions at the NLO (the factors in $ J $), and in a full NLO computation matching corrections at the NLO are also required. We would like now to point out a feature concerning the extension of the MR to the NLO. In the cases of top-top and charm-charm contributions, Eq.~\eqref{eq:SMloopfunctions}, LO and NLO short-distance corrections resum the following terms

\begin{equation}
( \alpha_s \cdot \log )^n \, , \qquad \alpha_s \cdot ( \alpha_s \cdot \log )^n \, , \qquad n \geq 0 \, ,
\end{equation}
respectively. In the charm-top case, however, there is a large logarithm in the loop function, Eq.~\eqref{eq:SMloopfunctions}, resulting in the counting

\begin{equation}
\log \cdot \, ( \alpha_s \cdot \log )^n \, , \qquad ( \alpha_s \cdot \log )^n \, , \qquad n \geq 0 \, .
\end{equation}

In this specific case, we extend the MR at NLO in the following way:

\begin{itemize}
	\item at LO, only the factor $ \log x_c $ contributes, whose important range of momenta is $ k^2 \in [m_c^2, M_W^2] $, while
	\item at NLO there is a contribution coming from the anomalous dimension matrices at NLO, namely the factors $ J $ in $ R^{NLO}_{\log} $, and
	\item there is a second contribution at NLO proportional to the non-logarithmic factor $ F (x_t) $ seen in Eq.~\eqref{eq:nonLogContribution}, whose important range of gluon momenta is given by $ k^2 \rightarrow m_c^2 $ (and therefore it is multiplied by the averaging function $ R_{1} $ instead of $ R_{\log} $).
\end{itemize}

\section{Calculation of the short-distance QCD corrections in the SM}\label{sec:comparisonMREFT}

We would like to check the validity of the MR in the SM where short-distance QCD corrections were computed in the EFT approach \cite{Gilman:1982ap,Herrlich:1993yv,Herrlich:1996vf,Brod:2010mj,Brod:2011ty}. Therefore, when moving to the LR Model in the next chapter, we will know when this method gives good estimates of the EFT, and when EFT is suitable instead in a more precise calculation.

In the SM, the short-distance corrections are summarized in the $ \eta $ parameters in the following expression

\begin{eqnarray}\label{eq:TheOldKnownSM}
&& H^{\rm SM} = \frac{G_F^2 M_W^2}{4 \pi^2}
\Bigg[ \lambda_c^{LL} \lambda_c^{LL} \eta_{cc} S^{LL}(x_c) \\
&& + \lambda_t^{LL} \lambda_t^{LL} \eta_{tt} S^{LL}(x_t) +2 \lambda_t^{LL} \lambda_c^{LL} \eta_{ct} S^{LL}(x_c,x_t) \Bigg] b (\mu_h) Q^{VLL}_1 + h.c. \nonumber
\end{eqnarray}
In this case, since one faces a unique operator $ Q^{VLL}_1 $ one usually factorizes out from the short-distance QCD corrections the dependence on the hadronization scale, defining a scale-independent quantity $ \eta $ related to $ \bar{\eta} $ by 

\begin{equation}
\bar{\eta} (\mu_h) = \eta b (\mu_h) \, ,
\end{equation}
where

\begin{equation}\label{eq:bFactormuh}
b (\mu_h) = \left( 1 + \frac{\alpha^{(3)}_s (\mu^2_h)}{4 \pi} J^{(3)}_{V} \right) \left( \alpha^{(3)}_s (\mu^2_h) \right)^{-d^{(3)}_{V}} \, .
\end{equation}
We now discuss the two ways to calculate $ \eta $.

\subsection{Method of Regions}




We first discuss the relevant set of diagrams. In the SM, the set of box diagrams including the $ W $ and its Goldstone forms a gauge invariant set. In the 't Hooft-Feynman gauge, one may distinguish the boxes $ WW $, $ WG $ and $ GG $, which are proportional to $ I_2 (x_U, x_V) $, $ x_U x_V I_1 (x_U, x_V) $ and $ x_U x_V I_2 (x_U, x_V) $ respectively, cf. Eq.~\eqref{eq:InamiLimSMcase} ($ I_{1,2} $ are the Inami-Lim functions). Then, these diagrams are dressed with gluons and identify the resulting two-loop diagrams contributing for the short-distance QCD corrections.

In the charm-charm case, contributions given by the exchange of Goldstones come at higher order in $ x_c $, and the loop function can be calculated simply by considering a $ W W $ box. Therefore, the method of regions gives


\begin{eqnarray}
\eta_{cc} & = & \sum_{r, \ell = \pm} \Bigg[ C_{\rm low} (k^{2}) \underbrace{ \left( a_{r, \ell}^{V (WW)} + \frac{\alpha^{(4)}_s (k^2)}{4 \pi} \left( \kappa_{r, \ell} \log \left( \frac{k^2}{m^{2}_{c}} \right) + \beta_{r, \ell} \right) \right) }_{matching ~ (4)} \\
&& C^{r}_{\rm high} (k^{2}) \, C^{\ell}_{\rm high} (k^{2}) \Bigg]\nonumber\\
& \times & \underbrace{ \left[ \left( 1 + \frac{\alpha^{(4)}_s (m^2_{c})}{4 \pi} J^{(4)}_{m} \right) \left( \frac{\alpha^{(4)}_s (k^2)}{\alpha^{(4)}_s (m^2_{c})} \right)^{d^{(4)}_{m}} \left( 1 - \frac{\alpha^{(4)}_s (k^2)}{4 \pi} J^{(4)}_{m} \right) \right]^{2} }_{running ~ (7)} , \nonumber
\end{eqnarray}

\begin{eqnarray}
C_{\rm low} (k^{2}) & = & \underbrace{ \left( \alpha^{(3)}_s (\mu^2_{4}) \right)^{d^{(3)}_{V}} \left( 1 - \frac{\alpha^{(3)}_s (\mu^2_{4})}{4 \pi} J^{(3)}_{V} \right) }_{running ~ (6)} \nonumber\\
& \times & \underbrace{ \left[ \left( 1 + \frac{\alpha^{(4)}_s (\mu^2_{4})}{4 \pi} J^{(4)}_{V} \right) \left( \frac{\alpha^{(4)}_s (k^2)}{\alpha^{(4)}_s (\mu^2_{4})} \right)^{d^{(4)}_{V}} \left( 1 - \frac{\alpha^{(4)}_s (k^2)}{4 \pi} J^{(4)}_{V} \right) \right] }_{running ~ (5)} \, , 
\end{eqnarray}

\begin{eqnarray}
C^{r}_{\rm high} (k^{2}) & = & \underbrace{ \left[ \left( 1 + \frac{\alpha^{(4)}_s (k^2)}{4 \pi} J^{(4)}_{r} \right) \left( \frac{\alpha^{(4)}_s (\mu^2_{5})}{\alpha^{(4)}_s (k^{2})} \right)^{d^{(4)}_{r}} \left( 1 - \frac{\alpha^{(4)}_s (\mu^2_{5})}{4 \pi} J^{(4)}_{r} \right) \right] }_{running ~ (3)} \nonumber\\
& \times & \underbrace{ \left[ \left( 1 + \frac{\alpha^{(5)}_s (\mu^2_{5})}{4 \pi} J^{(5)}_{r} \right) \left( \frac{\alpha^{(5)}_s (\mu^2_W)}{\alpha^{(5)}_s (\mu^2_{5})} \right)^{d^{(5)}_{r}} \left( 1 - \frac{\alpha^{(5)}_s (\mu^2_W)}{4 \pi} J^{(5)}_{r} \right) \right] }_{running ~ (2)} \nonumber\\
& \times & \underbrace{ \left( 1 + \frac{\alpha^{(5)}_s (\mu^2_{W})}{4 \pi} B_{r} \right) }_{matching ~ (1)} , \quad r = \pm , 
\end{eqnarray}
with $ k^{2} \rightarrow m^{2}_{c} $ and

\begin{eqnarray}
a_{r, \ell}^{V (WW)} & = & \frac{1 + r + \ell + 3 \cdot r \cdot \ell}{4} , \quad \sum_{r, \ell = \pm} a_{r, \ell}^{V (WW)} = 1 , \\
\kappa_{++} & = & a_{++}^{V (WW)} \, 3 \, (N-1) , \quad \kappa_{+-} = \kappa_{-+} = a_{+-}^{V (WW)} \, 3 \, (N+1) , \nonumber\\
\kappa_{--} & = & a_{--}^{V (WW)} \, 3 \, (N+3) , \\
\beta_{++} & = & (1-N) \left( \frac{N^2 - 6}{12 N} \pi^{2} + 3 \frac{- N^2 + 2 N + 13}{4 N} \right) , \\
\beta_{+-} & = & \beta_{-+} = (1-N) \left( \frac{- N^2 + 2 N - 2}{12 N} \pi^{2} + \frac{3 N^2 + 13}{4 N} \right) , \\
\beta_{--} & = & (1-N) \left( \frac{N^2 - 4 N + 2}{12 N} \pi^{2} - \frac{3 N^2 + 10 N + 13}{4 N} \right) \, .
\end{eqnarray}
Let us describe the many factors in the expression of $ \eta_{cc} $, enumerated from one to seven above: (1) is a NLO matching onto $ | \Delta F | = 1 $ operators at the scale $ \mu_W $, including the factor $ B_r $ calculated from \cite{Buchalla:1995vs}; (2) and (3) describe the running of $ | \Delta F | = 1 $ operators from the scale $ \mu_W^2 $ down to the scale $ \mu_5^2 $, and from the scale $ \mu_5^2 $ down to the scale $ k^2 $, respectively; (4) describes the contraction of two $ | \Delta F | = 1 $ operators resulting in a $ | \Delta F | = 2 $ operator, $ a_{r, \ell}^{V (WW)} $ arriving at the LO and $ \beta_{r \ell}, \kappa_{r \ell} $ at the NLO, the latter calculated in \cite{Herrlich:1993yv}; (5) and (6) describe the running of the SM $ | \Delta F | = 2 $ operator from the scale $ k^2 $ down to the scale $ \mu_4^2 $, and from the scale $ \mu_4^2 $ down to the scale $ \mu_h^2 $, respectively; in (6), the dependence on $ \mu_h^2 $ has been further factorized out into $ b(\mu_h) $ defined in Eq.~\eqref{eq:bFactormuh}; finally, (7) gives the running of the overall mass factor from $ k^2 $ (coming from the computation of the $ WW $ box diagram), to $ m^2_c $ (note that since $ k^2 $ is replaced by the relevant scale $ m^2_c $, this factor reduces to 1). Factors which are proportional to the many $ J $ come at the NLO, and are found in Appendix~\ref{app:anomdimgen}.



In the top-top case, contributions where Goldstones are exchanged in the 't Hooft-Feynman gauge are not suppressed by $ x_c $ as in the charm-charm case and must be considered.  
Note that $ \log x_t \simeq 1.5 $ and therefore we do not apply in such a case the $ R_{\log} $ averaging. Considering the three different contributions, $ WW, WG, GG $, the short-distance QCD correction for the top-top contribution to meson-mixing in the SM is given by

\begin{equation}
\eta_{tt} = \frac{(\eta_{tt}^{V (WW)} + \eta_{tt}^{V (GG)} x_t^2 / 4) I_2 (x_t,x_t,1) - \eta_{tt}^{V (WG)} 2 x_t^2 I_1 (x_t,x_t,1)}{(1 + x_t^2 / 4) I_2 (x_t,x_t,1) - 2 x_t^2 I_1 (x_t,x_t,1)} \, ,
\label{eq:etattMR}
\end{equation}
where we have

\begin{eqnarray}
\eta^{V (WW)}_{tt} & = & \sum_{r, \ell = \pm} \left[ C_{\rm low} (k^{2}) a_{r, \ell}^{V (WW)} C^{r}_{\rm high} (k^{2}) \, C^{\ell}_{\rm high} (k^{2}) \right] , \quad k^{2} \rightarrow m^{2}_{t}, \\
C_{\rm low} (k^{2}) & = & \left( \alpha^{(3)}_s (\mu^2_{4}) \right)^{d^{(3)}_{V}} \left( 1 - \frac{\alpha^{(3)}_s (\mu^2_{4})}{4 \pi} J^{(3)}_{V} \right) \\
& \times & \left[ \left( 1 + \frac{\alpha^{(4)}_s (\mu^2_{4})}{4 \pi} J^{(4)}_{V} \right) \left( \frac{\alpha^{(4)}_s (\mu^2_{5})}{\alpha^{(4)}_s (\mu^2_{4})} \right)^{d^{(4)}_{V}} \left( 1 - \frac{\alpha^{(4)}_s (\mu^2_{5})}{4 \pi} J^{(4)}_{V} \right) \right] \nonumber\\
& \times & \left[ \left( 1 + \frac{\alpha^{(5)}_s (\mu^2_{5})}{4 \pi} J^{(5)}_{V} \right) \left( \frac{\alpha^{(5)}_s (k^{2})}{\alpha^{(5)}_s (\mu^2_{5})} \right)^{d^{(5)}_{V}} \left( 1 - \frac{\alpha^{(5)}_s (k^{2})}{4 \pi} J^{(5)}_{V} \right) \right] \, , \nonumber\\
C^{r}_{\rm high} (k^{2}) & = & \left[ \left( 1 + \frac{\alpha^{(5)}_s (k^2)}{4 \pi} J^{(5)}_{r} \right) \left( \frac{\alpha^{(5)}_s (\mu^2_W)}{\alpha^{(5)}_s (k^2)} \right)^{d^{(5)}_{r}} \left( 1 - \frac{\alpha^{(5)}_s (\mu^2_W)}{4 \pi} J^{(5)}_{r} \right) \right] \nonumber\\
& \times & \left( 1 + \frac{\alpha^{(5)}_s (\mu^2_{W})}{4 \pi} B_r \right), \quad r = \pm ,
\end{eqnarray}
read as $ \eta_{cc} $, and

\begin{eqnarray}
\eta^{V (WG)}_{tt} & = & \sum_{r = \pm} \left[ C_{\rm low} (k^{2}) \left( a_{r}^{V (WG)} \cdot \overrightarrow{C}_{\rm high} (k^{2}) \right) C^{r}_{\rm high} (k^{2}) \right] \\
& \times & \left[ \left( 1 + \frac{\alpha^{(5)}_s (m^2_{t})}{4 \pi} J^{(5)}_{m} \right) \left( \frac{\alpha^{(5)}_s (k^2)}{\alpha^{(5)}_s (m^2_{t})} \right)^{d^{(5)}_{m}} \left( 1 - \frac{\alpha^{(5)}_s (k^2)}{4 \pi} J^{(5)}_{m} \right) \right]^{2} \nonumber\\
& \times & \left[ \left( 1 + \frac{\alpha^{(5)}_s (m^2_{t})}{4 \pi} J^{(5)}_{m} \right) \left( \frac{\alpha^{(5)}_s (\mu^2_{W})}{\alpha^{(5)}_s (m^2_{t})} \right)^{d^{(5)}_{m}} \left( 1 - \frac{\alpha^{(5)}_s (\mu^2_{W})}{4 \pi} J^{(5)}_{m} \right) \right]^{2} , \quad k^{2} \rightarrow m^{2}_{t} , \nonumber\\
\overrightarrow{C}_{\rm high} (k^{2}) & = & \left[ \left( 1_{2} + \frac{\alpha^{(5)}_s (k^2)}{4 \pi} \hat{J}^{(5)} \right) \hat{V} \left( \frac{\alpha^{(5)}_s (\mu^2_W)}{\alpha^{(5)}_s (k^2)} \right)^{\overrightarrow{d}^{(5)}} \hat{V}^{-1} \left( 1_{2} - \frac{\alpha^{(5)}_s (\mu^2_W)}{4 \pi} \hat{J}^{(5)} \right) \right] \overrightarrow{C}_{0} , \nonumber
\end{eqnarray}
\noindent
where

\begin{equation}
a_{r}^{V (WG)} = \begin{pmatrix}
- (1 + N \cdot r) & - (1 + r) \\
\end{pmatrix} , \quad \overrightarrow{C}^{\rm T}_{0} = \begin{pmatrix}
0 & - 1/2 \\
\end{pmatrix}, \quad \sum_{r = \pm} a_{r}^{V (WG)} \cdot \overrightarrow{C}_{0} = 1 ,
\end{equation}
\noindent
and

\begin{eqnarray}
\eta^{V (GG)}_{tt} & = & \left[ C_{\rm low} (k^{2}) \left( \overrightarrow{C}^{\rm T}_{\rm high} (k^{2}) \cdot \hat{a}^{V (GG)} \cdot \overrightarrow{C}_{\rm high} (k^{2}) \right) \right] \\
& \times & \left[ \left( 1 + \frac{\alpha^{(5)}_s (m^2_{t})}{4 \pi} J^{(5)}_{m} \right) \left( \frac{\alpha^{(5)}_s (\mu^2_{W})}{\alpha^{(5)}_s (m^2_{t})} \right)^{d^{(5)}_{m}} \left( 1 - \frac{\alpha^{(5)}_s (\mu^2_{W})}{4 \pi} J^{(5)}_{m} \right) \right]^{4} , \quad k^{2} \rightarrow m^{2}_{t} , \nonumber
\end{eqnarray}
\noindent
where

\begin{equation}
\hat{a}^{V (GG)} = 4 \begin{pmatrix}
N & 1 \\
1 & 1 \\
\end{pmatrix} , \quad \overrightarrow{C}^{\rm T}_{0} \cdot \hat{a}^{V (GG)} \cdot \overrightarrow{C}_{0} = 1 .
\end{equation}

The factors in $ \eta^{V (WG)}_{tt} $ and $ \eta^{V (GG)}_{tt} $ are analogous to those found in $ \eta^{V (WW)}_{tt} $ or $ \eta_{cc} $. Note that we include $ B_{r, \ell} $ corrections known from \cite{Buchalla:1995vs} for the matching of a $ \vert \Delta F \vert \, = 1 $ process with a $ W $ dynamic onto an EFT where the $ W $ is integrated out (and the top is left dynamic), matching onto a operator of the form $ \gamma^\mu P_L \otimes \gamma_\mu P_L $. Similar corrections onto $ P_L \otimes P_R $ operator structures were not considered, which should be anyways small since they are suppressed by $ \alpha_s (\mu_W) $.





The discussion of the $ \eta_{ct} $ short-distance QCD correction is different due to the $ \log x_c $ in its loop function. The following equation follows the procedure described at the end of Section~\ref{sec:MRforPedestrians} (ignoring thresholds, resulting in an approximation better than $ 1~\% $)

\begin{eqnarray}\label{eq:etactSMinMR}
\eta_{ct}&=&\frac{1}{-\log x_c + F(x_t)}\alpha_s(m_c)^{d_V} \sum_{r,\ell=\pm } a_{r\ell}
\left(\frac{\alpha_s(\mu_W)}{\alpha_s(m_c)}\right)^{d_\ell+d_r} 
 \\\nonumber
&&\times 
\biggl(-\log x_c \, R^{NLO}_{\log} \Bigg[-d_\ell-d_r+d_V+2 d_m,
 u_{r\ell}, j_{r\ell} ; m_c,M_W \Bigg] 
+ F(x_t)  \biggr)\,,\nonumber
\end{eqnarray}
where
\begin{equation}
a_{r\ell} = [1+r+\ell+N r\ell]/4
\end{equation}
and the sum $ \sum_{r,\ell} $ runs over the possible contractions $ \mathcal{O}_{\varepsilon_{R} \varepsilon_{L}} \equiv \mathbf{\rm T} \{ Q^L_{\varepsilon_{R}}, Q^L_{\varepsilon_{L}} \} $, $ \varepsilon_{R}, \varepsilon_{L} = \pm $, where the $ Q^L_{\varepsilon} $ operators are defined as

\begin{equation}\label{eq:plusMinusOps}
Q^L_{\varepsilon} = \bar{d} \gamma^{\lambda} P_{L} q' \cdot \bar{q} \gamma_{\lambda} P_{L} s \, \frac{(\mathbf{\hat{1}} + \varepsilon \mathbf{\tilde{\hat{1}}})}{2} \, .
\end{equation}
These are multiplicatively renormalizable operators (neglecting penguin operators), where $ \mathbf{\hat{1}} $ denotes a color singlet structure and $ \mathbf{\tilde{\hat{1}}} $ a color anti-singlet. The other factors seen in Eq.~\eqref{eq:etactSMinMR} include the NLO corrections from matching and running, and are defined by
\begin{eqnarray}
u_{r\ell}&=&1+ 2\frac{\alpha_s(m_c)}{4\pi} J_m-\frac{\alpha_s(\mu_W)}{4\pi}( J_\ell +J_r-B_\ell-B_r), \nonumber\\
j_{r\ell} &=& J_\ell+J_r-J_V-2J_m\,,
\end{eqnarray}
where the factors $ d, J, B $ are given in Appendix~\ref{app:anomdimgen}.



\subsection{EFT}

\begin{figure}
\begin{minipage}{0.3\textwidth}
	\hspace{1cm} \SetWidth{1.2}
	\begin{picture}(90,40)(-20,0)
		\ArrowLine(0,0)(30,20)
		\ArrowLine(30,20)(60,20)
		\ArrowLine(60,20)(90,0)
		\ArrowLine(90,60)(60,40)
		\ArrowLine(60,40)(30,40)
		\ArrowLine(30,40)(0,60)
		\Photon(30,20)(30,40){3}{5}
		\Photon(60,20)(60,40){3}{5}
		\LongArrow(140,30)(160,30)
		\Text(16,33)[t]{$ W $}
		\Text(75,33)[t]{$ W $}
		\Text(45,55)[t]{$ u, t $}
		\Text(45,12)[t]{$ u, t $}
		\Text(147,45)[t]{\textbf{integrate out}}
		\Text(147,20)[t]{$ t, W $ \textbf{@} $ \mu_W $}
		\Text(46,80)[t]{\textbf{full theory}}
	\end{picture}
\end{minipage}
\begin{minipage}{0.4\textwidth}
	\SetWidth{1.2}
	\vspace{-0.5cm} \begin{picture}(100,60)(-55,-5)
		\Text(100,80)[t]{\textbf{EFT: 5, 4, 3-quark theory}}
		\ArrowLine(60,0)(90,30)
		\ArrowLine(90,30)(120,60)
		\ArrowLine(90,30)(120,0)
		\ArrowLine(60,60)(90,30)
		\GCirc(90,30){3}{0}
		\Text(156,30)[r]{$ \vert \Delta F \vert \, = 2 $}
		\Text(103,5)[r]{$ Q^{VLL}_1 $}
	\end{picture}
\end{minipage}
\vspace{0.3cm}
\caption{\it Top-top contribution and its EFT description. The full theory matches onto a single local structure at $ \mu_W $ where the top and the $ W $ boson are integrated out.}\label{fig:TopTopEFT}
\end{figure}
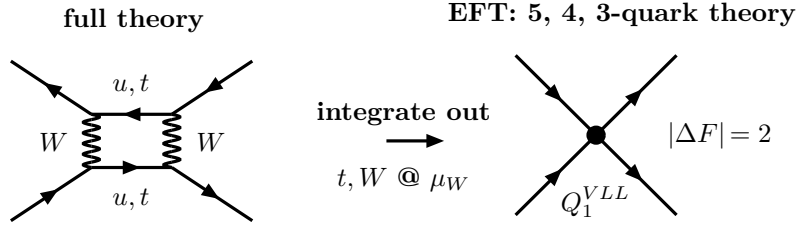


The detailed discussion of the EFT approach can be found in \cite{Buchalla:1995vs,Herrlich:1993yv,Herrlich:1996vf}. Here, we comment on its basic features. We start by discussing the simpler case, the top-top contribution represented in Figure~\ref{fig:TopTopEFT}. In this case, the top-quark and the $ W $ boson are integrated out at a single scale $ \mathcal{O} (M_W, m_t) $, and the matching of the full theory onto an EFT requires one single local operator $ Q^{VLL}_1 $. The anomalous dimension of this operator is calculated from its dressing with gluons. Then, one is able to evolve the EFT from the scale of the matching down to the scale $ \mu_h $ where non-perturbative methods are applied to take into account hadronization effects, by calculating the matrix element $ \langle Q^{VLL}_1 \rangle (\mu_h) $.



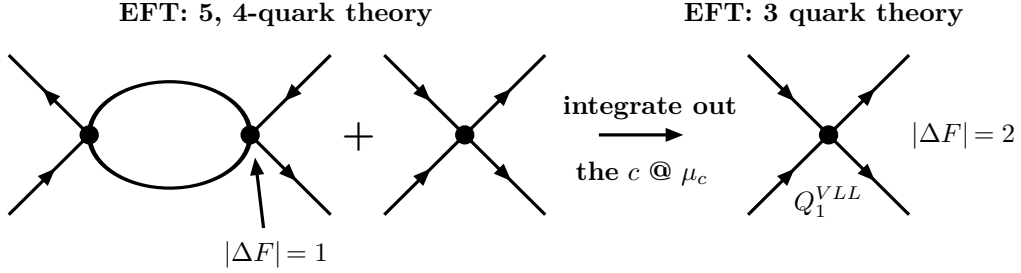
\begin{figure}
	\hspace{1cm} \SetWidth{1.2}
	\begin{picture}(100,60)(10,30)
		\ArrowLine(0,0)(30,30)
		\ArrowLine(30,30)(0,60)
		\ArrowLine(90,30)(120,0)
		\ArrowLine(120,60)(90,30)
		\Oval(60,30)(20,30)(0)
		\GCirc(30,30){3}{0}
		\GCirc(90,30){3}{0}
		\Text(100,-10)[t]{$ \vert \Delta F \vert \, = 1 $}
		\LongArrow(95,-5)(92,20)
		\Text(100,80)[t]{\textbf{EFT: 5, 4-quark theory}}
		\Text(131,35)[t]{\Large \textbf{+}}
		\ArrowLine(140,0)(170,30)
		\ArrowLine(170,30)(200,60)
		\ArrowLine(170,30)(200,0)
		\ArrowLine(140,60)(170,30)
		\GCirc(170,30){3}{0}
		\LongArrow(220,30)(250,30)
		\Text(240,45)[t]{\textbf{integrate out}}
		\Text(237,20)[t]{\textbf{the} $ c $ \textbf{@} $ \mu_c $}
		\Text(305,80)[t]{\textbf{EFT: 3 quark theory}}
	\end{picture}
	\SetWidth{1.2}
	\begin{picture}(60,60)(-160,30)
		\ArrowLine(0,0)(30,30)
		\ArrowLine(30,30)(60,60)
		\ArrowLine(30,30)(60,0)
		\ArrowLine(0,60)(30,30)
		\GCirc(30,30){3}{0}
		\Text(100,30)[r]{$ \vert \Delta F \vert \, = 2 $}
		\Text(43,5)[r]{$ Q^{VLL}_1 $}
	\end{picture}
\vspace{1.6cm}
\caption{\it Possible diagrams and operators found in the EFT description of the charm-charm and charm-top cases. Internal flavours include $ uu, uc, cu, cc $ (charm-charm case), or $ uu, uc, cu $ (charm-top case). Note that there is one first EFT between $ \mu_W $ and $ \mu_c $, described by two insertions of $ \vert \Delta F \vert \, = 1 $ operators and possibly local $ \vert \Delta F \vert \, = 2 $ ones, and another one below $ \mu_c $ described by a single local operator.}\label{fig:CharmCharmTopEFT}
\end{figure}

Moving to the charm-top and charm-charm cases, we discuss both at the same time since in both two cases the EFT built at $ \mu_W $ where the top-quark and the $ W $ boson are integrated out includes $ \vert \Delta F \vert \, = 1 $ operators, cf. Figure~\ref{fig:CharmCharmTopEFT}. Compared to the previous top-top case the computation is a bit more involved because it requires to consider a different EFT below $ \mu_c $, the scale where the charm-quark is integrated out.

There is an important simplification in the $ cc $ case compared to the $ ct $ one: thanks to the GIM mechanism, there is no divergence introduced in the computation coming from the double insertion of $ \vert \Delta F \vert \, = 1 $ operators, and therefore we do not need a $ \vert \Delta F \vert \, = 2 $ operator in the EFT between $ \mu_W $ and $ \mu_c $. On the other hand, due to the spread of masses of the top and the charm, the GIM mechanism does not operate in the same way in the $ ct $ case (a difference already seen from the $ \log (x_c) $ factor in its loop function), and local $ \vert \Delta F \vert \, = 2 $ operators are also present. Moreover, penguin operators are also present in the charm-top case \cite{Herrlich:1996vf}.

\subsection{Comparison}

A numerical comparison between the two methods, EFT and MR, is shown in Table~\ref{tab:comparisonEFTandMRinSM}. EFT expressions are taken from Ref.~\cite{Herrlich:1996vf}, and the numerical results are obtained using the same inputs found in there, namely $m_t(m_t) =167$~GeV, $m_c(m_c)=\mu_c=1.3$~GeV, $M_W=80$~GeV, $\Lambda^{(4)}=0.310$~GeV. The matchings onto the effective theories are performed at $\mu_b =4.8$ GeV, whereas the high scale $\mu_W$ is chosen differently depending on the
box considered: $\mu_W=130$ GeV when a $t$ quark is involved in order to take into account the fact that in the EFT approach the top quark and the $W$ boson are integrated out at the same time (hence $\mu_W$ is an average of the two masses), whereas  $\mu_W=M_W$ when only $c$ and $u$ quarks are involved and only the $W$ boson has to be integrated out in the diagram.

In Table~\ref{tab:comparisonEFTandMRinSM}, we do not show a numerical comparison for the charm-charm contribution between the two methods since they end up giving identical expressions once matching corrections, originally calculated in the EFT framework \cite{Herrlich:1993yv}, are considered in the MR calculation.

For the case of the top-top contribution, the small difference in the numerical values can be traced back to a different treatment of the top in the MR. Indeed, the top is not integrated out together with the $ W $ boson. Instead the method integrates out the $ W $ boson first and resums short-distance corrections between $ \mu_M \sim \mathcal{O} (M_W) $ and $ \mu_t \sim \mathcal{O} (m_t) $ with the RGE for the anomalous dimension matrix of $ \vert \Delta F \vert \, = 1 $ operators \cite{Datta:1995he}. Then, in the final step further (and more important) short-distance QCD corrections are resummed from $ \mu_t \sim \mathcal{O} (m_t) $ down to $ \mu_h $, with the RGE for the anomalous dimension matrix of $ \vert \Delta F \vert \, = 2 $ operators.

Note from Table~\ref{tab:comparisonEFTandMRinSM} that at the LO (the first numerical values given in this same table) there are differences smaller than $ 6~\% $ for both cases, charm-top and top-top. However, when moving to the NLO, a larger discrepancy of $ 30~\% $ is seen in the former case. This can be partly traced back to the presence of a large logarithm in the loop function in the charm-top case, which requires to take into account the anomalous dimension matrix describing the mixing of the local $ \vert \Delta F \vert \, = 2 $ operators with the double insertion of $ \vert \Delta F \vert \, = 1 $ operators, a feature missed in our implementation of the MR.

\begin{table}
\begin{center}
{\renewcommand{\arraystretch}{1.4}
\begin{tabular}{|c| c c |}
\hline
\textbf{SM} & $\eta_{tt}$ & $\eta_{ct}$ \\
\hline
Leading Order & $ (\alpha_s \cdot \log (x_c))^n $ & $ \log x_c \cdot (\alpha_s \cdot \log (x_c))^n $ \\
\hline
Next-to-LO & $ \alpha_s \cdot (\alpha_s \cdot \log (x_c))^n $ & $ (\alpha_s \cdot \log (x_c))^n $ \\
\hline
EFT {\scriptsize (LO, NLO)} & $ 0.612 - 0.038 = 0.574 $ & $ 0.368 + 0.099 = 0.467 $ \\
\hline
MR {\scriptsize (LO, NLO)} & $ 0.591 - 0.010 = 0.581 $ & $ 0.345 - 0.011 = 0.334 $ \\
\hline
\end{tabular}}
\end{center}
\caption{\it Numerical results: the first value corresponds to the LO, while the second one is the correction from the NLO.}\label{tab:comparisonEFTandMRinSM}
\end{table}

\section{Conclusion}

We have given the expressions of the many diagrams contributing to meson-mixing in LR Models. For a precise calculation of meson-mixing observables, we must take into account short-distance QCD corrections, thus the need for calculating the $ \bar{\eta} $ parameters. They can be calculated by two different approaches, Effective Field Theory and the less formal Method of Regions. Both consider Renormalization Groups Equations, which are unavoidable for resuming perturbative QCD corrections of the form $ \alpha_s^m \cdot \log^n $, where $ m, n $ depend on the order considered (LO, NLO, etc.). They have different philosophies though: the EFT approach builds a sequence of successive effective theories, and is the reference method for considering the computation of the $ \bar{\eta} $ factors; the MR inspects the possible diagrams and out of them determines the relevant factors $ \alpha_s^m \cdot \log^n $, from known anomalous dimension matrices and matching corrections.




\chapter{Short-distance QCD corrections in Left-Right Models}\label{ch:technicalEFT}


Over this chapter we are going to employ the two methods discussed in the previous chapter to calculate short-distance QCD corrections to meson-mixing. Over the literature, it is mostly common to find in the context of LR Models calculations based on the Method of Regions. Calculations made in the EFT approach can also be found \cite{Blanke:2011ry}, but they miss the effect of diagrams at low energies where the charm is still dynamical.

On the other hand, calculations in the SM are known up to the NLO in the top-top case and up to the NNLO in the charm-charm and charm-top cases within the EFT approach, showing a slow convergence in the charm-charm case. Higher-orders may shift considerably the numerical results and are important to control the size of the uncertainties one has from the residual dependence on the matching scales. They are also important to cancel the scheme dependence one has in the results of hadronic matrix elements calculated at low-energies by non-perturbative methods. In order to derive solid bounds on the LR Model (independent of the doublet or triplet specific realizations) structure when considering meson-mixing constraints, we therefore consider the calculation of short-distance QCD corrections at NLO. 




\section{The MR in the LR Model}


The short-distance QCD corrections effects have been addressed at LO in Ref.~\cite{Ecker:1985vv,Bigi:1983bpa} and are collected into the $ \bar{\eta} $ parameters defined in the following expressions:

\begin{eqnarray}\label{eq:boxSDQCDMR}
A^{({\rm box})} &=&\frac{G_F^2 M^2_W}{4\pi^2} 2 \beta h^2 \langle Q_2^{LR} \rangle \sum_{U,V=c,t}\lambda_U^{LR}\lambda_V^{RL} \sqrt{x_Ux_V} \\
& \times & [4 \bar{\eta}^{(W'W)}_{2, U V} I_1(x_U,x_V,\beta) - \bar{\eta}^{(W'G)}_{2, U V} I_2(x_U,x_V,\beta)] \,,
\nonumber
\end{eqnarray}

\begin{eqnarray}\label{eq:selfSDQCDMR}
A^{\rm (self)} =
-2\beta \sum^2_{i=1} \mathcal{F}^2_i \omega_i h^2\frac{G_F^2 M_W^2}{4\pi^2}  \langle Q_2^{LR}\rangle S_S(\omega_i) \sum_{U,V=c,t} \bar{\eta}^{(H)}_{2, U V} \lambda_U^{LR}\lambda_V^{RL}  \sqrt{x_U x_V} \,,
\end{eqnarray}

\begin{equation}\label{eq:vertSDQCDMR}
A^{\rm (vertex)} =
-32\beta \sum^2_{i=1} \mathcal{F}_i \omega_i h^2\frac{G_F^2 M_W^2}{4\pi^2} \langle Q_2^{LR}\rangle S_V(\omega_i)  \sum_{U,V=c,t} \bar{\eta}^{(H)}_{2, U V} \lambda_U^{LR}\lambda_V^{RL}  \sqrt{x_U x_V} \,,
\end{equation}

We also consider the short-distance corrections for the two other classes of contributions described in Section~\ref{sec:DiagramsSMLRM}: the box containing a charged Higgs

\begin{eqnarray}
A^{(H^{\pm} \, {\rm box})} &=& \frac{G_F^2 M_W^2}{4\pi^2} \frac{k^2}{1+r^2} \sum^2_{i=1} \tilde{G}^2_i 2 \omega_i \beta u \langle Q_2^{LR} \rangle \sum_{U,V=c,t}\lambda_U^{LR} \lambda_V^{RL} \sqrt{x_U x_V} \\
& \times & [\bar{\eta}^{(HG)}_{2, U V} x_U x_V I_1(x_U,x_V,\beta_H) - \bar{\eta}^{(HW)}_{2, U V} I_2(x_U,x_V,\beta_H)]\,,
\nonumber
\end{eqnarray}
and the tree level exchange of a neutral Higgs

\begin{eqnarray}
A^{(H^0)} &=& \biggl( - \frac{4 G_F \beta u}{\sqrt{2}} \frac{k^2}{1+r^2} \sum^2_{i=1} \omega_i \tilde G_i^2 \nonumber\\
&& + h^2 \frac{G_F^2 M_W^2}{2 \pi^2} \beta {\cal F}^r \sqrt{\omega_1 \omega_2}   S_S\bigl(\sqrt{\omega_1 \omega_2}\bigr) \biggr) \\
&& \sum_{U, V=c, t} \bar{\eta}^{(H)}_{2, U V} \lambda_U^{LR}\lambda_V^{RL} \sqrt{x_U x_V} \langle Q_2^{LR}\rangle \, . \nonumber
\end{eqnarray}

We have verified by an explicit inspection of the relevant range of internal momenta of the one-loop diagrams that the self-energy, the vertex, and the tree level Higgs exchange receive all the same short-distance corrections, indicated by $ \bar{\eta}^{(H)}_{2, U V} $.

Table~\ref{tab:characteristicsMRinLRM} summarizes some features of MR when applied to LR Model box diagrams: we indicate with a cross in the columns labeled as $ x_c, x_t, \beta $ the relevant energy scales, then in columns $ \sim $ and $ Range $ we show the dominant terms from the Inami-Lim functions (and therefore the counting to be performed) and the corresponding range over which $ \alpha_s (k^2) $ is considered. The final integration over the momentum of the EW diagram $ k^2 $ is performed accordingly to $ R_{\rm \log} $ or $ R_1 $. The cases $ cc $ and $ ct $ proportional to $ x_i x_j I_1(x_i,x_j,\beta) $ lead to suppressed contributions given the precision of the method, and are therefore not shown in Table~\ref{tab:characteristicsMRinLRM}. Masses coming from the coupling with Goldstone or Higgs fields are taken at $ \mathcal{O} (\mu_{W,W'}, \mu_H) $, while masses from the propagator are calculated at $ k^2 $.

\begin{table}

\begin{equation*}
\begin{array}{cccccccc}
         & ij & x_c & x_t & \beta & \sim & Range & R\\
\hline
            & cc& \times & & & \log x_c & [m_c,\mu_W] & R_{\log}\\ 
I_1(x_i,x_j,\beta) & ct & & \times & & \mathcal{O} (1) & m_t & R_1\\
          & tt & & \times & & \mathcal{O} (1) & m_t & R_1\\
\hline
        & cc&  & & \times & \log\beta & [\mu_W,\mu_{R,H}] &  R_{\log}\\ 
 I_2(x_i,x_j,\beta) & ct & & & \times & \log\beta & [\mu_W,\mu_{R,H}] & R_{\log}\\
          & tt & & & \times & \log\beta & [\mu_W,\mu_{R,H}] & R_{\log}\\
\hline 
x_i x_j I_1(x_i,x_j,\beta) & tt & & \times & & \mathcal{O} (1) & m_t & R_1       
\end{array}
\end{equation*}
\caption{\it Characteristics relevant for the MR for each individual contribution. $ x_i = m_i^2 / M_W^2, \beta = M_W^2 / M_{W'}^2 $. Contributions proportional to $ I_1(x_i,x_j,\beta) $ or $ x_i x_j I_1(x_i,x_j,\beta) $ come respectively from $ W W' $ and $ G H $ box diagrams, while a contribution proportional to $ I_2(x_i,x_j,\beta) $ comes from a $ W' G $ or a $ W H $ box. $ G G' $ come at a higher order in $ \beta $ and are thus not considered.}\label{tab:characteristicsMRinLRM}
\end{table}




Note that there is a large logarithm, $ \log x_c $, in the charm-charm case. This large logarithm comes from the $ W W' $ box in the 't Hooft-Feynman gauge and we now discuss the corresponding short-distance correction in detail (neglecting thresholds, which correspond to a small correction). Compared to the SM case, we have in the LR Model more operators, $ Q^{LR}_{1,2} $, which mix in the running. Therefore, we must calculate the two factors $\bar\eta^{(W'W)}_{a,UV}$, $ a=1,2 $, $U$ and $V$ denoting the quarks in the loop with $m_U \leq m_V$. In order to express the short-distance QCD correction, we start by defining
\begin{eqnarray}\nonumber
&&\xi^{(W'W)}_{a,UV} [R]=\sum_{r,l=\pm \, ,i=1,2}
 \left(\frac{\alpha_s(m_{V})}{\alpha_s(\mu_{h})}\right)^{-d_l-d_r+d_i+ d_m} 
  \left(\frac{\alpha_s(m_U)}{\alpha_s(\mu_{h})}\right)^{-d_m} \nonumber\\
&&\times \left(\frac{\alpha_s(\mu_W)}{\alpha_s(\mu_{h})}\right)^{d_l}
    \left(\frac{\alpha_s(\mu_R)}{\alpha_s(\mu_{h})}\right)^{d_r}
\left[\left(1+\frac{\alpha_s(\mu_{h})}{4\pi} \hat{K}\right) \hat{W}\right]_{ai}\nonumber\\
&& \quad\times 
 R^{NLO}\Bigg(-d_l-d_r+d_i+2d_m,\label{eq:etaLRW1}\\
&&  \qquad
  \Bigg[\hat{W}^{-1}\Bigg(1-\frac{\alpha_s(\mu_W)}{4\pi}[J_l-B_l] -\frac{\alpha_s(\mu_R)}{4\pi}[J_r-B_r] \nonumber\\
&&  \qquad\;\; +\frac{\alpha_s(m_U)+\alpha_s(m_V)}{4\pi}J_m\Bigg)
 \left(\begin{array}{c} \tau_{1}^{rl}\\\tau_{2}^{rl}\end{array}\right)
\Bigg]_i, \nonumber\\
&&   \qquad\quad
   \left[\hat{W}^{-1}\left(-\hat{K}+J_l+J_r-2J_m\right) \left(\begin{array}{c} \tau_{1}^{rl}\\\tau_{2}^{rl}\end{array}\right)
\right]_i , m_{V},\mu_W \Bigg)\,,\nonumber
\end{eqnarray} 
with $d_{l,r}$ determined from the anomalous dimension matrices of the $|\Delta F|=1$ current-current operators, $d_i$ from the corresponding $|\Delta F|=2$ local operator, $d_m$ from the evolution of the masses, $J_{l,r,i,m}, \hat{K}, J_{m}$ the respective terms from the anomalous dimension matrix at NLO, $\hat{W}$ being a diagonalisation matrix needed for solving the RGE, and finally the values of the Wilson coefficients coming from the matching between the bilocal operators ${\rm T} \{ Q^L_r, Q^R_l \}$ and the local $|\Delta F|=2$ operators are proportional to
\begin{eqnarray}
\tau_{1}^{rl}=\tau_{rl}/4 \, ,\qquad \qquad \tau_{2}^{rl}=1/4 \, , \qquad\qquad \tau_{rl}=-(r+l+Nrl)/2\,.\label{eq:taurl}
\end{eqnarray}
In Eq.~\eqref{eq:etaLRW1} the index $ i $ indicates the individual contributions from the operators $ Q^{LR}_{i} $ at $ m_V $, while at the ending of the running they are indexed by $ a $: as we have already commented on, the two operators mix through running, and their mixing is described by the $ 2 \times 2 $ matrices $\hat{W}$ and $ \hat{K} $. Their expressions, together with the others needed in Eq.~\eqref{eq:etaLRW1}, are found in Appendix~\ref{app:anomdimgen}.

Similarly to the SM case for $\eta_{ct}$ for which the counting in Eq.~\eqref{eq:etactSMinMR} was considered, we finally have the expression

\begin{eqnarray}
\bar \eta^{(W'W)}_{a,cc}
&=&\frac{1}{1+\log x_c} \biggl(\xi^{(W'W)}_{a,cc}\log(x_c)  +\sum_{r,l=\pm,i=1,2}
 \left(\frac{\alpha_s(m_c)}{\alpha_s(\mu_{h})}\right)^{-d_l-d_r+d_i}
\biggr.
\\\nonumber
&& \biggl.\times
    \left(\frac{\alpha_s(\mu_W)}{\alpha_s(\mu_{h})}\right)^{d_l}
    \left(\frac{\alpha_s(\mu_R)}{\alpha_s(\mu_{h})}\right)^{d_r}
\hat W_{ai}\left[\hat{W}^{-1}\left(\begin{array}{c} \tau_{1}^{rl}\\\tau_{2}^{rl}\end{array}\right)
\right]_i \biggr)\,.
\end{eqnarray}

For $\bar\eta_{a,ct}^{(W'W)}$ and $\bar\eta_{a,tt}^{(W'W)}$, there are no large logarithms in the contribution from $I_1$ in Eq.~\eqref{eq:InamiLimObetaLRM}, the integral is dominated by $k^2=\mathcal{O}(m_V^2)$  and  we have
\begin{equation}
\bar\eta_{a,ct}^{(W'W)}=\xi_{a,ct}^{(W'W)}[R^{NLO}\to R^{NLO}_1] , \qquad
\bar\eta_{a,tt}^{(W'W)}=\xi_{a,tt}^{(W'W)}[R^{NLO}\to R^{NLO}_1] , \qquad
\end{equation}
where
$R^{NLO}$ should be replaced by $R^{NLO}_1$ defined in Eq.~\eqref{eq:rnlo1} to express the fact that a single scale dominates the loop momentum.


The next step is to compute $ G W' $ contributions. Note from Table~\ref{tab:characteristicsMRinLRM} that they come with the logarithm $ \log \beta $. Its size for phenomenological interesting values of $ \beta $ is not largely dominant as for $ \log x_c $. We have therefore considered resumming (in the way just described above) or not this logarithm. As argued in Section~\ref{sec:running}, we do not expect large corrections coming from the resumming of $ \log \beta $, and we have indeed found a rather small modulation of the MR results for the combined short-distance QCD corrections $ \bar{\eta}^{(LR)}, \bar{\eta}^{(H^{\pm} \, {\rm box})} $ in Eqs.~\eqref{eq:Fullsetwneq0}, \eqref{eq:HiggsBoxFactEta} when resumming or not it. The analytical expressions are given in Appendix~\ref{lrregion}, while the numerical values are given in Tables~\ref{tab:noLargeLogBeta} and \ref{tab:LargeLogBeta}. Note that we only give the value of $ \bar{\eta}_{2,UV} $, the reason being that the numerical values of $ \bar{\eta}_{1,UV} $ are very much suppressed and therefore irrelevant for phenomenology. We further note that, for simplicity reasons, the numerical values for $ \bar{\eta}^{(HW)}_{2,UV} $ in Table~\ref{tab:LargeLogBeta} were calculated for the limiting case where $ w = 0 $ and therefore in the loop functions found in those tables $ F(\omega_1, \omega_2) $ should be replaced by a function depending only on $ \omega_1 $, which we have set to $ 0.1 $ and $ 0.8 $. In any case, the variation of $ \bar{\eta}^{(HW)}_{2,UV} $ with $ \omega_{1,2} $ is small compared to the uncertainties we will attribute to the final values and will be therefore neglected.

A similar discussion applies for $ W H, G H $ and the tree level diagram. The final numerical results will be discussed later in Section~\ref{sec:FinalResult}, together with the results for the EFT calculation in the charm-charm case.


\begin{table}
	\centering
	\begin{tabular}{|c|c|c|}
		\hline
		& $ \bar{\eta}_{2,tt} $ & loop functions \\
		\hline
		$ W W' $ & 4.65 + 0.99 = 5.64 & $ \frac{4 \log (x_t)}{(x_t - 1)^2} - \frac{4}{x_t - 1} $ \\
		\hline
		$ G W' $ & 4.66 + 0.98 = 5.64 & $ \frac{x_t^2 - 2 x_t}{(x_t - 1)^2} \log (x_t) + \frac{x_t}{x_t - 1} + \log (\beta) $ \\
		\hline
		$ ({\rm tree}), ({\rm vert}), ({\rm self}) $ & 4.66 + 1.00 = 5.66 & $ F (\omega_1, \omega_2) $ \\
		\hline
		$ W H $ & 4.66 + 0.98 = 5.64 & $ u \cdot \omega_1 \left( x_t \frac{x_t + (x_t - 2) \log (x_t) - 1}{(x_t - 1)^2} + \log (\beta \omega_1) \right) $ \\
		\hline
		$ G H $ & 4.66 + 1.00 = 5.66 & $ u \cdot \omega_1 \cdot x_t^2 \frac{- x_t + \log (x_t) + 1}{(x_t - 1)^2} $ \\
		\hline
		\hline
		& $ \bar{\eta}_{2,ct} $ & \\
		\hline
		$ W W' $ & 2.42 + 0.27 = 2.69 & $ - \frac{4 \log (x_t)}{x_t - 1} $ \\
		\hline
		$ G W' $ & 2.42 + 0.27 = 2.69 & $ \frac{x_t}{x_t - 1} \log (x_t) + \log (\beta) $ \\
		\hline
		$ ({\rm tree}), ({\rm vert}), ({\rm self}) $ & 2.42 + 0.28 = 2.70 & $ F (\omega_1, \omega_2) $ \\
		\hline
		$ W H $ & 2.42 + 0.27 = 2.69 & $ u \cdot \omega_1 \left( \frac{x_c + x_t}{x_t - 1} \log (x_t) + \log (\beta \omega_1) \right) $ \\
		\hline
		$ G H $ & - & higher order \\
		\hline
		\hline
		& $ \bar{\eta}_{2,cc} $ & \\
		\hline
		$ W W' $ & 1.46 + 0.16 - 0.28 = 1.34 & $ 4 \log (x_c) + 4 $ \\
		\hline
		$ G W' $ & 1.26 + 0.01 = 1.27 & $ \log (\beta) $ \\
		\hline
		$ ({\rm tree}), ({\rm vert}), ({\rm self}) $ & 1.26 + 0.02 = 1.28 & $ F (\omega_1, \omega_2) $ \\
		\hline
		$ W H $ & 1.26 + 0.02 = 1.28 & $ u \cdot \omega_1 \log (\beta \omega_1) $ \\
		\hline
		$ G H $ & - & higher order \\
		\hline
	\end{tabular}
	\caption{\it MR values for the LR Model when $ \log \beta $ is not resummed are indicated in the second column: the first value indicates the LO, and the second one the NLO correction. For the $ W W' $ contribution to the charm-charm case, the second correction comes from the NLO related to the $ \log x_c $, while the third is the LO correction related to the remaining $ \mathcal{O} (1) $ term. The different loop functions, indicated in the last column, are calculated in the 't Hooft-Feynman gauge. When not indicated, the dependence on $ \omega_1 $ is negligible.}\label{tab:noLargeLogBeta} 
\end{table}


\begin{table}
	\centering
	\begin{tabular}{|c|c|c|}
		\hline
		& $ \bar{\eta}_{2,tt} $ & loop functions \\
		\hline
		$ W W' $ & 4.68 + 0.96 = 5.64 & $ \frac{4 \log (x_t)}{(x_t - 1)^2} - \frac{4}{x_t - 1} $ \\
		\hline
		$ G W' $ & 4.86 + 7.32 - 5.26 = 6.92 & $ \frac{x_t^2 - 2 x_t}{(x_t - 1)^2} \log (x_t) + \frac{x_t}{x_t - 1} + \log (\beta) $ \\
		\hline
		$ ({\rm tree}), ({\rm vert}), ({\rm self}) $ & 4.66 + 0.98 = 5.64 & $ F (\omega_1, \omega_2) $ \\
		\hline
		$ W H, \omega_1 = 0.1 $ & 4.86 + 4.11 - 2.65 = 6.33 & $ u \cdot \omega_1 \left( x_t \frac{x_t + (x_t - 2) \log (x_t) - 1}{(x_t - 1)^2} + \log (\beta \omega_1) \right) $ \\
		\hline
		$ W H, \omega_1 = 0.8 $ & 4.84 + 6.70 - 4.76 = 6.79 & - \\
		\hline
		$ G H $ & 4.66 + 0.99 = 5.65 & $ u \cdot \omega_1 \cdot x_t^2 \frac{- x_t + \log (x_t) + 1}{(x_t - 1)^2} $ \\
		\hline
		\hline
		& $ \bar{\eta}_{2,ct} $ & \\
		\hline
		$ W W' $ & 2.43 + 0.26 = 2.69 & $ - \frac{4 \log (x_t)}{x_t - 1} $ \\
		\hline
		$ G W' $ & 2.52 + 1.91 - 1.51 = 2.92 & $ \frac{x_t}{x_t - 1} \log (x_t) + \log (\beta) $ \\
		\hline
		$ ({\rm tree}), ({\rm vert}), ({\rm self}) $ & 2.42 + 0.27 = 2.69 & $ F (\omega_1, \omega_2) $ \\
		\hline
		$ W H, \omega_1 = 0.1 $ & 2.53 + 1.17 - 0.86 = 2.83 & $ u \cdot \omega_1 \left( \frac{x_c + x_t}{x_t - 1} \log (x_t) + \log (\beta \omega_1) \right) $ \\
		\hline
		$ W H, \omega_1 = 0.8 $ & 2.52 + 1.77 - 1.40 = 2.89 & - \\
		\hline
		$ G H $ & - & higher order \\
		\hline
		\hline
		& $ \bar{\eta}_{2,cc} $ & \\
		\hline
		$ W W' $ & 1.55 + 0.16 - 0.31 = 1.40 & $ 4 \log (x_c) + 4 $ \\
		\hline
		$ G W' $ & 1.31 - 0.02 = 1.29 & $ \log (\beta) $ \\
		\hline
		$ ({\rm tree}), ({\rm vert}), ({\rm self}) $ & 1.26 + 0.02 = 1.28 & $ F (\omega_1, \omega_2) $ \\
		\hline
		$ W H, \omega_1 = 0.1 $ & 1.31 - 0.02 = 1.29 & $ u \cdot \omega_1 \log (\beta \omega_1) $ \\
		\hline
		$ W H, \omega_1 = 0.8 $ & 1.31 - 0.03 = 1.28 & - \\
		\hline
		$ G H $ & - & higher order \\
		\hline
	\end{tabular}
	\caption{\it Same as Table~\ref{tab:noLargeLogBeta}, with the difference of resumming $ \log \beta $. The first numerical values are the LO contributions while the second and possibly third are the NLO corrections.}\label{tab:LargeLogBeta}
\end{table}

\section{EFT calculation of the $ cc $ contribution in LR Models}\label{sec:MYSECTION1year}




In the SM case we have seen that a good comparison between the EFT approach and the MR one happens when there is no large logarithm in the loop function, and when that was not the case and a large $ \log x_c $ was present we have found a difference of $ 30~\% $ in the $ ct $ case. This is related to a more involved evolution of the relevant operators, because the double insertion of effective $ \vert \Delta F \vert \, = 1 $ operators has divergences that are renormalized by local counterterms, a feature missed in our implementation of the MR: in this case, we have not employed any information concerning the anomalous dimension matrix $ \gamma_{r \ell, a} $, $ r, \ell = \pm $ and $ a = 1,2 $, governing the mixing between double and single insertions of $ \vert \Delta F \vert \, = 1 $ and $ \vert \Delta F \vert \, = 2 $ operators. Note that the presence of $ \log x_c $ in the full theory announces the need to consider local operators in the renormalization of the effective theory: indeed, since $ \log q^2 + {\rm cnt} = \int \frac{d q^2}{q^2} $, once the propagator of the $ W $ is replaced by $ \frac{i}{M_W^2} $ one expects a worse control of divergences in the UV range.

A logarithm is also seen in the $ W W' $ box calculated in the 't Hooft-Feynman gauge in the charm-charm contribution seen in Eq.~\eqref{eq:boxWWprimecc}. We would like in this case to compute the short-distance correction in the EFT approach and compare it with the MR result. To this effect, we have already discussed the new energy scales $ M_{W'}, M_H $, which are expected to be found beyond $ M_W, m_t $, and we have thus argued that we can integrate out the $ W, W', H, t $ at the same scale. Then, in order to evolve the Wilson coefficients determined at $ \mu_W $ down to $ \mu_h $, we need the anomalous dimension matrices describing how the operators $ \vert \Delta F \vert \, = 1, 2 $ evolve without mixing, already known from the literature, and how the operators of the class $ \vert \Delta F \vert \, = 2 $ mix with double insertions of $ \vert \Delta F \vert \, = 1 $ operators, which we will determine. Table~\ref{tab:steps} summarizes the steps that will lead to the calculation of $ \bar{\eta}^{LR}_{cc} $.

\begin{table}
	\centering
	\begin{tabular}{lcc}
		\multicolumn{3}{c}{\textbf{Matching} at $ \mu_W = \mathcal{O} (M_W, M_{W'}, M_H) $} \\
		\hline
		Wilson coefficients at $ \mu_W $: & $ | \Delta F | = 1 $ & Ref.~\cite{Buras:1991jm,Buras:1992tc} \\
		                                  & $ | \Delta F | = 2 $ & calculated here \\
		\hline
		\\
		\multicolumn{3}{c}{\textbf{Running} from $ \mu_W $ to $ \mu_c $} \\
		\hline
		Anomalous dimension matrix & $ | \Delta F | = 1, 2 $ \textit{individually} & Ref.~\cite{Buras:2001ra} \\
		or tensor:                 & $ | \Delta F | = 1, 2 $ \textbf{mixing} & calculated here \\
		\hline
		\\
		\multicolumn{3}{c}{\textbf{Matching} at $ \mu_c = \mathcal{O} (m_c) $} \\
		\hline
		Wilson coefficients at $ \mu_c $: & $ | \Delta F | = 2 $ & calculated here \\
		\hline
		\\
		\multicolumn{3}{c}{\textbf{Running} from $ \mu_c $ to $ \mu_{h} $} \\
		\hline
		Anomalous dimension matrix: & $ | \Delta F | = 2 $ & Ref.~\cite{Buras:2001ra,Buras:2012fs} \\
		\hline
	\end{tabular}
	\caption{Full NLO calculation of $ \bar{\eta}^{LR}_{cc} $. Note that the mixing of $ | \Delta F | \; = 1, 2 $ operators is calculated here, while their individual running ($ | \Delta F | \; = 1 $ or $ | \Delta F | \; = 2 $ separately) is found in \cite{Buras:2001ra}.}\label{tab:steps}
\end{table}

\subsection{Basic elements}

The set of operators in the charm-charm case includes local $ \vert \Delta F \vert \, = 1 $ and $ \vert \Delta F \vert \, = 2 $ operators. The effective Lagrangian is

\begin{eqnarray}\label{eq:eqMyLagrangian}
\mathcal{L}^{(5)}_{\rm eff} (cc) & = & - \frac{4 G_F}{\sqrt{2}} \lambda^{LL}_c \sum_{i=\pm} C^L_i \sum_{j=\pm} Z^{-1}_{ij} Q^{L, {\rm bare}}_{j} \nonumber\\
&& - \frac{4 G_F}{\sqrt{2}} h^2 \beta \lambda^{RR}_c \sum_{i=\pm} C^R_i \sum_{j=\pm} Z^{-1}_{ij} Q^{R, {\rm bare}}_{j} \nonumber\\
&& - 2 G^2_F h^2 \beta \lambda^{LR}_c \lambda^{RL}_c \sum^2_{b=1} \left[ \sum_{k, l = \pm} C^L_k C^R_l \tilde{Z}^{-1}_{kl, b} + \sum^2_{a=1} C^r_a \tilde{Z}^{-1}_{ab} \right] \tilde{Q}^{LR, {\rm bare}}_{b} \nonumber\\
&& \quad + {\rm unphysical \;\; operators} ,
\end{eqnarray}
\noindent
where $ Z, \tilde{Z} $ are the renormalization matrices which absorb the divergences of the $ \vert \Delta F \vert \, = 1, 2 $ amplitudes of the bare Lagrangian \cite{Buras:1998raa} \cite{Herrlich:1996vf}

\begin{eqnarray}\label{eq:definitionInverseZ}
Z^{-1} & = & 1 + \frac{\alpha_s}{4 \pi} Z^{-1,(1)} + \ldots \, , \\
Z^{-1,(n)} & = & \sum^{n}_{r = 0} \frac{1}{\epsilon^r} Z^{-1,(n)}_r \, ,
\end{eqnarray}
and similarly for $ \tilde{Z} $.

Above, $ \vert \Delta F \vert \, = 1 $ operators seen in the first two lines involve the charm flavour only, and we have the same coefficient for both at the matching scale $ \mu_W $, i.e. $ C^L_i = C^R_i \equiv C_i $. Running effects being generated by strong interactions, $ C^{L}_{i} $ and $ C^{R}_{i} $ evolve in the same way below $ \mu_W $.

In the third line of Eq.~\eqref{eq:eqMyLagrangian}, the first term renormalizes the contraction of two $ \vert \Delta F \vert \, = 1 $ operators, while the second one is necessary in the matching to the full theory. For the renormalization, two local operators have been introduced:

\begin{eqnarray}
\tilde{Q}^{LR}_{1} & = & \frac{m^2_c}{g^2 \mu^{2 \varepsilon}} \gamma L \otimes \gamma R , \qquad {\rm and} \\
\tilde{Q}^{LR}_{2} & = & \frac{m^2_c}{g^2 \mu^{2 \varepsilon}} \; \; L \otimes R \, ,
\end{eqnarray}
whose normalization is chosen so that the mixing with two $ \vert \Delta F \vert \, = 1 $ operators is treated on the same footing as QCD radiative corrections and a common RGE framework can be applied to discuss the mixing of all the operators. The notation $ \gamma L \otimes \gamma R $, and $ L \otimes R $ avoids precising the quark flavours, which are in a singlet structure under color, and the obvious contraction of Lorentz indices.

The renormalization group equations describing the evolution of the Wilson coefficients are (the details of their derivation are given in Appendix~\ref{sec:toughRGE}):


\begin{equation}\label{eq:RGEDF1eq}
\sum_{j} \left[ \delta_{jk} \, \mu \frac{d}{d \mu} - \gamma_{jk} \right] C_j = 0 , \quad \gamma_{jk} \equiv \sum_{i} Z^{-1}_{ji} \mu \frac{d}{d \mu} Z_{ik} .
\end{equation}

\begin{equation}\label{eq:RGEDF2eq}
\sum^2_{a=1} \left[ \delta_{ac} \, \mu \frac{d}{d \mu} - \tilde{\gamma}_{ac} \right] C^r_{a} = \sum_{k,l=\pm} C_k C_l \gamma_{kl,c} , \quad \tilde{\gamma}_{ac} = \sum^2_{b=1} \tilde{Z}^{-1}_{ab} \mu \frac{d}{d \mu} \tilde{Z}_{bc} \, ,
\end{equation}
\noindent
and

\begin{equation}
\gamma_{kl,c} \equiv - \sum^2_{b=1} \left[ \sum_{k',l'=\pm} \left( \gamma_{kk'} \delta_{ll'} + \gamma_{ll'} \delta_{kk'} \right) \tilde{Z}^{-1}_{k'l',b} + \mu \frac{d}{d \mu} \tilde{Z}^{-1}_{kl,b} \right] \tilde{Z}_{bc} \, ,
\end{equation}
describing the mixing of $ \vert \Delta F \vert \, = 1, 2 $ operators and calculated below. We now move to the fourth line of Eq.~\eqref{eq:eqMyLagrangian}, containing evanescent operators.

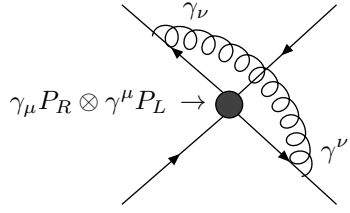
\begin{figure}
\hspace{3cm}
	\begin{picture}(300,85)(-50,-20)
	\ArrowLine(60,25)(20,60)
	\ArrowLine(20,-15)(60,20)
	\GlueArc(35,-3.7)(52,-0.9,94){4}{11}
	\ArrowLine(100,60)(62,25)
	\ArrowLine(62,20)(100,-15)
	\GCirc(60,22.5){5}{0.25}
	\Text(15,23)[]{$\gamma_\mu P_R \otimes \gamma^\mu P_L \, \rightarrow$}
	\Text(48,58)[]{$\gamma_\nu$}
	\Text(100,5)[]{$\gamma^\nu$}
	\end{picture}
	\caption{\it The simplification of the Dirac algebra leads to Eq.~\eqref{eq:illustrateFigure}, among other Lorentz structures.}\label{fig:illustrateEquation}
\end{figure}

\subsection{Evanescent operators}\label{sec:EObasics}

Apart the physical operators we have commented on above, we also need to include a set of non-physical operators when considering dimensional regularization, a common feature of higher order calculations. These are operators that vanish in $ D = 4 $ dimensions, but are present in $ D \neq 4 $ in order to close the Dirac algebra. As an example of such an operator, we consider the Dirac structure

\begin{equation}\label{eq:illustrateFigure}
\gamma_\nu \gamma_\mu P_R \otimes \gamma^\nu \gamma^\mu P_L \, ,
\end{equation}
that happens once dressing the operators $ Q^{LR}_{1} $ (or $ \tilde{Q}^{LR}_{1} $) with gluons, see Figure~\ref{fig:illustrateEquation}. In four dimensions, such a Dirac structure would simplify to $ 4 P_R \otimes P_L $, but when $ D = 4 - 2 \epsilon $ we define the following \textit{evanescent operator} (EO)

\begin{equation}\label{eq:simpleEvanescent}
E [Q^{LR}_{1}] = \bar{d} \gamma_\nu \gamma_\mu P_R s \cdot \bar{d} \gamma^\nu \gamma^\mu P_L s - (4 + a \epsilon) \bar{d} P_R s \cdot \bar{d} P_L s \, ,
\end{equation}
where the constant $ a $ is arbitrary. When defining $ E [\tilde{Q}^{LR}_{1}] $, the factor proportional to $ \epsilon $ may be chosen independently, in which case it is denoted with a tilde, i.e. $ \tilde{a} $. The choice we make for the value of the constants $ a $ and $ \tilde{a} $ makes part of the renormalization scheme and as so their values must be mentioned when providing the final results. For the example given above, we will choose $ a = \tilde{a} = 4 $, which is the most common choice (related to Fierz identities, \cite{Buras:1998raa}). This particular choice, and the ones for the other relevant evanescent operators, intend to match the same choice made when calculating Lattice matrix elements by non-perturbative methods. We do not give here the full set of EO we use: they can be found in Appendix~\ref{sec:setOfEOs}.

As shown in \cite{Buras:1989xd} \cite{Collins:1984xc}, Wilson coefficients of evanescent operators are not relevant in the matching because their matrix elements vanish in four dimensions under an appropriate choice of their finite renormalization. Moreover, \cite{Dugan:1990df} have shown that, under certain appropriate choices of EO and finite renormalization, EO do not mix into physical ones, and \cite{Herrlich:1994kh} have generalized this statement showing that any choice can be made. However, these operators should not be ignored, since the bare definitions of the EO are relevant for the evolution of the physical operators into themselves in the calculation of the two-loop anomalous dimension matrix.

\subsection{EFT between $ \mu_W $ and $ \mu_c $}

\subsubsection{Set of operators}


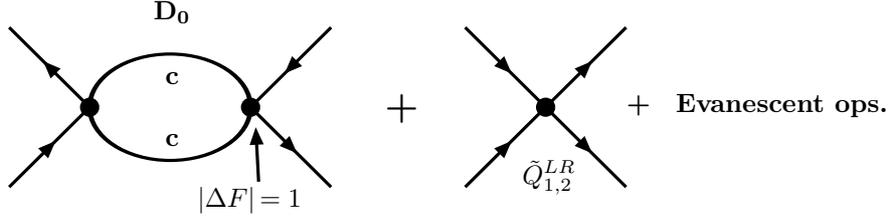
\begin{figure}
	\hspace{1cm} \SetWidth{1.2}
	\begin{picture}(100,60)(-10,0)
		\ArrowLine(0,0)(30,30)
		\ArrowLine(30,30)(0,60)
		\ArrowLine(90,30)(120,0)
		\ArrowLine(120,60)(90,30)
		\Oval(60,30)(20,30)(0)
		\GCirc(30,30){3}{0}
		\GCirc(90,30){3}{0}
		\Text(61,70)[t]{$ \mathbf{D_0} $}
		\Text(61,43)[t]{$ \mathbf{c} $}
		\Text(61,20)[t]{$ \mathbf{c} $}
		\Text(90,0)[t]{$ \vert \Delta F \vert \, = 1 $}
		\LongArrow(93,2)(92,21)
		\Text(147,35)[t]{\Large \textbf{+}}
		\ArrowLine(170,0)(200,30)
		\ArrowLine(200,30)(230,60)
		\ArrowLine(200,30)(230,0)
		\ArrowLine(170,60)(200,30)
		\GCirc(200,30){3}{0}
		\Text(202,10)[t]{$ \tilde{Q}^{LR}_{1,2} $}
		\Text(280,35)[t]{\textbf{+} $ \; $ \textbf{Evanescent ops.}}
	\end{picture}
	\vspace{0.3cm}
	\caption{\it Illustration of the set of operators required in the matching with the full theory. Local $ \vert \Delta F \vert \, = 1 $ operators give origin to the diagram $ D_0 $ represented in the left, which needs to be renormalized by physical and unphysical operators.}\label{fig:illlustrationLocals}
\end{figure}

We illustrate the need for local operators by calculating the diagram in the left of Figure~\ref{fig:illlustrationLocals}, referred to as $ D_0 $. In the LR Model $ D_0 $ diverges, thus introducing local counterterms of order $ \alpha^{0}_{s} $ in the strong coupling. On the other hand, in the case of the SM for $ cc $, there is no need for local counterterms, thanks to the GIM mechanism: the $ D_0 $ set of diagrams is finite after the combination of internal flavours $ cc - uc - cu + uu $ is taken.

The $ \vert \Delta F \vert \, = 1 $ operators are 

\begin{equation}
Q^X_{\varepsilon} = \bar{d} \gamma^{\lambda} P_{X} q' \cdot \bar{q} \gamma_{\lambda} P_{X} s \, \frac{(\mathbf{\hat{1}} + \varepsilon \mathbf{\tilde{\hat{1}}})}{2} \, , \quad X = L, R \, ,
\end{equation}
cf. Eq.~\eqref{eq:plusMinusOps}. The calculation of the $ D_0 $ diagram gives a kinematic structure proportional to

\begin{equation}\label{eq:eq30}
\frac{m^{2}_{c}}{(p^2 - m^{2}_{c})^2} \left( 2 D \, R \otimes L - \gamma^{\mu_1} \gamma^{\mu_2} R \otimes \gamma_{\mu_1} \gamma_{\mu_2} L \right) \, , \quad D = 4 - 2 \epsilon \, ,
\end{equation}
\noindent
and a color factor given by $ \frac{1}{4} \left( \mathbf{\hat{1}} - 2 \tau_{m n} \mathbf{\tilde{\hat{1}}} \right) $, with

\begin{equation}
\tau_{m n} = - (m + n + N m n) / 2 \, , \quad m, n = \pm \, .
\end{equation}
After integrating over the internal momentum $ p^{\mu} $ and a little bit of algebra the final result for the diagram $ D_0 $ reads

\begin{eqnarray}\label{eq:eq22}
D_0 & = & i \frac{m_c^2}{16 \pi^2} \frac{1}{\epsilon} \Bigg( Q^{LR}_{2} + Q^{LR}_{1} \tau_{m n} \nonumber\\
&& - 4 E^{LR}_{1} \tau_{m n} - \frac{1}{4} E^{LR}_{5} + \frac{1}{2} E^{LR}_{6} \Bigg) \nonumber\\
&& - i \frac{m_c^2}{16 \pi^2} \left[ 2 + \log \left( \frac{m^{2}_{c}}{\mu^{2}} \right) \right] \left( Q^{LR}_{2} + Q^{LR}_{1} \tau_{m n} \right) ,
\end{eqnarray}
where the set of evanescent operators $ E^{LR}_{1,5,6} $ is

\begin{eqnarray}
E^{LR}_{1} & = & \bar{d}^{\alpha} P_{R} s^{\beta} \cdot \bar{d}^{\beta} P_{L} s^{\alpha} + Q^{LR}_{1} / 2 , \\
E^{LR}_{5} & = & \bar{d}^{\alpha} \gamma^{\mu_1} \gamma^{\mu_2} P_{R} s^{\alpha} \cdot \bar{d}^{\beta} \gamma_{\mu_1} \gamma_{\mu_2} P_{L} s^{\beta} - (4 + a^{LR}_{2 \gamma} \epsilon) Q^{LR}_{2} , \\
E^{LR}_{6} & = & \bar{d}^{\alpha} \gamma^{\mu_1} \gamma^{\mu_2} P_{R} s^{\beta} \cdot \bar{d}^{\beta} \gamma_{\mu_1} \gamma_{\mu_2} P_{L} s^{\alpha} + (4 + a^{LR}_{2 \gamma} \epsilon) Q^{LR}_{1} / 2 ,
\end{eqnarray}
where we take $ a^{LR}_{2 \gamma} = 4 $. We do not keep evanescent operators coming as $ \epsilon^{0} \times E^{LR}_{1,5,6} $ in Eq.~\eqref{eq:eq22} since they play no role in the physical effective Lagrangian defined at $ D = 4 $ dimensions. However, evanescent operators that come as $ \epsilon^{-1} \times E^{LR}_{1,5,6} $ are necessary in the determination of the NLO anomalous dimension matrices.


Following Eq.~\eqref{eq:eq22}, we need to introduce in the effective Lagrangian counterterms proportional to the operators

\begin{eqnarray}\label{eq:eq19}
&& Q^{LR}_{2} + Q^{LR}_{1} \tau_{m n} - 4 \lambda E^{LR}_{1} \tau_{m n} - \frac{1}{4} \lambda E^{LR}_{5} + \frac{1}{2} \lambda E^{LR}_{6} \tau_{m n} \, ,
\end{eqnarray}
\noindent
where $ \lambda $ is included in order to keep track of the evanescent operators, which will be needed when discussing the calculation of the anomalous dimensions. On the other hand, the finite terms in Eq.~\eqref{eq:eq22}, necessary in the matching with the full theory, gives

\begin{eqnarray}\label{eq:eq18}
\langle \mathcal{O}_{m n} \rangle^{(0)} (\mu) & \stackrel{}{=} & \left[ 2 + \log \left( \frac{m^{2}_{c}}{\mu^{2}} \right) \right] \sum^{2}_{a=1} \tau_{a}^{m n} \frac{m^{2}_{c} (\mu)}{4 \pi^{2}} \langle Q^{LR}_{a} \rangle^{(0)} (\mu) \nonumber\\
& = & \sum^{2}_{a=1} r_{m n, a} (\mu) \frac{m^{2}_{c} (\mu)}{4 \pi^{2}} \langle Q^{LR}_{a} \rangle^{(0)} (\mu) ,
\end{eqnarray}
\noindent
where
\begin{equation}\label{eq:tauFactors}
\tau_{1}^{m n} = \frac{\tau_{m n}}{4} \, , \qquad\qquad \tau_{2}^{m n} = \frac{1}{4} \, ,
\end{equation}
and we have defined

\begin{eqnarray}
r_{ij,a} (\mu_c) = \left[ 2 + \log \left( \frac{m_{c}^2}{\mu_c^2} \right) \right] \tau_{a}^{i j} , \quad a = 1, 2 .
\end{eqnarray}

\subsubsection{Wilson coefficients at $ \mu_W $}

To leading order in $ \beta $ and in $ x_c $, the complete calculation performed in the full theory gives (quark masses are understood to be evaluated at $ \mu_W $)

\begin{eqnarray}
H^{(WW')} &=& \frac{G^{2}_{F} M^{2}_{W}}{4 \pi^2} \lambda^{LR}_{c} \lambda^{RL}_{c} 2 \beta h^{2} x_{c} Q^{LR}_{2} \\
&& \left[ 4 \log (m^2_c / \mu^2_W) + \left( 4 \log (\mu^2_W / M^2_W) + 4 + f_{(W',H)} \right) \right] + h.c. \, , \nonumber
\end{eqnarray}
\noindent
from Eqs.~\eqref{eq:finalLoopFunct}, where

\begin{equation}\label{eq:functionWpH}
f_{(W',H)} = \log (\beta) + F(\omega_1, \omega_2) ,
\end{equation}
\noi
that is going to show up repeatedly in the calculation.

In the EFT, one has Eq.~\eqref{eq:eqMyLagrangian}

\begin{eqnarray}
\mathcal{H}^{(5)}_{\rm eff} = 8 G^{2}_{F} \beta h^2 \lambda^{LR}_{c} \lambda^{RL}_{c} \left( \sum_{m, n = \pm} C_m (\mu_W) C_n (\mu_W) \langle \mathcal{O}_{m n} \rangle^{(0)} (\mu_W) \right. \nonumber\\
+ \left. \sum^2_{a=1} C^r_a (\mu_W) \langle \tilde{Q}^{LR}_a \rangle^{(0)} (\mu_W) \right) ,
\end{eqnarray}
\noindent
at order $ \alpha_s^0 $, including an overall normalization $ 8 G^{2}_{F} \beta h^2 \lambda^{LR}_{c} \lambda^{RL}_{c} $ factor for convenience.
Therefore

\begin{eqnarray}\label{eq:initialWilson}
C^r_2 (\mu_W) & = & \frac{\alpha_s (\mu_W)}{4 \pi} \left( 4 \log (\mu^2_W / M^2_W) - 4 + f_{(W',H)} \right) + \mathcal{O} (\alpha^2_s (\mu_W)) \, , \nonumber\\
C^r_1 (\mu_W) & = & \mathcal{O} (\alpha^2_s (\mu_W)) \, .
\end{eqnarray}


The initial conditions for the $ \vert \Delta F \vert \, = 1 $ Wilson coefficients are given by \cite{Buras:1991jm}

\begin{equation}\label{eq:initialWilsonDF1}
C_r (\mu_{W}) = 1 + \frac{\alpha_s (\mu_{W})}{4 \pi} \left( \log \left( \frac{\mu_{W}}{M_W} \right) \gamma^{(0)}_{m} + B_2 + r B_1 \right) \, , \; r=\pm \, ,
\end{equation}
\noindent
where $ B_{1,2} $ are given in Appendix~\ref{app:anomdimgen}, which have the same values at $ \mu_W $ since QCD is invariant under parity.

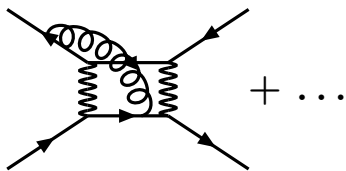
\begin{figure}
	\centering
	\SetWidth{1.2}
	\begin{picture}(90,60)(0,0)
		\ArrowLine(0,0)(30,20)
		\ArrowLine(30,20)(60,20)
		\ArrowLine(60,20)(90,0)
		\ArrowLine(90,60)(60,40)
		\ArrowLine(60,40)(30,40)
		\ArrowLine(30,40)(0,60)
		\Photon(30,20)(30,40){3}{5}
		\Photon(60,20)(60,40){3}{5}
		\GlueArc(20,20)(30,0,100){4}{6}
		\Text(110,35)[t]{\Large $ \textbf{+ \dots} $}
	\end{picture}
	\caption{\it Set of diagrams from the full theory at the order $ \alpha_s $. They match onto NNLO corrections to the Wilson coefficients, and therefore we do not need to consider them.}\label{fig:DoNotInterestUs}
\end{figure}

The Wilson coefficients we have above are the ones necessary for a full NLO computation. It is clear at this point that the diagram in Figure~\ref{fig:DoNotInterestUs} matches at the next order in perturbation theory, i.e. at the NNLO. 

\subsubsection{Anomalous dimension matrix at LO}

The set of anomalous dimension matrices for $ \vert \Delta F \vert \, = 1, 2 $ without mixing is given in Appendix~\ref{app:anomdimgen}. Here, we derive the anomalous dimension tensor describing their mixing. The anomalous dimension tensor is calculated from the divergences calculated in Eq.~\eqref{eq:eq19}. Writing the divergent part given in Eq.~\eqref{eq:eq22} in the $ \tilde{Q}^{LR}_a $ basis, one has

\begin{eqnarray}\label{eq:normalisationQtilde}
\sum_{m,n} \left( - 4 \frac{1}{\epsilon} \right) \tau_{a}^{m n} \frac{m^2_c}{16 \pi^2} \langle Q^{LR}_{a} \rangle^{(0)} (\mu_W) = \frac{\alpha_s}{4 \pi} \sum_{m,n} \left( - 4 \frac{1}{\epsilon} \right) \tau_{a}^{m n} \langle \tilde{Q}^{LR}_{a} \rangle^{(0)} (\mu_W) \, .
\end{eqnarray}

We now have the following definition, cf. Appendix~\ref{sec:toughRGE}

\begin{equation}
\gamma^{(0)}_{mn, a} = 2 \left[ \tilde{Z}^{-1,(1)}_{1} \right]_{mn, a} \, ,
\end{equation}
where $ \tilde{Z}^{-1,(1)}_{1} $ has been introduced in the renormalized effective Lagrangian of Eq.~\eqref{eq:eqMyLagrangian}, containing the necessary counterterms. Therefore,

\begin{equation}
\gamma^{(0)}_{mn, a} = - [ 2 \times ( - 4 \cdot \tau_{a}^{m n} ) ] = 8 \cdot \tau_{a}^{m n} \, .
\end{equation}

\subsubsection{Anomalous dimension matrix at NLO}

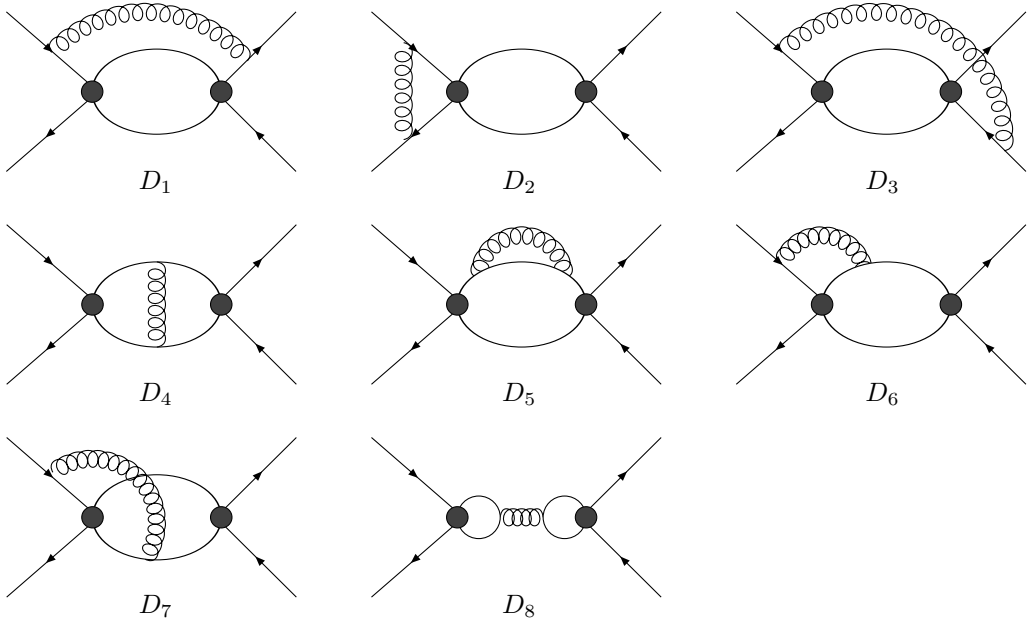
\begin{figure}[t]
\hspace{1.3cm}
\begin{picture}(550,250)(50,-200)
\SetScale{0.8}\setlength{\unitlength}{0.8pt}
\ArrowLine(40,60)(80,25)
\ArrowLine(80,20)(40,-15)
\GlueArc(105,0)(60,38,137){4}{15}
\GOval(110,22.5)(20,30)(360){1}
\GCirc(80,22.5){5}{0.25}
\ArrowLine(140,25)(175,60)
\ArrowLine(175,-15)(140,20)
\Text(110,-20)[]{$D_1$}
\GCirc(140,22.5){5}{0.25}

\ArrowLine(210,60)(250,25)
\ArrowLine(250,20)(210,-15)
\Gluon(225,45.5)(225,0){4}{6}
\GOval(280,22.5)(20,30)(360){1}
\GCirc(250,22.5){5}{0.25}
\ArrowLine(310,25)(345,60)
\ArrowLine(345,-15)(310,20)
\Text(280,-20)[]{$D_2$}
\GCirc(310,22.5){5}{0.25}

\ArrowLine(380,60)(420,25)
\ArrowLine(420,20)(380,-15)
\GlueArc(445,0)(60,-5,137){4}{20}
\GOval(450,22.5)(20,30)(360){1}
\GCirc(420,22.5){5}{0.25}
\ArrowLine(480,25)(515,60)
\ArrowLine(515,-15)(480,20)
\Text(450,-20)[]{$D_3$}
\GCirc(480,22.5){5}{0.25}

\ArrowLine(40,-40)(80,-75)
\ArrowLine(80,-80)(40,-115)
\GOval(110,-77.5)(20,30)(360){1}
\GCirc(80,-77.5){5}{0.25}
\Gluon(110,-57.5)(110,-97.5){4}{6}
\ArrowLine(140,-75)(175,-40)
\ArrowLine(175,-115)(140,-80)
\Text(110,-120)[]{$D_4$}
\GCirc(140,-77.5){5}{0.25}

\ArrowLine(210,-40)(250,-75)
\ArrowLine(250,-80)(210,-115)
\GlueArc(280,-65)(20,0,180){4}{10}
\GOval(280,-77.5)(20,30)(360){1}
\GCirc(250,-77.5){5}{0.25}
\ArrowLine(310,-75)(345,-40)
\ArrowLine(345,-115)(310,-80)
\Text(280,-120)[]{$D_5$}
\GCirc(310,-77.5){5}{0.25}

\ArrowLine(380,-40)(420,-75)
\ArrowLine(420,-80)(380,-115)
\GlueArc(420,-65)(20,0,160){4}{10}
\GOval(450,-77.5)(20,30)(360){1}
\GCirc(420,-77.5){5}{0.25}
\ArrowLine(480,-75)(515,-40)
\ArrowLine(515,-115)(480,-80)
\Text(450,-120)[]{$D_6$}
\GCirc(480,-77.5){5}{0.25}

\ArrowLine(40,-140)(80,-175)
\ArrowLine(80,-180)(40,-215)
\GOval(110,-177.5)(20,30)(360){1}
\GCirc(80,-177.5){5}{0.25}
\GlueArc(80,-180)(30,-35,128){4}{16}
\ArrowLine(140,-175)(175,-140)
\ArrowLine(175,-215)(140,-180)
\Text(110,-220)[]{$D_7$}
\GCirc(140,-177.5){5}{0.25}

\ArrowLine(210,-140)(250,-175)
\ArrowLine(250,-180)(210,-215)
\Gluon(270,-177.5)(290,-177.5){4}{4}
\ArrowLine(310,-175)(345,-140)
\ArrowLine(345,-215)(310,-180)
\Text(280,-220)[]{$D_8$}
\GCirc(260,-177.5){10}{1.}
\GCirc(300,-177.5){10}{1.}
\GCirc(250,-177.5){5}{0.25}
\GCirc(310,-177.5){5}{0.25}
\end{picture}
\vspace{-0.5cm}
\caption{\it Diagrams $D_i$, corresponding to the contributions at order $ \alpha_s $ from the double insertion of $ \vert \Delta F \vert \, = 1 $ operators.}\label{fig:diagsDi}
\end{figure}

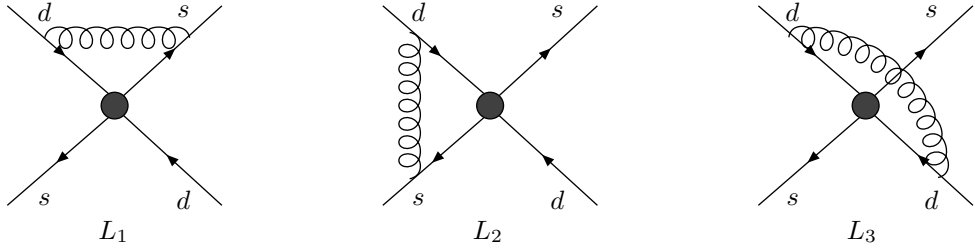
\begin{figure}[t]
\hspace{1.2cm} \begin{picture}(300,85)(50,0)
\ArrowLine(40,60)(80,25)
\Text(54,58)[]{$d$}
\ArrowLine(80,20)(40,-15)
\Text(54,-13)[]{$s$}
\Gluon(54,48)(108,48){4}{6}
\ArrowLine(82,25)(120,60)
\Text(106,58)[]{$s$}
\ArrowLine(120,-15)(82,20)
\Text(106,-13)[]{$d$}
\Text(80,-25)[]{$L_1$}
\GCirc(80,22.5){5}{0.25}

\ArrowLine(180,60)(220,25)
\Text(194,58)[]{$d$}
\ArrowLine(220,20)(180,-15)
\Text(194,-13)[]{$s$}
\Gluon(190,50)(190,-5){4}{7}
\ArrowLine(222,25)(260,60)
\Text(246,58)[]{$s$}
\ArrowLine(260,-15)(222,20)
\Text(246,-13)[]{$d$}
\Text(220,-25)[]{$L_2$}
\GCirc(220,22.5){5}{0.25}

\ArrowLine(320,60)(360,25)
\Text(334,58)[]{$d$}
\ArrowLine(360,20)(320,-15)
\Text(334,-13)[]{$s$}
\GlueArc(335,-3.7)(52,-0.9,94){4}{11}
\ArrowLine(362,25)(400,60)
\Text(386,58)[]{$s$}
\ArrowLine(400,-15)(362,20)
\Text(386,-13)[]{$d$}
\Text(360,-25)[]{$L_3$}
\GCirc(360,22.5){5}{0.25}
\end{picture}

\vspace{0.8cm}
\caption{\it Diagrams $L_i$, which correspond to the dressing of local $ \vert \Delta F \vert \, = 2 $ operators with gluons.}\label{fig:diagsLi}
\end{figure}

When going to the NLO, we need to dress the diagrams and operators in Figure~\ref{fig:illlustrationLocals} (including evanescent operators) with one gluon in order to compute the anomalous dimension tensor at this order. The full set of diagrams we have is given in Figures~\ref{fig:diagsDi} and \ref{fig:diagsLi}. Their divergent and finite contributions can be computed with the help of \textsc{Mathematica} and \textsc{TARCER} \cite{Mertig:1998vk},\footnote{See \cite{Jamin:1991dp} for calculations in the 't Hooft-Veltman scheme.} a package of \textsc{FeynCalc} \cite{Mertig:1990an} \cite{Shtabovenko:2016sxi} for the reduction of two-loop integrals, and using the integrals from \cite{Amoros:1999dp} or \cite{Davydychev:1992mt}. We use the $ \overline{{\rm MS}} $ scheme for extracting the divergences, where possible terms $ \log 4 \pi $ and $ \gamma_E $ are absorbed into the definition of the integral in $ D $ dimensions. External momenta are set to zero. At the same time, the IR divergences are controlled by external quark-masses $ m_{d,s} $, that are kept all long the calculations. We spare the reader from long intermediate results of the different classes of diagrams, which can be found in Ref.~\cite{Bernard:2015boz}.

Compared to the LO, new evanescent operators are present in the NLO. In our computation, they appear when many Dirac matrices of one quark line are contracted with the second line: for example, the insertion of $ \gamma_\nu \gamma_\mu P_R \cdot \gamma^\nu \gamma^\mu P_L $ seen in the definition of $ E [Q^{LR}_{1}] $ in Eq.~\eqref{eq:simpleEvanescent} in the diagram seen in Figure~\ref{fig:illustrateEquation} results in a structure with four Lorentz indices. Following this same example, consider

\begin{equation}
E^{LR}_{7} = \bar{d}^{\alpha} \gamma^{\mu_1} \gamma^{\mu_2} \gamma^{\mu_3} \gamma^{\mu_4} P_{R} s^{\alpha} \cdot \bar{d}^{\beta} \gamma_{\mu_1} \gamma_{\mu_2} \gamma_{\mu_3} \gamma_{\mu_4} P_{L} s^{\beta} - \left( (4 + a^{LR}_{2 \gamma} \epsilon)^2 + b \epsilon \right) Q^{LR}_{2} \, ,
\end{equation}
where we indicate the further arbitrarity in the choice of the $ \mathcal{O} (\epsilon) $ term by the inclusion of the parameter $ b $. Similarly to Section~\ref{sec:EObasics}, when considering the operators $ \tilde{Q}^{LR}_{a} $ with a different normalization we replace $ b $ with $ \tilde{b} $. The value of $ b $ for the same $ a^{LR}_{2 \gamma} = \tilde{a}^{LR}_{2 \gamma} = 4 $ is $ b = \tilde{b} = 96 $: we consider these values because then Fierz transformations can be applied in $ D \neq 4 $ dimensions. 

To determine the anomalous dimension tensor one computes $ \gamma^{(1)}_{r \ell, a} $ defined in

\begin{eqnarray}\label{eq:HowToCalculateGamma}
&& \gamma^{(1)}_{mn, a} = 4 \left( \left[ \tilde{Z}^{-1,(2)}_{1} \right]_{mn, a} + \beta_0 \left[ \frac{\tilde{Z}^{-1,(1)}_{0}}{2} \right]_{mn, a} \right. \nonumber\\
&& - \sum^{2}_{b=1} \left\{ \left[ \frac{\tilde{Z}^{-1,(1)}_{0}}{2} \right]_{mn, b} \left[ \tilde{Z}^{-1,(1)}_{1} \right]_{ba} + \left[ \tilde{Z}^{-1,(1)}_{1} \right]_{mn, b} \left[ \frac{\tilde{Z}^{-1,(1)}_{0}}{2} \right]_{ba} \right\} \nonumber\\
&& - \sum_{m', n' = \pm} \left\{ \left( \left[ \frac{Z^{-1,(1)}_{0} }{2} \right]_{m m'} \delta_{n n'} + \delta_{m m'} \left[ \frac{Z^{-1,(1)}_{0}}{2} \right]_{n n'} \right) \left[ \tilde{Z}^{-1,(1)}_{1} \right]_{m' n', a} \right. \nonumber\\
&& + \left. \left. \left( \left[ Z^{-1,(1)}_{1} \right]_{m m'} \delta_{n n'} + \delta_{m m'} \left[ Z^{-1,(1)}_{1} \right]_{n n'} \right) \left[ \frac{\tilde{Z}^{-1,(1)}_{0}}{2} \right]_{m' n', a} \right\} \right) \, ,
\end{eqnarray}
see Appendix~\ref{sec:toughRGE}. In this expression, the finite renormalization constants $ Z_0, \tilde{Z}_0 $, which go with an overall factor $ -1/2 $, absorb the contributions from evanescent operators which come as $ E / \epsilon $ (a precision is in order: the term proportional to $ \beta_0 $ does not contribute when indices $ k, n $ corresponding to physical operators are taken in $ [\tilde{Z}_0^{-1,(1)}]_{kn,1} $ or $ [\tilde{Z}_0^{-1,(1)}]_{kn,2} $, see Ref.~\cite{Buras:1998raa}). These same operators, namely $ E / \epsilon $, also contribute to $ \tilde{Z}^{-1, (2)}_1 $ in the computation of two-loop diagrams, which by its turn is not suppressed by a factor $ 1/2 $. All in all, evanescent contributions come suppressed by a factor $ 1/2 $.



One finally gets

\begin{equation}
\gamma^{(1)}_{rl,i}=-4 h^{rl,i}(1/2) \, ,
\end{equation}
with
\begin{eqnarray}\label{eq:firsthforgamma}
-h^{rl,1}(\lambda)&=&\frac{\lambda}{32 N}  \Bigg[(\tilde b-96) \left(N^2-2\right) \beta_{rl}+\left(8(\tilde b-48)
  -6 (\tilde b-96) N^2\right) \tau_{rl}\,
\nonumber \\
&&+ 6 N (\tilde b-80)\Bigg]-(\tilde b-280)
   \left(N^2-2\right)\frac{ \beta_{rl}}{64 N} \nonumber \\
&& +\left(3 \tilde b N^2-4\tilde b-152 N^2+48\right)
\frac{ \tau_{rl}}{32 N}+\frac{1}{32} (376-3 \tilde b) ,
\end{eqnarray}

\begin{eqnarray}\label{eq:secondhforgamma}
-h^{rl,2}(\lambda)&=& \frac{ \lambda}{8 N }  \left(3 \left(\tilde b-16 \left(N^2+4\right)\right)
   +\left(48 -\frac{\tilde b}{2}\right)N \beta_{rl}+\left(\tilde b +96\right) N
   \tau_{rl}\right)
\nonumber \\
&& +\frac{1}{16 N}(-3 \tilde b+72 N^2+304)+ 
( \tilde b-280) \frac{\beta_{rl}}{32}-\left(\frac{\tilde b}{8}+13\right) \frac{ \tau_{rl}}{2}\,, \nonumber \\
\end{eqnarray}
$ \beta_{rl}=l+r $, where the contributions from the evanescent operators are indicated with a factor $\lambda$. Therefore

\begin{eqnarray}\label{eq:eq25}
\gamma^{(1)}_{++, 1} = - \frac{251}{6} , \quad \gamma^{(1)}_{+-, 1} = \gamma^{(1)}_{-+, 1} = \frac{169}{2} , \quad \gamma^{(1)}_{--, 1} = - \frac{355}{6} , \nonumber\\
\gamma^{(1)}_{++, 2} = - \frac{41}{3} , \quad \gamma^{(1)}_{+-, 1} = \gamma^{(1)}_{-+, 1} = \frac{73}{3} , \quad \gamma^{(1)}_{--, 1} = \frac{223}{3} .
\end{eqnarray}

The above expression satisfies a series of checks: (1) we have performed our computations in an arbitrary QCD gauge $ \xi $, and no dependence on $ \xi $ is present in the final result (but it is present in intermediate steps, see \cite{Bernard:2015boz}); (2) similarly, we have regularized the IR divergences by considering masses $ m_{d,s} $ for the external states, which are not seen in the final expressions (but also present in the intermediate expressions, see \cite{Bernard:2015boz}); (3) as shown in Ref.~\cite{Herrlich:1994kh}, the $ \tilde{b} $ shown above have no effect in the calculation of $ \gamma^{(1)}_{rl,i} $, which is seen easily from Eqs.~\eqref{eq:firsthforgamma} and \eqref{eq:secondhforgamma} when $ \lambda = 1/2 $.

\subsection{EFT below $ \mu_c $}


At the energy scale $ \mu_c $ we match the EFT defined above $ \mu_c $ onto an EFT for which the charm is integrated out. The Wilson coefficients of this new EFT are determined by comparing the Green's functions in the two EFT, which is summarized in the following equation

\begin{equation}\label{eq:matchingNNLO}
\Bigg( \underbrace{ \frac{\pi}{\alpha_s (\mu_c)} C^r_a (\mu_c)  }_{\rm LO + NLO} + \sum_{i,j=\pm} \left[ \left( \underbrace{ r_{ij,a} }_{\rm NLO} + \underbrace{ \frac{\alpha_s}{4 \pi} C_a^{{\rm op}} }_{\rm NNLO} \right) C_i C_j \right] (\mu_c) \Bigg) \stackrel{\rm matching}{=} F_a (\mu_c) \, ,
\end{equation}
where the Wilson coefficients $ C^r_a, \; a = 1, 2 $ and $ C_i, \; i = \pm $, in the LHS are the Wilson coefficients of the EFT defined between the two energy scales $ \mu_W $ and $ \mu_c $, while the Wilson coefficients $ F_a, \; a = 1, 2 $, correspond to the  EFT defined below $ \mu_c $: this matching is illustrated in Figure~\ref{fig:CharmCharmTopEFT} seen in the previous chapter in the context of the SM. Note that the factor $ \pi / \alpha_s (\mu_c) $ seen above comes from the normalization of the operators $ \tilde{Q}^{LR}_{1,2} $ (which is the reason why the same factor does not multiply $ F_a (\mu_c) $). Also concerning this same normalisation factor, the LO term indicated in Eq.~\eqref{eq:matchingNNLO} gives a factor proportional to $ \log (\mu_c / \mu_W) $, as it should, since the running of the Wilson coefficient $ C^r $ results in a factor proportional to $ \alpha_s (\mu_c) \log (\mu_c / \mu_W) $. We will see in a while the values of the Wilson coefficients at $ \mu_c $, which are found from their evolution starting from $ \mu_W $, where they are given by Eqs.~\eqref{eq:initialWilson} and \eqref{eq:initialWilsonDF1}.

It was possible to check that  the gauge-dependent terms as well as the terms
involving small quark masses $m_s$ and $m_d$ are
 canceled at the matching scale $\mu_c$  for any choice of the coefficients at order $ \mathcal{O} (\epsilon) $ in the definition
of the evanescent operators. This provides additional powerful checks of the calculation and shows 
that our results are indeed independent of the choice of the QCD gauge and the infrared regularisation.

In addition, our results also provide  an 
estimate of the size of NNLO corrections. Indeed,
at NNLO several new contributions appear, one of them coming from the ${\cal O}(\alpha_s)$ corrections
to the operators shown in Eq.~\eqref{eq:matchingNNLO} and proportional to $ C_a^{{\rm op}} $.

Its final expression is
\begin{eqnarray}
8 C_1^{\rm op}&=&\log \left(\frac{m_c^2}{\mu ^2}\right) \left[\frac{11
   \left(N^2-2\right) \beta_{rl}}{2 N}-\frac{\left(N^2+12\right) \tau_{rl}}{N}+5\right]
\nonumber\\
&&+\log ^2\left(\frac{m_c^2}{\mu ^2}\right) \left(\left(\frac{3}{N}-\frac{3 N}{2}\right) \beta_{rl}-3 N \tau_{rl}-3\right) \nonumber\\
&&-\frac{3
   \left(N^2-2\right) \beta_{rl}}{8 N}+\left(\frac{95 N}{4}-\frac{73}{2 N}\right) \tau_{rl}-\frac{65}{4}\,,
\end{eqnarray}
\begin{eqnarray}
8 C_2^{\rm op}&=&\log ^2\left(\frac{m_c^2}{\mu ^2}\right) \left(-\frac{6}{N}+3 \beta_{rl}-6 \tau_{rl}\right)
\nonumber\\
&&+\log \left(\frac{m_c^2}{\mu ^2}\right) \left(\frac{10}{N}-11
   \beta_{rl}-26 \tau_{rl}\right)
\nonumber\\
&&+\frac{11 N}{2}-\frac{38}{N}+\frac{3 \beta_{rl}}{4}-\frac{51
   \tau_{rl}}{2}\,,
\end{eqnarray}
which were calculated for $ a^{LR}_{2 \gamma} = \tilde{a}^{LR}_{2 \gamma} = 4, b = \tilde{b} = 96 $ and the other choices seen in Eq.~\eqref{eq:choicesEVOPS} in Appendix~\ref{sec:setOfEOs}.


In the 3-quark effective theory, the relevant anomalous dimension matrix $ \hat{\gamma}_{LR} $ describing the evolution of the system $ \{ \gamma L \otimes \gamma R, L \otimes R \} $ in the three-quark EFT are already known at the NLO and are given by \cite{Buras:2001ra}, and found in Appendix~\ref{app:anomdimgen}.


\subsection{Short-distance corrections in EFT} \label{sec:result}

Combining Eq.~\eqref{eq:matchingNNLO} with the renormalisation equation for $C^r_a, C_i$, $ a = 1, 2 $ and $ i = \pm $, down to the low scale $\mu$ below $m_c$,
we obtain the final result for $\bar\eta_{a,cc}^{(LR)}$ at NLO in the EFT approach,
corresponding to the gauge-invariant combination of box, vertex and self-energy diagrams:
 \begin{eqnarray}\label{eq:etaeft}
&& \bar\eta_{a,cc}^{(LR)}=\frac{1}{S^{LR}(x_c (\mu_c),\beta,\omega)} \\
&& \sum_{j=1,2}\biggl( \bigl(1 +\frac{\alpha_s(\mu)}{4 \pi} K^{[3]}\bigr)
 \exp \biggl[ d^{[3]} \cdot \log\frac{\alpha_s(\mu_c)}{\alpha_s(\mu)}
\biggr] \bigl(1 -\frac{\alpha_s(\mu_c)}{4 \pi} K^{[3]}\bigr)\biggr)_{aj}
F_j(\mu_c)\,, \nonumber
\end{eqnarray}
with $S^{LR}(x_c,\beta,\omega)$ given by
\begin{eqnarray}\label{eq:ccSLR}
S^{LR}(x_c,\beta,\omega)&=&1 +\log(x_c) +\frac{1}{4} f_{(W',H)} \,,
\end{eqnarray}
and

\begin{eqnarray}
F_a(\mu_c)&=&\frac{\pi}{\alpha_s(\mu_c)}  C_a^r(\mu_c)+  \sum_{r,l=\pm} 
\left( r_{rl,a}(\mu_c) +\frac{\alpha_s(\mu_c)}{4 \pi} C_a^{\rm op}(\mu_c) \right)C_r(\mu_c) C_l(\mu_c) ,
\nonumber\\
&& \quad r_{rl,a}(\mu_c)= \bigl(2+\log(m_c^2/\mu_c^2)\bigr) \tau_a^{rl}  \, ,\qquad\qquad a=1,2\,,
 \label{eqmatch}
\end{eqnarray}
where  the values of $C_a^r(\mu_c)$, $C_r(\mu_c)$ and $C_l(\mu_c)$ are given by
\begin{eqnarray}
\begin{pmatrix}
C_r C_l  \\
C^r_1 \\
C^r_2 \\
\end{pmatrix} (\mu_c)&=&\left(1 +\frac{\alpha_s(\mu_c)}{4 \pi} \tilde J^{[4]}\right) \cdot \exp \biggl[ \tilde d^{[4]} \cdot \log\frac{\alpha_s(\mu_b)}{\alpha_s(\mu_c)}
\biggr] \\
&& \cdot \left(1 +\frac{\alpha_s(\mu_b)}{4 \pi} (\tilde J^{[5]}- \tilde J^{[4]})\right)
\nonumber\\
&& \cdot \exp \biggl[ \tilde d^{[5]} \cdot \log\frac{\alpha_s(\mu_W)}{\alpha_s(\mu_b)}\biggr]
\cdot \left(1 -\frac{\alpha_s(\mu_W)}{4 \pi} \tilde J^{[5]}\right)\cdot \begin{pmatrix}
C_r C_l  \\
C^r_1 \\
C^r_2 \\
\end{pmatrix} (\mu_W)\,,\nonumber
\label{eq:vecD}
\end{eqnarray}
following the running formalized in Eqs.~\eqref{eq:RGEDF1eq} and \eqref{eq:RGEDF2eq}. In order to get an estimate of the uncertainty due to neglected higher-order contributions, we have added in Eq.~\eqref{eqmatch} the contribution $C_a^{\rm op}$ which 
first appears at the next order.
The $C^r_{1,2} (\mu_W)$ are defined in Eq.~\eqref{eq:initialWilson} while $C_{\pm}(\mu_W)$
is defined in Eq.~\eqref{eq:initialWilsonDF1}.

Finally the matrices $\tilde d= \tilde d^{[f]}$, $\tilde J = \tilde J^{[f]}$ and $d=d^{[3]}$, $K=K^{[3]}$ encode respectively the 
$6 \times 6$  anomalous dimension matrix $\tilde \gamma$ and the $2 \times 2$ one
$\hat\gamma_{LR}$ defined in Appendix~\ref{sec:toughRGE}, with the additional definition
\begin{equation}
\tilde d=\frac{({\tilde\gamma}^{(0)})^T}{2 \beta_0} \, , \quad \quad \tilde{J} +[\tilde{d} ,\tilde{J}]=-\frac{(\tilde\gamma^{(1)})^T}{2 \beta_0}+\frac{\beta_1}{ \beta_0} \tilde{d}\, .
\end{equation}
Simplified expressions for $F_a(\mu_c)$ where effects from the 
five-flavour theory have been neglected and which are extremely good approximations
to the complete results read
\begin{eqnarray}
F_1&=&  
\frac{3}{104} \frac{\pi}{\alpha_s} \left(2 A^{--}-39 A^{+-}-26 A^{++}
+63 A_1\right)
\\
&&\biggl. -\frac{1}{8} \left(
\log \left(\frac{m_c{}^2 (\mu_c)}{\mu _c{}^2}\right)+2\right) \left(A^{--}-6 A^{+-}+5
   A^{++}\right) 
\biggr. 
\nonumber \\
&&
\biggl. +\frac{1}{4}\biggl( -\frac{1761281}{390000} A^{--}+\frac{587029}
{220000} A^{+-}+\frac{16120889}{1110000}A^{++}
-\frac{4789827}{260000}A_1+\frac{1737}{296} A_2
 \biggr. \biggr.
\nonumber\\
&& \biggl. \biggl. 
+A \left(A^{--} \left(-\frac{12}{13} \log \left(\frac{\mu
   _W}{M_W}\right)-\frac{10181}{16250}\right)+A^{+-} 
\left(\frac{9}{2} \log \left(\frac{\mu
   _W}{M_W}\right)+\frac{39993}{10000}\right)  \right.
\nonumber\\
&&  \biggl. \left. 
+A^{++} \left(-6 \log \left(\frac{\mu
   _W}{M_W}\right)-\frac{7031}{2500}\right)+A_1 \left(\frac{63}{26} \log \left(\frac{\mu
   _W}{M_W}\right)-\frac{974889}{1430000}\right)\right)
\biggr)\,,\nonumber
\end{eqnarray}
\begin{eqnarray}
F_2&=&
 \frac{3}{1924} \frac{\pi}{\alpha_s} \left(2590 A^{--}-481 A^{+-}-182 A^{++}+777 A_1-2704 A_2\right) \biggr.
\nonumber \\
&&\biggl. + \frac{1}{4} \left(\log \left(\frac{m_c{}^2 (\mu_c)}{\mu _c{}^2}\right)+2\right) \left(A^{--}+2
   A^{+-}+A^{++}\right)
\nonumber \\
&&
 +\frac{1}{4}\biggl(-\frac{101273 A^{--}}{9750}+\frac{3969529
   A^{+-}}{330000}+\frac{6590729 A^{++}}{555000}-\frac{5219109 A_1}{130000}+\frac{21963
   A_2}{3700}
 \biggr.
\nonumber\\
&&  \biggl. 
+A \left(-\frac{7}{1625} A^{--} \left(15000 \log \left(\frac{\mu
   _W}{M_W}\right)+10181\right)+A^{+-} \left(3 \log \left(\frac{\mu
   _W}{M_W}\right)+\frac{13331}{5000}\right)  \right.\biggr. 
\nonumber\\
&&  \biggl. \left. 
-\frac{7}{46250} \left(15000 \log \left(\frac{\mu
   _W}{M_W}\right)+7031\right) A^{++}
\right .\biggr.
\nonumber \\
&&\biggl. \biggl. \left.
+A_2 \left(2\log \left(\frac{M_W{}}{M_{W'}{}}\right) +F(\omega_1, \omega_2)+\frac{2600}{37} \log \left(\frac{\mu
   _W}{M_W}\right)+\frac{1318747}{22200}\right)\biggr. \biggr.\right.
\nonumber\\
&& \biggl. \left. +A_1 \left(\frac{21}{13} \log \left(\frac{\mu
   _W}{M_W}\right)-\frac{324963}{715000}\right)\right)
\biggr) \,,
\end{eqnarray}
with
\begin{eqnarray}
&& A=\frac{\alpha_s(\mu_W)}{\alpha_s(\mu_c)}\,, \qquad \qquad 
A_1=\biggl(\frac{\alpha_s(\mu_W)}{\alpha_s(\mu_c)}\biggr)^{\frac{2}{25}}\,, \quad \quad
A_2=\biggl(\frac{\alpha_s(\mu_W)}{\alpha_s(\mu_c)}\biggr)^{-1}\,,
\nonumber\\
&& A^{++}=\biggl(\frac{\alpha_s(\mu_W)}{\alpha_s(\mu_c)}\biggr)^{\frac{12}{25}}\,, \quad \quad
A^{+-}=\biggl(\frac{\alpha_s(\mu_W)}{\alpha_s(\mu_c)}\biggr)^{-\frac{6}{25}}\,, \\
&&A^{--}=\biggl(\frac{\alpha_s(\mu_W)}{\alpha_s(\mu_c)}\biggr)^{-\frac{24}{25}}\,. \nonumber
\end{eqnarray}

The value of $ \bar\eta_{1,cc}^{(LR)} $ is negligible and the one of $\bar\eta_{cc}^{(LR)}\equiv\bar\eta_{2,cc}^{(LR)}$ at the scale $\mu=1$~GeV is 
\begin{equation}
\left.\bar\eta_{cc}^{(LR)}\right|_{EFT}=\frac{1}{1-0.0294 \, F(\omega_1, \omega_2)}[1.562  +(0.604 -0.037 F(\omega_1, \omega_2))-0.473]\,,
\label{eq:etaEFT}
\end{equation}
where $F(\omega_1, \omega_2)$ is defined in Eq.~\eqref{eq:TheFunctioFomega1omega2} and we have
taken $M_{W'}=1$ TeV (for \linebreak $M_{W'}=\mathcal{O} (1-10)$~TeV, the dependence on this parameter is very weak). 
The first and second numerical values in the brackets are the LO and NLO contributions stemming from the first term in Eq.~\eqref{eq:matchingNNLO} or \eqref{eqmatch}, whereas the
last term comes from the $r_{rl,a}$ term in the same equation (the term $C_a^{\rm op}$ in Eq.~\eqref{eq:matchingNNLO} or \eqref{eqmatch} being of a higher order).

\begin{figure}[t!]
\centering
\includegraphics[width=6.5cm,angle=0]{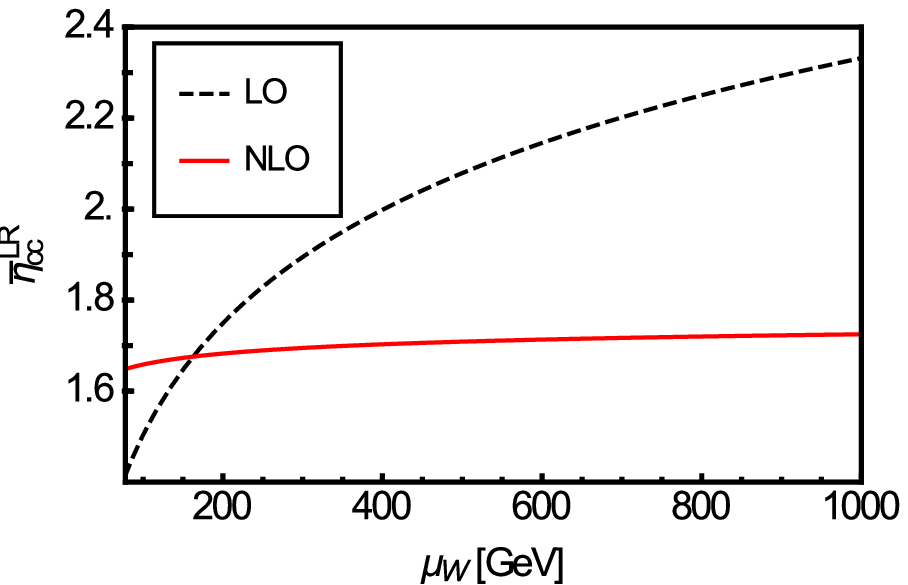}
\hspace{0.5cm}
\includegraphics[width=6.5cm,angle=0]{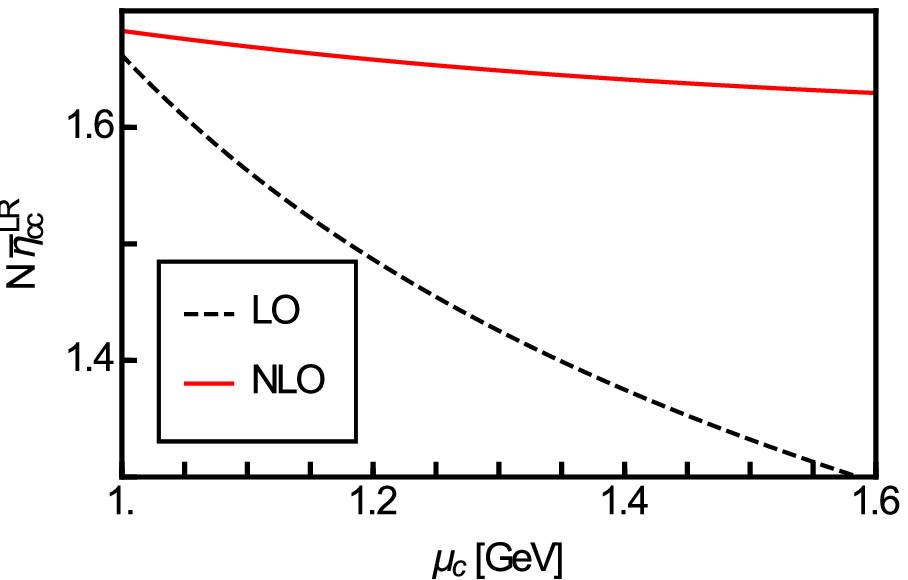}
\caption{\it Dependence of $\bar \eta_{cc}$ on the high (left panel) and on the low (right panel) scale  in the EFT approach for $M_{W_R}=1$ TeV and respectively for $\mu_c=m_c$ and
$\mu_W=M_W$. The other parameters are given in the text. The relevant quantity when $\mu_c \neq m_c$ is $ N \bar \eta_{cc}$ with $N$ defined in Eq.~\eqref{eq:noretamuc}. We also consider the limiting case $ \omega_1 = 0.1 $ and $ w=0 $.}
\label{figure:etaccmuw}
\end{figure}

\begin{figure}[t!]
\centering
\includegraphics[width=6.5cm,angle=0]{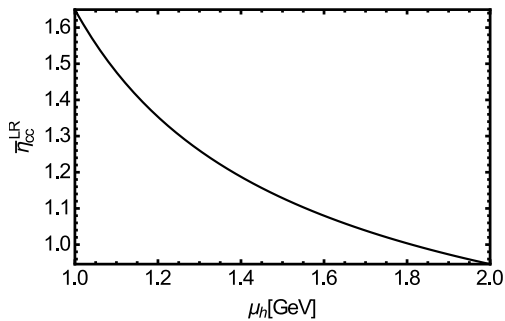}
\caption{\it Dependence of $\bar \eta^{LR}_{cc}$ on the hadronic scale  $\mu_h$ in the EFT approach. We also consider the limiting case $ \omega_1 = 0.1 $ and $ w=0 $.}\label{figure:etaccmuEFT}
\end{figure}

The dependence on the matching scales $\mu_W$
and $\mu_c$ is illustrated on Fig.~\ref{figure:etaccmuw}. This 
shows the  strong dependence of the LO result on the matching scales 
and the much milder dependence at NLO. 
This behaviour is similar to 
what is observed in the SM~\cite{Herrlich:1993yv,Buchalla:1995vs,Herrlich:1996vf} and it constitutes another significant check of our computation. In the case of the dependence on $\mu_c$, the relevant quantity is $N \bar \eta_{cc}$ with the normalisation factor given by
\begin{equation}
N= S^{LR}(x_c(\mu_c),\beta,\omega)/S^{LR}(x_c(m_c),\beta,\omega) \,,
\label{eq:noretamuc}
\end{equation}
considering that $S^{LR}(x_c(m_c),\beta,\omega)$ is the quantity multiplied by $\bar\eta_{cc}^{(LR)}$.
We also show the
dependence on the choice of the hadronic scale $\mu_h$ in Fig.~\ref{figure:etaccmuEFT} for typical values between $1 < \mu_h / {\rm GeV} < 2$ in the effective theory. 

\section{Discussion of the results}\label{sec:FinalResult}

We are now in a position to give our final results for the short-distance QCD corrections to $ K \bar{K} $ mixing
at NLO in LR Models. Adding up our results from the previous sections yields the effective Hamiltonian:
\begin{eqnarray}
H &=& H^{SM} +\frac{G_F^2 M^2_W}{4\pi^2} 8 \beta h^2 Q_2^{LR} \sum_{U,V = c,t} \lambda_U^{LR} \lambda_V^{RL} \bar\eta_{UV}^{(LR)} \sqrt{x_U x_V} S^{LR}(x_U,x_V,\beta,\omega) \nonumber\\
&& - \frac{4 G_F}{\sqrt{2}} u \beta \omega Q_2^{LR} \sum_{U,V = c,t} \lambda_U^{LR} \lambda_V^{RL} \bar\eta_{UV}^{(H)} \sqrt{x_U x_V} \nonumber\\
&& + \frac{G_F^2 M_W^2}{4\pi^2} Q_2^{LR} \sum_{U,V = c,t} \lambda_U^{LR} \lambda_V^{RL} \bar\eta_{UV}^{(H^\pm \rm box)} S^{H}_{LR}(x_U,x_V,\beta \omega) + h.c. ,
\label{eq:LRMHam}
\end{eqnarray}
where $H^{SM}$ is given in Eq.~\eqref{eq:TheOldKnownSM} and $ S^{LR} $ in Eq.~\eqref{eq:finalLoopFunct}. In all cases, the value of $ \bar\eta_{1, UV} $ is negligible, so that we will only consider $ \bar\eta_{2,UV} \equiv \bar\eta_{UV} $.

In the MR approach we add the individual contributions of Eqs.~\eqref{eq:boxSDQCDMR}, \eqref{eq:selfSDQCDMR}, \eqref{eq:vertSDQCDMR} (given in Tables~\ref{tab:noLargeLogBeta} and \ref{tab:LargeLogBeta}) for the three diagrams \ref{fig:tabNPdiagrams}(b), (c), (d) with the relevant weights and we normalize the result to 
$S^{LR}(x_U,x_V,\beta,\omega)$ in order to get the result in the appropriate form (the same applies to the charged Higgs in the
box which corresponds to the third line in Eq.~\eqref{eq:LRMHam}).

\subsection{Short-range contributions for the $cc$ box}

Since we computed $\bar\eta_{cc}^{(LR)} $ in both approaches, we can  compare the EFT result with the MR calculation.
We get from Eq.~\eqref{eq:etaEFT} and the MR value at the scale $\mu_h=1$ GeV for  $\omega_1=0.1$ ($\omega_1=0.8$) in the limiting case $ w=0 $
\begin{eqnarray}
\left.\bar\eta_{cc}^{(LR)}\right|_{EFT}&=&1.41 +0.67-0.43=1.65\qquad (3.41-0.17-1.03=2.21)\, , \nonumber\\
&& \\
\left.\bar\eta_{cc}^{(LR)}\right|_{MR}&=&1.16+0.13+0.03=1.32 \qquad (2.46+0.27-1.32=1.41)\, . \nonumber\\
\label{eq:etaccLRMRnum}
\end{eqnarray}
As in the SM case, we see that the central values from the MR are only in broad agreement (around 30\%) with the EFT approach in the presence of large logarithms, and in this sense we could quote a 30\% uncertainty in Eq.~\eqref{eq:etaccLRMRnum}. Including this uncertainty in our result and averaging with the values obtained with resummation of $\log \beta$, we have
\begin{equation}
\left. \bar\eta^{(LR)}_{cc} \right|_{MR} = 1.35 \pm 0.41 \pm 0.08 \qquad (1.48 \pm 0.44 \pm 0.10) ,
\end{equation}
where the first uncertainty comes from the comparison of MR and EFT, and the
second uncertainty is obtained by considering the values obtained with and without the resummation of $\log \beta$.

The EFT NLO central value will be taken as our final result. At the scale $\mu_h=1$ GeV and for $\omega_1=0.1$ ($\omega_1=0.8$)  in the limiting case $ w=0 $, we have:
\begin{equation}\label{eq:LRetacc}
\bar\eta_{cc}^{(LR)}= 1.65 \pm  0.50 \qquad  (2.21\pm 0.66)\,,
\end{equation}
where the conservative $30 \%$ error bar includes our estimate of higher-order terms, namely:
the contribution from $C_a^{op}$ (which turns out to be very small), contributions from the expansion of 
Eq.~\eqref{eq:etaeft} up to NNLO, an estimate of the NNLO term assuming a geometrical growth from  LO to NLO, the arbitrariness in the choice of $\mu_W$ when integrating out the $W$ and $W'$ bosons to match onto the four-flavour theory (we vary $\mu_W$ between the two high scales $M_W$ and $M_{W'}$), the dependence on the choice of the matching scales for the matching onto the three-flavour theory. Each of these uncertainties are of the order of a few percent. Furthermore we have not resummed the contributions $\log\beta$. This last uncertainty is clearly difficult to determine without an explicit calculation, however this logarithm $\log \beta$ is multiplied by a suppressing factor $\alpha_s(\mu_W)$, suggesting that the uncertainty should be smaller than our conservative estimate of $30 \%$.

\subsection{Short-range contributions  for the $ct$ and $tt$ boxes }

The short-distance contributions from the $ct$ and $tt$ boxes in the MR are: 
\begin{eqnarray}
\bar\eta_{ct}^{(LR)}&=&  2.74 \pm 0.82 \pm 0.05   \qquad (2.67 \pm 0.80 \pm 0.03)\,,\\
\bar\eta_{tt}^{(LR)}&=& 5.88 \pm 1.76 \pm 0.23 \qquad (5.55 \pm 1.67 \pm 0.11)\,,
\end{eqnarray}
where the central value and the second uncertainty are  obtained by considering the values obtained with or without a resummation of $\log\beta$. The first uncertainty is a conservative 30\% estimate of the uncertainty of the MR coming from our previous experience in the SM, in relation with the fact that the top quark is not treated on the same footing as other heavy degrees of freedom in this approach.
As indicated earlier, resumming or not $\log\beta$ yields a small uncertainty of a few percent in both cases (as expected, since the
potentially large logarithm $\log\beta$ is multiplied by a suppressing factor  $\alpha_s(\mu_W)$).

\subsection{Short-range contribution from neutral and charged Higgs exchange}
The values of the QCD short-distance corrections for the box containing a charged heavy Higgs (see Fig.~\ref{fig:tabNPdiagrams}) are 
\begin{eqnarray}
\bar\eta_{ct}^{(H^\pm \rm box)} & = & 2.76 \pm 0.83 \pm 0.07 \qquad (2.79 \pm 0.84 \pm 0.10) , \\
\bar\eta_{tt}^{(H^\pm \rm box)} & = & 5.85 \pm 1.76 \pm 0.20  \qquad (5.90 \pm 1.77 \pm 0.25) , \\
\bar\eta_{cc}^{(H^\pm \rm box)} & = & 1.29 \pm 0.39 \pm 0.01 ,
\end{eqnarray}
where the first uncertainty corresponds to a conservative 30\% uncertainty related to the MR method,\footnote{Note that we provide only one $ \bar\eta_{cc}^{(H^\pm \rm box)} $ since the dependence on $ \omega_{1,2} $ is negligible.} and the second uncertainty  corresponds to an average of the results with and without a resummation of $ \log\beta $. For the tree-level neutral Higgs exchange we have
\begin{eqnarray}
\bar\eta_{ct}^{(H)} & = & 2.70 \pm 0.09 , \\
\bar\eta_{tt}^{(H)} & = & 5.66 \pm 0.30 , \\
\bar\eta_{cc}^{(H)} & = & 1.28 \pm 0.04 ,
\end{eqnarray}
where the quoted uncertainty assesses conservatively the neglected NLO corrections coming from the matching at $ \mu_H $ and the NNLO corrections based on a geometrical progression of the perturbative series.

\subsection{Set of numerical values for different energy scales}

Our results need to be combined with hadronic matrix elements calculated at the low energy scale $ \mu_h $. In the literature, values for $ \mu_h = 2 $~GeV and $ 3 $~GeV are found, and in Chapter~\ref{ch:PHENO} we are going to combine our calculations at $ 3 $~GeV with bag parameters calculated at this same scale. The numerical results of the short-distance QCD corrections at these scales are given in Table~\ref{tab:differentEnergies}.

\begin{table}[h]
\centering
	\begin{tabular}{|c|ccc|}
	\hline
	@ $ 2 $~GeV & tt & ct & cc \\
	\hline
	$ \bar\eta_{ij}^{(LR)} $ & $\frac{3.42 - 1.05 \; F(\omega_1, \omega_2)}{1 - 0.329 F(\omega_1, \omega_2)} $ & $\frac{1.53 - 0.312 \; F(\omega_1, \omega_2)}{1 - 0.207 F(\omega_1, \omega_2)} $ & $ \frac{0.968 - 0.0227 \; F(\omega_1, \omega_2)}{1 - 0.0294 \; F(\omega_1, \omega_2)} $ \\
	& $ \pm 1.12 $ & $ \pm 0.51 $ & $ \pm 0.21 $ (EFT) \\
	\hline
	$ \bar\eta_{ij}^{(H^\pm \rm box)} $ & $ 3.29 \pm 1.09 $ & $ 1.54 \pm 0.51 $ & $ 0.72 \pm 0.24 $ \\
	\hline
	$ \bar\eta_{ij}^{(H)} $ & $ 3.19 \pm 0.36 $ & $ 1.51 \pm 0.36 $ & $ 0.71 \pm 0.19 $ \\
	\hline
	\hline
	@ $ 3 $~GeV & tt & ct & cc \\
	\hline
	$ \bar\eta_{ij}^{(LR)} $ & $\frac{2.80 - 0.862 \; F(\omega_1, \omega_2)}{1 - 0.329 F(\omega_1, \omega_2)} $ & $\frac{1.26 - 0.256 \; F(\omega_1, \omega_2)}{1 - 0.207 F(\omega_1, \omega_2)} $ & $ \frac{0.803 - 0.0191 \; F(\omega_1, \omega_2)}{1 - 0.0294 \; F(\omega_1, \omega_2)} $ \\
	& $ \pm 0.92 $ & $ \pm 0.44 $ & $ \pm 0.18 $ (EFT) \\
	\hline
	$ \bar\eta_{ij}^{(H^\pm \rm box)} $ & $ 2.71 \pm 0.89 $ & $ 1.27 \pm 0.42 $ & $ 0.59 \pm 0.19 $ \\
	\hline
	$ \bar\eta_{ij}^{(H)} $ & $ 2.61 \pm 0.31 $ & $ 1.24 \pm 0.23 $ & $ 0.58 \pm 0.16 $ \\
	\hline 
	\end{tabular}
	\caption{\it Short-distance QCD corrections for the different classes of diagrams we consider: diagrams containing the exchange of a pair $ WW' $ (indicated by the superscript LR), a pair $ W H $ in a box diagram, and a neutral Higgs in a tree level diagram. In the last two lines we have ignored the very small dependence compared to the uncertainties of the results on $\omega_i$. The function $F(\omega_1, \omega_2)$ in the first line is defined in Eq.~\eqref{eq:TheFunctioFomega1omega2}.}\label{tab:differentEnergies}
\end{table}

\section{$ B $ meson systems}

\begin{figure}
	\centering
	\includegraphics[scale=0.65]{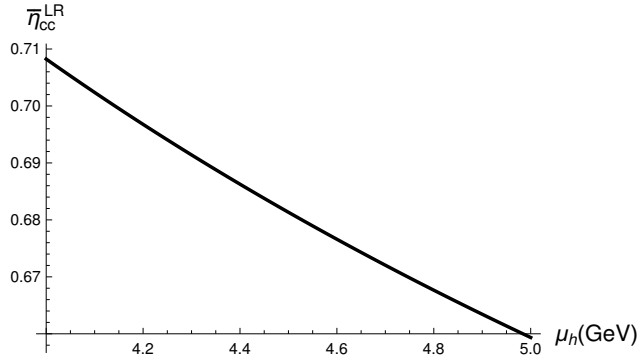}
	\caption{\it Dependence of $ \bar{\eta}_{cc} $ with the hadronic scale for the system of $ B $ mesons.}\label{fig:dependenceWithmuh}
\end{figure}

Similarly to the kaon system, we can also calculate short-distance QCD corrections for the $ B $ meson systems. As noted in Section~\ref{sec:setOfOpsEFT}, in the case where the right-handed mixing matrix has an arbitrary structure, we must calculate the contributions related to the $ tt $, $ ct $ and $ cc $ in the LR Model. In the EFT calculation, we have considered integrating out the heavy degree of freedom of the bottom at $ \mu_b = \mathcal{O} (m_b) $ and neglecting operators which are suppressed by $ 1 / m_b $. Therefore, the only difference we have compared to the kaon system calculation is the value of the low energy scale $ \mu_h $, which for the $ B $ meson systems is fixed at $ \mu_b = 4.3 $~GeV. As expected, similar dependences on the matching scales $ \mu_W $ and $ \mu_c $ are observed, while the dependence with $ \mu_h $ is shown in Figure~\ref{fig:dependenceWithmuh}. The numerical results for the EFT and the MR calculations are given in Table~\ref{tab:numericalValuesBBbar}.


\begin{table}[h]
\centering
	\begin{tabular}{|c|ccc|}
	\hline
	 & tt & ct & cc \\
	\hline
	$ \bar\eta_{ij}^{(LR)} $ & $\frac{2.44 - 0.75 \; F(\omega_1, \omega_2)}{1 - 0.329 F(\omega_1, \omega_2)} $ & $\frac{1.10 - 0.221 \; F(\omega_1, \omega_2)}{1 - 0.207 F(\omega_1, \omega_2)} $ & $ \frac{0.703 - 0.0169 \; F(\omega_1, \omega_2)}{1 - 0.0294 \; F(\omega_1, \omega_2)} $ \\
	& $ \pm 0.80 $ & $ \pm 0.35 $ & $ \pm 0.21 $ (EFT) \\
	\hline
	$ \bar\eta_{ij}^{(H^\pm \rm box)} $ & $ 2.36 \pm 0.80 $ & $ 1.12 \pm 0.36 $ & $ 0.52 \pm 0.18 $ \\
	\hline
	$ \bar\eta_{ij}^{(H)} $ & $ 2.28 \pm 0.31 $ & $ 1.07 \pm 0.21 $ & $ 0.50 \pm 0.16 $ \\
	\hline 
	\end{tabular}
	\caption{\it Short-distance QCD corrections at NLO for the LR contributions to $ B $ meson-mixing. Flavour thresholds are taken into account. The $\bar \eta$ are calculated at the hadronisation scale $\mu_{{h}}=4.3$~GeV with the other parameters as for the kaon system. In the last two lines we have ignored the very small dependence compared to the uncertainties of the results on $\omega_i$. The function $F(\omega_1, \omega_2)$ in the first line is defined in Eq.~\eqref{eq:TheFunctioFomega1omega2}.}\label{tab:numericalValuesBBbar}
\end{table}

\section{Conclusions}

Over this section we have considered two methods for computing short-distance QCD corrections at NLO for meson-mixing observables in LR Models. These are the Method of Regions, designed to include the most important corrections by inspecting all possible diagrams dressed with gluons, and the EFT approach, which builds effective descriptions of the full theory valid at low energies.

In the previous chapter, we have compared the two methods in the Standard Model where EFT results are available from the literature, and have obtained similar results using the MR except when large logarithms are present in the loop functions, i.e. $ \log x_c $. In the SM, this is the case for the charm-top contribution, requiring one to consider local $ \vert \Delta F \vert \, = 2 $ operators in the effective theory defined at the energy scale where the $ W $ boson and the top are integrated out in order to renormalize the divergences of the double insertion of $ \vert \Delta F \vert \, = 1 $ operators.

In the LR Model, we apply the same MR approach to compute short-distance QCD corrections for the diagrams shown in Figure~\ref{fig:tabNPdiagrams}, namely the set of contributions box, vertex and self-energy diagrams where a $ W' $ is exchanged, the box with a charged Higgs in place of the $ W' $ and a tree level diagram where a neutral Higgs is exchanged. In the first set, containing a $ W' $, the loop function of the charm-charm contribution has a large logarithm $ \log x_c $, and since there may be doubt concerning the MR result in this case, we have applied the EFT approach. This required the computation of the anomalous dimension matrix at NLO describing the mixing of (double insertions of) $ \vert \Delta F \vert \, = 1 $ operators and $ \vert \Delta F \vert \, = 2 $ operators. The other anomalous dimension matrices and matchings at NLO are known from the literature. We have as well computed a contribution coming at the NNLO in order to make a test of the good convergence of the perturbative series.

The whole interest of these calculations is to derive more trustful bounds coming from meson-mixing on the structure of LR Models, as for instance the scale of the masses of the $ W' $ and the exchanged scalar sector. We now consider the phenomenology of LR Models based on meson-mixing observables and the observables discussed in Chapter~\ref{ch:EWPO}.




\chapter{Global fits}\label{ch:PHENO}


We aim at having a clear picture of the doublet scenario of the LR Model based on experimental data. As we have discussed in previous chapters, this class of extensions of the SM introduces new energy scales $ \kappa_{1,2,L,R} $ from the spontaneous breakdown of its gauge symmetries, parameterized in terms of $ r = \kappa_2 / \kappa_1, w = \kappa_L / \kappa_1, \epsilon = \sqrt{\kappa_1^2 + \kappa_2^2 + \kappa_L^2} / \kappa_R, \kappa_R $. Together with the right-handed gauge coupling constant, rewritten in terms of $ c_\phi $, this set of parameters describes LR Model corrections at tree level to EWPO beyond the SM and the spectrum of masses of the new gauge bosons.

EWPO and direct bounds on the mass of the $ W' $ were already considered in Chapter~\ref{ch:EWPO}. Here, we would like to integrate the information coming from meson-mixing in a global fit. When considering meson-mixing observables, other parameters become relevant, which are the right-handed mixing matrix $ V^R $, together with the masses  of the neutral scalar sector, $ M_{H_1} $ and $ M_{H_2} = \delta M_{H_1} $. There is also an extra parameter coming from the Higgs potential, $ x = \mu'_1 / \mu'_2 $, appearing in the couplings of the scalars to fermions. The full set of parameters we have to constrain is then

\begin{equation}
\{ r, w, \epsilon, c_\phi, V^R, M_{H_1}, \delta, x \} \, .
\end{equation}
This is the set of parameters we consider, and we will be particularly interested in extracting lower bounds on the spectrum of LR particles. As a simplification, we consider here a particular structure of $ V^R $, namely $ V^R = V^L $, while more general structures will be considered in \cite{Bernard}.








\section{Meson-mixing observables}


Observables from meson-mixing have been extensively studied in the context of LR Models \cite{Beall:1981ze,Langacker:1989xa,Frere:1991db,Barenboim:1996nd,Barenboim:1996wz,Ball:1999mb,Kiers:2002cz,Zhang:2007da,Maiezza:2010ic,Blanke:2011ry,Bertolini:2014sua} (see also \cite{Isidori:2010kg,Charles:2013aka} for generic NP considerations), setting lower bounds of a few TeV on the mass of the $ W' $, and pushing the scalar masses beyond $ \mathcal{O} (10) $~TeV. These constraints were obtained in the case $ w = 0 $, and we would like to revisit these constraints in the more general case where $ w $ can be sizable, a possibility not allowed in the triplet case considered by these analyses.


We are going to focus on the better known observables in the context of the SM, namely $ \vert \epsilon_K \vert $, cf. for instance \cite{Bailey:2015wta} for its status, and the mass differences $ \Delta m_{d,s} $. Other observables have also been employed in the literature, such as the mass difference in the kaon system $ \Delta m_K $, and the direct CP violation quantity $ {\epsilon'}_K / \epsilon_K $. Due to the uncertainties coming from the long-distance effects in the prediction of these observables, we prefer at the moment to consider only the above mentioned observables which have a more solid status, though recent developments in $ {\epsilon'}_K / \epsilon_K $ have been reported in \cite{Buras:2015xba}. Similarly, we consider premature to employ $ D $ mixing observables, since theoretical uncertainties from the SM prediction limit a New Physics analysis.\footnote{See, however, Ref.~\cite{DDbarmixing} for bounds on LR Models from $ D \bar{D} $ mixing.}




The experimental values are given by Ref.~\cite{Agashe:2014kda}

\begin{eqnarray}
\vert \epsilon_K \vert &=& (2.228 \pm 0.011) \cdot 10^{-3} \, , \\
\Delta m_s &=& ( 17.757 \pm 0.021 ) \; {\rm ps}^{-1} \, , \\
\Delta m_d &=& ( 0.510 \pm 0.003 ) \; {\rm ps}^{-1} \, ,
\end{eqnarray}

\noindent
and we now introduce their theoretical expressions. The expression for the indirect CP violation $  \epsilon_K $ is \cite{Branco:1999fs} \cite{Anikeev:2001rk} \cite{Andriyash:2005ax}

\begin{eqnarray}\label{eq:epsilonDef}
&& \epsilon_K = \frac{G^2_F M^2_W}{8 \sqrt{2} \pi^2 \Delta m_K} {\rm e}^{i \phi_\epsilon} \Bigg[ \kappa_\epsilon \, {\rm Im} \{ \langle \bar{K} \vert Q^{VLL}_{1} \vert K \rangle (\mu_h) b(\mu_h) C^{SM} \} \nonumber\\
&& \qquad\qquad\qquad\qquad\qquad + {\rm Im} \{ \langle \bar{K} \vert Q^{LR}_{2} \vert K \rangle (\mu_h) C^{LR}_2 (\mu_h) \} \Bigg] \, ,
\end{eqnarray}
where $ C^{SM} $ is read from Eq.~\eqref{eq:TheOldKnownSM} by dropping $ \frac{G^2_F M^2_W}{4 \pi^2} Q^{VLL}_1 $ in that equation, and we focus first in this contribution. It includes the short-distance QCD factors $ \eta_{tt} $, $ \eta_{ct} $ and $ \eta_{cc} $, which are scale independent: the scale dependence has been factorized out in $ b (\mu_h) $, which combines with $ \langle Q^{VLL}_{1} \rangle (\mu_h) $ to give a scale invariant quantity. The value of $ C^{SM} $ is dominated by the top-top contribution ($ \sim 2 \cdot 10^{-7} $), but the charm-top and charm-charm ones have important sizes ($ \sim 6 \cdot 10^{-8} $ and $ \sim (- 3) \cdot 10^{-8} $, respectively).

Above, $ \kappa_\epsilon $ corresponds to long-distance effects, whose value has been calculated in the SM \cite{Buras:2008nn,Buras:2010pza,Lenz:2010gu}, thus giving


\begin{equation}
\kappa_\epsilon = 0.940 \pm 0.013 \pm 0.023 \, .
\end{equation}
Other long-distance effects are hidden in $ \Delta m_K $, which is precisely measured from experiment to be

\begin{equation}
\Delta m_K = 3.48392 \, \times \, 10^{-15} \; {\rm GeV} \, .
\end{equation}
Finally, since we are going to use the information coming from the modulus of $ \epsilon_K $, the value of $ \phi_\epsilon $ is not relevant for our purposes.

Concerning the LR Model corrections seen in Eq.~\eqref{eq:epsilonDef}, They consist in

\begin{eqnarray}
C^{LR}_2 = \Delta_{WW'} C^{LR}_2 + \Delta_{H^\pm \rm box} C^{LR}_2 + \Delta_{H} C^{LR}_2 \, ,
\end{eqnarray}
whose terms were introduced in Section~\eqref{sec:ExpressionsLRMwetaparameters}: 

\begin{eqnarray}
H^{(X)} = \frac{G^2_F M^2_W}{4 \pi^2} Q^{LR}_2 \Delta_{X} C^{LR}_2 \, ,
\end{eqnarray}
where $ X $ labels $ WW', H^\pm \, \rm box $ and $ H $. We are going to discuss in a while $ \langle \bar{K} \vert Q^{VLL}_{1} \vert K \rangle $ and $ \langle \bar{K} \vert Q^{LR}_{2} \vert K \rangle $.

Shifting to the $ B $ meson systems, the mass differences are given by

\begin{eqnarray}
&&\Delta M^{LR}_{q} = \frac{G^2_F M^2_W}{4 \pi^2} \Bigg| \langle \bar{B}_q \vert  Q^{VLL}_{1} \vert B_q \rangle (\mu_h) b(\mu_h) C^{SM} \\
&& \qquad\qquad\qquad\qquad + \langle \bar{B}_q \vert Q^{LR}_{2} \vert B_q \rangle (\mu_h) C^{LR}_2 (\mu_h) \Bigg| \, . \nonumber
\end{eqnarray}
The expressions of $ C^{SM} $ and the many contributions to $ C^{LR}_2 $ can also be read from Eq.~\eqref{eq:TheOldKnownSM} and Section~\ref{sec:ExpressionsLRMwetaparameters}, by replacing the mixing-matrix elements and short-distance QCD corrections from the system of kaons to the system of mesons $ B $. We further note that in the systems of $ B $ mesons, the individual terms in the SM are largely dominated by the top-top contribution (at least a factor $ 10^3 $ larger than the charm-top and charm-charm contributions).

Short-distance QCD corrections necessary in the calculation of the LR Model corrections to $ \epsilon_K $ and $ \Delta m_{d,s} $ observables were calculated in Chapter~\ref{ch:technicalEFT} and can be read from Tables~\ref{tab:differentEnergies} and \ref{tab:numericalValuesBBbar}.

\subsection{Bag parameters}

Apart from short-distance QCD corrections, other theoretical inputs include quantities parameterizing long-distance QCD corrections, or hadronic effects, which we now describe. While in the SM we find only the local operator $ Q^{VLL}_1 $ at low energy scales, LR Model requires a different operator $ Q^{LR}_{2} $ having a different chiral structure. This is the only new structure we need to consider, since we have already argued that $ Q^{LR}_{1} $ gives suppressed contributions.

Their matrix elements are related to the bag parameters $ B^M_{1,4} (\mu_h) $ as follows

\begin{eqnarray}\label{eq:BagParametersEnhancement}
&& \langle \overline{M} \vert Q^i_a \vert M \rangle (\mu_h) = \frac{2}{3} m_M F^2_M P^i_a (\mu_h) \, , \\
&& P^{VLL}_{1} (\mu_h) = B^M_{1} (\mu_h) \, , \quad P^{LR}_{2} (\mu_h) = \frac{3}{4} \left[ \left( \frac{m_M}{m_{q_1} + m_{q_2}} \right)^2 + \frac{1}{6} \right] B^M_{4} (\mu_h) \, , \nonumber
\end{eqnarray}
where the factor $1/6$ is subleading and thus often neglected, and $ B^M_{1} (\mu_h) $ and $ B^M_{4} (\mu_h) $ combine respectively with $ b (\mu_h) $ and the various $ \bar{\eta}_{UV} $, $ U, V = c, t $, to give a scale invariant quantity. At this point, note that there is a chiral enhancement of LR contributions to $ \epsilon_K $: indeed, $ m_K^2 / (m_s + m_d)^2 \simeq 25 $, while the same enhancement is not shared with LR contributions to the $ B- $meson observables.



The values of $ \hat{B}_{B_s} \equiv B^s_1 $, $ \hat{B}_{B_s}/\hat{B}_{B_d} \equiv B^s_1 / B^d_1 $ and $ \hat{B}_K \equiv B^K_1 $ are given in Table~\ref{tab:bagLRK}. As can be seen from 
Table~\ref{tab:bagLRK} the lattice results for $B_4^K$ for the three different 
collaborations show sizable deviations. 
The  SWME \cite{Jang:2015sla} and the preliminary RBC-UKQCD  
\cite{Hudspith:2015wev} ones which use the intermediate RI-SMOM schemes  are 
consistent with each other, but significantly different from those using the 
intermediate RI-MOM scheme \cite{Carrasco:2015pra,Boyle:2012qb,Bertone:2012cu}. 
The source of discrepancy
seems to be the different intermediate renormalization scheme used to 
match from the matrix elements to those in the continuum $\overline{MS}$
scheme, see \cite{Jang:2015sla}, and more discussions
can be found in \cite{Aoki:2016frl}. Without further understanding of the source of uncertainty, it is premature then to consider their average.

For definiteness, the results we show in the global fit take into account only the values indicated in boldface numbers, except when otherwise stated. For $ B^K_4 $ the value calculated with $ 2+1+1 $ dynamical flavours on the lattice, Ref.~\cite{Carrasco:2015pra} (but we will also comment on the larger value given in Ref.~\cite{Jang:2015sla}). The computation done by this reference considered the matrix element of local, dimension 6 operators. Therefore, dimension eight operators where the charm quark is exchanged in a loop, as in Figure~\ref{fig:illlustrationLocals}, are not considered. In this way, we can employ directly the short-distance QCD corrections $ \bar{\eta}^{(LR)}_{cc} \vert_{EFT} $ that we have computed in an effective theory described by local dimension 6 operators. The combination of the different values will be considered in a more complete analysis \cite{Bernard}.






\begin{table}[!t]
\begin{center}
\begin{tabular}{|c|c |c |c |}
\hline
&$N_f$ &$\mu_h$ [GeV]  &   \\
\hline
&2+1+1 \cite{Carrasco:2015pra} &3 & $ \mathbf{0.78(4)(3)} $\\
&2+1 \cite{Hudspith:2015wev} &3 &$0.92(2)$ \\
$B_4^K$&2+1 \cite{Jang:2015sla}& 3 &$0.981(3)(62)$\\
&2+1 \cite{Boyle:2012qb} &3 & $ 0.69(7)$\\
&2+1  \cite{Bertone:2012cu} &3 & $ 0.76(3)$\\
&2 \cite{Bertone:2012cu} &2 & $ 0.78(3)$\\
\hline
\hline
&2+1 \cite{Bazavov:2016nty} & 4.18 & $ 1.040(75)(45) $\\
$B_4^d$&2 \cite{Carrasco:2013zta} & 4.29 &$\mathbf{0.95(4)(3)}$\\
&quenched \cite{Becirevic:2001xt} & 4.6  &$1.15(3)(^{+5}_{-7})$\\
\hline
\hline
&2+1 \cite{Bazavov:2016nty} & 4.18 & $ 1.022(57)(34) $\\
$B_4^s$&2 \cite{Carrasco:2013zta} & 4.29 &$\mathbf{0.93(4)(1)}$\\
&quenched \cite{Becirevic:2001xt} & 4.6  &$1.16(2)(^{+5}_{-7})$\\
\hline
\end{tabular}
\end{center}
\caption{\it Values of the bag parameters $B_4$ in the $\overline{MS}$
scheme for various numbers of dynamical quarks. Various intermediate renormalization
scheme are used to match from the matrix elements obtained on the lattice to the 
continuum ones. When two uncertainties are given, the first corresponds to a statistical uncertainty and the second to a theoretical one. Note that among the references given here only \cite{Carrasco:2015pra}, \cite{Jang:2015sla} and \cite{Bertone:2012cu} meet the required quality criteria of the last update of the Flag working group
\cite{Aoki:2016frl} for $ K $ meson-mixing. In boldface, we indicate the values we use in our analysis.}\label{tab:bagLRK}
\end{table}


\section{Perturbativity of the potential parameters}\label{sec:perturbativeRequirements}


We will consider in the global fit of LR Model parameters the impact of perturbativity bounds on the Higgs potential parameters, which we now detail. The Higgs mass spectrum depends on three parameters from the scalar potential

\begin{equation}
\alpha_{34} = \alpha_{3} - \alpha_{4} \, , \quad \rho = \rho_{3} / 2 - \rho_{1} \, , \quad x = \mu'_1 / \mu'_2 \, ,
\end{equation}
as seen from Appendix~\ref{sec:spectrumScalars}. Conversely, we can express $ \alpha_{34} $ and $ \rho $ as functions of the masses

\begin{equation}\label{eq:rhoPotential}
\rho = \frac{M_{H_1}^2}{\kappa_R^2} \frac{\left(1+\delta^2\right) (1-X)}{2 \left(1+w^2 \beta (x)\right)} \geqslant 0
\end{equation}
(one can verify that $ 0 \leqslant X \leqslant 1 $), and

\begin{equation}\label{eq:alpha34Potential}
\frac{\alpha_{34}}{\rho} = \frac{4 \nu (x)  \left(x^2 \left(r^2
    X-1\right)+r^2-X\right)+2 \left(r^2-1\right) (X+1) (r
    x+1)}{\left(r^2+1\right) (X-1) (r x+1)} \, ,
\end{equation}
where $ X, \nu (x), \beta (x) $ are found in Eqs.~\eqref{eq:Xnubetak2defs}. Note that from the stability conditions of the vacuum, Appendix~\ref{sec:stableConds}, $ \rho $ and $ \mu'_2 $  are related by the following expression:

\begin{equation}\label{eq:mup2potential}
\mu'_2 / \kappa_R = - \frac{\sqrt{2} \rho w}{1 + rx} \, .
\end{equation}




One does not expect the parameters from the potential to be too large, otherwise one could approach non-perturbative regimes. Perturbativity bounds then require $ \alpha_{34}, \rho, \mu'_{1,2} / \kappa_R $ to be inferior than $ \mathcal{O} (4 \pi) $. Therefore, we assume in a first exploratory analysis

\begin{eqnarray}
&& \alpha_{34}, \mu'_{1,2} / \kappa_R \in [-10, 10] \, , \\
&& \rho \in [0, 10] \, .
\end{eqnarray}
These requirements translate into bounds for $ M_{H_1} $, $ r $ and $ w $ through Eqs.~\eqref{eq:rhoPotential}, \eqref{eq:alpha34Potential}, \eqref{eq:mup2potential}.





Note that, Refs.~\cite{Senjanovic:1979cta,Olness:1985bg} have derived unitarity bounds relating the masses of the heavy scalars and the new gauge bosons, based on vector boson scattering amplitudes. These bounds were derived in the triplet case, under particular assumptions such as $ g_L = g_R $. Therefore, they do not directly apply in our case, and the translation into our case remains to be worked out \cite{Bernard}.

\section{Global fit}

\subsection{Inputs and ranges of definition}

For completeness, we list the set of EWPO we use in the global fit:

\begin{eqnarray}
&& M^{light}_H \, , \quad m_{top}^{pole} \, , \quad M_Z \, , \quad \alpha_s (M_Z) \, , \\
&& \Gamma_Z \, , \quad \sigma_{had} \, , \nonumber\\ 
&& R_{b, c, e, \mu, \tau} \, , \nonumber\\
&& A^{b, c, e, \mu, \tau}_{FB} \, , \nonumber\\
&& A_{b, c} \, , \quad A^{SLD}_{e, \mu, \tau} \, , \quad A_{e, \tau} (P_{\tau}) \, , \nonumber\\
&& M_W \, , \quad \Gamma_W \, , \nonumber\\
&& Q_W (Cs) \, , \quad Q_W (Tl) \, , \nonumber
\end{eqnarray}
whose inputs are found in Table~\ref{tab:tab2}. As indicated in the same table and discussed in Chapter~\ref{ch:EWPO}, we also take into account the following direct bound on the mass of the $ W' $ boson

\begin{equation}\label{eq:boundmWR2}
M_{W'} > 2 \, {\rm TeV} \, .
\end{equation}
Together with meson-mixing observables $ \epsilon_K, \Delta m_{d,s} $, this is the full set of observables we consider.

Concerning the ranges of definition of the parameters of the fit, as we have observed in Section~\ref{sec:EWSBSMLRM} we choose as a simplification the VEVs $ \kappa_2, \kappa_L $ to be positive, by setting their complex phases $ \alpha, \theta_L $ to $ 0 \, {\rm mod} \, 2 \pi $. Therefore, we have allowed $ r $ to vary over the range $ [0, 1] $ like in Chapter~\ref{ch:EWPO}. On the other hand, $ w $ is allowed to be larger than one. Having values of $ w $ larger than one implies that the EWSB scale $ \sqrt{\kappa_1^2 + \kappa_2^2 + \kappa_L^2} $, up to $ \epsilon^2 $ corrections, is dominated by $ \kappa_L $. Now, once the mass of the top is given by $ \kappa_1 \lambda_1 + \kappa_2 \lambda_2 $, where $ \lambda_{1,2} $ come from the Yukawa coupling matrices, suppressing $ \kappa_{1,2} $ may push $ \lambda_{1,2} $ to non-perturbative values. We have therefore considered varying $ w $ over the rather conservative range $ [0, 3] $. Therefore


\begin{equation}
r \in [0, 1] \, , \qquad w \in [0, 3] \, .
\end{equation}

We also include the perturbativity bounds on $ c_\phi $ derived from $ g_R^2 , g_{B-L}^2 < 4 \pi $, thus implying

\begin{equation}
\vert c_\phi \vert \in [0.1, 0.99] \, .
\end{equation}

Shifting to the scalar sector, particles are labeled in such a way that $ \delta \geqslant 1 $, and $ M_{H_1} $, which is the mass of the scalars $ H^0_1, A^0_1, H^\pm_1 $ (degenerate up to $ \mathcal{O} (\epsilon^2) $ corrections), will be always smaller than $ M_{H_2} $, the masses of $ H^0_2, A^0_2, H^\pm_2 $ (also degenerate up to $ \mathcal{O} (\epsilon^2) $ corrections). Note that the inclusion of the sector $ H^0_2, A^0_2, H^\pm_2 $ in the analysis of meson-mixing observables, which decouples in the limiting case $ w = 0 $, is a novelty of the analysis shown in here.



\subsection{Right-handed mixing matrix}

We have argued in Section~\ref{sec:VRStructure} that one may expect the structure of $ V^R $ to be very different from the structure of the left-handed mixing matrix $ V^L $. However, we study here the most simple and constrained case where $ V^R = V^L $. A more general analysis will be the object of future work, where (semi-)leptonic decays will be integrated in the global fit in order to better constrain different structures of $ V^R $ \cite{Bernard}.

The matrix $ V^L $ is equal to the CKM matrix of the SM up to $ \mathcal{O} (\epsilon^2) $ corrections. Therefore, $ V^L $ has the same hierarchical structure of the CKM matrix, and can be similarly parameterized by Wolfestein parameters $ \tilde{A}, \tilde{\lambda}, \tilde{\rho}, \tilde{\eta} $. Lacking of a more complete analysis, we are going to set the Wolfenstein parameters to the ones found for the CKM matrix extracted in the SM framework. Their values are given in Eq.~\eqref{eq:resultsCKMparams}, and in the following we allow $ \tilde{A}, \tilde{\lambda}, \tilde{\rho}, \tilde{\eta} $ to vary over the $ 1 \times \sigma $ intervals we have derived before for the CKM matrix, namely

\begin{eqnarray}
\tilde{A} = 0.819 \pm 0.010 \, , & \qquad &
\tilde{\lambda} = 0.22549 \pm 0.00037 \, , \\
\tilde{\bar \rho} = 0.154 \pm 0.009 \, , & \qquad &
\tilde{\bar \eta} = 0.3535 \pm 0.0075 \, , \nonumber
\end{eqnarray}
which were symmetrized compared to Eq.~\eqref{eq:resultsCKMparams}. (We have considered varying the sizes of the intervals shown above without much effect on the results from the global fit.)

\subsection{Results}

We combine the observables using \texttt{CKMfitter}, employing the Rfit scheme for treating theoretical uncertainties. The best fit point gives a $ \chi^2_{min} $ of $ 22.24 $ and a p-value of $ \sim 27~\% $, for $ 19 $ degrees of freedom (though some of the parameters are not constrained by the fit, and the real number of degrees of freedom can be actually lower). 


At the best fit point we have

\begin{eqnarray}
&& c_\phi = 0.78 \, , \quad
\epsilon = 0.0007 \, , \quad
r = 0.33 \, , \quad
w = 2.4 \, , \nonumber\\
&& M_{H_1} = 1.3 \, \cdot \, 10^{3} \; {\rm TeV} \, , \quad
\delta = 1.2 \, , \quad
x = 0.19 \, , \\
&& \alpha_{34} = 4.6 \, , \quad
\rho = 2.5 \, , \quad
\mu'_1 = -1.5 \, , \quad
\mu'_2 = -7.7 \, . \nonumber
\end{eqnarray}
Table~\ref{tab:individualContrs} shows the individual impacts coming from the different contributions discussed in Chapter~\ref{ch:generalEFT} at the best fit point. In all the cases, contributions to the SM are destructive. Note that $ W H $ contributions are far too small at the best fit point, while the tree level diagram $ H $ dominates, but $ W W' $ also gives sizable contributions. 

\begin{table}
\centering
\begin{tabular}{c|c|c|c}
& $ | \epsilon_K | \cdot 10^3 $ & $ \Delta m_d \; ({\rm ps}^{-1}) $ & $ \Delta m_s \; ({\rm ps}^{-1}) $ \\
\hline
SM: $ W W $ box &	2.305	&	0.5102	&	17.764	\\
\hline
$ W W' $ &	-0.021	&	$ -4 \, \cdot \, 10^{-5} $	&	$ -0.001 $ \\
\hline
$ W H $ box &	-0.002	&	$ -10^{-6} $	&	$ -10^{-4} $ \\
\hline
$ H $ tree &	-0.054	&	-0.0002	&	-0.005 \\
\hline
All &	2.228	&	0.510	&	17.757 \\
\end{tabular}
\caption{\it Numerical impact of the different LR Model contributions to $ | \epsilon_K | $ and $ \Delta m_{d,s} $ at the best fit point. Above, $ W W' $ refers to the full set of contributions which form a gauge invariant set.}\label{tab:individualContrs}
\end{table}

We now discuss their confidence level intervals and correlations.

\subsubsection{One-dimensional constraints}

The outcome of the global fit does not show any preference at $ 1 \, \sigma $ for the values of the parameters $ r, w, c_\phi, x, \alpha_{34}, \rho, \mu'_{1,2} $. On the other hand, we derive the following bounds for $ M_{H_1} $, $ \epsilon $, $ \delta $ at $ 68~\% $ CL ($ 95~\% $ CL)

\begin{eqnarray}\label{eq:results1}
M_{H_1} &>& 28.1 \, {\rm TeV} \qquad (26.6 \, {\rm TeV}) \, , \\
\epsilon &<& 0.0131 \qquad\quad\;\; (0.0135) \, , \\
\delta &<& 33.7 \qquad\quad\quad\quad\; (34.8) \, .
\end{eqnarray}
The upper bound on $ \delta $ is due to perturbativity requirements of Section~\ref{sec:perturbativeRequirements}, which limit the scale of $ M_{H_2} = \delta M_{H_1} $ compared to $ M_{H_1}, M_{W'} $.

Considering Ref.~\cite{Jang:2015sla} instead of Ref.~\cite{Carrasco:2015pra} for the value of $ B^K_4 $ shifts the results for $ M_{H_{1,2}}, 1/\epsilon $ upwards as expected due to the higher magnitude of $ B^4_K $ obtained in \cite{Jang:2015sla}

\begin{eqnarray}\label{eq:results2}
M_{H_1} &>& 31.5 \, {\rm TeV} \qquad (30.1 \, {\rm TeV}) \, , \\
\epsilon &<& 0.0116 \qquad\quad\;\; (0.0120) \, , \\
\delta &<& 29.9 \qquad\quad\quad\quad\; (30.2) \, .
\end{eqnarray}

\begin{figure}
	\centering
	\includegraphics[scale=0.35]{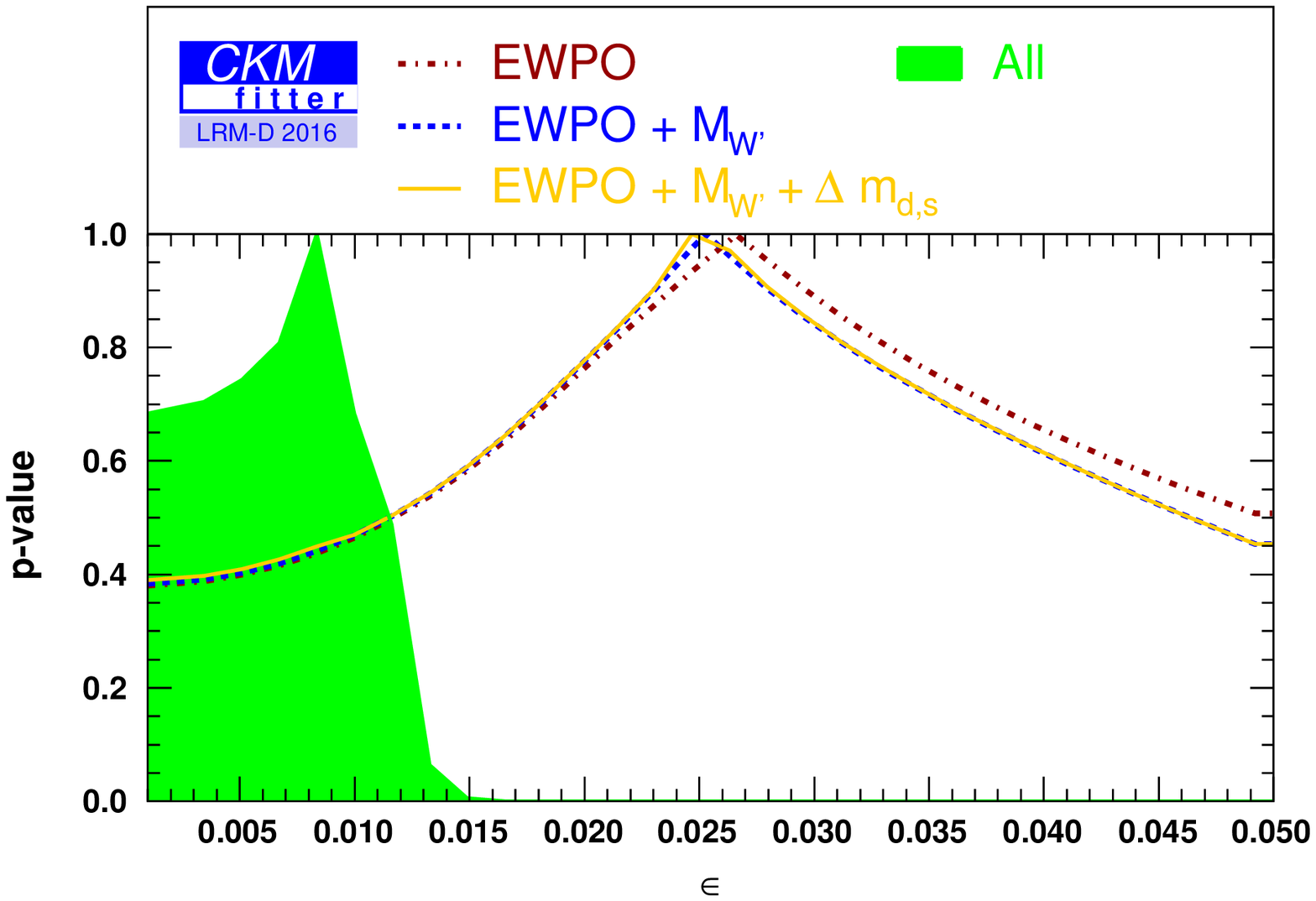}
	\includegraphics[scale=0.35]{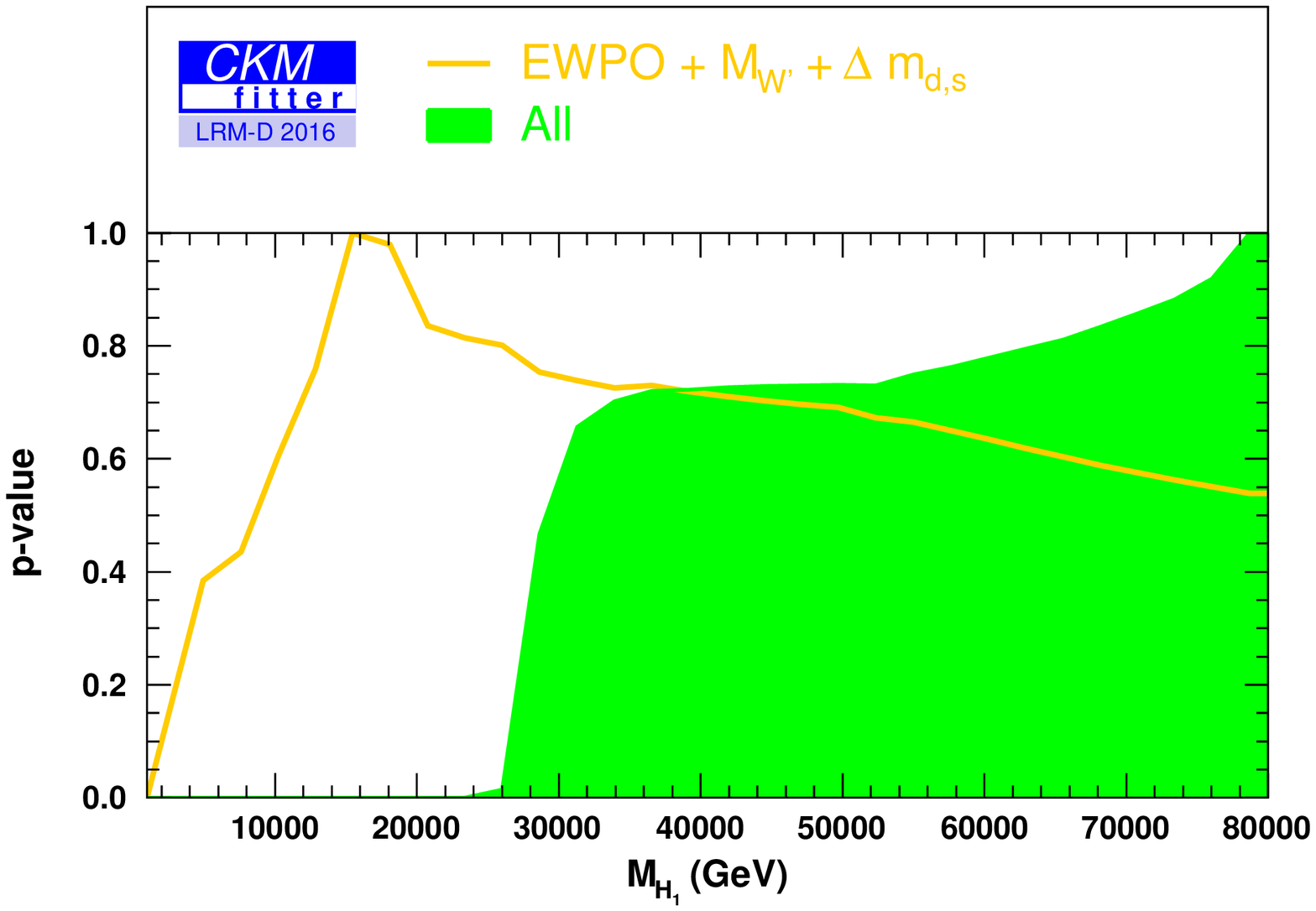}
	\caption{\it (Left) Impact on the LR Model parameter $ \epsilon $ of EWPO, the bound on the mass of the $ W' $ resulting from direct searches, and the meson-mixing observables $ \Delta m_{d,s} $ and $ \epsilon_K $. The full set of the constraints is indicated by the label ``All." (Right) Impact of $ \Delta m_{d,s} $ and $ \epsilon_K $ on the lower bound of the Higgs mass $ M_{H_1} $.}\label{fig:setOfConstraints}
\end{figure}

The individual bounds for $ \epsilon $ and $ M_{H_1} $ are indicated in Figure~\ref{fig:setOfConstraints}. Note that the constraint from $ \vert \epsilon_K \vert $ plays a particularly important role when added to the fit. Since only the imaginary part of the LR contribution enters in the expression of $ \vert \epsilon_K \vert $, it will be certainly important to reconsider the global fit if $ V^R $ is allowed to be different from $ V^L $. In particular, in the case where $ V^R = (V^L)^* $ there is no contribution from the LR Model to $ \vert \epsilon_K \vert $, as discussed in \cite{Maiezza:2010ic}. Also note that the constraint on $ M_{H_1} $ is significantly stronger than the constraint found in literature in Ref.~\cite{Blanke:2011ry}, which is due to a non-manifest structure of the $ V^R $ mixing-matrix in the latter reference.

\begin{figure}
\centering
	\includegraphics[scale=0.35]{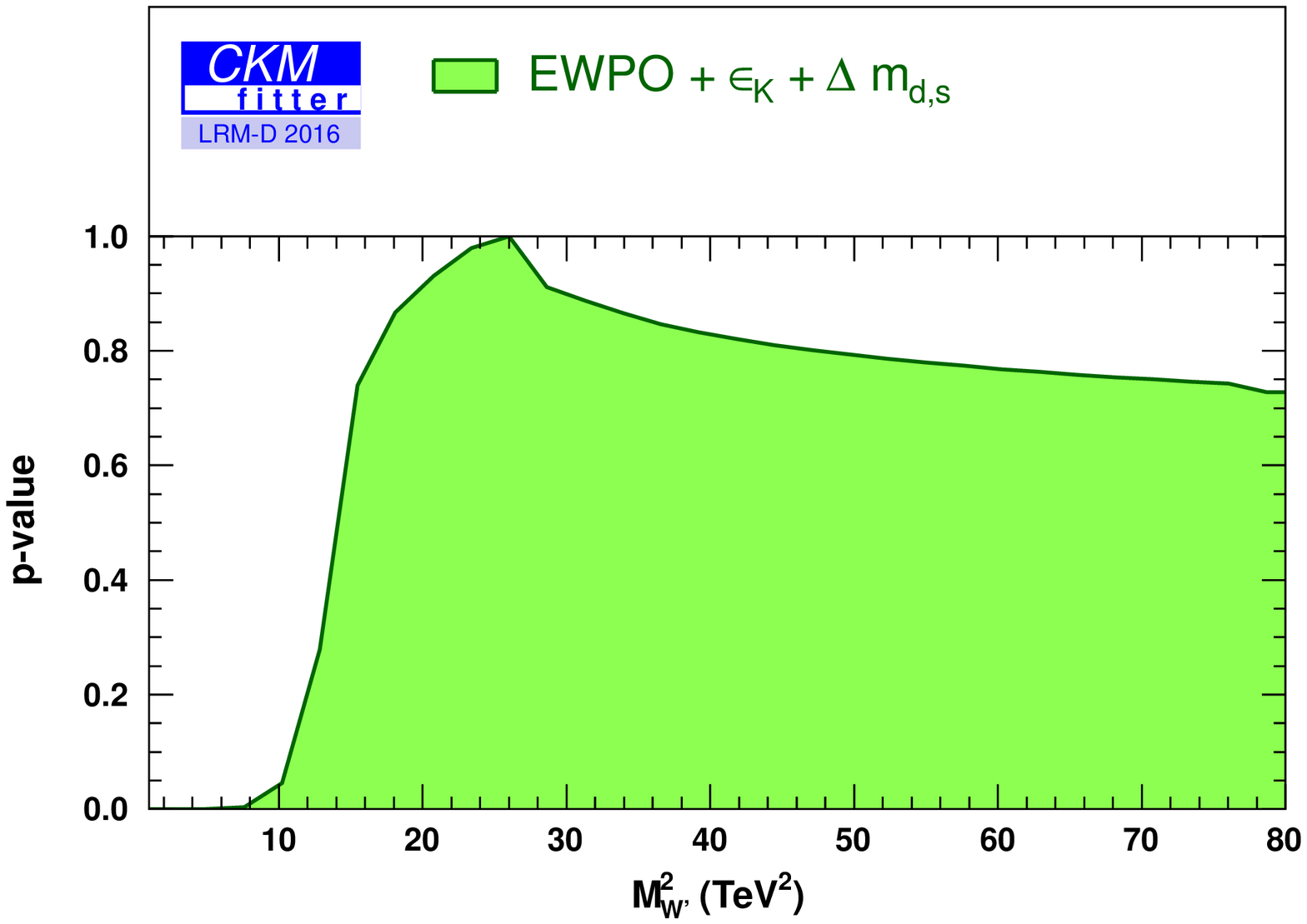}
	\includegraphics[scale=0.35]{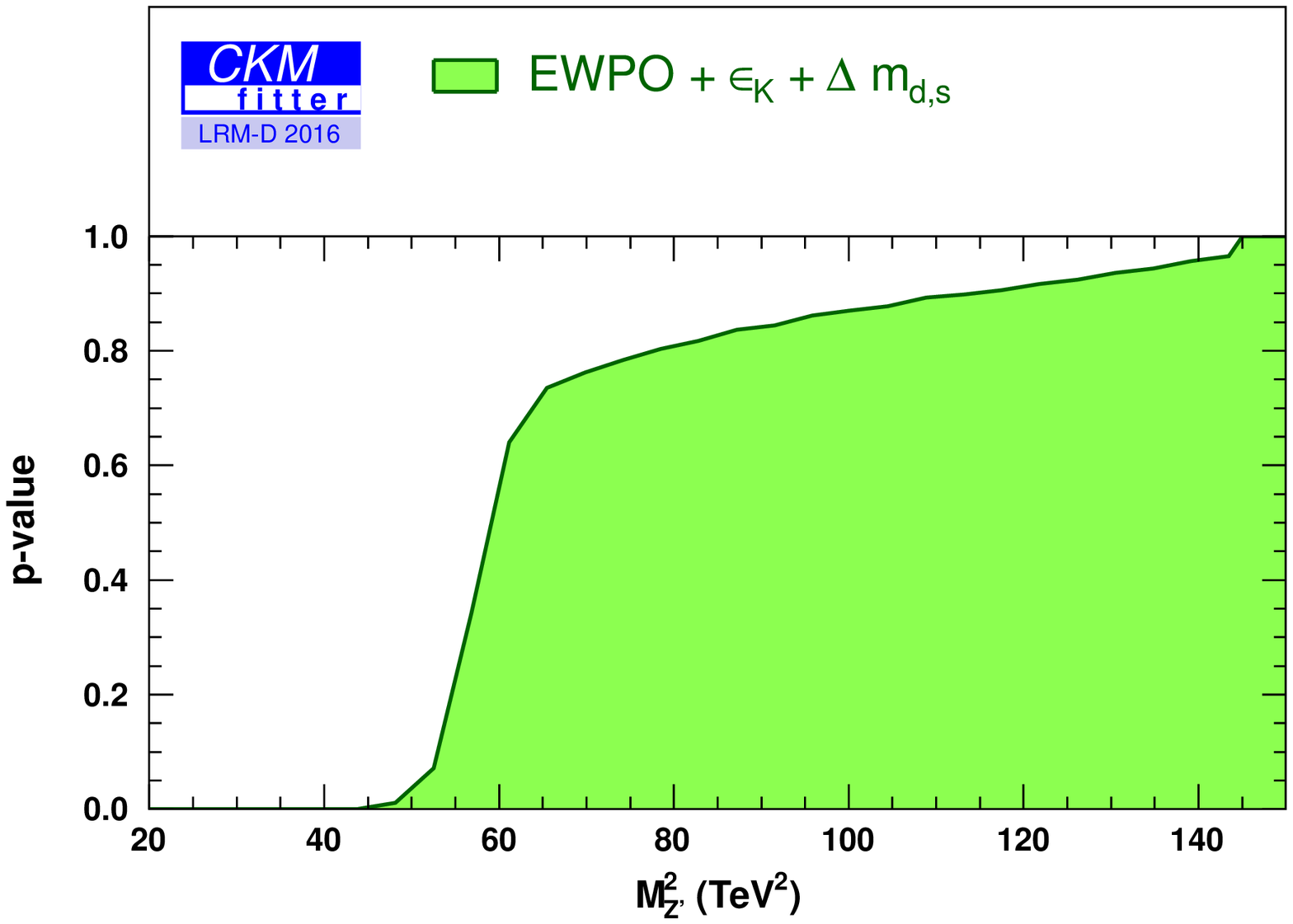}
	\caption{\it Indirect bounds on the masses of the $ W', Z' $ coming from EWPO and meson-mixing observables (masses are squared).}\label{fig:MWR2}
\end{figure}

Based on the constraints from EWPO and meson-mixing, it is also possible to set stronger bounds on the mass of the $ W' $ compared to Eq.~\eqref{eq:boundmWR2}:

\begin{eqnarray}\label{eq:results3}
M_{W'} > 3.6 \; {\rm TeV} \quad (3.2 \; {\rm TeV}) \, , \quad\quad {\rm and} \quad M_{Z'} > 7.5 \; {\rm TeV} \quad (7.2 \; {\rm TeV})
\end{eqnarray}
at $ 68~\% $ CL ($ 95~\% $ CL), cf. Figure~\ref{fig:MWR2}, while for the value of $ B^K_4 $ given in Ref.~\cite{Jang:2015sla} we have

\begin{eqnarray}\label{eq:results4}
M_{W'} > 4.0 \; {\rm TeV} \quad (3.7 \; {\rm TeV}) \, , \quad\quad {\rm and} \quad M_{Z'} > 8.5 \; {\rm TeV} \quad (8.2 \; {\rm TeV}) \, .
\end{eqnarray}


\subsubsection{Correlations among the parameters}

\begin{figure}
	\centering
	\includegraphics[scale=0.35]{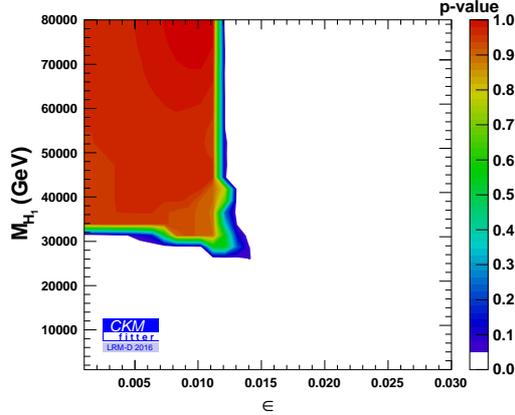}
	\caption{\it Correlation among $ \epsilon = \sqrt{\kappa^2_1 + \kappa^2_2 + \kappa^2_L} / \kappa_R $ and the Higgs mass $ M_{H_{1}} $.}\label{fig:epsilonMH1}
\end{figure}

\begin{figure}
	\centering
	\includegraphics[scale=0.35]{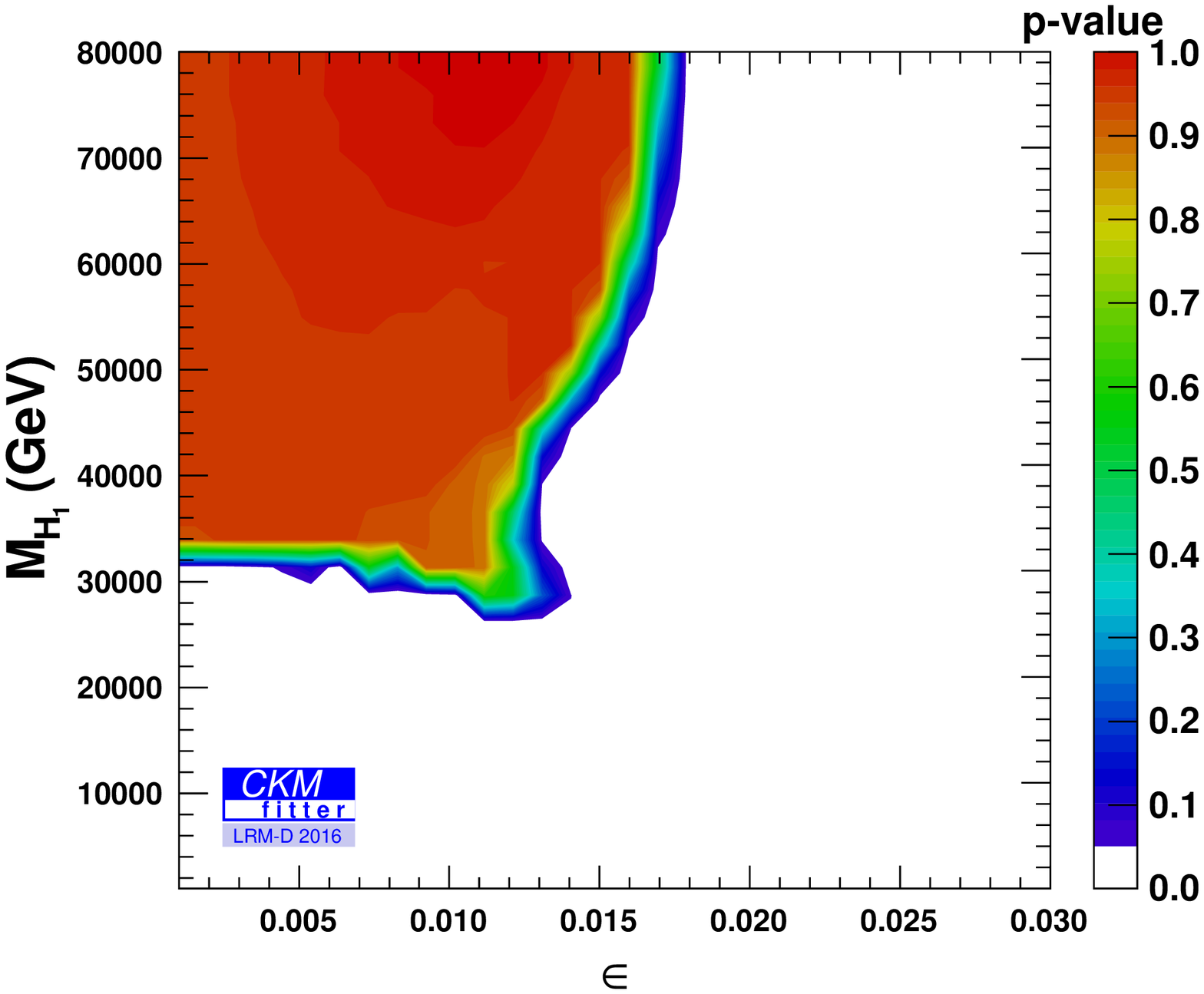}
	\includegraphics[scale=0.35]{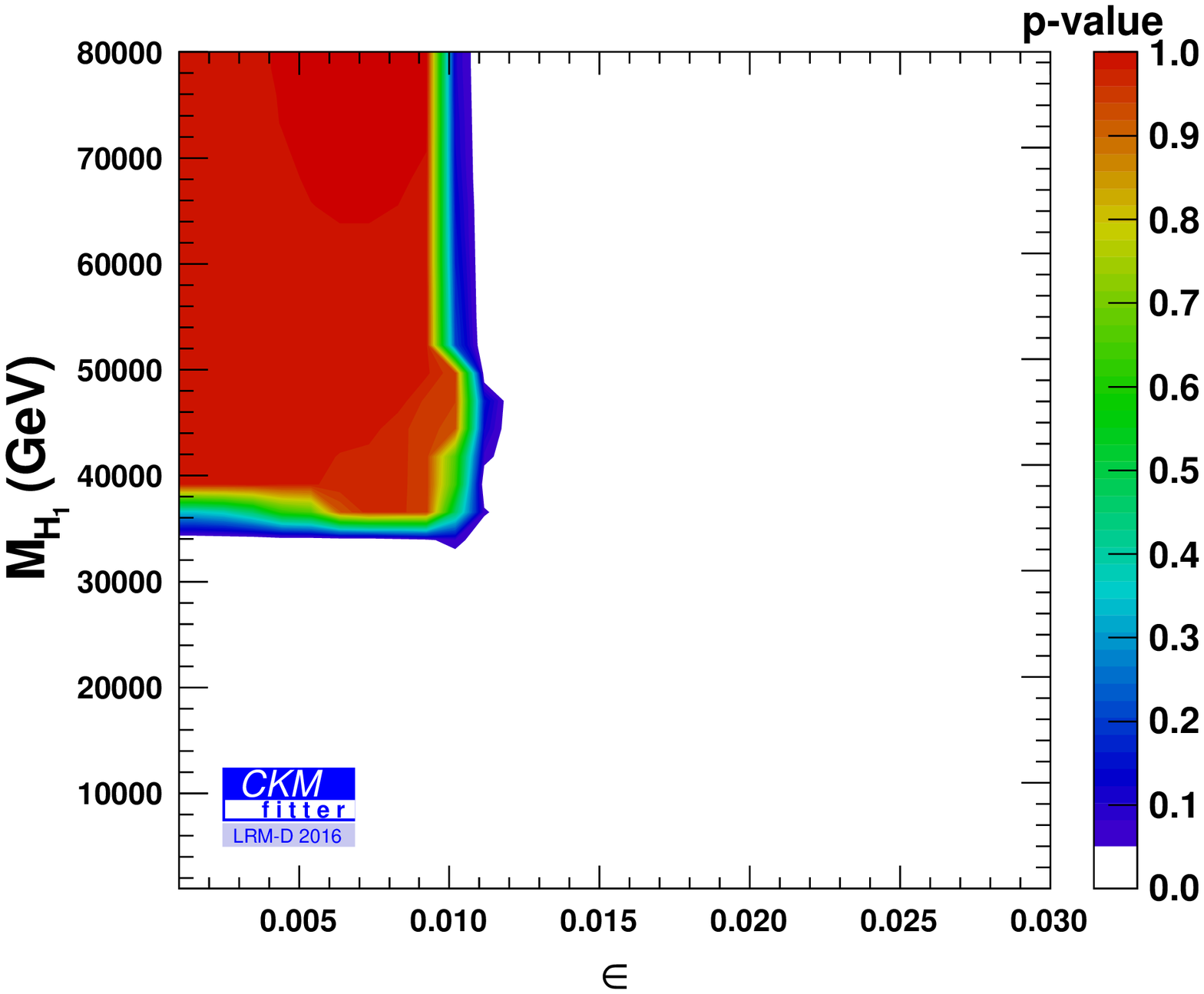}
	\caption{\it (Left) The same fit as in Figure~\ref{fig:epsilonMH1} is considered except that perturbative bounds on $ \alpha_{34}, \rho $ and $ \mu'_{1,2}/\kappa_R $ are not required. (Right) The errors of the short-distance QCD corrections calculated in Chapter~\ref{ch:technicalEFT} are considered to be four times smaller.}\label{fig:epsilonMH1pertAndSD}
\end{figure}

In Figure~\ref{fig:epsilonMH1} we show the effect of the different classes of constraints on the plan $ ( \epsilon, M_{H_1} ) $. It is therefore clear that the scenario $ V^R= V^L $ excludes a large region of the parameter space. Note that there is some impact on the same plan $ ( \epsilon, M_{H_1} ) $ from perturbativity bounds: in Figure \ref{fig:epsilonMH1pertAndSD} (Left), we consider a global fit without the effects of the bounds on $ \alpha_{34}, \rho $ and $ \mu'_{1,2} / \kappa_R $. Indeed, $ \alpha_{34} $ and $ \mu'_2 / \kappa_R $ are both proportional to $ M_{H_1}^2 \times \epsilon^2 $, thus cutting off simultaneous large values of $ M_{H_1} $ and $ \epsilon $.

The effect of the size of the uncertainties coming from the short-distance QCD corrections calculated in Chapter~\ref{ch:technicalEFT} is seen in Figure \ref{fig:epsilonMH1pertAndSD} (Right), showing a reduction of the allowed region for smaller uncertainties when they are all divided by a factor of four. This therefore illustrates the need for precise short-distance QCD correction calculations, thus further justifying the attention we have dedicated to them in Chapters~\ref{ch:generalEFT} and \ref{ch:technicalEFT}.


\begin{figure}
	\centering
	\includegraphics[scale=0.35]{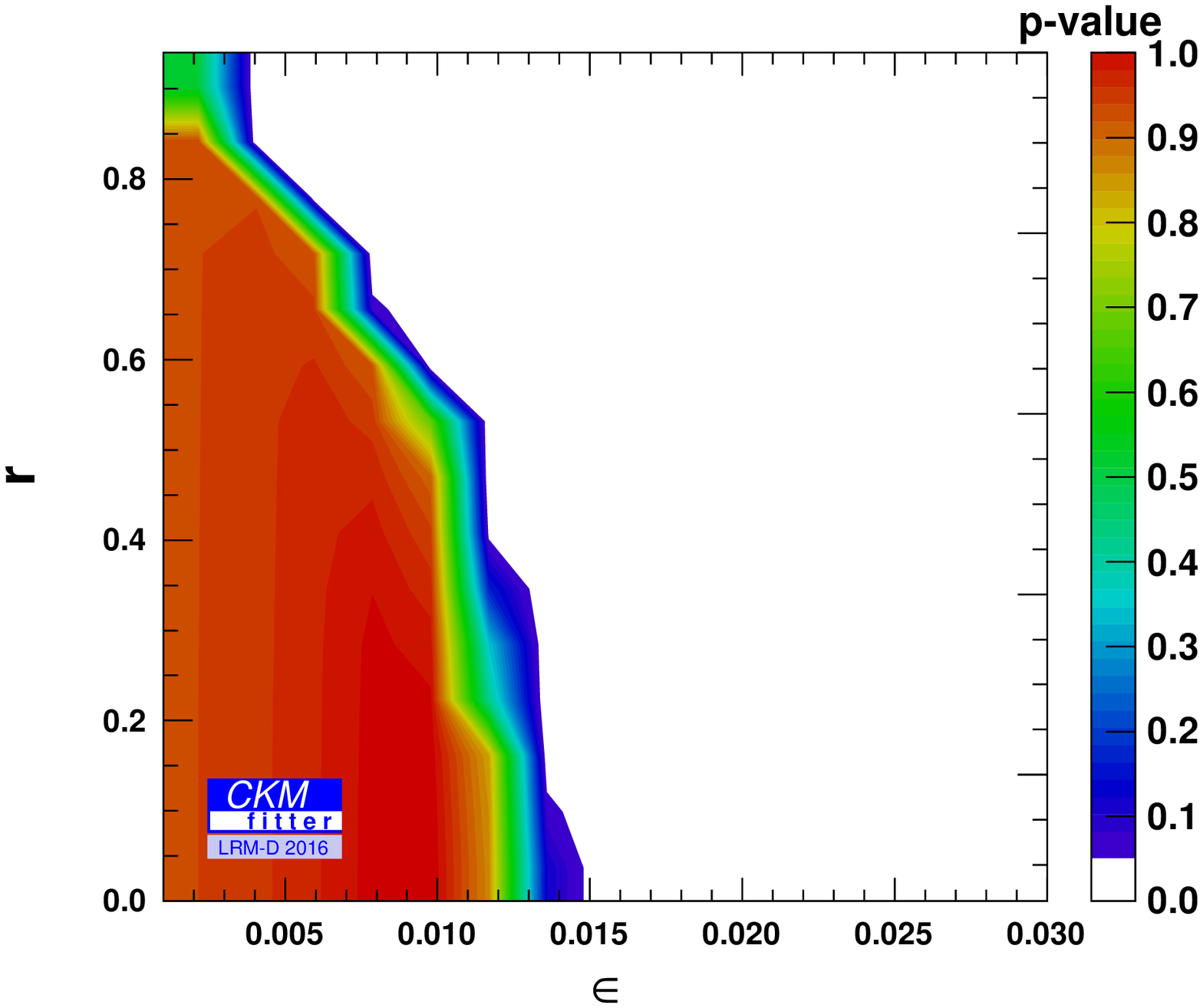}
	\includegraphics[scale=0.35]{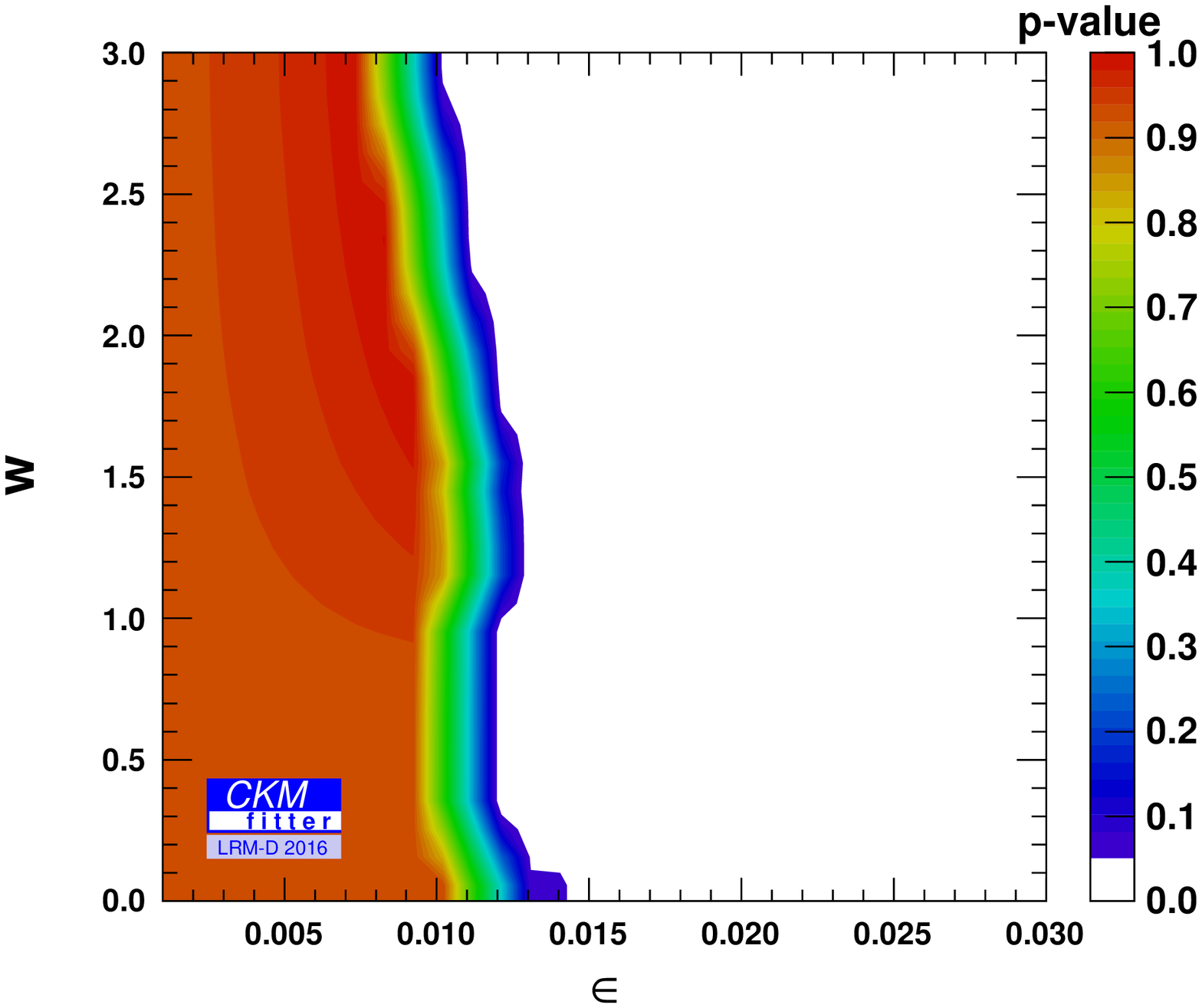}
	\caption{\it (Left) Correlation among $ \epsilon $ and $ r $. (Right) Correlation among $ \epsilon $ and $ w $.}\label{fig:correlationsWepsilon}
\end{figure}

\begin{figure}
	\centering
	\includegraphics[scale=0.35]{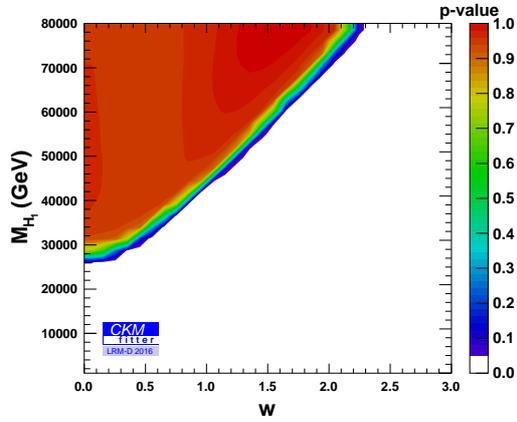}
	\caption{\it Correlation among $ M_{H_{1}} $ and $ w $.}\label{fig:correlationsWMH1}
\end{figure}

\begin{figure}
	\centering
	\includegraphics[scale=0.35]{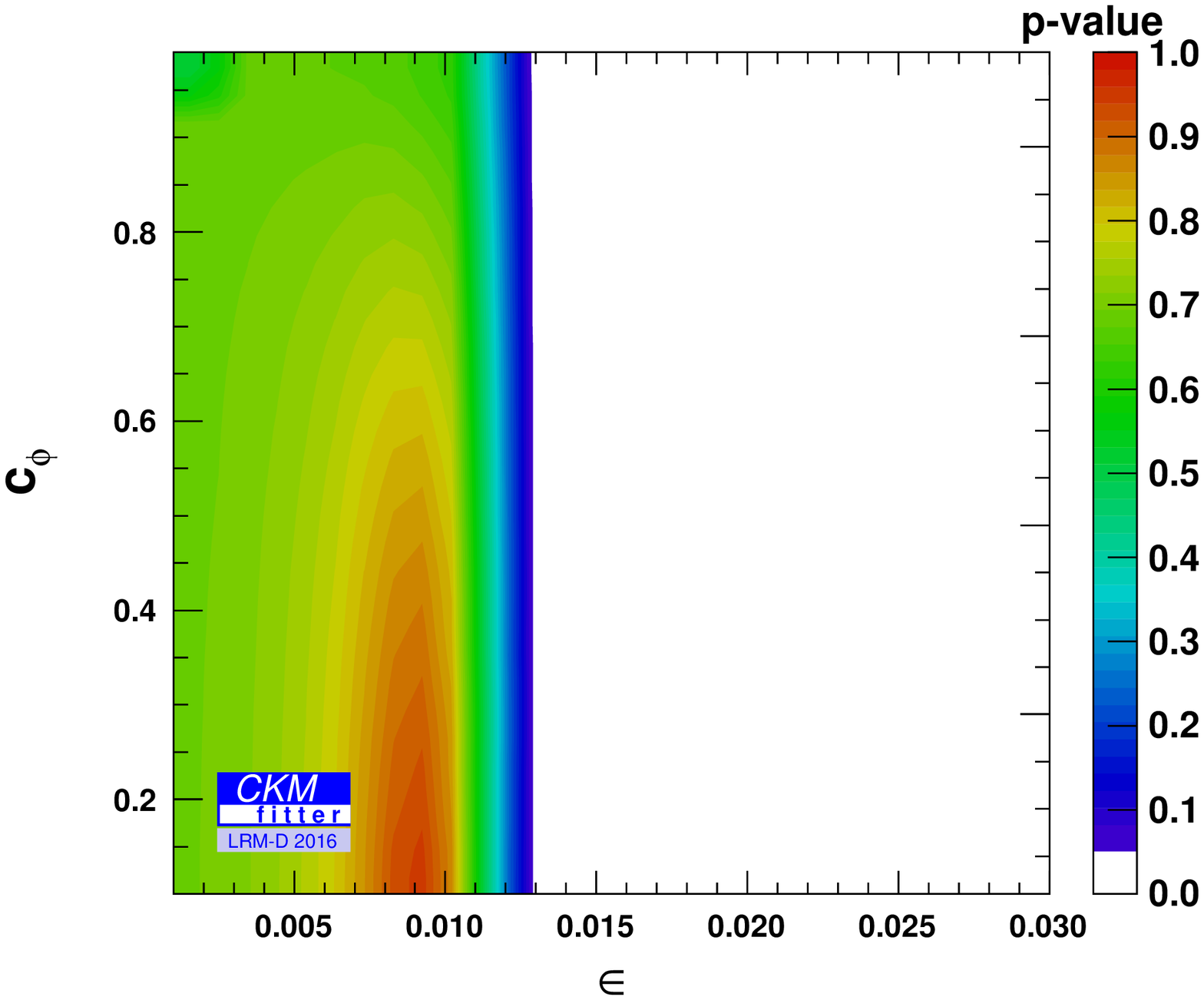}
	\includegraphics[scale=0.35]{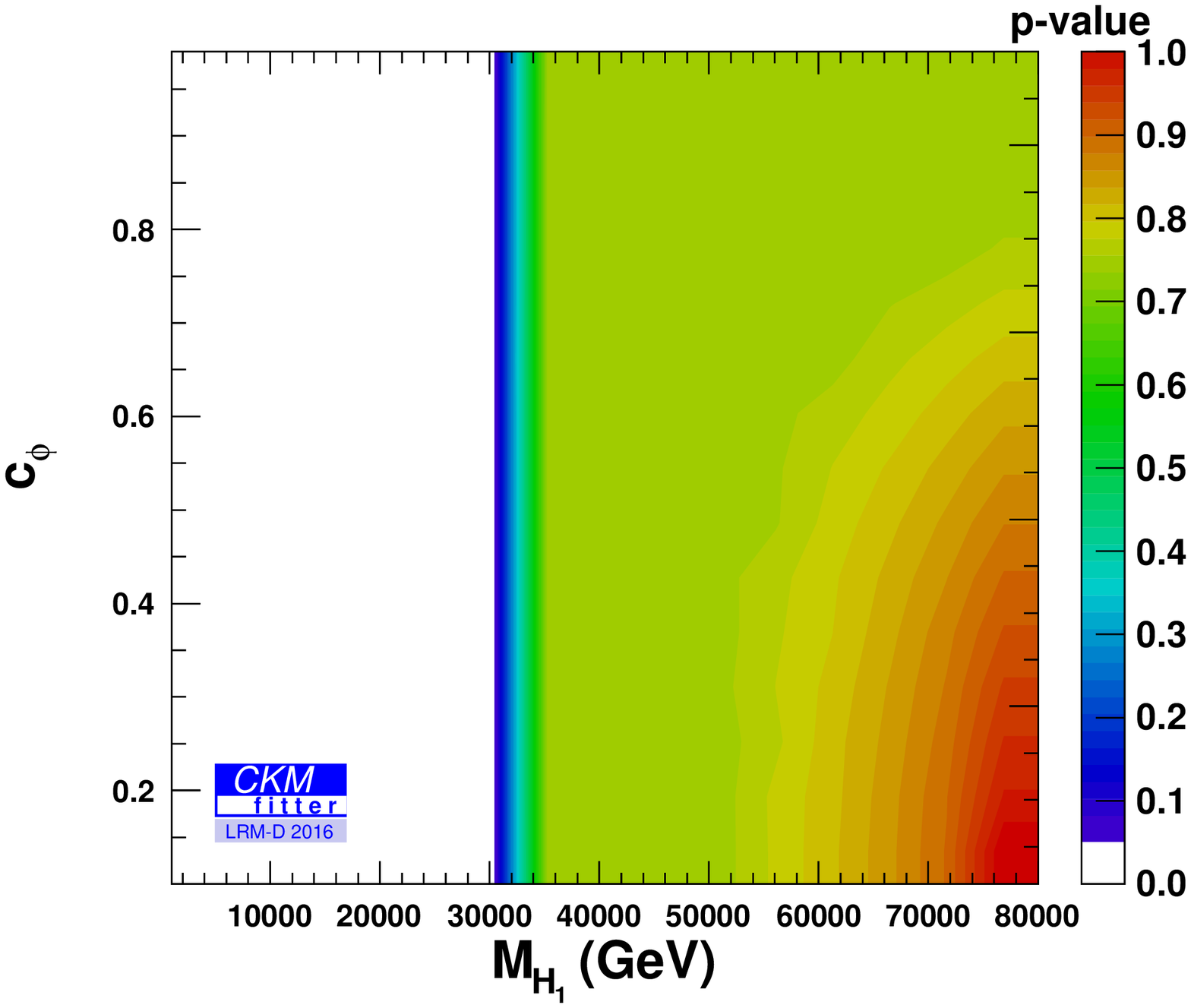}
	\caption{\it (Left) Correlation among $ c_\phi $ and $ \epsilon $. (Right) Correlation among $ c_\phi $ and $ M_{H_1} $. In both two plots, the p-value surface has been calculated by considering 1 degree of freedom, since there is essentially no correlation among the quantities, as the figures show.}\label{fig:correlationsWithcR}
\end{figure}

Though no bounds are set on $ r, w $ and $ c_\phi $, we can still have access to their correlations. In Figure~\ref{fig:correlationsWepsilon} we show the correlation of $ \epsilon $ with $ r, w $. We see that larger values of $ \epsilon $ require smaller values of $ r $. Then, in Figure~\ref{fig:correlationsWMH1} we show the correlation of $ M_{H_1} $ with $ w $. We see that smaller values of $ M_{H_1} $ require $ w $ small. Concerning correlations with $ c_\phi $, we do not observe any correlation in the plan $ ( \epsilon, c_\phi ) $ or $ ( M_{H_1}, c_\phi ) $, see Figure~\ref{fig:correlationsWithcR}. 




\section{Conclusion}



We have considered constraints on the parameters of the LR Models, including the masses of the extended Higgs sector, based on EWPO, direct searches for the $ W' $ and meson-mixing observables. We have studied here the simplest and most constrained possible structure of the mixing matrix in the right-handed sector, namely $ V^R = V^L $, called manifest scenario. We have therefore derived the bounds seen in Eq.~\eqref{eq:results1}-\eqref{eq:results4} in this specific case, showing that the indirect bounds on the new gauge bosons and new scalars are stronger than the bounds from collider physics.


We have not been able to constrain the other parameters of the LR Model. Therefore, the precise way in which the LR gauge symmetries are broken still offers different pictures, claiming for the inclusion of more observables. On the other hand, we were able to extract their correlations with other quantities, thus concluding in particular that large values for $ r $ ($ w $) do not favor large (small) values for $ \epsilon $ ($ M_{H_1} $). 

\chapter{Modeling theoretical uncertainties}\label{ch:theo}


When dealing with experimental results, one is usually interested in extracting some information about the underlying theory in charge of explaining them. From both experimental and theoretical sides, systematic or theoretical uncertainties may be present, which difficults a straightforward comparison between experiment and theory. From a flavour perspective, the problem of dealing with theoretical uncertainties is of central importance, since in the flavour global fit discussed in Chapter~\ref{ch:SM} many observables suffer from large theoretical uncertainties.

Our goal here is to compare the properties of the Rfit approach, which we have employed in Chapters~\ref{ch:SM}, \ref{ch:EWPO} and \ref{ch:PHENO}, and has shown good properties in the context of flavour physics \cite{CKMfitterStandard,Charles:2015gya}, with other possible models of theoretical uncertainty. The Rfit scheme was developed to deal with the vary large uncertainties one had when Lattice QCD results had not achieved the degree of accuracy they have nowadays. At the present stage of development in particle physics, we would like to know if other procedures offer interesting alternatives.



The problem of extracting the fundamental parameters of a model (given certain hypotheses) is going to be discussed in detail in Section \ref{sec:toolbox}. After we make clear the way we understand on general grounds the inference of the values of the fundamental parameters of a model, we move to a more original discussion and include in Section \ref{sec:theoProblem} theoretical uncertainties. This class of uncertainties does not fit straightforwardly in the framework and we will face the problem concerning its interpretation. We will therefore propose different possibilities for modeling theoretical uncertainties: a common method called \textit{random} approach, the \textit{Rfit} scheme used by the \CKM $ \, $ Collaboration, a method called \textit{external} approach (close to what experimentalists do when dealing with systematic uncertainties), and different \textit{nuisance} approaches (fixed and adaptive ranges). Then, in Sections \ref{sec:lambdaCholesky} and \ref{sec:globalFit} we are going to discuss the problem of combining many measurements under the perspective of these different approaches.


\section{Statement of the problem}

Given a model, such as the SM theory or a NP model, which is formulated in terms of a set of fundamental parameters $ \mu $, suppose we are able to calculate an observable $ x $ as a function of $ \mu $, $ x (\mu) $. By ``fundamental parameters," we refer here to parameters which are free in the framework of the model, and must have their values extracted from the observation of nature.

From the experimental side, we project then an apparatus able to probe the true value $ x_t $ of the observable, i.e. its value in nature. Since repeating the same experiment twice may not imply the same outcome, an experiment designed to measure $ x_t $ may be thought of as a ``generator'' of values of a random variable $ X $

\begin{eqnarray}\label{eq:apparatus0}
&& X \sim g (X ; x_t) \, , \quad {\rm (``\sim " \; means \; ``distributed \; as")} 
\end{eqnarray} 
where the probability of measuring $ X $ in the subspace $ \mathcal{S} \subseteq \mathcal{U} $, $ \mathcal{U} $ containing all possible values, is

\begin{equation}
\int_{\mathcal{S}} dX \, g (X ; x_t) = \mathcal{P} [X \in \mathcal{S} | x_t] \, ,
\end{equation}
where $ g (X ; x_t) $ is normalized so that $ \int_{\mathcal{U}} dX \, g (X ; x_t) = \mathcal{P} [X \in \mathcal{U} | x_t] = 1 $. Interpreted in terms of an underlying model, $ x_t $ is a function of its true fundamental parameters $ \mu_t $, i.e. $ x_t = x (\mu_t) $, and therefore we replace systematically $ g (X ; x_t) \rightarrow g (X ; \mu_t) $.

Now, if the outcome of the experimental analysis gives $ X_0 $, the following result is quoted

\begin{equation}\label{eq:outcome}
X_0 \pm \sigma \, ,
\end{equation}
where $ \sigma $ is the precision or the accuracy of the measurement. However, this is not the full story: the next step is to extract the values of the fundamental parameters $ \mu_t $. To this effect, based on $ X_0 \pm \sigma $ we aim at extracting constraints on the value of $ \mu_t $.

\section{Frequentist toolbox}\label{sec:toolbox}

We introduce in this section a set of concepts which will help us in the extraction of physical parameters and in the comparison of different physical models. We intend to provide the minimal elements necessary for the understanding of what follows: more complete discussions on the subject are found in \cite{Lyons:1986em,James:2006zz,Geyer,Demortier,Feldman:1997qc}. For the time being, we will not be concerned about theoretical uncertainties: the way to include them in our statistical framework will be discussed later in Section \ref{sec:theoProblem}.

The first object we introduce is called \textit{test statistic}, which is a positive definite function of the experimental value $ X_0 $, $ T (X_0 ; \mu) $. It is expected basically to, given the outcome $ X_0 \pm \sigma $, indicate whether it is in ``good" or ``bad" agreement with the hypothesis $ \mathcal{H}_\mu $

\begin{eqnarray}\label{eq:null}
&& \mathcal{H}_\mu : \mu_t = \mu 
\end{eqnarray}
of $ \mu_t $ being given by $ \mu $: the larger the value of $ T (X_0 ; \mu) $, the worse the agreement of data with $ \mathcal{H}_\mu $. Seen as a function of the random variable $ X $, it is a random variable itself whose distribution is given by

\begin{equation}
h (T \vert \mathcal{H}_\mu) = \int dX \, \delta [T - T (X ; \mu)] g (X ; \mu) \, .
\end{equation}
\noindent
From the distribution $ h $, one can compute the probability of finding a value of the test statistic smaller than a given $ T (X_0 ; \mu) $

\begin{eqnarray}
&&\mathcal{P} [T < T (X_0 ; \mu)] = \int^{T (X_0 ; \mu)}_0 dT \, h (T \vert \mathcal{H}_\mu) \, , 
\end{eqnarray}
and then

\begin{equation}
\mathcal{P} [T < T (X_0 ; \mu)] = \int dX \, \theta [T(X_0 ; \mu) - T (X ; \mu)] g (X ; \mu) \, ,
\end{equation}
where $ \theta [y] = 1 $, when $ y > 0 $, and $ \theta [y] = 0 $, when $ y \leq 0 $.

The criterion we use to distinguish a good from a bad agreement is based on the p-value defined from the test statistic as

\begin{equation}\label{eq:pvalueFirstDef}
1 - p (X_0 ; \mu) = \mathcal{P} [T < T (X_0 ; \mu)] \, , \qquad p (X_0 ; \mu) = \mathcal{P} [T \geq T (X_0 ; \mu)] \, ,
\end{equation}
as we will see in detail in Section \ref{sec:pvalueCoverage}. Therefore, large values of $ p (X_0 ; \mu) $ tend to indicate that the hypothesis $ \mathcal{H}_\mu $ is in good agreement with the data $ X_0 $.

Seen as a function of $ X $, the p-value follows the distribution

\begin{equation}
\int dX \, \delta [p - p (X ; \mu)] g (X ; \mu) \, ,
\end{equation}
implying that the probability of having a p-value which is smaller than or equal to a given $ \alpha $ is (using $ \delta (f(y)) = \sum_i \delta (y - y_i) / \vert f' (y_i) \vert $ for a continuously differentiable function of roots $ y_i $)


\begin{equation}\label{eq:exactCoverage}
\mathcal{P} [p \leq \alpha \vert \mathcal{H}_\mu] = \alpha \, , \quad \mathcal{P} [p > \alpha \vert \mathcal{H}_\mu] = 1 - \alpha \, ,
\end{equation}
meaning that the distribution of the p-value is uniform, i.e. the distribution of values of $ p $ is flat between 0 and 1.

\subsection{Confidence intervals from the p-value}\label{sec:pvalueCoverage}

\begin{figure}[t]
\begin{center}
\includegraphics[scale=0.45]{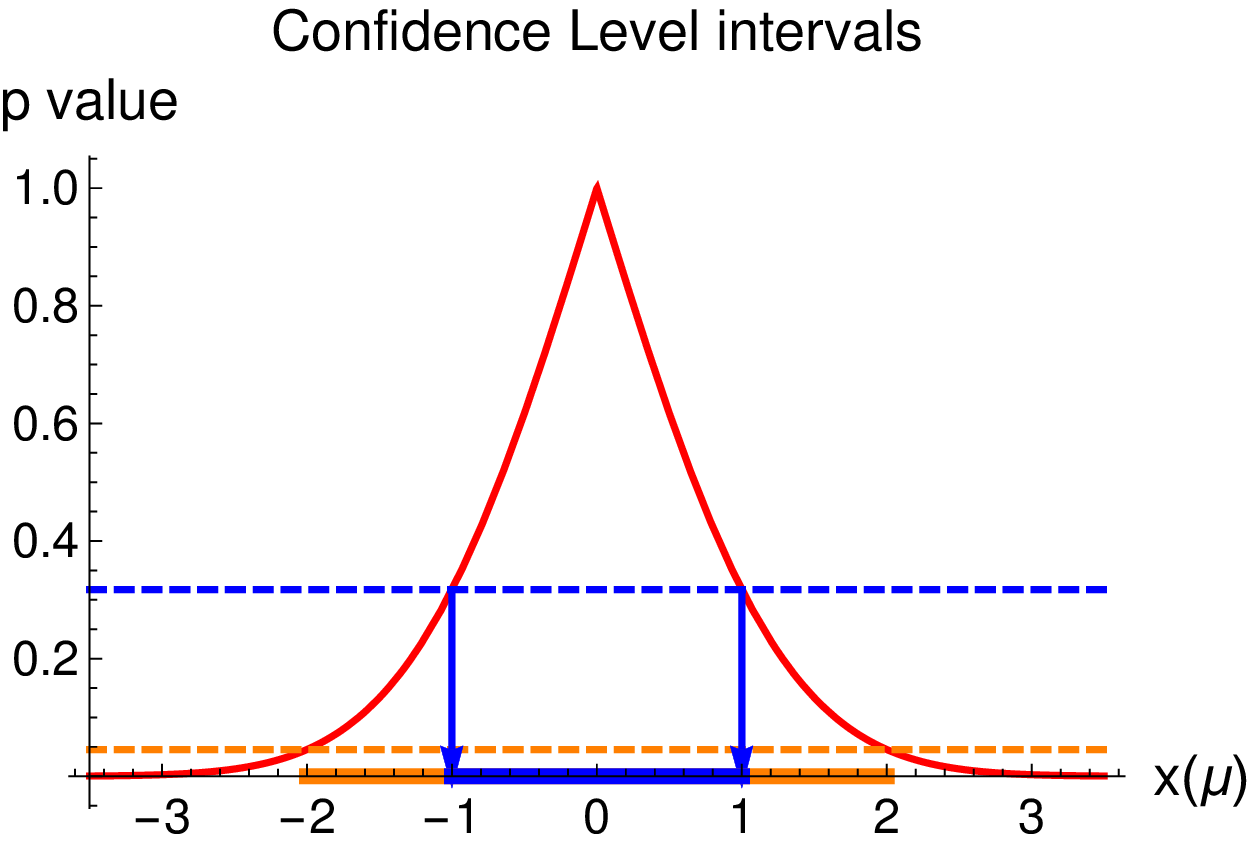}
\hspace{0.6cm}
\includegraphics[scale=0.45]{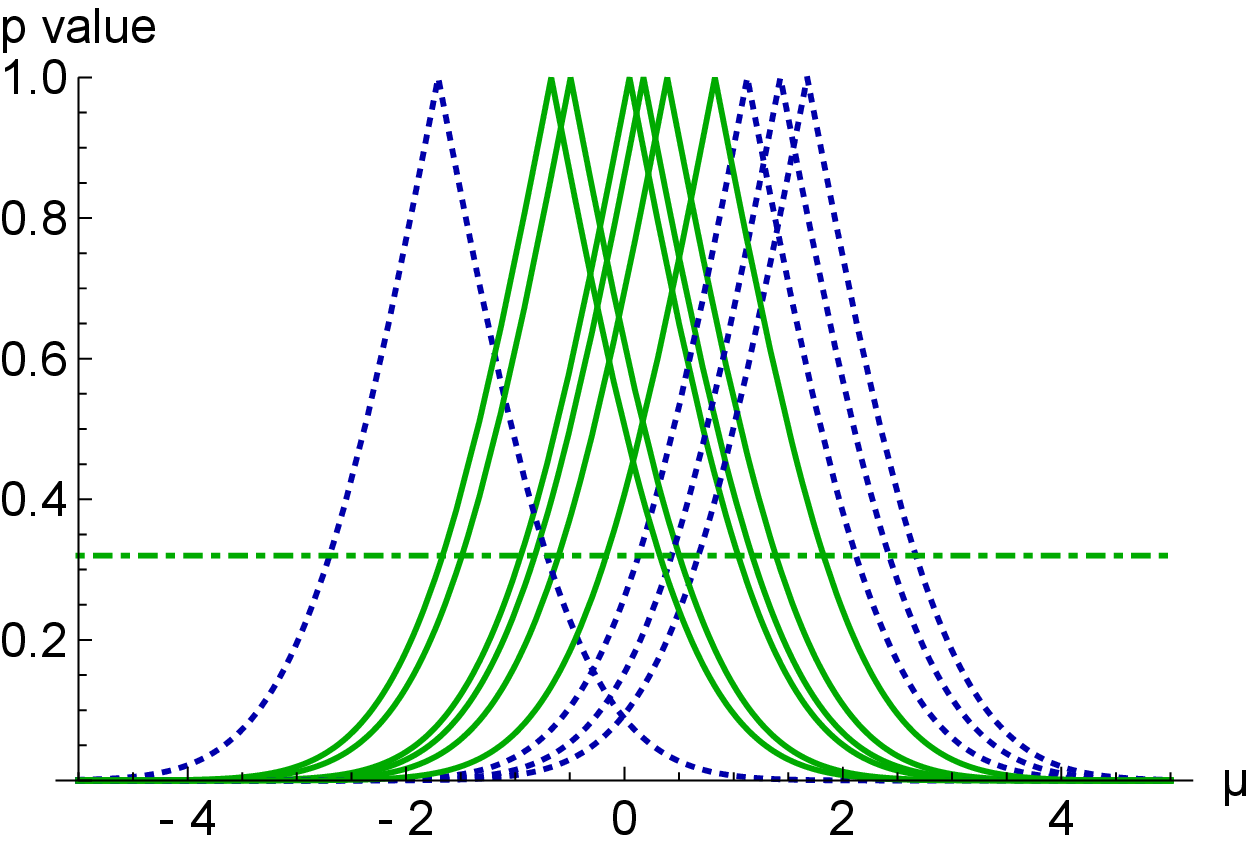}
\end{center}
\caption{\it (Left) p-value curves built out of a random variable modeling a measurement distributed normally. We show on top of the $ x(\mu) $ axis the intervals determined for $p=0.32$ for the dashed-blue line, or $p=0.05$ for the dashed-orange line. (Right)  A $\alpha$ CL interval built from a $p$-value with exact coverage has a probability of $\alpha$ of containing the true value. This is illustrated in the simple case of a quantity which has a true value $\mu_t=0$ but is measured with an uncertainty $\sigma=1$.}
\label{fig:coverageexample}
\end{figure}

Before discussing the content of Eq.~\eqref{eq:exactCoverage}, we first define the notion of confidence level interval. For the Gaussian example of Figure~\ref{fig:coverageexample} (Left), we have considered a measurement $ X_0 $ and calculated the curve of the p-value as a function of the numerical hypothesis $ x(\mu) $ from Eq.~\eqref{eq:pvalueFirstDef}. Then we show in this figure the intervals over the space of values of $ x(\mu) $ built from the requirement p-value $ > 0.32 $, or p-value $ > 0.05 $, which correspond to Confidence Levels of $ 68~\% $ and $ 95~\% $, respectively.

The meaning and usefulness of this overall procedure becomes clear from Eq.~\eqref{eq:exactCoverage}. We illustrate the property stated in Eq.~\eqref{eq:exactCoverage} in the following way, simplifying the notation by setting $ x(\mu) $ to $ \mu $: consider generating a set of \textit{toy} events $ \{ X^{(1)}_0, \ldots, X^{(n)}_0 \} $ respecting a certain distribution $ g $ determined from a chosen $ \mu_t $ (in practice both are unknown). This will imply a set of p-values $ \{ p (X^{(1)}_0 ; \mu), \ldots, p (X^{(n)}_0 ; \mu) \} $ determined from the same $ g $. If $ \mu $ is chosen so that $ \mu = \mu_t $, then for large enough $ n $ there will be a fraction $ \alpha $ of the set of p-values $ p (X^{(i)}_0 ; \mu_t) $ which will be inferior than or equal to $ \alpha $.

Yet, we may state the previous paragraph in a different way. Consider building p-value curves as a function of $ \mu $ for each given $ X^{(i)}_0 $. Pick a value $ 0 \leq \alpha \leq 1 $ and consider the set of values of $ \mu $ for which $ p (X^{(i)}_0 ; \mu) \leq \alpha $. We call this set of the parameter space the \textit{exclusion} interval $ C_\alpha (X^{(i)}_0) $, which depends on the value of $ \alpha $ and on the realization $ X^{(i)}_0 $. Then, what Eq.~\eqref{eq:exactCoverage} tells us is that for large enough $ n $ a fraction $ \alpha $ of the sets $ C_\alpha (X^{(i)}_0) $ will contain $ \mu_t $.

Finally then, we arrive to the following \textit{frequentist} procedure to extract the value of $ \mu_t $: consider that $ g $ correctly describes a certain outcome $ X_0 $, and consider the p-value curve as a function of $ \mu $. Then the complementary set of the interval $ C_\alpha (X^{(i)}_0) $ has the $ 1 - \alpha $ chance of containing the true value $ \mu_t $, and we quote for, say $ 1 - \alpha = 68~\% $, an interval which contains the true value $ \mu_t $ with $ 68~\% $ probability. This is the precise meaning of a $ 1 - \alpha $ Confidence Level (CL) interval in the frequentist sense. 

Let us now discuss Figure~\ref{fig:coverageexample} (Right) in detail to better clarify this discussion. Each time a measurement is performed, it will yield a different value and thus a different $p$-value curve as a function of the hypothesis tested $\mu_t=\mu$. We assume these measurements to be distributed normally, leading to the shapes seen in the figure. From each measurement $ X_0 $, a $ 68~\% $ CL interval can be determined by considering the part of the curve above the line $p=0.32$, but this interval may or may not contain the true value $\mu_t=0$. We show an example with ten measurements in the figure: the curves corresponding to the first case (second case), (not) containing the true value, are indicated with 6 green solid lines (4 blue dotted lines). Asymptotically, if the $p$-value has exact coverage, $ 68~\% $ of these confidence intervals will contain the true value.

Following what we have said above, we say that a set of hypotheses $ \mathcal{H}_\mu $ is excluded with $ 1 - \alpha $ CL if they fall in the exclusion interval $ C_\alpha (X_0) $ of a given measurement $ X_0 $: \textbf{for a given $ \alpha $ and an outcome $ X_0 $, the hypothesis $ \mathcal{H}_\mu $ is excluded at $ 1 - \alpha $ CL if $ p (X_0 ; \mu) \leq \alpha $}. 

If the property announced in Eq.~\eqref{eq:exactCoverage} is satisfied, one says that the p-value has \textit{exact coverage}. p-values are meaningful tool only if they have good coverage properties, otherwise the confidence intervals have not the interpretation we have stated in the previous paragraphs. By ``good" we mean \textit{exact} coverage, Eq.~\eqref{eq:exactCoverage}, or slight \textit{over} coverage if exact coverage cannot be assured:

\begin{eqnarray}\label{eq:defAgressiveConservative}
\mathcal{P} [p \leq \alpha \vert \mathcal{H}_\mu] &=& \alpha \quad : \quad {\rm exact,} \\
\mathcal{P} [p \leq \alpha \vert \mathcal{H}_\mu] & < & \alpha \quad : \quad {\rm conservative \;\quad\quad\quad\, (overcoverage),} \\
\mathcal{P} [p \leq \alpha \vert \mathcal{H}_\mu] & > & \alpha \quad : \quad {\rm liberal/aggressive \;\;\; (undercoverage),}
\end{eqnarray}
which is a property dependent on the value of $ \alpha $.



In the case of over (under) coverage, the CL intervals tend to be broader (smaller). This is reproduced, for example, by quoting an uncertainty $ \sigma $ in the pure statistical case bigger (smaller) than necessary (and sufficient) for exact coverage.

\subsection{Likelihood: comparison of hypotheses}\label{sec:LikelihoodDefinition}

\begin{figure}
	\centering
	\includegraphics[scale=0.7]{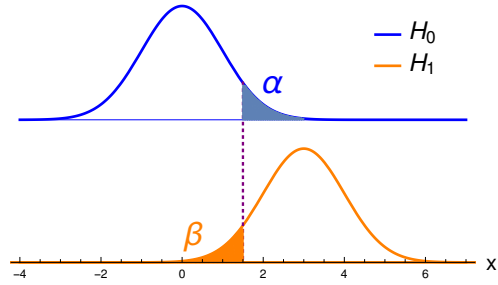}
	\caption{\it Illustration of the likelihood ratio discriminant, where the blue and orange functions have Gaussian shapes (normalized to one). If a measure of $ x $ falls to the right of the dotted purple line, chosen with the help of the \NP $ \, $ lemma, one may exclude the hypothesis $ H_0 \equiv \mathcal{H}_{\mu_0} $ in favour of the hypothesis $ H_1 \equiv \mathcal{H}_{\mu_1} $, which may lead to the exclusion of $ H_0 $ when it is true in $ \alpha $ percent of the cases. Conversely, if a measure of $ x $ falls to the left of the dotted purple line, one may exclude the hypothesis $ H_1 $ when it is true in $ \beta $ (related to the power of the method) percent of the cases.}\label{fig:NeymanPearson}
\end{figure}

We consider the following product of \textit{probability distribution functions} (pdf), seen as a function of the value of the fundamental parameters $ \mu $ (non correlated case)

\begin{equation}\label{eq:definitionLikelihood}
\mathcal{L}_{X_0} (\mu) = \prod^n_{i = 1} g_i (X^{(i)}_0 ; \mu) \, ,
\end{equation}
called the Likelihood. This is an interesting object because the ratio of Likelihoods under different hypotheses can be used to compare hypotheses as follows: if

\begin{equation}
\mathcal{L}_{X_0} (\mu_1) / \mathcal{L}_{X_0} (\mu_2) < t_\alpha
\end{equation}
\noindent
for a given $ t_\alpha $ then we exclude the hypothesis $ \mathcal{H}_{\mu_1} $, called null hypothesis, in favour of the hypothesis $ \mathcal{H}_{\mu_2} $. It can be shown (\NP $ \, $ lemma, see Ref.~\cite{Metzger}) that among the possible exclusion tests which suffer from an error $ \alpha $ of excluding $ \mathcal{H}_{\mu_1} $ if $ \mathcal{H}_{\mu_1} $ is true (error of Type-I, false negative), the Likelihood ratio has the smallest possible error for accepting $ \mathcal{H}_{\mu_1} $ if $ \mathcal{H}_{\mu_2} $ is true (error of Type-II, false positive). In this sense, we say that the test of the Likelihood ratio is \textit{optimal}: given a Type-I error of \textit{size} $ \alpha $, the Likelihood ratio minimizes the Type-II error, thus maximizing the so-called \textit{power} of the comparison, see Figure~\ref{fig:NeymanPearson}.

Obviously, the logarithm

\begin{center}
$ - 2 \log ( \mathcal{L}_{X_0} (\mu_1) / \mathcal{L}_{X_0} (\mu_2) ) = - 2 \log \mathcal{L}_{X_0} (\mu_1) + 2 \log \mathcal{L}_{X_0} (\mu_2) $
\end{center}

\noindent
inherits the properties of the Likelihood ratio, and we build a test statistic to compare both hypotheses out of that

\begin{equation}\label{eq:testfromLikelihood}
T = - 2 \log \mathcal{L}_{X_0} (\mu_1) + 2 \log \mathcal{L}_{X_0} (\mu_2) \, .
\end{equation}
Then, low values of $ T $ show a preference for the hypothesis $ \mu_1 $, thus being able to discriminate $ \mu_1 $ and $ \mu_2 $.

\subsubsection{Composite hypotheses}\label{sec:compositeStat}

Consider, however, a case where the hypothesis depends on parameters we would like to extract $ \mu $, and additional parameters $ \delta $: in this case, we cannot formulate a \textit{simple} hypothesis $ \mathcal{H}_\mu $ because something must be said about the value of $ \delta $. One talks about \textit{composite} hypotheses, for which the optimality property of the ratio of Likelihoods is not guaranteed any more.

By analogy with the principle of maximizing Likelihood ratios discussed above, which leads to the most \textit{powerful} situation, one usually considers

\begin{eqnarray}\label{eq:MLRtestComposite}
T_\mu &=& - 2 \log \frac{\max_{\delta} \mathcal{L}_{X_0} (\mu, \delta)}{\max_{\mu, \delta} \mathcal{L}_{X_0} (\mu, \delta)} \\
&=& \min_\delta [- 2 \log \mathcal{L}_{X_0} (\mu, \delta)] - \min_{\mu, \delta} [- 2 \log \mathcal{L}_{X_0} (\mu, \delta)] \nonumber
\end{eqnarray}
for comparing hypotheses, i.e. we minimize over the parameters $ \delta $, called in this context \textit{nuisance} parameters. 

\subsection{Gaussian case without theoretical uncertainties}\label{sec:candy} 



We would like now to discuss the simplest case where all the random variables are independent, or decorrelated, and follow a normal distribution. We assume a Gaussian law motivated by the Central-Limit theorem, which implies that the specific way in which we model an experiment does not (asymptotically) matter. Therefore

\begin{eqnarray}\label{eq:apparatus}
&& g (X ; \mu_t) = \frac{1}{\sqrt{2 \pi} \sigma} \exp \left[ - \frac{1}{2} \left( \frac{X - x (\mu_t)}{\sigma} \right)^2 \right] \, ,
\end{eqnarray}
\noindent
where $ \sigma $ is the accuracy of the experimental technique.\footnote{Though it may sound doubtful that one may be able to model an apparatus far away from the range where it is specially designed to work at, i.e. far away $ X_0 \simeq x(\mu) $ and deep in the ``tail'' region (borrowing the image of a Gaussian), this is going to be admitted and $ X_0 \gg x(\mu) $ or $ X_0 \ll x(\mu) $ will thus cast doubt on the interpretation of $ x_t $ as $ x (\mu) $ -- or the very modeling of the apparatus (cf. \cite{Lyons:2013yja}).}


We now build the test statistic from the Likelihood, cf. Eq~\eqref{eq:testfromLikelihood}


\begin{equation}
T (X_0 ; \mu) = \sum^n_{i=1} \left( \frac{X^{(i)}_0 - x(\mu)}{\sigma} \right)^2 \, .
\end{equation}
Beyond the property of optimality of Type-II errors, one can use the Likelihood to build estimators of the fundamental parameters, by the minimization of $ T $, whose minimum follows a $ \chi^2 $ distribution of $ n - \| \mu \| $ degrees of freedom (d.o.f.), where $ n $ is the number of measurements and $ \| \mu \| $ is the number of fundamental parameters estimated, i.e. $ T_{min} \sim \chi^2 (n - \| \mu \|) $. In principle, however, this is only true for linear models where $ x(\mu) $ depends linearly on $ \mu $, but for large enough $ n $ this distribution is asymptotically valid (Wilks' theorem, see Ref.~\cite{Cowan:1998ji}). 



Building the p-value as described previously, we have

\begin{equation}
p (X_0 ; \mu) = \frac{\Gamma (N_{dof}/2, T (X_0 ; \mu) / 2)}{\Gamma (N_{dof}/2)} = {\rm Prob} (T (X_0 ; \mu) , N_{dof}) \, ,
\end{equation}
where $ {\rm Prob} $ is the well-known routine from the CERN library, which calculates the probability to find a value for a variable distributed as $ \chi^2 (N_{dof}) $ at least as large as $ T (X_0 ; \mu) $.\footnote{$ \Gamma (a , y) $ ($ \Gamma (a) $) is the incomplete (usual, Euler) gamma function

\begin{equation}
\Gamma (a , y) = \int^{\infty}_y t^{a - 1} \exp (- t) \, dt \, , \qquad \Gamma (a) = \int^{\infty}_0 t^{a - 1} \exp (- t) \, dt \, ,
\end{equation}
and thus $ \Gamma (1/2) = \sqrt{\pi} $, $ \Gamma (1) = 1 $, $ \Gamma (3/2) = \sqrt{\pi} / 2 $, etc.} For further illustration, consider the case of a single random variable. We have the following expression

\begin{equation}
1 - p (X_0 ; \mu) = \int^{t_0}_{-t_0} \frac{1}{\sqrt{2 \pi} \sigma} \exp \left[ - \frac{1}{2} \left( \frac{y}{\sigma} \right)^2 \right] \, dy \, , \qquad t_0 = T (X_0 ; \mu) \, ,
\end{equation}
then\footnote{The function $ {\rm Erf} (y) $ is defined by

\begin{equation}
{\rm Erf} (y) = \frac{2}{\sqrt{\pi}} \int^y_0 \exp (- t^2) \, dt \, .
\end{equation}
It is also useful to define the complementary of the error function

\begin{equation}
{\rm Erfc} (y) = \frac{2}{\sqrt{\pi}} \int^{\infty}_y \exp (- t^2) \, dt \, .
\end{equation}

}

\begin{equation}\label{eq:numberofsigma1D}
\frac{t_0}{\sigma} = \sqrt{2} {\rm Erf}^{-1} (1 - p (X_0 ; \mu)) = \sqrt{2} {\rm Erfc}^{-1} (p (X_0 ; \mu)) \, ,
\end{equation}
and $ t_0 / \sigma \equiv k_\sigma (p) $ gives the ``number of units of $ \sigma $" of a given p-value:

\begin{center}
$ p (X_0 ; \mu) \simeq 0.32 \leftrightarrow t_0 = 1 \times \sigma $, $ p (X_0 ; \mu) \simeq 0.05 \leftrightarrow t_0 = 2 \times \sigma $, etc.
\end{center}
\noindent
which correspond to test statistic values $ 1^2 $, $ 2^2 $, etc. In words, the numbers of $ \sigma $ tell how far away the predicted value $ x(\mu) $ is from the best fit point. 

\section{Theoretical uncertainties}\label{sec:theoProblem} 

A common problem in the determination of an observable is that theoretical uncertainties are usually present. This is indeed a problem because, by their own nature, theoretical uncertainties do not decrease with the amount of data: even if the limit where the sample has an infinite size is taken, the extraction of $ x_t $ cannot be done with an absolute certainty, and one quotes a systematic effect $ \pm \Delta $ instead. From a different perspective, it may happen that the technology we dispose to predict the value of the observable is limited: examples are given by the need to extrapolate a calculation, or to truncate a perturbative series in theoretical works.

We are going to discuss possible ways to circumvent the difficulty introduced by theoretical uncertainties in the next subsections. It should be kept in mind through our discussion that, no matter what their modeling is, the presence of theoretical uncertainties will imply a worse capacity of extracting the true values of the fundamental parameters of a model.




Some comments about the terminology are relevant. We consider the test statistic and the p-value as functions of $ X $, a random variable of variance $ \sigma^2 $. Moreover, since we are only interested in the comparison $ X - x (\mu) $, we can (for practical reasons here) quote the theoretical uncertainty altogether with the experimental value

\begin{equation}
X_0 \pm \sigma \pm \Delta \, .
\end{equation}
One may represent the theoretical uncertainty $ \Delta $ as coming from an additional unknown parameter $ \delta $: if the true value $ \delta_t $ of $ \delta $ was known, we would quote $ X_0 + \delta_t $ for the central value of the measurement. It then follows that the meaning of $ \Delta $ is not the same as the one for $ \sigma $, which parameterizes the probability distribution function of a random variable, and we will need a way to interpret $ \Delta $ or $ \delta $. Starting from this section, we are going to discuss different models for $ \delta $.



For definiteness, we discuss the case of only one random variable $ X $, and for simplicity we consider only symmetric cases over this chapter (a longer discussion is found in Ref.~\cite{Charles}). Also for simplicity reasons, we take $ x(\mu) \rightarrow \mu $, i.e. the observable is the fundamental parameter itself (or it is linearly related to the single parameter $ \mu $ of the theory).

\subsection{Random approach: naive Gaussian}



In principle, $ \delta $ comes from a non-statistical source and has no reason to be a random variable. However, for simplicity one can model it as a random variable of mean $ 0 $ and dispersion $ \Delta $. Then, we say that what we measure are the realizations of the random variable

\begin{equation}
X = X' + \delta \, ,
\end{equation}
where

\begin{equation}
X' \sim \mathcal{N}_{(\mu_t, \sigma)}
\end{equation}
models the case where no theoretical uncertainties are present.

We then convolute the distribution of $ X' $ with that of $ \delta $, resulting in the distribution of the random variable $ X $. In what we call ``naive Gaussian," we assume that the theoretical uncertainty is distributed as a Gaussian

\begin{equation}
\delta \sim \mathcal{N}_{(0,\Delta)} \, ,
\end{equation}
and we end up having the commonly used distribution

\begin{equation}\label{eq:apparatusnG}
X \sim \mathcal{N}_{(\mu_t, \sqrt{\sigma^2 + \Delta^2})} \, .
\end{equation}

Note that we are able to formulate the simple hypothesis $ \mathcal{H}_\mu : \mu_t = \mu $, since we have gotten rid of $ \delta $ in some sense. The test statistic is then built from the Likelihood Eq.~\eqref{eq:testfromLikelihood} as

\begin{equation}
T (X ; \mu) = \frac{(X - \mu)^2}{\sigma^2 + \Delta^2} \, ,
\end{equation}
resulting in the p-value

\begin{eqnarray}
p^{nG} (X ; \mu) = 1 - {\rm Erf} \left[ \frac{\vert X - \mu \vert}{\sqrt{2} \sqrt{\sigma^2 + \Delta^2}} \right] = {\rm Erfc} \left[ \sqrt{ \frac{T (X ; \mu)}{2} } \right] \, .
\end{eqnarray}

\begin{figure}[H]
	\centering
	\includegraphics[scale=0.5]{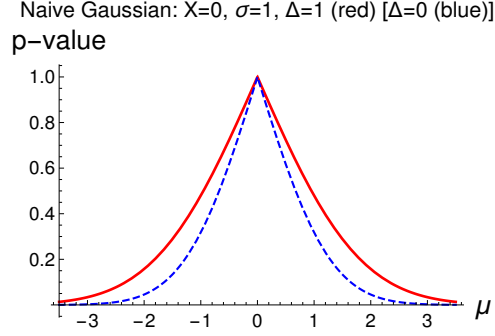}
	\caption{\it Naive Gaussian: p-value as a function of $ \mu $.}\label{fig:simplenG}
\end{figure}

In Figure \ref{fig:simplenG}, we illustrate that the inclusion of theoretical uncertainties (solid, red) increases the size of the CL intervals compared to the case where no theoretical uncertainties are present (dashed, blue).

\subsection{External approach: Scan method}

The external approach intends to be close in philosophy to what experimentalists often do to estimate systematic uncertainties. In this case, the theoretical uncertainty $ \delta $ is seen as an \textit{external} parameter, just like $ \sigma $, whose value one admits knowing. The apparatus is modeled as

\begin{equation}\label{eq:apparatusExternal}
X \sim \mathcal{N}_{(\mu_t + \delta, \sigma)} \, ,
\end{equation}
and under the simple hypothesis $ \mathcal{H}^{(\delta)}_{\mu} : \mu_t = \mu + \delta $ we have the following test statistic from the Likelihood Eq.~\eqref{eq:testfromLikelihood}

\begin{equation}\label{eq:testExternal}
T (X ; \mu) = \frac{( X - \mu - \delta )^2}{\sigma^2} \, .
\end{equation}
\noindent
Note that, given \eqref{eq:apparatusExternal} and \eqref{eq:testExternal}, the estimator of the observable built from the minimization of the test statistic converges to $ \mu $ with the size of the sample, thus implying that this is not a biased estimator.

The test statistic in Eq.~\eqref{eq:testExternal} implies the following p-value



\begin{equation}
p (X ; \mu) = 1 - {\rm Erf} \left[ \frac{\vert X - \mu - \delta \vert}{\sqrt{2} \sigma} \right] = {\rm Erfc} \left[ \sqrt{\frac{T (X ; \mu)}{2}} \right] \, ,
\end{equation}
which depends explicitly in the (external) value of $ \delta $. We therefore need a way to combine the different p-values corresponding to the different values of $ \delta $. To this effect, we maximize the p-value over the range $ \Omega = [- \Delta , \Delta] $:

\begin{equation}\label{eq:externalpv}
p^{Scan} (X_0 ; \mu) = \max_{\delta \in [- \Delta , \Delta]} p (X_0 ; \mu, \delta) \, ,
\end{equation}

\begin{figure}[H]
	\centering
	\includegraphics[scale=0.5]{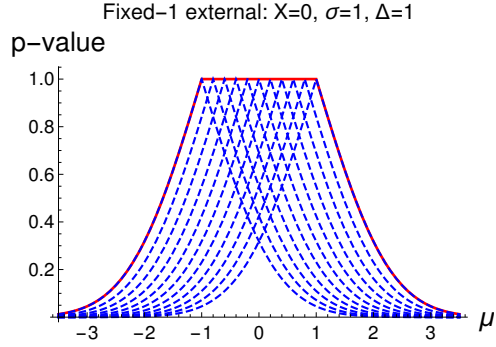}
	\caption{\it Envelope (red) of p-values for $ \delta \in [- \Delta, \Delta] $.}\label{fig:simpleExternal}
\end{figure}

\noindent
thus implying

\begin{eqnarray}\label{eq:externalpvSimplified}
p^{Scan} (X ; \mu) &=& 1 \qquad \qquad \qquad \qquad \qquad \quad \; \; , \quad {\rm if} \; \vert X - \mu \vert \leq \Delta \, , \\
&=& 1 - {\rm Erf} \left( \frac{\vert X - \mu \pm \Delta  \vert}{\sqrt{2} \sigma} \right) \, , \quad {\rm if} \; \mp ( X - \mu ) > \Delta \, . \nonumber
\end{eqnarray}

The idea behind the maximization is to consider the ``envelope" of all admissible values of $ \delta $: the various p-value curves shown in Figure \ref{fig:simpleExternal} (dashed, blue) fit into an envelope (solid, red), which by its turn is used to determine confidence intervals. The range we have chosen above, namely $ \Omega = [- \Delta , \Delta] $, makes part of the model and for simplicity reasons we are not going to discuss different possibilities, such as $ \Omega = 3 [- \Delta , \Delta] $, which would correspond to an envelope in Figure~\ref{fig:simpleExternal} three times broader.

The external approach, called fixed-1 external or 1-external in what follows, is fundamentally the same as the Scan method discussed in Ref.~\cite{Scan}. Among the differences, the Scan method relies on the test statistic $T = - 2 \log {\cal L} (\mu,\delta)$ which is interpreted assuming that $T$ follows a $\chi^2$ distribution law, including both the parameters of interest and nuisance parameters. The $1-\alpha$ confidence region is then determined by varying the nuisance parameters in given intervals (typically a $1 \, \sigma$ range), and accepting only points for which $T \leq T_c$, $T_c$ being a critical value so that $P(T \geq T_c ; N | H_0) \geq \alpha$ (typically $\alpha=0.05$), where $ H_0 $ is the hypothesis of the Standard Model and $ N $ is the number of degrees of freedom.

\subsection{Nuisance approach}

In this case, the theoretical uncertainty is considered as a nuisance parameter in the lines of Section \ref{sec:compositeStat}, and the apparatus is modeled as

\begin{equation}\label{eq:apparatusNuisance}
X \sim \mathcal{N}_{(\mu_t + \delta_t, \sigma)} \, .
\end{equation}

In principle, we can choose any test statistic that tests correctly the null hypothesis $ \mathcal{H}_\mu : \mu_t = \mu $, and we start from the following quadratic test statistic

\begin{equation}\label{eq:quadraticTestNuisance}
T (X ; \mu) = \frac{(X - \mu)^2}{\sigma^2 + \Delta^2} \, .
\end{equation}
This ansatz is motivated by the following procedure

\begin{equation}\label{eq:quadraticTestNuisanceMotivation}
T (X_0 ; \mu) = \min_{\delta \in \Re} \left[ \left( \frac{X_0 - \mu - \delta}{\sigma} \right)^2 + \left( \frac{\delta}{\Delta} \right)^2 \right] \, ,
\end{equation}
already illustrated in Eq.~\eqref{eq:MLRtestComposite}, where the minimum above is found at $ (X_0 - \mu) \frac{\Delta^2}{\sigma^2 + \Delta^2} $, and can assume any value. The advantage of the last ansatz is that, when a larger number of nuisance parameters is present, this procedure always leads to a quadratic form like the one of Eq.~\eqref{eq:quadraticTestNuisance}.


Though $ T (X ; \mu) $ does not depend on $ \delta $, its distribution does due to Eq.~\eqref{eq:apparatusNuisance} and the p-value inherits this dependence

\begin{eqnarray}\label{eq:pvalueNuisance}
&& \mathcal{P}_\delta [T < T (X ; \mu)] = \frac{1}{2} \left[ {\rm Erf} \left( \frac{\delta + \vert X - \mu \vert}{\sqrt{2} \sigma} \right) - {\rm Erf} \left( \frac{\delta - \vert X - \mu \vert}{\sqrt{2} \sigma} \right) \right] \, , \nonumber\\
&& \qquad\qquad 1 - p_\delta (X ; \mu) = \mathcal{P}_\delta [T < T (X ; \mu)] \, ,
\end{eqnarray}
where the subscript $ \delta $ indicates that we do not treat $ \delta $ and $ \mu $ on the same footing. The above expression states that the p-value depends on the unknown bias parameter $ \delta $. To build Confidence Level (CL) intervals, we consider the following procedure


\begin{equation}
\max_{\delta \in \Omega} p_\delta (X_0 ; \mu) \, .
\end{equation}
Taking the maximum in this case as we have done previously in the external approach will them imply broader CL intervals, expectedly leading to conservative situations.


The question now is what to take for $ \Omega $, and in particular suppose we are interested in quoting confidence intervals at different significances (i.e. ``numbers of $ \sigma $"): should we take always the same interval $ \Omega $ where the true value of $ \delta $ is supposed to be found, or should we adapt $ \Omega $ correspondingly to the size of the Confidence Level interval of $ \mu $ we would like to quote? To answer to this question, we compare two procedures to define the p-value

\begin{itemize}
	\item Fixed range $ \Omega = r [- \Delta , \Delta] $: $ p^{fixed} (X ; \mu) = p_{\pm r \Delta} (X ; \mu) $, where $ r $ is fixed. We are going to refer to this method as ``fixed$ -r $ nuisance approach," or simply $ r- $nuisance.
	\item Adaptive range $ \Omega = k_\sigma (p) [- \Delta , \Delta] $: $ p^{adapt} (X ; \mu) = p_{\pm k_\sigma (p) \Delta} (X ; \mu) $, where $ k_\sigma (p) $ is the ``number of $ \sigma $" of a given p-value. The adaptive procedure then means that: if one is interested to quote a significance $ n \times \sigma $, one maximizes $ p_\delta (X ; \mu) $ over the range $ \Omega = n [- \Delta , \Delta] $, i.e. for each $ n $ one calculates the corresponding p-value and determines from it the CL interval for the physical parameter $ \mu $.
\end{itemize}

\begin{figure}[H]
\centering
\includegraphics[scale=0.45]{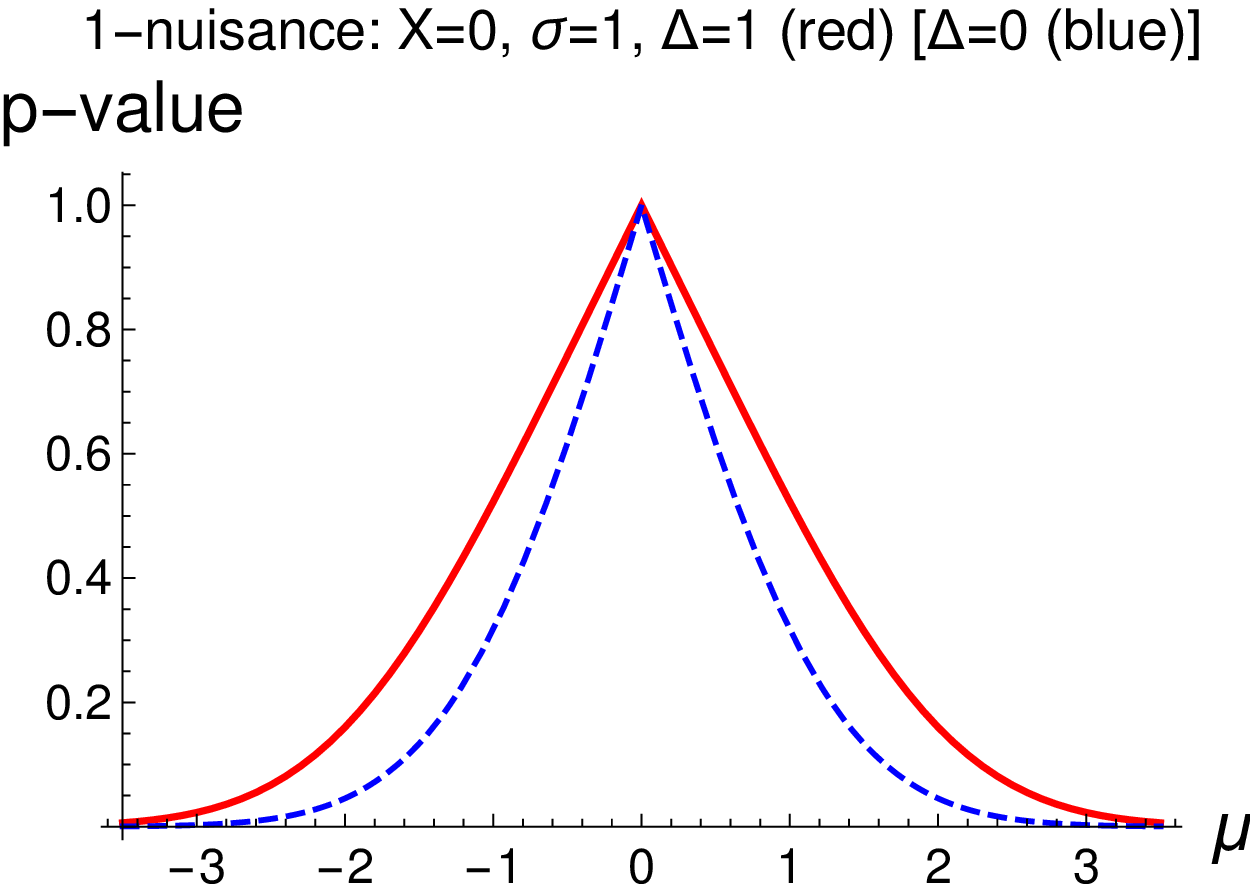}
\includegraphics[scale=0.45]{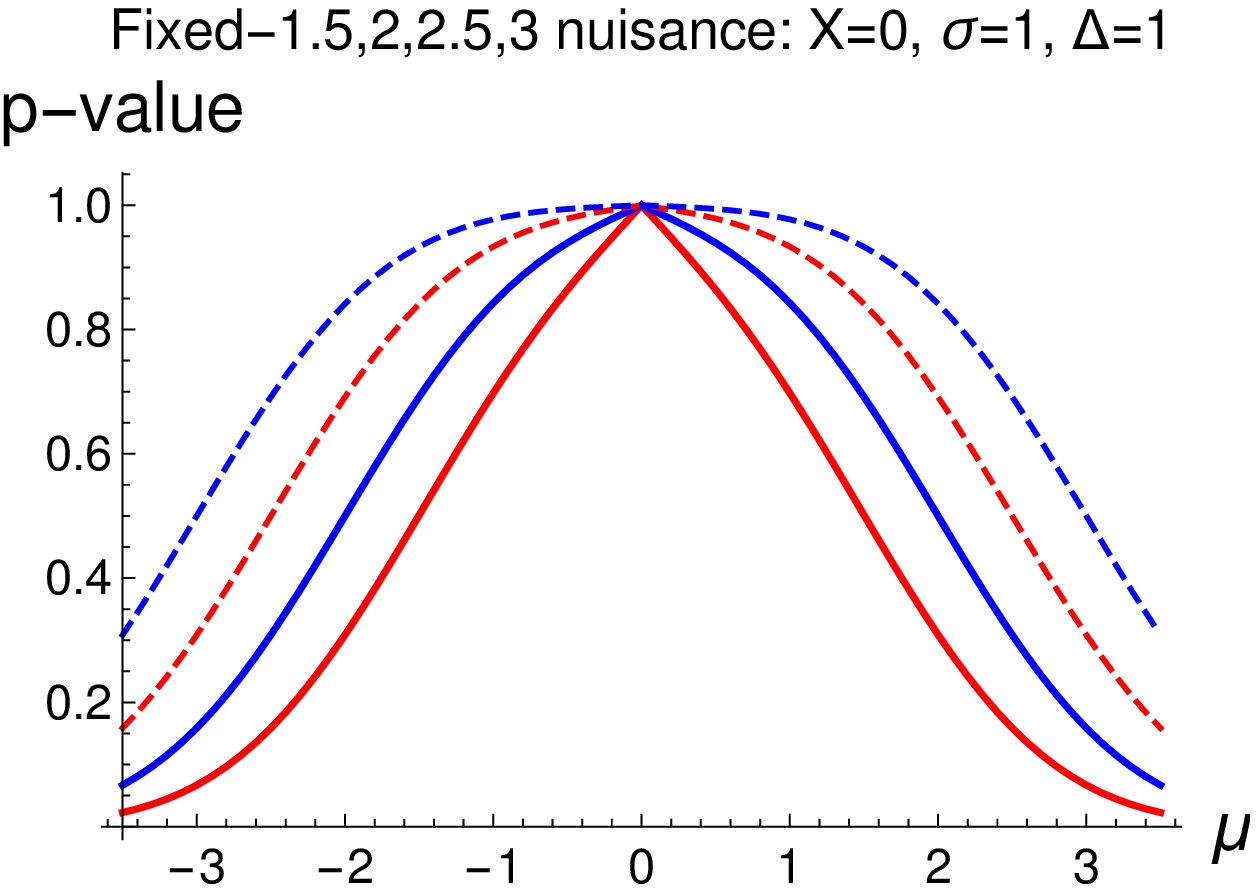} \\
\vspace{3mm}
\includegraphics[scale=0.45]{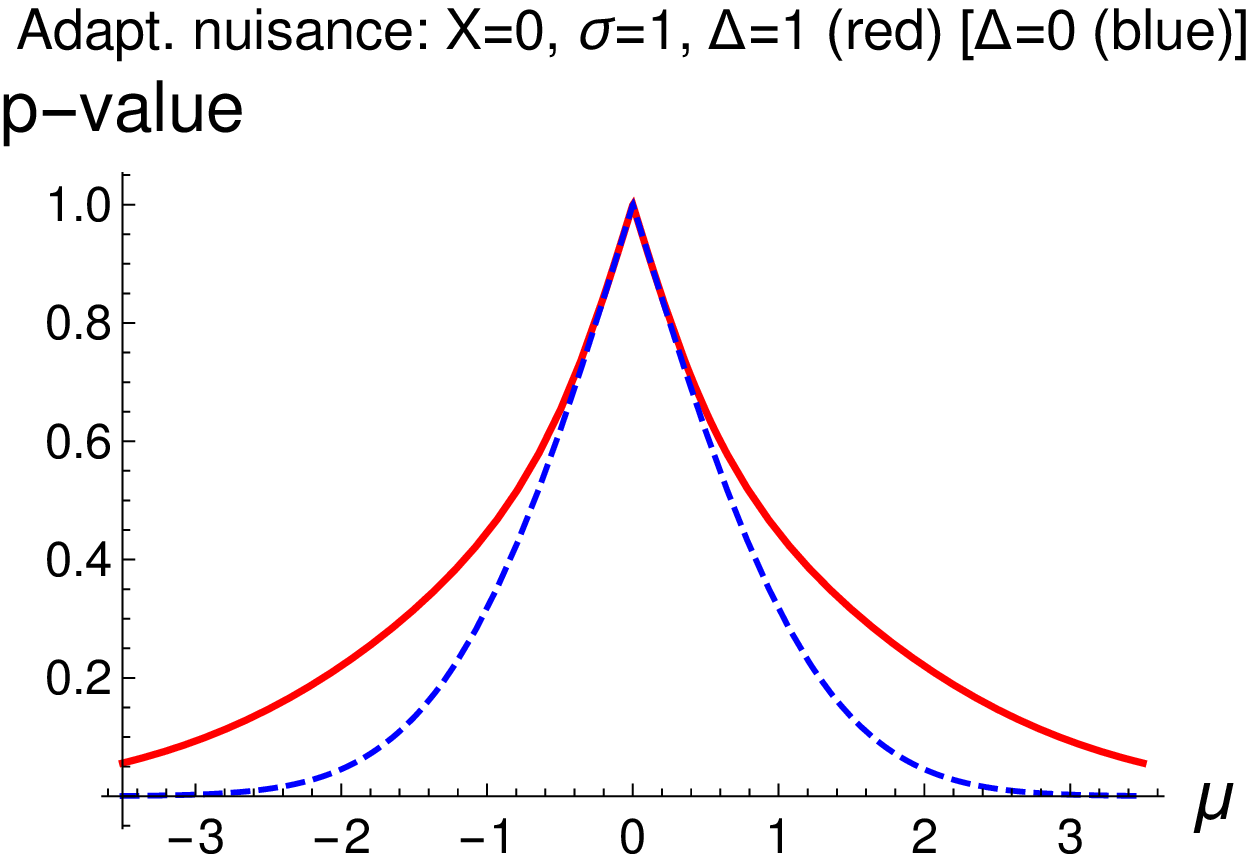}
\caption{\it Illustration of the nuisance approach: (Left) Fixed-1 nuisance (solid, red) compared to the case of no theoretical uncertainty (dashed, blue). (Right) comparison of fixed-1.5 (solid, red), fixed-2 (solid, blue), fixed-2.5 (dashed, red), and fixed-3 (dashed, blue) nuisance, where one notes that a smooth \textit{plateau} shows up for increasingly larger intervals $ \Omega $. (Bottom) Adaptive nuisance (solid, red) compared to the case of no theoretical uncertainty (dashed, blue).}\label{fig:differentOmegaNuisance}
\end{figure}

The choice for $ \Omega $ makes part of our way to model theoretical uncertainties: a graphical comparison between the different choices is given in Figure \ref{fig:differentOmegaNuisance}. The different graphs illustrate that the two nuisance approaches depicted above lead to very different p-value curves, and consequently to possibly very different confidence level intervals.

\subsection{Rfit}

The Rfit procedure follows the nuisance philosophy discussed in the previous section, and considers that the theoretical parameter $ \delta $ strictly relies inside the interval $ [-\Delta, \Delta] $. In practice, one considers a theoretical Likelihood that is a constant over the fixed range $ \delta \in [- \Delta , \Delta] $ and zero outside, and then combines it with a pure statistical Likelihood. This procedure has been briefly exemplified in Chapter~\ref{ch:SM} and is adopted by the \CKM $ \, $ Collaboration \cite{Hocker:2001xe}, \cite{CKMfitterStandard}.

As said previously, we can choose any test statistic that models correctly the null hypothesis (the hypothesis we want to test), and we consider the following test statistic
\begin{eqnarray}\label{eq:WellShape}
T (X_0 ; \mu) & = & 0 \, , \quad {\rm if} \; \vert X_0 - \mu \vert \leq \Delta \\
& = & \left( \frac{X_0 - \mu + \Delta}{\sigma} \right)^2 \, , \quad {\rm if} \; \mu - X_0 > \Delta \, , \nonumber\\
& = & \left( \frac{X_0 - \mu - \Delta}{\sigma} \right)^2 \, , \quad {\rm if} \; X_0 - \mu > \Delta \, , \nonumber
\end{eqnarray}
which we have already found in Chapter~\ref{ch:SM}. This expression means that we consider at the same footing values of $ \mu $ at the flat bottom of $ T (X_0 ; \mu) $, of size given by $ 2 \Delta $.

In the previous nuisance approaches, one would build a p-value from this test statistic and would maximize it over a given range for $ \delta $. Though this is straightforward to do in one-dimension, in multi-dimensional cases the computation is much more involved in the Rfit framework. Therefore, it is usual to assume that the distribution of $ T (X ; \mu) $ is well approximated by a $ \chi^2 $ distribution, in which case we can use the expressions given in Section~\ref{sec:candy}. In the one-dimensional case, we have:

\begin{eqnarray}
p^{Rfit} (X ; \mu) &=& 1 \qquad \qquad \qquad \qquad \qquad \quad \; \; , \quad {\rm if} \; \vert X - \mu \vert \leq \Delta \\
&=& 1 - {\rm Erf} \left( \frac{\vert X - \mu \pm \Delta  \vert}{\sqrt{2} \sigma} \right) \, , \quad {\rm if} \; \mp ( X - \mu ) > \Delta \, . \nonumber
\end{eqnarray}

\begin{figure}[H]
\centering
\includegraphics[scale=0.45]{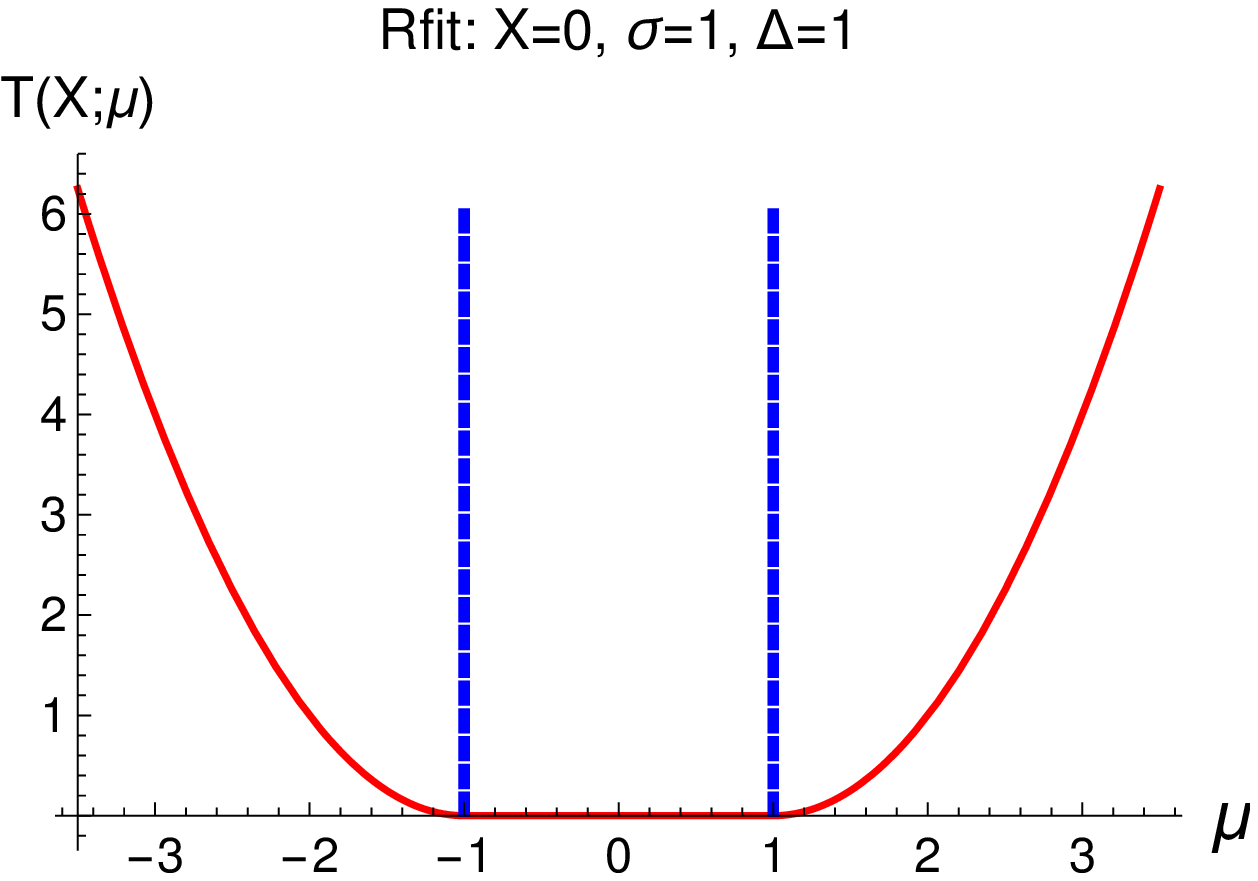}
\hspace{0.5cm}
\includegraphics[scale=0.45]{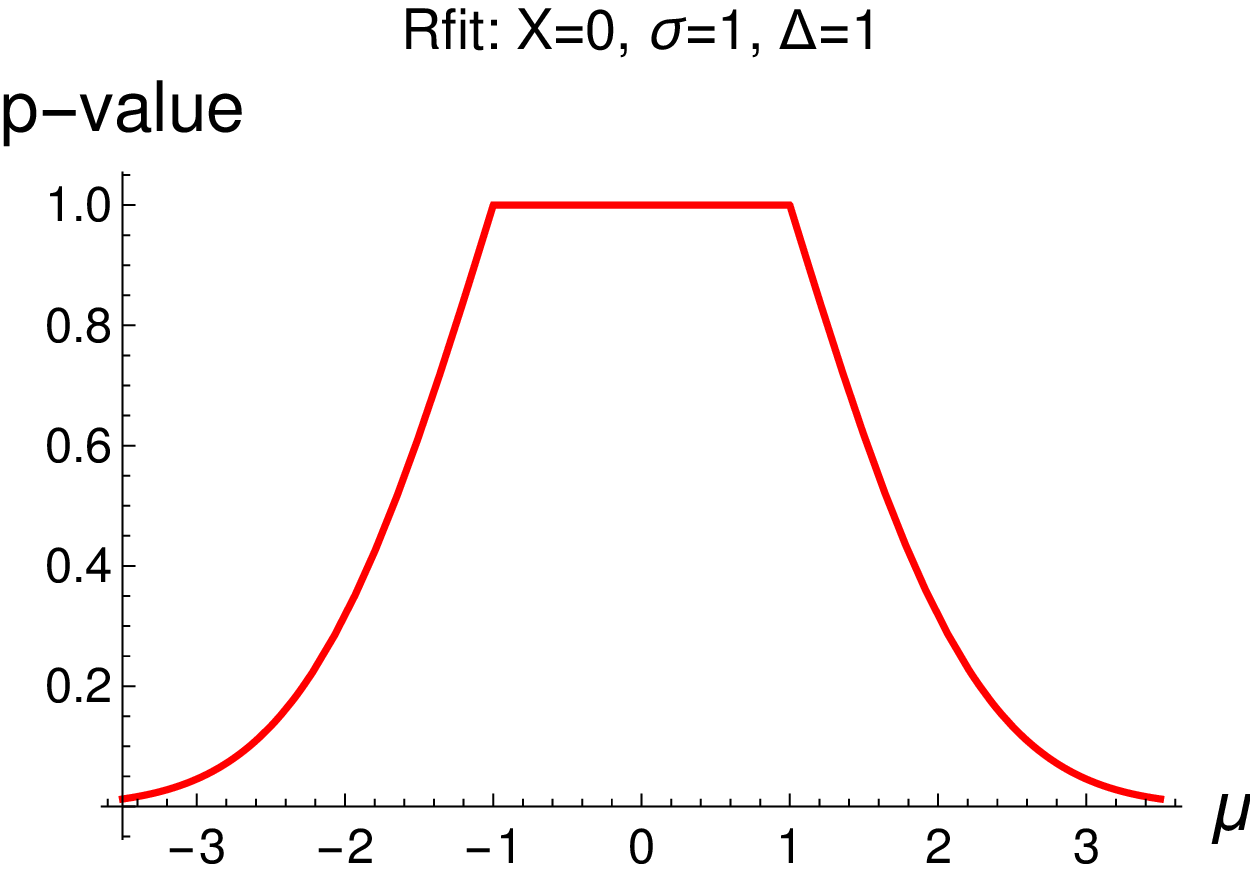}
\caption{\it (Left) The test statistic given in Eq.~\eqref{eq:WellShape} is null over the interval indicated by the two dashed-blue lines, and follows a quadratic shape outside. This ansatz corresponds to a true value of the theoretical uncertainty $ \delta $ bounded inside the range $ [-\Delta, \Delta] $. (Right) At higher dimensions, it is preferable to approximate the test statitistic by a known distribution. It is usual to consider a $ \chi^2 $ distribution, whose shape has a flat top, or a \textit{plateau}, following the assumption $ -\Delta \leq \delta_t \leq \Delta $. Note that other approaches discussed beforehand also show a flat top or \textit{plateau}.}\label{fig:RfitApproach}
\end{figure}

Note that this expression ends up being identical to Eq.~\eqref{eq:externalpvSimplified} in the 1-dimensional case. This, however, is not a property found in higher dimensional cases. The ``well'' shape function and the p-value in the Rfit approach are seen in Figure~\ref{fig:RfitApproach}.

\subsection{Impact of the modeling of theoretical uncertainties}\label{sec:impactmodelling}





We now move to the comparison of the different approaches discussed so far. We consider then a measurement

\begin{equation}
X_0 = 0 \pm \sigma \pm \Delta \, , \quad \sigma^2 + \Delta^2 = 1 \, .
\end{equation}

As said previously, 1-external and Rfit lead to the same p-value curve in the 1-dimensional case, and in what follows we are going to consider both indistinctly, except when otherwise stated.


\subsubsection{Confidence intervals}\label{sec:ConfidenceIntervals1D}

\begin{figure}
\centering
\includegraphics[scale=0.47,trim={1.2cm 0 0 0},clip]{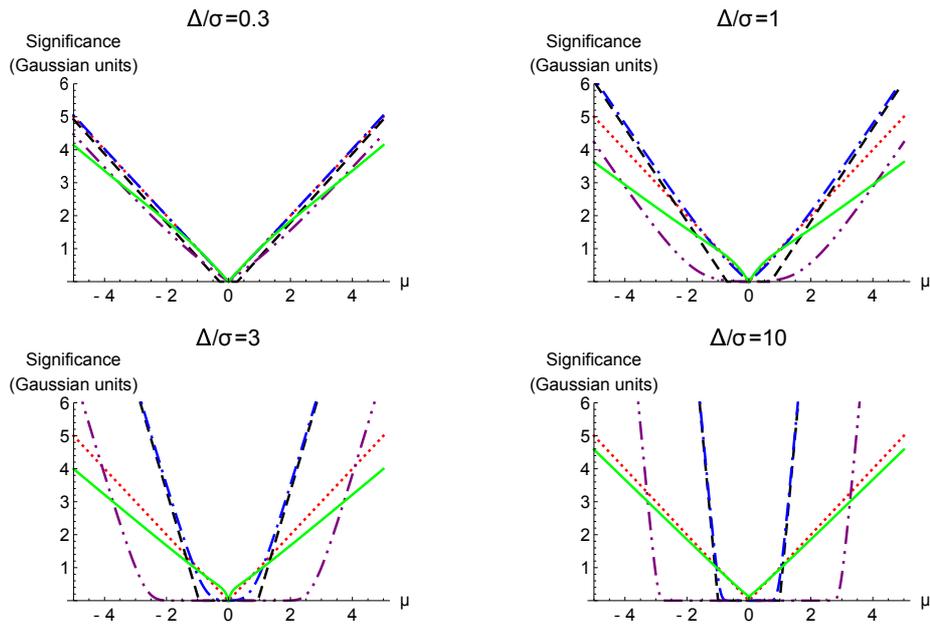}
\caption{\it Comparison of different methods for modeling theoretical uncertainties for the measurement $ X_0 = 0 \pm \sigma \pm \Delta \, , \sigma^2 + \Delta^2 = 1 $, for different ratios $ \Delta / \sigma $. Plots are shown in ``units of $ \sigma $." The various p-values are: (dotted, \textbf{{\color{red} red}}) naive Gaussian (nG); (dashed, \textbf{black}) fixed-1 external $ \delta \in [- \Delta , \Delta] $, or Rfit; (dotted-dashed, \textbf{{\color{blue} blue}}) fixed-1 nuisance $ \delta \in [- \Delta , \Delta] $; (dotted-dotted-dashed, \textbf{{\color{magenta} purple}}) fixed-3 nuisance $ \delta \in 3 [- \Delta , \Delta] $; (solid, \textbf{{\color{green} green}}) adaptive-nuisance. Beyond being independent of the ratio $ \Delta / \sigma $, by definition, note that at $ n \times \sigma $ the nG method always provides a CL whose size is $ n $.}\label{fig:graphicCI}
\end{figure}

The fundamental usefulness of p-values is to quote confidence intervals in the extraction of fundamental parameters. Therefore, the first comparison to be made concerns the different properties of the CL intervals of the different approaches we have discussed. In practice, we would like to find a method which does not quote too large $ 68~\% $ CL intervals, or otherwise we would not learn much from data, but on the other hand is rather ``conservative" for signalizing tensions with the Standard Model.

Figure \ref{fig:graphicCI} shows the p-value curves for different models of theoretical uncertainty, and the same information under a more quantitative form can be seen in Table \ref{tab:compar1Derrors}. Note the following characteristics, independent of the ratio $ \Delta / \sigma $:

\begin{itemize}
	\item At $ 1 \sigma $ (important for quoting the metrology): fixed-3 nuisance provides broader CL, while the other methods provide similar CL. (By definition, fixed-1 nuisance and adaptive nuisance give exactly the same $ 1 \, \sigma $ CL.)
	\item At $ 3 \sigma $ (evidence of tension): fixed-3 nuisance and adaptive nuisance provide broader CL than the others methods (and identical $ 3 \, \sigma $ CL in these cases, by definition). 
	\item At $ 5 \sigma $ (threshold for discovery): broader CL are given by adaptive nuisance, followed by nG or fixed-3 nuisance.
\end{itemize}

\noindent
As one increases $ \Delta / \sigma $, note the following:

\begin{itemize}
	\item Globally, the different methods give very similar answers for small $ \Delta / \sigma $, as expected, and for intermediate or large $ \Delta / \sigma $ the differences between the confidence intervals may scale as a factor 2 for large or intermediate confidence levels.
	\item Fixed-1 external (or Rfit) and fixed-1,3 nuisance show flat bottoms, and ``saturate" for  large $ \Delta / \sigma $, meaning that the size of the confidence intervals grows slowly with the number of ``units of $ \sigma $" compared to the other methods.
	\item nG and adaptive-nuisance are similar for low and large $ \Delta / \sigma $, and in the examples shown more important differences happen for $ \Delta / \sigma \simeq 1 $. 
\end{itemize}

\begin{table}
\begin{center}
\begin{tabular}{cc|cccc}
& & nG & fixed-1 nuisance & adapt. nuisance & fixed-1 ext./Rfit \\
\hline
&$1\sigma$ & 1.0 & 1.0 & 1.0 & 1.2 \\
$\frac{\Delta}{\sigma}=0.3$&$3\sigma$ & 3.0 & 3.0 & 3.5 & 3.2 \\
&$5\sigma$ & 5.0 & 5.0 & 6.1 & 5.1 \\
\hline
&$1\sigma$ & 1.0 & 1.1 & 1.1 & 1.4 \\
$\frac{\Delta}{\sigma}=1$&$3\sigma$ & 3.0 & 2.7 & 4.1 & 2.8 \\
&$5\sigma$ & 5.0 & 4.1 & 7.0 & 4.2 \\
\hline
&$1\sigma$ & 1.0 & 1.1 & 1.1 & 1.3 \\
$\frac{\Delta}{\sigma}=3$&$3\sigma$ & 3.0 & 1.8 & 3.7 & 1.9 \\
&$5\sigma$ & 5.0 & 2.5 & 6.3 & 2.5 \\
\hline
&$1\sigma$ & 1.0 & 1.0 & 1.0 & 1.1 \\
$\frac{\Delta}{\sigma}=10$&$3\sigma$ & 3.0 & 1.3 & 3.3 & 1.3 \\
&$5\sigma$ & 5.0 & 1.5 & 5.5 & 1.5 \\
\hline
\end{tabular}
\caption{\it Comparison of the size of one-dimensional confidence intervals at $1,3,5 \sigma$ for various methods and various values of $\Delta/\sigma$.
\label{tab:compar1Derrors}}
\end{center}
\end{table}

Therefore, none of the methods gives systematically the broadest CL interval for all significances. At large significances, however, adaptive nuisance is always ``more conservative" than the other methods, and for low and large $ \Delta / \sigma $ nG gives similar intervals. Whether or not the methods are aggressive for large ratios and large significances in the precise meaning of Eq.~\eqref{eq:defAgressiveConservative} is going to be discussed in Section \ref{sec:coverageComparison}.


\subsubsection{Significance}

When comparing a prediction such as $ \mu = 0 $ with a measurement $ X_0 \neq 0 $ one would like to quantify the tension between both, which surely depends on the quoted values for the statistical variance, $ \sigma^2 $, and systematic uncertainty, $ \Delta $. 

The comparison of significances can be qualitatively seen from Figure \ref{fig:graphicCI}: if the size of the uncertainty is fixed, or in other words the size of the CL interval, the different approaches quote different significances for this same CL interval. One can take the observations made previously in a reverse way: if confidence intervals for a method are broader, then a possible tension becomes less significant. A more quantitative comparison can be seen in Table \ref{tab:sigthresholds1}. In this table, built for the special case $\Delta/\sigma=1$, one clearly sees that if for instance the random approach is employed to quote a significance of $ 5 \, \sigma $ (last table), the adaptive nuisance approach would be more ``cautious," quoting $ 3.6 \, \sigma $ instead, while the other methods seen in table would quote even higher significances compared to nG. 

\begin{table}
\begin{center}
\begin{tabular}{c|cccc|}
\cline{2-5}
1 $\sigma$ signif. threshold  & nG & 1-nuisance & adaptive nuisance & 1-external/Rfit \\
\hline
nG & \textbf{1} & 0.9 & 1.0 & 0.4 \\\cline{2-5}
fixed-1 nuisance & 1.1 & \textbf{1} & 1.0 & 0.5 \\\cline{2-5}
adaptive nuisance & 1.1 & 1.0 & \textbf{1} & 0.5 \\\cline{2-5}
fixed-1 external/Rfit & 1.4 & 1.4 & 1.2 & \textbf{1}
\\\cline{1-5}
\\\cline{2-5} 3 $\sigma$  signif. threshold  & nG & $1$-nuisance & adaptive nuisance & $1$-external/Rfit \\
\hline
nG & \textbf{3} & 3.4 & 2.3 & 3.2 \\\cline{2-5}
fixed-1 nuisance & 2.7 & \textbf{3} & 2.0 & 2.8 \\\cline{2-5}
adaptive nuisance & 4.1 & 4.9 & \textbf{3} & 4.8 \\\cline{2-5}
fixed-1 external/Rfit & 2.8 & 3.2 & 2.1 & \textbf{3}
\\\cline{1-5}
\\\cline{2-5} 5 $\sigma$  signif. threshold  & nG & $1$-nuisance & adaptive nuisance & $1$-external/Rfit \\
\hline
nG & \textbf{5} & 6.2 & 3.6 & 6.1 \\\cline{2-5}
fixed-1 nuisance & 4.1 & \textbf{5} & 3.0 & 4.9 \\\cline{2-5}
adaptive nuisance & 7.0 & $ > $8 & \textbf{5} & $ > $8 \\\cline{2-5}
fixed-1 external/Rfit & 4.2 & 5.1 & 3.1 & \textbf{5} \\
\hline
\end{tabular}
\end{center}
\caption{\it Comparison of 1D $1,3,5 \sigma$ significance thresholds for $\Delta/\sigma=1$. For instance, the first line should read: if with nG a p-value=1 $\sigma$ is found, then the corresponding values for the three other methods are 0.9/1.0/0.4 $\sigma$.}\label{tab:sigthresholds1}
\end{table}

As a practical example, consider the anomalous magnetic moment of the muon

\begin{equation}
a_{muon} \equiv \frac{g_{muon} - 2}{2} \, ,
\end{equation}
whose measurement, $ a_{muon}^{exp} $, compared to the Standard Model prediction, $ a_{muon}^{SM} $, is

\begin{figure}[H]
\hspace{0.cm}
\begin{minipage}{0.5\textwidth}
	\begin{center}
		\begin{eqnarray}
			(a_{muon}^{exp} - a_{muon}^{SM}) \times 10^{11} = 288 \pm 63_{exp} \pm 49_{SM} \nonumber
		\end{eqnarray}
	\end{center}
\end{minipage}
\hspace{1.5cm}
\begin{minipage}{0.5\textwidth}
	\begin{tabular}{c|c}
	\multicolumn{2}{c}{significance ($ \mathcal{H}_0 : \mu_t = 0 $)} \\
	\hline
	nG & $ 3.6 \, \sigma $ \\
	\hline
	1-external/Rfit & $ 3.8 \, \sigma $ \\
	\hline
	1-nuisance & $ 3.9 \, \sigma $ \\
	\hline
	adapt. nuisance & $ 2.7 \, \sigma $ \\
	\hline
	\end{tabular}
\end{minipage}
\end{figure}
\noindent
where the first uncertainty comes from the experimental results and is treated as a statistical uncertainty, while the second uncertainty comes from the SM prediction and is treated as a theoretical uncertainty. Note that for the given uncertainties, the three different approaches quote very different significances: while nG, 1-external (or Rfit) and 1-nuisance quote an evidence for New Physics, the adaptive nuisance procedure does not imply a significance beyond the $ 3 \, \sigma $ threshold. This example clarifies that, when large tensions are claimed and the prediction or/and the measurements presents large theoretical uncertainties, it is fundamental to pay some special attention to the way this class of uncertainties is interpreted, since it may change the quantitative information about the size of the deviation.




\subsubsection{Coverage}\label{sec:coverageComparison}

p-values only have a meaning if the confidence intervals they quote have good coverage properties, i.e. exact coverage or not excessive overcoverage. This guarantees the robustness of a test statistic, since uncontrolled levels of undercoverage may be too risky for excluding a hypothesis when it is true, and large overcoverage limits too much our capacity of extracting the true value of fundamental quantities.

Based in the discussion made in Section \ref{sec:pvalueCoverage}, for given $ \mu_t $ and $ \delta_t $, we generate a set of $ n $ values $ \{ X^{(1)}_0 , \ldots , X^{(n)}_0 \} $ from the distribution of $ X $. For each of these values, we calculate the p-values $ p (X^{(i)}_0 ; \mu_t) $, $ i = 1, \ldots, n $, for the same $ \mu_t $ and $ \delta_t $ and consider for a given $ \alpha $ the fraction of p-values which is higher than $ \alpha $ itself. For exact coverage, this fraction must be equal to $ 1 - \alpha $ when $ n $ goes to infinity, and for over (under) coverage this fraction is higher (lower) than $ 1 - \alpha $. The values of $ \mathcal{P} [p > \alpha \vert \mathcal{H}_\mu] $, i.e. the fraction of p-values covering the true value $ \mu_t $, are shown in Tables~\ref{tab:coverage2} and \ref{tab:coverage1} for different methods and different CL intervals:

\begin{center}
$ \alpha = 0.3173, 0.0455, 0.0027 \quad $ for $ \quad 68.27~\%, 95.45~\%, 99.73~\% $~CL.
\end{center}

Notice that when the true value of $ \delta $ is included in the expected interval, called ``fortunate'' cases, all approaches present exact or overcoverage, except for a slight undercoverage of nG when $ \delta \neq 0 $ in the example. Then, when moving to the ``unfortunate'' cases where the true value of $ \delta $ is outside the interval $ \Omega $, where it was supposed to be found, the approaches systematically undercover. An exception is the adaptive approach, for which $ \Omega $ grows proportionally to the confidence interval we want to build, thus showing good coverage properties as the size of the confidence interval grows.

\begin{table}
\begin{center}

\begin{tabular}{cccc}
& 68.27\% CL & 95.45\% CL & 99.73\% CL \\
\hline
$\Delta / \sigma = 1, \; \delta / \Delta = 0$ & \phantom{68.27\% CL} & \phantom{95.45\% CL} & \phantom{99.73\% CL} \\
nG & 84.1\% & 99.5\% & 100.0\% \\
$1$-nuisance & 86.5\% & 99.3\% & 100.0\% \\
adaptive nuisance &86.4\% & 100.0\% & 100.0\% \\
$1$-external/Rfit & 95.4\% & 99.7\% & 100.0\% \\
$1$-ext./Rfit (excl. $p\equiv 1$) & 85.5\% & 99.1\% & 100.0\% \\
\end{tabular} 

\begin{tabular}{cccc}
\hline
$\Delta / \sigma=3, \; \delta / \Delta=0$ & \phantom{68.27\% CL} & \phantom{95.45\% CL} & \phantom{99.73\% CL} \\
nG & 99.8\% & 100.0\% & 100.0\% \\
$1$-nuisance & 100.0\% & 100.0\% & 100.0\% \\
adaptive nuisance &99.9\% & 100.0\% & 100.0\% \\
$1$-external/Rfit & 100.0\% & 100.0\% & 100.0\% \\
$1$-ext./Rfit (excl. $p\equiv 1$) & 98.5\% & 100.0\% & 100.0\% \\
\end{tabular} 

\begin{tabular}{cccc}
\hline
$\Delta/\sigma=3, \; \delta/\Delta=1$ & \phantom{68.27\% CL} & \phantom{95.45\% CL} & \phantom{99.73\% CL} \\
nG & 56.3\% & 100.0\% & 100.0\% \\
$1$-nuisance & 68.1\% & 95.5\% & 99.7\% \\
adaptive nuisance & 68.2\% & 100.0\% & 100.0\% \\
$1$-external/Rfit & 84.1\% & 97.7\% & 99.9\% \\
$1$-ext./Rfit (excl. $p\equiv 1$) & 68.2\% & 95.4\% & 99.7\% \\
\hline
\end{tabular} 
\end{center}
\caption{\it Coverage properties of the various methods in ``fortunate'' cases where the true value of $\delta/\Delta$ is contained in (or at the border of) the volume $\Omega$ (for all confidence intervals). Since the  1-external (or Rfit) approach produces clusters of $p$-values equal to 1 due to a \textit{plateau}, the coverage values excluding this \textit{plateau} are also considered.}\label{tab:coverage2}
\end{table}


\begin{table}
\begin{center}

\begin{tabular}{cccc}
& 68.27\% CL & 95.45\% CL & 99.73\% CL \\
\hline
$\Delta/\sigma=1, \; \delta/\Delta=1$ & \phantom{68.27\% CL} & \phantom{95.45\% CL} & \phantom{99.73\% CL} \\
nG & 65.2\% & 96.6\% & 99.9\% \\
$1$-nuisance & 68.2\% & 95.4\% & 99.7\% \\
adaptive nuisance & 68.3\% & 99.6\% & 100.0\% \\
$1$-external  & 83.9\% & 97.8\% & 99.9\% \\
$1$-external (excl. $p\equiv 1$) & 69.2\% & 95.7\% & 99.8\% \\
\end{tabular} 

\begin{tabular}{cccc}
\hline
$\Delta/\sigma=1, \; \delta/\Delta=3$ & \phantom{68.27\% CL} & \phantom{95.45\% CL} & \phantom{99.73\% CL} \\
nG & 5.76\% & 43.2\% & 89.1\% \\
$1$-nuisance & 6.60\% & 38.0\% & 78.4\% \\
adaptive nuisance & 6.53\% & 75.4\% & 99.8\% \\
$1$-external & 16.0\% & 50.3\% & 84.2\% \\
$1$-external (excl. $p\equiv 1$) & 14.0\% & 49.1\% & 83.8\% \\
\end{tabular} 

\begin{tabular}{cccc}
\hline
$\Delta/\sigma=3, \; \delta/\Delta=3$ & \phantom{68.27\% CL} & \phantom{95.45\% CL} & \phantom{99.73\% CL} \\
nG & 0.00\% & 0.35\% & 68.7\% \\
$1$-nuisance & 0.00\% & 0.00\% & 0.07\% \\
adaptive nuisance & 0.00\% & 9.60\% & 99.8\% \\
$1$-external & 0.00\% & 0.00\% & 0.13\% \\
$1$-external (excl. $p\equiv 1$) & 0.00\% & 0.00\% & 0.13\% \\
\hline
\end{tabular} 
\end{center}
\caption{\it Coverage properties of the various methods  in ``unfortunate'' cases where the true value of $\delta/\Delta$ is not contained in (or at the border of) the volume $\Omega$ (for all the confidence intervals). Since the 1-external approach produces clusters of $p$-values equal to 1 due to a \textit{plateau}, the coverage values excluding this \textit{plateau} are also considered.}\label{tab:coverage1}
\end{table}



\section{Combining data}\label{sec:lambdaCholesky}

In this section we consider the combination of many possibly correlated measures of the same observable, first in the pure statistical case, and then in the case where theoretical uncertainties are present. 








\subsection{Pure statistical case}

We first introduce the definitions of covariance and correlation matrices in the case where no theoretical uncertainties are present. In Appendix~\ref{sec:cumberInverse}, we study the cumbersome case where the inverse of the covariance matrix, necessary in order to build the usual test statistic, is not (naively) defined. 

Consider now a set of random variables $ X^{(1)}, X^{(2)}, X^{(3)}, \ldots, X^{(n)} $ equally distributed $ X^{(i)} \sim \mathcal{N}_{(\mu_t, \sigma_i)} $ with same mean $ \mu_t $. We want to study the estimator $ \widehat{\mu} $ of $ \mu_t $ under the case where these random variables are correlated. Following the comments in Section~\ref{sec:candy}, we minimize the quadratic test statistic to estimate $ \mu $

\begin{equation}
   T(X,\mu) =(X-\mu U)^T \cdot C_s^+ \cdot (X-\mu U) \, ,
\end{equation}
where $ U $ is a column of entries ``1" $ n $ times, $ X $ is the column vector

\begin{center}
$ (X^{(1)}, X^{(2)}, X^{(3)}, \ldots, X^{(n)})^T $,
\end{center}
\noindent
and $ C_s^+ $ is the inverse of the covariance matrix $ C_s $, which is a (semi-)positive definite matrix defined as


\begin{equation}
( C_s )_{i j} = E[(X^{(i)} - \mu_t) (X^{(j)} - \mu_t)] \equiv \rho_{i j} \sigma_i \sigma_j \, ,
\end{equation}
where the correlation matrix $ \rho_{i j} = E[(X^{(i)} - \mu_t) (X^{(j)} - \mu_t)] / (\sigma_i \sigma_j) $ is squared and symmetric.

Following the minimization of $ T $, $ \widehat\mu $ and $ \sigma_\mu $ are given as follows

\begin{eqnarray}\label{eq:estimatorNoTheo}
&& \widehat\mu=\frac{U^T \cdot C_s^+ \cdot X}{U^T \cdot C_s^+ \cdot U}= w^T \cdot X \, , \qquad 
\sigma_\mu^2=\frac{U^T \cdot C_s^+ \cdot C_s \cdot C_s^+ \cdot U}{(U^T \cdot C_s^+ \cdot U)^2}=w^T \cdot C_s \cdot w \, , \nonumber\\
&& \qquad w=\frac{ C_s^{+} \cdot U }{U^T \cdot C_s^+ \cdot U} \, , \qquad \sum_{i=1}^n w_i = 1 \, ,
\end{eqnarray}
where $ w_i $ are the weights, and $ \sigma^2_\mu $ is the variance. The above average is efficient in the sense that it is the unbiased estimator of the true value $ \mu_t $ which minimizes the variance.

\subsection{Theoretical uncertainties}


We now consider the case where correlated theoretical uncertainties are present. To make things simpler, we focus specifically in the nuisance approach case. The external approach has a similar discussion if an \textit{ad hoc} overall normalization for Eq.~\eqref{eq:testExternal} is assumed, in order to balance the contributions to the average according to the theoretical uncertainties. 



\begin{figure}
	\centering
	\includegraphics[scale=0.4]{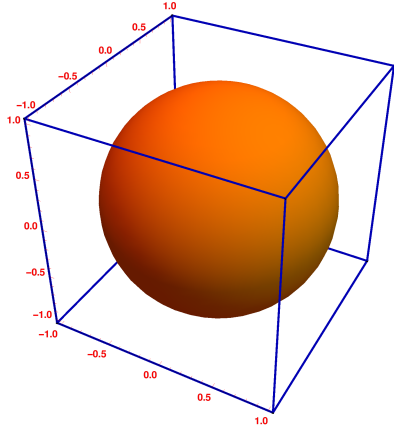}
	\caption{\it Illustration in a three-dimensional example of the (hyper-)space over which we maximize the p-value in the two different cases discussed in the text: hypercube and hyperball. In the former, the three normalized theoretical uncertainties $ \tilde\delta_{1,2,3} $ can assume any value inside the interval $ [-1, 1] $, including $ | \tilde\delta_{1,2,3} | = 1 $ simultaneously, while for the hyperball the phase space is reduced, and avoids these ``fine-tuned" possibilities where all the uncertainties have their extreme admited value.}\label{fig:sphereInABox}
\end{figure}


We start with the following test statistic, cf. Eq.~\eqref{eq:quadraticTestNuisanceMotivation}

\begin{equation}\label{eq:startingTest}
T_{\delta} (X, \mu) = (X-\mu U-\Delta\tilde\delta)^T\cdot W_s \cdot (X-\mu U-\Delta\tilde\delta)+\tilde\delta^T \cdot\widetilde{W}_t \cdot \tilde\delta \, ,
\end{equation}
where $ W_s $ is the inverse $ C^{-1}_s $ (or the generalized inverse $ C^+_s $, see Appendix~\ref{sec:cumberInverse}) of the $ n \times n $ statistical covariance matrix, $ \widetilde{W}_t $ is the inverse $ \widetilde{C}^{-1}_t $ (or the generalized inverse $ \widetilde{C}^+_t $, see Appendix~\ref{sec:correlatedTheo}) of the $ m \times m $ theoretical correlation matrix, and the true values of the theoretical uncertainties $ \tilde{\delta}_1, \ldots, \tilde{\delta}_m $ are normalized, which is indicated by the use of ``tilded" symbols. The $ n \times m $ matrix $ \Delta $ tells the dependence of the $ i- $th measurement on the $ j- $th theoretical uncertainty, where $ i $ runs from 1 to $ n $, and $ j $ runs from 1 to $ m $, i.e. the quoted measure is $ X^{(i)} \pm \sigma_i \pm \Delta_{i 1} \pm \ldots \pm \Delta_{i m} $.

The term ``correlation matrix'' for theoretical uncertainties is in fact an abuse: they are not random variables, so that ``correlation" does not apply as in the statistical sense. Note, however, that the way we have introduced them in Eq.~\eqref{eq:startingTest} is quite symmetric with respect to the statistical case.

We then minimize over the set of theoretical uncertainties, resulting in the following test statistic

\begin{equation}
T(X, \mu) = (X-\mu)^T \cdot \overline{W} \cdot (X-\mu) \, ,
\end{equation}
where

\begin{equation}
\overline{W} = W_s - B^T\cdot A^{-1}\cdot B \, , \qquad\qquad
B = ( W_s \cdot \Delta)^T \, , \qquad\qquad
A = \widetilde{W}_t +B \cdot \Delta \, .
\end{equation}

\noindent
The above test statistic results then in the following estimator

\begin{eqnarray}\label{eq:averagesCorrelated}
&& \widehat\mu = \sum_i w_i X^{(i)} \, , \qquad\qquad
\sigma_\mu^2 = \sum_{i,j} w_iw_j (C_s)_{ij} \, , \nonumber\\
&& w_i = \sum_j \overline{W}_{ij} \times \left( \sum_{i,j} \overline{W}_{ij} \right)^{-1} \, ,
\end{eqnarray}
which is biased with bias given by

\begin{equation}\label{eq:biasCorrelatedCase}
\delta_\mu=\sum_i w_i \Delta_i \tilde\delta_i \, .
\end{equation}

We now discuss the way of varying the different $ \tilde\delta_i $ according to the last expression. The case without correlations is trivial to understand, and it corresponds to letting $ \tilde\delta_i $ vary independently in the space we want them to be defined: (\textit{linear}) one could understand that they should be varied over a hypercube when maximizing the p-value, which implies in particular that they can all assume at the same time their extreme values, i.e. the corners of the hypercube in Figure~\ref{fig:sphereInABox} are accessible; or (\textit{quadratic}) over a hyperball, avoiding the ``fine-tuned" corners of the last case, see Figure~\ref{fig:sphereInABox}. Therefore, depending on the approach we have

\begin{equation}\label{eq:nDtheo}
\Delta_\mu = \sum_i \left| w_i \Delta_i \right| \ \ \quad (\mathrm{linear}) \, , \qquad\qquad
\Delta_\mu = \sqrt{\sum_i \left( w_i \Delta_i \right)^2} \ \ \quad (\mathrm{quadratic}) \, ,
\end{equation}
recovering the ansatz of \cite{Porter:2008uw} in the linear case.

Now, when correlations are present in Eq.~\eqref{eq:biasCorrelatedCase}, it should mean that they cannot be varied independently: in particular, when they are all totally correlated, one would expect to have only a particular combination of them that can be varied. This is formalized and illustrated in Appendix~\ref{sec:correlatedTheo}.

\begin{table}
\begin{center}
{\small
\begin{tabular}{cccccc}
Reference     &   & Mean &  Stat &  Theo\\
\hline
Exclusive & \cite{Amhis:2014hma} & 3.28 & $\pm$ 0.15 &$\pm$  0.26\\ 
Inclusive & \cite{Amhis:2014hma} & 4.359 & $\pm$ 0.180 & $\pm 0.013 \pm 0.027 \pm 0.037\pm 0.161 \pm 0.200$\\ 
\end{tabular}

\vspace{0.4cm}

\begin{tabular}{ccccc}
Method & Average & 1 $\sigma$ CI & 3 $\sigma$ CI &5 $\sigma$ CI \\
\hline
nG & $3.79\pm0.22   \pm0 $ & $ 3.79 \pm0.22$ & $ 3.79 \pm0.65$ & $ 3.79 \pm1.1$\\
naive Rfit & $3.70\pm0.12   \pm0$ & $3.70 \pm0.12$ & $3.70 \pm0.35$ & $3.70 \pm0.58$\\
educ Rfit & $3.70\pm0.11   \pm0.26$ & $3.70 \pm0.38$ & $3.70 \pm0.61$ & $3.70 \pm0.84$\\
1-hypercube & $ 3.79\pm0.12   \pm0.34$ & $ 3.79 \pm0.40$ & $ 3.79 \pm0.67$ & $ 3.79 \pm0.91$\\
adapt hyperball & $ 3.79\pm0.12   \pm0.18 $ & $ 3.79 \pm0.24$ & $ 3.79 \pm0.88$ & $ 3.79 \pm1.49$
\end{tabular}}
\end{center}
\caption{\it Top: determinations of $|V_{ub}|\cdot 10^3$ from semileptonic decays. Note that we decompose the different sources of theoretical uncertainty for the inclusive determination of $|V_{ub}|$, while in Chapter~\ref{ch:SM} they were combined linearly accordingly to the Rfit approaches discussed in this context. Bottom: averages according to the various methods, and corresponding confidence intervals for various significances. The uncertainties are distinguished at $ 1 \, \sigma $ according to Eqs.~\eqref{eq:averagesCorrelated} and \eqref{eq:biasCorrelatedCase}.}\label{tab:inputsVub}
\end{table}

\begin{table}
\begin{center}
{\small
\begin{tabular}{cccccc}
Reference     &   & Mean &  Stat &  Theo\\
\hline
Exclusive & \cite{Amhis:2014hma} & 38.99 & $\pm 0.49$ & $\pm 0.04\pm  0.21\pm  0.13\pm  0.39\pm  0.17\pm  0.04\pm 0.19$\\ 
Inclusive & \cite{Amhis:2014hma} & 42.42 & $\pm 0.44$ & $\pm 0.74$\\ 
\end{tabular}

\vspace{0.4cm}

\begin{tabular}{ccccc}
Method & Average & 1 $\sigma$ CI & 3 $\sigma$ CI &5 $\sigma$ CI \\
\hline
nG & $40.41\pm0.55  \pm0$ & $40.41\pm0.55$ & $40.41 \pm1.66$ & $40.41 \pm2.77$\\
naive Rfit & $41.00\pm0.33 \pm0 $ & $41.00\pm0.32$ & $41.00 \pm0.98$ & $41.00\pm1.64$\\
educ Rfit & $41.00\pm0.33 \pm0.74$ & $41.00 \pm1.07$ & $41.00 \pm1.72$ & $41.00 \pm2.38$\\
1-hypercube & $40.41\pm0.34  \pm0.99$ & $40.41 \pm1.15$ & $40.41 \pm1.94$ & $40.41\pm2.65$\\
adapt hyperball & $40.41\pm0.34 \pm0.44$ & $40.41 \pm0.60$ & $40.41 \pm2.26$ & $40.41 \pm3.84$
\end{tabular}}
\end{center}
\caption{\it Same as Table~\ref{tab:inputsVub} for $|V_{cb}|\cdot 10^3$.}\label{tab:inputsVcb}
\end{table}

\subsection{Examples}

\begin{figure}
	\centering
	\includegraphics[scale=0.55]{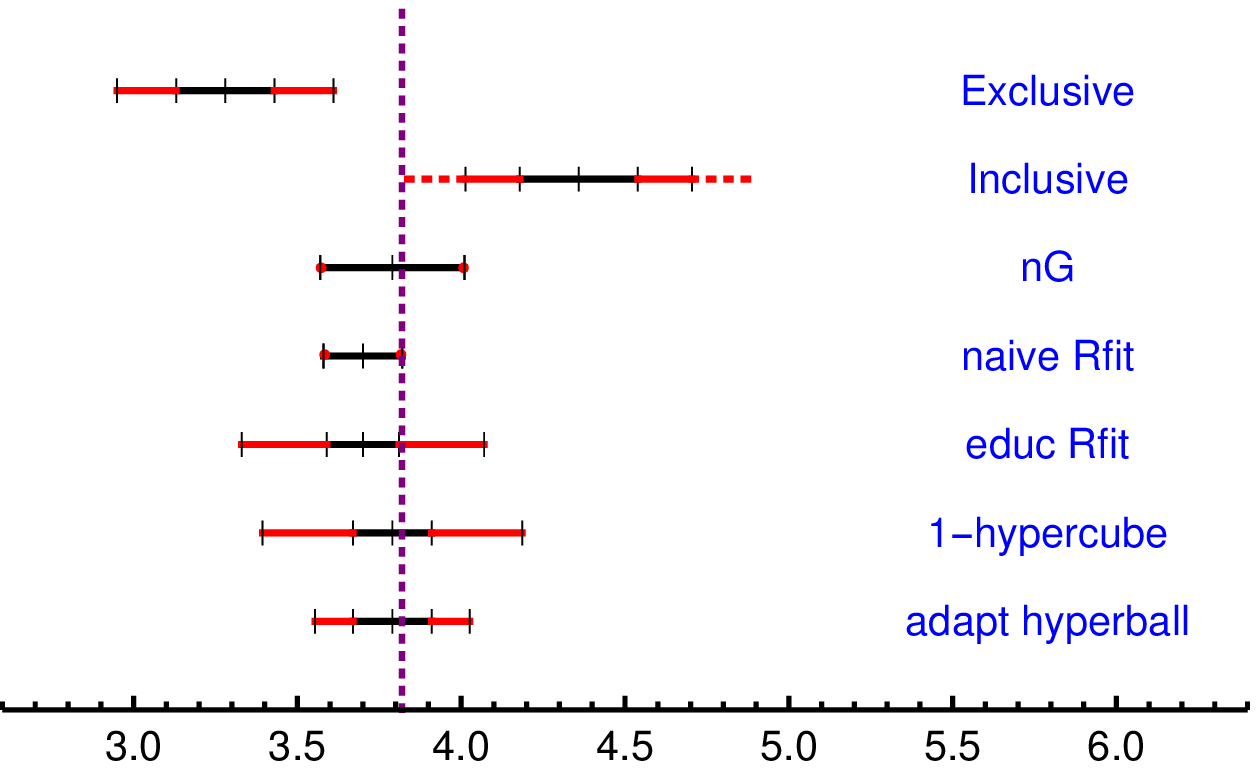}
	\hspace{0.2cm}
	\includegraphics[scale=0.55]{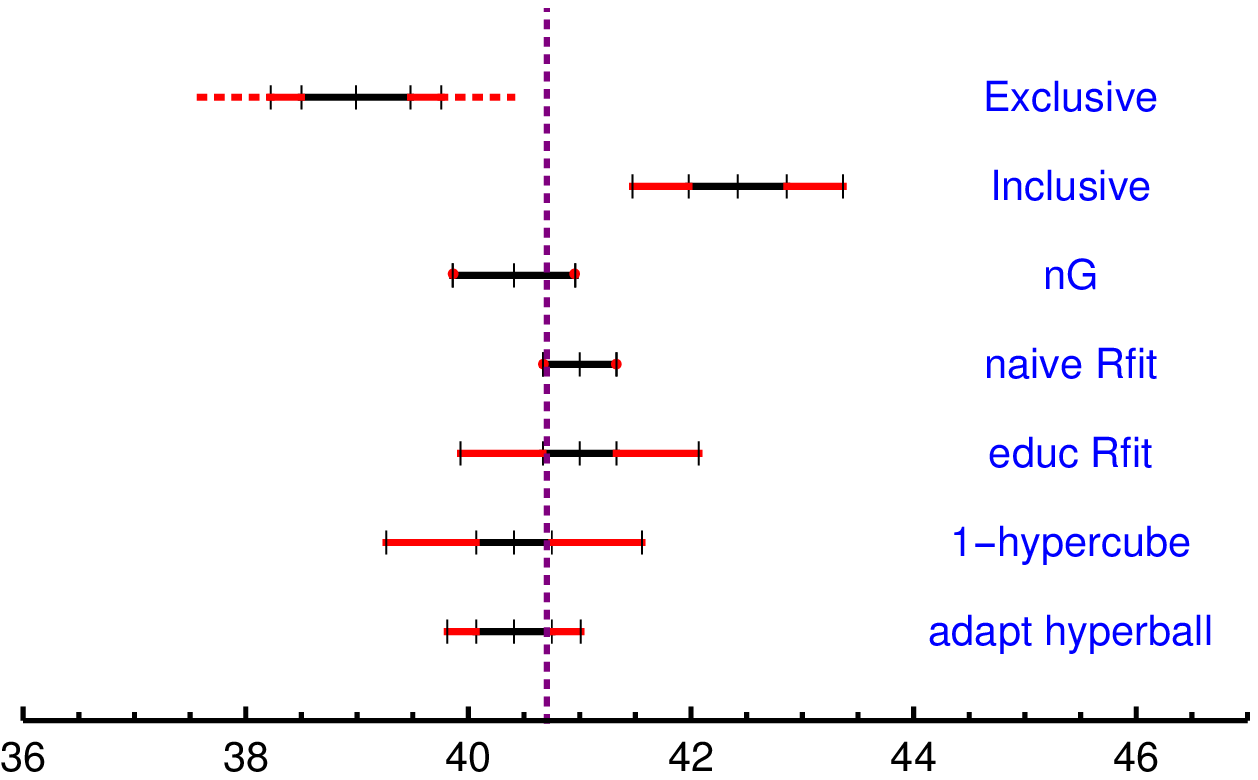}
	\caption{\it Comparison between the different approaches for $|V_{ub}|$ (left) and $|V_{cb}|$ (right) at $ 1 \, \sigma $, where the purple dotted-line indicates the simple average of the two classes on inputs, namely inclusive and exclusive extractions. The black intervals indicate the size of the statistical uncertainties, while in solid red we indicate the remaining theoretical uncertainty. Instead, the dotted red lines indicate the linear combination of the individual uncertainties seen in Tables~\ref{tab:inputsVub} and \ref{tab:inputsVcb} for the inclusive extraction of $|V_{ub}|$ (left) and the exclusive extraction of $|V_{cb}|$ (right): in each of the two cases, the red intervals give $ 68~\% $ CL intervals. The incompatibility between the inclusive and exclusive input is clearly seen.}\label{fig:graphicalRepVubVcb}
\end{figure}

We consider the same physical-oriented example already discussed in Section~\ref{sec:CKMfitterPhilo}, namely the case of incompatible measurements. The two different procedures we call naive Rfit and educated Rfit were already discussed at that point, where it has been noted that the former implies no theoretical uncertainty in the average of incompatible measurements (no flat \textit{plateau}) while the latter quotes the minimum of the two theoretical uncertainties as the resulting theoretical uncertainty. Here we compare these two methods with the pure statistical case (naive Gaussian) and the nuisance approach with fixed hypercube and adaptive hyperball volumes (though the p-values for the nuisance and external approaches have different interpretations, they lead to the same expressions for the combined variance and theoretical uncertainty, once the test statistic in the external approach is defined suitably). The results can be seen numerically in Tables~\ref{tab:inputsVub} and \ref{tab:inputsVcb}, or graphically at $ 1 \, \sigma $ in Figure~\ref{fig:graphicalRepVubVcb}.

The first difference to highlight is that the different methods quote different central values: in the comparisons we have made in Section~\ref{sec:impactmodelling} this was not possible because we had a single measurement $ X_0 $. Now, concerning the uncertainties, also note that at $ 1 \, \sigma $ the 1-hypercube approach quotes larger uncertainties, while naive Gaussian and naive Rfit are more aggressive in their averages. At larger confidence intervals, the adaptive hyperball becomes the more conservative, while the naive Rfit model is the more aggressive for the reason already mentioned.

\section{Global fit of flavour observables}\label{sec:globalFit}


So far, we have considered the impact of the modeling of theoretical uncertainties in one-dimensional cases where there is only one parameter $ \mu $ that we would like to extract. In this section we are going to consider an example of utmost interest for flavour physics, which is the extraction of the Wolfenstein parameters $ A, \lambda, \bar{\rho}, \bar{\eta} $. For our purposes, it is sufficient to consider a subset of the observables given in Chapter~\ref{ch:SM}, Table~\ref{tab:expinputs}.

\begin{table}
\begin{center}
\begin{tabular}{cc}
Observable & Input\\
$|V_{ud}|_{\rm nucl}$ & $0.97425\pm 0\pm 0.00022$\\
$|V_{ub}|$ & $(3.70\pm 0.12\pm 0.26)\times 10^{-3}$\\
$|V_{cb}|$ & $(41.00\pm 0.33\pm 0.74)\times 10^{-3}$ \\
$\Delta m_{d}$ & $(0.510\pm 0.003)$ ps$^{-1}$ \\
$\Delta m_{s}$ & $(17.757\pm 0.021)$ ps$^{-1}$ \\
$\hat{B}_{B_s}/\hat{B}_{B_d}$ & $1.023\pm 0.013\pm0.014$\\
$\hat{B}_{B_s}$ & $1.320\pm 0.017\pm0.030$\\
$f_{B_s}/f_{B}$ & $1.205\pm 0.004\pm 0.007$\\
$f_{B_s}$ & $225.6\pm 1.1\pm 5.4$ MeV\\
$\eta_B$ & $0.5510\pm 0\pm 0.0022$\\
$\bar{m}_t$ & $165.95\pm 0.35 \pm 0.64$ GeV\\
\end{tabular}
\caption{\it Set of inputs considered over this section for the extraction of $ \{ A, \lambda, \bar{\rho}, \bar{\eta} \} $, consisting of observables dominated by theoretical uncertainties. The numerical values correspond to those used by the \texttt{CKMfitter} Collaboration as of Summer 14 (therefore $|V_{ub}|$, $\hat{B}_{B_s}$, $f_{B_s}/f_{B}$ and $f_{B_s}$ are not exactly the same as in Chapter~\ref{ch:SM}).}\label{tab:inputssystdominated}
\end{center}
\end{table}


Following the discussion we have had so far, where the nuisance has been considered for definiteness, cf. Eq~\eqref{eq:startingTest}, we consider minimizing the quadratic form


\begin{equation}\label{eq:testForGlobalFit}
T_\delta (X_0 ; \mu) = \sum^{N}_{i=1} \left( \frac{X_{i,0} - x_i (\mu, \delta)}{\Sigma_i} \right)^2 \, ,
\end{equation}
where the measurements $ X_{i,0} $ have respective uncertainties $ \Sigma_i $ (which we suppose non-correlated over this section for simplicity). Note that the true values of the theoretical uncertainties have been absorbed into the notation of the SM prediction, namely the set of parameter $ \delta $ in the expression for the observables $ x_i (\mu, \delta) $. The sum runs over the total number $ N $ of observables we dispose, see Table~\ref{tab:inputssystdominated}, and $ \mu $ is a short-cut for the parameters we would like to extract: $ \{ A, \lambda, \bar{\rho}, \bar{\eta} \} $ in the example we consider.



The best-fit point $ \hat{\chi} \equiv \{ \hat{\mu}, \hat{\delta} \} $ can be determined using the full quadratic form above, Eq.~\eqref{eq:testForGlobalFit}. In order to derive analytically the intervals of the parameters we want to extract, $ \{ A, \lambda, \bar{\rho}, \bar{\eta} \} $, we are going to consider the following linear model

\begin{equation}
x_i (\chi_j) = \hat{x}_i + \sum_k a_{ik} (\chi_k - \hat{\chi}_k) + \mathcal{O} [ (\chi - \hat{\chi})^2 ] \, ,
\end{equation}
with

\begin{equation}
\hat{x}_i = x_i (\hat{\chi}_j) \, , \qquad a_{ik} = \frac{\partial x_i}{\partial \chi_k} \Bigg|_{\chi = \hat{\chi}} \, ,
\end{equation}
where the index $ j $ runs over the full set of parameters, $ \{ \mu_1, \ldots, \mu_m, \delta_1, \ldots, \delta_n \} $. We stress that over this section our goal is to perform an illustrative comparison between the different theoretical models, and that in actual global fits produced by the \texttt{CKMfitter} Collaboration this simplifying step is not taken (but in any case the differences are not important even for large confidence intervals). The advantage of considering the linear model is that we can compute analytical expressions for the best values of $ \{ A, \lambda, \bar{\rho}, \bar{\eta} \} $ and their related statistical and theoretical uncertainties.

Then one can rewrite the minimum of Eq.~\eqref{eq:testForGlobalFit} under the matrix form
\begin{equation}\label{eq:linsol}
\sum_{i=1}^{N} \sum_{j=1}^{m+n} \frac{a_{ij}a_{ik}}{\Sigma_i^2} \hat\chi_j =
\sum_{i=1}^{N} \frac{a_{ik}}{\Sigma_i^2}\left( X_{i,0}-\hat x_i\right)  
\Leftrightarrow M\cdot\hat\chi = A+B \cdot X_0 \, ,
\end{equation}
where we define implicitly $ M, A $ and $ B $. Since measurements $X_i$ and parameters are linearly related, the statistical uncertainty for $\hat\chi_k$ can readily be obtained from the variances $\sigma_{X_i}$
\begin{equation} \label{eq:minlinstat}
\sigma_{\hat\chi_k}^2 = \sum_{i,j=1}^n\left[(M^{-1})_{kj}\right]^2\times
\left[\frac{a_{ij}}{\Sigma_i^2}\right]^2\times \left(\sigma_{X_i}\right)^2 \, .
\end{equation}
\noindent
The theoretical uncertainty on  $\hat\chi_k$ is obtained similarly, depending on the type of combination considered
\begin{eqnarray} \label{eq:minlintheo}
\Delta_{\hat\chi_k} &=& \sqrt{\sum_{i,j}\left[(M^{-1})_{kj}\right]^2\times
\left[\frac{a_{ij}}{\Sigma_i^2}\right]^2\times \left(\Delta_{X_i}\right)^2} \ \ \;\; ({\rm quadratic})\, , \\
\Delta_{\hat\chi_k} &=& \sum_{i,j}\left[(M^{-1})_{kj}\right]\times
\left[\frac{a_{ij}}{\Sigma_i^2}\right]\times \Delta_{X_i} \ \ \qquad\quad\quad ({\rm linear}) \, .
\end{eqnarray}





In the case of the nuisance approach, we can use the above expressions to determine the confidence intervals. In the other cases, we extend the one-dimensional discussion of Section~\ref{sec:theoProblem}, determining the relevant p-values by assuming that the test statistic in each case follows a known distribution (in the Rfit case) or by determining its actual distributing and computing the p-value numerically. We provide in Table~\ref{tab:scenarioA} below the set of results. Note from this table that Rfit and 1-hypercube give similar results at $ 1 \sigma $, being more conservative than the other methods at this confidence level. As one increases the significance, the adaptive hyperball approach gives systematically more conservative intervals.

\begin{figure}
	\centering
	\includegraphics[scale=0.3]{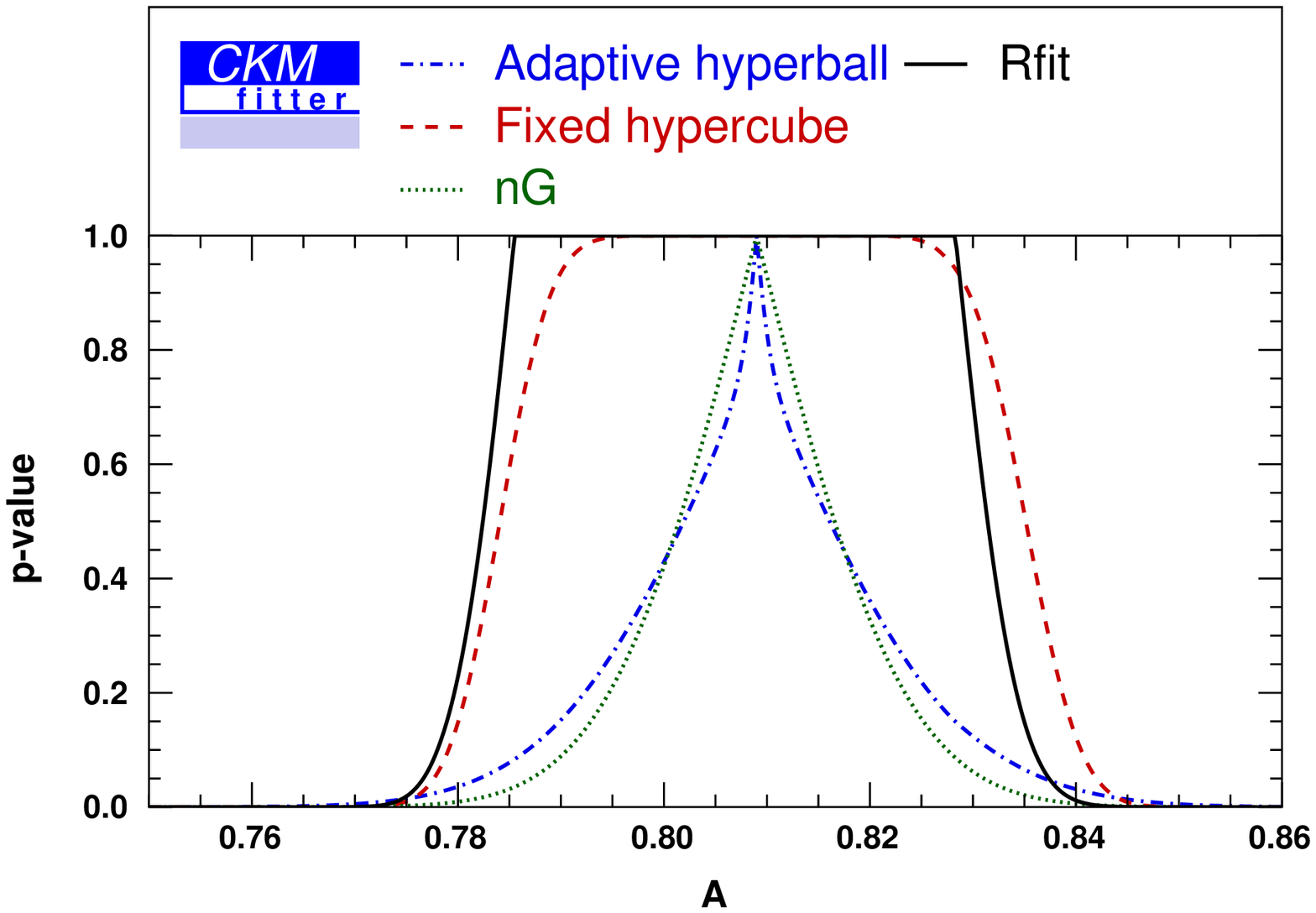}\includegraphics[scale=0.3]{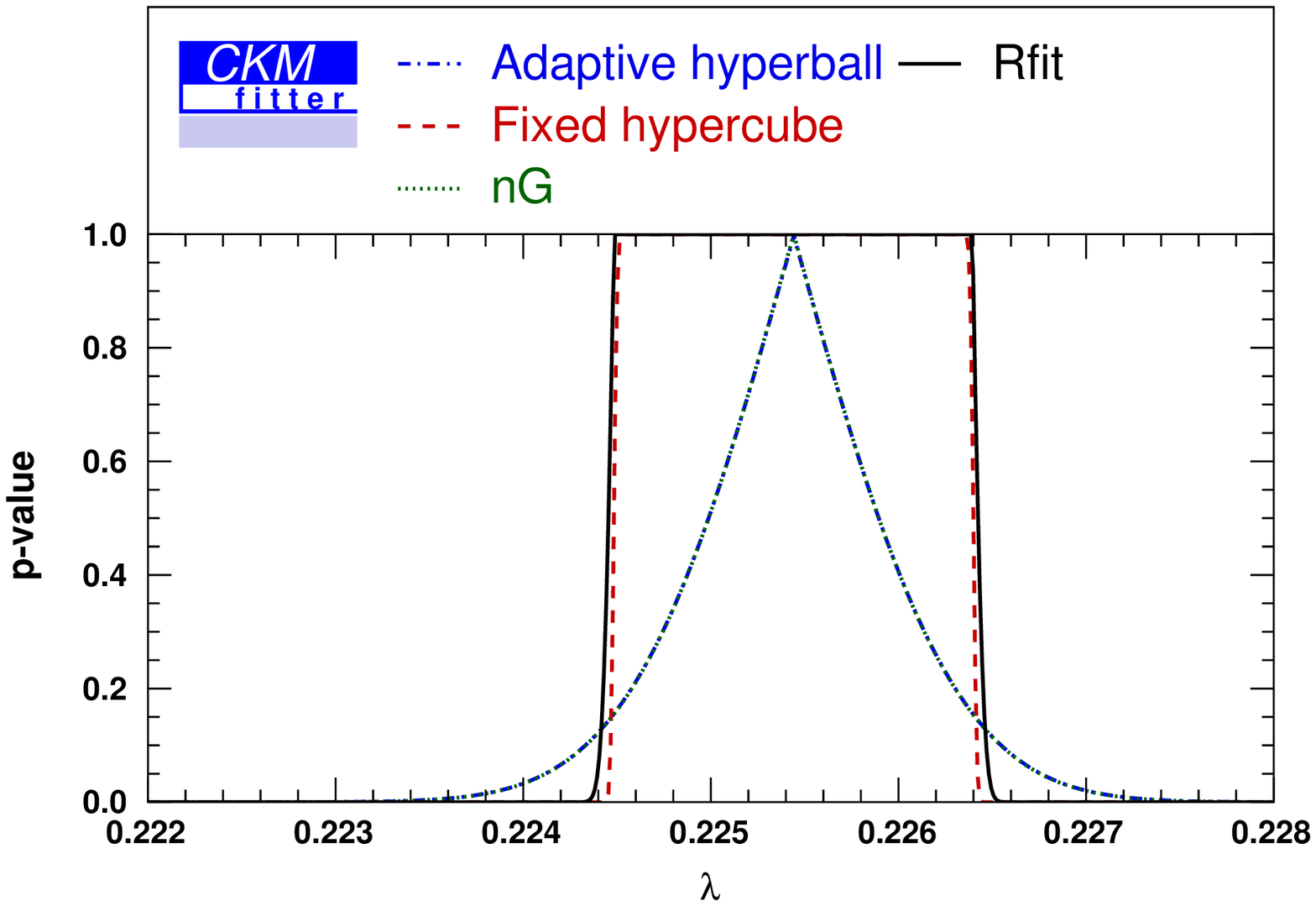}
	\includegraphics[scale=0.3]{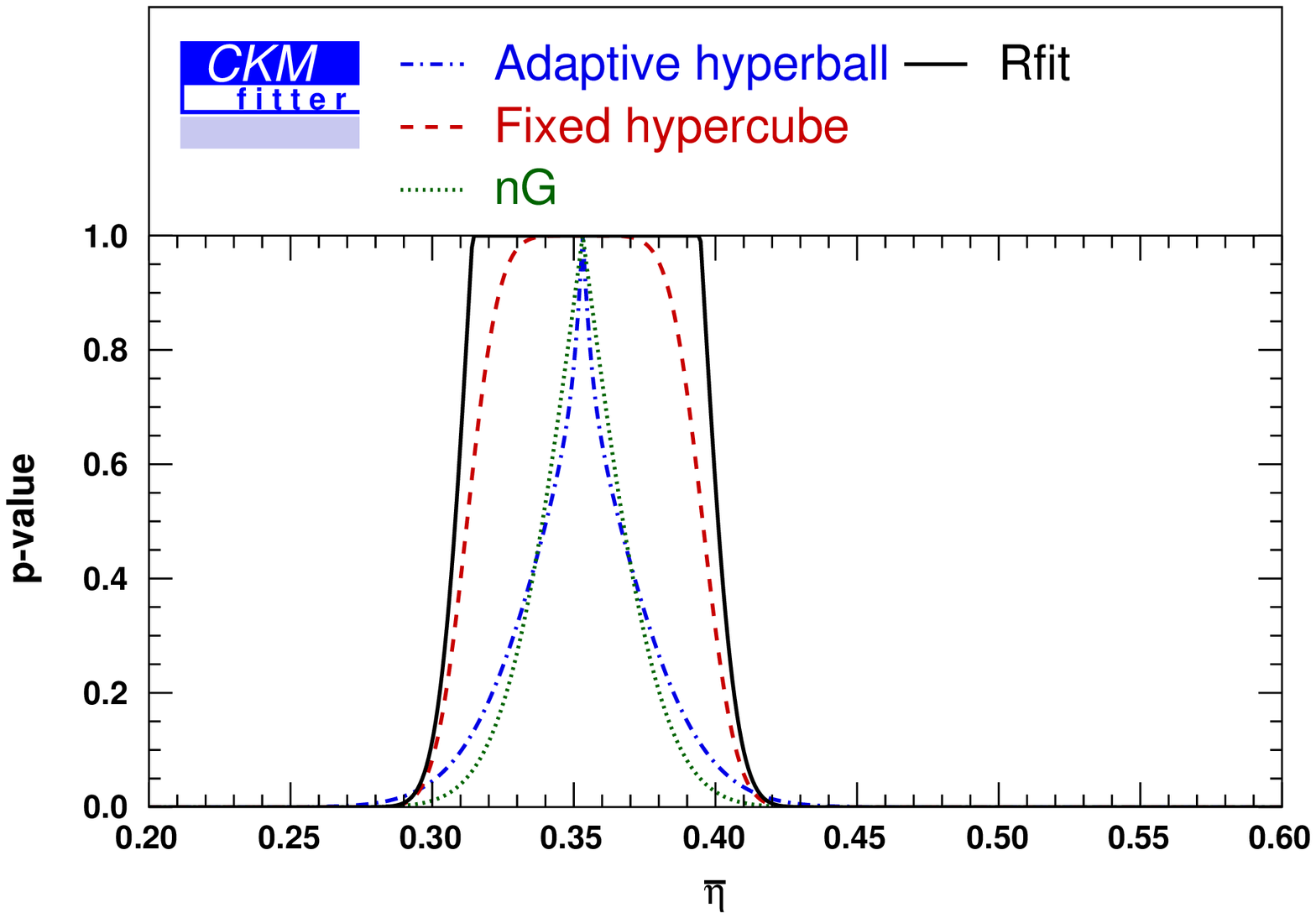}\includegraphics[scale=0.3]{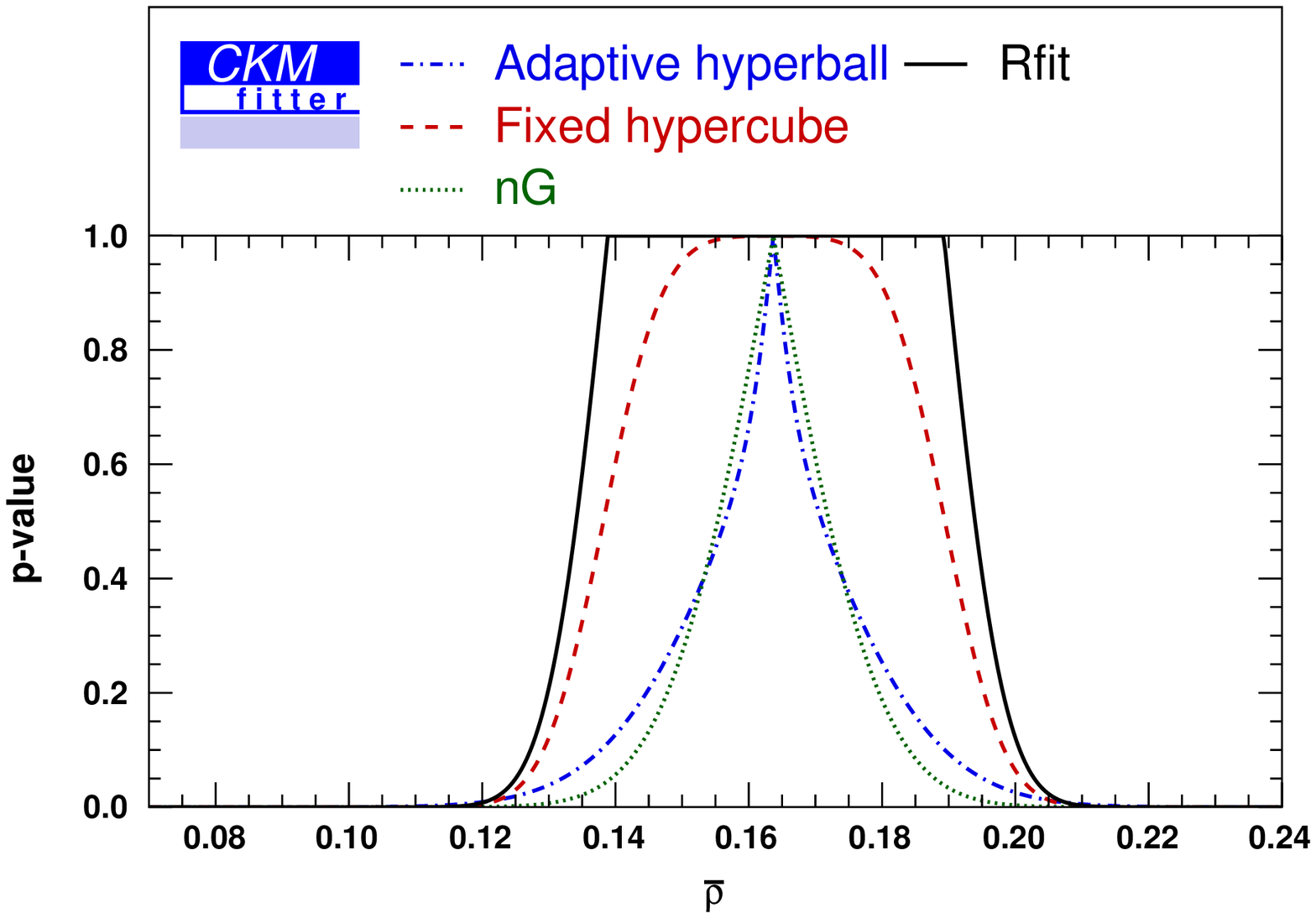}
	\caption{\it Shapes of the p-values for the different treatements of theoretical uncertainties discussed in the text. We consider the extraction of $ \{ A, \lambda, \bar{\rho}, \bar{\eta} \} $ for the set of inputs given in Table~\ref{tab:inputssystdominated}, and as a simplification we consider linearizing the SM predictions around the best-fit point for these fundamental parameters.}\label{fig:scenarioA}
\end{figure}

Note from Table~\ref{tab:scenarioA} that we do not separate the resulting uncertainties in the Rfit scheme into a statistical and a theoretical parts. Indeed, in more general situations the p-value curves have more complex shapes than seen in Figure~\ref{fig:scenarioA}, and in those situations a \textit{plateau} corresponding to the size of the theoretical uncertainty may not be present.




\pagebreak

\section{Conclusions}

We have considered the problem of extracting some information from the measurement

\begin{equation}
X_0 \pm \sigma \pm \Delta \nonumber
\end{equation}
of an observable $ x_t $ predicted to have the value $ x (\mu) $ in the framework of a given model, which has as free parameters a set represented by $ \mu $. In the measurement above, $ X_0 $ is called the central value, while $ \sigma $ is the statistical uncertainty and $ \Delta $ is the size of the theoretical uncertainty. 

Compared with statistical uncertainties, which by definition are modeled by random variables following a certain distribution law, theoretical (or systematic) uncertainties form a very different class of uncertainties. Indeed, their origin can be traced back, for instance, to calculations where one truncates a perturbative series, and where the remaining corrections $ \delta_t $, in principle unknown, are estimated and a theoretical uncertainty $ \Delta $ is quoted.

In the comparison between data and the predictions made by a model for the extraction of the underlying parameters $ \mu $, this class of uncertainties must be taken into account and they diminish our ability for the extraction of $ \mu $, just like statistical uncertainties. The interpretation of theoretical uncertainties is somewhat arbitrary and we have considered throughout this chapter different modelings. Ideally, we would like to find a method which is not too ``conservative" or too ``aggressive" when quoting confidence level intervals, among other properties. The underlying issue is to extract useful information from data, without claiming a tension with the Standard Model which may be only an artifact of the modeling of theoretical uncertainties. For each of the models, we have considered the corresponding p-values in order to build confidence intervals for the extraction of the values of $ \mu $ (the precise way to build p-values and the interpretation of confidence intervals are discussed in the main text). The following methods were considered:

\begin{itemize}
	\item In the naive Gaussian approach, we have considered treating them as a source of statistical uncertainty, and modeled $ \delta $ by a random variable distributed normally, with variance $ \Delta^2 $. Though widely employed, it corresponds to the awkward situation where the full result of perturbative calculation, in the example commented above, has a random behaviour.
	\item In the external approach, we have considered that they should be treated as an \textit{external} parameter, i.e. supposing first that its value is known. Then, when confidence intervals of the parameters of interest $ \mu $ are considered, we vary the external parameter $ \delta $ over a fixed range, typically $ [-\Delta, \Delta] $.
	\item In the nuisance approach, we consider maximizing the p-value, usually carrying a dependence on the value of $ \delta $, over a certain interval. We have then considered two different cases: a fixed interval, and an interval which grows with the size of the confidence interval for $ \mu $ we want to have (adaptive approach).
	\item The Rfit approach also treats $ \delta $ as a nuisance parameter and models the interval $ [-\Delta, \Delta] $ with a \textit{plateau} in the p-value or test statistic.
\end{itemize}

We have then studied the properties of these different methods, such as: confidence intervals (including significances and coverage), averaging different measurements of the same quantity, and a global fit of the CKM matrix for illustration. These questions are not purely mathematical and have a clear interest for physical problems, as some of the examples discussed in the main text attest. As we have seen over this chapter, the sizes of the quoted intervals, apart from depending on the method chosen, are sensitive to the relative size of statistical and theoretical uncertainties and the significance at which one wants to build the confidence intervals. Moreover, the combination of different extractions of the same quantity, and the outcome of global fits also depend on the scheme in use.

At higher significances, the adaptive approach has broader confidence level intervals and has good coverage properties. The adaptive approach also has the interesting property of decomposing statistical and theoretical uncertainties in global analyses, a property not found in the Rfit. \textit{It is then an encouraging possibility to be further investigated.}

A longer analysis and more comments are found in \cite{Charles}. The comparison between different models for dealing with theoretical uncertainties, plus the way to treat singular cases when combining correlated measurements correspond to a prospective study to be considered by future \CKM $ \, $ analyses.






\begin{sidewaystable}
\begin{center}
\begin{tabular}{ccccc}
A \\
Method & Fit result & 1 $\sigma$ & 2 $\sigma$ & 3 $\sigma$ \\
nG & $0.809\pm 0.011$ & $0.809\pm 0.011$ & $0.809 \pm 0.023$ & $0.809\pm 0.034$ \\
Rfit & $0.807 \pm 0.026$ & $0.807 \pm 0.026$ & $0.807\pm 0.031$ & $0.807 \pm 0.035$\\
1-hypercube & $0.809 \pm 0.004\pm 0.025$ & $0.809\pm 0.028$ & $0.809\pm 0.033$ & $0.809\pm 0.037$\\
adaptive hyperball & $0.809 \pm 0.004\pm 0.010$ & $0.809\pm 0.012$ & $0.809\pm 0.029$ & $0.809\pm 0.043$\\
\hline
$\lambda$ \\
Method & Fit result & 1 $\sigma$ & 2 $\sigma$ & 3 $\sigma$ \\
nG & $0.2254\pm 0.0007$ & $0.2254\pm 0.0007$ & $0.225\pm 0.0013$ & $0.2254\pm 0.0020$ \\
Rfit &  $0.2254 \pm 0.0010$& $0.2254 \pm 0.0010$ & $0.2254 \pm 0.0010$ & $0.2254 \pm 0.0010$\\
1-hypercube & $0.2254 \pm 0.0000\pm 0.0010$ & $0.2254\pm 0.0010$ & $0.2254\pm 0.0010$ & $0.2254\pm 0.0010$ \\
adaptive hyperball & $0.2254 \pm 0.0000 \pm 0.0007$ &  $0.2254\pm 0.0007$ & $0.2254\pm 0.0014$ & $0.2254\pm 0.0020$ \\
\hline
$\bar\rho$\\
Method & Fit result & 1 $\sigma$ & 2 $\sigma$ & 3 $\sigma$ \\
nG & $0.164\pm 0.012$ & $0.164\pm 0.012$ & $0.164\pm 0.025$ & $0.164\pm 0.037$ \\
Rfit & $0.164 \pm 0.032$ & $0.164 \pm 0.032$ & $0.164 \pm 0.039$ & $0.164 \pm 0.046$\\
1-hypercube & $0.164 \pm 0.007\pm 0.026$ & $0.164\pm 0.029$ & $0.164\pm 0.038$ & $0.164\pm 0.045$\\
adaptive hyperball & $0.164 \pm 0.007\pm 0.010$ & $0.164\pm 0.014$ & $0.164\pm 0.032$ & $0.164\pm 0.051$\\
\hline
$\bar\eta$\\
Method & Fit result & 1 $\sigma$ & 2 $\sigma$ & 3 $\sigma$ \\
nG & $0.353\pm 0.021$  & $0.353\pm 0.021$ & $0.353\pm 0.042$ & $0.353\pm 0.063$ \\
Rfit & $0.354^{+0.050}_{-0.049}$ & $0.354^{+0.050}_{-0.049}$ & $0.354^{+0.059}_{-0.058}$ & $0.354^{+0.068}_{-0.067}$\\
1-hypercube & $0.353 \pm 0.009 \pm 0.041$ & $0.353\pm 0.046$ & $0.353\pm 0.057$ & $0.353\pm 0.067$\\
adaptive hyperball & $0.353  \pm 0.009 \pm 0.019$ & $0.353 \pm 0.023$ & $0.353\pm 0.054$ & $0.353\pm 0.083$\\
\end{tabular}
\caption{\it Numerical results for different confidence intervals for the models of theoretical uncertainty discussed in the text: naive Gaussian (nG), Rfit (usually employed by the \texttt{CKMfitter} Collaboration), nuisance approach with theoretical uncertainties varying in a hypercube (1-hypercube), and adaptive intervals in the nuisance approach with theoretical uncertainties varying in a hyperball (adaptive hyperball).}\label{tab:scenarioA}
\end{center}
\end{sidewaystable}

\chapter*{Conclusion}
\addcontentsline{toc}{chapter}{Conclusion}

We have at hand nowadays a very successful theory to describe a wide variety of phenomena involving particles and their interactions, which is for this reason called \textit{Standard} Model. We have reviewed in Chapter~\ref{ch:SM} two classes of observables for testing the SM: EWPO, consisting of very precise measurements, and flavour observables used in the extraction of the elements of the CKM matrix. In each context, we have considered a global fit based on the \texttt{CKMfitter} statistical framework for combining observables. The results show an overall successful description of these observables by the SM.

Though generally very successful, the SM does not explain some features of nature but rather include them into its framework. A good example is the chiral structure of the weak interactions, or in other words the violation of parity symmetry (and charge conjugation). We would like to have a better understanding of this aspect of the SM, and we have therefore considered a class of extensions of the SM called Left-Right Models \cite{Pati:1974yy,Mohapatra:1974hk,Senjanovic:1975rk,Senjanovic:1978ev}, a framework where the symmetry between left and right is restored at a high energy scale, similarly to the way in which the electroweak symmetry is restored at high enough energies.

We have revisited the LR Models realized with doublet representations, which was the first scalar content to be considered in the old literature on the subject. Later on, more attention has been given to a model containing triplet scalars, due to the possibility of explaining the smallness of neutrino masses with a see-saw mechanism. However, the model with triplets has been very much constrained, justifying the study of other realizations of the LR Models (in our case a simpler one).

From the point of view of the phenomenology of this class of models, new gauge bosons $ Z', W' $ are introduced. The $ W' $ couples to the right-handed fields with strengths described by a mixing-matrix analogous to the CKM matrix of the SM. This is particularly interesting due to the possibility of introducing new sources of $ \mathcal{C P} $ violation. In full generality, other $ \mathcal{C P}- $violating phases could also come from the VEV triggering the spontaneous breaking of the LR gauge group down to the electromagnetism at low energies; for simplicity reasons though, we have preferred to work in the case where these phases vanish. A different interesting phenomenological aspect is that there is a whole new scalar sector, with the new neutral scalar particles introducing flavour changing neutral currents at the tree level. 

We have discussed in Chapter~\ref{ch:LRM} the specific way in which the gauge symmetries of the LR Models are spontaneously broken. As we have seen at that moment, the energy scale of the EW Symmetry Breaking is described in full generality by three different Vacuum Expectation Values, $ \kappa_{1,2,L} $, which combine as $ \sqrt{\kappa^2_{1} + \kappa^2_{2} + \kappa^2_{L}} $ to set up the scale of the EWSB.  We stress that this is one of the novelties of the work we show in here, namely the consideration of the VEV $ \kappa_{L} $, which is constrained to be very much suppressed in the triplet case (due to the value of the $ \rho $ parameter).

A sizable value of $ \kappa_{L} $ would imply a richer pattern of the EWSB, triggered by scalar fields of different representations. To probe this aspect, we have performed a global fit including EW Precision Observables and the lower direct bound on the mass of the $ W' $. The results of the fit for the predictions of the different EWPO are very similar if compared to the SM global fit, and we were not able to solve the tensions found in the SM context (such as $ A^b_{FB} $). As seen in Chapter~\ref{ch:EWPO}, the scale $ \kappa_R $ at which parity is restored ends up being very high, due to the direct bound on the mass of the $ W' $, limiting in part the sensitivity of the fit to the new gauge coupling $ g_R $ and the VEVs $ \kappa_{1,2,L} $. We have however noted a preference for high values of $ \kappa_L $.

We need new observables to constrain the LR Model, and more specifically observables sensitive to new physics effects even for large values of $ \kappa_R $. Meson-mixing observables are sensitive to new energy scales much beyond the reach of modern particle colliders, and are then good candidates for constraining LR Models, including the couplings of the $ W' $ and the new scalar sector. However, in order to learn something accurate from this class of observables, we need dedicated calculations of the effects introduced by QCD, which are unavoidable when processes involving quarks are considered.

While the parameters describing long-distance QCD effects in meson-mixing observables in LR Models have been addressed by many different collaborations, short-distance QCD effects have not received the same attention. We have calculated in Chapter~\ref{ch:technicalEFT}, after a brief introduction of the necessary tools in Chapter~\ref{ch:generalEFT}, the effects from short-distance QCD corrections in meson-mixing in LR Models. These effects were computed by exploiting two different methods: one exact approach based on a successive set of EFT, and another one giving an estimate of the latter exact approach which we have called Method of Regions. The novelty of our calculation in the LR Model context was the integration of effects coming from dynamical charms in the EFT approach, and the effort to extend the MR calculation up to the NLO.

Having achieved the computation of short-distance QCD corrections necessary for the LR Model predictions of meson-mixing observables, we integrated in Chapter~\ref{ch:PHENO} meson-mixing observables together with EWPO and direct searches for the $ W' $ in a global fit, using the \texttt{CKMfitter} statistical framework in the context of the LR Models. Compared to EWPO, the new set of parameters includes the masses of the scalar sector and the mixing-matrix of the right-handed quarks $ V^R $, analogous to the CKM matrix of the SM. For the latter, however, we have assumed the simplified case $ V^R = V^L $ in our analysis, a case called \textit{manifest} scenario.

We were then able to set lower bounds on the masses of the gauge bosons $ W' $ and $ Z' $ at $ \sim 3.6 - 4.0 $~TeV and $ \sim 7.5 - 8.5 $~TeV ($ 68~\% $ CL), respectively, more restrictive or at least competitive when compared to direct search programs. We have as well been able to set bounds on the masses of the extended scalar sector beyond $ \sim 25 $~TeV. On the other hand, we were not able to extract bounds for $ \kappa_L, \kappa_1, \kappa_2, g_R $, but we have extracted some of their correlations.

Some aspects of the global fit can be certainly improved or generalized. We could still refine the information coming from direct searches, which is usually made under specific assumptions concerning the couplings of the $ W', Z' $ gauge bosons (such as $ g_L = g_R $). Moreover, different structures of the mixing-matrix $ V^R $ could be probed, requiring new observables, in particular tree level processes where a charged gauge boson is exchanged in the SM. Their inclusion could point towards features of the $ V^R $ mixing-matrix such as new $ \mathcal{C P}- $violating phases, and in this context of (semi-)leptonic processes it would be interesting to include observables which have recently shown tensions with the SM, e.g. $ R_K, R_{D^{}}, R_{D^{*}} $. This would lead us to questionings related to the leptonic sector of the model, a challenging but quite exciting perspective if one considers that LR Models provide a $ Z' $ boson as currently hinted at by $ b \rightarrow s \ell \ell $ observables, see e.g. \cite{Descotes-Genon}.


Shifting to a different issue, we have compared in Chapter~\ref{ch:theo} different modelings of theoretical uncertainties, a class of uncertainties that is specially important in flavour physics. We have shown that the size of a tension between experiment and prediction, or the outcome of the combination of different extractions of the same quantity, or yet the results in terms of confidence intervals of a global fit, depend on the way we understand theoretical errors. The underlying interest of investigating different models of theoretical uncertainties is not only for the exercise of illustrating their differences or similarities: aiming to be up-to-date with the present landscape of uncertainties in experimental data and theoretical inputs in flavour physics, much more accurate than ten years ago, we would like to improve the analysis performed by the \texttt{CKMfitter} Collaboration. We have therefore looked for alternatives to the Rfit scheme, presently used  in the modeling of theoretical errors, and have found a promising candidate, called \textit{adaptive} nuisance approach, which shows interesting properties from the point of view of coverage and decomposition of statistical and theoretical uncertainties in a global fit.

\newpage


\section*{Acknowledgments}
\addcontentsline{toc}{chapter}{Acknowledgments}

I would like first to thank Véronique Bernard and Sébastien Descotes-Genon for their coordination during the complexion of this thesis, without which this work would not have been possible. We had an intensive collaboration with much effort from all sides, punctuated by many and long discussions. Moreover, Véronique and Sébastien were always ready to help me for pursuing my career, and I will always be thankful for that.

\vspace{1.5mm}

As a member of the \texttt{CKMfitter} Collaboration, I had the opportunity to discuss my work with Jérôme Charles, Olivier Deschamps, Heiko Lacker, Andreas Menzel, Stéphane Monteil, Valentin Niess, José Ocariz, Jean Orloff, Alejandro Perez, Wenbin Qian, Vincent Tisserand, Stéphane T'Jampens, Karim Trabelsi and Philip Urquijo. I am therefore glad to have had both their collaboration and their questions over the last three years.

\vspace{1.5mm}

I am very grateful for the careful reading and precious comments of Ulrich Nierste and Renata Zukanovich Funchal, who have kindfully accepted to review my thesis.

\vspace{1.5mm}

Science is made of a series of questionings and some people have clarified a bunch of specific issues, making the path easier to track. I thank Andrzej Buras, Otto Eberhardt, Ulrich Ellwanger, Alberto Escalante del Valle, Adam Falkowski, Marc Knecht, Emi Kou, Alexander Lenz, Miha Nemev\v{s}ek, Tord Riemann, Hagop Sazdjian, Goran Senjanovi\'{c} and Jiang-Hao Yu for sharing their knowledge.

\vspace{1.5mm}

Special thanks go to Asmaa Abada, Giorgio Arcadi, Damir Be\v{c}irevi\'{c} and Pradipta Ghosh for their advices, in particular those beyond the scope of this thesis.

\vspace{1.5mm}

I also thank Jean-Marie Frère, Jernej Kamenik and Fabrizio Nesti for their invitation to give talks in their home institutes and the organizers of \textit{Flavorful Ways to New Physics}, \textit{RPP 15} and \textit{RPP 16} for the same reason, thanks to which I have had many useful insights to improve this manuscript.

\vspace{1.5mm}

It is a pleasure to thank Philippe Boucaud, Olivier Brand-Foissac, Mireille Calvet, Patricia Dubois-Violette, Odile Heckenauer, Henk Hilhorst, Jean-Pierre Leroy, Yvette Mamilonne, Philippe Molle, Marie-Agnes Poulet, Jocelyne Puech and Jocelyne Raux for their technical support and constant help. 

\vspace{1.5mm}

I am also very grateful to the people who have made \textit{le Laboratoire de Physique Théorique} a joyful place to work, namely, Andrei Angelescu, Renaud Boussarie, Ma\'{i}ra Dutra, Gabriel Jung, Kyle Kalutkiewicz, Sarah Klein, Antoine Lehébel, Mathias Pierre, Timothé Poulain, Mat\'{i}as Rodr\'{i}guez-V\'{a}zquez and Olcyr Sumensari de Lima.

\vspace{1.5mm}

I cannot forget to mention the \textit{triad} Pedro Francisco Baraçal de Mecê, Juan Raphael Diaz Simões and Rafael Sampaio de Rezende, and Brahim Ait Haddou, Edouard Benoit, Jana Biedov\'{a}, Guillaume Binet, Arindam Biswas, Natalia Borek, Mathieu Bouyrie, Frédéric Dreyer, Anne-Sophie de Groër, Simona Ricci and Simon Rioult, who have been very present in my life all over my Ph.D. and cannot therefore be disentangled from the work required to complete this thesis.

\vspace{1.5mm}

{\it Finalmente, eu gostaria de agradecer a algumas pessoas que tiveram uma participaç\~{a}o importante na bagagem cient\'{i}fica e metodol\'{o}gica que eu trouxe para esta tése, Gustavo de Medeiros Azevedo, Daniel Augusto Turolla Vanzella e S\'{e}rgio Lu\'{i}s Zani. N\~{a}o posso esquecer o suporte da minha fam\'{i}lia ao longo do meu percurso, meu pai Francisco de Assis Filho, minha primeira professora Maria de F\'{a}tima Alves Vale Silva, o membro mais ``jovem" da fam\'{i}lia Marcos Napole\~{a}o Rabelo e agradecimentos especiais v\~{a}o para Luciana Vale Silva Rabelo, quem primeiro me direcionou à carreira cient\'{i}fica, ainda que n\~{a}o intencionalmente, e a quem eu gostaria de dedicar esta tése.}

\appendix


\chapter{EWPO tree level expressions in the SM}\label{sec:EWPOTreeLevel}

\section{Partial widths}

Partial widths of the $ Z $ boson, $ Z \rightarrow f \bar{f} $, are calculated in the following way

\begin{equation}
\Gamma_{f \bar{f}} = C_f \frac{G_F}{6 \pi \sqrt{2}} [(g^f_V)^2 + (g^f_A)^2] M^3_Z \, ,
\end{equation}
where $ C_f $ is the number of colours, thus implying

\begin{eqnarray}
\Gamma_{\nu \bar{\nu}} &=& \frac{G_F}{12 \pi \sqrt{2}} M^3_Z \, , \\
\Gamma_{\ell \bar{\ell}} &=& \frac{1}{2} \Gamma_{\nu \bar{\nu}} [(1 - 4 \sin^2 \theta)^2 + 1] \, , \\
\Gamma_{U \bar{U}} &=& \frac{3}{2} \Gamma_{\nu \bar{\nu}} [(1 - \frac{8}{3} \sin^2 \theta)^2 + 1] \, , \\
\Gamma_{D \bar{D}} &=& \frac{3}{2} \Gamma_{\nu \bar{\nu}} [(1 - \frac{4}{3} \sin^2 \theta)^2 + 1] \, ,
\end{eqnarray}
where $ \ell = e, \mu, \tau $, $ U = u, c $, $ D = d, s, b $, and we have neglected the masses of the fermions for illustration (which is an extremely good approximation except for the bottom-quark). Therefore, the total width is given by\footnote{In principle, $ \Gamma_{total} $ as defined above can be different from $ \Gamma_Z $. In our SM and LRM analyses we assume that they are the same, i.e. there is no extra invisible channel apart from the three neutrino flavours.}

\begin{eqnarray}
\Gamma_{total} &=& 3 \Gamma_{\nu \bar{\nu}} + 3 \Gamma_{\ell \bar{\ell}} + 2 \Gamma_{U \bar{U}} + 3 \Gamma_{D \bar{D}} \, , \\
\Rightarrow \Gamma_{total} &=& \frac{3}{2} \Gamma_{\nu \bar{\nu}} \left[ 14 - \frac{80}{3} \sin^2 \theta + \frac{320}{9} \sin^4 \theta \right] .
\end{eqnarray}

From the previous expressions one has (and for the numerical exercise only, we take $ G_F \simeq 1.166 \cdot 10^{-5} \, {\rm GeV}^{-2} $, $ \sin^2 \theta \simeq 0.23 $, $ M_Z \simeq 91.2 $~GeV) 

\begin{eqnarray}
\Gamma_{\nu \bar{\nu}} &\simeq& 166~{\rm MeV} \, , \qquad \Gamma_{\ell \bar{\ell}} \simeq 83~{\rm MeV} \, , \\
\Gamma_{U \bar{U}} &\simeq& 286~{\rm MeV} , \qquad \Gamma_{D \bar{D}} \simeq 368~{\rm MeV} \, , \\
\Gamma_{total} &\simeq& 2.4~{\rm GeV} \, ,
\end{eqnarray}
from where one has the following branching ratios

\begin{eqnarray}
Br(\nu \bar{\nu}) &\simeq& 6.8~\% , \qquad Br(\ell \bar{\ell}) \simeq 3.4~\% \, , \\
Br(U \bar{U}) &\simeq& 12~\% , \qquad Br(D \bar{D}) \simeq 15~\% \, , \\
Br(hadrons) &\simeq& 69~\% \, .
\end{eqnarray}
As a thumb rule, $ Z $ decays twice more into neutrinos than into charged leptons, and $ 2/3 $ of the time into hadrons.

From the partial-widths, one can define the following ratios

\begin{equation}
R_\ell = \frac{\Gamma_{had}}{\Gamma_{\ell \bar{\ell}}} \, , \qquad R_q = \frac{\Gamma_{q \bar{q}}}{\Gamma_{had}} \, ,
\end{equation}
measured for $ \ell = e, \mu, \tau $ and $ q = b, c $.

\section{Cross sections}

Consider a $ e^+ e^- $ collision\footnote{Of course, hadron collisions are also useful to investigate the lineshape of the $ Z $ boson resonance, though they suffer from initial-quark PDF uncertainties.} with a center-of-mass energy $ \sqrt{s} $ in the center-of-mass frame; the non-polarized differential cross section at tree level is given by (neglecting the masses of the fermions)

\begin{eqnarray}
&& \frac{d \sigma_f}{d \cos \theta^*} (\sqrt{s}) = \frac{C_f}{8 \pi} \left( \frac{g_L^2}{8 \cos^2 \theta} \right)^2 \frac{s}{(s - M^2_Z)^2 + \Gamma^2_Z M_Z^2} \\
&& \left\{ [(g^e_V)^2 + (g^e_A)^2] [(g^f_V)^2 + (g^f_A)^2] (1 + \cos^2 \theta^*) + 8 g^e_V g^e_A g^f_V g^f_A \cos \theta^* \right\} \, , \nonumber
\end{eqnarray}
where $ \theta^* $ is the emission angle in the center-of-mass frame (angle between the positron (electron) and the final (anti-)particle). Note that the last term shows an asymmetric dependence on the emission angle $ \theta^* $, which is an explicit violation of parity (proportional to the asymmetry $ A_{FB} (f) $ defined below).

Above, a Breit-Wigner approximation was employed, i.e. for a virtual $ Z^* $ exchanged in a $ s- $channel\footnote{Other channels have a non-resonant character but still contribute to the line-shape of $ e^+ e^- \rightarrow Z^* \rightarrow e^+ e^- $.} at tree level ($ k^2 = s $)

\begin{equation}
- i \frac{g_{\mu \nu} - k_\mu k_\nu / M^2_Z}{k^2 - M^2_Z + i \Gamma_Z M_Z} \, ,
\end{equation}
in the unitary gauge. (Experimentally, however, one employs a width which has an energy dependence on the energy scale, $ \propto (k^2 - M^2_Z + i k^2 \Gamma_Z / M_Z)^{-1} $.)

The differential cross-section implies the following total cross-section

\begin{equation}
\sigma_f (\sqrt{s}) = \frac{s \Gamma^2_Z}{(s - M^2_Z)^2 + \Gamma^2_Z M_Z^2} \sigma_f \, ,
\end{equation}
where $ \sigma_f \equiv \sigma_f (M_Z) $. In terms of partial-widths at tree level:

\begin{equation}
\sigma_f = \frac{12 \pi}{M^2_Z} \frac{\Gamma_{e \bar{e}} \Gamma_{f \bar{f}}}{\Gamma^2_Z} \, .
\end{equation}
Finally, it is useful in order to better control systematic errors to write

\begin{equation}
\sigma_f (\sqrt{s}) = \frac{s \Gamma^2_Z}{(s - M^2_Z)^2 + \Gamma^2_Z M_Z^2} \sigma_{had} \frac{\Gamma_{f \bar{f}}}{\Gamma_{had}} \, ,
\end{equation}
where $ \sigma_{had} $ corresponds to the sum over hadronic channels of $ \sigma_f $, i.e. $ \sigma_{had} = \sum_{f \in \{ u,d,s,c,b \}} \sigma_f $.

\section{Asymmetries}

Parity violation in neutral weak interactions in the SM is at the origin of a different class of observables. Though the asymmetries of the $ Z $ couplings are not as large as the asymmetries of the $ W $ couplings, they have their own interest due to the precision measurements at the center-of-mass energy $ \sqrt{s} \sim 91 $~GeV (and they offer a privileged way to determine $ \sin^2 \theta $, by writing $ g^f_{V,A} $ in terms of $ \sin^2 \theta $, see Table \ref{tab:SMquantumNumbers} in Chapter~\ref{ch:EWPO}). One of the possible asymmetries to measure is the Forward-Backward asymmetry, i.e. the angular asymmetry on $ Z $ decays

\begin{equation}
A_{FB} (f) = \frac{n (\theta^* < 90^{\rm o}) - n (\theta^* > 90^{\rm o})}{n (\theta^* < 90^{\rm o}) + n (\theta^* > 90^{\rm o})} \, ,
\end{equation}
where $ n $ is the number of events, and $ f= e, \mu, \tau, b, c $. Note that parity conserving effects, such as QED and QCD, cancel in the ratio. From the theoretical side, $ A_{FB} (f) $ is given as follows

\begin{equation}
A_{FB} (f) = \frac{3}{4} \mathcal{A}_e \mathcal{A}_f \, , \qquad \mathcal{A}_f = 2 \frac{g^f_V g^f_A}{(g^f_V)^2 + (g^f_A)^2} \, .
\end{equation}

The asymmetry of $ Z $ decays in polarised final-states, measured for $ \tau $ leptons by the observation of their decay products, is

\begin{equation}
P^\tau_{-} = \frac{n (\tau_R) - n (\tau_L)}{n (\tau_R) + n (\tau_L)} = - \mathcal{A}_\tau \, ,
\end{equation}
and more generally one has a function of the emission angle, $ P^\tau_{-} (\cos \theta^*) $.

Another asymmetry to measure is the Left-Right asymmetry, i.e. the asymmetry of the cross-section with incident polarized electron beams

\begin{equation}
A_{LR} = \frac{\sigma (e_L) - \sigma (e_R)}{\sigma (e_L) + \sigma (e_R)} = \mathcal{A}_e \, .
\end{equation}

\section{Atomic Parity Violation (APV)}


Apart from these Z-lineshape and asymmetries we have low-energy	measures coming from Atomic Parity Violation (APV), see e.g. \cite{Bardin:2001ii} (and \cite{Diener:2011jt,Deandrea:1997wk} for NP studies), which is defined from the parity violating effects related to the exchanges of $ Z $ bosons between the atomic nucleus and the atomic electrons. The most important contribution comes from the axial coupling to electrons and to the vectorial couplings to quarks from the nucleons (for more details, see the ``EW model and constraints on NP'' review in \cite{Beringer:1900zz}):

\begin{eqnarray}
Q_{W} (q) &=& 2 g^{e}_{A} g^{q}_{V} \, , \\
Q_{W} (p) &=& 2 \, Q_{W} (u) + Q_{W} (d) \, , \\
Q_{W} (n) &=& Q_{W} (u) + 2 \, Q_{W} (d) \, ,
\end{eqnarray}
\noindent
and we have considered $ ^{133}Cs_{55} $ and $ ^{205,\,203}Tl_{81} $ in our fit, whose APV magnitudes are given by

\begin{equation}
Q_{W} (n) = -2 (Z Q_{W} (p) + N Q_{W} (n)) \, .
\end{equation}

\section{W boson partial widths and cross sections}

We also provide useful bounds from the $ W $ boson properties. In the limit of massless neutrinos we have

\begin{equation}
\Gamma_{\nu \bar{\ell}} = \frac{G_F}{6 \pi \sqrt{2}} M^3_W \, ,
\end{equation}
and for the quarks there is additionally a dependence on the CKM mixing matrix ($ q_1 $ ($ q_2 $) is the up-type (down-type) flavour)

\begin{equation}
\Gamma_{q_1 \bar{q}_2} = 3 \frac{G_F}{6 \pi \sqrt{2}} \vert V^{L}_{q_1 q_2} \vert^2 M^3_W = 3 \vert V^{L}_{q_1 q_2} \vert^2 \Gamma_{\nu \bar{\ell}} \, .
\end{equation}

The above two equations imply

\begin{equation}
\Gamma_{total} = 3 \Gamma_{\nu \bar{\ell}} + 3 \sum_{q_1 , q_2} \vert V^{L}_{q_1 q_2} \vert^2 \Gamma_{\nu \bar{\ell}} = 9 \Gamma_{\ell \bar{\nu}_\ell} \, ,
\end{equation}
where in $ \sum_{q_1 , q_2} $ the kinematically allowed flavours are $ q_1 = u, c $ and $ q_2 = d, s, b $. Therefore, one has the following approximate values ($ G_F \simeq 1.166 \cdot 10^{-5} \, {\rm GeV}^{-2} $, $ M_W \simeq 80.4 $~GeV)

\begin{eqnarray}
&& \Gamma_{\nu \bar{\ell}} \simeq 227~{\rm MeV} \, , \\
&& {\rm from} \;\;\; \Gamma_{u \bar{b}} \simeq 6~{\rm keV} \, , \; {\rm to} \;\;\; \Gamma_{u \bar{d}, \, c \bar{s}} \simeq 642~{\rm MeV} \, , \\
&& \Gamma_{total} \simeq 2~{\rm GeV} \, ,
\end{eqnarray}
and the following branching ratios

\begin{equation}
Br(\nu \bar{\ell}) \simeq 1/9 , \qquad Br(hadrons) \simeq 6/9 \, .
\end{equation}

Though we are not going to exploit the following expressions in our analysis, it is interesting to note the parity asymmetry in them. These are: $ q \bar{q'} \rightarrow W^* \rightarrow \nu_\ell \bar{\ell} $ cross-sections

\begin{eqnarray}
&& \frac{d \bar{\sigma}_{\nu \overline{\ell}}}{d \cos \theta^*} (\sqrt{\overline{s}}) = \frac{1}{8 \pi} \left( \frac{g^2_L}{8} \right)^2 \frac{\overline{s} \vert V^{L}_{q q'} \vert^2}{(\overline{s} - M^2_W)^2 + \Gamma^2_W M^2_W} (1 + \cos \theta^*)^2 \, , \nonumber\\
\end{eqnarray}
and $ q \overline{q'} \rightarrow W^* \rightarrow q_1 \overline{q}_2 $

\begin{eqnarray}
&& \frac{d \overline{\sigma}_{q_1 \overline{q}_2}}{d \cos \theta^*} (\sqrt{\overline{s}}) = \frac{3}{8 \pi} \left( \frac{g^2_L}{8} \right)^2 \frac{\overline{s} \vert V^{L}_{q q'} \vert^2 \vert V^{L}_{q_1 q_2} \vert^2}{(\overline{s} - M^2_W)^2 + \Gamma^2_W M^2_W} (1 + \cos \theta^*)^2 \, . \nonumber\\
\end{eqnarray}

In the above expressions, the measured cross section is given by

\begin{equation}
\sigma_f (\sqrt{s}) = \int \int d x_1 \, d x_2 \, F(x_1) F(x_2) \overline{\sigma}_f (\sqrt{\overline{s}}) \, ,
\end{equation}
where $ \overline{s} = s x_1 x_2 $ and $ F(x_{1,2}) $ are the relevant PDFs.

\chapter{Parameterization of the EWPO}\label{sec:AppParameters}

In practice, we want to combine the set of EWPO in order to constrain the SM or the LRM using \texttt{CKMfitter}. Asking \texttt{CKMfitter} to call \texttt{Zfitter} for each realization of $ \mathcal{S} $ would be too time-consuming. Therefore, we prefer to parameterize the observables as a function of $ \mathcal{S} $ before using \texttt{CKMfitter}. A simple (Fortran) code makes calls to \texttt{Zfitter} for certain realizations of $ \mathcal{S} $, which gives as output the corresponding values for the set of observables we want to study. Then, with a second program (written in Mathematica), we parameterize these observables as a function of $ \mathcal{S} $. It is this parameterization that is used by \texttt{CKMfitter} to produce global fits. The following ``chart graph" resumes the task

\begin{eqnarray}
&& {\rm \mathtt{Fortran} \; code} \rightarrow {\rm \mathtt{Zfitter}} \rightarrow {\rm Observables} \rightarrow {\rm \mathtt{Mathematica} \; notebook} \qquad \nonumber\\
&& \qquad \qquad \rightarrow {\rm parameterization} \rightarrow {\rm \mathtt{CKMfitter}} \rightarrow {\rm constraints} \nonumber
\end{eqnarray}

When using \texttt{Zfitter}, we have left the value of $ \alpha^{-1} (0) = 137.0360 $ fixed. The values of other relevant parameters are: $ m_{s} = 0.3 $~GeV, $ m_{c} = 1.5 $~GeV, $ m_{b} = 4.7 $~GeV, and $ G_{\mu} = 1.1664 \cdot 10^{-5} \operatorname{GeV^{-2}} $, which were all kept fixed during the analysis. Both of the references, \texttt{Zfitter} \cite{Arbuzov:2005ma} and Freitas in \cite{Freitas:2014hra}, calculate loop corrections in the on-shell scheme. In this scheme, the mass of the top is the pole mass, $ m^{pole}_{top} $.

\begin{sidewaystable}
\begin{center}

\begin{tabular}{|c|c|c|c|c|c|c|c|c|c|}
\hline
Obs. & $ X_{0} $ & $ c_{1} $ & $ c_{2} $ & $ c_{3} $ & $ c_{4} $ & $ c_{5} $ & $ c_{6} $ & $ c_{7} $ & max. dev. \\
\hline
\hline
$ \Gamma_{Z} $ [MeV] & 2494.24 & -2.0 & 19.7 & 58.60 & -4.0 & 8.0 & -55.9 & 9267 & $ < $ 0.01 \\
\hline
\hline
$ \sigma_{had} $ [pb] & 41488.4 & 3.0 & 60.9 & -579.4 & 38 & 7.3 & 85 & -86027 & $ < $ 0.1 \\
\hline
\hline
$ R_{b} $ * & 215.80 & 0.031 & -2.98 & -1.32 & -0.84 & 0.035 & 0.73 & -18 & $ < $ 0.01 \\
\hline
$ R_{c} $ * & 172.23 & -0.029 & 1.0 & 2.3 & 1.3 & 0.38 & -1.2 & 37 & $ < $ 0.01 \\
\hline
$ R_{\ell} $ * & 20750.9 & -8.1 & -39 & 732.1 & -44 & 5.5 & -358 & 11702 & $ < $ 0.1 \\
\hline
\end{tabular}
\caption{Parametrization from \cite{Freitas:2014hra}, $ * = 10^{3} $. We do not use its parameterization for $ R_{\ell} $ since it is the result of a combination of $ R_{e}, R_{\mu}, R_{\tau} $.} 

\vspace{3mm}

\begin{tabular}{|c|c|c|c|c|c|c|c|c|c|}
\hline
Obs. & $ X_{0} $ & $ c_{1} $ & $ c_{2} $ & $ c_{3} $ & $ c_{4} $ & $ c_{5} $ & $ c_{6} $ & $ c_{7} $ & max. dev. \\
\hline
\hline
$ \Gamma_{Z} $ [MeV] & 2495.22 & -2.4 & 20.1 & 63.48 & -3.2 & -1.8 & -54.4 & 9225 & $ < $ 0.006 \\
\hline
\hline
$ \sigma_{had} $ [pb] & 41478.9 & 1.3 & 52.8 & -630.9 & 134 & -3.4 & 82 & -86323 & $ < $ 0.5 \\
\hline
\hline
$ R_{b} $ * & 215.81 & 0.039 & -3.12 & -0.0285 & -0.74 & 0.070 & 0.81 & -19. & $ < $ 0.0006 \\
\hline
$ R_{c} $ * & 172.24 & -0.032 & 1.0 & 2.3 & 1.4 & 0.38 & -1.2 & 37. & $ < $ 0.0006 \\
\hline
$ R_{e} $ * & 20739.5 & -7.7 & -32. & 791.2 & -78. & 0.70 & -361. & 11950. & $ < $ 0.3 \\
\hline
$ R_{\mu} $ * & 20739.5 & -6.8 & -32. & 791.5 & 31. & -0.17 & -362. & 11880. & $ < $ 0.3 \\
\hline
$ R_{\tau} $ * & 20786.5 & -8.4 & -32. & 792.9 & -40. & -0.37 & -363. & 11399. & $ < $ 0.3 \\
\hline
\hline
$ A_{FB} (b) $ * & 103.1 & -2.7 & 15.2 & -2.2 & -4.1 & 0.09 & -115.5 & 3725.2 & $ < $ 0.03 \\
\hline
$ A_{FB} (c) $ * & 73.7 & -2.1 & 11.9 & -1.7 & -2.9 & -0.01 & -89.4 & 2877.5 & $ < $ 0.02 \\
\hline
$ A_{FB} (\ell) $ * & 16.2 & -0.8 & 4.8 & -0.7 & -0.7 & -0.09 & -35.9 & 1. & $ < $ 0.02 \\
\hline
\hline
$ \mathcal{A}_{b} $ * & 934.6 & -0.3 & 0.4 & -0.1 & -0.5 & 0.5 & -13.2 & 429.6 & $ < $ 0.03 \\
\hline
$ \mathcal{A}_{c} $ * & 667.9 & -1.6 & 9.5 & -1.4 & -1.5 & 0.05 & -71.5 & 2260.2 & $ < $ 0.01 \\
\hline
$ \mathcal{A}_{\ell} $ * & 147.1 & -3.7 & 21.6 & -3.1 & -5.9 & 0.05 & -162.7 & 5246.9 & $ < $ 0.03 \\
\hline
\hline
$ M_{W} $ [GeV] & 80.361 & -0.058 & 0.522 & -0.073 & -0.034 & 0.0002 & -1.069 & 114.885 & $ < $ 0.0002 \\
\hline
$ \Gamma_{W} $ [MeV] & 2090.6 & -4.4 & 41.1 & 48.2 & -2.6 & 0.9 & -83.0 & 8953.8 & $ < $ 0.02 \\
\hline
\hline
$ Q_{W} (Cs) $ & -72.98 & -0.09 & 0.05 & -0.21 & 0.83 & 0.12 & -5.17 & 172.91 & $ < $ 0.006 \\
\hline
$ Q_{W} (Tl) $ & -116.48 & -0.13 & 0.02 & -0.33 & 1.31 & 0.17 & -7.7 & 257.02 & $ < $ 0.008 \\
\hline
\end{tabular}
\caption{The numeric values (truncated here for illustration) for the parameters of different observables, $ * = 10^{3} $.}

\end{center}

\end{sidewaystable}

\vspace*{5mm}

When doing the parameterization, we have divided the intervals of $ \mathcal{S} $ by the following numbers of points :

\vspace*{5mm}

\begin{center}
\begin{tabular}{|c|c|c|c|}
\hline
variable & interval(s) & nb. of points & parameter(s) fitted \\
\hline
$ \Delta \alpha^{(5)}_{had} (M_{Z}) $ & $ 0.02757 \pm 0.00050 $ & 81 & $ c_{6} $ \\
\hline
$ M_{Z} $ & $ 91.1876 \pm 0.0042 $ & 21 & $ c_{7} $ \\
\hline
$ m^{pole}_{top} $ & $ 173.2 \pm 2.0 $ & 71 & $ c_{2} $ \\
\hline
$ M_H $ & $ 125.7 \pm 2.5 $ & 51 & $ c_{1} $ \\
\hline
$ \alpha_{s} (M_{Z}) $ & $ 0.1184 \pm 0.0050 $ & 111 & $ c_{3} $ \\
\hline
All & the same & $ 5^5 $ & $ c_{4}, c_{5} $ \\
\hline
\end{tabular}
\end{center}

\vspace*{5mm}

On determining $ c_{3} $ from $ \alpha_{s} (M_{Z}) $, we have set $ c_{4} $ to zero (the dependence of $ c_{3} $ on $ c_{4} $ is very small): the resulting value for $ c_{3} $ is the same as the value found for a conjoint fit of $ c_{3} $ and $ c_{4} $ from varying $ \alpha_{s} (M_{Z}) $; the difference on the values of this $ c_{4} $ and the one determined from the variation of all the variables together was generally not big. $ \mathcal{X}_{0} $ was determined exactly when setting $ \Delta \alpha^{(5)}_{had} (M_{Z}) = 0.02757 $, $ M_{Z} = 91.1876 $, $ m^{pole}_{top} = 173.2 $, $ M_H = 125.7 $, and $ \alpha_{s} (M_{Z}) = 0.1184 $. Varying all the variables was as well used to determine the maximum deviations of the parameterization, thus showing that extra quadratic dependences are not necessary.

\chapter{$ \rho $ parameter}\label{sec:rhoParameter}

The problem of the $ \rho $ parameter value for a bi-doublet and two triplets is considered by \cite{Branco:1999fs}, so we are going to list some of the results. The important mixing matrices are, where we take $ g_L = g_R = g $ for simplicity

\begin{equation}
\frac{1}{2} \begin{pmatrix}
\frac{| \kappa_{1} |^{2} + | \kappa_{2} |^{2}}{2} + | \kappa_L |^{2} & - \kappa_{1} \kappa^{}_{2} \\
- \kappa^{}_{1} \kappa_{2} & \frac{| \kappa_{1} |^{2} + | \kappa_{2} |^{2}}{2} + | \kappa_R |^{2}
\end{pmatrix}
\end{equation}
\noindent
in the $ W_L, W_R $ basis and

\begin{equation}
\frac{1}{2} \begin{pmatrix}
\frac{| \kappa_{1} |^{2} + | \kappa_{2} |^{2}}{4} + | \kappa_L |^{2} & - \frac{| \kappa_{1} |^{2} + | \kappa_{2} |^{2}}{4} + \frac{s^{2}_{\theta}}{c^{2}_{\theta} - s^{2}_{\theta}} | \kappa_L |^{2} \\
- \frac{| \kappa_{1} |^{2} + | \kappa_{2} |^{2}}{4} + \frac{s^{2}_{\theta}}{c^{2}_{\theta} - s^{2}_{\theta}} | \kappa_L |^{2} & \frac{| \kappa_{1} |^{2} + | \kappa_{2} |^{2}}{4} + \frac{s^{4}_{\theta} | \kappa_L |^{2} + c^{4}_{\theta} | \kappa_R |^{2}}{(c^{2}_{\theta} - s^{2}_{\theta})^{2}}
\end{pmatrix}
\end{equation}
\noindent
in the $ X_1, X_2 $ basis. Here the weak angles are given by

\begin{equation}
c^{2}_{\theta} \equiv \frac{g_{B-L}^{2} + g^{2}}{2 g_{B-L}^{2} + g^{2}} \, , \quad s^{2}_{\theta} \equiv \frac{g_{B-L}^{2}}{2 g_{B-L}^{2} + g^{2}} \, .
\end{equation}

On searching for parity breaking, $ | \kappa_L | \neq | \kappa_R | $, and supposing $ | \kappa_R | $ much bigger than $ | \kappa_L |, | \kappa_{1,2} | $, one has the following light masses, to leading order in $ 1/| \kappa_R |^{2} $:

\begin{eqnarray}
&& M^{2}_{W} \approx \frac{g^{2}}{2} \left( \frac{| \kappa_{1} |^{2} + | \kappa_{2} |^{2}}{2} + | \kappa_L |^{2} \right) \, , \\
&& M^{2}_{Z} \approx \frac{g^{2}}{2 c^{2}_{\theta}} \left( \frac{| \kappa_{1} |^{2} + | \kappa_{2} |^{2}}{2} + 2 | \kappa_L |^{2} \right) \approx \frac{M^{2}_{W}}{c^{2}_{\theta}} + g^{2} \frac{| \kappa_L |^{2}}{2 c^{2}_{\theta}} \, .
\end{eqnarray}

Thus

\begin{equation}
\rho \approx (1 + g^{2} | \kappa_L |^{2} / (2 M^{2}_{W}))^{-1}
\end{equation}
is close to 1 only when $ | \kappa_L | $ is much smaller than $ | \kappa_{1,2} | $.

Following \cite{Branco:1999fs}, we calculate explicitly the case where the triplets are replaced by doublets. We find

\begin{equation}
\frac{1}{2} \begin{pmatrix}
\frac{| \kappa_{1} |^{2} + | \kappa_{2} |^{2} + | \kappa_L |^{2}}{2} & - \kappa_{1} \kappa^{}_{2} \\
- \kappa^{}_{1} \kappa_{2} & \frac{| \kappa_{1} |^{2} + | \kappa_{2} |^{2} + | \kappa_R |^{2}}{2}
\end{pmatrix}
\end{equation}
\noindent
in the $ W $ sector and

\begin{equation}
\frac{1}{2} \begin{pmatrix}
\frac{| \kappa_{1} |^{2} + | \kappa_{2} |^{2} + | \kappa_L |^{2}}{4} & - \frac{| \kappa_{1} |^{2} + | \kappa_{2} |^{2}}{4} + \frac{s^{2}_{\theta}}{c^{2}_{\theta} - s^{2}_{\theta}} \frac{| \kappa_L |^{2}}{4} \\
- \frac{| \kappa_{1} |^{2} + | \kappa_{2} |^{2}}{4} + \frac{s^{2}_{\theta}}{c^{2}_{\theta} - s^{2}_{\theta}} \frac{| \kappa_L |^{2}}{4} & \frac{| \kappa_{1} |^{2} + | \kappa_{2} |^{2}}{4} + \frac{s^{4}_{\theta} | \kappa_L |^{2} + c^{4}_{\theta} | \kappa_R |^{2}}{4(c^{2}_{\theta} - s^{2}_{\theta})^{2}}
\end{pmatrix}
\end{equation}
\noindent
in the neutral sector, which implies $ \rho = 1 $ to leading order in $ 1/| \kappa_R |^{2} $.

\chapter{Stability conditions}\label{sec:stableConds}


The extreme conditions (first order derivatives of the potential in Eq.~\eqref{eq:potentialSymPdoublet}, giving the conditions for the stability of the vacuum state) with respect to $ \{ \varphi^{0r}_{1}, \varphi^{0r}_{2}, \chi^{0r}_{R}, \chi^{0r}_{L}, \varphi^{0i}_{2}, \chi^{0i}_{L} \} $ are

\begin{eqnarray}
0 &=& \kappa_{2} \left( B + 4 \lambda_{4} \kappa^{2}_{1} \right) \cos \alpha + \kappa_{1} \left[ A + \alpha_{4} (\kappa^{2}_{R} + \kappa^{2}_{L}) + \left( 4 \lambda_{3} + 8 \lambda_{2} \cos (2 \alpha) \right) \kappa^{2}_{2} \right] \nonumber\\
&& + \sqrt{2} {\mu'}_{2} \kappa_L \kappa_R \cos \theta_{L} + \alpha_{2} \kappa_{2} \left[ \kappa^{2}_{R} \cos (\alpha + \delta_{2}) + \kappa^{2}_{L} \cos (\alpha - \delta_{2}) \right] , \\
0 &=& \kappa_{1} \left( B + 4 \lambda_{4} \kappa^{2}_{2} \right) \cos \alpha + \kappa_{2} \left[ A + \alpha_{3} (\kappa^{2}_{R} + \kappa^{2}_{L}) + \left( 4 \lambda_{3} + 8 \lambda_{2} \cos (2 \alpha) \right) \kappa^{2}_{1} \right] \nonumber\\
&& + \sqrt{2} {\mu'}_{1} \kappa_L \kappa_R \cos (\alpha - \theta_{L}) + \alpha_{2} \kappa_{1} \left[ \kappa^{2}_{R} \cos (\alpha + \delta_{2}) + \kappa^{2}_{L} \cos (\alpha - \delta_{2}) \right] , \nonumber\\
&& \\
0 &=& \frac{C}{\kappa^{2}_{R}} + \frac{\kappa^{2}_{L}}{\kappa^{2}_{R}} 2 \rho + \sqrt{2} \frac{\kappa_L}{\kappa_R} \left( \frac{{\mu'}_{1}}{\kappa_R} \frac{\kappa_{2}}{\kappa_R} \cos ( \alpha - \theta_{L} ) + \frac{{\mu'}_{2}}{\kappa_R} \frac{\kappa_{1}}{\kappa_R} \cos \theta_{L} \right) \nonumber\\
&& + 2 \alpha_{2} r \left( \frac{\kappa_{1}}{\kappa_R} \right)^{2} \cos (\alpha + \delta_{2}) , \\
0 &=& \frac{\kappa_L}{\kappa_R} \left( \frac{C}{\kappa^{2}_{R}} + 2 \rho \right) + \sqrt{2} \left( \frac{{\mu'}_{1}}{\kappa_R} \frac{\kappa_{2}}{\kappa_R} \cos ( \alpha - \theta_{L} ) + \frac{{\mu'}_{2}}{\kappa_R} \frac{\kappa_{1}}{\kappa_R} \cos \theta_{L} \right) \nonumber\\
&& + 2 \alpha_{2} r w \left( \frac{\kappa_{1}}{\kappa_R} \right)^{2} \cos (\alpha - \delta_{2}) , \\
0 &=& \kappa_{1} \left( B + 16 \lambda_{2} \kappa_{1} \kappa_{2} \cos \alpha \right) \sin \alpha + \sqrt{2} {\mu'}_{1} \kappa_L \kappa_R \sin (\alpha - \theta_{L}) \nonumber\\
&& + \alpha_{2} \kappa_{1} \left[ \kappa^{2}_{R} \sin (\alpha + \delta_{2}) + \kappa^{2}_{L} \sin (\alpha - \delta_{2}) \right] , \label{eq:eqStabDoublets5}\\
0 &=& {\mu'}_{1} \kappa_{2} \sin ( \alpha - \theta_{L} ) - {\mu'}_{2} \kappa_{1} \sin \theta_{L} , \label{eq:eqStabDoublets6}
\end{eqnarray}
where $ \rho \equiv \rho_3 / 2 - \rho_1 $, $ w \equiv \frac{\kappa_L}{\kappa_{1}} $, $ r \equiv \frac{\kappa_{2}}{\kappa_{1}} $ and

\begin{eqnarray}
A &\equiv& -2 \mu^{2}_{1} + \alpha_{1} (\kappa^{2}_{R} + \kappa^{2}_{L}) + 2 \lambda_{1} (\kappa^{2}_{1} + \kappa^{2}_{2}) , \\
B &\equiv& -4 \mu^{2}_{2} + 2 \lambda_{4} (\kappa^{2}_{1} + \kappa^{2}_{2}) , \\
C &\equiv& -2 \mu^{2}_{3} + 2 \rho_{1} (\kappa^{2}_{R} + \kappa^{2}_{L}) + \alpha_{1} (\kappa^{2}_{1} + \kappa^{2}_{2}) + \alpha_{4} \kappa^{2}_{1} + \alpha_{3} \kappa^{2}_{2} .
\end{eqnarray}
These six equations provide relations among the VEV values $ \kappa_{1,2} $, $ \kappa_{L,R} $, $ \alpha $ and $ \theta_{L} $, and the underlying parameters of the potential, $ \mu^2_{1,2,3} $, $ \alpha_{1,2,3,4} $, $ \mu'_{1,2} $, $ \rho_{1,3} $, $ \lambda_{1,2,3,4} $ and $ \delta_2 $.

In order to analyze the limit where $ \kappa_R $ is much bigger than the other VEVs, we define $ \epsilon \equiv \sqrt{1 + r^2 + w^2} \frac{\kappa_{1}}{\kappa_R} $. It is useful to further define 

\begin{eqnarray}
A' &\equiv & -2 \mu^{2}_{1} + \alpha_{1} \kappa^{2}_{R} , \\
B' &\equiv & -4 \mu^{2}_{2} , \\
C' &\equiv & -2 \mu^{2}_{3} + 2 \rho_{1} \kappa^{2}_{R} .
\end{eqnarray}
The extreme conditions are now given under the following form, to leading order in $ \epsilon \ll 1 $:

\begin{eqnarray}
0 &\simeq& r \frac{B'}{\kappa^{2}_{R}} \cos \alpha + \left( \frac{A'}{\kappa^{2}_{R}} + \alpha_{4} \right) + \sqrt{2} \frac{{\mu'}_{2}}{\kappa_R} w \cos \theta_{L} + \alpha_{2} r \cos (\alpha + \delta_{2}) , \label{eq:equation1simpDoub}\\
0 &\simeq& \frac{B'}{\kappa^{2}_{R}} \cos \alpha + r \left( \frac{A'}{\kappa^{2}_{R}} + \alpha_{3} \right) + \sqrt{2} \frac{{\mu'}_{1}}{\kappa_R} w \cos (\alpha - \theta_{L}) + \alpha_{2} \cos (\alpha + \delta_{2}) , \nonumber\\
&& \label{eq:equation2simpDoub}\\
0 &\simeq& \frac{C'}{\kappa^{2}_{R}} , \label{eq:equation3simpDoub}\\
0 &\simeq& w 2 \rho_{} + \sqrt{2} \left( r \frac{{\mu'}_{1}}{\kappa_R} \cos ( \alpha - \theta_{L} ) + \frac{{\mu'}_{2}}{\kappa_R} \cos \theta_{L} \right) , \label{eq:equation4simpDoub}\\
0 &\simeq& \frac{B'}{\kappa^{2}_{R}} \sin \alpha + \sqrt{2} \frac{{\mu'}_{1}}{\kappa_R} w \sin (\alpha - \theta_{L}) + \alpha_{2} \sin (\alpha + \delta_{2}) , \label{eq:equation5simpDoub}\\
&& \frac{{\mu'}_{1}}{\kappa_R} r \sin ( \alpha - \theta_{L} ) = \frac{{\mu'}_{2}}{\kappa_R} \sin \theta_{L} , \label{eq:equation6simpDoub}
\end{eqnarray}
where in the fourth equation (Eq. (\ref{eq:equation4simpDoub})) we have used the third one (Eq. (\ref{eq:equation3simpDoub})). When two very different energy scales are present, such as $ \kappa_R \gg \kappa_{1,2,L} $, one may face a certain amount of tuning as we now explain (an exception would be the situation where two equations reduce to the same one). In our case, we have six equations (corrected by higher orders in $ \epsilon $) and five parameters $ \{ r, w, \kappa_R, \theta_L, \alpha \} $ related to the VEVs (\cite{Barenboim:2001vu}, \cite{Kiers:2005gh}, \cite{Ellwanger:2011mu}). We can imagine solving for the VEVs using five out of the six equations, and plugging the solutions into the sixth. In other words a pure combination of parameters from the potential, say $ f(\mu^2_{1,2} / \mu^2_{3}, \alpha_{1,2,3,4}, \rho_{1,3}, \mu'_{1} / \mu'_{2}) $, is zero up to corrections suppressed by $ \epsilon $, $ f = \mathcal{O} (\epsilon) $. Whether this resulting combination is stable under radiative corrections, thus stating or not a certain amount of tuning, remains to be verified. 



The equations given so far hold in the most general situation, and for simplicity we are going to compute the mass spectrum in the particular case where $ \sin \theta_{L} $, and consequently $ \sin \alpha $, goes to zero. For this limit, it is also necessary to have $ \sin \delta_{2} \rightarrow 0 $, where $ \delta_{2} $ is the $ \mathcal{C P}- $violating phase of the Higgs potential: note from Eq.~\eqref{eq:eqStabDoublets6} or \eqref{eq:equation6simpDoub} that in the special case where $ \sin \theta_L \rightarrow 0 $,

\begin{equation}
\mu'_1 \kappa_2 \sin \alpha \rightarrow 0 \, ,
\end{equation}
and in this case we ask for $ \sin \alpha \rightarrow 0 $; then, Eq.~\eqref{eq:eqStabDoublets5} or \eqref{eq:equation5simpDoub} would imply

\begin{equation}
\alpha_2 \kappa_1 (v^2_R - v^2_L) \sin \delta_2 \rightarrow 0 \, ,
\end{equation}
and we ask for $ \sin \delta_2 \rightarrow 0 $.

To deal with the limit $ \sin \alpha \rightarrow 0 $, a new parameter is introduced by the relative speed $ s_{\delta,\alpha} \equiv \lim_{\sin \alpha \rightarrow 0} \frac{\sin \delta_{2}}{\sin \alpha} $. For further discussion, the eigenvalues of the mass matrix are given in Table~\ref{tab:ExtraTable1}, up to order $ \mathcal{O} (\epsilon) $.

\begin{table}[H]
\centering
\def\arraystretch{2.0}
\begin{tabular}{|c|c|}
  \hline
  Name & Mass$ ^2 $ / $ \kappa^{2}_{R} $ \\
  \hline
  \hline
  $ h^0 $ & $ 0 $ \\
  \hline
  $ H^0_3 $ & $ 2 \rho_{1} $ \\
  \hline
  $ H^0_1, A^0_1 $ & $ \frac{1}{4 r^{2}} (b - \sqrt{b^{2} - 8 r^{2} \Delta} ) $ \\
  \hline
  $ H^0_2, A^0_2 $ & $ \frac{1}{4 r^{2}} (b + \sqrt{b^{2} - 8 r^{2} \Delta} ) $ \\
  \hline
\end{tabular}
\caption{Neutral scalar spectrum.} \label{tab:ExtraTable1}
\end{table}
\noindent
Above, we have employed the definitions

\begin{eqnarray}
b &=& \alpha_{2} s_{\delta,\alpha} r (1 + r^{2}) + 2 \rho \{ c^{2} w^{2} + r^{2} [ 1 + (1 - c)^{2} w^{2} ] \} , \\
\Delta &=& \alpha_{2} s_{\delta,\alpha} \rho r (1 + r^{2} + w^{2}) \, ,
\end{eqnarray}
where $ c $ is defined to be $ \frac{C}{1+C} $, $ C \equiv \frac{{\mu'}_{1}}{{\mu'}_{2}} r $, and $ \rho \equiv (\frac{\rho_{3}}{2} - \rho_{1}) $ (we take $ \cos \alpha = \cos \theta_{L} = 1 $). Therefore, we see that when $ s_{\delta,\alpha} = 0 $ two heavy particles become light, i.e. massless up to order $ \mathcal{O} (\epsilon) $.  In other words, if $ \sin \delta_2 = 0 $ from the beginning, while $ \sin \alpha , \sin \theta_L \neq 0 $, the spectrum would have many light physical scalars.

If one considers solving for $ \{ \mu^{2}_{1,2,3}, {\mu'}_{1,2}, \alpha_{2} \} $, to leading order in $ \epsilon $:

\begin{eqnarray}
\frac{\mu^{2}_{1}}{\kappa^{2}_{R}} &\simeq& \frac{(\alpha_{1} + \alpha_{3})}{2} - \frac{\alpha_{34} + 2 (1 - 2 c) \rho w^{2}}{2 (1 - r^{2})} , \\
\frac{\mu^{2}_{2}}{\kappa^{2}_{R}} &\simeq& \frac{\alpha_{2}}{4} + \frac{\alpha_{34} r^{2} - 2 (c - (1 - c) r^{2}) \rho w^{2}}{4 r (1 - r^{2})} , \\
\frac{\mu^{2}_{3}}{\kappa^{2}_{R}} &\simeq& \rho_{1} , \\
\frac{{\mu'}_{1}}{\kappa_R} &\simeq& - \frac{\sqrt{2} c \rho w}{r} , \quad \frac{{\mu'}_{2}}{\kappa_R} \simeq \sqrt{2} (c - 1) \rho w , \label{eq:equation9limit}\\
\alpha_{2} s_{\delta, \alpha} &\simeq& \frac{-2 c^{2} \rho  w^{2} + r^{2} (\alpha_{34} + 2 (1 - c)^{2} \rho w^{2})}{r (1 - r^{2})} , \label{eq:equation8limit}
\end{eqnarray}
where $ \alpha_{34} \equiv \alpha_{3} - \alpha_{4} $. Equation (\ref{eq:equation8limit}) should be seen as the definition of $ s_{\delta, \alpha} $, and Eqs.~(\ref{eq:equation9limit}) are not independent. Therefore we have four equations relating the three remaining VEVs $ r, w, \kappa_R $.

To conclude, though we are discussing the limit $ \sin \delta_{2}, \sin \theta_{L}, \sin \alpha \rightarrow 0 $, we have verified that this case is equivalent to the special case where $ \sin \delta_{2} = \sin \theta_{L} = \sin \alpha = 0 $. The interesting point to note, however, is the interplay between the masses of the scalars and the $ \mathcal{C P}- $violating phases: when setting $ \sin \delta_2 $ to zero we also require real VEVs or otherwise there would be extra light scalars (of masses of the order of the EWSB energy scale), a case not considered here, due to the danger related to potentially too large Flavour-Changing Neutral Currents (FCNC) amplitudes.

\chapter{Spectrum of the scalar particles}\label{sec:spectrumScalars}

When diagonalizing neutral the mass matrix, we find the following two massless combinations (up to $ \mathcal{O} (\epsilon^{3}) $ corrections) eigenstates, orthonormal up to order $ \mathcal{O} (\epsilon^2) $

\begin{eqnarray} \label{eq:equation2}
G^{0}_{1} &=& \frac{1}{\sqrt{1 + r^{2} + w^{2}}} (-\varphi^{0i}_{1} + r \varphi^{0i}_{2} + w \chi^{0i}_{L}) , \nonumber\\
G^{0}_{2} &=& \chi^{0i}_{R} - \epsilon (-\varphi^{0i}_{1} + r \varphi^{0i}_{2} - \frac{1 + r^{2}}{w} \chi^{0i}_{L}) \frac{w^{2}}{1 + r^{2} + w^{2}}
\end{eqnarray}
(apart from the light SM-like Higgs, built out of $ \varphi^{0r}_{1,2} $, $ \chi^{0r}_{L,R} $). We could in principle quote any orthogonal combination out of them, and we need to determine the one which gives the would-be Goldstone bosons. From the couplings of these scalar fields to the Gauge boson $ Z' $, we determine then the linear term of the form $ Z' \partial G^0_{Z'} $, and similarly for $ G^0_Z $. The neutral Goldstones are finally given by

\begin{eqnarray}
G^{0}_{Z} &=& G^{0}_{1} - \epsilon \sqrt{1 + r^{2} + w^{2}} \left[ s^{2}_{R} - \left( 1 + \frac{w^{2}}{1 + r^{2}} \right)^{-1} \right] G^{0}_{2} , \\
G^{0}_{Z'} &=& G^{0}_{2} + \epsilon \sqrt{1 + r^{2} + w^{2}} \left[ s^{2}_{R} - \left( 1 + \frac{w^{2}}{1 + r^{2}} \right)^{-1} \right] G^{0}_{1} .
\end{eqnarray}

Similarly, we find the following expressions for the charged Goldstones (massless up to order $ \mathcal{O} (\epsilon^{3}) $)

\begin{eqnarray}\label{eq:GoldstonesUpToEpsilon}
G^{\pm} &=& \frac{1}{\sqrt{1+r^{2}+w^{2}}} (- \varphi^{\pm}_{1} + r \varphi^{\pm}_{2} + w \chi^{\pm}_{L}) + 2 \epsilon \frac{r}{\sqrt{1+r^{2}+w^{2}}} \chi^{\pm}_{R} \nonumber\\
{G'}^{\pm} &=& \chi^{\pm}_{R} + \epsilon (r \varphi^{\pm}_{1} - \varphi^{\pm}_{2}) .
\end{eqnarray}

In Table~\ref{tab:ExtraTable2}, we give the neutral Higgses, up to order $ \mathcal{O} (\epsilon^{1}) $.

\vspace{5mm}

\begin{table}[H]
\centering
\def\arraystretch{2.0}
\begin{tabular}{|c|c|c|}
  \hline
  Name & Mass$ ^2 $ / $ \kappa^{2}_{R} $ & Vectorial space (not normalized) \\
  \hline
  \hline
  $ h^{0} $ & $ 0 $ & $ \varphi^{0r}_{1} + r \varphi^{0r}_{2} + w \chi^{0r}_{L} $ \\
  \hline
  $ H^{0}_{3} $ & $ 2 \rho_{1} $ & $ \chi^{0r}_{R} $ \\
  \hline
  $ H^{0}_{1} $ & $ \frac{1}{4 r^{2} (1 - r^{2})} \left( b - \sqrt{b^{2} - 8 r^{2} (1 - r^{2}) \Delta} \right) $ & $ \tilde{p} \varphi^{0r}_{1} + \varphi^{0r}_{2} + p \chi^{0r}_{L} $ \\
  \hline
  $ H^{0}_{2} $ & $ \frac{1}{4 r^{2} (1 - r^{2})} \left( b + \sqrt{b^{2} - 8 r^{2} (1 - r^{2}) \Delta} \right) $ & $ \tilde{q} \varphi^{0r}_{1} + \varphi^{0r}_{2} + q \chi^{0r}_{L} $ \\
  \hline
  \hline
  $ A^{0}_{1} $ & $ \frac{1}{4 r^{2} (1 - r^{2})} \left( b - \sqrt{b^{2} - 8 r^{2} (1 - r^{2}) \Delta} \right) $ & $ - \tilde{p}  \varphi^{0i}_{1} + \varphi^{0i}_{2} + p \chi^{0i}_{L} $ \\
  \hline
  $ A^{0}_{2} $ & $ \frac{1}{4 r^{2} (1 - r^{2})} \left( b + \sqrt{b^{2} - 8 r^{2} (1 - r^{2}) \Delta} \right) $ & $ - \tilde{q} \varphi^{0i}_{1} + \varphi^{0i}_{2} + q \chi^{0i}_{L} $ \\
  \hline
\end{tabular} \caption{Neutral mass spectrum.}\label{tab:ExtraTable2}
\end{table}

\vspace{5mm}

Above, we have employed the following definitions

\begin{eqnarray}
b &\equiv& \alpha_{34} r^{2} (1 + r^{2}) + 2 \rho r^{2} \{ 2 ( 1 - 2 c ) w^{2} + 1 - r^{2} \} , \label{eq:equation6}\\
\Delta &\equiv& \rho (1 + r^{2} + w^{2}) \{ \alpha_{34} r^{2} + 2 \rho w^{2} [r^{2} (1 - c)^{2} - c^{2}] \} , \label{eq:equation7}
\end{eqnarray}
\noindent
where $ c = \frac{C}{1+C} $ and $ C = \frac{{\mu'}_{1}}{{\mu'}_{2}} r $. At this order then, the only relevant parameters are $ \alpha_{34} $, $ \rho_3 / 2 - \rho_1 $ and $ \mu'_1 / \mu'_2 $. $ p, \tilde{p} $ and $ q, \tilde{q} $ are combinations of parameters of the potential and VEV values. They can be related to each other since the eigenvectors shown above are orthogonal, leaving a dependence on only one function:



\begin{equation}
p=-\frac{
   k^2 (1+r x)(1-\delta^2 +\left(1 + \delta^2\right)X))+2 w^2 r(r-x)}{2 w  \left(k^2 (r - x\delta^2  \right)- r^2 (r-x))} \, .
\label{eq:parp}
\end{equation}




For completeness, the light Higgs mass is given by

{\def\arraystretch{2.0}
\begin{eqnarray}
&& \{ - \frac{1}{\rho_{1}} \{ [ (\alpha_{1} + \alpha_{3}) r^{2} (1 + r^{2}) + \alpha_{2} r (2 r^{2} - (1 - r^{2}) s_{\delta, \alpha}) ] \nonumber\\
&& + w^{2} [ 2 c \rho_{1} (2 r^{2} + (1 - r^{2}) c) + \rho_{3} (-c^{2} + (1 - c)^{2} r^{2}) ] \}^{2} \nonumber\\
&& + 4 r^{2} \{ r^{2} [ \lambda_{1} (1 + r^{2})^{2} + 8 \lambda_{2} r^{2} + 4 \lambda_{3} r^{2} + 4 \lambda_{4} r (1 + r^{2}) ] \nonumber\\
&& + w^{2} [ (\alpha_{1} + \alpha_{3}) r^{2} (1 + r^{2}) + \alpha_{2} r (2 r^{2} - (1 - r^{2}) s_{\delta, \alpha}) ] \nonumber\\
&& + w^{4} [ \rho_{1} (2 c^{2} (1 + r^{2}) - (1 - 2 c)^{2} r^{2}) + \rho_{3} (-c^{2} + (1 - c)^{2} r^{2}) ] \} \} \nonumber\\
&& \frac{1}{2 r^{4} (1 + r^2 + w^2)} \epsilon^{2} \kappa^{2}_{R} ,
\end{eqnarray}
}
up to order $ \mathcal{O} (\epsilon^{3}) $.



To conclude, for the charged Higgses, one has:

\begin{center}
\begin{tabular}{|c|c|c|}
  \hline
  Name & Mass$ ^2 $ / $ \kappa^{2}_{R} $ & Vectorial space (not normalized) \\
  \hline
  \hline
  $ H^{\pm}_{1} $ & $ \frac{1}{4 r^{2} (1 - r^{2})} (b - \sqrt{b^{2} - 8 r^{2} (1 - r^{2}) \Delta} ) $ & $ p \chi^{\pm}_{L} - \tilde{p} \varphi^{\pm}_{1} + \varphi^{\pm}_{2} $ \\
  \hline
  $ H^{\pm}_{2} $ & $ \frac{1}{4 r^{2} (1 - r^{2})} (b + \sqrt{b^{2} - 8 r^{2} (1 - r^{2}) \Delta} ) $ & $ q \chi^{\pm}_{L} - \tilde{q} \varphi^{\pm}_{1} + \varphi^{\pm}_{2} $ \\
  \hline
\end{tabular}
\end{center}



\section{Tables of couplings}\label{sec:tableOfCouplings}

\begin{table}[H]
\begin{center}
\def\arraystretch{2.5}
\begin{tabular}{c|c|c|c}
 & $A^0_\mu$ & $Z^0_\mu$ & ${Z'}^0_\mu$\\
 \hline
$\bar{u}^i_L u^i_L$ 
& $\frac{2}{3}e\gamma^\mu$ 
& $\frac{e}{c_Ws_W}\gamma^\mu\left(\frac{1}{2}-\frac{2}{3}s_W^2+\frac{1}{6}\epsilon^2[(1+r^2)s_R^2-k^2s_R^4]\right)$ 
& $\frac{es_R}{c_Rc_W}\left(-\frac{1}{6}\right)\gamma^\mu$\\
$\bar{u}^i_R u^i_R$ 
& $\frac{2}{3}e\gamma^\mu$ 
& $\frac{e}{c_Ws_W}\gamma^\mu\left(-\frac{2}{3}s_W^2+\frac{(4c_R^2-1)}{6}\epsilon^2[-w^2+c_R^2k^2]\right)$ 
& $\frac{e}{s_Rc_Rc_W}\left(\frac{1}{2}-\frac{2}{3}s_R^2\right)\gamma^\mu$\\
$\bar{d}^i_L d^i_L$ 
& $-\frac{1}{3}e\gamma^\mu$ 
& $\frac{e}{c_Ws_W}\gamma^\mu\left(-\frac{1}{2}+\frac{1}{3}s_W^2-\frac{1}{6}\epsilon^2[(1+r^2)s_R^2-k^2s_R^4]\right)$ 
& $\frac{es_R}{c_Rc_W}\left(-\frac{1}{6}\right)\gamma^\mu$\\
$\bar{d}^i_R d^i_R$ 
& $-\frac{1}{3}e\gamma^\mu$ 
& $\frac{e}{c_Ws_W}\gamma^\mu\left(\frac{1}{3}s_W^2-\frac{(2c_R^2+1)}{6}\epsilon^2[-w^2+c_R^2k^2]\right)$ 
& $\frac{e}{s_Rc_Rc_W}\left(-\frac{1}{2}+\frac{1}{3}s_R^2\right)\gamma^\mu$\\
$\bar{\nu}^i_L \nu^i_L$ 
& $0$ 
& $\frac{e}{c_Ws_W}\gamma^\mu\left(\frac{1}{2}+\frac{1}{2}\epsilon^2[(1+r^2)s_R^2-k^2s_R^4]\right)$ 
& $\frac{es_R}{c_Rc_W}\left(\frac{1}{2}\right)\gamma^\mu$\\
$\bar{\nu}^i_R \nu^i_R$ 
&$0$ 
& $\frac{e}{c_Ws_W}\gamma^\mu\left(\frac{1}{2}+\frac{1}{2}\epsilon^2[(1+r^2)s_R^2-k^2s_R^4]\right)$ 
& $\frac{es_R}{c_Rc_W}\left(\frac{1}{2}\right)\gamma^\mu$\\
$\bar{\ell}^i_L \ell^i_L$ 
& $-e\gamma^\mu$ 
& $\frac{e}{c_Ws_W}\gamma^\mu\left(-\frac{1}{2}+s_W^2+\epsilon^2[(1+r^2)s_R^2-k^2s_R^4]\right)$ 
& $\frac{es_R}{c_Rc_W}\left(\frac{1}{2}\right)\gamma^\mu$\\
$\bar{\ell}^i_R \ell^i_R$ 
& $-e\gamma^\mu$ 
& $\frac{e}{c_Ws_W}\gamma^\mu\left(\frac{1}{2}s_W^2-\frac{(2c_R^2-1)}{2}\epsilon^2[-w^2+c_R^2k^2]\right)$ 
& $\frac{e}{s_Rc_Rc_W}\left(-\frac{1}{2}+\frac{1}{2}s_R^2\right)\gamma^\mu$
\end{tabular}
\end{center}
\caption{Couplings of the fermions to the neutral gauge bosons.}
\end{table}

\begin{table}[H]
\begin{center}
\def\arraystretch{2.5}
\begin{tabular}{c|c|c}
 & $W^+_\mu$ & ${W'}^+_\mu$\\
 \hline
$\bar{u}^i_L d^j_L$ &  $\frac{e}{\sqrt{2}s_W} V^L_{ij} \gamma_\mu $ & 0\\
$\bar{u}^i_R d^j_R$ & $\frac{e\sqrt{2}}{s_W}  r \epsilon^2 V^R_{ij} \gamma_\mu $ & $\frac{e}{\sqrt{2}s_Rc_W}V^R_{ij} \gamma_\mu$
\end{tabular}
\end{center}
\caption{Couplings of the fermions to the charged gauge bosons. For leptons, $ V^{L,R} \rightarrow V^{L,R}_{lept} $, and similarly for the masses.}
\end{table}

\begin{table}[H]
\begin{center}\small
\def\arraystretch{2.5}
\begin{tabular}{c|c|c}
 & $H_1^0$ & $H_2^0$  \\
 \hline
$\bar{u}^i_R u^j_L$ & $\frac{-(2r+wp)m_u^i \delta^{ij}+(1+r^2+rwp) V^R_{ia}m^a_d V^L_{ja*}}{(1-r^2)\kappa_1\sqrt{1+p^2+(r+wp)^2}}$ 
& $\frac{(w-r^2 w -2pr -2prw^2)m_u^i \delta^{ij}+pk^2 V^R_{ia}m^a_d V^L_{ja*}}{(1-r^2)\kappa_1 k \sqrt{1+p^2+(r+wp)^2}}$ \\
$\bar{u}^i_L u^j_R$ & $\frac{-(2r+wp)m_u^i \delta^{ij}+((1+r^2+rwp) V^L_{ia}m^a_d V^R_{ja*}}{(1-r^2)\kappa_1\sqrt{1+p^2+(r+wp)^2}}$
& $\frac{(w-r^2 w -2pr -2prw^2)m_u^i \delta^{ij}+pk^2 V^L_{ia}m^a_d V^R_{ja*}}{(1-r^2)\kappa_1 k \sqrt{1+p^2+(r+wp)^2}}$ \\
$\bar{d}^i_R d^j_L$ & $\frac{(1+r^2+rwp)V^R_{*ai}m^a_u V^L_{aj}- (2r+wp) m^i_d \delta^{ij}}{(1-r^2)\kappa_1\sqrt{1+p^2+(r+wp)^2}}$ 
& $\frac{pk^2 V^R_{*ai}m^a_u V^L_{aj}+ (w-r^2w-2pr-w^2pr)m^i_d \delta^{ij}}{(1-r^2)\kappa_1k\sqrt{1+p^2+(r+wp)^2}}$ \\
$\bar{d}^i_L d^j_R$ &  $\frac{(1+r^2+rwp)V^L_{*ai}m^a_u V^R_{aj}- (2r+wp)m^i_d \delta^{ij}}{(1-r^2)\kappa_1\sqrt{1+p^2+(r+wp)^2}}$ 
& $\frac{pk^2 V^L_{*ai}m^a_u V^R_{aj}+ (w-r^2w-2pr-w^2pr)m^i_d \delta^{ij}}{(1-r^2)\kappa_1k\sqrt{1+p^2+(r+wp)^2}}$
\end{tabular}
\vspace{0.3cm}
\caption{Couplings of the fermions to the heavy $ \mathcal{C P}- $even physical scalars. $H_0^3$ does not couple to fermions at this order.} 
\end{center}
\end{table}

\begin{table}[H]
\begin{center}\small
\def\arraystretch{2.5}
\begin{tabular}{c|c|c}
 & $A_1^0$ & $A_2^0$  \\
 \hline
$\bar{u}^i_R u^j_L$ & $i\frac{-(2r+wp)m_u^i \delta^{ij}+(1+r^2+rwp) V^R_{ia}m^a_d V^L_{ja*}}{(1-r^2)\kappa_1\sqrt{1+p^2+(r+wp)^2}}$ 
& $i\frac{(w-r^2w-2pr-w^2pr)m_u^i \delta^{ij}+pk^2 V^R_{ia}m^a_d V^L_{ja*}}{(1-r^2)\kappa_1k\sqrt{1+p^2+(r+wp)^2}}$ \\
$\bar{u}^i_L u^j_R$ & $-i\frac{-(2r+wp)m_u^i \delta^{ij}+(1+r^2+rwp) V^L_{ia}m^a_d V^R_{ja*}}{(1-r^2)\kappa_1\sqrt{1+p^2+(r+wp)^2}}$
& $-i\frac{(w-r^2w-2pr-w^2pr)m_u^i \delta^{ij}+pk^2 V^L_{ia}m^a_d V^R_{ja*}}{(1-r^2)\kappa_1k\sqrt{1+p^2+(r+wp)^2}}$ \\
$\bar{d}^i_R d^j_L$ & $-i\frac{(1+r^2+rwp)V^R_{*ai}m^a_u V^L_{aj}- (2r+wp) m^i_d \delta^{ij}}{(1-r^2)\kappa_1\sqrt{1+p^2+(r+wp)^2}}$ 
& $-i\frac{pk^2 V^R_{*ai}m^a_u V^L_{aj}+ (w-r^2w-2pr-prw^2)m^i_d \delta^{ij}}{(1-r^2)\kappa_1k\sqrt{1+p^2+(r+wp)^2}}$ \\
$\bar{d}^i_L d^j_R$ &  $i\frac{(1+r^2+rwp)V^L_{*ai}m^a_u V^R_{aj}- (2r+wp)m^i_d \delta^{ij}}{(1-r^2)\kappa_1\sqrt{1+p^2+(r+wp)^2}}$ 
& $i\frac{pk^2 V^L_{*ai}m^a_u V^R_{aj}+ (w-r^2w-2pr-prw^2)m^i_d \delta^{ij}}{(1-r^2)\kappa_1k\sqrt{1+p^2+(r+wp)^2}}$ 
\end{tabular}
\end{center}
\caption{Couplings of the fermions to the $ \mathcal{C P}- $odd physical scalars.}
\end{table}

\begin{table}[H]
\begin{center}
\def\arraystretch{2.5}
\begin{tabular}{c|c}
 & $H_1^+$ \\
 \hline
$\bar{u}^i_R d^j_L$ & $-\sqrt{2}\times \frac{-(2r+wp) m_u^i V_{ij}^L+(1+r^2+rwp) V^R_{ij} m^j_d }{(1-r^2)\kappa_1\sqrt{1+p^2+(r+wp)^2}}$ \\
$\bar{u}^i_L d^j_R$ &  $\sqrt{2}\times\frac{(1+r^2+rwp)V^R_{ij} m^i_u - (2r+wp) m^j_d V_{ij}^L}{(1-r^2)\kappa_1\sqrt{1+p^2+(r+wp)^2}}$ \\
 & $H_2^+$ \\
 \hline
$\bar{u}^i_R d^j_L$ & $-\sqrt{2}\frac{(w-r^2 w -2pr - prw^2) m_u^i V_{ij}^L+pk^2 V^R_{ij} m^j_d}{(1-r^2)\kappa_1 k \sqrt{1+p^2+(r+wp)^2}}$ \\
$\bar{u}^i_L d^j_R$ &  $\sqrt{2}\frac{pk^2 V^R_{ij} m^i_u+ (w-r^2w-2pr-w^2pr) m^j_d V_{ij}^L}{(1-r^2)\kappa_1k\sqrt{1+p^2+(r+wp)^2}}$
\end{tabular}
\end{center}
\caption{Couplings of the fermions to the charged physical scalars.}\label{tab:chargedHiggsContributions}
\end{table}

\begin{table}[H]
\begin{center}
\def\arraystretch{2.5}
\begin{tabular}{c|c}
 & $G^+_W$\\
 \hline
$\bar{u}^i_R d^j_L$ & $\frac{-i \sqrt{2}}{\kappa_1k}
\left[m^{i}_u V^L_{ij}\left(1+\frac{2r^2}{k^2}\epsilon^2\right)-2r\epsilon^2 V^R_{ij}m^j_d\right]$ \\
$\bar{u}^i_L d^j_R$ &  $\frac{i \sqrt{2}}{\kappa_1k}
\left[V^L_{ij}m^j_d\left(1+\frac{2r^2}{k^2}\epsilon^2\right)-2r\epsilon^2 m^{i}_u V^R_{ij}\right]
$ \\
\end{tabular}
\end{center}
\caption{Couplings of the fermions to the charged Goldstone corresponding to the $ W^+ $ boson.}
\end{table}

\begin{table}[H]
\begin{center}
\def\arraystretch{2.5}
\begin{tabular}{c|c|c}
 & $G^0_Z$ & $h^0$\\
 \hline
$\bar{u}^i_L u^i_R$ & $\frac{-i m^i_u}{\kappa_1k}
\left[1+\frac{w^4-c_R^4k^2}{2k^4}\epsilon\right]
$ & $\frac{m^i_u}{\kappa_1k} +O(\epsilon^2)$ \\
$\bar{d}^i_L d^i_R$ & $\frac{i m^i_d}{\kappa_1k}\left[1+\frac{w^4-c_R^4k^2}{2k^4}\epsilon\right]$ & $\frac{m^i_d}{\kappa_1 k}+O(\epsilon^2)$
\end{tabular}
\end{center}
\caption{Couplings of the fermions to the neutral Goldstone corresponding to the $ Z^0 $ boson and the SM-like Higgs particle.}
\end{table}

\begin{table}[H]
\begin{center}
\begin{tabular}{c| c| c }
 & $H_i^0$& $ A_i^0$ \\
\hline
&&\\
$ W W'$ & $i g_R M_W F_i(r,w,t_i)$ &$ - g_R M_W F_i(r,w,t_i)$ \\
&&\\
$W^\pm G_{W'}^\mp$ & $i \frac{g_R}{2} \frac{M_W}{M_W'} F_i(r,w,t_i) (p'-p)_\mu$  & $ \frac{g_R}{2} \frac{M_W}{M_W'} F_i(r,w,t_i) (p'-p)_\mu$    \\
&&\\
$G^\pm_{W} {W'}^\mp$ & $i \frac{g_R}{2}  F_i(r,w,t_i) (p'-p)_\mu$  & $ \frac{g_R}{2}  F_i(r,w,t_i) (p'-p)_\mu $  \\
&&\\
$G_W^\pm G_{W'}^\mp$ & $- i \frac{g_R}{2} \frac{{M^2_{H{_i^0}}}}{M_W'} F_i(r,w,t_i)$  &$ \frac{g_R}{2} \frac{{M^2_{A{_i^0}}}}{M_W'} F_i(r,w,t_i)$    \\
&&\\
$Gh \; Gh'$ &$ -i \xi  \frac{g_R}{\sqrt 2} M_W F_i(r,w,t_i)$ &$ \xi \frac{g_R}{\sqrt 2} M_W F_i(r,w,t_i)$ \\
&&\\
\end{tabular}
\end{center}
\caption{Feynman rules relevant for meson mixing in the Left-Right model. $t_1 \equiv p$ and 
$t_2\equiv q$. The last line gives the couplings of the ghost particles ($Gh$, $Gh'$) to the heavy Higgses.}
\label{tab:feyn}
\end{table}

\chapter{Potential and scalar spectrum in the triplet case}\label{sec:tripletCaseStability}

In the triplet case, the most general scalar potential $ V $ symmetric under $ \mathcal{P} $ is\footnote{Ref.~\cite{Branco:1999fs} has a different expression for $ V $, with different operators, but one can show that they are the same. Indeed, $ V $ contains all possible operators, constituting a basis for any potential one could consider.}

\begin{equation}\label{eq:equation51}\begin{split}
V &= -\mu^{2}_{1} \operatorname{Tr} (\phi^{\dag} \phi) -\mu^{2}_{2} [\operatorname{Tr} (\tilde{\phi} \phi^{\dag}) + \operatorname{Tr} (\tilde{\phi}^{\dag} \phi)] -\mu^{2}_{3} [\operatorname{Tr} (\Delta_{L} \Delta^{\dag}_{L}) + \operatorname{Tr} (\Delta_{R} \Delta^{\dag}_{R})] \\
&+\lambda_{1}[\operatorname{Tr} (\phi \phi^{\dag})]^{2} +\lambda_{2} \{ [\operatorname{Tr} (\tilde{\phi} \phi^{\dag})]^{2} + [\operatorname{Tr} (\tilde{\phi}^{\dag} \phi)]^{2} \} \\
&+\lambda_{3} [\operatorname{Tr} (\tilde{\phi} \phi^{\dag}) \operatorname{Tr} (\tilde{\phi}^{\dag} \phi)] +\lambda_{4} \operatorname{Tr} (\phi \phi^{\dag}) [\operatorname{Tr} (\tilde{\phi} \phi^{\dag}) + \operatorname{Tr} (\tilde{\phi}^{\dag} \phi)] \\
&+\rho_{1} \{ [\operatorname{Tr} (\Delta_{L} \Delta^{\dag}_{L})]^{2} + [\operatorname{Tr} (\Delta_{R} \Delta^{\dag}_{R})]^{2} \} \\
&+\rho_{2} [\operatorname{Tr} (\Delta_{L} \Delta_{L}) \operatorname{Tr} (\Delta^{\dag}_{L} \Delta^{\dag}_{L}) + \operatorname{Tr} (\Delta_{R} \Delta_{R}) \operatorname{Tr} (\Delta^{\dag}_{R} \Delta^{\dag}_{R})] \\
&+\rho_{3} [\operatorname{Tr} (\Delta_{L} \Delta^{\dag}_{L}) \operatorname{Tr} (\Delta_{R} \Delta^{\dag}_{R})] \\
&+\rho_{4} [\operatorname{Tr} (\Delta_{L} \Delta_{L}) \operatorname{Tr} (\Delta^{\dag}_{R} \Delta^{\dag}_{R}) + \operatorname{Tr} (\Delta^{\dag}_{L} \Delta^{\dag}_{L}) \operatorname{Tr} (\Delta_{R} \Delta_{R})] \\
&+\alpha_{1} \operatorname{Tr} (\phi \phi^{\dag}) [\operatorname{Tr} (\Delta_{L} \Delta^{\dag}_{L}) + \operatorname{Tr} (\Delta_{R} \Delta^{\dag}_{R})] \\
&+ \alpha_{2} \{ \operatorname{e}^{i \delta_{2}} [\operatorname{Tr} (\phi \tilde{\phi}^{\dag}) \operatorname{Tr} (\Delta_{R} \Delta^{\dag}_{R}) + \operatorname{Tr} (\phi^{\dag} \tilde{\phi}) \operatorname{Tr} (\Delta_{L} \Delta^{\dag}_{L})] \\
& \quad + \operatorname{e}^{-i \delta_{2}} [\operatorname{Tr} (\phi^{\dag} \tilde{\phi}) \operatorname{Tr} (\Delta_{R} \Delta^{\dag}_{R}) + \operatorname{Tr} (\tilde{\phi}^{\dag} \phi) \operatorname{Tr} (\Delta_{L} \Delta^{\dag}_{L})] \} \\
&+\alpha_{3} [\operatorname{Tr} (\phi \phi^{\dag} \Delta_{L} \Delta^{\dag}_{L}) + \operatorname{Tr} (\phi^{\dag} \phi \Delta_{R} \Delta^{\dag}_{R})] +\beta_{1} [\operatorname{Tr} (\phi \Delta_{R} \phi^{\dag} \Delta^{\dag}_{L}) + \operatorname{Tr} (\phi^{\dag} \Delta_{L} \phi \Delta^{\dag}_{R})] \\
&+\beta_{2} [\operatorname{Tr} (\tilde{\phi} \Delta_{R} \phi^{\dag} \Delta^{\dag}_{L}) + \operatorname{Tr} (\tilde{\phi}^{\dag} \Delta_{L} \phi \Delta^{\dag}_{R})] +\beta_{3} [\operatorname{Tr} (\phi \Delta_{R} \tilde{\phi}^{\dag} \Delta^{\dag}_{L}) + \operatorname{Tr} (\phi^{\dag} \Delta_{L} \tilde{\phi} \Delta^{\dag}_{R})].
\end{split}\end{equation}
Because of Hermitian conjugation and $ \mathcal{P} $-symmetry, all the parameters above ($ \mu^{2}_{1,2,3} $, $ \lambda_{1,2,3,4} $, $ \rho_{1,2,3,4} $, $ \alpha_{1,2,3} $, and $ \beta_{1,2,3} $) are real.\footnote{For a potential symmetric under $ \mathcal{C} $, some of the parameters of the potential acquire a phase relatively to the $ \mathcal{P} $ case: thus

\begin{eqnarray}
\mu^{2}_{2} [\operatorname{tr} (\tilde{\phi} \phi^{\dag}) + \operatorname{tr} (\tilde{\phi}^{\dag} \phi)] \, ,
\end{eqnarray}
becomes

\begin{equation}
\tilde{\mu}^{2}_{2} \operatorname{tr} (\tilde{\phi} \phi^{\dag}) + (\tilde{\mu}^{2}_{2})^{*} \operatorname{tr} (\tilde{\phi}^{\dag} \phi) \, ,
\end{equation}
and similarly for $ \lambda_{2,4} $ and $ \beta_{1,2,3} $. The $ \alpha_{2} $-term has instead a different structure:

\begin{eqnarray}
\alpha_{2} e^{i \delta_{2}} [\operatorname{tr} (\phi \tilde{\phi}^{\dag}) \operatorname{tr} (\Delta_{R} \Delta^{\dag}_{R}) + \operatorname{tr} (\phi^{\dag} \tilde{\phi}) \operatorname{tr} (\Delta_{L} \Delta^{\dag}_{L})] + h.c.
\end{eqnarray}
becomes

\begin{equation}
\tilde{\alpha}_{2} e^{-i \tilde{\delta}_{2}} \operatorname{tr} (\tilde{\phi} \phi^{\dag}) [ \operatorname{tr} (\Delta_{L} \Delta^{\dag}_{L}) + \operatorname{tr} (\Delta_{R} \Delta^{\dag}_{R})] + h.c.
\end{equation}
The final budget is that a general potential symmetric under $ \mathcal{C} $ has 6 extra phases compared to the $ \mathcal{P} $-symmetric one because $ \tilde{\mu}^{2}_{2}, \tilde{\lambda}_{2,4}, \tilde{\beta}_{1,2,3} $ are generic complex numbers. Similar comments would hold for the doublet case. 

It is not difficult to verify that when one asks for $ \mathcal{P} $ \textit{and} $ \mathcal{C} $ invariance, all the complex phases vanish. Instead, in the general case where neither $ \mathcal{P} $ or $ \mathcal{C} $ is considered, many more phases are present in the potential.}


In the triplet case, the neutral fields expanded around their VEVs are

\begin{equation}\begin{split}
\varphi^{0}_{1} &= (\varphi^{0r}_{1} + i \varphi^{0i}_{1} + \kappa_{1}) / \sqrt{2}, \\
\varphi^{0}_{2} &= (\varphi^{0r}_{2} + i \varphi^{0i}_{2} + \kappa_{2}) \operatorname{e}^{i \alpha} / \sqrt{2}, \\
\delta^{0}_{R} &= (\delta^{0r}_{R} + i \delta^{0i}_{R} + \kappa_R) / \sqrt{2}, \\
\delta^{0}_{L} &= (\delta^{0r}_{L} + i \delta^{0i}_{L} + \kappa_L) \operatorname{e}^{i \theta_{L}} / \sqrt{2},
\end{split}\end{equation}
\noindent
together with $ \varphi^{+}_{1,2} $, $ \delta^{+}_{L,R} $, and $ \delta^{++}_{L,R} $. The extreme conditions with respect to $ \{ \varphi^{0r}_{1}, \varphi^{0r}_{2}, \varphi^{0i}_{2}, \delta^{0r}_{R}, \delta^{0r}_{L}, \delta^{0i}_{L} \} $ are ($ \kappa_{2} \neq \kappa_{1} $)

\begin{equation}\label{eq:equation4triplet}\begin{split}
\frac{\mu^{2}_{1}}{\kappa^{2}_{R}} &= \frac{\alpha_{1}}{2} \left( 1 + \frac{\kappa^{2}_{L}}{\kappa^{2}_{R}} \right) - {\frac{\alpha_{3}}{2} \frac{\kappa^{2}_{2}}{\kappa^{2}_{-}} \left( 1 + \frac{\kappa^{2}_{L}}{\kappa^{2}_{R}} \right)} + \left( \lambda_{1} + {2 \lambda_{4} \frac{\kappa_{1} \kappa_{2}}{\kappa^{2}_{+}} \cos \alpha} \right) \frac{\kappa^{2}_{+}}{\kappa^{2}_{R}} \\
&+ \left[ \beta_{2} \frac{\kappa^{2}_{1}}{\kappa^{2}_{-}} \cos \theta_{L} - \beta_{3} \frac{\kappa^{2}_{2}}{\kappa^{2}_{-}} \cos (\theta_{L} - 2 \alpha) \right] \frac{\kappa_L}{\kappa_R},
\end{split}\end{equation}

\begin{equation}\label{eq:equation5triplet}\begin{split}
\frac{\mu^{2}_{2}}{\kappa^{2}_{R}} &= \frac{\alpha_{2}}{2 \cos \alpha} \left[ \cos (\alpha + \delta_{2}) + \cos (\alpha - \delta_{2}) \frac{\kappa^{2}_{L}}{\kappa^{2}_{R}} \right] + \frac{\alpha_{3}}{4 \cos \alpha} \frac{\kappa_{1} \kappa_{2}}{\kappa^{2}_{-}} \left( 1 + \frac{\kappa^{2}_{L}}{\kappa^{2}_{R}} \right) \\
&+ \left\{ [2 \lambda_{2} \cos (2 \alpha) + \lambda_{3} ] \frac{\kappa_{1} \kappa_{2}}{\kappa^{2}_{+}} \frac{1}{\cos \alpha} + \frac{1}{2} \lambda_{4} \right\} \frac{\kappa^{2}_{+}}{\kappa^{2}_{R}} \\
&+ \left\{ \beta_{1} \cos (\theta_{L} - \alpha) - 2 [ \beta_{2} \cos \theta_{L} - \beta_{3} \cos (\theta_{L} - 2 \alpha)] \frac{\kappa_{1} \kappa_{2}}{\kappa^{2}_{-}} \right\} \frac{\kappa_L/\kappa_R}{4 \cos \alpha},
\end{split}\end{equation}

\begin{equation}\label{eq:equation6triplet}\begin{split}
\frac{\mu^{2}_{3}}{\kappa^{2}_{R}} &= \rho_{1} \left( 1 + \frac{\kappa^{2}_{L}}{\kappa^{2}_{R}} \right) + \frac{1}{2} \left( \alpha_{1} + \alpha_{3} \frac{\kappa^{2}_{2}}{\kappa^{2}_{+}} \right) \frac{\kappa^{2}_{+}}{\kappa^{2}_{R}} \\
&+ 2 \alpha_{2} \left[ \cos (\alpha + \delta_{2}) - \cos (\alpha - \delta_{2}) \frac{\kappa^{2}_{L}}{\kappa^{2}_{R}} \right] \frac{\kappa_{1} \kappa_{2} / \kappa^{2}_{R}}{1 - \kappa^{2}_{L}/\kappa^{2}_{R}},
\end{split}\end{equation}

\begin{equation}\label{eq:equation7triplet}\begin{split}
&\left[ (2 \rho_{1} - \rho_{3}) - \frac{8 \alpha_{2} \sin \alpha \sin \delta_{2}}{1 - \kappa^{2}_{L}/\kappa^{2}_{R}} \frac{\kappa_{1} \kappa_{2}}{\kappa^{2}_{R}} \right] \frac{\kappa_L}{\kappa_{+}} = \\
&\left[ {\beta_{1} \frac{\kappa_{1} \kappa_{2}}{\kappa^{2}_{+}} \cos (\theta_{L} - \alpha)} + \beta_{2} \frac{\kappa^{2}_{1}}{\kappa^{2}_{+}} \cos \theta_{L} + \beta_{3} \frac{\kappa^{2}_{2}}{\kappa^{2}_{+}} \cos (\theta_{L} - 2 \alpha) \right] \frac{\kappa_{+}}{\kappa_R},
\end{split}\end{equation}

\begin{eqnarray} \label{eq:equation8triplet}
0 = \beta_{1} \frac{\kappa_{1} \kappa_{2}}{\kappa^{2}_{+}} \sin (\theta_{L} - \alpha) + \beta_{2} \frac{\kappa^{2}_{1}}{\kappa^{2}_{+}} \sin \theta_{L} + \beta_{3} \frac{\kappa^{2}_{2}}{\kappa^{2}_{+}} \sin (\theta_{L} - 2 \alpha),
\end{eqnarray}

\begin{equation}\label{eq:equation9triplet}\begin{split}
&2 \alpha_{2} \left( 1 - \dfrac{\kappa^{2}_{L}}{\kappa^{2}_{R}} \right) \sin \delta_{2} = \frac{\kappa_{1} \kappa_{2}}{\kappa^{2}_{-}} \sin \alpha \left[ \alpha_{3} \left( 1 + \dfrac{\kappa^{2}_{L}}{\kappa^{2}_{R}} \right) - 4 (2 \lambda_{2} - \lambda_{3}) \kappa^{2}_{-}/\kappa^{2}_{R} \right] \\
&+ \left\{ {2 \frac{\kappa_{1} \kappa_{2}}{\kappa^{2}_{-}} \sin (\theta_{L} - \alpha) (\beta_{2} + \beta_{3})} + \left[ \frac{\kappa^{2}_{1}}{\kappa^{2}_{-}} \sin \theta_{L} + \frac{\kappa^{2}_{2}}{\kappa^{2}_{-}} \sin (\theta_{L} - 2 \alpha) \right] \beta_{1} \right\} \dfrac{\kappa_L}{\kappa_R}.
\end{split}\end{equation}
\noindent
We now define $ \varepsilon \equiv \kappa_{1}/\kappa_R $, which is a natural expansion parameter, and the extreme conditions shown above become much simpler when expanded in $ \varepsilon $. Considering that the parameters of the potential are of order $ \mathcal{O} (\varepsilon^{0}) $, except for $ \mu^2_{1,2,3} $, and solving for $ \mu^2_{1,2,3} $, we have

\begin{eqnarray} \label{eq:equation11triplet}
\frac{\mu^{2}_{1}}{\kappa^{2}_{R}} \simeq \frac{\alpha_{1}}{2} - \frac{\alpha_{3}}{2} \frac{\kappa^{2}_{2}}{\kappa^{2}_{-}} ,
\end{eqnarray}

\begin{eqnarray} \label{eq:equation12triplet}
\frac{\mu^{2}_{2}}{\kappa^{2}_{R}} \simeq \frac{\alpha_{2}}{2} \frac{\cos (\alpha + \delta_{2})}{\cos \alpha} + \frac{\alpha_{3}}{4 \cos \alpha} \frac{\kappa_{1} \kappa_{2}}{\kappa^{2}_{-}} ,
\end{eqnarray}

\begin{eqnarray} \label{eq:equation13triplet}
\frac{\mu^{2}_{3}}{\kappa^{2}_{R}} \simeq \rho_{1} ,
\end{eqnarray}

\begin{eqnarray} \label{eq:equation14triplet}
&& (2 \rho_{1} - \rho_{3}) \frac{\kappa_L}{\kappa_{+}} \simeq \left[ \beta_{1} \frac{\kappa_{1} \kappa_{2}}{\kappa^{2}_{+}} \cos (\theta_{L} - \alpha) + \beta_{2} \frac{\kappa^{2}_{1}}{\kappa^{2}_{+}} \cos \theta_{L} \right. \\
&& \left. + {\beta_{3} \frac{\kappa^{2}_{2}}{\kappa^{2}_{+}} \cos (\theta_{L} - 2 \alpha)} \right] \frac{\kappa_{+}}{\kappa_R} , \nonumber
\end{eqnarray}

\begin{eqnarray} \label{eq:equation16triplet}
2 \alpha_{2} \sin \delta_{2} \simeq \frac{\kappa_{1} \kappa_{2}}{\kappa^{2}_{-}} \alpha_{3} \sin \alpha ,
\end{eqnarray}
\noindent
plus Eq.~(\ref{eq:equation8triplet}), which does not change.

Equations (\ref{eq:equation11triplet})-(\ref{eq:equation13triplet}) are very similar to those found in the doublet case, and constitute equations for $ \mu^{2}_{1,2,3} $. Equation (\ref{eq:equation14triplet}) in the triplet case corresponds to a see-saw mechanism: $ \frac{\kappa_L}{\kappa_{+}} $ in one side, $ \frac{\kappa_{+}}{\kappa_R} $ in the other, see Section~\ref{sec:TripletModelDiscussion}.

We now shift to the physical spectrum of LRM in the triplet case. We have two charged and two neutral Goldstone bosons (giving masses to the vector gauge bosons), one light Higgs (the SM Higgs), and many other massive scalars: five neutral, two singly-charged, and two doubly-charged Higgses, the former being given in Table~\ref{table:tbl1}.

\begin{table}[H]
\begin{center}
\def\arraystretch{2.0}
\begin{tabular}{|c|c|c|}
  \hline
  State & $ \operatorname{Mass}^{2} $ \\
  \hline
  \hline
  $ \varphi^{0r}_{1} + r \varphi^{0r}_{2} - \varepsilon \frac{\alpha_{1}}{2 \rho_{1}} \delta^{0r}_{R} $ & $ \left( 4 \lambda_{1} - \frac{\alpha^{2}_{1}}{\rho_{1}} \right) \kappa^{2}_{1} / 2 $ \\
  $ \varphi^{0r}_{2} - r \varphi^{0r}_{1} + \varepsilon \frac{4 \alpha_{2}}{\alpha_{3} - 4 \rho_{1}} \delta^{0r}_{R} $ & $ \alpha_{3} \kappa^{2}_{R} / 2 + \mathcal{O} (\varepsilon^{2},r^{2},\varepsilon r) \kappa^{2}_{R} $ \\
  $ \delta^{0r}_{R} + \varepsilon \frac{\alpha_{1}}{2 \rho_{1}} \varphi^{0r}_{1} - \varepsilon \frac{4 \alpha_{2}}{\alpha_{3} - 4 \rho_{1}} \varphi^{0r}_{2} $ & $ 2 \rho_{1} \kappa^{2}_{R} + \mathcal{O} (\varepsilon^{2},r^{2},\varepsilon r) \kappa^{2}_{R} $ \\
  $ \delta^{0r}_{L} $ & $ \rho \kappa^{2}_{R} $ \\
  $ \varphi^{0i}_{2} + r \varphi^{0i}_{1} $ & $ \alpha_{3} \kappa^{2}_{R} / 2 + \mathcal{O} (\varepsilon^{2},r^{2},\varepsilon r) \kappa^{2}_{R} $ \\
  $ \delta^{0i}_{L} $ & $ \rho \kappa^{2}_{R} $ \\
  \hline
  $ - \varphi^{0i}_{1} + r \varphi^{0i}_{2} $ & 0 \\
  $ \delta^{0i}_{R} $ & 0 \\
  \hline
\end{tabular}
\end{center}\caption{Neutral spectrum in the triplet case. $ r \equiv \frac{\kappa_{2}}{\kappa_{1}} $ is taken small for simplicity. We take $ \sin \alpha = \sin \theta_{L} = 0 $, $ \cos \alpha = \cos \theta_{L} = 1 $; one also asks for $ \beta_{1,2,3} \ll \mathcal{O}(1) $, a usual requirement when implementing a see-saw mechanism for neutrinos. Moreover, the $ \rho- $parameter implies $ \kappa_L \sim 0 $.}\label{table:tbl1}
\end{table}

\chapter{RGE formulae}\label{sec:toughRGE}

We now exploit the fact that the Lagrangian is independent on the renormalization scale. The equation $ \mu \frac{d}{d \mu} \mathcal{L}^{(5)}_{\rm eff} \vert_{\vert \Delta F \vert = 1} = 0 $ (i.e. the net effect of comparing the renormalized Lagrangian at two different scales is zero up to a certain order) then implies

\begin{equation}
\sum_{i} \mu \frac{d}{d \mu} C_i (\mu) Z^{-1}_{ij} (\mu) = 0 \Rightarrow \sum_{i,j} \left( \mu \frac{d}{d \mu} C_i (\mu) Z^{-1}_{ij} (\mu) \right) Z_{jk} (\mu) = 0 ,
\end{equation}
\noindent
and it follows that

\begin{equation}\label{eq:RGEDF1operator}
\sum_{j} \left[ \delta_{jk} \, \mu \frac{d}{d \mu} - \gamma_{jk} \right] C_j = 0 , \quad \gamma_{jk} \equiv \sum_{i} Z^{-1}_{ji} \mu \frac{d}{d \mu} Z_{ik} .
\end{equation}

The matrix $ \gamma $ is called the anomalous dimension of the $ \vert \Delta F \vert = 1 $ local operator $ Q_i^L $, or $ Q_i^R $ defined as

\begin{equation}
Q^{X}_{i} = \bar{d} \gamma^{\lambda} P_{X} q' \cdot \bar{q} \gamma_{\lambda} P_{X} s \, \frac{(\mathbf{\hat{1}} + i \mathbf{\tilde{\hat{1}}})}{2} \, , \; X = L, R \, , \; i = \pm \, ,
\end{equation}
where $ \mathbf{\hat{1}} $ denotes a color singlet and $ \mathbf{\tilde{\hat{1}}} $ a color anti-singlet.


We further have, from $ \mu \frac{d}{d \mu} \mathcal{L}^{(5)}_{\rm eff} \vert_{\vert \Delta F \vert = 2} = 0 $,

\begin{eqnarray}
&& \mu \frac{d}{d \mu} \left[ \sum_{k, l = \pm} C_k C_l \tilde{Z}^{-1}_{kl, b} + \sum^2_{a=1} C^r_a \tilde{Z}^{-1}_{ab} \right] = 0 \nonumber\\
& \Rightarrow & \sum^2_{b=1} \mu \frac{d}{d \mu} \left[ \sum_{k, l = \pm} C_k C_l \tilde{Z}^{-1}_{kl, b} + \sum^2_{a=1} C^r_a \tilde{Z}^{-1}_{ab} \right] \tilde{Z}_{bc} = 0 ,
\end{eqnarray}
\noindent
that implies, from the Eq.~\eqref{eq:RGEDF1operator}

\begin{equation}
\sum^2_{a=1} \left[ \delta_{ac} \, \mu \frac{d}{d \mu} - \sum^2_{b=1} \tilde{Z}^{-1}_{ab} \mu \frac{d}{d \mu} \tilde{Z}_{bc} \right] C^r_{a} = \sum_{k,l=\pm} C_k C_l \gamma_{kl,c} ,
\end{equation}
\noindent
where

\begin{equation}\label{eq:definitionGammaMixing}
\gamma_{kl,c} \equiv - \sum^2_{b=1} \left[ \sum_{k',l'=\pm} \left( \gamma_{kk'} \delta_{ll'} + \gamma_{ll'} \delta_{kk'} \right) \tilde{Z}^{-1}_{k'l',b} + \mu \frac{d}{d \mu} \tilde{Z}^{-1}_{kl,b} \right] \tilde{Z}_{bc} .
\end{equation}
Note from Eq.~\eqref{eq:definitionInverseZ} that the renormalization matrices include by definition powers in $ 1/\epsilon $. Therefore, when expanding the previous equation defining $ \gamma_{kl,c} $ we include $ \epsilon $ corrections in the $ \mu $ derivative as follows

\begin{equation}
\mu \frac{d}{d \mu} = - 2 a (\mu) [ \epsilon + \beta_0 a (\mu) ] \frac{d}{d a} \, , \qquad a (\mu) \equiv \frac{\alpha_s (\mu)}{4 \pi} \, .
\end{equation}
Following this precision, we have the following expressions for $ \gamma_{kl,c} $, \cite{Buras:1998raa} \cite{Herrlich:1996vf}

\begin{equation}
\gamma^{(0)}_{mn, a} = 2 \left[ \tilde{Z}^{-1,(1)}_{1} \right]_{mn, a} \, ,
\end{equation}
at the LO, and

\begin{eqnarray}
&& \gamma^{(1)}_{mn, a} = 4 \left( \left[ \tilde{Z}^{-1,(2)}_{1} \right]_{mn, a} + \beta_0 \left[ \frac{\tilde{Z}^{-1,(1)}_{0}}{2} \right]_{mn, a} \right. \nonumber\\
&& - \sum^{2}_{b=1} \left\{ \left[ \frac{\tilde{Z}^{-1,(1)}_{0}}{2} \right]_{mn, b} \left[ \tilde{Z}^{-1,(1)}_{1} \right]_{ba} + \left[ \tilde{Z}^{-1,(1)}_{1} \right]_{mn, b} \left[ \frac{\tilde{Z}^{-1,(1)}_{0}}{2} \right]_{ba} \right\} \nonumber\\
&& - \sum_{m', n' = \pm} \left\{ \left( \left[ \frac{Z^{-1,(1)}_{0} }{2} \right]_{m m'} \delta_{n n'} + \delta_{m m'} \left[ \frac{Z^{-1,(1)}_{0}}{2} \right]_{n n'} \right) \left[ \tilde{Z}^{-1,(1)}_{1} \right]_{m' n', a} \right. \nonumber\\
&& + \left. \left. \left( \left[ Z^{-1,(1)}_{1} \right]_{m m'} \delta_{n n'} + \delta_{m m'} \left[ Z^{-1,(1)}_{1} \right]_{n n'} \right) \left[ \frac{\tilde{Z}^{-1,(1)}_{0}}{2} \right]_{m' n', a} \right\} \right) \, ,
\end{eqnarray} 
at the NLO, where

\begin{eqnarray}
Z & = & 1 - \frac{\alpha_s}{4 \pi} Z^{-1,(1)} - \left( \frac{\alpha_s}{4 \pi} \right)^2 (Z^{-1,(2)} - Z^{-1,(1)} \times Z^{-1,(1)}) + \ldots \, , \nonumber\\
&&
\end{eqnarray}
has been employed. Of course, the divergent terms that one may face in the definition in Eq.~\eqref{eq:definitionGammaMixing} vanish by suitable relations satisfied by the renormalization matrices.

Finally, we have

\begin{equation}
\mu \frac{d}{d \mu} \begin{pmatrix}
C^r_1 \\
C^r_2 \\
\end{pmatrix} = \tilde{\gamma}^{\rm T} \begin{pmatrix}
C^r_1 \\
C^r_2 \\
\end{pmatrix} + \sum_{k,l=\pm} \begin{pmatrix}
C_k C_l \, \gamma_{kl,1} \\
C_k C_l \, \gamma_{kl,2} \\
\end{pmatrix} ,
\end{equation}
\noindent
where $ \tilde{\gamma} \equiv \tilde{Z}^{-1} \mu \frac{d}{d \mu} \tilde{Z} $, and

\begin{equation}
\tilde{\gamma}_{LR} = \tilde{\gamma} - 2 ( \gamma_m - \beta ) \cdot \mathbf{1}_{2}
\end{equation}
\noindent
is the matrix necessary to evolve the system $ \{ \gamma L \otimes \gamma R, L \otimes R \} $, while in here we are concerned with the system $ \{ \tilde{Q}^{LR}_{1}, \tilde{Q}^{LR}_{2} \} $ with a different normalization.



An equivalent way to write the last equation is


\begin{eqnarray}\label{eq:headacheGamma}
&& \mu \frac{d}{d \mu} \begin{pmatrix}
C^2_{+} \\
C_{+} C_{-} \\
C^2_{-} \\
C^r_1 \\
C^r_2 \\
\end{pmatrix} = \begin{pmatrix}
2 \cdot \gamma_{+} && 0 && 0 && 0 \quad 0 \\
0 && \gamma_{+} + \gamma_{-} && 0 && 0 \quad 0 \\
0 && 0 && 2 \cdot \gamma_{-} && 0 \quad 0 \\
\gamma_{+ +,1} && 2 \cdot \gamma_{+ -,1} && \gamma_{- -,1} && \gamma^{\rm T} \\
\gamma_{+ +,2} && 2 \cdot \gamma_{+ -,2} && \gamma_{- -,2} && \\
\end{pmatrix} \begin{pmatrix}
C^2_{+} \\
C_{+} C_{-} \\
C^2_{-} \\
C^r_1 \\
C^r_2 \\
\end{pmatrix} \nonumber
\end{eqnarray}
or

\begin{equation}
\mu \frac{d}{d \mu} \overrightarrow{D} = \Gamma^{\rm T} \overrightarrow{D} \, .
\end{equation}

The solution of the above differential equations is standard and can be found in \cite{Buras:1998raa} \cite{Adams:2007tk}: first we define $ V $ diagonalizing the $ N \times N $ matrix $ \Gamma^{(0) {\rm T}} $

\begin{equation}
\Gamma^{(0)}_D = V^{-1} \Gamma^{(0) {\rm T}} V = {\rm diag} \left[ \Gamma^{(0)}_1, \Gamma^{(0)}_2, \ldots, \Gamma^{(0)}_N \right] ,
\end{equation} 
and then we define matrices $ G, H, J $ in the following way

\begin{eqnarray}
G &=& V^{-1} \Gamma^{(1) {\rm T}} V , \\
H_{i j} &=& \delta_{ij} \Gamma^{(0)}_i \frac{\beta_1}{2 \beta^2_0} - \frac{G_{i j}}{2 \beta_0 + \Gamma^{(0)}_i - \Gamma^{(0)}_j} , \quad (2 \beta_0 + \Gamma^{(0)}_i - \Gamma^{(0)}_j \neq 0) \nonumber\\
&&\\
J &=& V H V^{-1} .
\end{eqnarray}
Finally, one has

\begin{eqnarray}
\overrightarrow{D} (\mu_1) &=& \left(\mathbf{1}_{N} + \frac{\alpha_s (\mu_1)}{4 \pi} J \right) V U(\mu_1 , \mu_2) V^{-1} \nonumber\\
&& \times \left(\mathbf{1}_{N} - \frac{\alpha_s (\mu_2)}{4 \pi} J \right) \overrightarrow{D} (\mu_2) , \\
U(\mu_1 , \mu_2) &=& {\rm diag} \left[ \left( \frac{\alpha_s (\mu_2)}{\alpha_s (\mu_1)} \right)^{d_1} , \ldots , \left( \frac{\alpha_s (\mu_2)}{\alpha_s (\mu_1)} \right)^{d_N} \right] ,
\end{eqnarray}
where $ d_i = \Gamma^{(0)}_i / (2 \beta_0) $.

\chapter{Set of evanescent operators}\label{sec:setOfEOs}


\section{Operators $ |\Delta F| = 2 $}

We define

\begin{eqnarray}
Q^{VLL} & = & \bar{d}^{\alpha} \gamma^{\mu} P_{L} s^{\alpha} \cdot \bar{d}^{\beta} \gamma_{\mu} P_{L} s^{\beta} , \\
Q^{LR}_{1} & = & \bar{d}^{\alpha} \gamma^{\mu} P_{R} s^{\alpha} \cdot \bar{d}^{\beta} \gamma_{\mu} P_{L} s^{\beta} , \\
Q^{LR}_{2} & = & \bar{d}^{\alpha} P_{R} s^{\alpha} \cdot \bar{d}^{\beta} P_{L} s^{\beta} , \\
Q^{SLL}_{1} & = & \bar{d}^{\alpha} P_{L} s^{\alpha} \cdot \bar{d}^{\beta} P_{L} s^{\beta} ,
\end{eqnarray}
\noindent
and similarly for $ Q^{VRR} $ and $ Q^{SRR}_{1} $. The evanescent operators are ($ f_{( 1 \rightarrow 2 )} = \frac{1}{f_{( 2 \rightarrow 1 )}} = - 2 $)

\begin{eqnarray}
E^{VLL}_{1} & = & \bar{d}^{\alpha} \gamma P_{L} s^{\beta} \cdot \bar{d}^{\beta} \gamma P_{L} s^{\alpha} - Q^{VLL}_{} , \\
E^{VLL}_{2} & = & \bar{d}^{\alpha} \gamma \gamma \gamma P_{L} s^{\alpha} \cdot \bar{d}^{\beta} \gamma \gamma \gamma P_{L} s^{\beta} - (16 - a^{VLL}_{3 \gamma} \epsilon) Q^{VLL}_{} , \\
E^{VLL}_{3} & = & \bar{d}^{\alpha} \gamma \gamma \gamma P_{L} s^{\beta} \cdot \bar{d}^{\beta} \gamma \gamma \gamma P_{L} s^{\alpha} - (16 - a^{VLL}_{3 \gamma} \epsilon) Q^{VLL}_{} , \\
E^{VLL}_{4} & = & \bar{d}^{\alpha} \gamma \gamma \gamma \gamma \gamma P_{L} s^{\alpha} \cdot \bar{d}^{\beta} \gamma \gamma \gamma \gamma \gamma P_{L} s^{\beta} - \left( (16 - a^{VLL}_{3 \gamma} \epsilon)^2 + \tilde{b}_{3V} \epsilon \right) Q^{VLL}_{} , \nonumber\\
\\
E^{VLL}_{5} & = & \bar{d}^{\alpha} \gamma \gamma \gamma \gamma \gamma P_{L} s^{\beta} \cdot \bar{d}^{\beta} \gamma \gamma \gamma \gamma \gamma P_{L} s^{\alpha} - \left( (16 - a^{VLL}_{3 \gamma} \epsilon)^2 + \tilde{b}_{3V} \epsilon \right) Q^{VLL}_{} , \nonumber\\
\end{eqnarray}

\begin{eqnarray}
E^{LR}_{1} & = & \bar{d}^{\alpha} P_{R} s^{\beta} \cdot \bar{d}^{\beta} P_{L} s^{\alpha} - f_{(2 \rightarrow 1)} Q^{LR}_{1} , \\
E^{LR}_{2} & = & \bar{d}^{\alpha} \gamma P_{R} s^{\beta} \cdot \bar{d}^{\beta} \gamma P_{L} s^{\alpha} - f_{(1 \rightarrow 2)} Q^{LR}_{2} , \\
E^{LR}_{3} & = & \bar{d}^{\alpha} \gamma \gamma \gamma P_{R} s^{\alpha} \cdot \bar{d}^{\beta} \gamma \gamma \gamma P_{L} s^{\beta} - (4 + a^{LR}_{3 \gamma} \epsilon) Q^{LR}_{1} , \\
E^{LR}_{4} & = & \bar{d}^{\alpha} \gamma \gamma \gamma P_{R} s^{\beta} \cdot \bar{d}^{\beta} \gamma \gamma \gamma P_{L} s^{\alpha} - f_{(1 \rightarrow 2)} (4 + a^{LR}_{3 \gamma} \epsilon) Q^{LR}_{2} , \\
E^{LR}_{5} & = & \bar{d}^{\alpha} \gamma \gamma P_{R} s^{\alpha} \cdot \bar{d}^{\beta} \gamma \gamma P_{L} s^{\beta} - (4 + a^{LR}_{2 \gamma} \epsilon) Q^{LR}_{2} , \\
E^{LR}_{6} & = & \bar{d}^{\alpha} \gamma \gamma P_{R} s^{\beta} \cdot \bar{d}^{\beta} \gamma \gamma P_{L} s^{\alpha} - f_{(2 \rightarrow 1)} (4 + a^{LR}_{2 \gamma} \epsilon) Q^{LR}_{1} , \\
E^{LR}_{7} & = & \bar{d}^{\alpha} \gamma \gamma \gamma \gamma P_{R} s^{\alpha} \cdot \bar{d}^{\beta} \gamma \gamma \gamma \gamma P_{L} s^{\beta} - \left( (4 + a^{LR}_{2 \gamma} \epsilon)^2 + \tilde{b}_{2LR} \epsilon \right) Q^{LR}_{2} , \\
E^{LR}_{8} & = & \bar{d}^{\alpha} \gamma \gamma \gamma \gamma P_{R} s^{\beta} \cdot \bar{d}^{\beta} \gamma \gamma \gamma \gamma P_{L} s^{\alpha} - f_{(2 \rightarrow 1)} \left( (4 + a^{LR}_{2 \gamma} \epsilon)^2 + \tilde{b}_{2LR} \epsilon \right) Q^{LR}_{1} , \nonumber\\
\\
E^{LR}_{9} & = & \bar{d}^{\alpha} \gamma \gamma \gamma \gamma \gamma P_{R} s^{\alpha} \cdot \bar{d}^{\beta} \gamma \gamma \gamma \gamma \gamma P_{L} s^{\beta} - \left( (4 + a^{LR}_{3 \gamma} \epsilon)^2 + \tilde{b}_{3LR} \epsilon \right) Q^{LR}_{1} , \nonumber\\
&& \\
E^{LR}_{10} & = & \bar{d}^{\alpha} \gamma \gamma \gamma \gamma \gamma P_{R} s^{\beta} \cdot \bar{d}^{\beta} \gamma \gamma \gamma \gamma \gamma P_{L} s^{\alpha} - f_{(1 \rightarrow 2)} \left( (4 + a^{LR}_{3 \gamma} \epsilon)^2 + \tilde{b}_{3LR} \epsilon \right) Q^{LR}_{2} , \nonumber\\
&&
\end{eqnarray}
\noindent
and similarly for right-right operators, where $ \bar{d} \gamma \gamma L s \cdot \bar{d} \gamma \gamma R s $ means $ \bar{d} \gamma^{\mu_1} \gamma^{\mu_2} L s \cdot \bar{d} \gamma_{\mu_1} \gamma_{\mu_2} R s $, etc. We take

\begin{eqnarray}\label{eq:choicesEVOPS}
&& a^{SLL}_{2 \gamma} = a^{SRR}_{2 \gamma} = 12 \, , \\
&& a^{VLL}_{3 \gamma} = a^{VRR}_{3 \gamma} = 4 \, , \nonumber\\
&& a^{SLL}_{4 \gamma} = a^{SRR}_{4 \gamma} = 224 \, , \nonumber\\
&& \tilde{b}_{3V} = - 96 \, , \nonumber\\
&& a^{LR}_{2 \gamma} = 4 \, , \quad a^{LR}_{3 \gamma} = 4 \, , \nonumber\\
&& \tilde{b}_{2LR} = 96 \, , \quad \tilde{b}_{3LR} = 96 \, , \nonumber
\end{eqnarray}
except when otherwise stated. 

For $ E^{LR}_{5,6} $, compared to \cite{Buras:2000if}, our definitions employ the equations 

\begin{eqnarray}
\{ \gamma_{\mu} , \gamma_{\nu} \} & = & 2 g_{\mu \nu} \cdot \mathbf{1} , \label{eq:gamma1}\\
\gamma_{\mu} \gamma^{\mu} & = & {g_{\mu}}^{\mu} \cdot \mathbf{1} = D \cdot \mathbf{1} = (4 - 2 \epsilon) \cdot \mathbf{1} . \label{eq:gamma2}
\end{eqnarray}
found in NDR.

It is interesting to note that, for example,

\begin{eqnarray}
\tilde{E}^{LR}_{4} & = & \bar{d}^{\alpha} \gamma \gamma \gamma P_{R} s^{\beta} \cdot \bar{d}^{\beta} \gamma \gamma \gamma P_{L} s^{\alpha} - (4 + 4 \epsilon) \bar{d}^{\alpha} \gamma P_{R} s^{\beta} \cdot \bar{d}^{\beta} \gamma P_{L} s^{\alpha} \nonumber\\
& = & \bar{d}^{\alpha} \gamma \gamma \gamma P_{R} s^{\beta} \cdot \bar{d}^{\beta} \gamma \gamma \gamma P_{L} s^{\alpha} - (4 + 4 \epsilon) ( E^{LR}_{2} + f_{(1 \rightarrow 2)} Q^{LR}_{2} ) \nonumber\\
& = & E^{LR}_{4} - (4 + 4 \epsilon) E^{LR}_{2} = E^{LR}_{4} - 4 E^{LR}_{2} ,
\end{eqnarray}
\noindent
and therefore it is the combination of two EO already defined, illustrating that the set of evanescent operators given above is a sufficient set.

\section{Operators $ |\Delta F| = 1 $}


The relevant operators are

\begin{eqnarray}
Q^{VLL}_{1} = \tilde{Q}_{V_{L} V_{L}} & = & \bar{d}^{\alpha} \gamma^{\mu} P_{L} q^{\beta} \cdot \bar{q'}^{\beta} \gamma_{\mu} P_{L} s^{\alpha} , \\
Q^{VLL}_{2} = Q_{V_{L} V_{L}} & = & \bar{d}^{\alpha} \gamma^{\mu} P_{L} q^{\alpha} \cdot \bar{q'}^{\beta} \gamma_{\mu} P_{L} s^{\beta} , \\
Q^{SLR}_{1} = \tilde{Q}_{LR} & = & \bar{d}^{\alpha} P_{L} q^{\beta} \cdot \bar{q'}^{\beta} P_{R} s^{\alpha} , \\
Q^{SLR}_{2} = Q_{LR} & = & \bar{d}^{\alpha} P_{L} q^{\alpha} \cdot \bar{q'}^{\beta} P_{R} s^{\beta} ,
\end{eqnarray}
\noindent
and the following definitions of evanescent operators are made

\begin{eqnarray}
E^{SLR}_{2} & = & \bar{d}^{\alpha} \gamma \gamma P_{R} q^{\alpha} \cdot \bar{q'}^{\beta} \gamma \gamma P_{L} s^{\beta} - (4 + 4 \epsilon) Q^{SLR}_{2} , \\
E^{SLR}_{1} & = & \bar{d}^{\alpha} \gamma \gamma P_{R} q^{\beta} \cdot \bar{q'}^{\beta} \gamma \gamma P_{L} s^{\alpha} - (4 + 4 \epsilon) Q^{SLR}_{1} , \\
E^{VLL}_{5} & = & \bar{d}^{\alpha} \gamma \gamma \gamma P_{L} q^{\alpha} \cdot \bar{q'}^{\beta} \gamma \gamma \gamma P_{L} s^{\beta} - (16 - 4 \epsilon) Q^{VLL}_{1} , \\
E^{VLL}_{6} & = & \bar{d}^{\alpha} \gamma \gamma \gamma P_{L} q^{\beta} \cdot \bar{q'}^{\beta} \gamma \gamma \gamma P_{L} s^{\alpha} - (16 - 4 \epsilon) Q^{VLL}_{2} .
\end{eqnarray}

For $ E^{SLR}_{1,2} $, compared to \cite{Buras:2000if} our definitions employ the equations (\ref{eq:gamma1}) and (\ref{eq:gamma2}) found in NDR.

\chapter{Operators and anomalous dimensions}
\label{app:anomdimgen}


\section{$|\Delta S|=1$ operators}\label{app:anomdimgenS1}

We have the $|\Delta S|=1$ vector operators for the SM case~\cite{Herrlich:1996vf,Buras:2000if}
\begin{eqnarray}
O_1^{VLL}&=&(\bar{d}^\alpha \gamma_\mu P_L s^\beta)(\bar{V}^\beta\gamma^\mu P_L U^\alpha)\,,
\qquad O_2^{VLL}=(\bar{d}\gamma_\mu P_L s)(\bar{V}\gamma^\mu P_L U)\,,\\
O_1^{VLR}&=&(\bar{d}\gamma_\mu P_L s)(\bar{V}\gamma^\mu P_R U)\,,
\qquad\qquad O_2^{VLR}=(\bar{d}^\alpha\gamma_\mu P_L s^\beta)(\bar{V}^\beta\gamma^\mu P_R U^\alpha)\,,
\end{eqnarray}
where $U$ and $V$ can be any up-type fermions.
As discussed in Ref.~\cite{Buras:2000if}, Fierz identities hold for these operators up two loops in 
the NDR-$\overline{MS}$ scheme.
The anomalous dimensions for the vector-vector operators is simpler for~\cite{Buras:2000if}
\begin{equation}
O_\pm=\frac{O_1\pm O_2}{2}\,,
\end{equation}
which are the following
\begin{eqnarray}
&&\gamma_\pm^{(0)} = \pm 6\frac{N\mp 1}{N}\,,\qquad  \gamma_\pm^{(1)} = \frac{N \mp 1}{2 N} \biggl(- 21 \pm 
\frac{57}{N} \mp 19\frac {N}{3} \pm \frac{4}{3} f \biggr)\,,
\nonumber
\\
&&\gamma_m^{(0)}=6 C_F\,, \qquad\qquad \qquad \gamma_m^{(1)}= C_F \biggl( 3 \,  C_F +\frac{97}{3}N - \frac{10}{3} f \biggr)\,,
\end{eqnarray}
where the second line corresponds to the anomalous dimensions for masses with \linebreak $C_F=(N^2-1)/2N$, and for $N=3$, $\gamma_+^{(0)}=4, \gamma_-^{(0)}=-8,\gamma_m^{(0)}=8$.

We introduce the correction of the anomalous dimensions
\begin{eqnarray}
J_\pm &=&\frac{d_\pm \beta_1}{\beta_0} -\frac{\gamma^{(1)}_\pm}{2 \beta_0}\,, \qquad d_\pm=\frac{\gamma^{(0)}_\pm}{2 \beta_0}\,,\\
J_m &=& \frac{d_m \beta_1}{\beta_0} -\frac{\gamma^{(1)}_m}{2 \beta_0}\,, \qquad d_m=\frac{\gamma^{(0)}_m}{2 \beta_0}\,,
\end{eqnarray}
and the value of the Wilson coefficients at the high scale
$C_{\pm}(\mu_W)$
defined in 
Ref.~\cite{Buchalla:1995vs}
\begin{equation}\label{eq:cpmuw}
C_{\pm}(\mu_W)=1 + \frac{\alpha_s (\mu_W)}{4 \pi} \biggl ( \log \frac{\mu_W}{M_W} \gamma_\pm^{(0)} + B_{\pm} \biggr) +{\cal O}(\alpha_s^2)\,,
\end{equation}
with 
\begin{equation}
B_{\pm}= -\frac{11}{2 N} \pm \frac{11}{2}\,,
\end{equation}
leading to the evolution
\begin{eqnarray}
C^{NLO}_\pm(\mu;\mu_0)=\left(1 + \frac{\alpha_s(\mu)}{4 \pi} J_\pm \right) \biggl( \frac{\alpha_s(\mu_0)}{\alpha_s(\mu)}\biggr)^{d_\pm}\left(1 - \frac{\alpha_s(\mu_0)}{4 \pi}[J_\pm-B_\pm] \right)\,, \\
C^{NLO}_m(\mu;\mu_0)=\left(1 + \frac{\alpha_s(\mu)}{4 \pi} J_m \right) \biggl( \frac{\alpha_s(\mu_0)}{\alpha_s(\mu)}\biggr)^{d_m} \left(1 -\frac{\alpha_s(\mu_0)}{4 \pi} J_m \right) \,.
\end{eqnarray}
We have 
\begin{equation}
d_m=4/\beta_0 \qquad d_+=2/\beta_0 \qquad d_-=-4/\beta_0\,.
\end{equation}
The same equations can be written for $O_i^{VRR}$ which will be useful for the discussion of the LRM, with identical results for the anomalous dimensions.

One may also consider the running of the $|\Delta S|=1$ local operators VLR. In the basis $O_1^{VLR},O_2^{VLR}$, the anomalous dimensions are
{\small\begin{eqnarray}
\hat\gamma^{(0)}_{VLR}&=&\left[\begin{array}{cc} 6/N & -6\\ 0 & -6N+6/N\end{array}\right]\,,\\
\hat\gamma^{(1)}_{VLR}&=&\left[\begin{array}{cc} \frac{137}{6}+\frac{15}{2N^2}-\frac{22}{3N} f & - \frac{100N}{3}+\frac{3}{N}+ \frac{22}{3} f\\\nonumber
-\frac{71}{2} N -\frac{18}{N}+4f & - \frac{203}{6} N^2+\frac{479}{6}+\frac{15}{2N^2}+\frac{10}{3} Nf -\frac{22}{3N} f  \end{array}\right]\,.
\end{eqnarray}}
Introducing
\begin{eqnarray}
\hat{V}&=&\left(\begin{array}{cc} 3/2 & 0 \\ -1/2 & -1/2 \end{array}\right)\,,\\
\hat\gamma^{(0)}_D&=& \hat{V}^{-1} \hat\gamma^{(0)T}_{VLR} \hat{V}
   =\left(\begin{array}{cc} 6/N & 0\\ 0 & -6N+6/N\end{array}\right)\,,\qquad \gamma^{(0)}_{1}=2\,,\qquad \gamma^{(0)}_{2}=-16\,,\\
\hat{G}&=& \hat{V}^{-1} \hat\gamma^{(1)T}_{VLR} \hat{V}\,,\\
\hat{H}_{ij}&=& \delta_{ij} \gamma^{(0)}_i \frac{\beta_1}{2\beta_0^2} - \frac{\hat{G}_{ij}}{2\beta_0 + \gamma^{(0)}_i - \gamma^{(0)}_j} \qquad\qquad ({2\beta_0 + \gamma^{(0)}_i - \gamma^{(0)}_j} \neq 0)\,,\\
\hat{J} &=& \hat{V} \hat{H} \hat{V}^{-1}\,,
\end{eqnarray}
one can write down the evolution
\begin{eqnarray}
\vec{C}^{LR}(\mu;\mu_0)&=&
   \left(1+\frac{\alpha_s(\mu)}{4\pi} \hat{J}\right) \hat{V} D(\mu;\mu_0) \hat{V}^{-1} 
   \left(1-\frac{\alpha_s(\mu_0)}{4\pi} \hat{J}\right)
   \vec{C}^{LR}(\mu_0)\,,\\
 D(\mu;\mu_0)&=&\left(\begin{array}{cc} (\alpha_s(\mu_0)/\alpha_s(\mu))^{d_1} & 0\\
 0&  (\alpha_s(\mu_0)/\alpha_s(\mu))^{d_2} \end{array}\right)\,,
\end{eqnarray}
with $d_i=\gamma^{(0)}_i/(2\beta_0)$.

\section{$|\Delta S|=2$ operators}\label{app:anomdimgenS2}

For $|\Delta S|=2$ operators, we recall the anomalous dimensions associated with the operator $Q_{V}$
\begin{equation}
Q_{V}=(\bar{s}^\alpha \gamma_\mu P_L d^\alpha)(\bar{s}^\beta \gamma^\mu P_L d^\beta)\,,
\end{equation}
with
\begin{eqnarray}
\gamma^{(0)}_V&=&6-6/N\,, \\
\gamma^{(1)}_V&=&-19/6N-22/3+39/N-57/(2N^2)+2/3f-2/(3N)f\,,\\
J_V &=&\frac{d_V \beta_1}{\beta_0} -\frac{\gamma^{(1)}_V}{2 \beta_0}\,,
\qquad\qquad\qquad d_V=\frac{\gamma^{(0)}_V}{2\beta_0}\,,
\end{eqnarray}
and we can write down a similar evolution for the $|\Delta S|=2$ local operators $Q_1^{LR},Q_2^{LR}$ 
\begin{eqnarray}
Q_1^{LR}&=&(\bar{s}^\alpha \gamma_\mu P_L d^\alpha)(\bar{s}^\beta \gamma_\mu P_R d^\beta)\,,
\qquad
Q_2^{LR}=(\bar{s}^\alpha P_L d^\alpha)(\bar{s}^\beta P_R d^\beta)\,,
\end{eqnarray}
with the anomalous dimensions
{\small
\begin{eqnarray}\label{eq:anoQLR}
\hat\gamma^{(0)}_{LR}&=&\left[\begin{array}{cc} 6/N & 12\\ 0 & -6N+6/N\end{array}\right]\,,\\
\hat\gamma^{(1)}_{LR}&=&\left[\begin{array}{cc}  \frac{137}{6}+\frac{15}{2N^2}-\frac{22}{3N} f & \frac{200N}{3}-\frac{6}{N}-\frac{44}{3} f\\\nonumber
\frac{71}{4} N + \frac{9}{N}-2f & - \frac{203}{6} N^2+ \frac{479}{6}+\frac{15}{2N^2}+\frac{10}{3} Nf -\frac{22}{3N} f  \end{array}\right]\,.
\end{eqnarray}
}
Introducing
\begin{eqnarray}
\hat{W}&=&\left(\begin{array}{cc} 3/2 & 0 \\ 1 & 1 \end{array}\right)\,,\\
\hat\gamma^{(0)}_D&=& \hat{W}^{-1} \hat\gamma^{(0)T}_{LR}\hat{W}
   =\left(\begin{array}{cc} 6/N & 0\\ 0 & -6N+6/N\end{array}\right)\qquad \gamma^{(0)}_{1}=2\qquad \gamma^{(0)}_{2}=-16\,,\\
\hat{G}&=& \hat{W}^{-1} \hat\gamma^{(1)T}_{LR}\hat{W}\,,\\
\hat{H}_{ij}&=& \delta_{ij} \gamma^{(0)}_i \frac{\beta_1}{2\beta_0^2} - \frac{\hat{G}_{ij}}{2\beta_0 + \gamma^{(0)}_i - \gamma^{(0)}_j} \qquad\qquad ({2\beta_0 + \gamma^{(0)}_i - \gamma^{(0)}_j} \neq 0)\,,\\
\hat{K} &=& \hat{W} \hat{H} \hat{W}^{-1}\,,
\end{eqnarray}
one can write down the evolution 
\begin{eqnarray}
\vec{C}^{LR}(\mu;\mu_0)&=&
   \left(1+\frac{\alpha_s(\mu)}{4\pi} \hat{K}\right) \hat{W} D(\mu;\mu_0) \hat{W}^{-1} 
   \left(1-\frac{\alpha_s(\mu_0)}{4\pi} \hat{K}\right)
   \vec{C}^{LR}(\mu_0)\,,\\
 D(\mu;\mu_0)&=&\left(\begin{array}{cc} (\alpha_s(\mu_0)/\alpha_s(\mu))^{d_1} & 0\\
 0&  (\alpha_s(\mu_0)/\alpha_s(\mu))^{d_2} \end{array}\right)\,,
\end{eqnarray}
with $d_i=\gamma^{(0)}_i/(2\beta_0)$.
The associated LO anomalous dimensions are 
\begin{equation}
 \gamma^{(0)}_{1}=2\,,\qquad \gamma^{(0)}_{2}=-16\,,\\
\end{equation}
and we have 
\begin{equation}
d_1=1/\beta_0\,, \qquad d_2=-8/\beta_0\,,\qquad d_V=2/\beta_0\,.
\end{equation}

\chapter{Case at NLO with the method of regions}\label{lrregion}

\section{Contributions with $\log\beta$}

{Following Ref.~\cite{Ecker:1985vv},} if we consider the box with the Goldstone boson associated to $W$ together with $W'$,
the masses stem from the Goldstone boson coupling (evaluated at the scale $\mu_W$), whereas the largest contribution to $I_2$ comes  from the range between $\mu_W$ and $\mu_R$.
We obtain
\begin{eqnarray}\nonumber
&&\xi^{(W'G)}_{a,UV}[R]=\sum_{r=\pm,i,j=1,2}
 \left(\frac{\alpha_s(\mu_W)}{\alpha_s(\mu_{h})}\right)^{-d_r+d_i+2d_m} 
  \left(\frac{\alpha_s(m_U)}{\alpha_s(\mu_{h})}\right)^{-d_m} 
    \left(\frac{\alpha_s(m_V)}{\alpha_s(\mu_{h})}\right)^{-d_m} 
    \left(\frac{\alpha_s(\mu_R)}{\alpha_s(\mu_{h})}\right)^{d_r}\\\nonumber
&& \quad\times 
\left[\left(1+\frac{\alpha_s(\mu_{h})}{4\pi} \hat{K}\right) \hat{W}\right]_{ai}\\\nonumber
&& \quad\times 
 R^{NLO}\Bigg(-d_r+d_i-d_j,\\\nonumber
&&  \qquad\qquad\qquad
  \left[\hat{W}^{-1}\hat{a}^{(W'G)}_r \hat{V}\right]_{ij} [\hat{V}^{-1}\vec{C}_0]_j  \\\nonumber
&&  \qquad\qquad\qquad\qquad\qquad
\times  \left(1 -\frac{\alpha_s(\mu_R)}{4\pi}[J_r-B_r]-\frac{\alpha_s(\mu_W)}{4\pi}2J_m
   +\frac{\alpha_s(m_U)+\alpha_s(m_V)}{4\pi}J_m\right)\\\nonumber
&&   \qquad\qquad\qquad\qquad  -\frac{\alpha_s(\mu_W)}{4\pi}
   \left[\hat{W}^{-1}\hat{a}^{(W'G)}_r \hat{V}\right]_{ij} [\hat{V}^{-1}\hat{J}\vec{C}_0]_j 
   , \\\nonumber
&&   \qquad\qquad\qquad
   \left[\hat{W}^{-1}[\hat{a}^{(W'G)}_r \hat{J}-\hat{K}\hat{a}^{(W'G)}_r]\hat{V}\right]_{ij} [\hat{V}^{-1}\vec{C}_0]_j 
   + \left[\hat{W}^{-1}\hat{a}^{(W'G)}_r \hat{V}\right]_{ij} [\hat{V}^{-1}\vec{C}_0]_j  J_r,
      \\
&&  \qquad\qquad\qquad\qquad
 \mu_W, \mu_R\Bigg)\,,
\end{eqnarray}
with the initial conditions for the evolution of the operators $O_{1,2}^{VLR}$ and the
coefficients for the matching from the two-point function of $O_\pm^{VRR}$  and $O_{1,2}^{VLR}$  to the local operators $Q_{1,2}^{LR}$ at $\mu=k^2$.
\begin{equation}
\vec{C}_0=\left(\begin{array}{c} 0\\ -1/2\end{array}\right)\,,
 \qquad
C_a^{LR}  \leftrightarrow \sum_{r,i} (\hat{a}^{(W'G)}_{r})_{ai} C_i^{VLR} C_r^{VRR}\,,
\qquad
\hat{a}^{(W'G)}_r=\left(\begin{array}{cc} (3r+1)/2 & r/2\\ 0 & -1\end{array}\right) \,.
\end{equation}

If we consider the box with $W$ and a charged Higgs boson $H$,
the masses stem from the Higgs couplings (to be evaluated at a high scale $\mu_H$), whereas the largest contribution to $I_2$ comes  from the range between $\mu_W$ and $M_H$. We obtain
\begin{eqnarray}\nonumber
\xi^{(HW)}_{a,UV}[R]\!\!\!&=&\!\!\!\!\!\!\sum_{l=\pm,i,j=1,2}
 \left(\frac{\alpha_s(\mu_W)}{\alpha_s(\mu_{h})}\right)^{d_i-d_j} 
  \left(\frac{\alpha_s(m_U)}{\alpha_s(\mu_{h})}\right)^{-d_m} 
    \left(\frac{\alpha_s(m_V)}{\alpha_s(\mu_{h})}\right)^{-d_m} 
    \left(\frac{\alpha_s(\mu_H)}{\alpha_s(\mu_{h})}\right)^{d_j+2d_m}\\\nonumber
&& \quad\times 
\left[\left(1+\frac{\alpha_s(\mu_{h})}{4\pi} \hat{K}\right) \hat{W}\right]_{ai}\\\nonumber
&& \quad\times 
 R^{NLO}\Bigg(-d_l+d_i-d_j,\\\nonumber
&&  \qquad\qquad\qquad
  \left[\hat{W}^{-1}\hat{a}^{(HW)}_l \hat{V}\right]_{ij} [\hat{V}^{-1}\vec{C}_0]_j  \\\nonumber
&&  \qquad\qquad\qquad
\times  \left(1 -\frac{\alpha_s(\mu_W)}{4\pi}[J_l-B_l]-\frac{\alpha_s(\mu_H)}{4\pi}2J_m
   +\frac{\alpha_s(m_U)+\alpha_s(m_V)}{4\pi}J_m\right), \\\nonumber
&&   \qquad\qquad\qquad
   \left[\hat{W}^{-1}[\hat{a}^{(HW)}_l \hat{J}-\hat{K}\hat{a}^{(HW)}_l ]\hat{V}\right]_{ij} [\hat{V}^{-1}\vec{C}_0]_j 
   - \left[\hat{W}^{-1}\hat{a}^{(HW)}_l \hat{V}\right]_{ij} [\hat{V}^{-1}\hat{J}\vec{C}_0]_j  \\
&&  \qquad\qquad\qquad\qquad
   + \left[\hat{W}^{-1}\hat{a}^{(HW)}_l \hat{V}\right]_{ij} [\hat{V}^{-1}\vec{C}_0]_j  J_l, \mu_W, \mu_H\Bigg)\,,
\end{eqnarray}
with the same initial conditions for the evolution of the operators $Q_{1,2}^{VLR}$  and the
coefficients for the matching from the two-point function of $O_\pm^{VLL}$ and $O_{1,2}^{VRL}$  to the local operators $Q_{1,2}^{LR}$ at $\mu=k^2$.
\begin{equation}
C_a^{LR}  \leftrightarrow \sum_{l,j} (\hat{a}^{(HW)}_{l})_{ai} C_j^{VRL} C_l^{VLL}\,,
\qquad
\hat{a}^{(HW)}_l=\hat{a}^{(W'G)}_{r=l}\,.
\end{equation}
One can check that the expressions from Ref.~\cite{Ecker:1985vv} are recovered at leading order.

If we consider $\log\beta$ as small (``small $\log\beta$ approach''), we see that the diagrams are dominated by the region $k^2=\mathcal{O}(m_t^2,\mu_W^2)$ in all cases: this is obvious for $tt$ and $ct$ boxes, whereas the $cc$ box receives only suppressed contributions from the region $k^2=\mathcal{O}(m_c^2)$. We obtain thus expressions involving the averaging weight for constant terms $R^{NLO}_1$
\begin{equation}
\bar\eta^{(W'G)}_{a,UV}=\xi^{(W'G)}_{a,UV}[R^{NLO}_1]\,,
\qquad\qquad \bar\eta^{(HW)}_{a,UV}=\xi^{(HW)}_{a,UV}[R^{NLO}_1]\,,
\end{equation}
where we have identified the two scales for the integration $\mu_W=\mu_R$ to a common average value (this is similar to the treatment of the region between $m_t$ and $M_W$ in the SM case).

In the case of a large $\log\beta$ (``large $\log\beta$ approach''), we want to perform the resummation of the large $\log\beta$ with $R^{NLO}_{\log}$ and consider the rest of the contribution as dominated by the region $k^2=\mathcal{O}(m_t^2,\mu_W^2)$. In the case of $(W'G)$ we obtain
\begin{eqnarray}
\bar\eta^{(W'G)}_{a,UV} &=& \Bigg[F^{(W'G)}_{UV} \\
& \times & \sum_{r=\pm, \, i,j=1,2}
\left(\frac{\alpha_s(\mu_W)}{\alpha_s(\mu_{h})}\right)^{-d_r+d_i+2d_m}
  \left(\frac{\alpha_s(m_U)}{\alpha_s(\mu_{h})}\right)^{-d_m} 
    \left(\frac{\alpha_s(m_V)}{\alpha_s(\mu_{h})}\right)^{-d_m} 
    \left(\frac{\alpha_s(\mu_R)}{\alpha_s(\mu_{h})}\right)^{d_r} \nonumber\\
\qquad & \times & \hat{W}_{ai} \left[\hat{W}^{-1}\hat{a}^{(W'G)}_r \hat{V}\right]_{ij} [\hat{V}^{-1}\vec{C}_0]_j
+\log(\beta) \times \xi^{(W'G)}_{a,UV}[R^{NLO}_{\log}]
     \Bigg] \frac{1}{\log(\beta) + F^{(W'G)}_{UV}}\qquad\qquad \nonumber
\end{eqnarray}
with the contributions from the constant term
\begin{equation}
F^{(W'G)}_{tt} =  \frac{x_t^2 - 2 x_t}{(x_t - 1)^2} \log (x_t) + \frac{x_t}{x_t - 1} \,,\quad
F^{(W'G)}_{ct}=  \frac{x_t}{x_t - 1} \log (x_t) \,,\quad
F^{(W'G)}_{cc} = 0 \,,
\end{equation}
and similarly for $(HW)$
\begin{eqnarray}
\bar\eta^{(HW)}_{a,UV} &=& \Bigg[  F^{(HW)}_{UV} \\
& \times & \sum_{l=\pm, \, i,j=1,2}
\left(\frac{\alpha_s(\mu_W)}{\alpha_s(\mu_{h})}\right)^{d_i-d_j}   \left(\frac{\alpha_s(m_U)}{\alpha_s(\mu_{h})}\right)^{-d_m} 
    \left(\frac{\alpha_s(m_V)}{\alpha_s(\mu_{h})}\right)^{-d_m} 
    \left(\frac{\alpha_s(\mu_H)}{\alpha_s(\mu_{h})}\right)^{d_j+2d_m} \nonumber\\
\qquad &\times &    \hat{W}_{ai} \left[\hat{W}^{-1}\hat{a}^{(HW)}_l \hat{V}\right]_{ij} [\hat{V}^{-1}\vec{C}_0]_j
   +\log(\beta \omega) \times  \xi^{(HW)}_{a,UV}[R^{NLO}_{\log}]   \Bigg] \frac{1}{\log(\beta \omega) + F^{(HW)}_{UV}}\qquad\qquad \nonumber
\end{eqnarray}
with the contributions from the constant term
\begin{equation}
F^{(HW)}_{tt} =  x_t \frac{x_t + (x_t - 2) \log (x_t) - 1}{(x_t - 1)^2} \,,\qquad
F^{(HW)}_{ct}  =  \frac{x_t}{x_t - 1} \log (x_t)\,,\qquad
F^{(HW)}_{cc}  =  0 \,.
\end{equation}

\section{Contributions without $\log\beta$}

If we consider the box with the Goldstone associated with $W$ and a charged Higgs boson $H$, 
the masses stem from the Higgs couplings, the Goldstone boson couplings and the propagator, whereas the largest contribution to $I_1$ comes  from the range between $m_V$ and $\mu_W$. We obtain
\begin{eqnarray}\nonumber
\!\!\!\!\!\!&&\bar\eta^{(HG)}_{a,UV}\!\!\!=\!\!\!\!\!\!\!\!\!\sum_{b,i,j,j',k,k'=1,2}
  \left(\frac{\alpha_s(m_U)}{\alpha_s(\mu_{h})}\right)^{-3d_m} \!\!\!
    \left(\frac{\alpha_s(m_V)}{\alpha_s(\mu_{h})}\right)^{d_i-d_k-d_{k'}-d_m} \!\!\!
     \left(\frac{\alpha_s(\mu_W)}{\alpha_s(\mu_{h})}\right)^{d_k+2dm} \!\!\!
    \left(\frac{\alpha_s(\mu_H)}{\alpha_s(\mu_{h})}\right)^{d_{k'}+2d_m}\\\nonumber
&& \quad\times 
\bar{a}^{(HG)}_{b,jj'}
\left[\left(1+\frac{\alpha_s(\mu_{h})}{4\pi} \hat{K}\right) \hat{W}\right]_{ai}
\left[\hat{V}^{-1}\left(1-\frac{\alpha_s(\mu_W)}{4\pi} \hat{J}\right) \vec{C}_0\right]_k
\left[\hat{V}^{-1}\left(1-\frac{\alpha_s(\mu_H)}{4\pi} \hat{J}\right) \vec{C}_0\right]_{k'}
\\\nonumber
&& \quad\times 
 R^{NLO}\Bigg(d_i-d_k-d_{k'}+2d_m,\\\nonumber
&&  \qquad\qquad\qquad
 \hat{W}^{-1}_{ib} \hat{V}_{jk} \hat{V}_{j'k'} 
 \times  \left(1 -\frac{\alpha_s(\mu_W)}{4\pi}2J_m-\frac{\alpha_s(\mu_H)}{4\pi}2J_m
   +\frac{\alpha_s(m_U)+\alpha_s(m_V)}{4\pi}3J_m\right), \\\nonumber
&&   \qquad\qquad\qquad
 -2J_m \hat{W}^{-1}_{ib} \hat{V}_{jk} \hat{V}_{j'k'}
 - (\hat{W}^{-1}\hat{K})_{ib} \hat{V}_{jk} \hat{V}_{j'k'}
 +\hat{W}^{-1}_{ib} (\hat{J}\hat{V})_{jk} \hat{V}_{j'k'}
 + \hat{W}^{-1}_{ib} \hat{V}_{jk} (\hat{J}\hat{V})_{j'k'},\\
&&   \qquad\qquad\qquad
   m_V,\mu_W\Bigg)\,,
\end{eqnarray}
where $\bar{a}^{(HG)}_{a,ij}$ provides the
coefficients for the matching from the two-point function of $O_{1,2}^{VLR}$  to the local operators $Q_{1,2}^{LR}$ at $\mu=k^2$:
\begin{equation}
C^{LR}_a  \leftrightarrow \sum_{ij} \bar{a}^{(HG)}_{a,ij} C_i^{VLR} C_j^{VRL}\,,
\end{equation}
with the non-vanishing entries
\begin{equation}
 \bar{a}^{(HG)}_{1,12}=-2\,, \qquad  \bar{a}^{(HG)}_{1,21}=-2\,, \qquad  \bar{a}^{(HG)}_{1,11}=-6\,,
 \qquad  \bar{a}^{(HG)}_{2,22}=4\, .
\end{equation}
The only relevant case is $tt$, where $R^{NLO}$ can be replaced by $R^{NLO}_1$.

If we consider tree-level $H^0$ exchanges, we have
\begin{eqnarray}\nonumber
\bar\eta^{(H)}_{a,UV}&=& \left(\frac{\alpha_s(m_U)}{\alpha_s(\mu_{h})}\right)^{-d_m} 
    \left(\frac{\alpha_s(m_V)}{\alpha_s(\mu_{h})}\right)^{-d_m} 
    \left(\frac{\alpha_s(\mu_H)}{\alpha_s(\mu_{h})}\right)^{2d_m} \\
&&\qquad\times   \left(1-\frac{\alpha_s(\mu_H)}{4\pi}2J_m
   +\frac{\alpha_s(m_U)+\alpha_s(m_V)}{4\pi}J_m\right) \\
&&\qquad\times 
 \left[\left(1+\frac{\alpha_s(\mu_{h})}{4\pi} \hat{K}\right) \hat{W}
 \left(\frac{\alpha_s(\mu_H)}{\alpha_s(\mu_{h})}\right)^{\vec{d}}
 \hat{W}^{-1}\left(1-\frac{\alpha_s(\mu_H)}{4\pi} \hat{K}\right)\vec{C}_0
 \right]_a\,, \nonumber
\end{eqnarray}
where the matching yields the value of the Wilson coefficients for the $|\Delta S|=2$ operators at the high scale. One can check that the expressions from Ref.~\cite{Ecker:1985vv} are recovered at leading order.

\chapter{Inverse of the covariance matrix}\label{sec:cumberInverse}

In practice, we may not know the correlation among the random variables $ X $, and it is common to suppose a correlation of $ 100~\% $ when two random variables are suspect to be strongly correlated, as a way to try to be conservative. In such a case, however, the (naive) inverse of the covariance matrix may not be well defined: in cases it is not possible, we would like to introduce a generalized inverse $ C^+_s $ of the covariance matrix $ C_s $, which is not an inverse in the usual sense, $ C_s^{-1} $, but makes the singular case treatable.\footnote{Note in particular the term $ C_s^+ \cdot C_s \cdot C_s^+ $ that appears in Eq.~\eqref{eq:estimatorNoTheo}: it cannot be simplified to $ C_s^+ $ because the condition $ C_s^+ \cdot C_s \cdot C_s^+ = C_s^+ $ is not always satisfied by the generalized inverse.}

We start by working out the form of $ C_s $. Since every symmetric matrix is orthogonally diagonalizable, we write

\begin{equation}
C_s=\Sigma \cdot \Gamma \cdot \Sigma=\Sigma \cdot R \cdot D \cdot R^T \cdot \Sigma \, ,
\end{equation}
where $ \Sigma $ collects the statistical uncertainties $ \sigma_1, \sigma_2, \sigma_3, \ldots, \sigma_n $ of the individual measurements, and the correlation matrix $ \Gamma $ (a symmetric matrix itself) is orthogonally diagonalized into $ R $, an orthogonal matrix, and $ D = {\rm diag} (d_1, \ldots, d_n) $, a diagonal matrix whose elements, without loss of generality, are ordered

\begin{equation}
d_1 \geq d_2 \geq \ldots \geq d_m > 0 = d_{m+1} = \ldots = d_n \, .
\end{equation}

We now ask $ C^+_s $ to satisfy the following condition characterizing what we call a generalized inverse\footnote{We can consider the following conditions to further characterize other notions of generalized inverse: given $ A \in \Re^{n \times m} $ and $ A^+ \in \Re^{m \times n} $, let a subset of the following equalities be satisfied

\begin{eqnarray}
&& [1] \qquad \qquad A \cdot A^+ \cdot A = A \, , \nonumber\\
&& [2] \qquad \qquad A^+ \cdot A \cdot A^+ = A^+ \, , \nonumber\\
&& [3] \qquad \qquad (A \cdot A^+)^T = A \cdot A^+ \, , \nonumber\\
&& [4] \qquad \qquad (A^+ \cdot A)^T = A^+ \cdot A \, . \nonumber
\end{eqnarray}
We say that: if $ [1] $ is satisfied, then $ A^+ $ is a ``generalized inverse" of $ A $; if $ [1,2] $ are satisfied, then $ A^+ $ is a ``generalized reflexive inverse" of $ A $; if $ [1,2,3,4] $ are satisfied, then $ A^+ $ is the ``Moore-Penrose pseudoinverse" of $ A $. In all three cases, a matrix $ A^+ $ can always be defined for any matrix $ A $, but only the Moore-Penrose pseudoinverse is always unique. Therefore, if an inverse exists, it is equal to the Moore-Penrose pseudoinverse.
}

\begin{equation}
C_s \cdot C_s^+ \cdot C_s=C_s \, ,
\end{equation}
which is fulfilled by the following symmetric choice

\begin{equation}
C_s^+=\Sigma^{-1} \cdot R \cdot D^+ \cdot R^T \cdot \Sigma^{-1} \, ,
\end{equation}
where $ D^+ $ is the generalized inverse of $ D $

\begin{equation}
D \cdot D^+ \cdot D=D \, ,
\end{equation}
considered to have the form

\begin{equation}
D^+=\left[
\begin{array}{c|c} 1/d & A\\
\hline
A^T & B
\end{array}
\right] \, .
\end{equation}

We would like to simplify the form $ D^+ $ may assume in order to simplify as much as possible the computation of $ C_s^+ $. For this reason we assume $A=0$ and $B=\lambda\times \mathbf{1}_{(n-m)\times (n-m)}$, with $ \lambda $ a constant which we will now fix. A condition we ask our generalized inverse to satisfy is:

\begin{center}
if one of the uncertainties is sent to zero, \textit{viz.} $ \sigma_a \rightarrow 0^+ $,

then the weight $ w_a $ must be the dominant one.
\end{center}
\noindent
This condition is not easily satisfied: for example, the Moore-Penrose pseudoinverse, which is somewhat closer to the usual inverse, and has the property of uniqueness, does not satisfy this condition in the singular case (see Ref.~\cite{Charles}).

Let us see the implications of this condition by considering the case where an uncertainty $ \sigma_a $ is much smaller than the others. In this case

\begin{equation}
w_i \rightarrow \frac{1}{U^T \cdot C_s^+ \cdot U} \frac{1}{\sigma_i\sigma_a} (R \cdot D^+ \cdot R^T)_{ia} \, ,
\end{equation}
where $ U^T \cdot C_s^+ \cdot U $ is a common factor, and $ R \cdot D^+ \cdot R^T $ is independent of the uncertainties $ \sigma_i $. Then, in order to have the dominance of $ w_a $, it is enough to satisfy

\begin{equation}\label{eq:motivationLambdaChoice}
0\neq (R \cdot D^+ \cdot R^T)_{aa}=\sum_{j=1}^m (R_{aj})^2 \frac{1}{d_j} + \lambda \sum_{j=m+1}^n (R_{aj})^2 
  = \lambda  + \sum_{j=1}^m (R_{aj})^2 \left(\frac{1}{d_j}-\lambda\right) \, ,
\end{equation}
which is always fulfilled for $0<\lambda\leq 1/d_1$. We now make a last choice and pick
\begin{equation}
\lambda=1/d_1 \, ,
\end{equation}
recovering the ansatz of \cite{Schmelling:1994pz}. Note that this choice is efficient (i.e. $ \sigma_\mu $ is minimal) over the range $ \lambda \in (0, 1/d_1] $ when $ 100~\% $ correlations are present among all the measurements, in which case

\begin{equation}
\sigma_\mu^2
 =\frac{(\sum_i 1/\sigma_i)^2/n^2}{\left\{(\sum_i 1/\sigma_i)^2/n^2
    +\lambda \left[\sum_i 1/\sigma^2_i-(\sum_i 1/\sigma_i)^2/n\right]\right\}^2} \, ,
\end{equation}
where $ \sum_i 1/\sigma^2_i \geq (\sum_i 1/\sigma_i)^2/n $ follows from the Cauchy-Schwarz inequality.

We refer to the resulting generalized inverse as ``generalized $ \lambda- $inverse"


\begin{equation}
D^+= {\rm diag} (1/d_1, \ldots, 1/d_m, \underbrace{ 1/d_1, \ldots, 1/d_1 }_{n - m \; \rm times} ) \, , \quad \lambda-{\rm inverse} \, .
\end{equation}
It should be stressed that, in cases where the inverse is defined, the index $ m $ above is equal to zero, and the generalized $ \lambda- $inverse reduces to the (usual) inverse.


Though there is arbitrariness in the choices made above, the generalized $ \lambda- $inverse satisfies to the basic property we would like the average to satisfy, namely the more precise measurement dominates the average, i.e. its weight is more relevant. Moreover, for the specific examples discussed in the next section, it is efficient in the sense that the variance is roughly given by $ \sigma^2_a $ when $ \sigma_a $ is much smaller than the other uncertainties.



\section{Examples}

\subsection{Two measurements}

In the case of two uncorrelated measurements, there is no problem with the inversion of the statistical covariance matrix, and we get for all methods
\begin{eqnarray}
&& C_s^{-1}=\left(\begin{array}{ccc}
\frac{1}{\sigma_1^2} & 0 \\
0 & \frac{1}{\sigma_2^2}
\end{array}\right) \, ,
\qquad \sigma_\mu^2=\frac{\sigma_1^2\sigma_2^2}{\sigma_1^2+\sigma_2^2} \rightarrow \sigma_1^2 \, , \nonumber\\
&& w=\frac{1}{\sigma_1^2+\sigma_2^2}\left(\begin{array}{c}
\sigma_2^2 \\
\sigma_1^2
\end{array}\right) \rightarrow \left(\begin{array}{c}
1 \\
\sigma_1^2/\sigma_2^2 
\end{array}\right) \, .
\end{eqnarray}

In the case of two fully correlated measurements, we have
\begin{equation}
C_s=\left(\begin{array}{cc} \sigma_1^2 & \sigma_1\sigma_2 \\
\sigma_1\sigma_2 & \sigma_2^2 
\end{array}
\right) \, ,
\end{equation}
with $d_1=2$, $d_2=0$. The $\lambda$-inverse for $C_s$ yields
\begin{eqnarray}
&& C_s^+=\left(\begin{array}{ccc}
\frac{1}{2 \sigma_1^2} & 0 \\
0 & \frac{1}{2 \sigma_2^2}
\end{array}\right) \, ,
\qquad \sigma_\mu^2=\frac{\sigma_1^2\sigma_2^2[\sigma_1+\sigma_2]^2}{[\sigma_1^2+\sigma_2^2]^2}\rightarrow \sigma_1^2 \, , \nonumber\\
&& w=\frac{1}{\sigma_1^2+\sigma_2^2}\left(\begin{array}{c}
\sigma_2^2 \\
\sigma_1^2
\end{array}\right) \rightarrow \left(\begin{array}{c}
1 \\
\sigma_1^2/\sigma_2^2 
\end{array}\right) \, ,
\end{eqnarray}
where we indicate the limit when $\sigma_1\to 0$.

\subsection{$n$ fully correlated measurements}

We have a correlation matrix with unit entries everywhere.
This yields $d_1=n$, $d_{i>1}=0$. 
The $\lambda$-inverse yields
\begin{eqnarray}
&& C_s^+=\left(\begin{array}{ccc}
\frac{1}{n \sigma_1^2} & \cdots & 0\\
\vdots & \ddots & \vdots  \\
0 & \cdots & \frac{1}{n \sigma_n^2}
\end{array}\right) \, ,
\qquad \sigma_\mu^2=\frac{(\sum_i 1/\sigma_i)^2}{(\sum_i 1/\sigma_i^2)^2}
\rightarrow \sigma_1^2 \, , \nonumber\\
&& w=\frac{1}{\sum_i 1/\sigma_i^2}
\left(\begin{array}{c}
1/\sigma_1^2 \\
\vdots \\
1/\sigma_N^2 
\end{array}\right) \rightarrow \left(\begin{array}{c}
1 \\
\sigma_1^2/\sigma_2^2\\
\vdots \\
\sigma_1^2/\sigma_n^2
\end{array}\right) \, ,
\end{eqnarray}
where we indicated the limit when $\sigma_1\to 0$. This is the ansatz found in Ref.~\cite{Schmelling:1994pz}.

\subsection{Two fully correlated measurements with an uncorrelated measurement }

Let us consider
\begin{equation}
C_s=\left(\begin{array}{ccc} \sigma_1^2 & \sigma_1\sigma_2 & 0\\
\sigma_1\sigma_2 & \sigma_2^2 & 0\\
0 & 0 & \sigma_3^2
\end{array}
\right) \, ,
\end{equation}
with $d_1=2$, $d_2=1$, $d_3=0$.

The $\lambda$-inverse for $C_s$ yields
\begin{eqnarray}
&& C^+=\left(\begin{array}{ccc}
\frac{1}{2 \sigma_1^2} & 0 & 0\\
0 & \frac{1}{2 \sigma_2^2} & 0 \\
0 & 0 & \frac{1}{\sigma_3^2}
\end{array}\right) \, , \nonumber\\
&& \sigma_\mu^2=\frac{\sigma_1^2\sigma_2^2\sigma_3^2[2\sigma_1\sigma_2\sigma_3^2+4\sigma_1^2\sigma_2^2+\sigma_1^2\sigma_3^2+\sigma_2^2\sigma_3^2]}{[2\sigma_1^2\sigma_2^2+\sigma_1^2\sigma_3^2+\sigma_2^2\sigma_3^2]^2} \rightarrow \sigma_1^2 \, , \\
&& w=\frac{1}{2\sigma_1^2\sigma_2^2+\sigma_1^2\sigma_3^2+\sigma_2^2\sigma_3^2}\left(\begin{array}{c}
\sigma_2^2\sigma_3^2 \\
\sigma_1^2\sigma_3^2 \\
2\sigma_1^2\sigma_2^2
\end{array}\right) \rightarrow \left(\begin{array}{c}
1 \\
\sigma_1^2/\sigma_2^2 \\
2\sigma_1^2/\sigma_3^2
\end{array}\right) \, . \nonumber
\end{eqnarray}

\chapter{Correlated theoretical uncertainties}\label{sec:correlatedTheo}



Apart from the difficulty of taking the inverse of a singular matrix, which led us to introduce the generalized $ \lambda- $inverse in the previous appendix, we need a way to characterize the range of variation of correlated theoretical uncertainties, particularly in the case where $ 100~\% $ theoretical correlations are assumed.

We start by introducing the Cholesky decomposition of a symmetric positive-definite matrix (which always exists and is unique)

\begin{equation}
\widetilde{C}_t=L\cdot L^T \, ,
\end{equation}
for a lower triangular matrix $ L $ with strictly positive diagonal elements. We are going to see that the Cholesky decomposition has the good properties to describe correlated theoretical uncertainties in singular cases.

When $ \widetilde{C}_t $ is positive-definite, $ L^{-1} $ is always well defined, and from Eq.~\eqref{eq:startingTest} this decomposition leads to the decorrelation of theoretical uncertainties $ L^{-1} \tilde{\delta} $ in the non-singular case, where $ \widetilde{C}_t^{-1} = (L^{-1})^T \cdot L^{-1} $:

\begin{equation}
\tilde\delta^T \cdot \widetilde{C}_t^{-1} \cdot \tilde\delta = \tilde\delta^T \cdot (L^{-1})^T \cdot L^{-1} \cdot \tilde\delta = (L^{-1} \cdot \tilde\delta)^T \cdot (L^{-1} \cdot \tilde\delta) \, .
\end{equation}

Therefore, we have the following prescription for the combination of correlated theoretical uncertainties:

\begin{eqnarray}
&& \Delta_\mu = \sum_j \left|\sum_i w_i \Delta_i L_{ij} \right| \ \ \quad (\mathrm{linear}) \, , \\
&& \Delta_\mu = \sqrt{\sum_j \left(\sum_i w_i \Delta_i L_{ij} \right)^2} \ \ \quad (\mathrm{quadratic}) \, . \nonumber
\end{eqnarray}
Note that in the above expression we only need to know $ L $. In the case where the matrix $ \widetilde{C}_t $ is only semi-definite positive, the Cholesky decomposition still exists if the diagonal elements of $ L $ are allowed to be zero, but it is not unique. To prescribe a triangular matrix $ L $ when $ \widetilde{C}_t $ is not positive-definite,\footnote{Meaning that there are directions $ v $ for which $ \widetilde{C}_t \cdot v = 0 $} we consider $\widetilde{C}_t + \epsilon\times \mathbf{1}_{m \times m}$, and then take the limit $ \epsilon \rightarrow 0 $.

\begin{figure}[t]
\begin{center}
\includegraphics[scale=0.4]{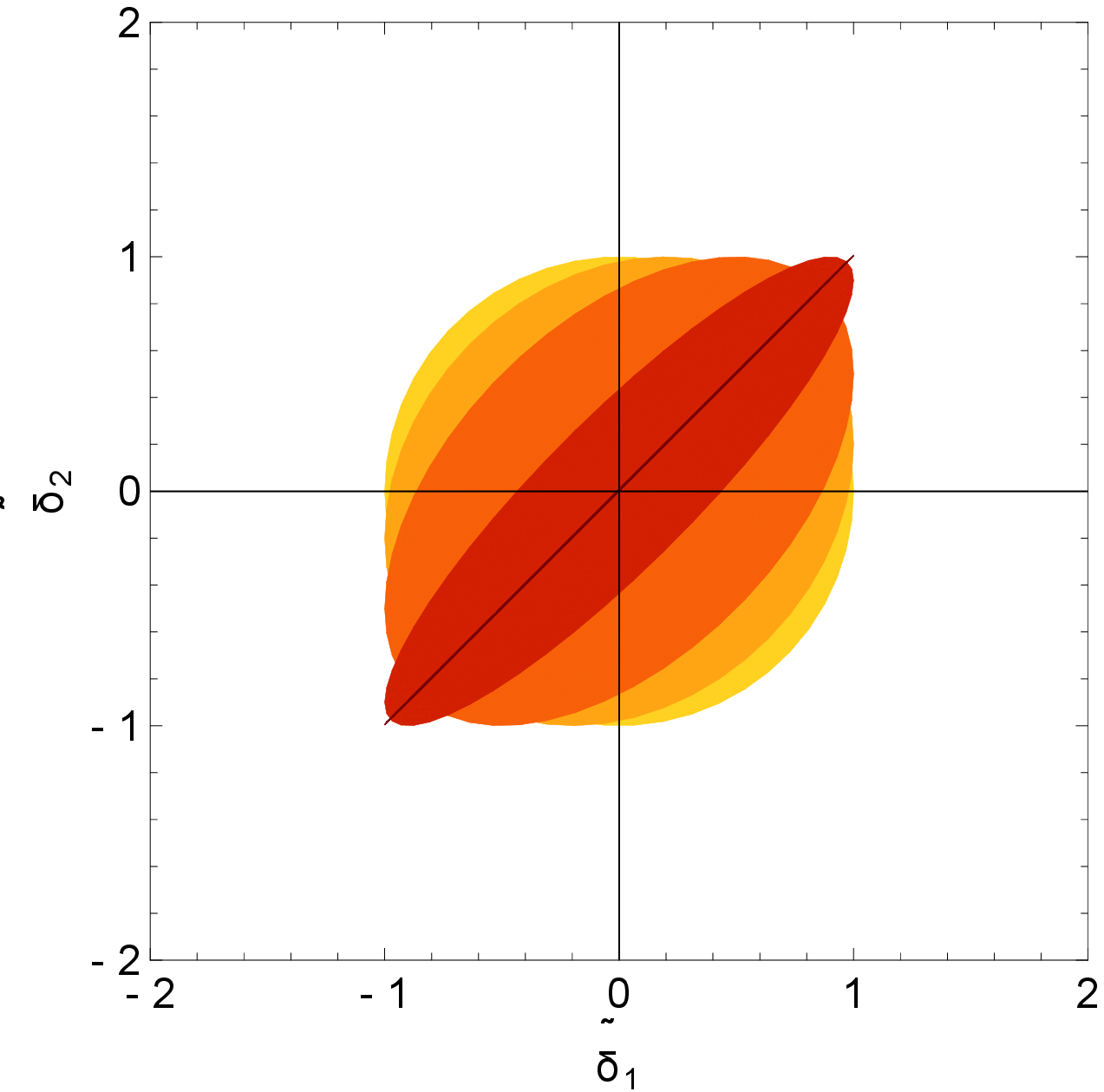} \hspace{5mm}
\includegraphics[scale=0.4]{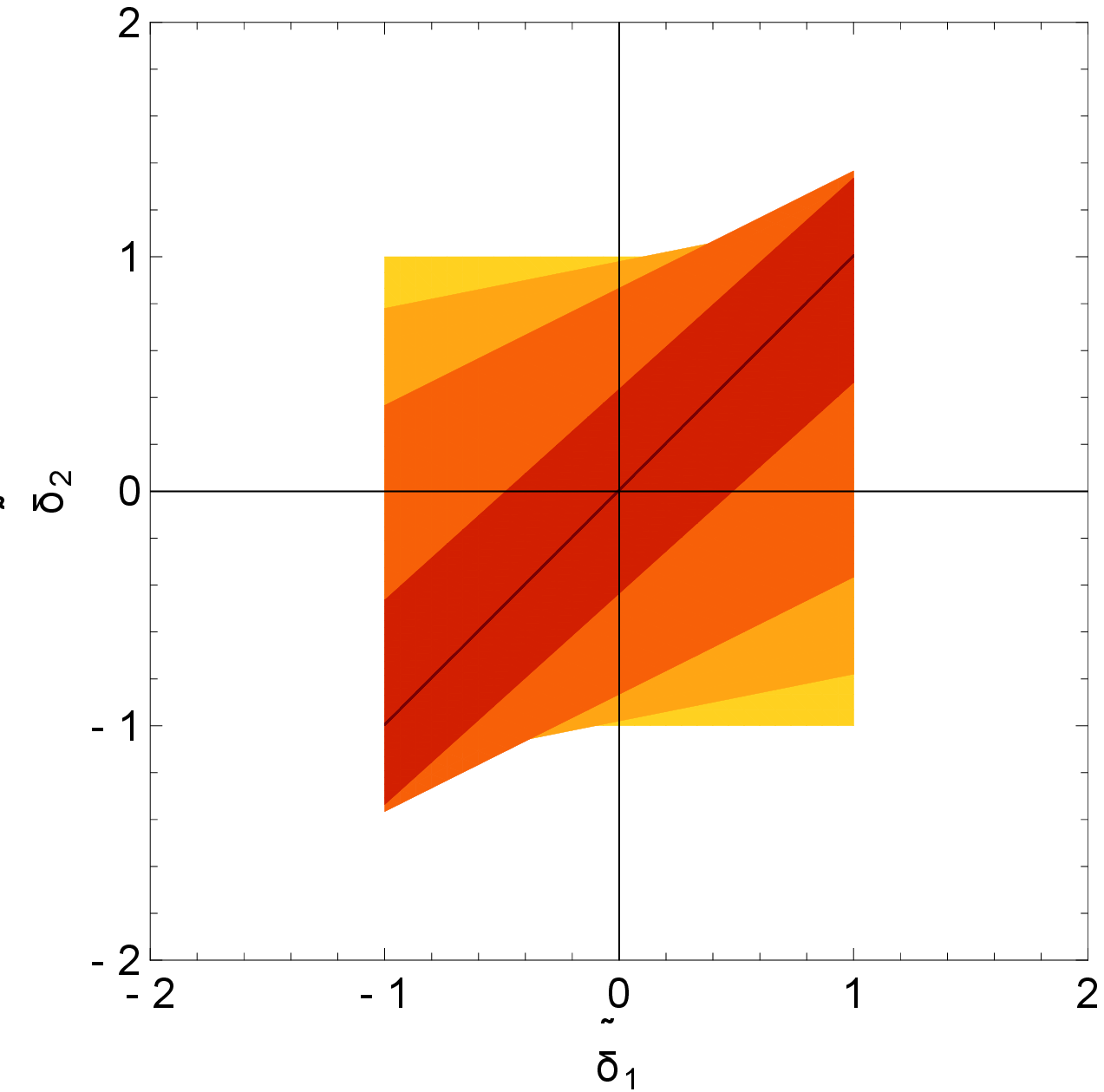}
\caption{Ranges of variation for $\tilde\delta_1$ and $\tilde\delta_2$ for $\rho=0,0.2,0.5,0.9,1$, going from light (yellow) to dark (red). The variation over a hyperball (left) or a hypercube (right) is considered.}\label{fig:theocorrrange}
\end{center}
\end{figure}

In 2 dimensions (see Ref.~\cite{Charles} for other examples), if two theoretical uncertainties $ \tilde{\delta}_{1,2} $ are totally (anti-) correlated, i.e. $ \tilde{\delta}_1 = 1 $ $ \Leftrightarrow \tilde{\delta}_2 = 1 $ ($ \tilde{\delta}_1 = 1 $ $ \Leftrightarrow \tilde{\delta}_2 = -1 $), we expect them to vary over a diagonal. Therefore, in intermediate cases a hypercube or a hyperball should continuously deform into a diagonal, which is seen in Figure \ref{fig:theocorrrange}. Note that the hypercube shows the unpleasant feature of not treating symmetrically $ \tilde{\delta}_1 $ and $ \tilde{\delta}_2 $. This is a property found in more general situations, including different modelings of theoretical correlations. On the other hand, the hyperball case does not suffer from the same problem in correlated situations.

\begin{thebibliography}{99}
\bibitem{Riemann}
  I would like to thank T. Riemann for advices concerning this version.
\end{thebibliography}
\end{document}